%% file: pdfs.tex
\documentclass[11pt]{cernrep}
\usepackage{graphicx,amsmath,amssymb}
\usepackage{cite,./mcite}
\usepackage{epsfig}

\usepackage{amssymb,amscd}
\usepackage[figuresright]{rotating}
\usepackage{array}
\usepackage{graphics}
\usepackage{multirow}
\usepackage{supertabular}
\usepackage{longtable}
\usepackage{dcolumn}
\usepackage{xspace}
\usepackage{tabularx}
\usepackage{color}
\usepackage{calc}
\usepackage{floatflt}

\newcommand{\F}{$ F_{2}(x,Q^2)\:$} 
\newcommand{\FL}{$ F_{L}(x,Q^2)\:$}

\newcommand{\Fc}{$ F_{2}\,$}

\newcommand{\amz}{$\alpha_s(M_Z^2)\,$} 
\newcommand{\bs}{\overline{s}}
\newcommand{\bcc}{\overline{c}}
\newcommand{\bu}{\overline{u}}
\newcommand{\bd}{\overline{d}}
\newcommand{\bU}{\overline{U}}
\newcommand{\bD}{\overline{D}}

\newcommand{\FLc}{$ F_{L}\,$} 
 
\newcommand{\xg}{$xg(x,Q^2)\,$}

\newcommand{\ipb}{pb$^{-1}\,$} 
\newcommand{\pdff}{$\partial F_{2} / \partial \ln Q^{2}\,$ }

\newcommand{\asmz}{\alpha_s(M_Z^2)}
\newcommand{\msbar}{\mbox{$\overline{\rm{MS}}$}\ }
\newcommand{\be}{\begin{equation}}
\newcommand{\ee}{\end{equation}}
\newcommand{\bea}{\begin{eqnarray}}
\newcommand{\eea}{\end{eqnarray}}

\newcommand{\la}{\left\langle}
\newcommand{\ra}{\right\rangle}
\newcommand{\lc}{\left[}
\newcommand{\rc}{\right]}
\newcommand{\lp}{\left(}
\newcommand{\rp}{\right)}

\newcommand{\aq}{\alpha_s\lp Q^2\rp}
\newcommand{\aqq}{\alpha_s\lp Q_0^2\rp}
\newcommand{\bc}{\begin{center}}
\newcommand{\ec}{\end{center}}

\newcommand{\bi}{\begin{itemize}}
\newcommand{\ei}{\end{itemize}}
\def\epm#1#2{\hbox{${\lower1pt\hbox{$\scriptstyle +~#1$}}
\atop {\raise1pt\hbox{$\scriptstyle -~#2$}}$}}
\newcommand{\dat}{\mathrm{dat}}
\newcommand{\art}{\mathrm{art}}
\newcommand{\net}{\mathrm{net}}
\newcommand{\rep}{\mathrm{rep}}
\newcommand{\rmexp}{\mathrm{exp}}

\newcommand{\as}{\alpha_{\rm s}}
\newcommand{\gsim}{\raisebox{-0.07cm}{$\:\:\stackrel{>}{{\scriptstyle
 \sim}}\:\: $} }
\newcommand{\lsim}{\raisebox{-0.07cm}{$\:\:\stackrel{<}{{\scriptstyle
 \sim}}\:\: $} }

\newcommand{\Li}{{\rm Li}}
\newcommand{\GeV}{{\rm GeV}}

\def\beq{\begin{equation}}
\def\eeq{\end{equation}}
\def\beqa{\begin{eqnarray}}
\def\eeqa{\end{eqnarray}}

\def\Eq#1{Eq.~(\ref{#1})}
\def\ifm{\ifmmode}
\def\msb{\ifm \overline{\rm MS}\,\, \else $\overline{\rm MS}\,\, $\fi} 

\begin{document}

\title{\sc 
  WORKING GROUP I: Parton Distributions\\
  Summary Report for the HERA - LHC Workshop proceedings}

\include{author-list}
\begin{abstract}
We provide an assessment of the impact of parton distributions on the determination
of LHC processes,  
 and of the accuracy with which parton distributions (PDFs)
can be extracted from data, in particular from current and forthcoming
HERA experiments. We give an overview of reference LHC processes and their
associated PDF uncertainties, and study in detail $W$ and $Z$
production at the LHC. We discuss the precision which may be
obtained from the analysis of existing HERA data,  
tests of consistency of HERA
data from different experiments, and the combination of these data. We determine
further improvements on PDFs  which may be obtained from
future HERA data (including measurements of $F_L$), 
and from combining present and future HERA data with
present and future hadron collider data. We review the current status
of knowledge of higher (NNLO) QCD corrections to perturbative
evolution and deep-inelastic
scattering, and provide reference results
for their impact on parton
evolution, and we briefly examine non-perturbative models for parton
distributions.  We discuss the state-of-the art in global parton fits,
we assess the impact on them of various kinds of data and of
theoretical corrections, by providing 
benchmarks of Alekhin and MRST parton distributions and a CTEQ
analysis of parton fit stability, and we briefly
presents proposals for alternative approaches to parton fitting.
We summarize the status of large and small $x$  resummation, by
providing estimates of the impact of large $x$ resummation on parton
fits, and a comparison of different approaches to small $x$
resummation, for which we also discuss numerical techniques.
\end{abstract}
\eject
\pagestyle{plain}
\tableofcontents
\include{genintro}

\include{processes}

\include{pdfexp}

\include{precisionlimits}
\include{comparisonh1zeus}

\include{averaging}
\include{longitudinal}

\include{mkbrsea}

\include{futureimpact}

\include{jetsinfits}

\include{evpfits}

\include{nnloprecision}

\include{mellinmath}

\include{evolution}

\include{nonperturbativeshape}
\include{alekhin}

\include{thorne}

\include{cteq}

\include{nnpdf}

\include{resumsec}

\include{acknowledgments}

\bibliographystyle{heralhc} 
{\raggedright
\bibliography{pdfs}
}

\end{document}

%% file: author-list.tex
\author{
{\sc Conveners:}
\\
M.~Dittmar$~^{1}$,
S.~Forte$~^{2}$,
A.~Glazov$~^{3}$,
S.~Moch$~^{4}$
\\
{\sc Contributing authors:}
\\
S.~Alekhin$~^{5}$,
G.~Altarelli$~^{6,7}$,
J.~Andersen$~^{8}$,
R.~D.~Ball$~^{9}$,
J.~Bl\"umlein$~^{4}$,
H.~B\"ottcher$~^{4}$,
T.~Carli$~^{10}$,
M.~Ciafaloni$~^{11}$,
D.~Colferai$~^{11}$,
A.~Cooper-Sarkar$~^{12}$,
G.~Corcella$~^{6}$,
L.~Del Debbio$~^{6,9}$, 
G.~Dissertori$~^{1}$,
J.~Feltesse$^{13}$,
A.~Guffanti$~^{4}$,
C.~Gwenlan$~^{12}$,
J.~Huston$~^{14}$,
G.~Ingelman$~^{15}$,
M.~Klein$~^{4}$,
J.~I.~Latorre$~^{16}$,
T.~La\v{s}tovi\v{c}ka$~^{10}$,
G.~La\v{s}tovi\v{c}ka-Medin$~^{17}$,
L.~Magnea$~^{18}$,
A.~Piccione$~^{18}$,
J.~Pumplin$~^{14}$,
V.~Ravindran$~^{19}$,
B.~Reisert~$^{20}$,
J.~Rojo$~^{16}$,
A.~Sabio Vera$~^{21}$,
G.~P.~Salam$~^{22}$,
F.~Siegert$~^{10}$,
A.~Sta\'sto$~^{23}$,
H.~Stenzel$~^{24}$,
C.~Targett-Adams$~^{25}$,
R.S.~Thorne$~^{8}$,
A.~Tricoli$^{12}$,
J.A.M.~Vermaseren$~^{26}$,
A.~Vogt$~^{27}$
}

\institute{
$^{1}$~Institute for Particle Physics, ETH-Z\"urich H\"onggerberg, CH 8093 Z\"urich, Switzerland
\\
$^{2}$~Dipartimento di Fisica, Universit\'a di Milano, INFN Sezione di Milano, Via Celoria 16, I 20133 Milan, 
        \\ $^{\phantom{2}}$~Italy
\\
$^{3}$~DESY, Notkestrasse 85, D 22603 Hamburg, Germany 
\\
$^{4}$~DESY, Platanenallee 6, D 15738 Zeuthen, Germany 
\\
$^{5}$~Institute for High Energy Physics, 142284 Protvino, Russia
\\
$^{6}$~CERN, Department of Physics, Theory Division,
CH 1211 Geneva 23, Switzerland 
\\
$^{7}$~Dipartimento di Fisica ``E.Amaldi'', 
Universit\`a Roma Tre and INFN, Sezione di Roma Tre,
\\ $^{\phantom{7}}$~via della Vasca Navale 84, I 00146 Roma, Italy
\\
$^{8}$~Cavendish Laboratory, University of Cambridge, Madingley Road, Cambridge, CB3 0HE, UK 
\\
$^{9}$~School of Physics, University of Edinburgh, Edinburgh EH9 3JZ, UK
\\
$^{10}$~CERN, Department of Physics, 
CH 1211 Geneva 23, Switzerland 
\\
$^{11}$~Dipartimento di Fisica, Universit\`a di Firenze and INFN,
Sezione di Firenze, I 50019 \\ $^{\phantom{11}}$~Sesto Fiorentino, Italy
\\
$^{12}$~Department of Physics, Nuclear and Astrophysics Lab., Keble Road, Oxford, OX1 3RH, UK 
\\
$^{13}$~DSM/DAPNIA, CEA, Centre d'Etudes de Saclay, F 91191
Gif-sur-Yvette, France
\\
$^{14}$~Department of Physics and Astronomy, Michigan State University, E.~Lansing, MI 48824, USA
\\
$^{15}$~High Energy Physics, Uppsala University, Box 535, SE 75121 Uppsala, Sweden
\\
$^{16}$~Departament d'Estructura i Constituents de la Mat\`eria, Universitat de Barcelona, Diagonal 647, 
        \\ $^{\phantom{15}}$~E~08028 Barcelona, Spain
\\
$^{17}$~University of Podgorica, Cetinjski put bb, CNG 81000 Podgorica, Serbia and Montenegro
\\
$^{18}$~Dipartimento di Fisica Teorica, Universit\`a di Torino and INFN Sezione di Torino, Via P. Giuria 1, \\ $^{\phantom{17}}$~I 10125 Torino, Italy
\\
$^{19}$~Harish-Chandra Research Institute, Chhatnag Road, Jhunsi,  Allahabad,  India 
\\
$^{20}$~FNAL, Fermi National Accelerator Laboratory, Batavia, IL 60126, USA
\\
$^{21}$~II. Institut f\"ur Theoretische Physik, Universit\"at Hamburg, Luruper Chaussee 149, D 22761 Hamburg, 
        \\ $^{\phantom{20}}$~Germany
\\
$^{22}$~LPTHE, Universities of Paris VI and VII and CNRS, F 75005, Paris, France 
\\
$^{23}$~H.~Niewodnicza\'nski Institute of Nuclear Physics, PL 31-342 Krak\'ow, Poland
\\
$^{24}$~II. Physikalisches Institut, Universit\"at Giessen, Heinrich-Buff-Ring 16, D 35392 Giessen, Germany
\\
$^{25}$~Department of Physics and Astronomy, UC London, Gower Street, London, WC1E 6BT, UK 
\\
$^{26}$~NIKHEF Theory Group, Kruislaan 409, NL 1098 SJ Amsterdam, The Netherlands 
\\
$^{27}$~IPPP, Department of Physics, Durham University, Durham DH1 3LE, UK 
\\
}

\maketitle

%% file: genintro.tex
\section{Introduction}
The physics of parton distributions, especially within the context of 
deep-inelastic scattering (DIS), has been an active subject of
detailed theoretical and experimental investigations since the origins
of perturbative quantum chromodynamics (QCD), which, thanks to
asymptotic freedom, allows one to determine perturbatively their
scale dependence~\cite{Gross:1973id,Politzer:1973fx,Gross:1973ju,Georgi:1974sr,Gross:1974cs}.

Since the advent of HERA, much progress has been made in determining 
the Parton Distribution Functions (PDFs) of the proton. 
A good knowledge of the PDFs is vital in order to make predictions for 
both Standard Model and beyond the Standard Model processes at
hadronic colliders, specifically the LHC.  
Furthermore, PDFs must be known as precisely as possible in order to maximize 
the discovery potential for new physics at the LHC. Conversely, LHC
data will lead to an imporvement in the knowledge of PDFs.

The main aim of this document is to provide a state-of-the art
assessment of the impact of parton distributions on the determination
of LHC processes,  and of the accuracy with which parton distributions
can be extracted from data, in particular current and forthcoming HERA
data.

In Section~\ref{section:processes} we will set the stage by
providing an overview of
relevant LHC processes and a discussion of their experimental and
theoretical accuracy.  In Section~\ref{section:exppdf} we will turn
to the experimental determination of PDFs, and in particular examine
the improvements to be expected from forthcoming measurements at HERA,
as well as from analysis methods which allow one to combine HERA data 
with each other, and also with data from existing (Tevatron) and
forthcoming (LHC) hadron colliders.
In Section~\ref{section:evpfits} we will discuss the state of the art
in the extraction of parton distributions of the data, by first
reviewing recent progress in higher-order QCD corrections and their
impact on the extraction of PDFs, and then discussing and comparing
the determination of PDFs from global fits. Finally, in
Section~\ref{sec:pdf,res} we will summarize the current status of
resummed QCD computations which are  not yet used in parton
fits, but could lead to an improvement in the theoretical precision of
PDF determinations.

Whereas we will aim at summarizing the state of the art, we will also
provide several new results, benchmarks and predictions, which have
been obtained within the framework of the HERA-LHC workshop.

%% file: processes.tex
\section{LHC final states and their potential experimental and
  theoretical accuracies\protect\footnote{Subsection coordinator: Michael Dittmar}}
\label{section:processes}
\subsection{Introduction}

Cross section calculations and experimental simulations for many LHC reactions, 
within the Standard Model and for many new physics 
scenarios have been performed during the last 20 years.
These studies demonstrate how various final states might eventually be selected above Standard Model backgrounds and 
indicate the potential statistical significance of such measurements. In general, these studies assumed that the uncertainties 
from various sources, like the PDF uncertainties, the experimental uncertainties and the 
various signal and background Monte Carlo simulations will eventually be controlled with uncertainties small  
compared to the expected statistical significance. This is the obvious approach for many so called 
discovery channels with clean and easy signatures and relatively small cross sections.

However, during the last years many new and more complicated signatures,  which require 
more sophisticated selection criteria, have been discussed. These studies indicate the possibility to perform    
more ambitious searches for new physics and for precise Standard Model tests, which would  
increase the physics potential of the LHC experiments. Most of these studies 
concentrate on the statistical significance only and potential systematic limitations are rarely discussed.

In order to close this gap from previous LHC studies,  questions related to the systematic limits of  
cross section measurements from PDF uncertainties, from imperfect Standard Model Monte Carlo simulations, from 
various QCD uncertainties and from the efficiency and luminosity uncertainties were discussed 
within the PDF working group of this first HERA-LHC workshop.
The goal of the studies presented during the subgroup meetings during the 2004/5 HERA LHC workshop
provide some answers to questions related to these systematic limitations.
In particular, we have discussed potential experimental and theoretical uncertainties for 
various Standard Model signal cross sections at the LHC.
Some results on the experimental systematics, on experimental and theoretical uncertainties for the inclusive W, Z and for diboson production, especially 
related to uncertainties from PDF's and from higher order QCD calculations
are described in the following sections. 

While it was not possible to investigate the consequences for various aspects of the LHC physics potential in detail,
it is important to keep in mind that  many of these Standard Model reactions are also important backgrounds in the search for the Higgs and other 
exotic phenomena. Obviously, the consequences from these unavoidable systematic uncertainties need to be investigated 
in more detail during the coming years.

\newpage

\subsection{Measuring and interpreting cross sections at the LHC\protect\footnote{
Contributing author: Michael Dittmar}}

The LHC is often called a machine to make discoveries. However, 
after many years of detailed LHC simulations, it seems clear that relatively few 
signatures exist, which do not involve cross section measurements for signals and 
the various backgrounds.   
Thus, one expects that cross section measurements for a large variety of well defined reactions 
and their interpretation within or perhaps beyond the Standard Model 
will be one of the main task of the LHC physics program.

While it is relatively easy to estimate the statistical precision of a particular measurement as a function of the luminosity, estimates  
of potential systematic errors are much more complicated. Furthmore, as almost nobody wants to know about systematic limitations
of future experiments, detailed studies are not rewarding. 
Nevertheless, realistic estimates of such systematic errors are relevant, 
as they might allow the LHC community to concentrate their efforts 
on the areas where current systematic errors, like the ones which are related to uncertainties from Parton Distribution Functions (PDF) 
or the ones from missing higher order QCD calculations, can still be improved during the next years.

In order to address the question of systematics, it is useful to start with the 
basics of cross section measurements. Using some clever criteria a particular signature
is separated from the data sample and the surviving N$_{\mathrm{observed}}$ events can be counted. 
Backgrounds, N$_{\mathrm{background}}$, from various sources have to be estimated either using the data or some Monte Carlo estimates.
The number of signal events, N$_{\mathrm{signal}}$, is then obtained from the difference.
In order to turn this experimental number of signal events into a measurement one has to apply a correction  
for the efficiency. This experimental number can now be compared with the product of the theoretical production cross section 
for the considered process and the corresponding Luminosity. 
For a measurement at a hadron collider, like the LHC, processes are calculated on the basis of quark and gluon luminosities which 
are obtained from the proton-proton luminosity ``folded'' with the PDF's.    

In order to estimate potential systematic errors one needs to examine carefully the 
various ingredients to the cross section measurement and their interpretation.
First, a measurement can only be as good as the impact from of the background uncertainties, which depend on 
the optimized signal to background ratio. Next, the experimental efficiency uncertainty depends 
on many subdetectors and their actual real time performance. While this can only be known exactly from real data,
one can use the systematic error estimates from previous experiments in order to guess the size of similar error sources 
for the future LHC experiments.
We are furthermore confronted with uncertainties from the PDF's and from the proton-proton luminosity.
If one considers all these areas as essentially experimental, then one should assign
uncertainties originating from imperfect knowledge of signal and background cross sections
as theoretical. 

Before we try to estimate the various systematic errors in the following subsections, we believe that  
it is important to keep in mind that particular studies need not to be much more detailed 
than the largest and limiting uncertainty, coming from either the experimental or the theoretical area.
Thus, one should not waste too much time, in order to achieve the smalled possible uncertainty in one particular 
area. Instead, one should try first to reduce the most important error sources and 
if one accepts the ``work division'' between experimental and theoretical contributions, then 
one should simply try to be just a little more accurate than either the theoretical 
or the experimental colleagues.  

\subsubsection{Guessing experimental systematics for ATLAS and CMS}

In order to guess experimental uncertainties, without doing lengthy and detailed Monte Carlo studies,
it seems useful to start with some simple and optimistic assumptions about ATLAS and CMS\footnote{Up to date performance of the ATLAS and CMS detectors and further detailed references can be found 
on the corresponding homepages {\tt http://atlas.web.cern.ch/Atlas/}
and {\tt http://cmsinfo.cern.ch/Welcome.html/}}.  

First of all, one should assume that both experiments can actually operate 
as planned in their proposals. As the expected performance goals are 
rather similar for both detectors the following list of measurement capabilities 
looks as a reasonable first guess.   

\begin{itemize} 
\item Isolated electrons, muons and photons with a transverse momentum 
above 20 GeV and a pseudorapidity $\eta$ with $|\eta| \leq 2.5$ are 
measured with excellent accuracy and high (perhaps as large as 95\% for some reactions) ``homogeneous'' efficiency.  
Within the pseudo rapidity coverage one can assume that experimentalists will perhaps be able, using 
the large statistics from leptonic W and Z decays, to control the efficiency for electrons and muons with a 1\% accuracy.
For simplicity, one can also assume that these events will allow to control measurements with 
high energy photons to a similar accuracy.
For theoretical studies one might thus assume that high $p_{t}$ electrons, muons and photons 
and $|\eta| \leq 2.5$ are measured with a systematic uncertainty of $\pm$ 1\% for each lepton (photon).
 
\item Jets are much more difficult to measure. Optimistically one could assume that jets can be seen with good efficiency 
and angular accuracy if the jet transverse momentum is larger than 30 GeV
and if their pseudo rapidity fulfills $|\eta| \leq 4.5$. The jet energy resolution is not easy to quantify, 
but numbers could be given using some ``reasonable'' assumptions like $\Delta E/E \approx 100-150\%/\sqrt{E}$.
For various measurements one want to know the uncertainty of the absolute jet energy scale. Various tools,
like the decays of $W \rightarrow q\bar{q}$ in $t\bar{t}$ events or the photon-jet final state,  
might be used to calibrate either the mean value or the maximum to reasonably good accuracy. 
We believe that only detailed studies of the particular signature will allow 
a quantitative estimate of the uncertainties related to the jet energy scale measurements.

\item 
The tagging of b--flavoured jets can be done, but the efficiency 
depends strongly on the potential backgrounds. Systematic efficiency uncertainties
for the b--tagging are difficult to quantify but it seems that, in the absence of a new method,
relative b-tagging uncertainties below $\pm$ 5\% are almost impossible to achieve.
\end{itemize}

With this baseline LHC detector capabilities,
it seems useful to divide the various high $q^{2}$ LHC reactions
into essentially five different non overlapping categories.
Such a devision can be used to make some reasonable accurate estimates of 
the different systematics.

\begin{itemize} 
\item Drell--Yan type lepton pair final states. 
This includes on-- and off--shell W and Z decays.
\item $\gamma$--jet and $\gamma \gamma X$ final states.
\item Diboson events of the type $WW,~WZ,~ZZ,~W\gamma$ with leptonic decays 
of the $W$ and $Z$ bosons. One might consider to include the Standard Model Higgs signatures 
into this group of signatures.
\item Events with top quarks in the final state, identified with at least one isolated lepton.
\item Hadronic final states with up to n(=2,3 ..) Jets and different $p_{t}$ and mass.
\end{itemize}

With this ``grouping'' of experimental final states, one can now start to analyze the different potential 
error sources. Where possible, one can try to define and use relative measurements of various reactions such 
that some systematic errors will simply cancel. 

Starting with the resonant W and Z production with leptonic decays, several million of
clean events will be collected quickly, resulting in relative statistical errors well below $\pm$1\%.
Theoretical calculations for these reactions are well advanced and these reactions are among the best understood LHC final states
allowing to build the most accurate LHC Monte Carlo generators. 
Furthermore, some of the experimental uncertainties can be reduced 
considerably if ratio measurements of cross section, such as $W^{+}/W^{-}$ and $Z/W$, are performed. 
The similarities in the production 
mechanism should also allow to reduce theoretical uncertainties for such ratios.
The experimental counting accuracy of W and Z events, 
which includes background and efficiency corrections, might achieve eventually uncertainties of 1\% or slightly better
for cross section ratios. 

\begin{figure}[htb]
\begin{center}
\includegraphics[width=7cm]{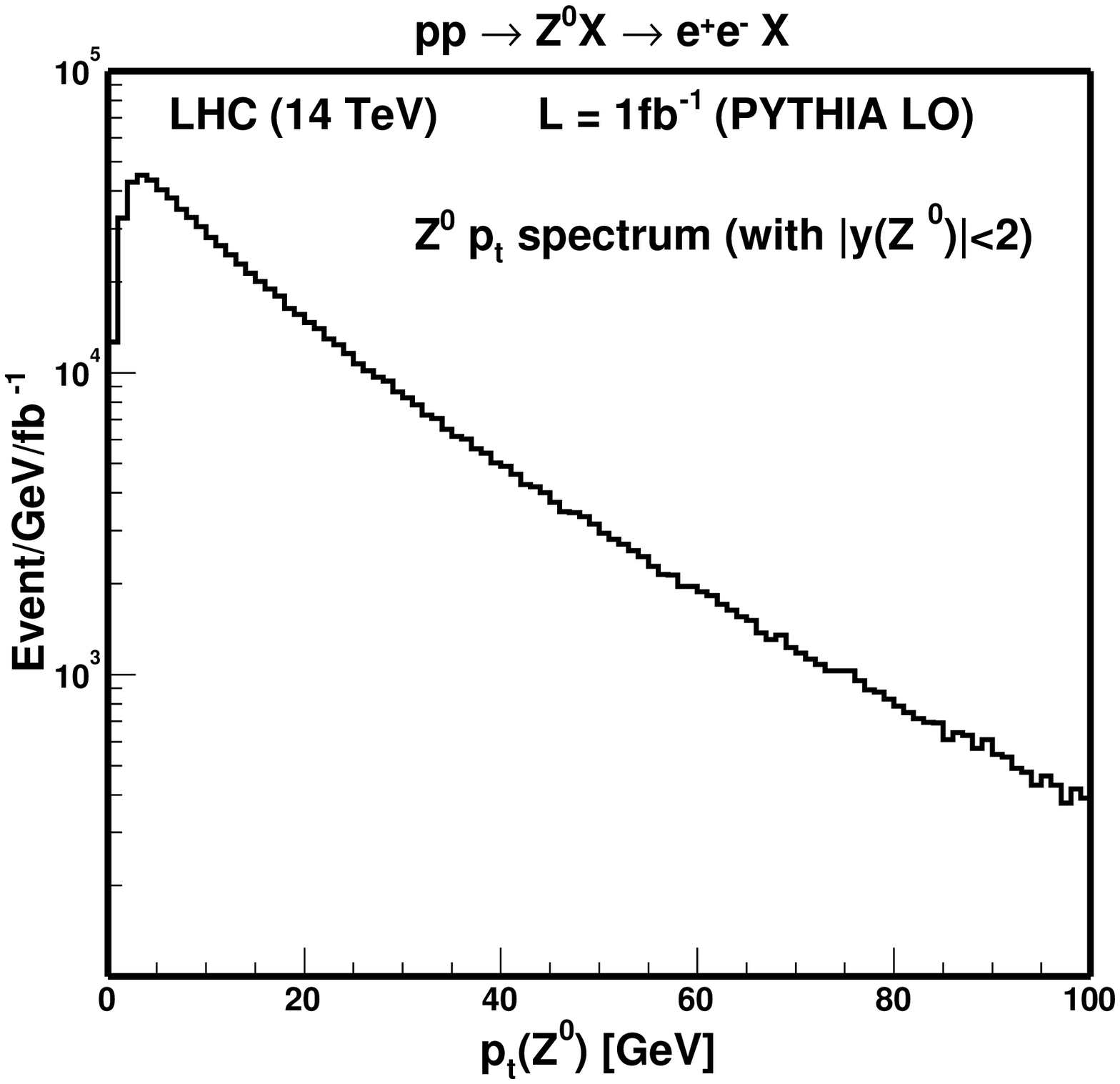}
\includegraphics[width=7cm]{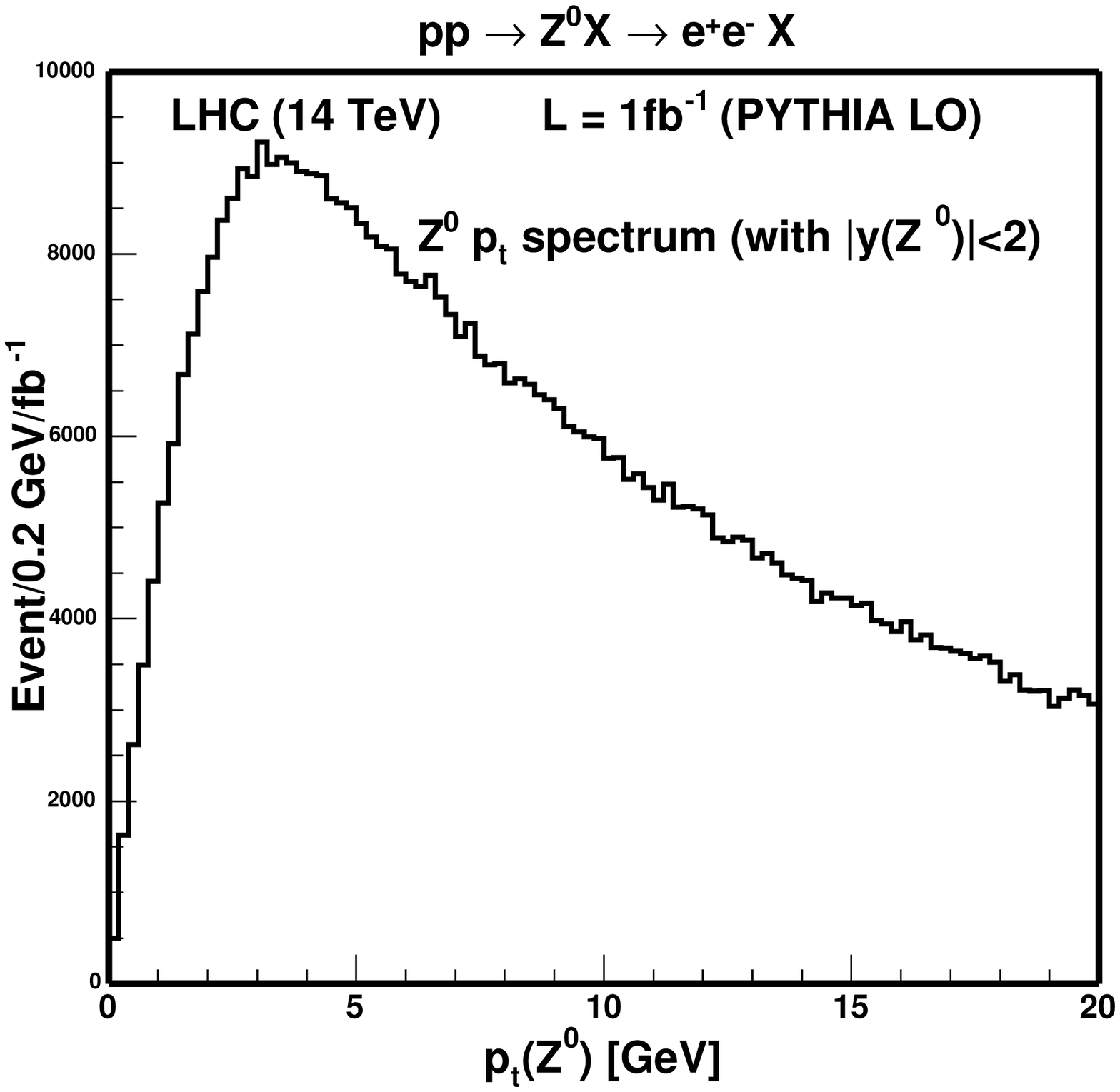}
\end{center}
\caption{
Simple simulation of a potential measurement of the Z  $p_{t}$ spectrum, possible with a luminosity of only 1 fb$^{-1}$.
Who will be able to predict this $p_{t}$ spectrum in all its beauty and with similar accuracy?} 
\label{z0ptdis}
\end{figure}

Furthermore, it seems that the shape of the $p_{t}$ distribution of the Z, using the decay into electron pairs
($pp \rightarrow ZX \rightarrow e{+}e^{-}X$),
can be determined with relative accuracies of much less than 1\%. This distribution, shown in figure~\ref{z0ptdis}, can be used to tune 
the Monte Carlo description of this particular process. 
This tuning of the Monte Carlo can than be used almost directly to predict theoretically also the W $p_{t}$ spectrum,
and the $p_{t}$ spectrum for high mass Drell-Yan lepton pair events. Once an accurate model description of these Standard Model
reactions is achieved one might use these insights also to predict the $p_{t}$ spectrum of other well defined final states. 

From all the various high $q^{2}$ reactions, 
the inclusive production of W and Z events is known to be the theoretically best understood
and best experimentally measurable LHC reaction. Consequently, the idea to use these simple well defined final states
as the LHC cross section normalisation tool, or standard candle was described  
first in reference \cite{Dittmar:1997md}. This study indicated that the W and Z production might result in 
a precise and simple parton luminosity monitor. In addition, these reactions can also be used to improve the 
relative knowledge of the PDF's.
In fact, if one gives up on the idea to measure absolute cross sections,
the relative parton luminosity can in principle be determined with
relative uncertainties well below $\pm$5\%, the previously expected possible limit for any absolut 
proton-proton luminosity normalisation procdure.  

In summary, one can estimate that it should be possible to reduce experimental uncertainties 
for Drell-Yan processes to systematic uncertainties below $\pm$5\%, optimistically one might envisage an event counting accuracy of perhaps $\pm$1\%, 
limited mainly from the lepton identification efficiency.    
   
The next class of final states, which can be measured exclusively with leptons, are the diboson pair events 
with subsequent leptonic decays. Starting with the ZZ final state, we expect that the statistical accuracy 
will dominate the measurement for several years. Nevertheless, the systematic uncertainties of the measurement, 
based on four leptons, should in principle be possible with relative errors of a few \% only.

The production of WZ and WW involves unmeasurable neutrinos. 
Thus, experimentally only an indirect and incomplete determination of the kinematics of the final states is possible
and very detailed simulations with precise Monte Carlo generators are required for the interpretation of these final state. 
It seems that a measurement of the event counting with an accuracy below $\pm$5\%, 
due to efficiency uncertainties from the selection alone,  to be highly non trivial. 
Nevertheless, if the measurements and the interpretations can be done relative to the W and Z resonance production, 
some uncertainties from the lepton identification efficiency, from the  
PDF and from the theoretical calculation can perhaps be reduced.
Without going into detailed studies for each channel, one could 
try to assume that a systematic uncertainty of $\pm$5\% might be defined as a goal. 
Similar characteristics and thus limitations can be expected for other diboson signatures.

The production cross section of top antitop quark pairs is large and several million of semileptonic tagged and relatively clean 
events ($pp \rightarrow t \bar{t} \rightarrow Wb Wb$ identified with one leptonic $W$ decay) can be expected. 
However, the signature involves several jets, some perhaps tagged as b--flavoured, and missing transverse momentum from the neutrino(s).
The correct association of the various jets to the corresponding top quark is known to be extremely difficult, 
leading to large combinatorial backgrounds. Thus, it seems that, 
even if precise Monte Carlo generators will become eventually available,
that systematic uncertainties smaller than 5-10\% should not be expected.
Consequently,we assume that top antitop backgrounds for a wide class of signals can not be determined 
with uncertainties smaller than 5-10\%.

Measurements of so called ``single'' top quarks are even more difficult, as
the cross section is smaller and larger backgrounds exist. Systematic errors 
will therefore always be larger than the one guessed for top-antitop pair production.

Finally, we can address the QCD jet production. Traditionally one measures and interprets the so called jet cross section as a function of 
$p_{t}$ jet and the mass of the multi jet system using various rapidity intervals. 
With the steeply falling $p_{t}$ jet spectrum and essentially no background, one will determine the differential spectrum 
such that only the slope has to be measured with good relative accuracy.
If one is especially interested into the super high mass or high $p_{t}$ events, then  
we expect that migrations due to jet mis-measurements and non Gaussian tails in the jet energy measurements will limit 
any measurement. A good guess might be that the LHC experiments can expect absolut normalisation uncertainties 
similar to the ones achieved with CDF and D0, corresponding to  
uncertainties of about $\pm$ 10-20\%.

Are the above estimated systematic limits for the various measurements pessimistic, optimistic or simply realistic? 
Of course, only real experiments will tell during the coming LHC years. However, while some of these estimates 
will need perhaps some small modification, they could be used as a limit waiting to be improved
during the coming years. Thus, some people full of ideas might take these numbers as a challenge, and discover and develop 
new methods that will improve these estimates. This guess of systematic limitations for LHC experiments could thus be considered 
as a ``provocation'', which will stimulate activities to prove them wrong.
In fact, if the experimental and theoretical communities could demonstrate why some of these ``pessimistic'' numbers are wrong 
the future real LHC measurements will obviously benefit from the required efforts to 
develope better Monte Carlo programs and better experimental methods. 

The following summary from a variety of experimental results from previous high energy collider experiments  
might help to quantify particular areas of concern for the LHC measurements.
These previous measurements can thus be used as a starting point for an LHC experimenter, who can study and explain 
why the corresponding errors at LHC will be smaller or larger. 

\subsubsection{Learning from previous collider experiments}

It is broadly accepted, due to the huge hadronic interaction rate and the short 
bunch crossing time, that the experimental conditions at the LHC will be similar or worse than 
the ones at the Tevatron collider.
One experimental answer was to improve the granularity, speed and accuracy 
of the different detector elements accordingly. 
Still, no matter how well an experiment can be realized, the LHC conditions to do experiments 
will be much more difficult than at LEP or any hypothetical future high energy $e^{+}e^{-}$ collider.
One important reason is the large theoretical uncertainty, which prevents to make 
signal and background Monte Carlos with accuracies similar to the ones which were used at LEP.  

Thus, we can safely expect that systematic errors at LHC experiments will be larger than 
the corresponding ones from LEP and that the Tevatron experience can be used as a first guess.

\begin{itemize}
\item Measurement of $\sigma \times$BR for W and Z production from CDF\cite{Acosta:2004uq} and D0\cite{Bellavance:2005rg}:

The CDF collaboration has presented a high statistics measurement with electrons and muons. Similar systematic errors of about 
$\pm$ 2\% were achieved for efficiency and thus the event counting with electrons and muons. 
The error was reduced to $\pm$ 1.4\% for the ratio measurement where some lepton identification efficiencies cancel.
Similar errors about $\times$ 1.5-2 larger have been obtained by the corresponding measurements from the D0 experiment.  

\item Measurement of the cross section for $p\bar{p} \rightarrow Z \gamma(\gamma)$ from D0\cite{Abazov:2005ea}: 

A total of 138 $ee\gamma$ and 152 $\mu\mu \gamma$ candidate events were selected. 
The background was estimated to be about 10\% with a systematic uncertainty of $\pm$ 10-15\%, mainly from $\gamma$-jet misidentification.
Using Monte Carlo and a large sample of inclusive Z events, the efficiency uncertainty has been estimated to be $\approx$ 5\%
and when the data were used in comparison with the Standard Model prediction another uncertainty of 3.3\% originating from PDF's
was added. 

\item Measurement of the $ p \bar{p} \rightarrow t\bar{t}$ production cross section from CDF\cite{Taffard:2004qy}

A recent CDF measurement, using 197 pb$^{-1}$, obtained a cross section (in pb) 
of 7.0 +2.4 (-2.1) from statistics. This should be comapred with +1.7 (-1.2) from systematics, which includes $\pm 0.4$ from the luminosity
measurement. 
Thus, uncertainties from efficiency and background are roughly $\pm$20\%. It is expected that
some of the uncertainties can be reduced with the expected 10 fold luminosity increase such that  
the systematic error will eventually decrease to about $\pm$ 10\%, sufficient to be better than the 
expected theoretical error of $\pm$ 15\%. 

\item A search for Supersymmetry with b-tagged jets from CDF\cite{Bortoletto:2004ev}:  

This study, using single and double b-tagged events was consistent with background only.
However, it was claimed that the background uncertainty was dominated by the systematic error, which 
probably originated mostly from the b tagging efficiency and the misidentification of b-flavoured jets.
The numbers given were 16.4$\pm$ 3.7 events (3.15 from systematics) for the single b-tagged events
and 2.6$\pm$0.7 events (0.66 from systematics) for the double b-tagged events.
These errors originate mainly from the b-tagging efficiency uncertainties, which 
are found to roughly $\pm$ 20-25\% for this study of rare events.

\item Some ``random'' selection of recent $e^{+}e^{-}$ measurements:

A recent measurement from ALEPH (LEP) of the 
W branching ratio to $q\bar{q}$ estimated a systematic uncertainty 
of about $\pm$ 0.2\% \cite{Heister:2004wr}. This small uncertainty was possible because 
many additional constraints could be used.
 
OPAL has reported a measurement of $R_{b}$ at LEP II energies, with a systematic uncertainty of 
$\pm$ 3.7\%. Even though this uncertainty could in principle be reduced with higher statistics, 
one can use it as an indication on how large efficiency uncertainties from b-tagging 
are already with clean experimental conditions\cite{Abbiendi:2004vw}

Recently, ALEPH and DELPHI have presented cross section measurements for $e^{+}e^{-} \rightarrow \gamma \gamma$
with systematic errors between 2.2\% (ALEPH)\cite{Heister:2002ut} and 1.1\% (DELPHI)\cite{Abdallah:2004rc}. 
In both cases, the efficiency uncertainty, 
mainly from conversions, for this in principle easy signal was estimated to be roughly 1\%. 
In the case of ALEPH an uncertainty of  
about $\pm$0.8\% was found for the background correction. 

\end{itemize}

Obviously, these measurements can only be used,
in absence of anything better, as a most optimistic guess for possible systematic limitations at a hadron collider.  
One might conclude that the systematics from LEP experiments 
give (1) an optimistic limit for comparable signatures at the LHC and (2) that the 
results from CDF and D0 should indicate systematics which might be obtained 
realistically during the early LHC years. 

Thus, in summary the following list might be used as a first order guess on 
achievable LHC systematics\footnote{Reality will hopefully show new brilliant ideas, which  
combined with hard work will allow to obtain even smaller uncertainties.}.

\begin{itemize}
\item ``Isolated'' muons, electrons and photons can be measured with a small momentum (energy) uncertainty
and with an almost perfect angular resolution. The 
efficiency for $p_{t} \geq 20$ GeV and $|\eta| \leq 2.5$ will be ``high'' and
can be controlled optimistically to $\pm$ 1\%. 
Some straight forward selection criteria should reduce 
jet background to small or negligible levels.
\item ``Isolated'' jets with a $p_{t} \geq 30$ GeV and $|\eta| \leq 4.5$ 
can be seen with high (veto) efficiency and a small uncertainty from the jet direction measurement.
However, it will be very difficulty to measure the absolute jet energy scale and 
Non-Gaussian tails will limit the systematics if the jet energy scale is important.
\item Measurements of the missing transverse momentum depend on the final state but will 
in general be a sum of the errors from the lepton and the jet accuracies. 
\end{itemize}
 
Using these assumptions, the following ``optimistic'' experimental systematic errors can be used as a guideline:

\begin{enumerate}
\item Efficiency uncertainties for isolated leptons and photons with a $p_{t}$ above 20 GeV 
can be estimated with a $\pm$1\% accuracy. 
\item Efficiencies for tagging jets will be accurate to a few percent and the efficiency to 
tag b-flavoured jets will be known at best within $\pm$5\%. 
\item Backgrounds will be known, combining theoretical uncertainties and 
some experimental determinations, at best with a $\pm$5-10\% accuracy. 
Thus, discovery signatures without narrow peaks require signal to background ratios larger than 0.25-0.5, if  
5 $\sigma$ discoveries are claimed. Obviously, for accurate cross section measurements, the signal to background 
ratio should be much larger. 
\item In case of ratio measurements with isolated leptons, like $pp \rightarrow W^{+}/pp \rightarrow W^{-}$,    
relative errors between 0.5-1\% should be possible. Furthermore, it seems that the measurement of the shape of 
Z $p_{t}$ spectrum, using Z$\rightarrow e^{+}e^{-}$, will be possible with a systematic error much smaller than 1\%. 
As the Z cross section is huge and clean we expect that this signature will become the best measurable final state
and should allow to test a variety of production models with errors below $\pm$ 1\%, thus challenging future QCD 
calculations for a long time. 
\end{enumerate}

\newpage
\subsection{Uncertainties on $W$ and $Z$ production at the LHC\protect\footnote{Contributing authors: 
Alessandro Tricoli, Amanda Cooper-Sarkar, Claire Gwenlan}}

\subsubsection{Introduction}
\label{sec:intro}

\begin{figure}[htbp]
\centerline{
\epsfig{figure=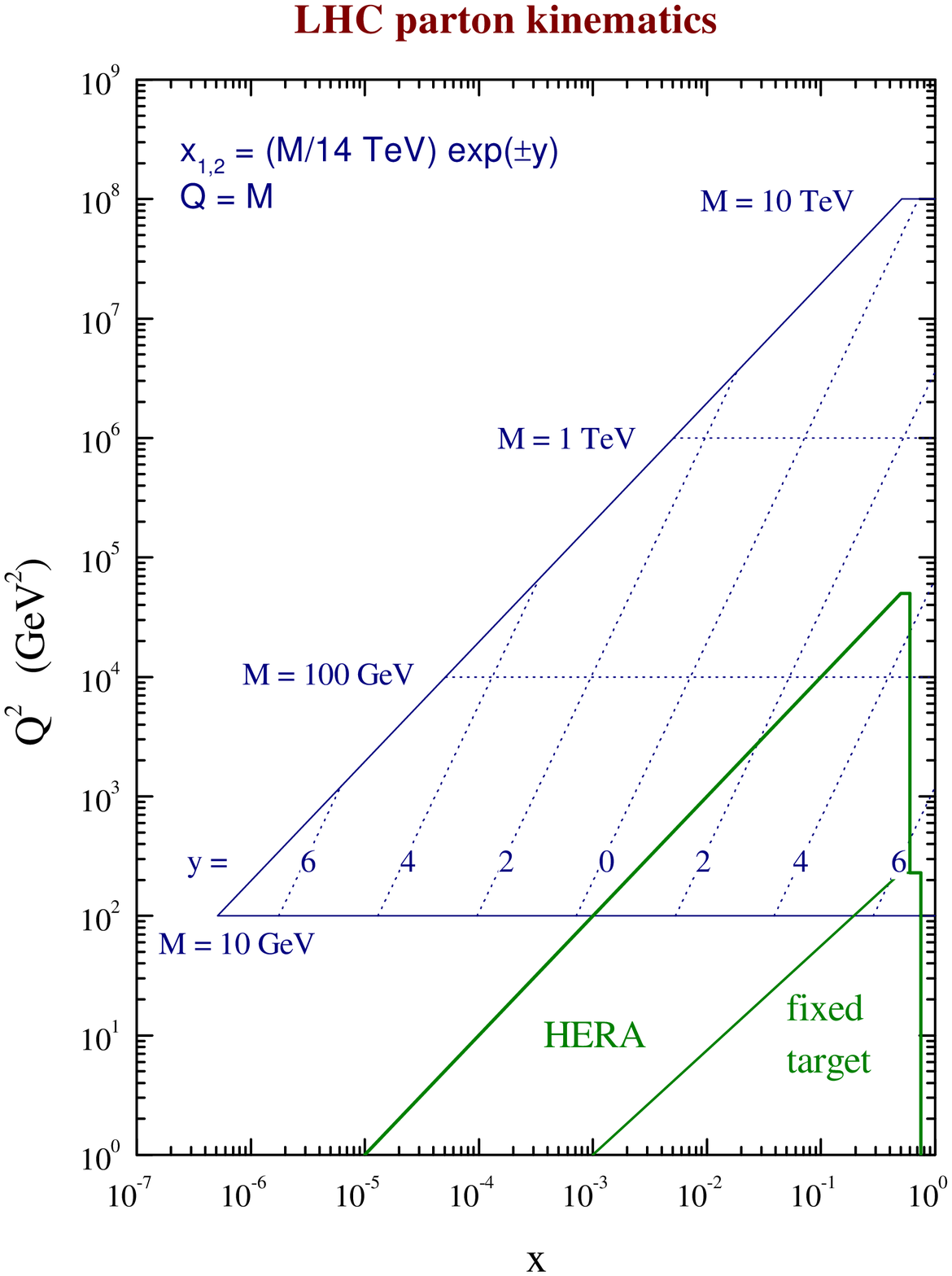,width=0.4\textwidth}
\epsfig{figure=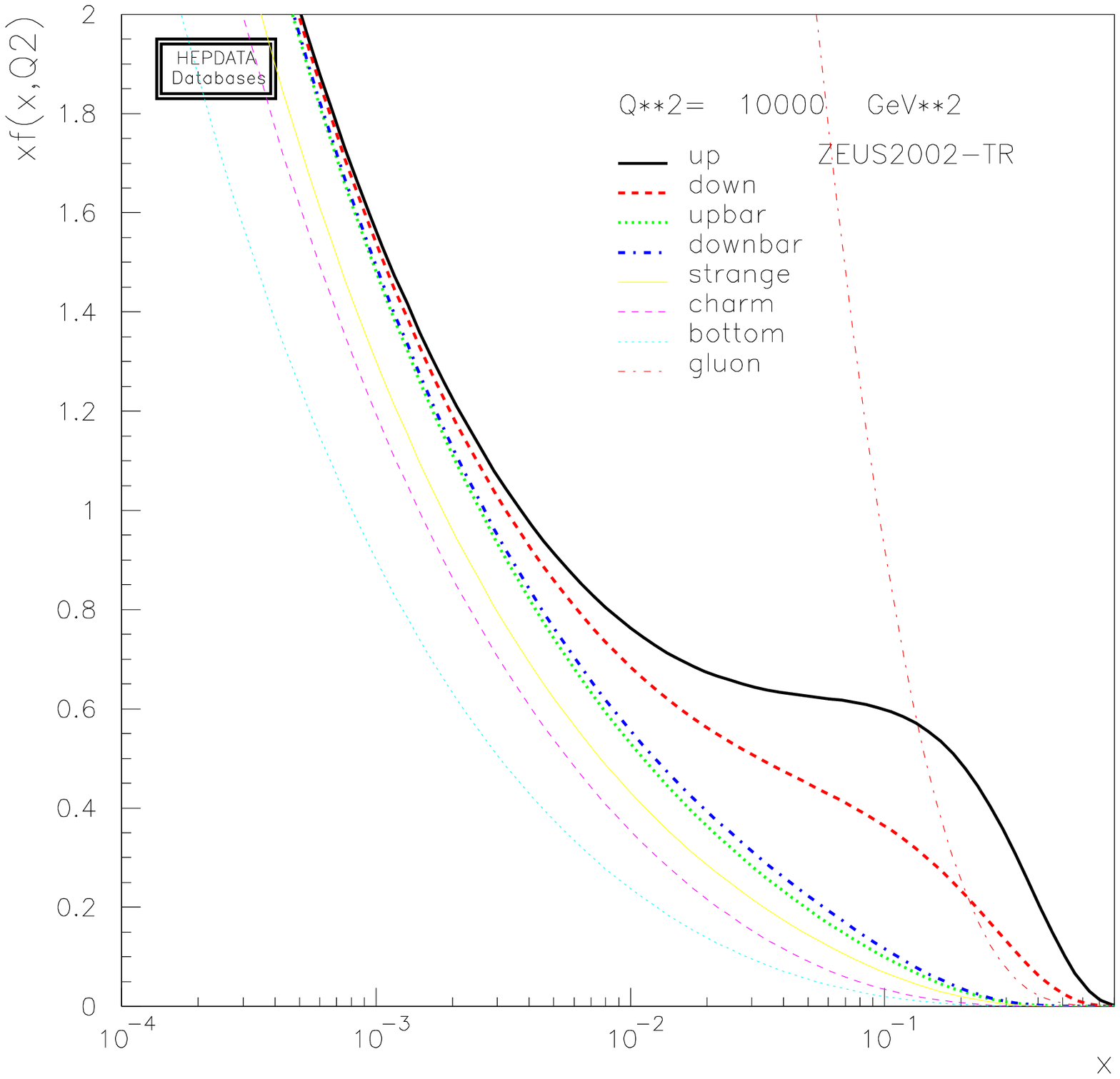,width=0.45\textwidth}}
\caption {Left plot: The LHC kinematic plane (thanks to James Stirling).
Right plot: PDF distributions at $Q^2 = 10,000$~GeV$^2$.}
\label{fig:kin/pdfs}
\end{figure}

 At leading order (LO), $W$ and $Z$ production occur by the process, $q \bar{q} \rightarrow W/Z$, 
and the momentum fractions of the partons 
participating in this subprocess are given by $x_{1,2} = \frac{M}{\surd{s}} \exp (\pm y)$, where 
$M$ is the centre of mass energy of the subprocess, $M = M_W$ or $M_Z$, $\surd{s}$ is the centre of
 mass energy of the reaction  ($\surd{s}=14$ TeV at the LHC) and 
$y = \frac{1}{2} \ln{\frac{(E+p_{l})}{(E-p_{l})}}$ gives the parton rapidity. 
The kinematic plane for LHC parton kinematics is shown in Fig.~\ref{fig:kin/pdfs}. 
Thus, at central rapidity, the participating partons have small momentum fractions, $x \sim 0.005$.
Moving away from central rapidity sends one parton to lower $x$ and one 
to higher $x$, but over the measurable rapidity range, $|y| < 2.5$, 
$x$ values remain in the range, 
$10^{-4} < x < 0.1$. Thus, in contrast to the situation at the Tevatron, valence quarks are not 
involved, the scattering is happening between sea quarks. Furthermore, the high 
scale of the process $Q^2 = M^2 \sim 10,000$~GeV$^2$ ensures that the gluon is the dominant 
parton, see Fig.~\ref{fig:kin/pdfs}, so that these sea quarks have mostly 
been generated by the flavour blind $g \to q \bar{q}$ splitting process. Thus the precision of 
our knowledge of $W$ and $Z$ cross-sections at the LHC is crucially dependent on the uncertainty on 
the momentum distribution of the gluon. 

\begin{figure}[tbp] 
\centerline{
\epsfig{figure=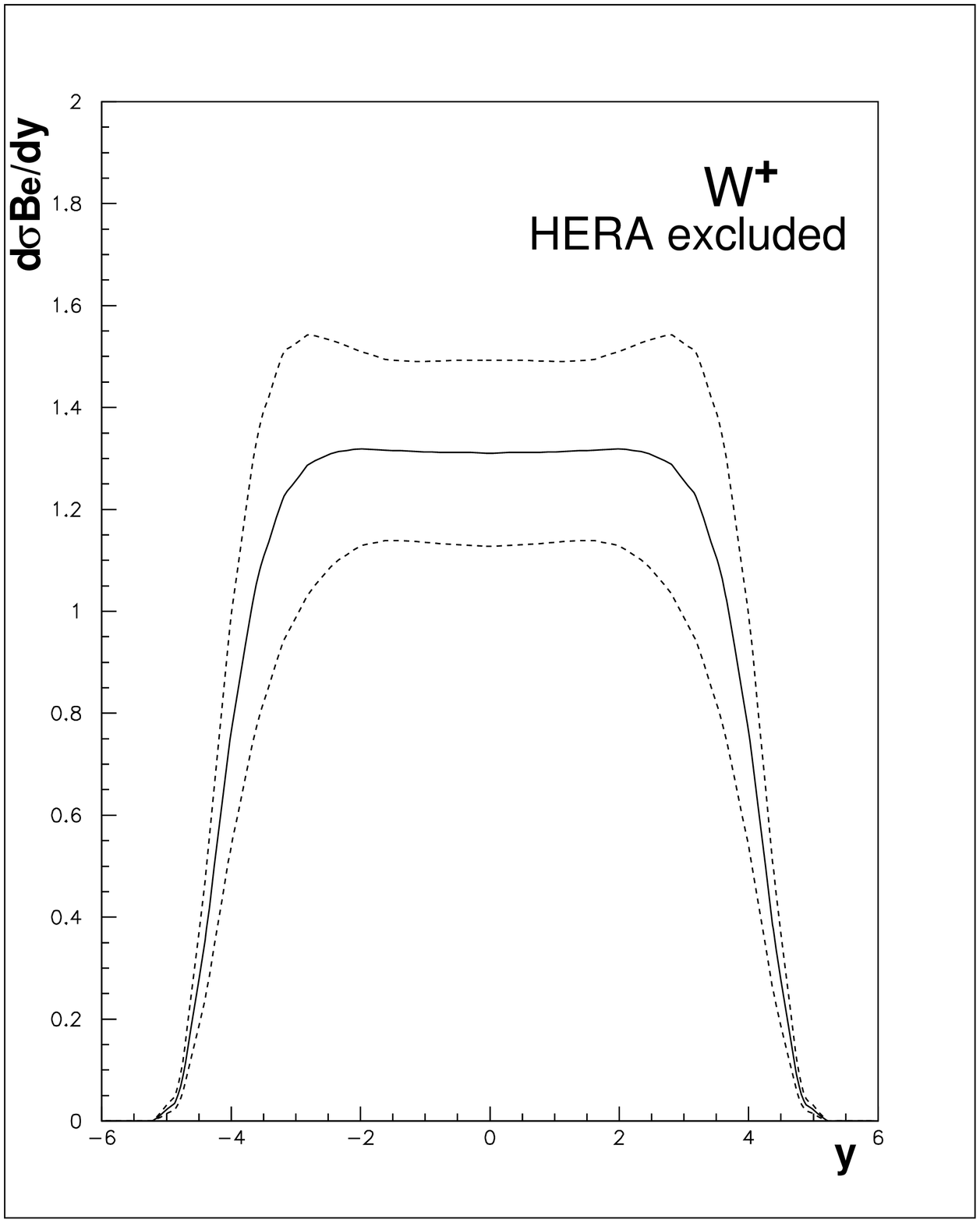,width=0.3\textwidth,height=5cm}
\epsfig{figure=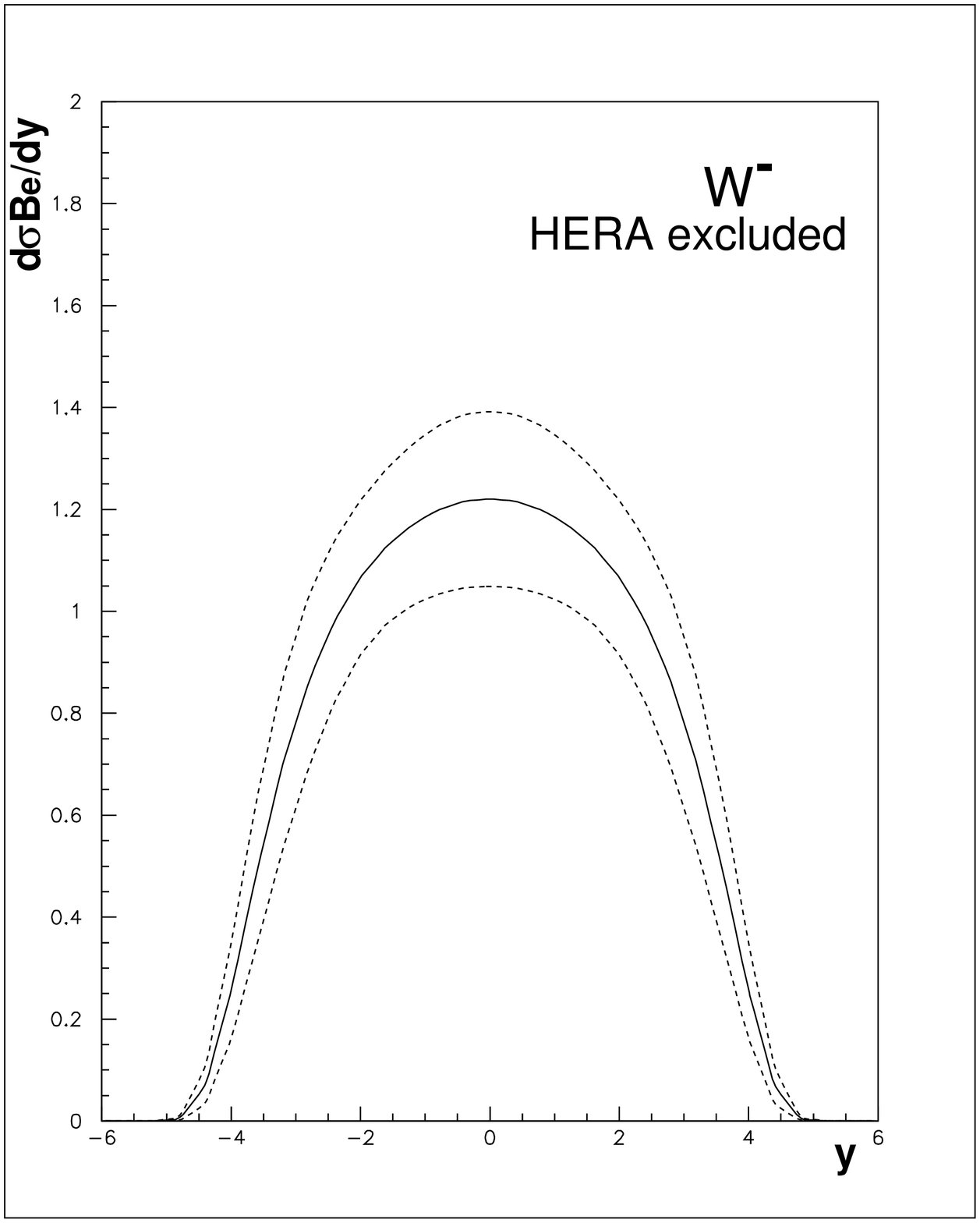,width=0.3\textwidth,height=5cm}
\epsfig{figure=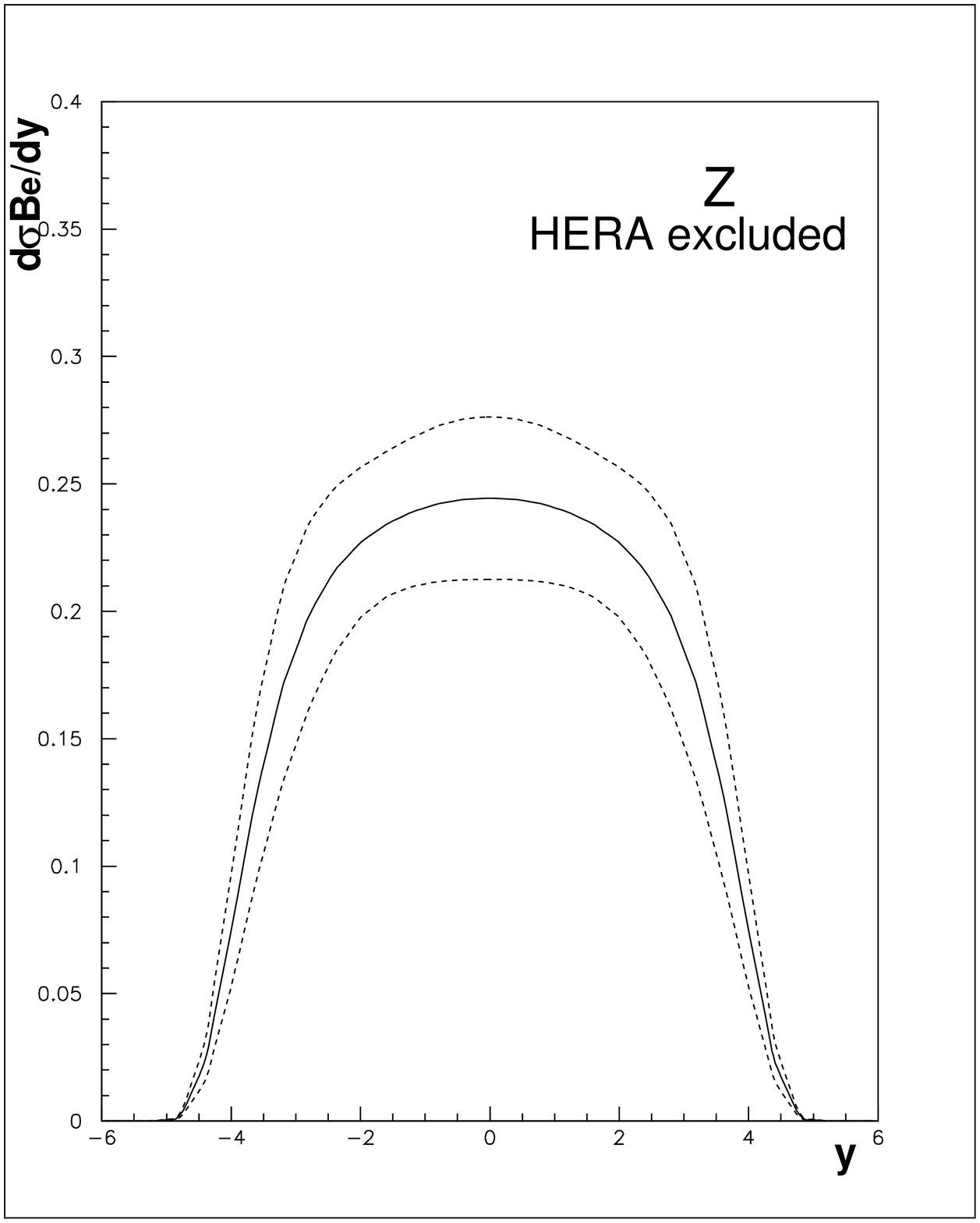,width=0.3\textwidth,height=5cm} 
}
\centerline{
\epsfig{figure=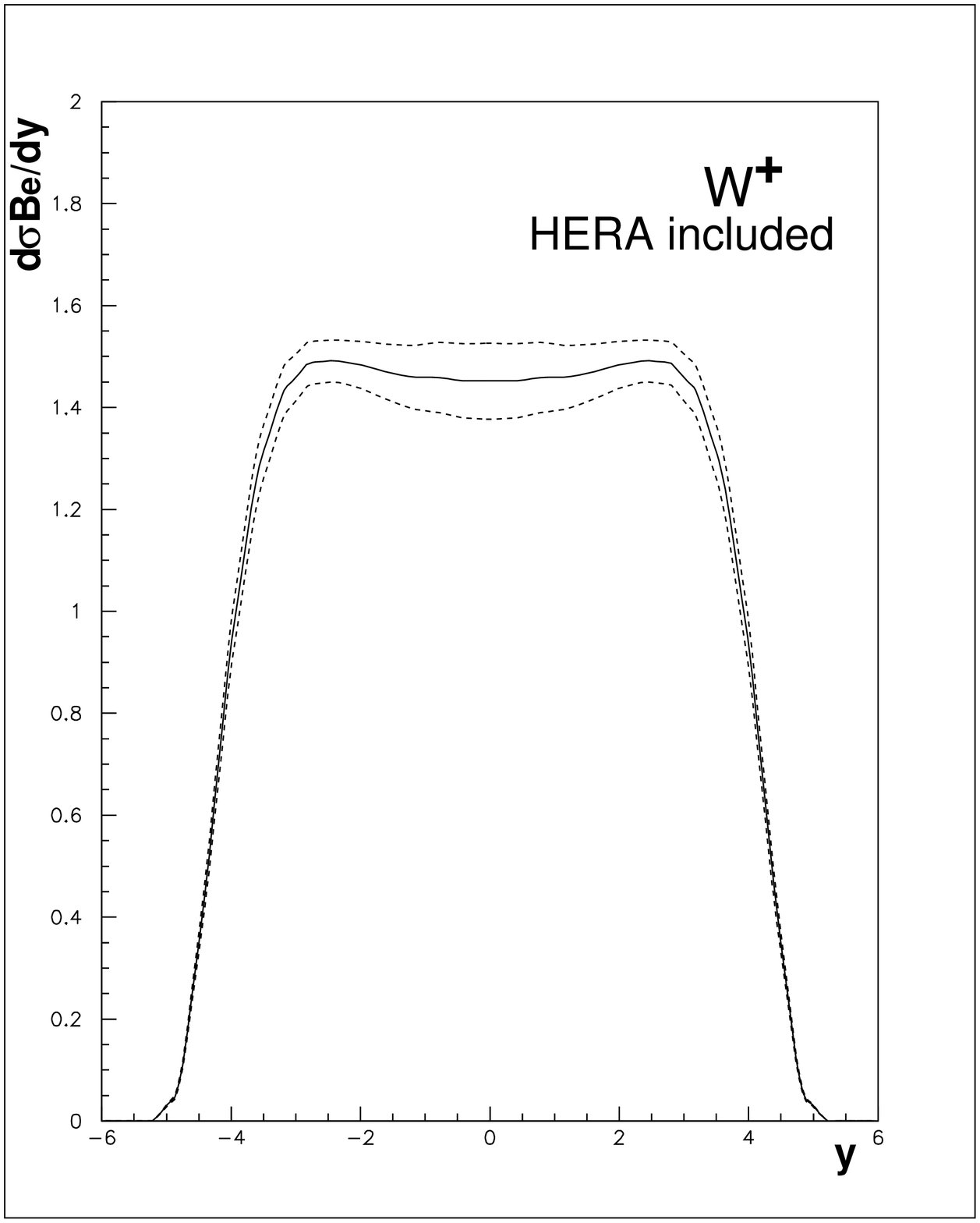,width=0.3\textwidth,height=5cm}
\epsfig{figure=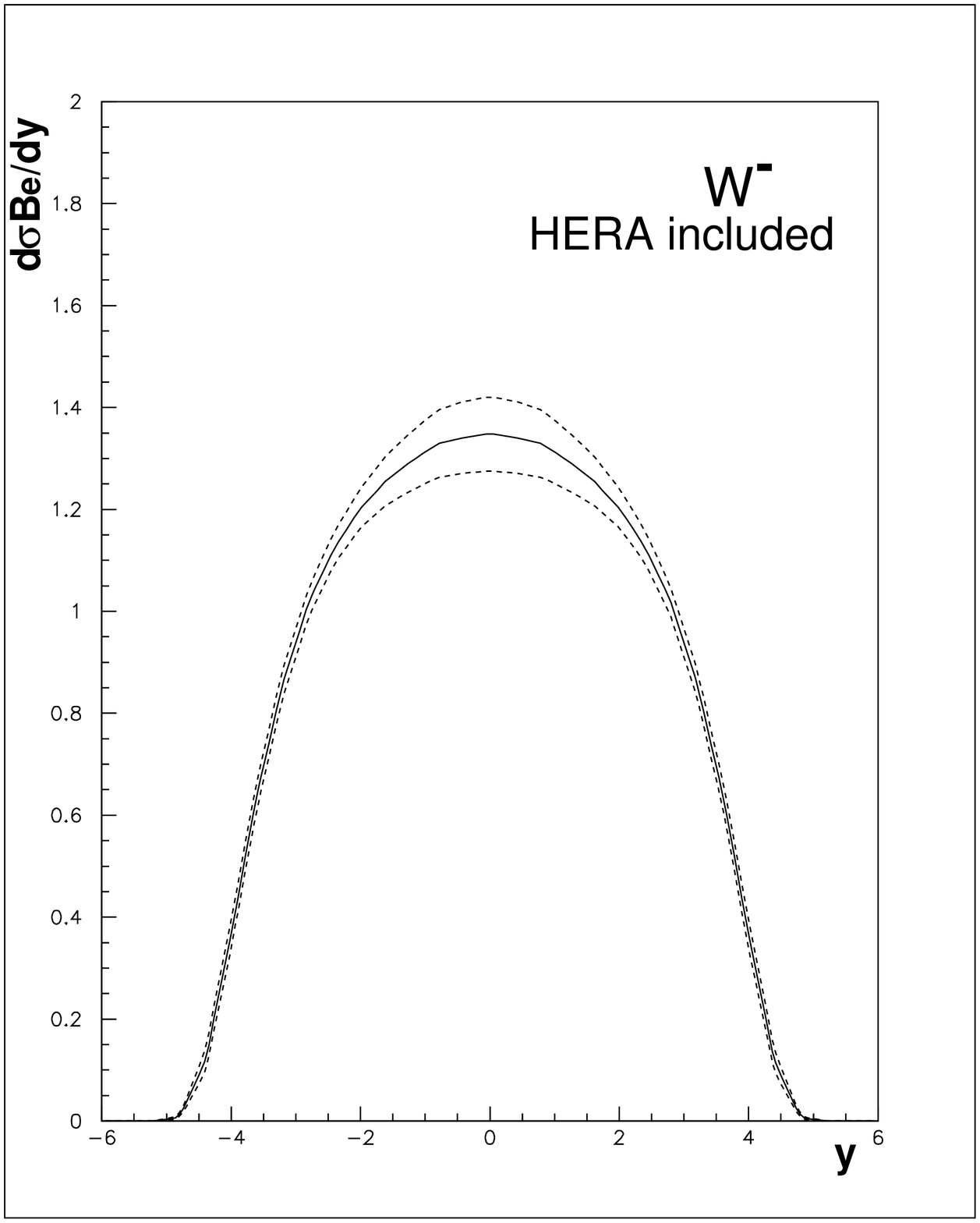,width=0.3\textwidth,height=5cm}
\epsfig{figure=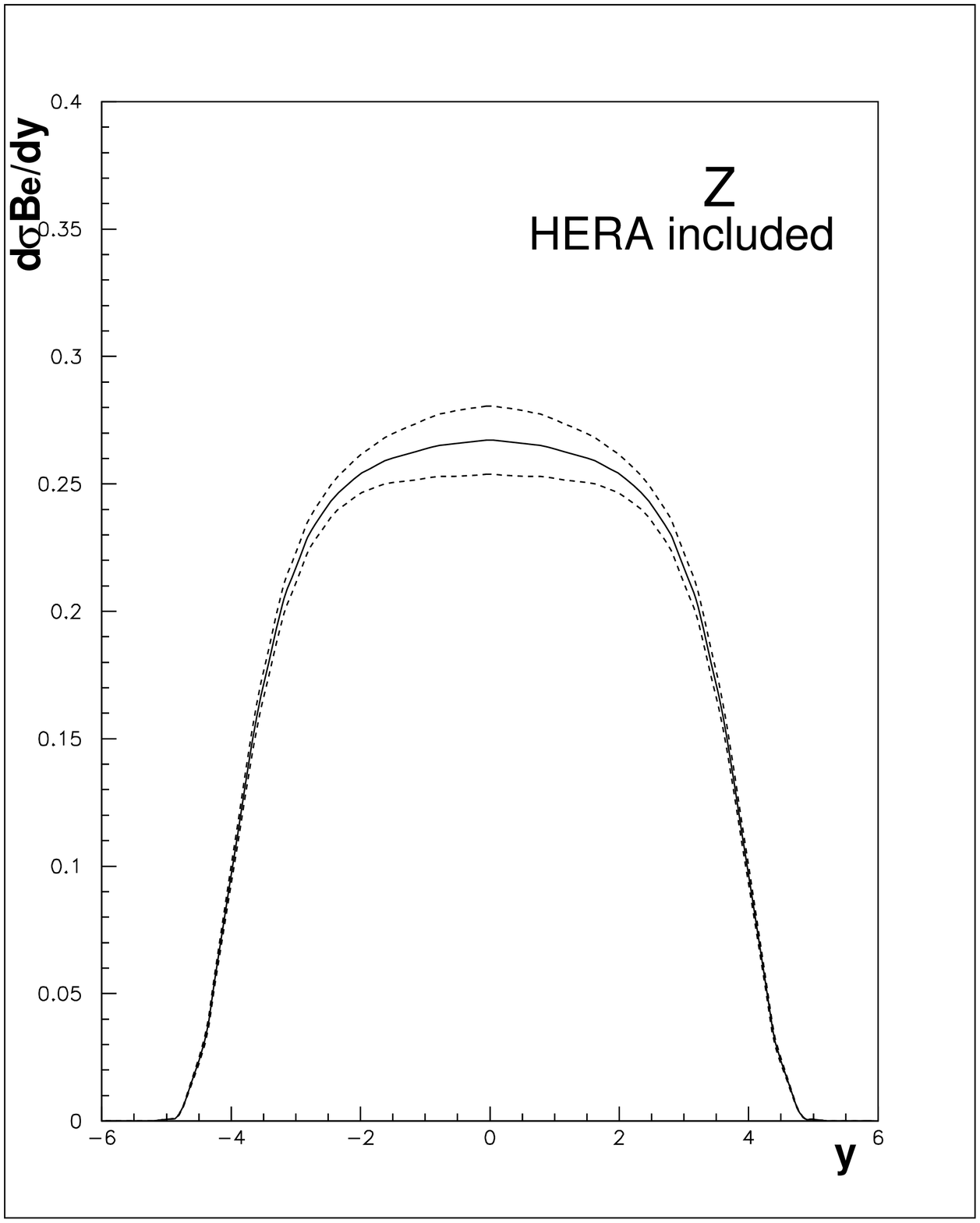,width=0.3\textwidth,height=5cm}
}
\caption {LHC $W^+,W^-,Z$ rapidity distributions and their PDF uncertainties 
(the full line shows the central value and the dashed lines show the spread of 
the uncertainty): 
Top Row: from the ZEUS-S global PDF analysis
not including HERA data; left plot $W^+$; middle plot $W^-$; right plot $Z$: Bottom Row: from the ZEUS-S
global PDF analysis including HERA data; left plot $W^+$; middle plot $W^-$; right plot $Z$}
\label{fig:WZrapFTZS13}
\end{figure}

\begin{figure}[tbp] 
\vspace{-1.0cm}
\centerline{
\epsfig{figure=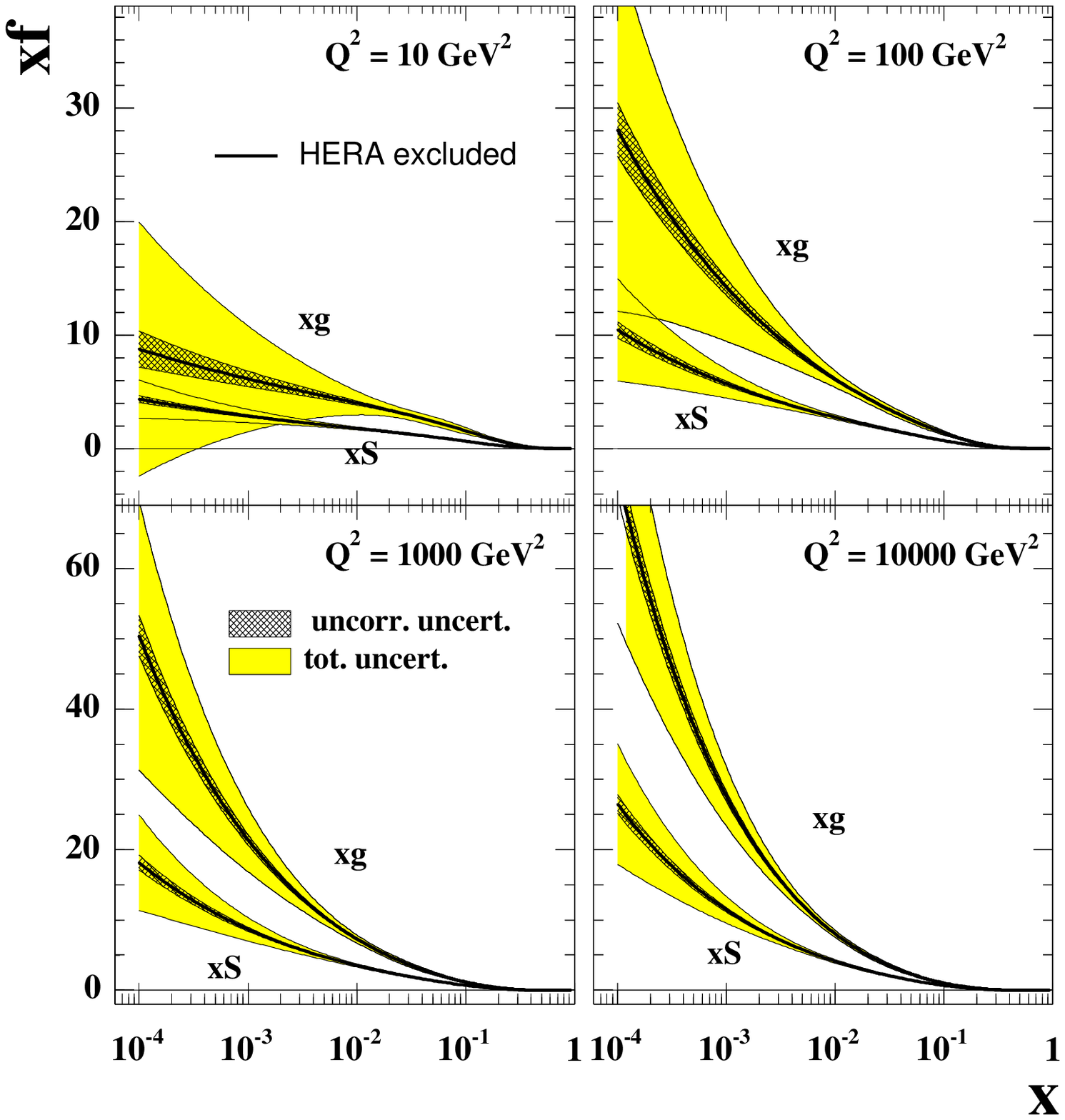,width=0.5\textwidth}
\epsfig{figure=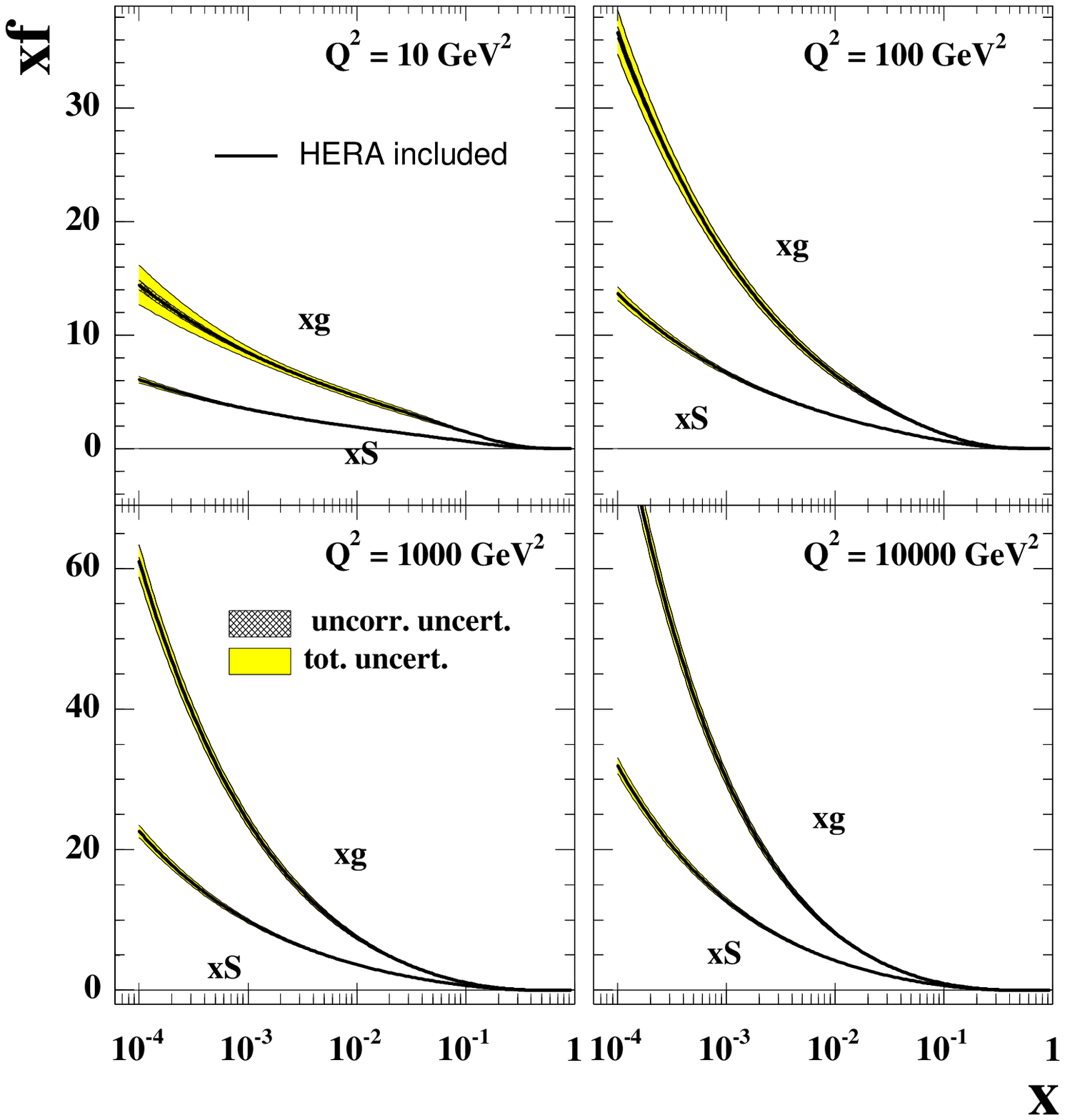,width=0.5\textwidth}}
\caption {Sea ($xS$) and gluon ($xg$) PDFs at various $Q^2$: left plot; 
from the ZEUS-S global PDF analysis
 not including HERA data; right plot: from the ZEUS-S global PDF analysis 
including HERA data. The inner cross-hatched error bands show the statistical 
and uncorrelated systematic uncertainty, the outer error bands show the total 
uncertainty including experimental correlated systematic uncertainties, 
normalisations and model uncertainty.}
\label{fig:pre/postPDFs}
\end{figure}

HERA data have dramatically improved our knowledge of the gluon, as illustrated in 
Fig.~\ref{fig:WZrapFTZS13}, 
which shows $W$ and $Z$ rapidity spectra predicted from a 
global PDF fit which does not include the HERA data, compared to a fit including HERA data. 
The latter fit is the ZEUS-S global fit~\cite{Chekanov:2002pv}, whereas the former is 
a fit using the same fitting analysis but leaving out the ZEUS data. The full 
PDF uncertainties for both fits are calculated from the error PDF sets of the ZEUS-S analysis
using LHAPDF~\cite{LHAPDF} (see the contribution of M.Whalley to these proceedings). 
The predictions for the $W/Z$ cross-sections, decaying to the lepton decay mode, are summarised in 
Table~\ref{tab:datsum}.
\begin{table}[t]
\centerline{\small
\begin{tabular}{llllcccc}\\
 \hline
PDF Set  & $\sigma(W^+).B(W^+ \rightarrow l^+\nu_l)$ & $\sigma(W^-).B(W^- \rightarrow l^-\bar{\nu}_l)$ & 
$\sigma(Z).B(Z \rightarrow l^+ l^-)$\\
 \hline
 ZEUS-S no HERA  & $10.63 \pm 1.73 $~nb & $7.80 \pm 1.18 $~nb & $1.69 \pm 0.23$~nb \\
 ZEUS-S  & $12.07 \pm 0.41 $~nb & $8.76 \pm 0.30 $~nb & $1.89 \pm 0.06$~nb\\
 CTEQ6.1 & $11.66 \pm 0.56 $~nb & $8.58 \pm 0.43 $~nb & $1.92 \pm 0.08$~nb\\
 MRST01 & $11.72 \pm 0.23 $~nb & $8.72 \pm 0.16 $~nb & $1.96 \pm 0.03$~nb\\
 \hline\\
\end{tabular}}
\caption{LHC $W/Z$ cross-sections for decay via the lepton mode, for various PDFs}
\label{tab:datsum}
\end{table}
The uncertainties 
in the predictions for these cross-sections have decreased from $\sim 16\%$ pre-HERA to $\sim 3.5\%$ 
post-HERA. The reason for this can be seen clearly in Fig.~\ref{fig:pre/postPDFs}, 
where the sea and gluon 
distributions for the pre- and post-HERA fits are shown for several different $Q^2$ bins, together 
with their uncertainty bands. It is the dramatically increased precision in the low-$x$ gluon PDF,
feeding into increased precision in the low-$x$ sea quarks, 
which has led to the increased precision on the predictions for $W/Z$ production at the LHC. 

\begin{figure}[tbp] 
\centerline{
\epsfig{figure=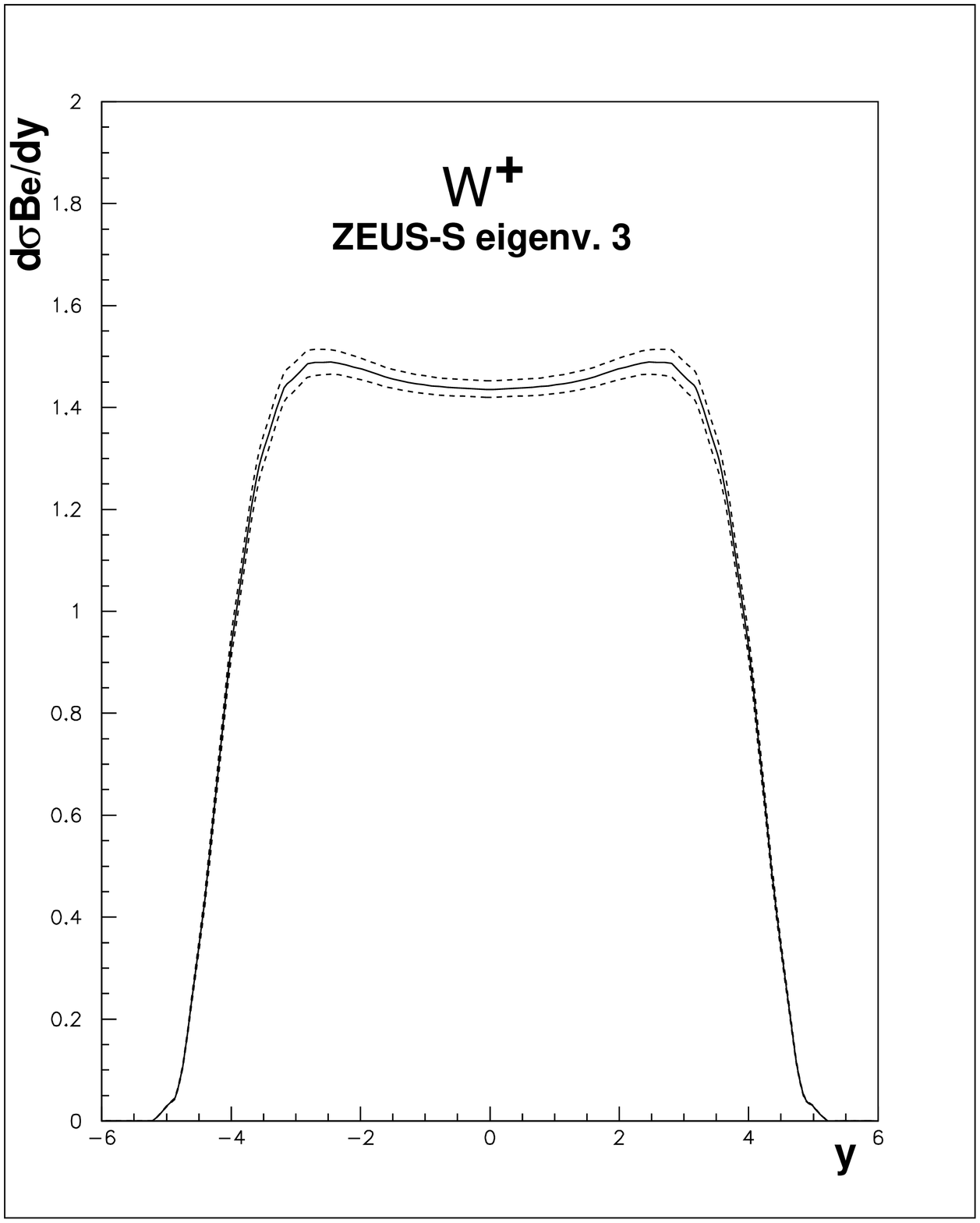,width=0.3\textwidth,height=5cm} 
\epsfig{figure=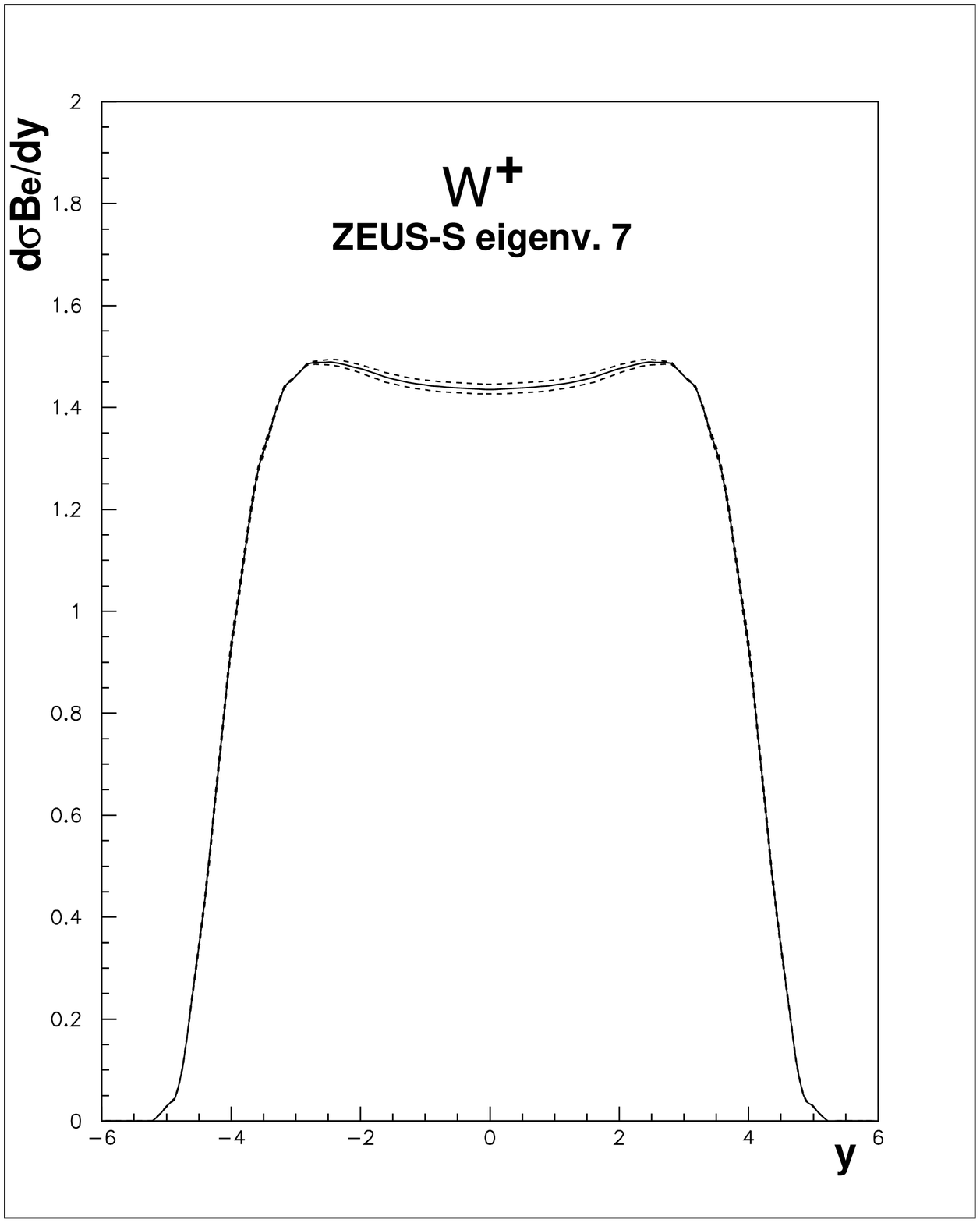,width=0.3\textwidth,height=5cm}
\epsfig{figure=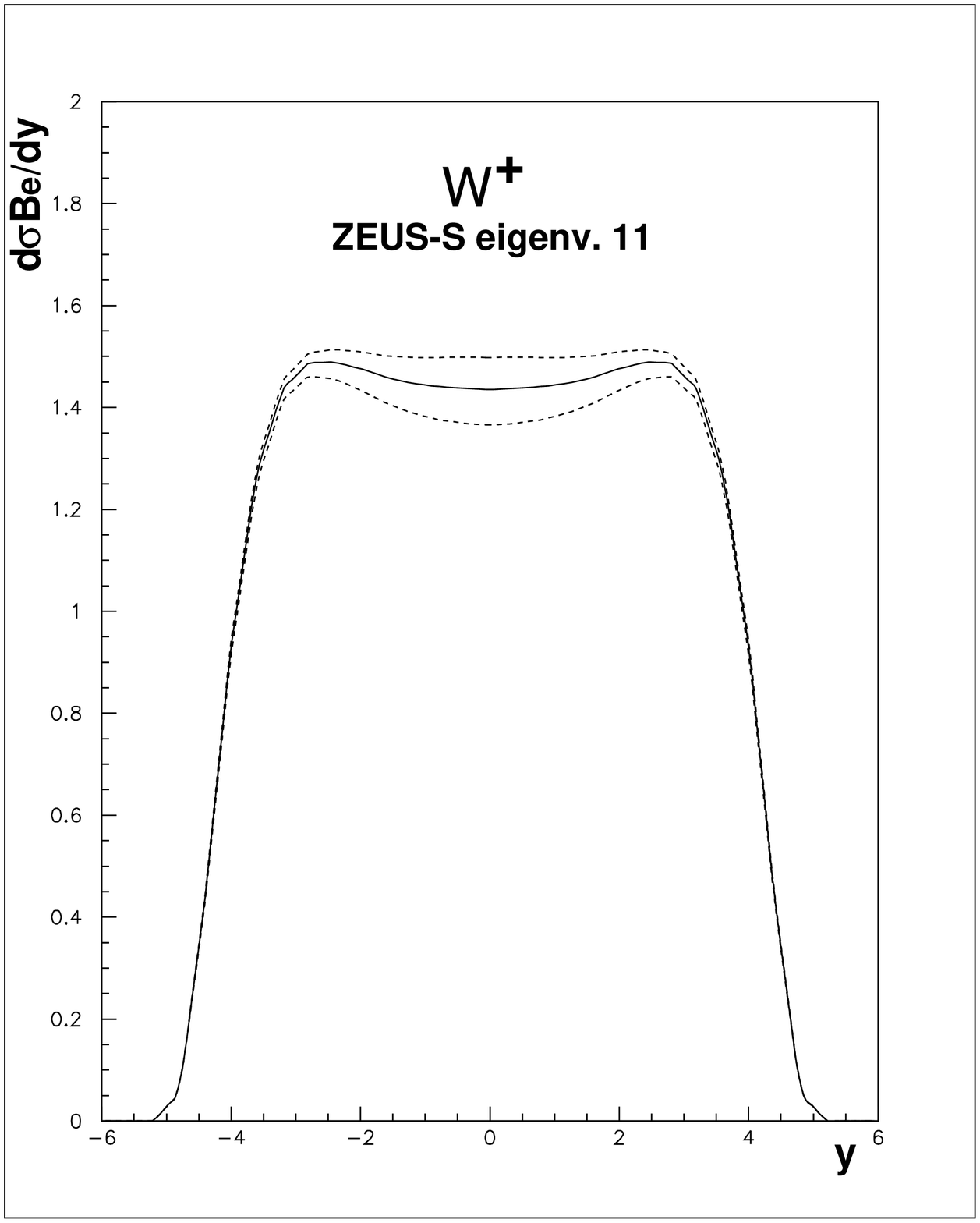,width=0.3\textwidth,height=5cm}
}
\caption {LHC $W^+$ rapidity distributions and their PDF uncertainties due to the eigenvectors 3,7 and 11 of the 
ZEUS-S analysis.}
\label{fig:eigen}
\end{figure}

Further evidence for the conclusion that the uncertainties on the gluon PDF at the input scale ($Q^2_0=7$~GeV$^2$,
 for ZEUS-S) are the major 
contributors to the uncertainty on the $W/Z$ cross-sections at $Q^2= M_W(M_Z)$, comes from decomposing the 
predictions down into their contributing eigenvectors. Fig~\ref{fig:eigen} shows the dominant contributions to the total 
uncertainty from eigenvectors 3, 7, and 11 which are eigenvectors which are
dominated by the parameters which control the low-$x$, mid-$x$ and high-$x$, gluon respectively. 

\begin{figure}[tbp] 
\centerline{
\epsfig{figure=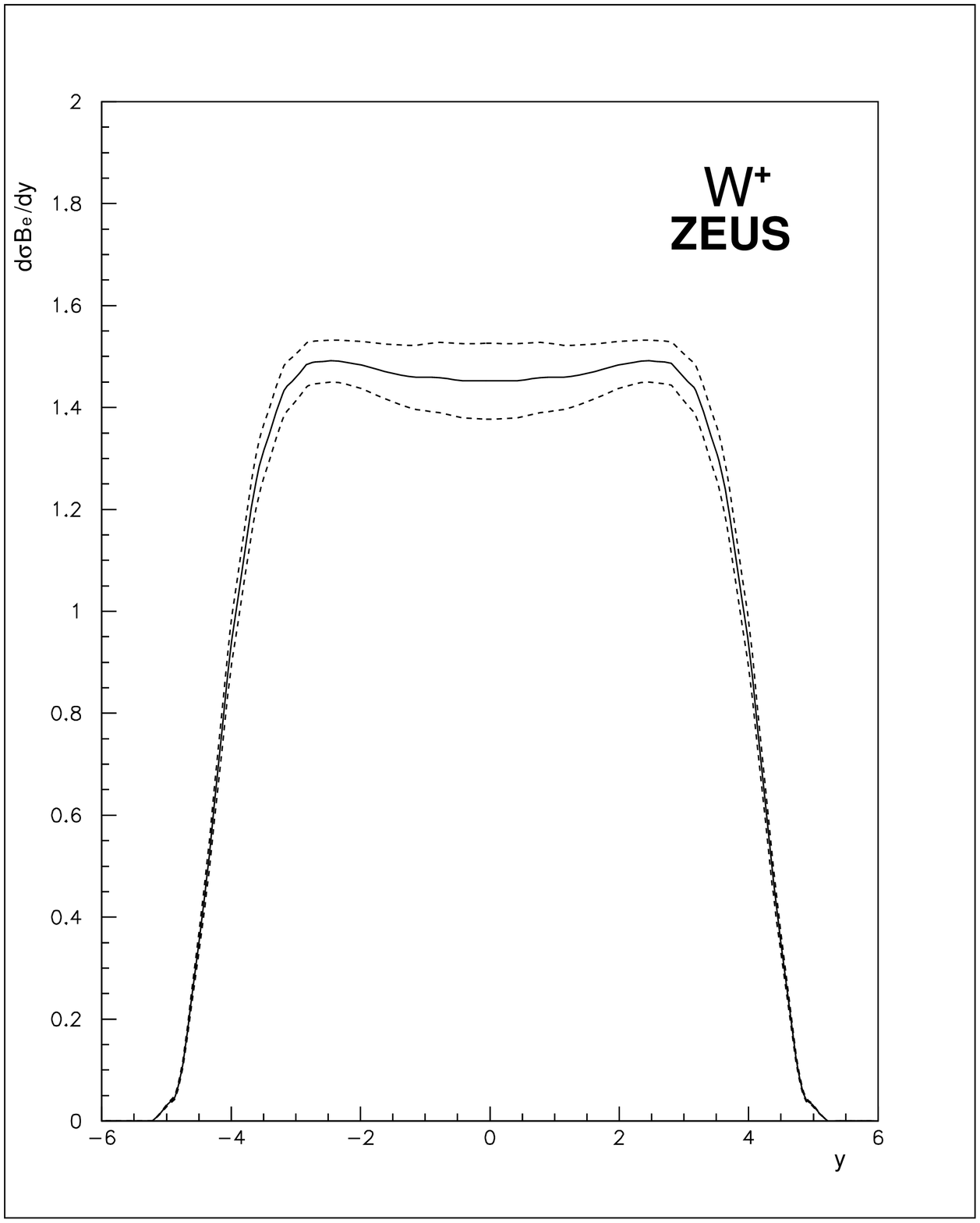,width=0.3\textwidth,height=4cm}
\epsfig{figure=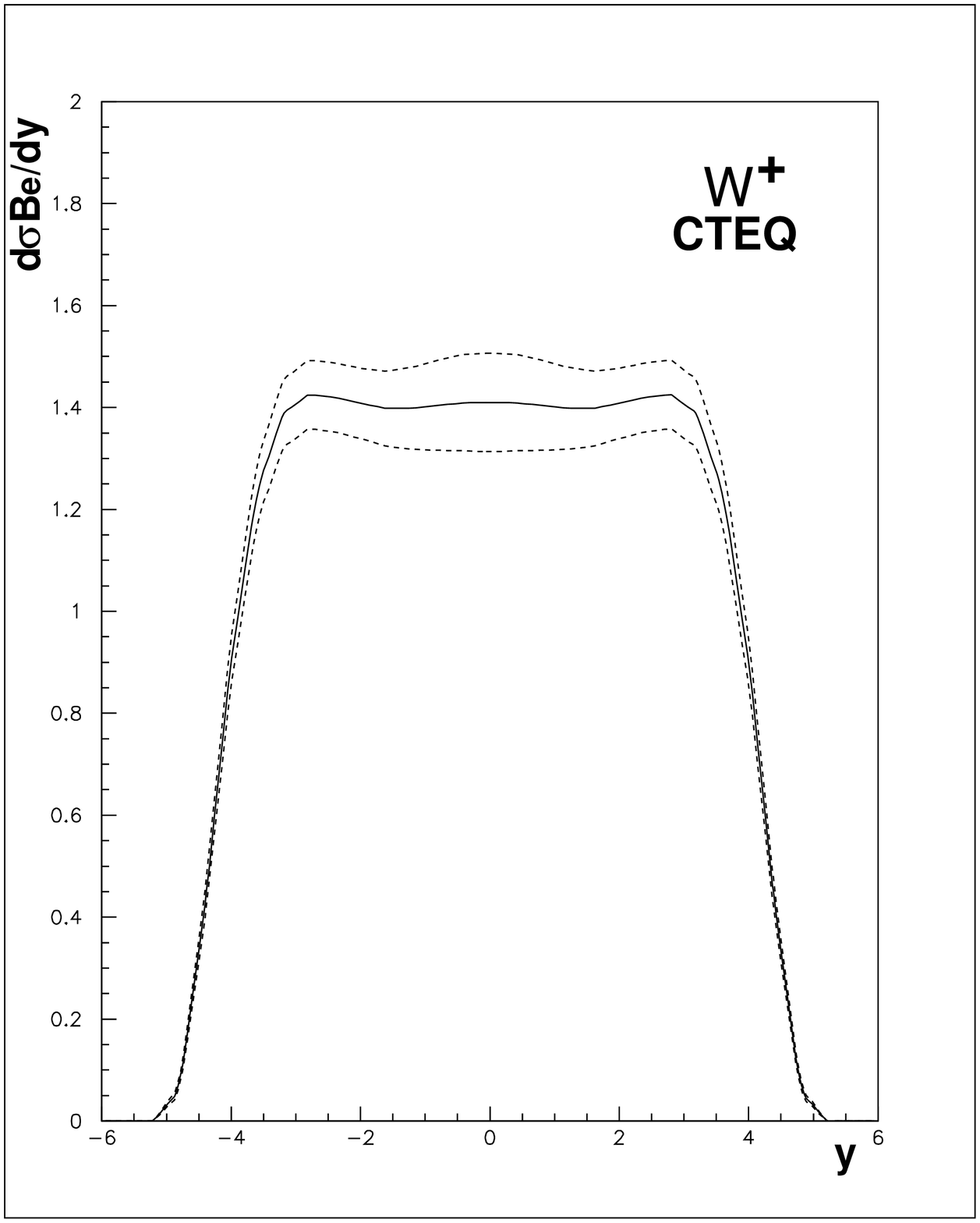,width=0.3\textwidth,height=4cm}
\epsfig{figure=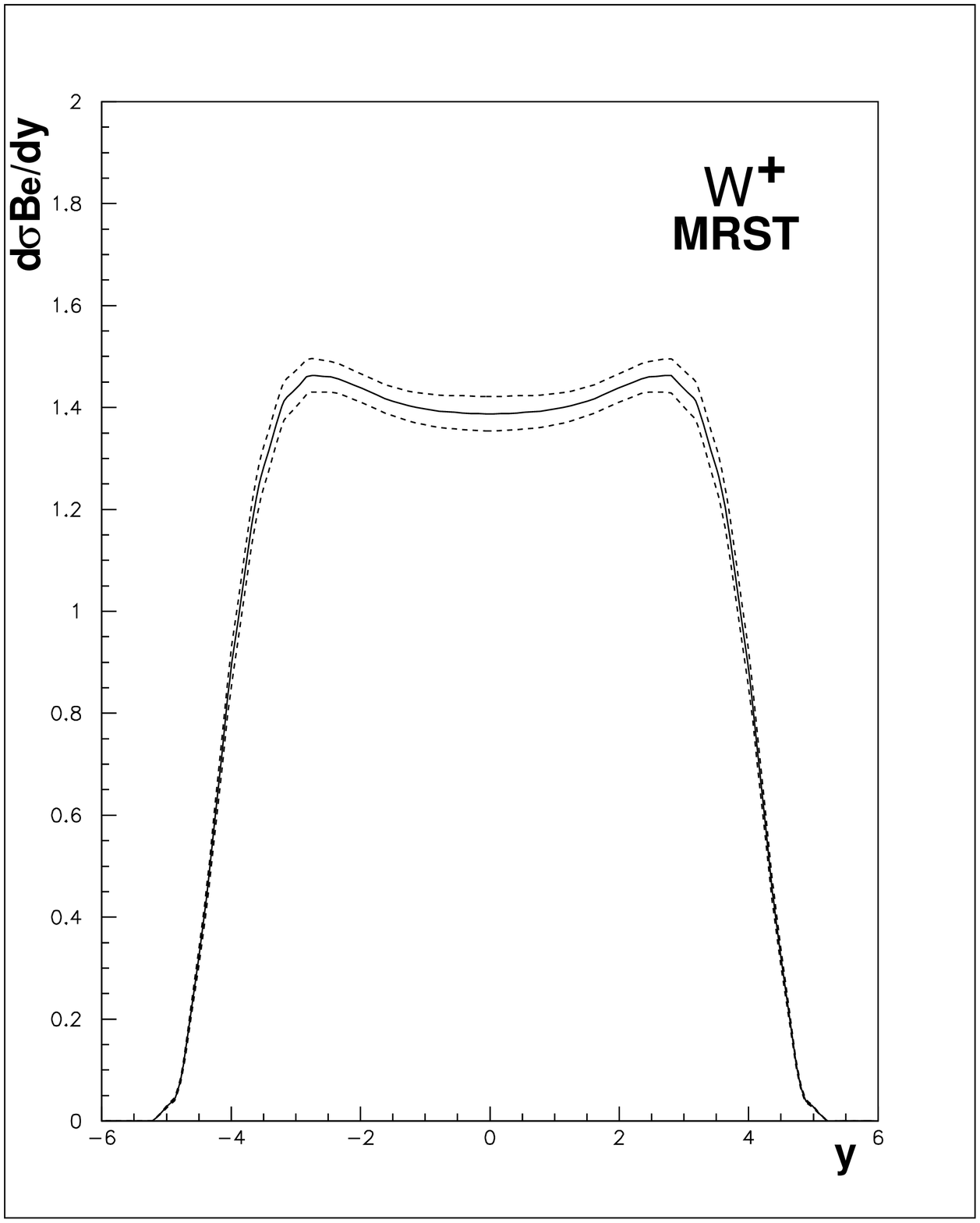,width=0.3\textwidth,height=4cm}
}
\caption {LHC $W^+$ rapidity distributions and their PDF uncertainties:
 left plot, ZEUS-S PDFs; middle plot, CTEQ6.1 PDFs;
right plot: MRST01 PDFs.}
\label{fig:mrstcteq}
\end{figure}

The post-HERA level of precision illustrated in Fig.~\ref{fig:WZrapFTZS13} 
is taken for granted in modern analyses, such that $W/Z$ production have 
been suggested as `standard-candle' processes for luminosity measurement. However, when 
considering the PDF uncertainties on the Standard Model (SM) predictions it is necessary not 
only to consider the uncertainties of a particular PDF analysis, but also to compare PDF 
analyses. Fig.~\ref{fig:mrstcteq} compares the predictions for $W^+$ production for the ZEUS-S PDFs 
with those of 
the CTEQ6.1\cite{cteq} PDFs and the MRST01\cite{mrst} PDFs\footnote{MRST01 PDFs are used because the 
full error analysis is available only for this PDF set.}. 
The corresponding $W^+$ cross-sections, for decay to leptonic mode are 
given in Table~\ref{tab:datsum}.
Comparing the uncertainty at central rapidity, rather 
than the total cross-section, we see that the uncertainty estimates are rather larger: $~5.2\%$ for ZEUS-S; 
$~8.7\%$ 
for CTEQ6.1M and about $~3.6\%$ for MRST01. The difference in the central value between 
ZEUS-S and CTEQ6.1 is $~3.5\%$. Thus the spread in the predictions of the different PDF sets is 
comparable to the uncertainty estimated by the individual analyses. Taking all of these 
analyses together the uncertainty at central rapidity is about $~8\%$. 

\begin{figure}[tbp] 
\vspace{-2.0cm}
\centerline{
\epsfig{figure=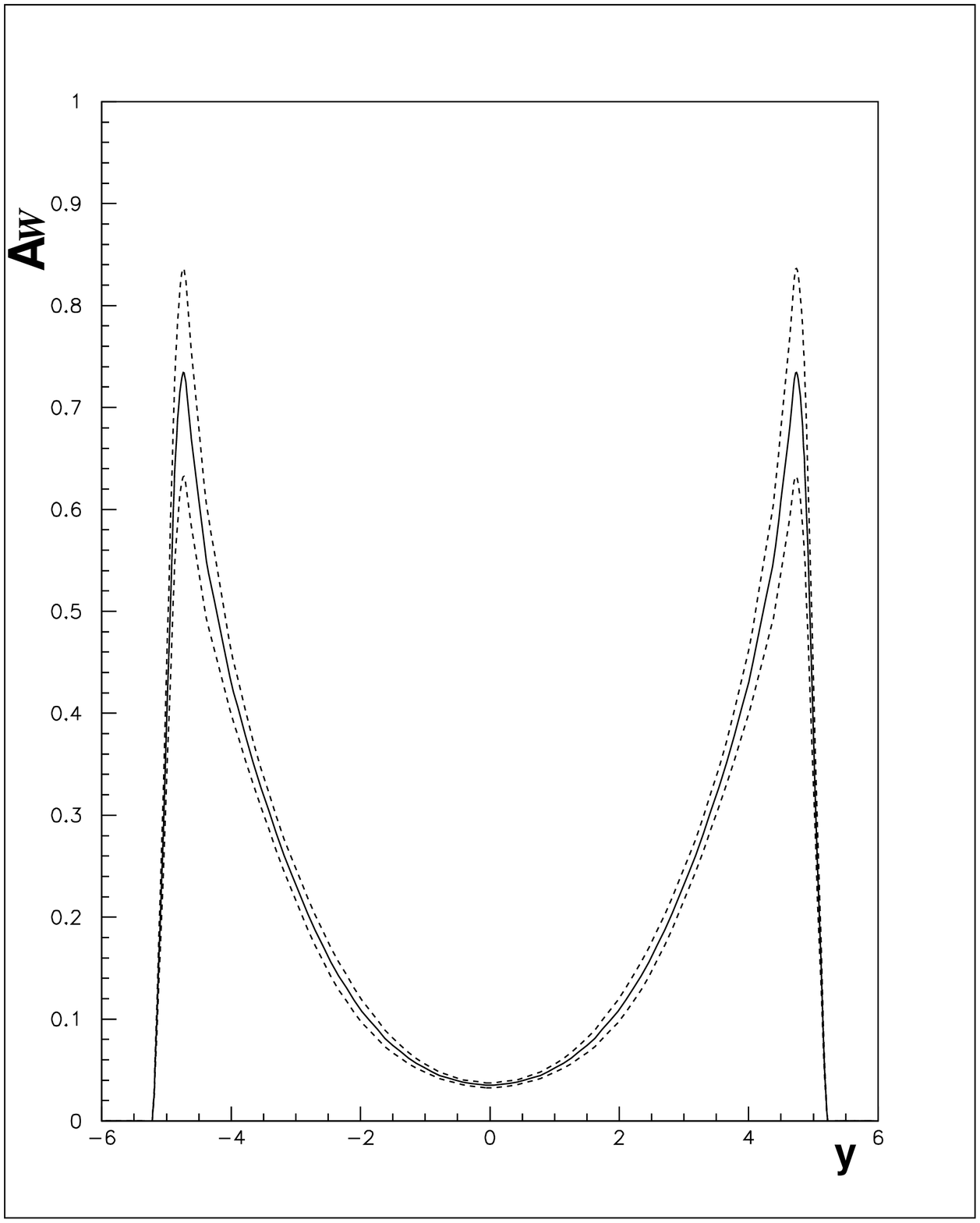,width=0.3\textwidth,height=4cm}
\epsfig{figure=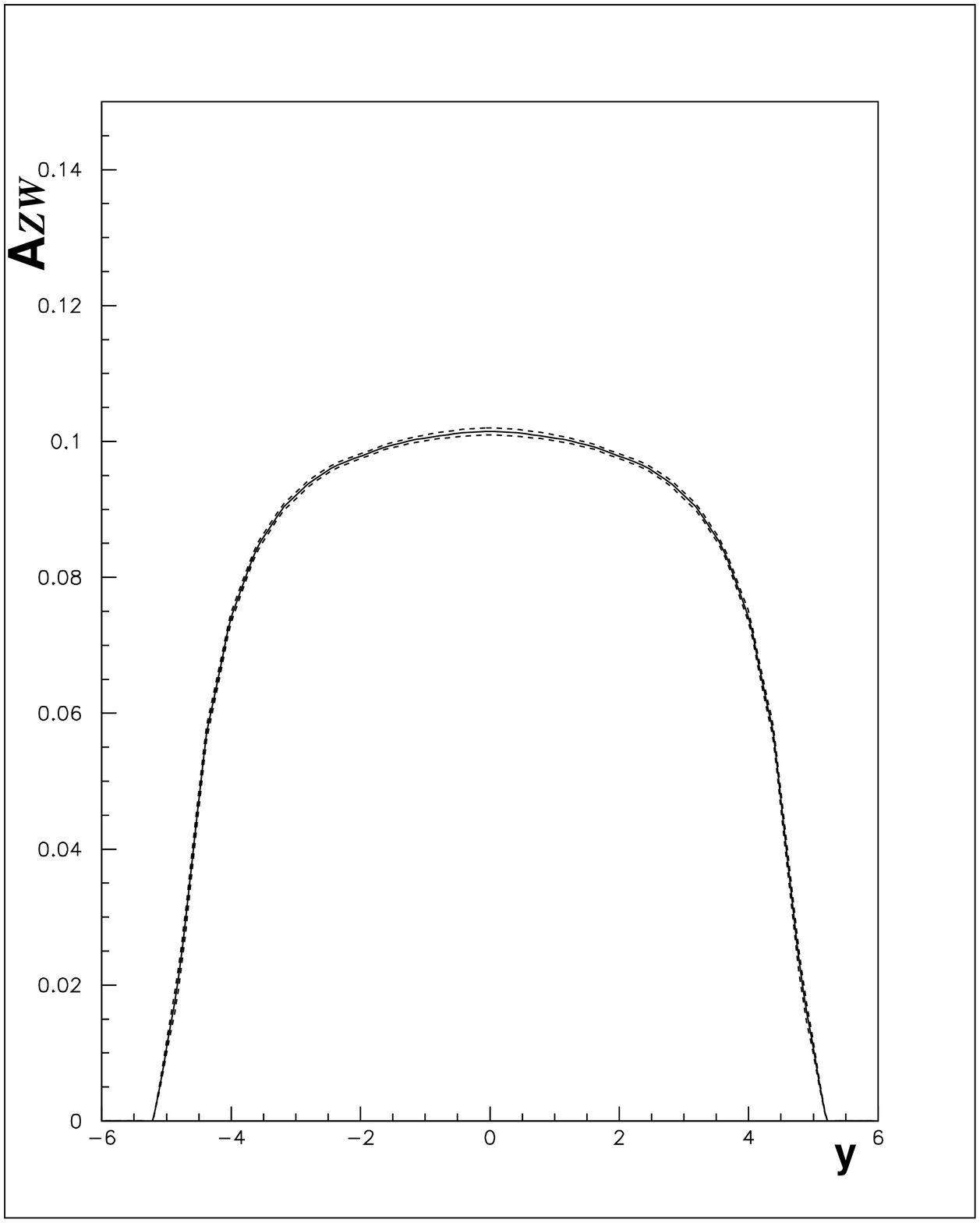,width=0.3\textwidth,height=4cm}
}
\centerline{
\epsfig{figure=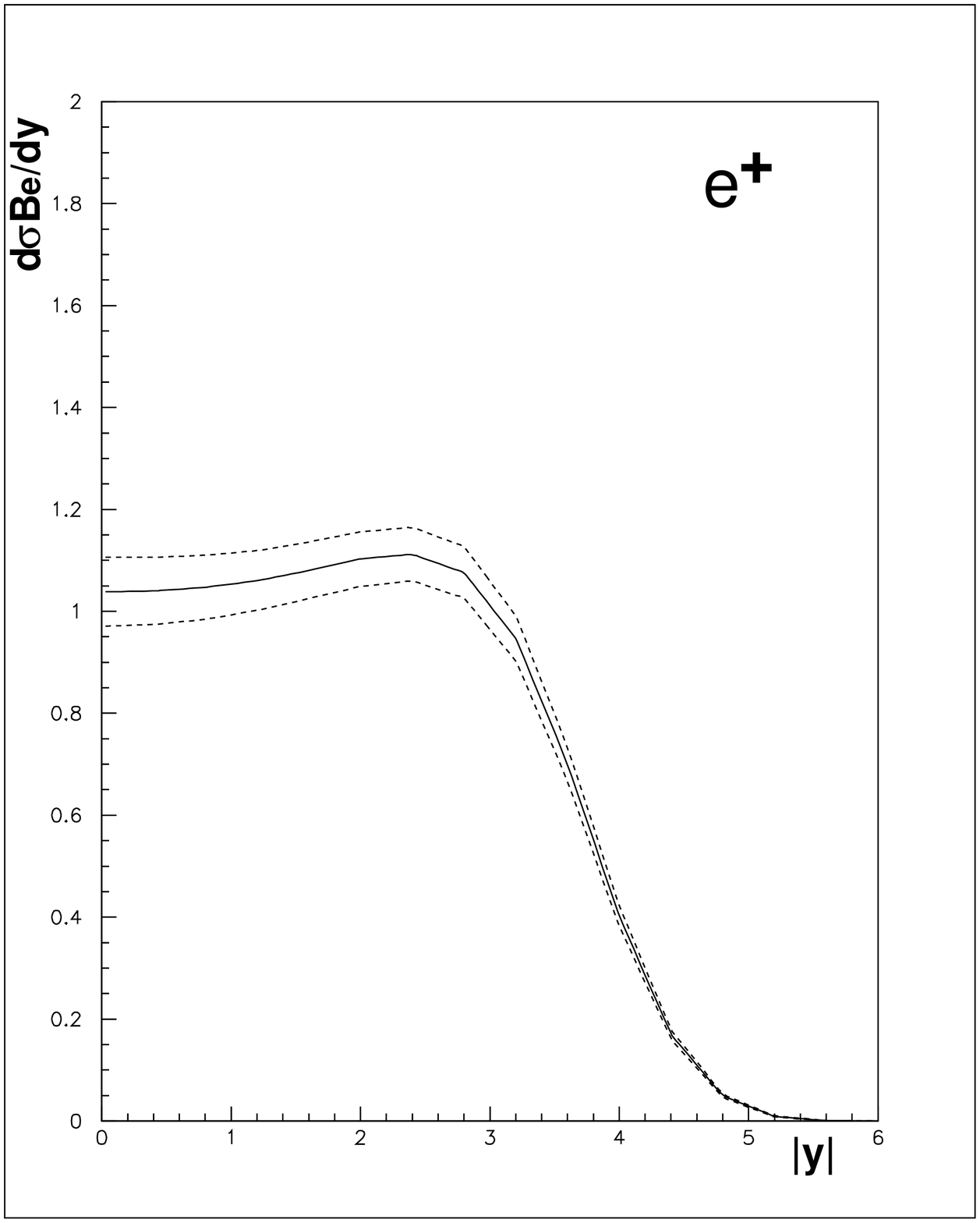,width=0.3\textwidth,height=4cm}
\epsfig{figure=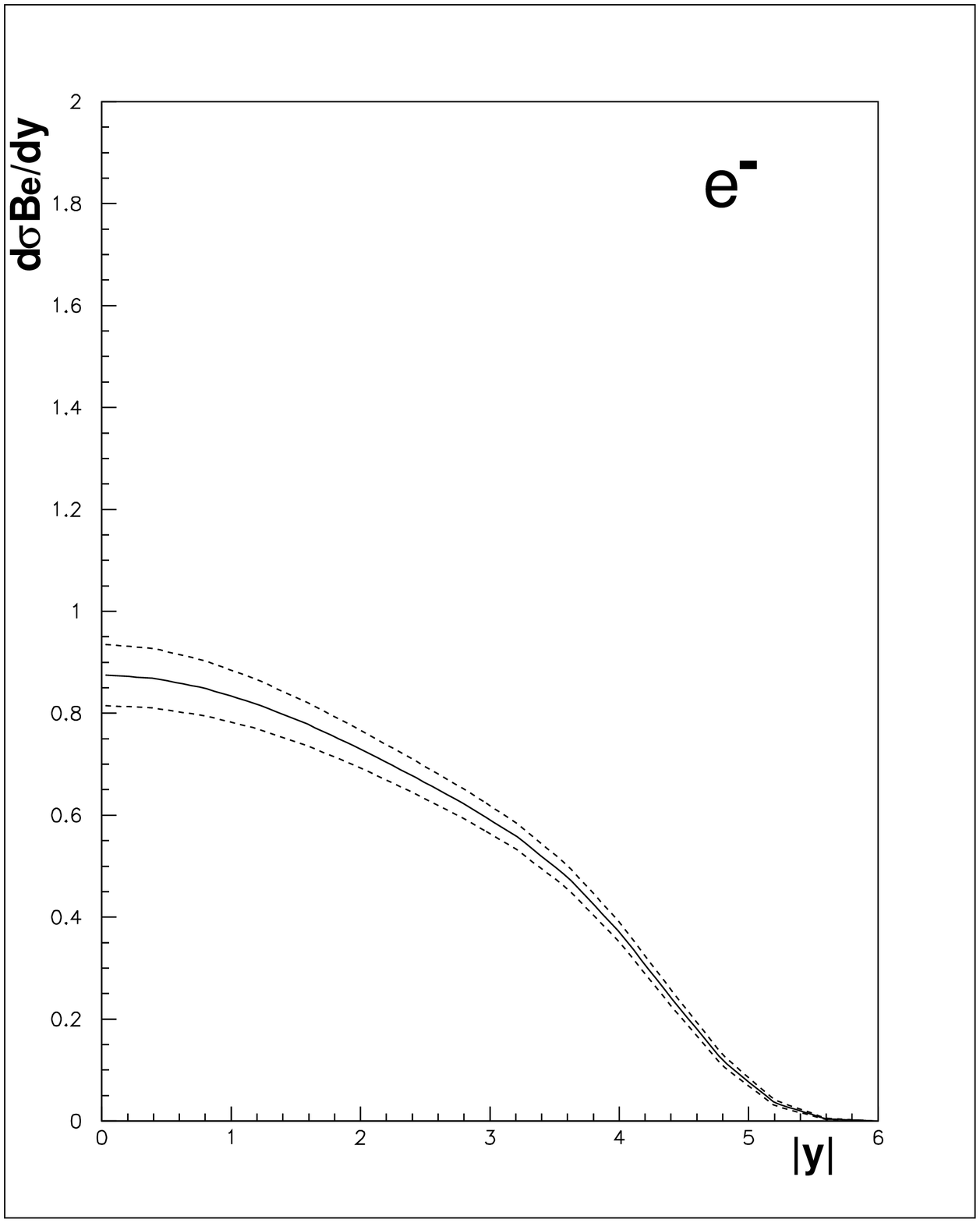,width=0.3\textwidth,height=4cm}
\epsfig{figure=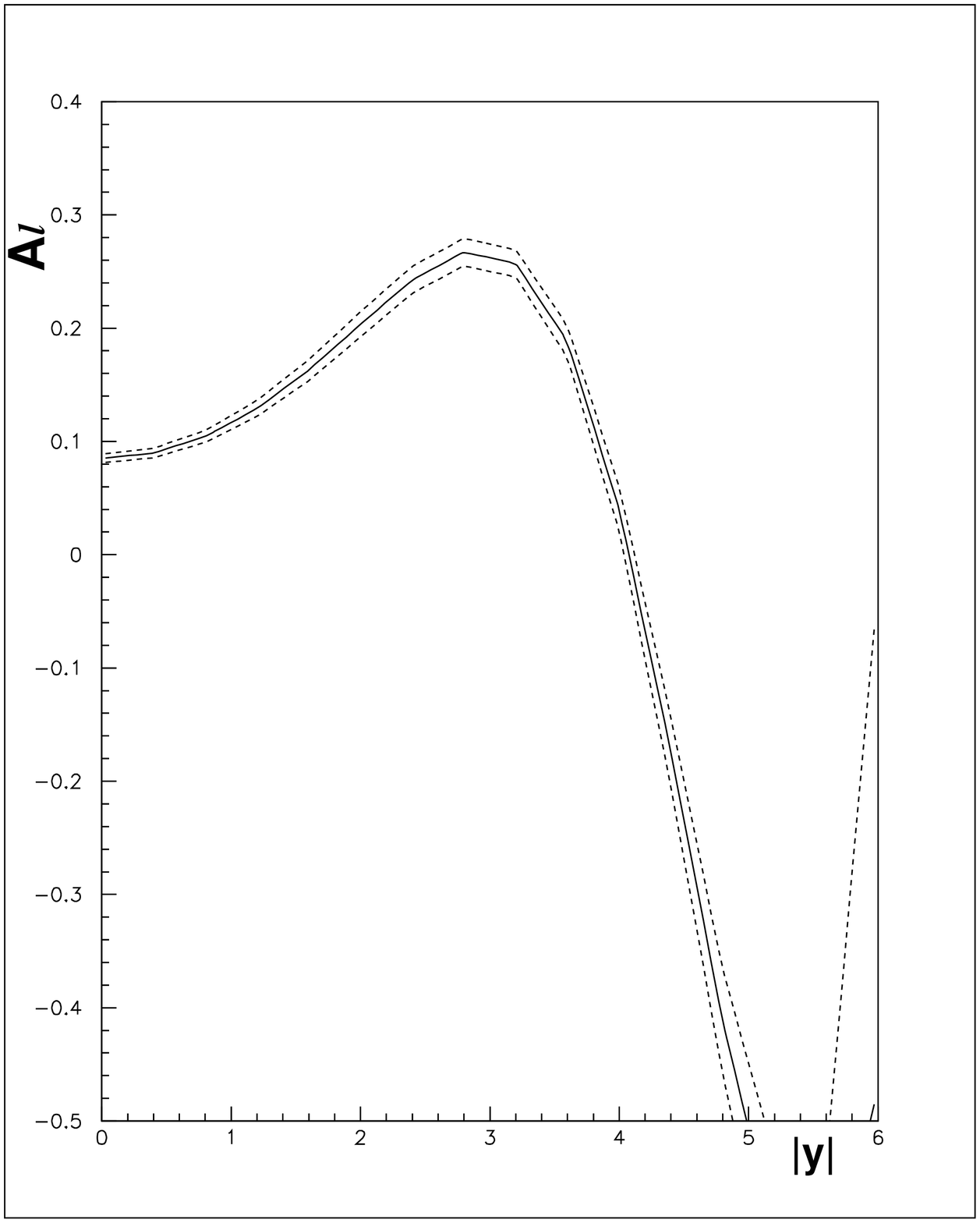,width=0.3\textwidth,height=4cm}
}
\caption {Predictions for $W,Z$ production at the LHC from the CTEQ6.1 PDFs. Top row: left plot, 
the $W$ asymmetry, $A_W$; 
right plot, the ratio, $A_{ZW}$: Bottom row: left plot, 
decay $e^+$ rapidity spectrum; middle plot, decay $e^-$ rapidity spectrum; right plot, 
lepton asymmetry, $A_e$}
\label{fig:awzwlepton}
\end{figure}

Since the PDF uncertainty feeding into the $W^+, W^-$ and $Z$ production is mostly coming from the
gluon PDF, for all three processes, there is a strong correlation in their uncertainties, which can be 
removed by taking ratios. Fig.~\ref{fig:awzwlepton} shows the $W$ asymmetry 
\[A_W = (W^+ - W^-)/(W^+ + W^-).\] for CTEQ6.1 PDFs, which have the largest uncertainties of published 
PDF sets. The PDF uncertainties on the asymmetry are very small in the measurable rapidity range. 
An eigenvector decomposition indicates that 
sensitivity to high-$x$ $u$ and $d$ quark flavour distributions is now evident at large $y$. 
Even this residual flavour 
sensitivity can be removed by taking the ratio \[A_{ZW} = Z/(W^+ +W^-)\] as also shown in 
Fig.~\ref{fig:awzwlepton}. 
This quantity is almost independent of PDF uncertainties. These quantities have been suggested as 
benchmarks for our understanding of Standard Model Physics at the LHC. However, 
whereas the $Z$ rapidity distribution can be fully reconstructed from its decay leptons, 
this is not possible for the $W$ rapidity distribution, because the leptonic decay channels 
which we use to identify the $W$'s have missing neutrinos. Thus we actually measure the $W$'s 
decay lepton rapidity spectra rather than the $W$ rapidity spectra. The lower half of  Fig.~\ref{fig:awzwlepton} 
shows the rapidity spectra for positive and 
negative leptons from $W^+$ and $W^-$ decay and the lepton asymmetry, \[A_l = (l^+ - l^-)/(l^+ + l^-).\] 
A cut of, $p_{tl} > 25$~GeV, has been applied on the decay lepton, since it will not be possible to
trigger on 
leptons with small $p_{tl}$. A particular lepton rapidity can be fed from a range 
of $W$ rapidities so that the contributions of partons at different $x$ values is smeared out 
in the lepton spectra, but the broad features of the $W$ spectra and the sensitivity to the gluon parameters 
remain. The lepton asymmetry shows the change of sign at large $y$ which is characteristic of the $V-A$ 
structure of the lepton decay. The cancellation of the 
uncertainties due to the gluon PDF is not so 
perfect in the lepton asymmetry as in the $W$ asymmetry. Nevertheless in the 
measurable rapidity range sensitivity to PDF parameters is small. Correspondingly, the PDF uncertainties are 
also small ($~4\%$) and this quantity provides a suitable Standard Model benchmark. 

In summary, these preliminary investigations indicate that PDF uncertainties on 
predictions for the $W,Z$ rapidity 
spectra, using standard PDF sets which describe all modern data, have 
reached a precision of $\sim 8\%$. This may be good enough to consider 
using these processes as luminosity monitors. The predicted precision on ratios such as the 
lepton ratio, $A_l$,  is better ($\sim 4\%$) and this measurement may be used as a SM benchmark. 
It is likely that this current level of uncertainty will have improved before the LHC turns on-
see the contribution of C. Gwenlan (section~\ref{sec:prec})
 to these proceedings. 
The remainder of this contribution will be concerned with the question: how accurately can we  
measure these quantities and can we use the early LHC data 
to improve on the current level of uncertainty?

\subsubsection{k-factor and PDF re-weighting} 
To investigate how well we can really measure $W$ production we need to generate samples 
of Monte-Carlo (MC) data and pass them through a simulation of a detector. 
Various technical problems 
arise. Firstly, many physics studies are done with HERWIG (6.505)\cite{Corcella:2000bw}, 
which generates events at LO with parton showers to account for higher order effects. 
Distributions can be corrected from LO to NLO by k-factors which are applied as a function of the 
variable of interest. The use of HERWIG is gradually being superceded by MC@NLO (2.3)\cite{MCatNLO} 
but this is 
not yet implemented for all physics processes. Thus it is necessary to 
investigate how much bias is introduced by using HERWIG with k-factors. Secondly, to simulate the 
spread of current PDF uncertainties, it is necessary to run the MC with all of the eigenvector 
error sets of the PDF of interest. This would be unreasonably time-consuming. Thus the technique 
of PDF reweighting has been investigated.

\begin{figure}[tbp] 
\vspace{-1.0cm}
\centerline{
\epsfig{figure=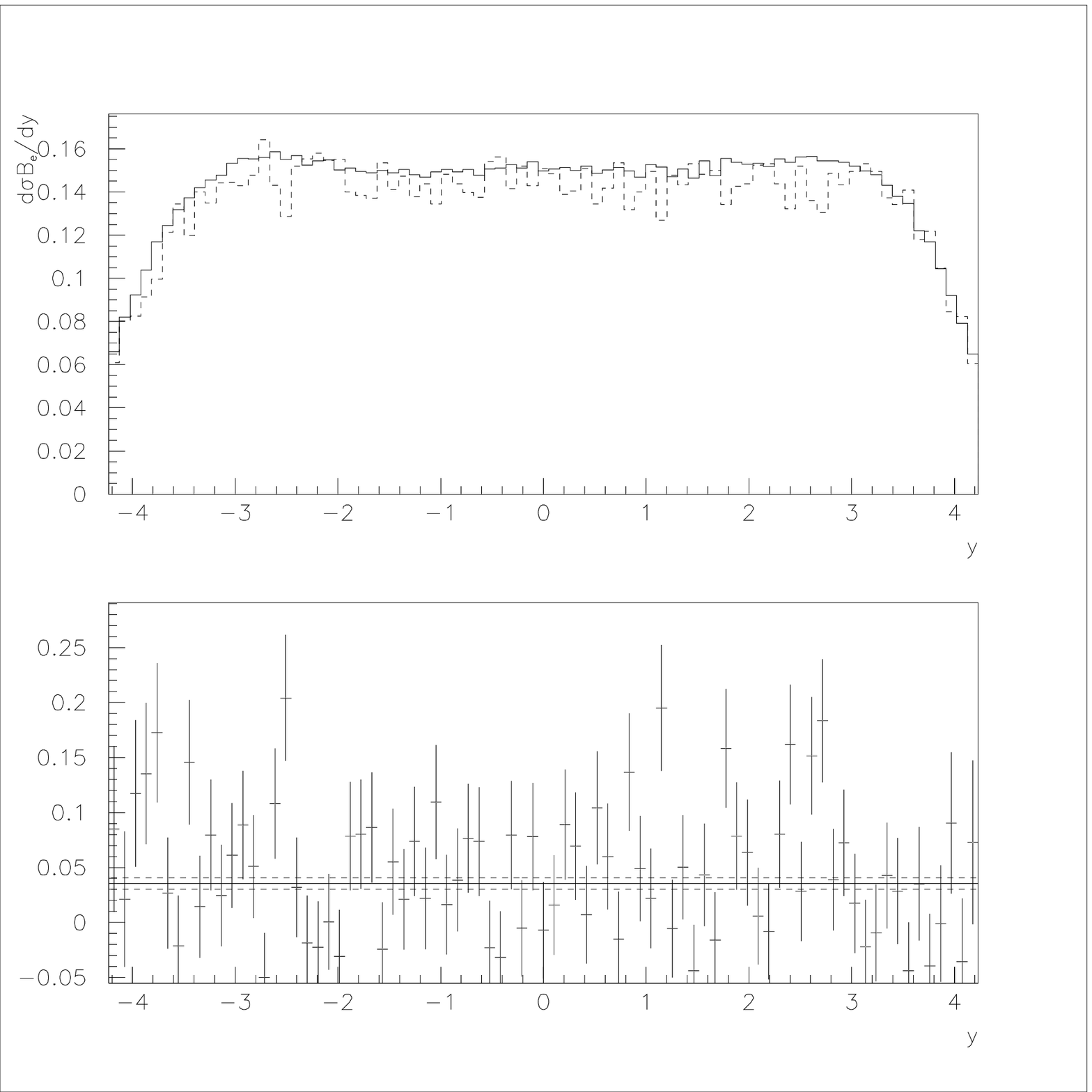,width=0.3\textwidth}
\epsfig{figure=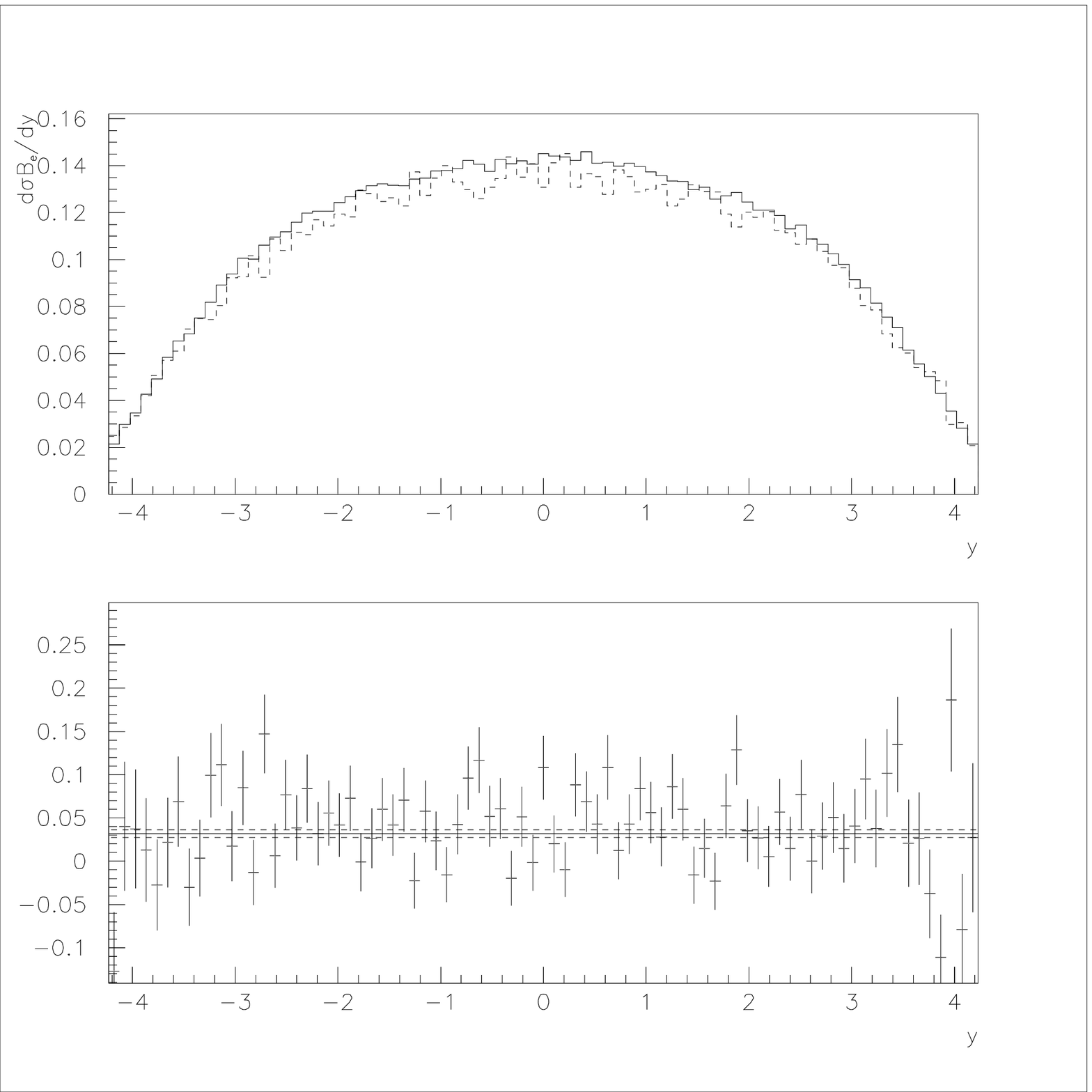,width=0.3\textwidth}
\epsfig{figure=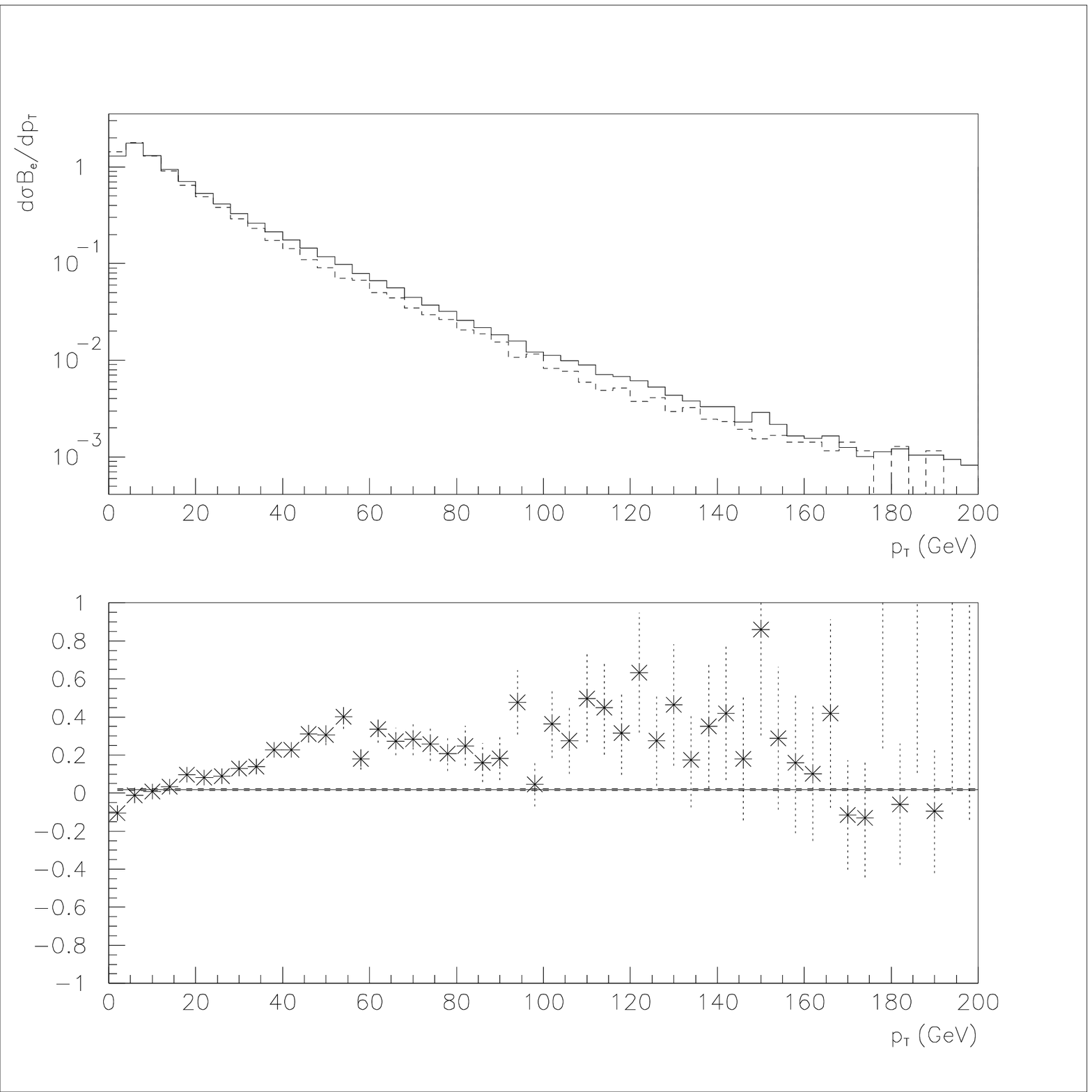,width=0.3\textwidth}
}
\caption {Top Row: $W$ rapidity and $p_t$ spectra for events generated with HERWIG + k-Factors (full line), 
compared to those generated by MC@NLO (dashed line); left plot $W^+$ rapidity; middle plot $W^-$ rapidity; 
right plot $W^-$ $p_t$. Bottom row:
the fractional differences of the spectra generated by HERWIG + k-factors and those generated by MC@NLO. 
The full line represents the weighted mean of these difference spectra and the dashed lines show 
its uncertainty }
\label{fig:kfactor}
\end{figure}

One million $W \to e \nu_e$ events were generated using HERWIG (6.505). This corresponds to 43 hours of LHC running at low 
luminosity, $10 fb^{-1}$. The events are split into $W^+$ and $W^-$ events according to 
their Standard Model 
cross-section rates, $58\%$: $42\%$ (the exact split depends on the input PDFs). These events are then 
weighted with k-factors, which are analytically calculated as the ratio of the NLO to LO 
cross-section as a function of rapidity for the same input PDF~\cite{Stirling}. 
The resultant rapidity 
spectra for $W^+, W^-$ are compared to rapidity spectra for $\sim 107,700$ events 
generated using MC@NLO(2.3) in Fig~\ref{fig:kfactor}\footnote{In MC@NLO the hard emissions are treated 
by NLO 
computations, whereas soft/collinear emissions are handled by the MC simulation. In the matching procedure a 
fraction of events with negative weights is generated to avoid double counting. The event  weights must be 
applied to the generated number of events 
before the effective number of events can be converted to an equivalent luminosity. The figure given is the 
effective number of events.}. The MRST02 PDFs were used for this investigation. 
The accuracy of this study is limited by the statistics 
of the MC@NLO generation. Nevertheless it is clear that HERWIG with k-factors does a good job of mimicking 
the NLO rapidity spectra. However, the normalisation is too high by $3.5\%$. This is not suprising since, 
unlike the analytic code, HERWIG is not a purely LO calculation, parton showering is also included. 
This normalisation difference is not too crucial since
in an analysis on real data the MC will only be used to correct data from the detector level to the 
generator level. For this purpose, 
it is essential to model the shape of spectra to understand the effect of experimental cuts and smearing
but not essential to model the overall normalisation perfectly. However, one should note that 
HERWIG with k-factors is not so successful in modelling the shape of the $p_t$ spectra, as shown in the 
right hand plot of
Fig.~\ref{fig:kfactor}. This is hardly surprising, since at LO the $W$ have no $p_t$ and non-zero $p_t$ 
for HERWIG is generated by parton showering, whereas for MC@NLO non-zero $p_t$ originates from additional higher order processes which cannot be scaled from LO, where they are not present. 

\begin{figure}[tbp] 
\vspace{-1.0cm}
\centerline{
\epsfig{figure=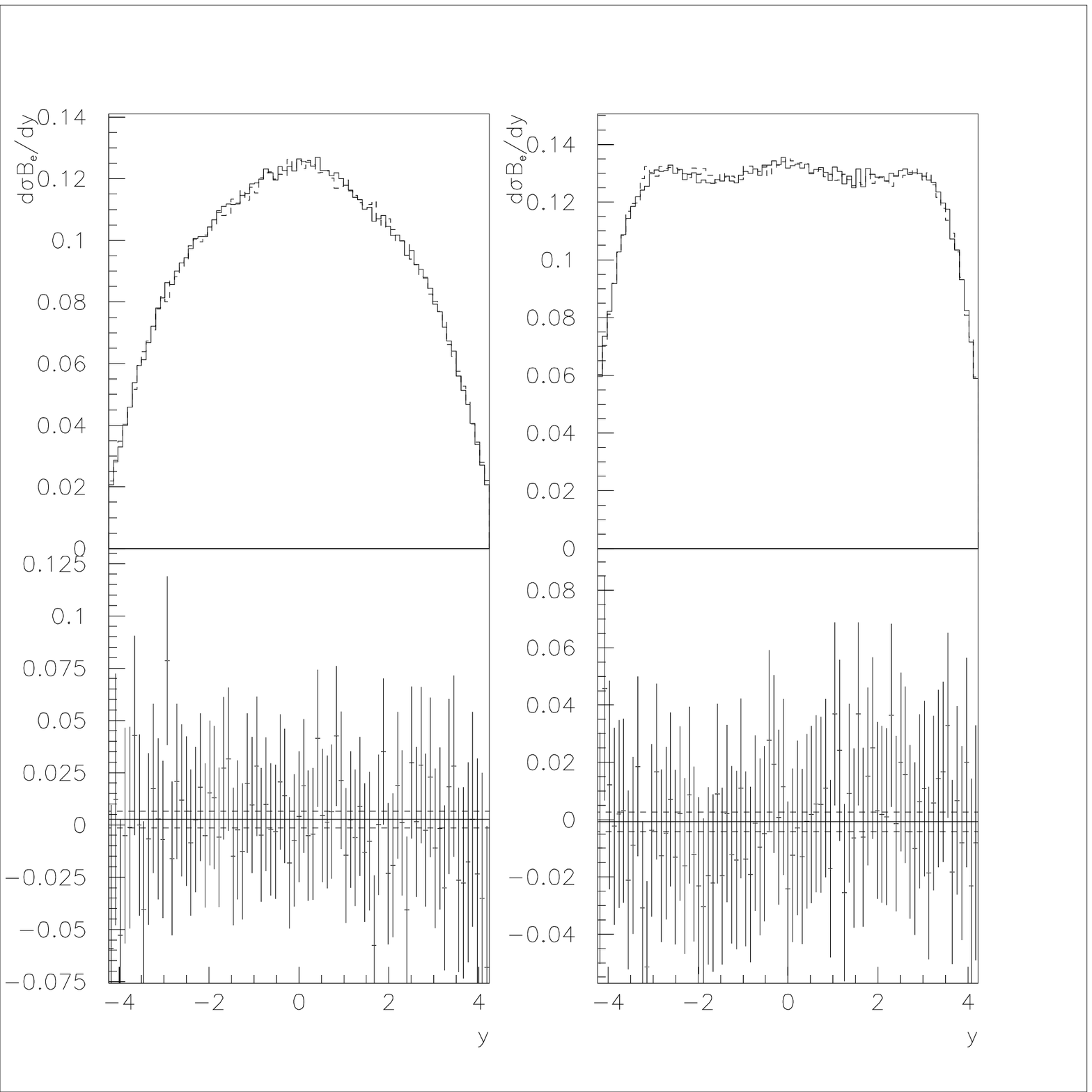,height=7cm}
\epsfig{figure=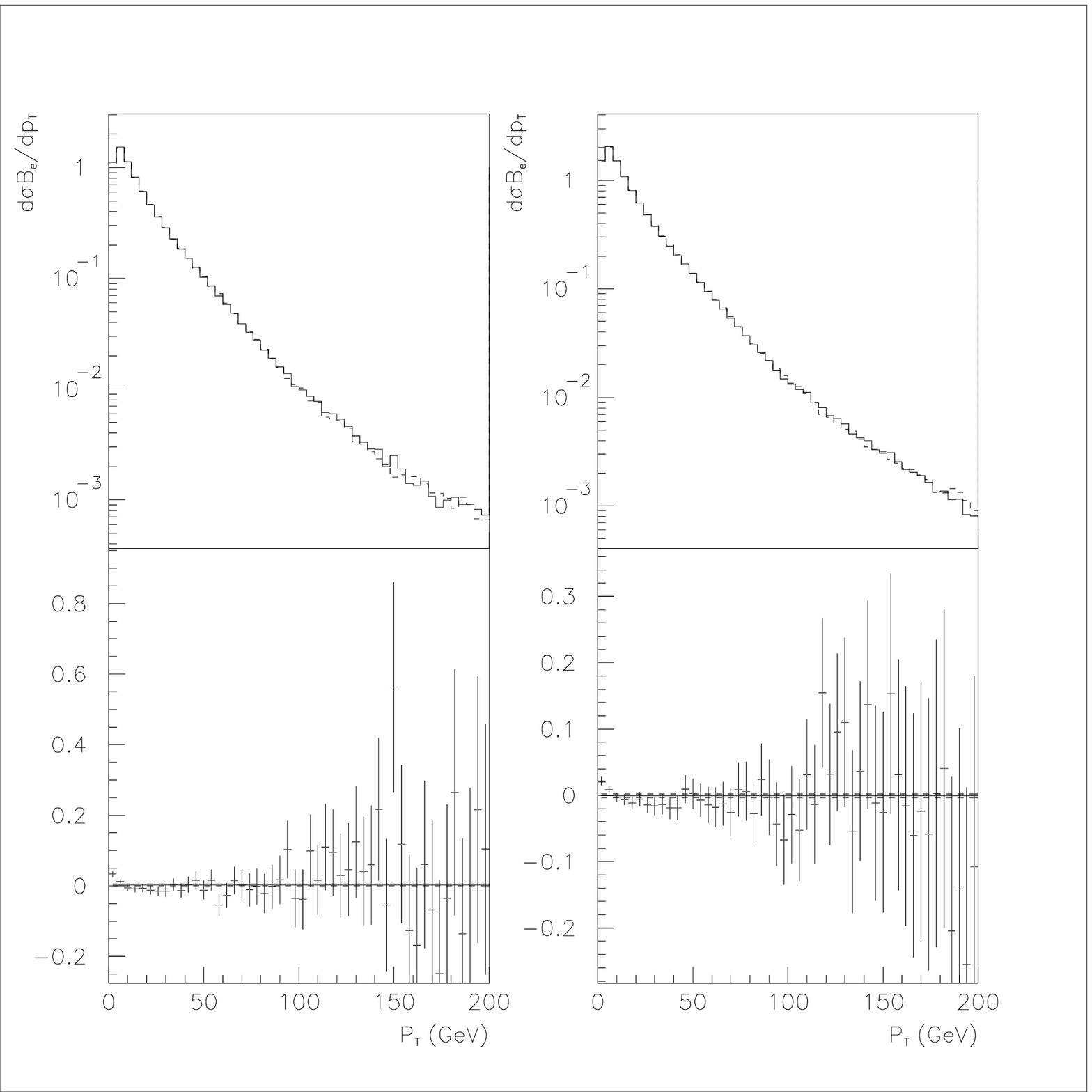,height=7cm}
}
\caption {Left side: 
$W^-$ (left) and $W^+$ (right) rapidity spectra, for events generated with MRST02 PDFs reweighted to 
CTEQ6.1 PDFs (full line), compared to events generated directly with CTEQ6.1 PDFs (dashed line). 
The fractional difference between these spectra are also shown
beneath the plots. The full line represents the weighted mean of these difference spectra and the dashed 
lines show its uncertainty. Right side: the same for $p_t$ spectra.}
\label{fig:pdfreweight}
\end{figure}

Suppose we generate $W$ events with a particular PDF set:
PDF set 1. Any one event has the hard scale, $Q^2 = M_{W}^{2}$, and two primary partons of 
flavours $flav_1$ and $flav_2$, 
with momentum fractions $x_1, x_2$ according to the distributions of PDF set 1. These momentum 
fractions are applicable to the hard process before the parton showers are implemented in backward
evolution in the MC. One can then evaluate the probability of picking up the same flavoured 
partons with the same momentum fractions from an alternative PDF set, PDF set 2, 
at the same hard scale. Then the event weight is given by
\begin{equation}
  \rm{PDF(re-weight)} = \frac{f_{PDF_2}(x_1,flav_1, Q^2). f_{PDF_2} (x_2, flav_2, Q^2)}
{ f_{PDF_1}(x_1,flav_1, Q^2). f_{PDF_1} (x_2, flav_2, Q^2)}
\end{equation}
where $xf_{PDF}(x,flav,Q^2)$ is the parton momentum distribution for flavour, $flav$, 
at scale, $Q^2$, and momentum fraction, $x$.
Fig.~\ref{fig:pdfreweight} compares the $W^+$ and $W^-$ spectra for a million events 
generated using MRST02 as PDF set 1 and re-weighting  to CTEQ6.1 as PDF set 2, 
with a million events which are directly generated with CTEQ6.1.  Beneath the spectra the fractional difference
between these distributions is shown. These difference spectra show that the reweighting is good 
to better than $1\%$, and there is no evidence of a $y$ dependent bias. 
This has been checked for reweighting between MRST02, CTEQ6.1 and ZEUS-S 
PDFs. Since the uncertainties of any one analysis are similar in size to the differences between 
the analyses it is clear that the technique can be used to produce spectra for the eigenvector 
error PDF sets of each analysis and thus to simulate the full PDF uncertainties from a single 
set of MC generated events.  
Fig.~\ref{fig:pdfreweight} also shows a similar comparison for $p_t$ spectra. 

\subsubsection{Background Studies}
\label{sec:bgd}
To investigate the accuracy with which $W$ events can be measured at the LHC
it is necessary to make an estimate of the importance of 
background processes. We focus on $W$ events which are identified through their decay to the 
$W \rightarrow e~\nu_e$ channel. There are several processes which can be misidentified as
$W \rightarrow e \nu_e$. These are: $W \rightarrow \tau \nu_\tau$, with $\tau$ decaying to the electron 
channel; $Z \rightarrow \tau^+ \tau^-$ with at least one $\tau$ decaying to the electron channel 
(including the case when both $\tau$'s decay to the electron channel, but one 
electron is not identified); $Z \rightarrow e^+ e^-$ with one electron not identified. We have generated one 
million events for each of these background processes, 
using HERWIG and CTEQ5L, and compared them to one million signal events generated with CTEQ6.1. 
We apply event selection criteria designed to eliminate the background preferentially. These criteria are:
\begin{itemize}
\item ATLFAST cuts (see Sec.~\ref{sec:gendet})
\item pseudorapidity, $|\eta| <2.4$, to avoid bias at the edge of the measurable rapidity range
\item  $p_{te} > 25$ GeV, high $p_t$ is necessary for electron triggering 
\item  missing $E_t > 25$ GeV, the $\nu_e$ in a signal event will have a correspondingly large missing $E_t$
\item  no reconstructed jets in the event with $p_t > 30$ GeV, to discriminate against QCD background 
\item  recoil on the transverse plane $p_t^{recoil} < 20$ GeV, to discriminate against QCD background
\end{itemize}
Table~\ref{tab:bgd} gives the percentage of background with respect to signal, calculated using the known 
relative cross-sections of these processes, as each of these cuts is 
applied. 
After, the cuts have been applied the background from these processes is negligible. However,
there are limitations on this study from the fact that in real data there will be further QCD backgrounds 
from $2 \rightarrow 2$ processes involving $q,\bar{q},g$ in which a final state $\pi^{0} \rightarrow \gamma
\gamma$ decay mimics a single electron. A preliminary study applying the selection criteria to MC 
generated QCD events suggests 
that this background is negligible, but the exact level of QCD background cannot be accurately estimated 
without passing a very large 
number of events though a full detector simulation, which is beyond the scope of the current contribution.
\begin{table}[tbp]
\vspace{-1.0cm}
\caption{Reduction of signal and background due to cuts}
\centerline{\small
\begin{tabular}{c|cc|cc|cc|cc}\\
 \hline
Cut & ~~~$W \rightarrow e \nu_e$ &  & 
~~~$Z \rightarrow \tau^+ \tau^-$ &  &
~~~$Z \rightarrow e^+ e^- $ &  &
~~~$W \rightarrow \tau \nu_\tau$ &  
\\
 & $e^+$ & $e^-$ &$e^+$ & $e^-$ &$e^+$ & $e^-$ &$e^+$ & $e^-$ 
\\ 
 \hline
 ATLFAST cuts& 382,902 & 264,415  & $5.5\%$ & $7.9\%$ & $34.7\%$ & $50.3\%$ & $14.8\%$ & $14.9\%$\\
 $|\eta| < 2.4$ & 367,815 & 255,514  & $5.5\%$ & $7.8\%$ & $34.3\%$ & $49.4\%$ & $14.7\%$ & $14.8\%$\\
 $p_{te} > 25$ GeV& 252,410 & 194,562  & $0.6\%$ & $0.7\%$ & $12.7\%$ & $16.2\%$ & $2.2\%$ & $2.3\%$\\
 $p_{tmiss} > 25$ GeV& 212,967 & 166,793  & $0.2\%$ & $0.2\%$ & $0.1\%$ & $0.2\%$ & $1.6\%$ & $1.6\%$\\
 No jets with $P_t > 30$ GeV& 187,634 & 147,415  & $0.1\%$ & $0.1\%$ & $0.1\%$ & $0.1\%$ & $1.2\%$ & $1.2\%$\\
$p_t^{recoil} < 20$ GeV& 159,873 & 125,003  & $0.1\%$ & $0.1\%$ & $0.0\%$ & $0.0\%$ & $1.2\%$ & $1.2\%$\\
 \hline\\
\end{tabular}}
\label{tab:bgd}
\end{table}

\subsubsection{Charge misidentification}

\begin{figure}[tbp] 
\vspace{-0.8cm}
\centerline{
\epsfig{figure=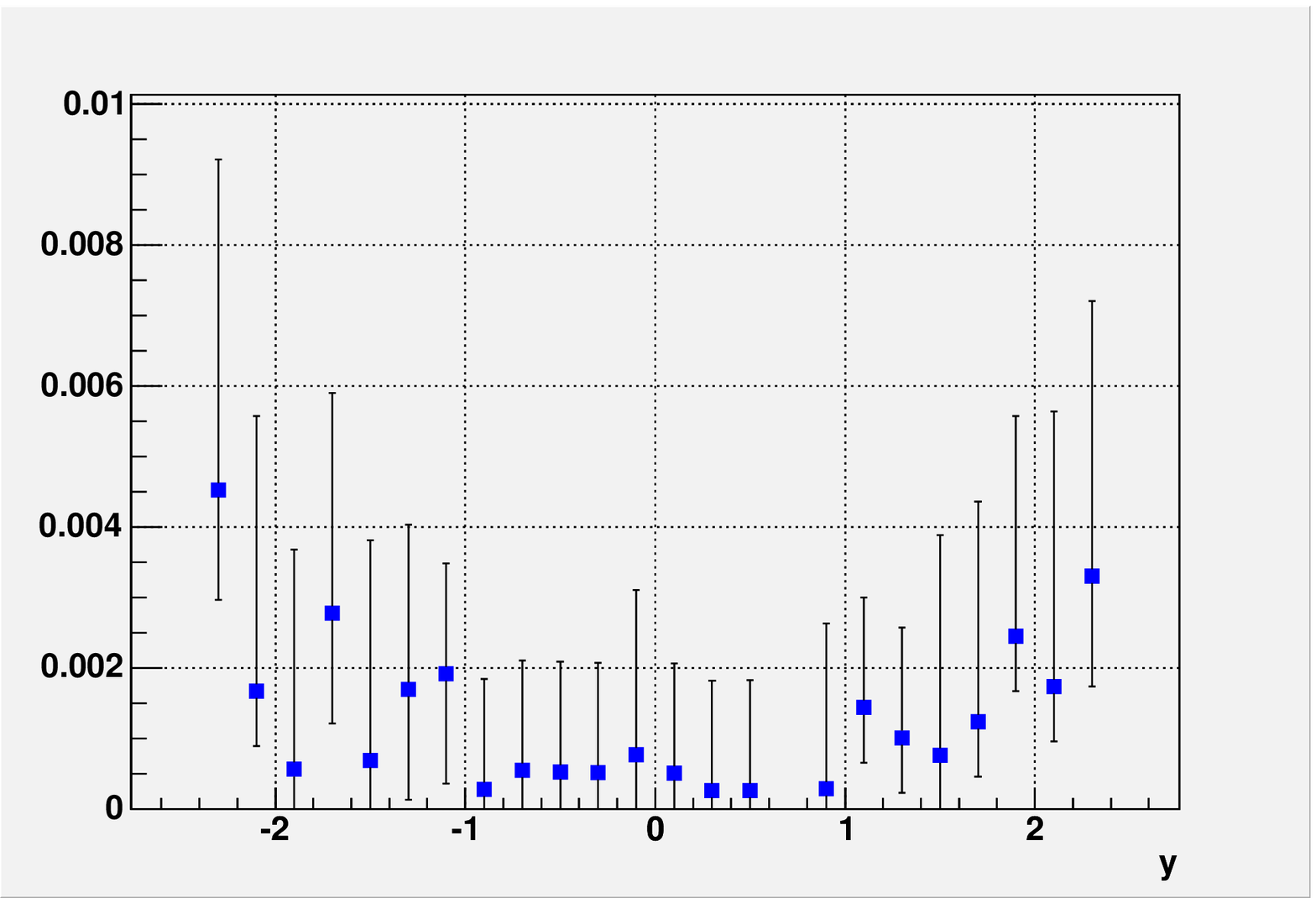,width=0.4\textwidth}
\epsfig{figure=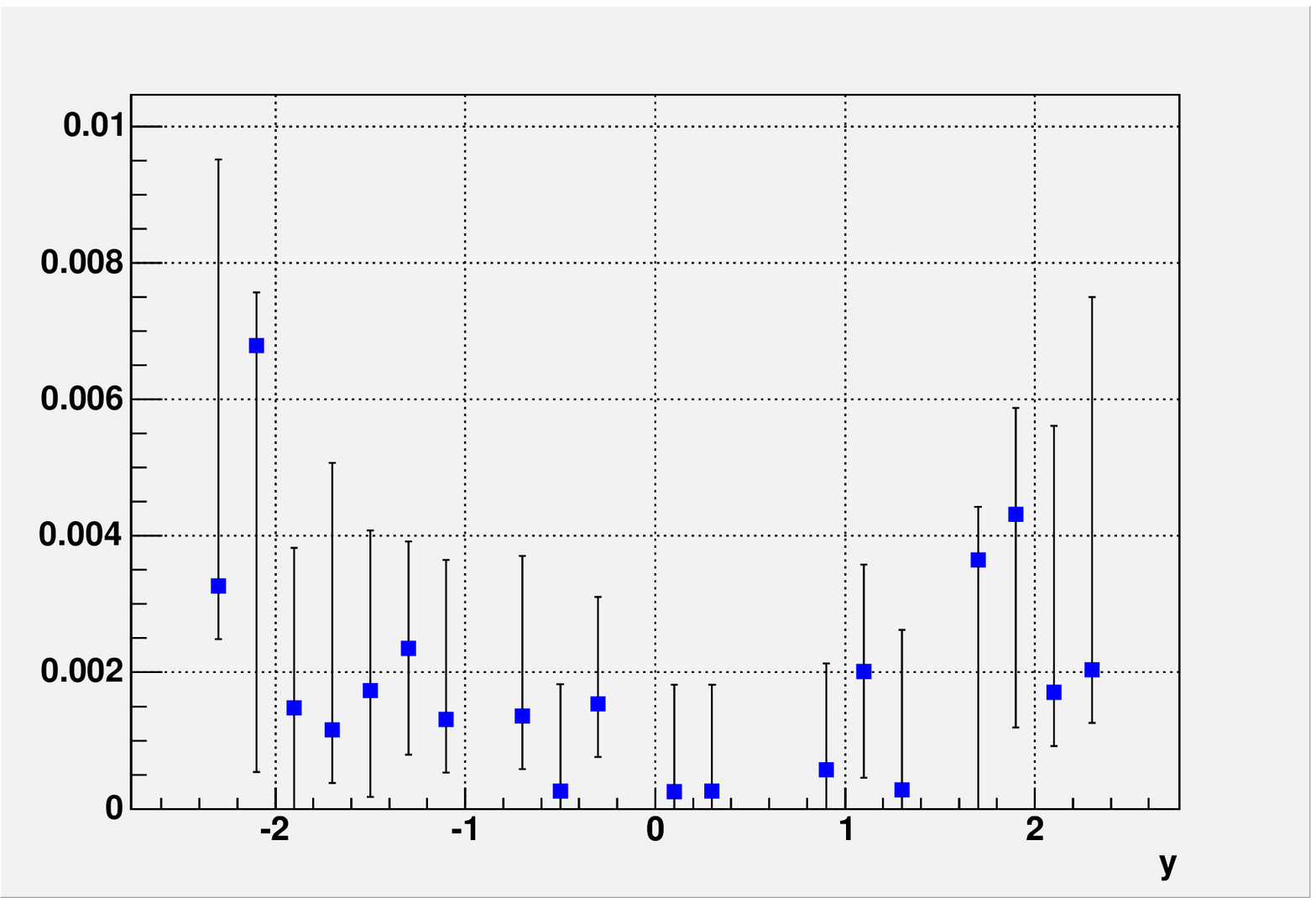,width=0.4\textwidth}
}
\caption {The rates of charge misidentification as a function of rapidity for $e^-$ misidentified 
as $e^+$ (left), $e^+$ misidentifed as $e^-$ (right).
}
\label{fig:misid}
\end{figure}

Clearly charge misidentification could distort the lepton rapidity spectra and dilute the asymmetry $A_l$.
\[
 A_{true} = \frac{ A_{raw} - F^+ + F^-}{1 -F^- - F^+}
\]
where $A_{raw}$ is the measured asymmetry, $A_{true}$ is the true asymmetry, $F^-$ is the rate of true $e^-$
misidentified as $e^+$ and $F^+$ is the rate of true $e^+$ misidentified as $e^-$. To make an estimate of 
the importance of charge misidentification we use a sample of $Z \rightarrow e^+ e^-$ events generated by 
HERWIG with CTEQ5L and passed through a full simulation of the ATLAS detector. Events with two
or more charged electromagnetic objects in the EM calorimeter are then selected and subject to the cuts;
$|\eta| < 2.5$, $p_{te} > 25$ GeV, as 
usual and, $E/p < 2$, for bremsstrahlung rejection. We then look for the charged electromagnetic pair with 
invariant mass closest to $M_Z$ and impose the cut, $60 < M_Z < 120$ GeV. 
Then we tag the charge of the better 
reconstructed lepton of the pair and check to see if the charge of the second lepton is the
 same as the first. Assuming that the pair really came from the decay of the $Z$ this gives us a measure of 
charge misidentification. Fig~\ref{fig:misid} show the misidentification rates $F^+$, $F^-$ as functions of 
pseudorapidity\footnote{These have been corrected for the small possibility that the better reconstructed lepton 
has had its charge misidentified as follows. In the central region, $|\eta| < 1$, assume the same probability 
of misidentification of the first and second leptons, in the more forward regions assume the same rate of 
first lepton misidentification as in the central region.}.
 These rates are very small. The quantity $A_l$, can be corrected 
for charge misidentification applying Barlow's method for combining asymmetric errors~\cite{Barlow:2004wj}. 
The level of correction is $0.3\%$ in the central region 
and $0.5\%$ in the more forward regions. 

\subsubsection{Compare events at the generator level to events at the detector level}
\label{sec:gendet}
\begin{figure}[tbp] 
\vspace{-1.0cm}
\centerline{
\epsfig{figure=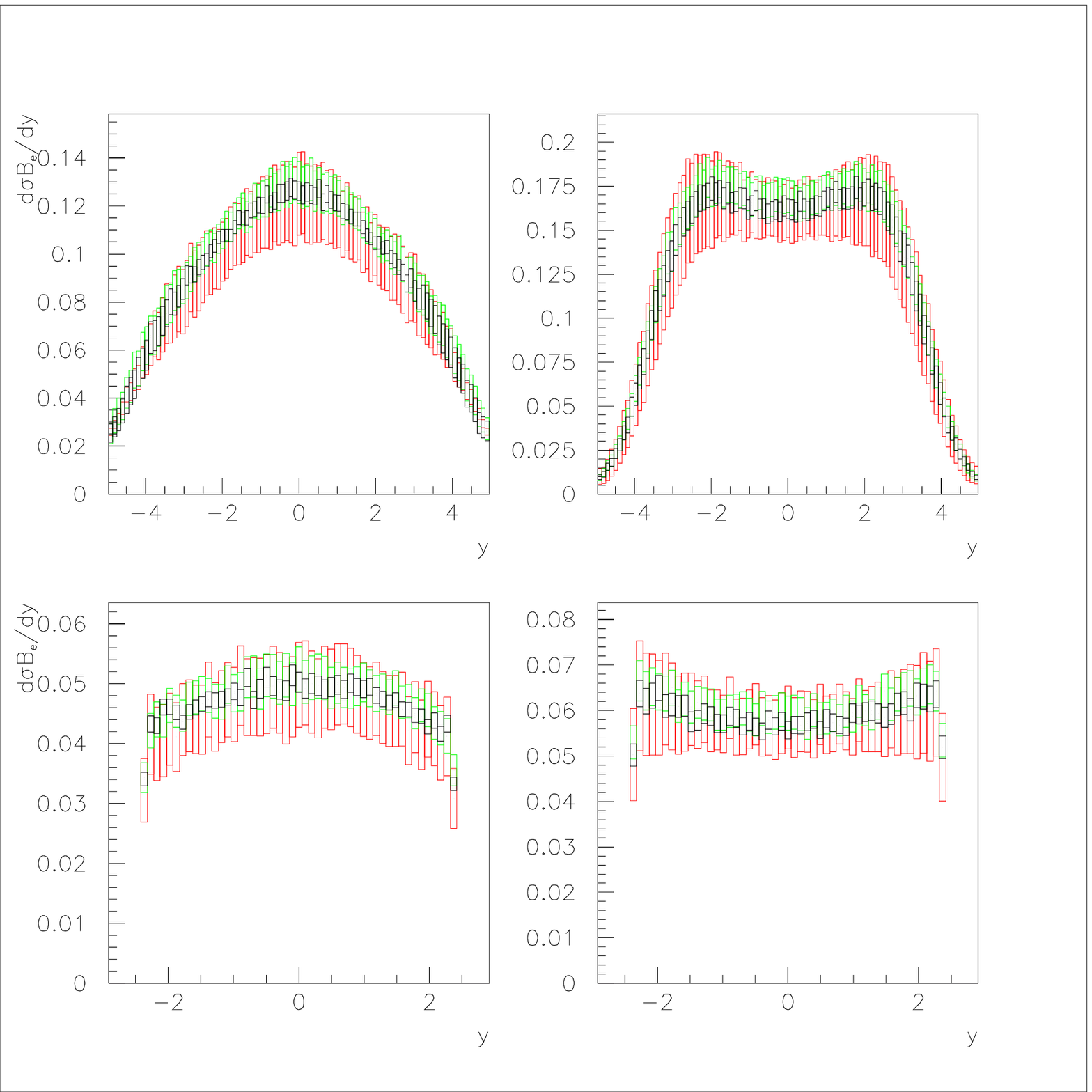,width=0.66\textwidth,height=8cm }
\epsfig{figure=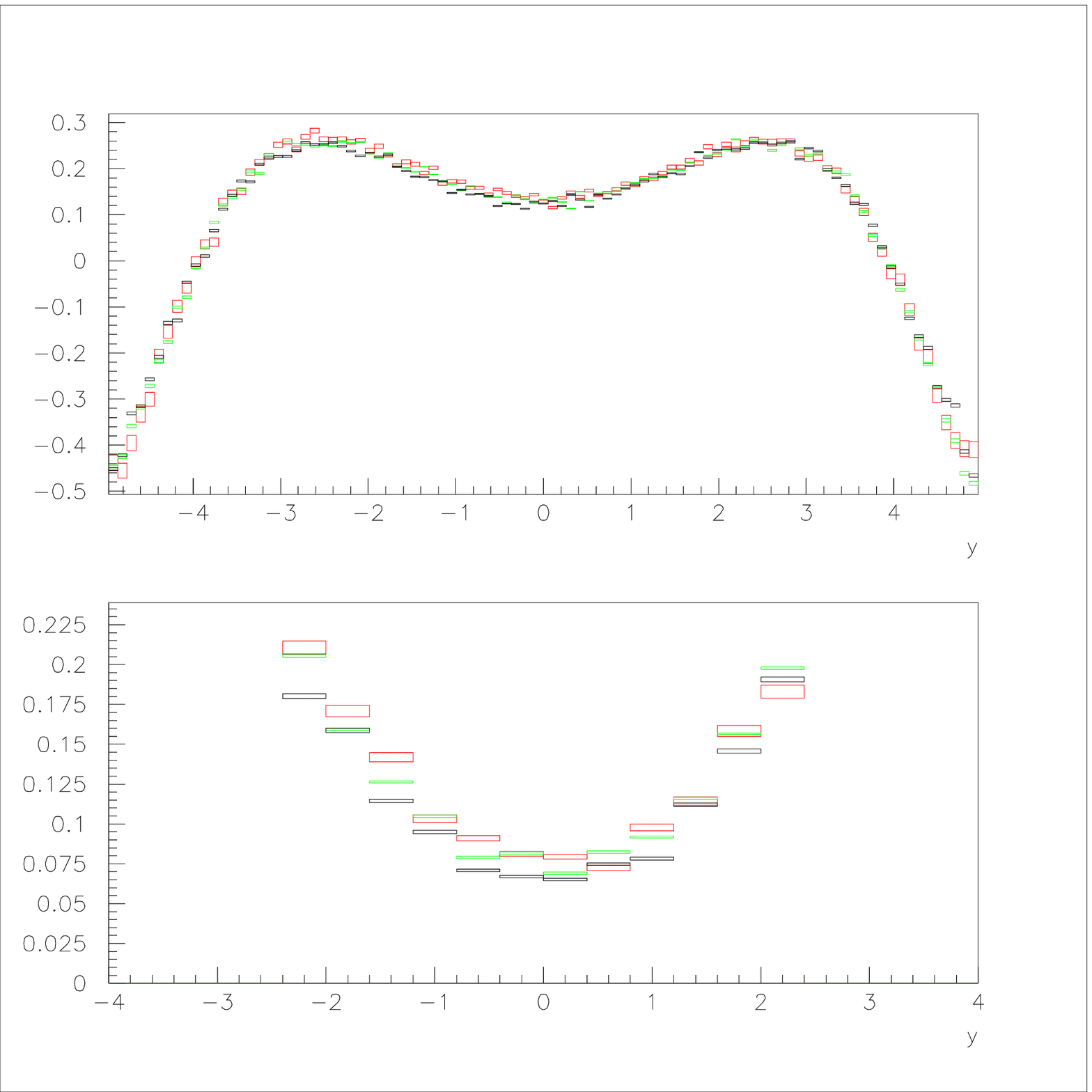,width=0.33\textwidth,height=8cm}
}
\caption {Top row: $e^-$, $e^+$ and $A_e$ rapidity spectra for the lepton from the $W$ decay, 
generated using HERWIG + k factors and CTEQ6.1 (red),
ZEUS-S (green) and MRST2001 (black) PDF sets with full uncertainties. Bottom row: the same spectra after passing 
through the ATLFAST detector simulation and selection cuts.}
\label{fig:gendet}
\end{figure}

We have simulated one million signal, $W \rightarrow e \nu_e$, events for each of the PDF sets CTEQ6.1, 
MRST2001 and ZEUS-S using HERWIG (6.505). 
For each of these PDF sets the eigenvector error PDF sets have been simulated 
by PDF reweighting and k-factors have been applied to approximate an NLO generation.
The top part of Fig.~\ref{fig:gendet} shows the $e^{\pm}$ and $A_l$ 
spectra at this generator level, for all of the PDF sets superimposed.
The events are then passed through the ATLFAST fast simulation of the ATLAS detector. This applies
loose kinematic cuts: $|\eta| < 2.5$, $p_{te} > 5$ GeV, and electron isolation criteria. 
It also smears the 4-momenta of the 
leptons to mimic momentum dependent detector resolution. We then apply the selection cuts described in 
Sec.~\ref{sec:bgd}. The lower half of
Fig.~\ref{fig:gendet} shows the $e^{\pm}$ and $A_l$ spectra at the detector level after application 
of these cuts, for all of the PDF sets superimposed. 
The level of precision of each PDF set, seen in the analytic calculations of Fig.~\ref{fig:mrstcteq},
 is only slightly degraded  at detector level, so that a net level of PDF 
uncertainty at central rapidity of $\sim 8\%$ is maintained. The anticipated cancellation of PDF 
uncertainties in the asymmetry spectrum is also observed, within each PDF set, and the spread between PDF sets
suggests that measurements which are accurate to better than $\sim 5\%$ could discriminate between PDF sets.

\subsubsection{Using LHC data to improve precision on PDFs}

\begin{figure}[tbp] 
\vspace{-1.5cm}
\centerline{
\epsfig{figure=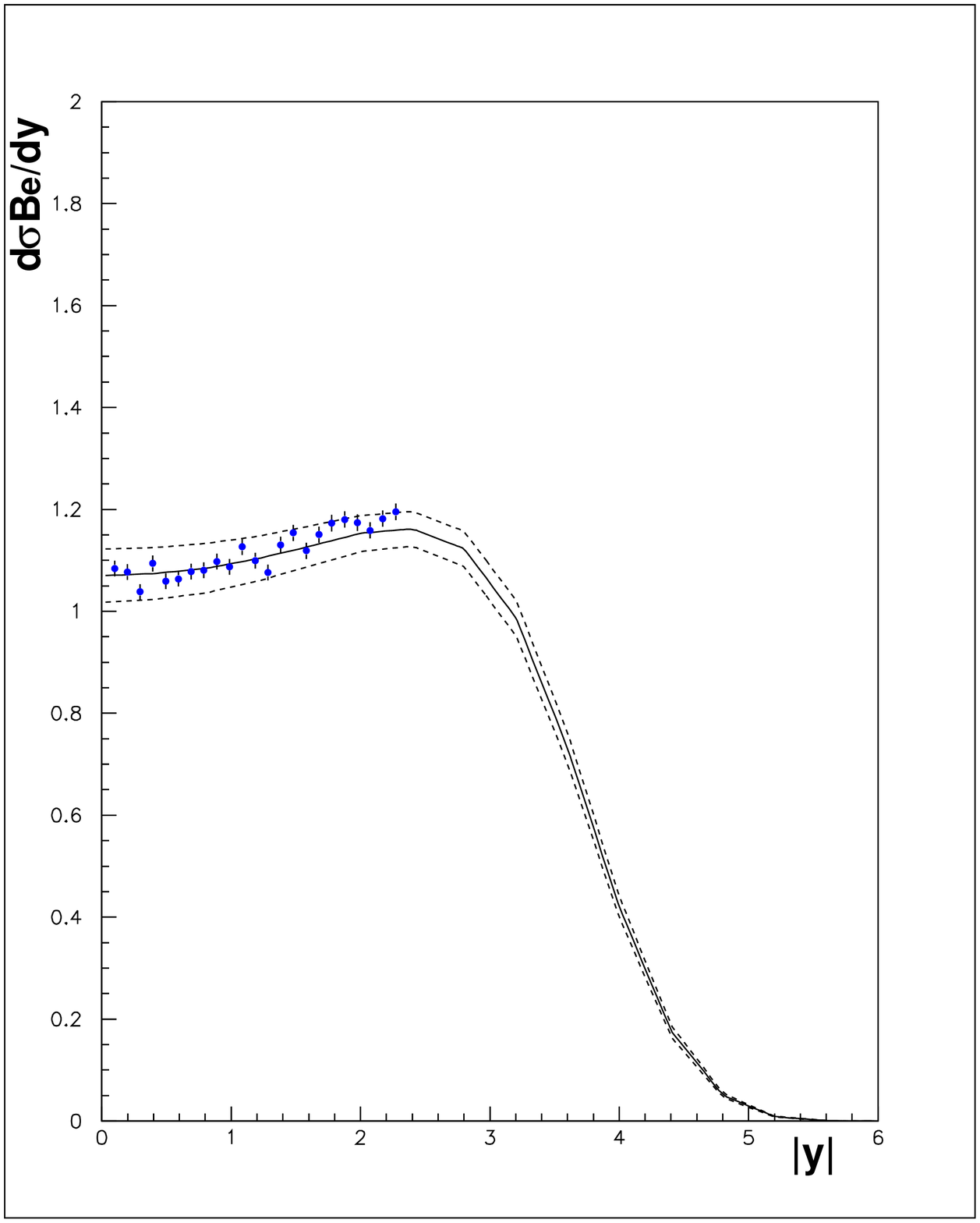,width=0.3\textwidth,height=4cm}
\epsfig{figure=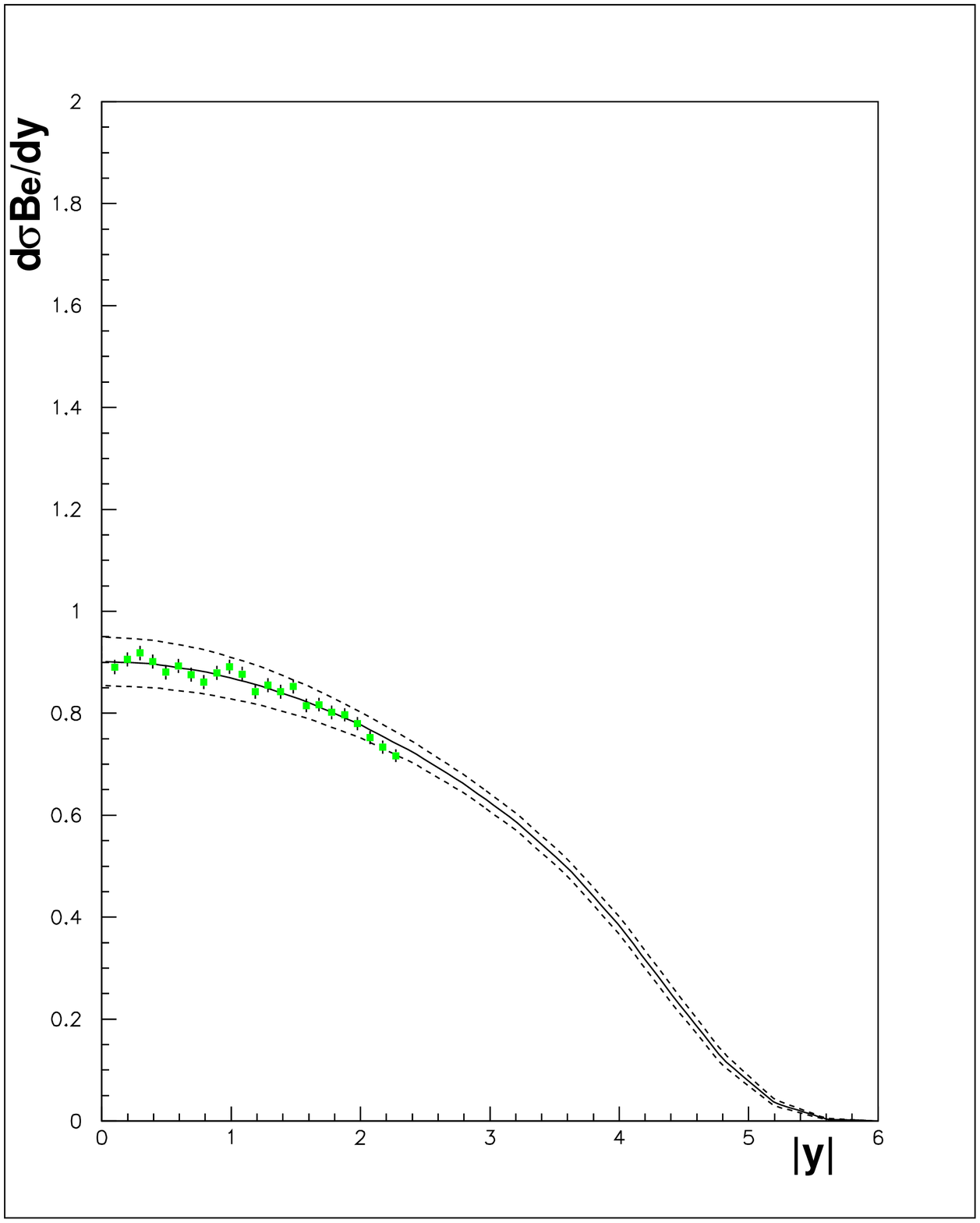,width=0.3\textwidth,height=4cm}
}
\centerline{
\epsfig{figure=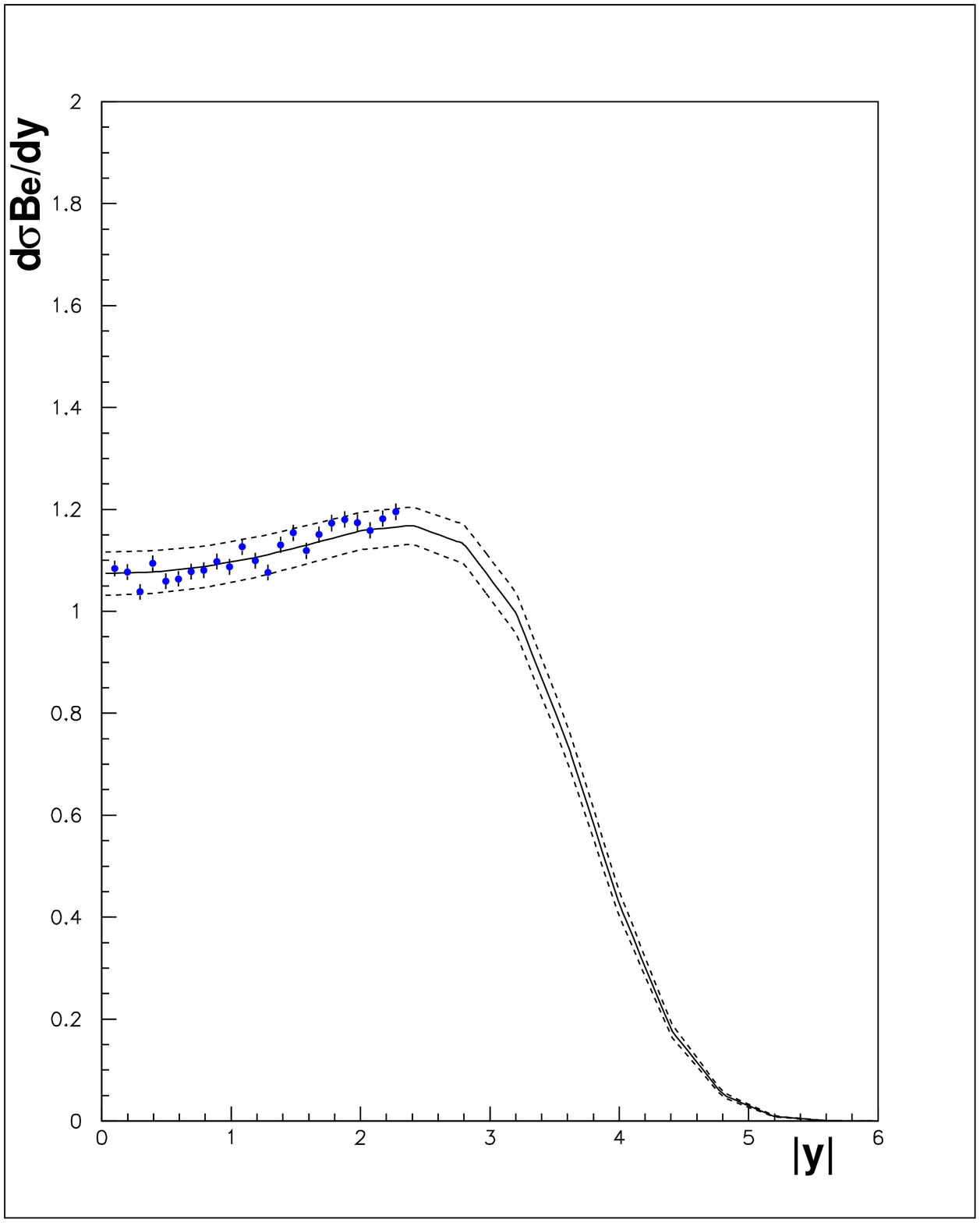,width=0.3\textwidth,height=4cm}
\epsfig{figure=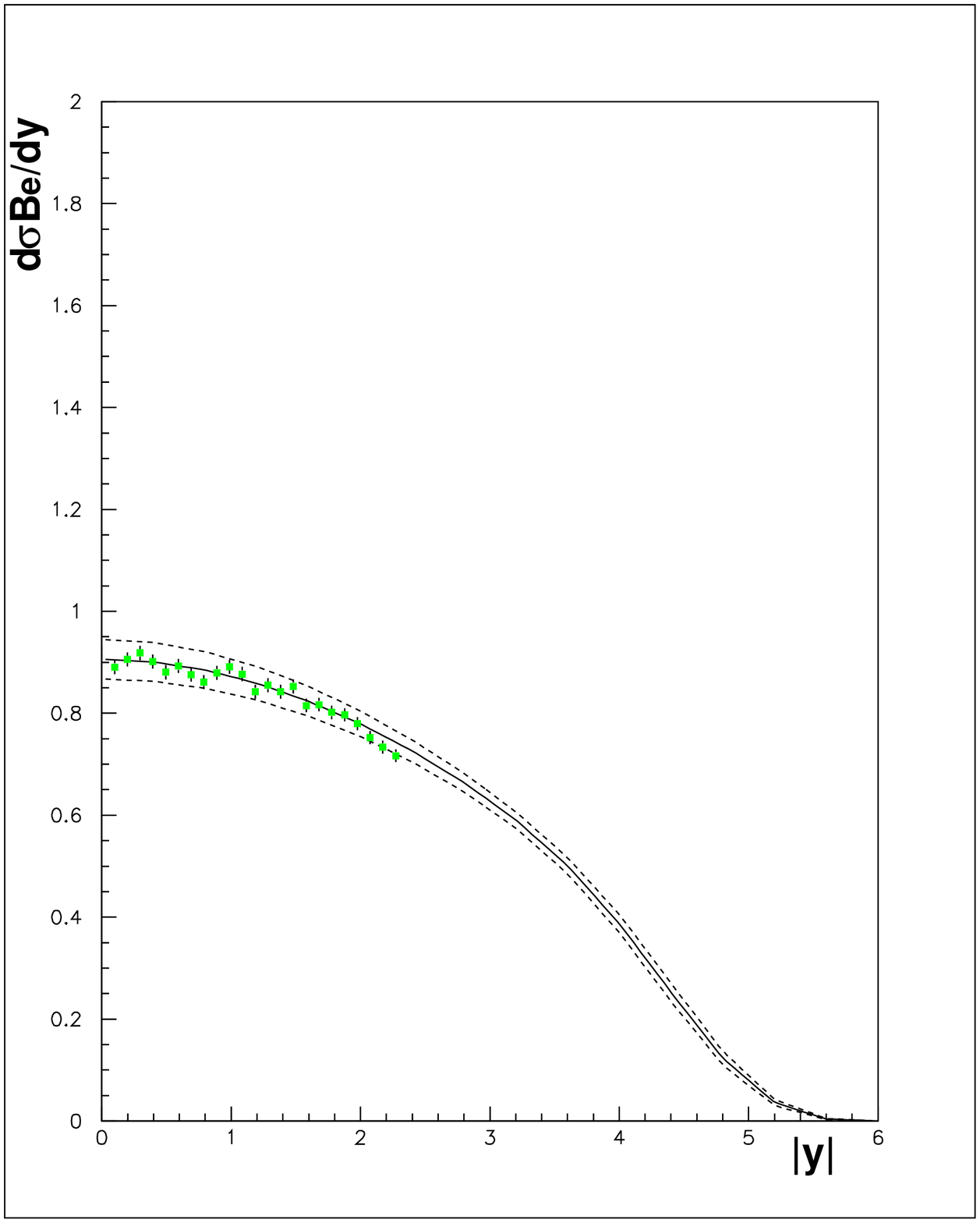,width=0.3\textwidth,height=4cm}
}
\caption {Top row: $e^+$ and $e^-$ rapidity spectra generated from ZEUS-S PDFs compared to the analytic prediction
using ZEUS-S PDFs. Bottom row: the same lepton rapidity spectra compared to the analytic 
prediction AFTER including these lepton pseudo-data in the ZEUS-S PDF fit.}
\label{fig:zeusfit}
\end{figure}

\begin{figure}[tbp] 
\vspace{-1.5cm}
\centerline{
\epsfig{figure=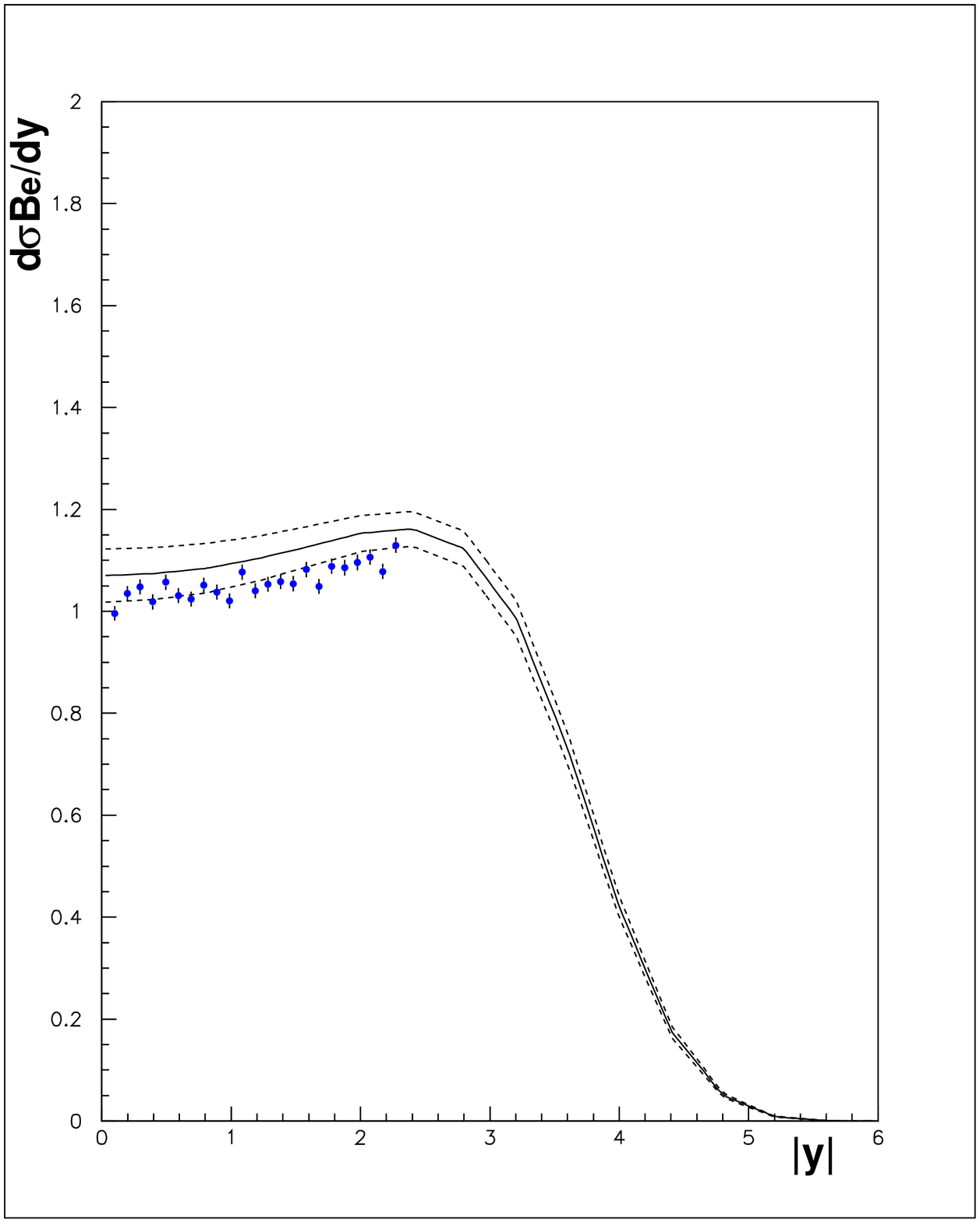,width=0.3\textwidth,height=4cm}
\epsfig{figure=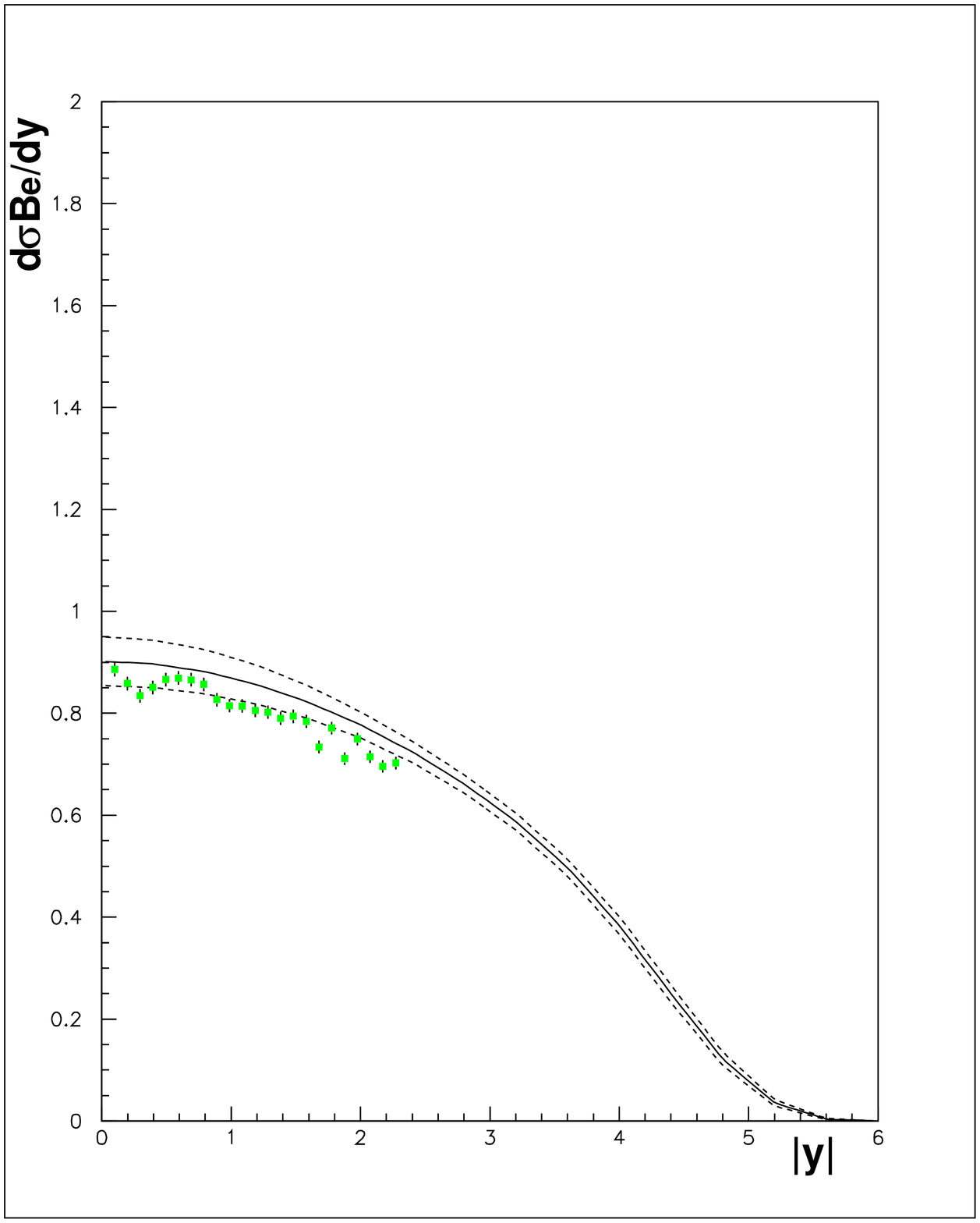,width=0.3\textwidth,height=4cm}
}
\centerline{
\epsfig{figure=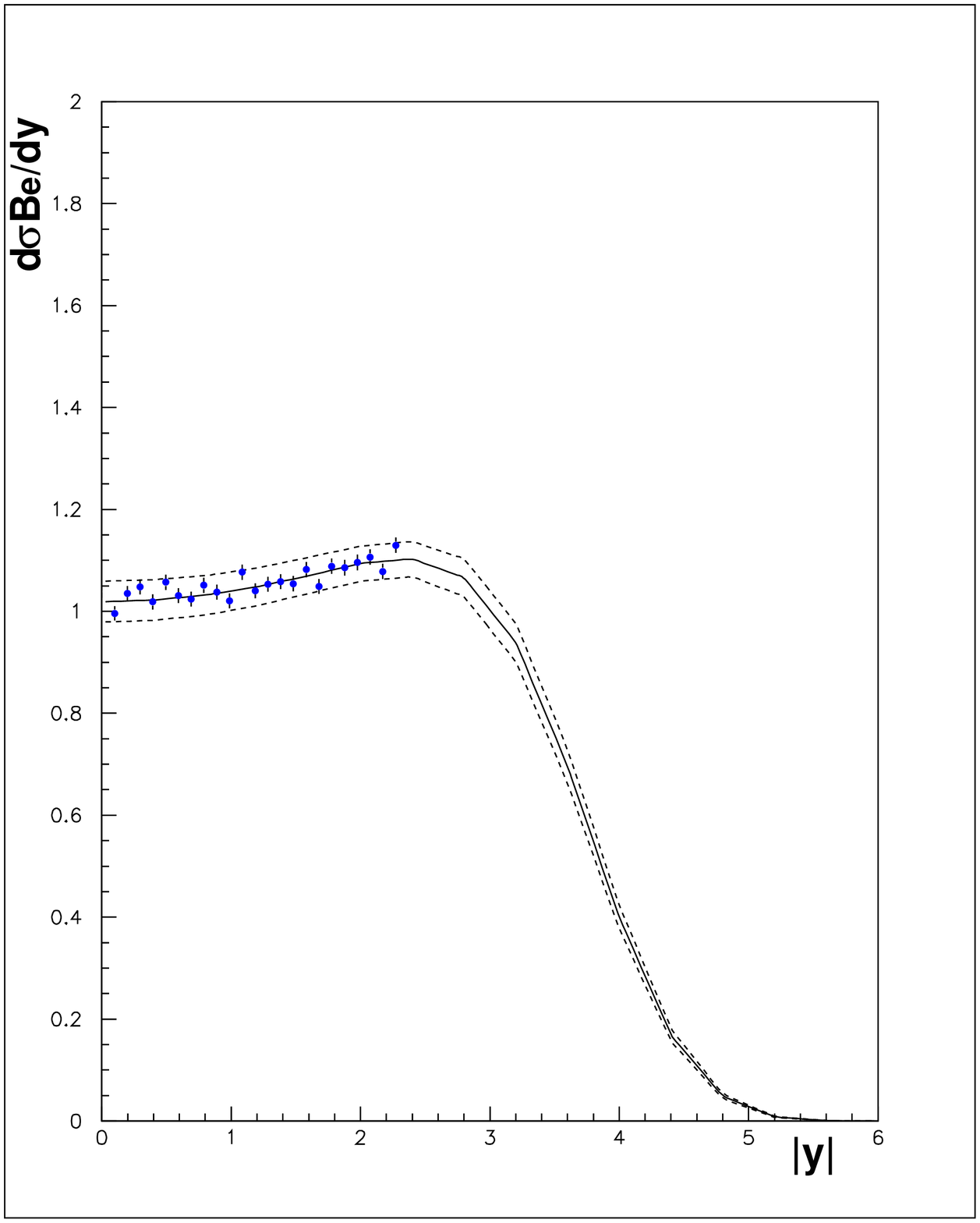,width=0.3\textwidth,height=4cm}
\epsfig{figure=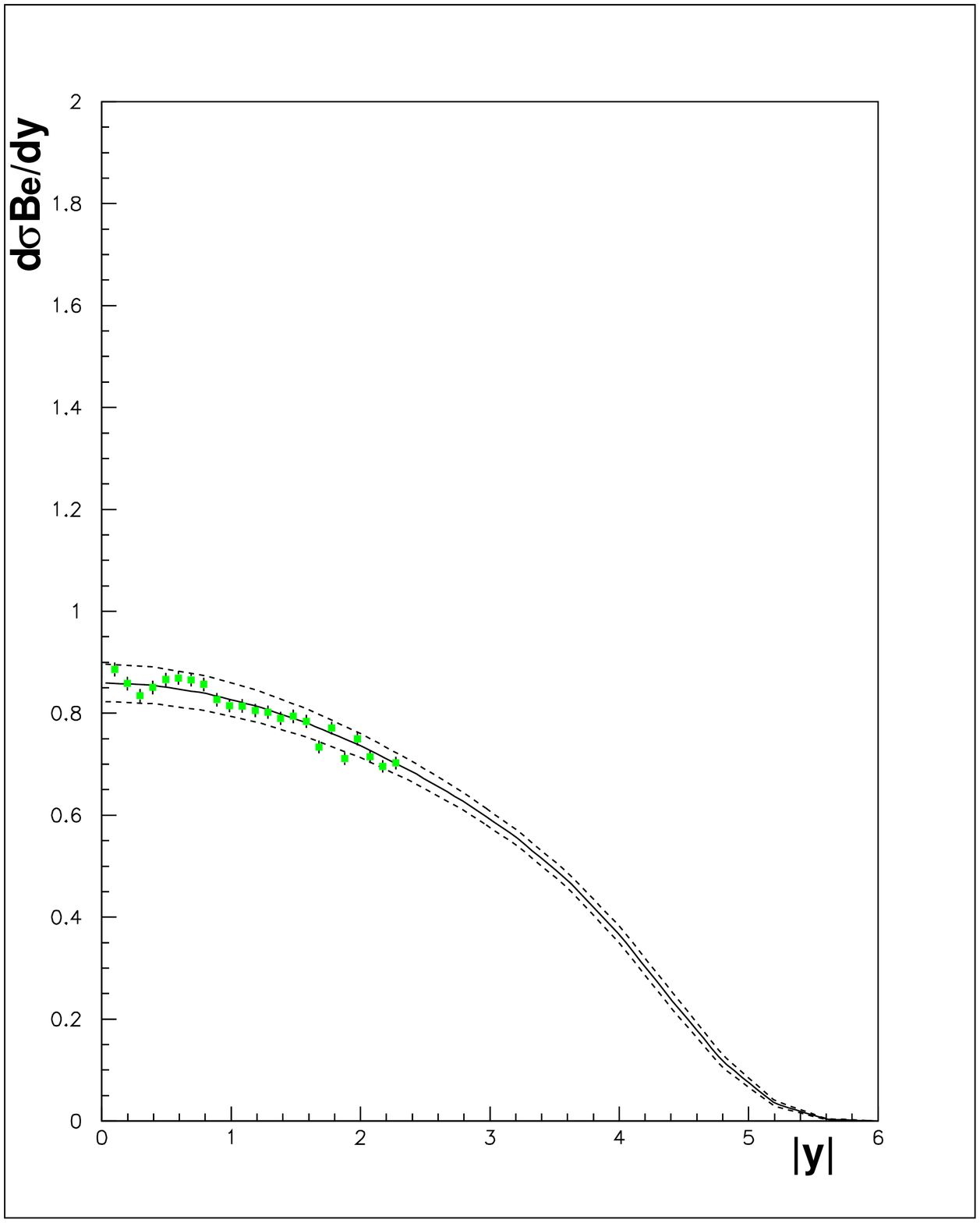,width=0.3\textwidth,height=4cm}
}
\caption {Top row: $e^+$ and $e^-$ rapidity spectra generated from CTEQ6.1 PDFs compared to the analytic prediction
using ZEUS-S PDFs. Bottom row: the same lepton rapidity spectra compared to the analytic 
prediction AFTER including these lepton pseudo-data in the ZEUS-S PDF fit.}
\label{fig:ctqfit}
\end{figure}

\begin{figure}[tbp] 
\vspace{-0.5cm}
\centerline{
\epsfig{figure=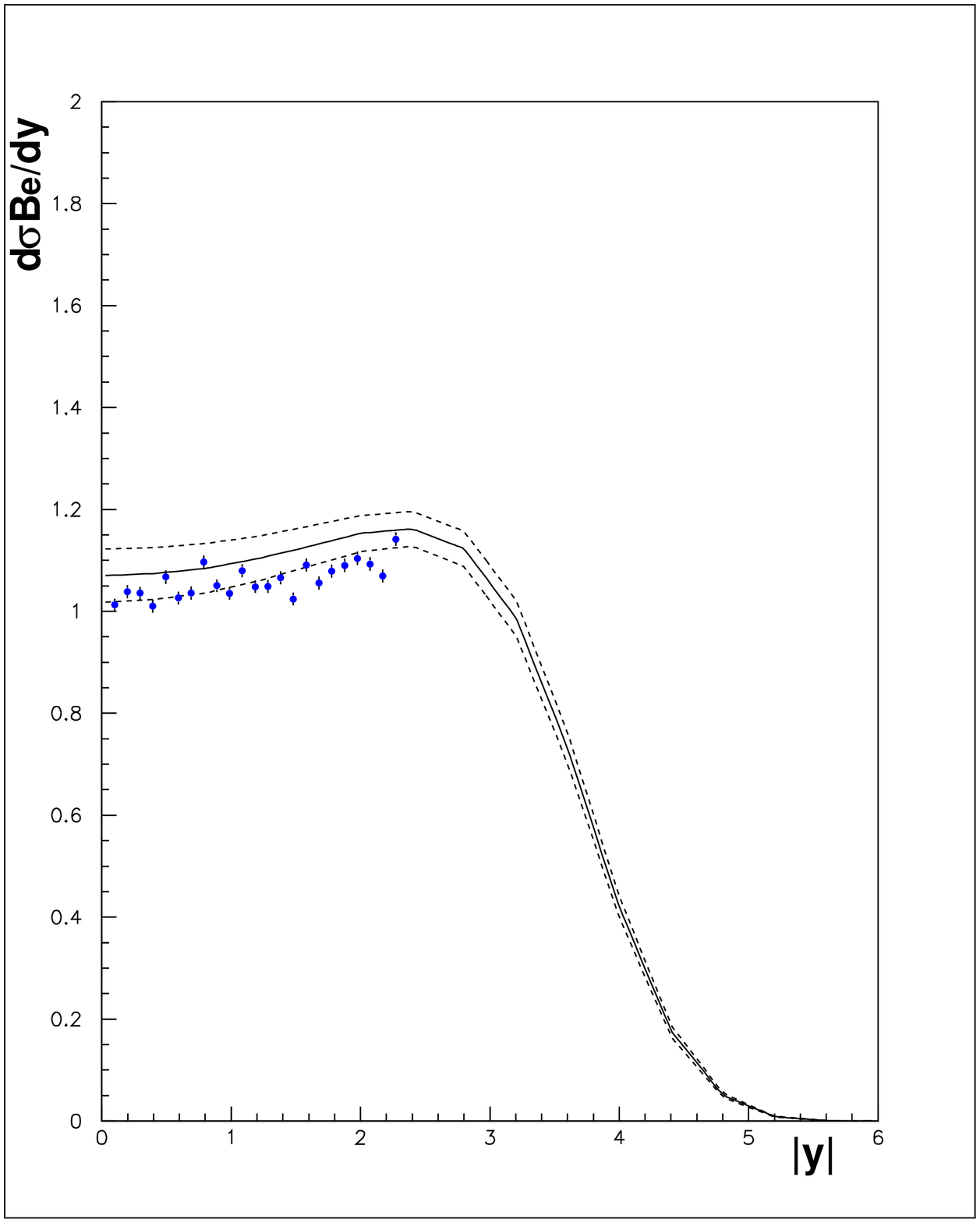,width=0.3\textwidth,height=4cm}
\epsfig{figure=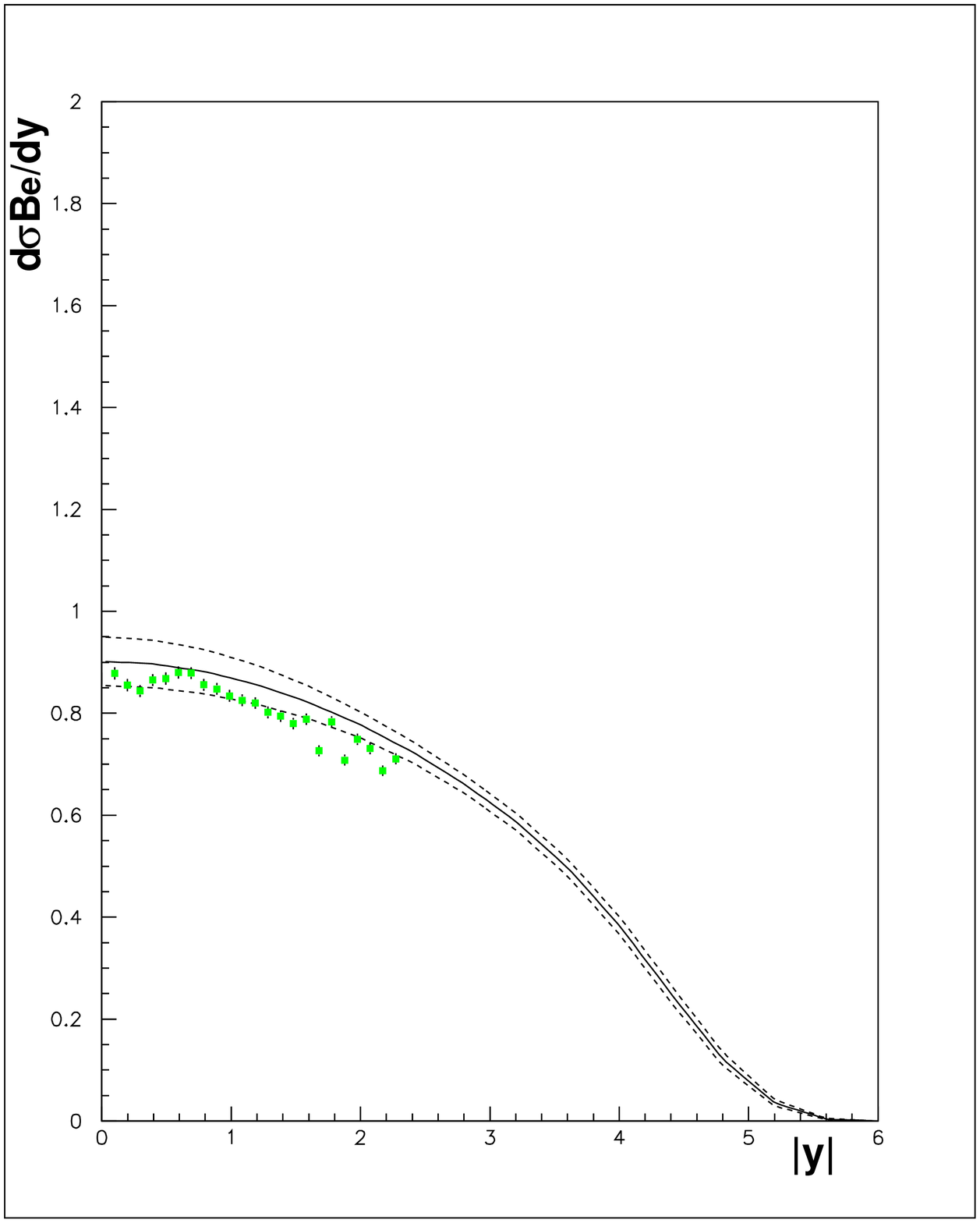,width=0.3\textwidth,height=4cm}
}
\centerline{
\epsfig{figure=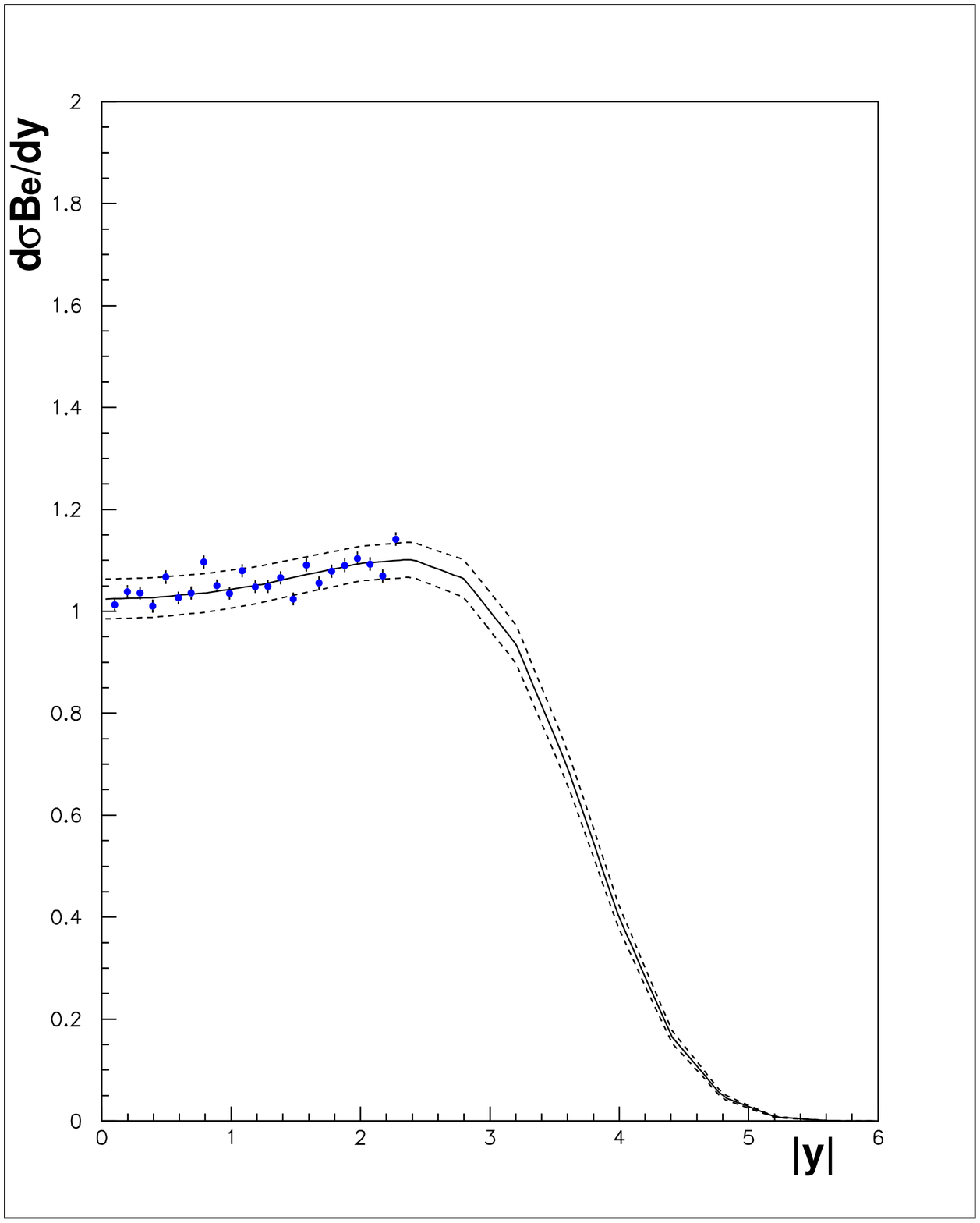,width=0.3\textwidth,height=4cm}
\epsfig{figure=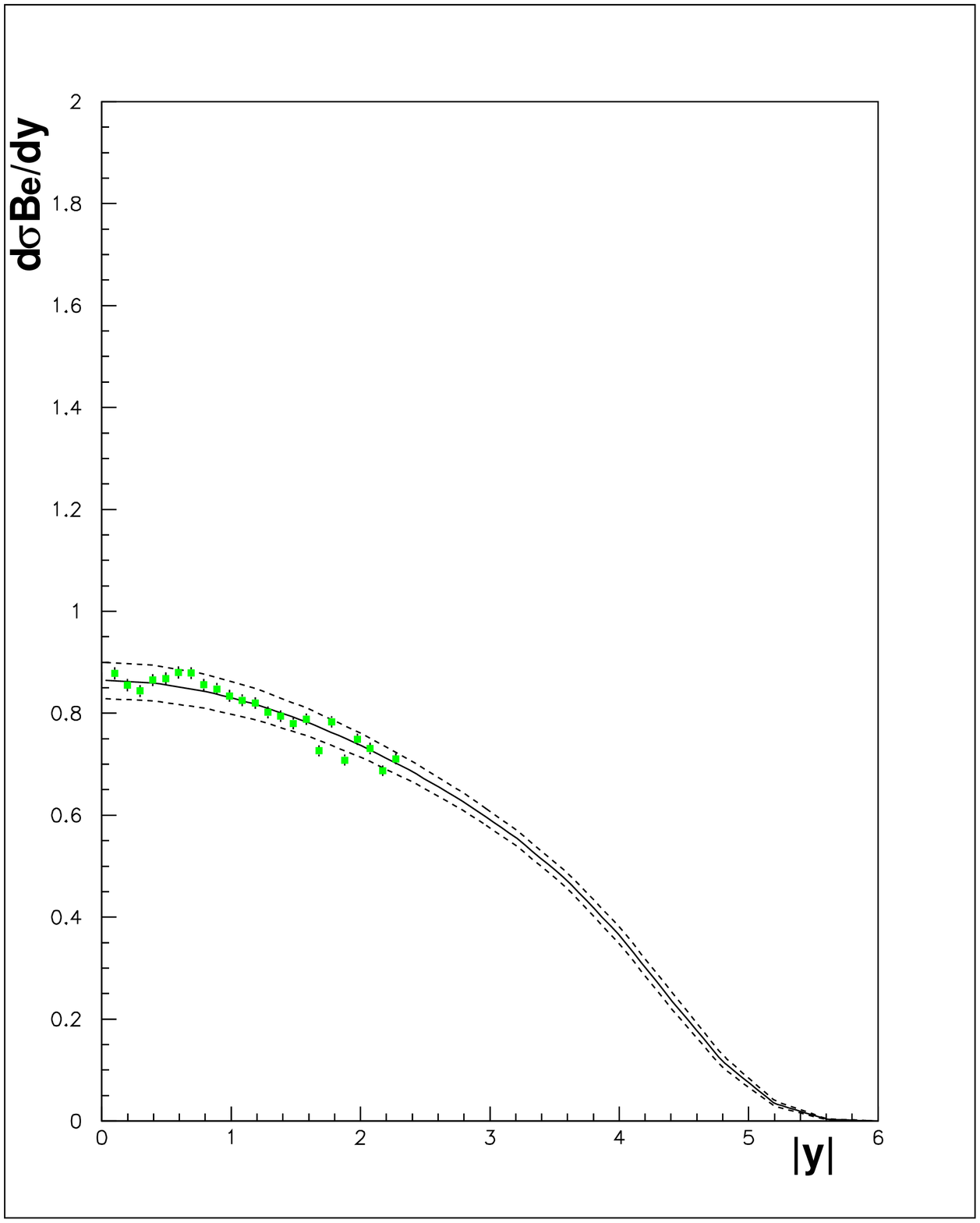,width=0.3\textwidth,height=4cm}
}
\caption {Top row: $e^+$ and $e^-$ rapidity spectra generated from CTEQ6.1 PDFs, which have been passed through 
the ATLFAST detector simulation and corrected back to generator level using ZEUS-S PDFs, 
compared to the analytic prediction
using ZEUS-S PDFs. Bottom row: the same lepton rapidity spectra compared to the analytic 
prediction AFTER including these lepton pseudo-data in the ZEUS-S PDF fit.}
\label{fig:ctqcorfit}
\end{figure}

The high cross-sections for $W$ production at 
the LHC ensure that it will be the experimental systematic errors, rather than the statistical errors, which 
are determining. We have imposed a random  $4\%$
scatter on our samples of one million $W$ events, generated using different PDFs,
in order to 
investigate if measurements at this level of precision will improve PDF uncertainties at central rapidity 
significantly if they 
are input to a global PDF fit. Fig.~\ref{fig:zeusfit} shows the $e^+$ and $e^-$ rapidity 
spectra for events generated from the ZEUS-S PDFs ($|\eta| < 2.4$) compared to the analytic 
predictions for these same
ZEUS-S PDFs. The lower half of this figure illustrates the result if these events are then 
included in the ZEUS-S PDF fit. The size of the PDF uncertainties, at $y=0$, 
decreases from $5.8\%$ to $4.5\%$.  
The largest improvement is in the PDF parameter $\lambda_g$ controlling the 
low-$x$ gluon at the input scale, $Q^2_0$: $xg(x) \sim x^{\lambda_g}$ at low-$x$, $\lambda_g = -0.199 \pm 0.046$, 
before the input of the LHC pseudo-data, compared to, $\lambda_g = -0.196 \pm 0.029$, after input. 
Note that whereas the relative normalisations of the $e^+$ and $e^-$ spectra are set by the PDFs, 
the absolute normalisation of the data is free in the fit so that no assumptions are made on 
our ability to measure luminosity.
Secondly, we repeat this procedure for events generated using the CTEQ6.1 PDFs. 
As shown in Fig.~\ref{fig:ctqfit}, the cross-section for these events is on the lower edge of the uncertainty 
band of the ZEUS-S predictions. If these events are input to the fit the central value shifts and the 
uncertainty decreases. The value of the parameter $\lambda_g$ becomes, $\lambda_g = -0.189 \pm 0.029$, 
after input of these pseudo-data.
Finally to simulate the situation which really faces experimentalists we generate events with CTEQ6.1, 
and pass them through the ATLFAST detector simulation and cuts. We then correct back from detector level 
to generator level using a different PDF set- in this case the ZEUS-S PDFs- since in practice we will not know 
the true PDFs. Fig.~\ref{fig:ctqcorfit} shows that the resulting corrected data look 
pleasingly like CTEQ6.1, but they are more smeared. When these data are input to the PDF fit the 
central values shift and 
errors decrease just as for the perfect CTEQ6.1 pseudo-data. The value of $\lambda_g$ becomes,
 $\lambda = -0.181 \pm 0.030$, after input of these pseudo-data. Thus we see that the bias introduced by the 
correction procedure from detector to generator level is small compared to the PDF uncertainty. 

\subsubsection{Conclusions and a warning: problems with the theoretical predictions at small-$x$?}

We have investigated the PDF uncertainty on the predictions for $W$ and $Z$ production at the LHC, taking 
into account realistic expectations for measurement accuracy and the cuts on data which will be needed to 
identify signal events from background processes. We conclude that at the present level of PDF uncertainty
the decay lepton asymmetry, $A_l$, will be a useful standard model benchmark measurement, and that the 
decay lepton spectra can be used as a luminosity monitor which will be good to $\sim 8\%$. However, 
we have also investigated the measurement accuracy 
necessary for early measurements of these decay lepton spectra to be useful in further constraining the 
PDFs. A systematic measurement error of less than $\sim 4\%$ would provide useful extra constraints.

However, a caveat is that the current study has been performed using 
standard PDF sets which are extracted using NLO QCD in the DGLAP~\cite{ap,gl,l,d} formalism. 
The extension to NNLO is 
straightforward, giving small corrections $\sim 1\%$. PDF analyses at NNLO 
including full accounting of the PDF 
uncertainties are not extensively available yet, so this small correction is not pursued here. However, 
there may be much larger  
uncertainties in the theoretical calculations because the kinematic region involves  
low-$x$.  There may be a need to account for $ln(1/x)$ resummation (first considered in the 
BFKL~\cite{Lipatov:1976zz,Kuraev:1977fs,Balitsky:1978ic} formalism) 
or high gluon density effects. See reference~\cite{dcs} for a review. 

\begin{figure}[tbp] 
\vspace{-1.0cm}
\centerline{
\epsfig{figure=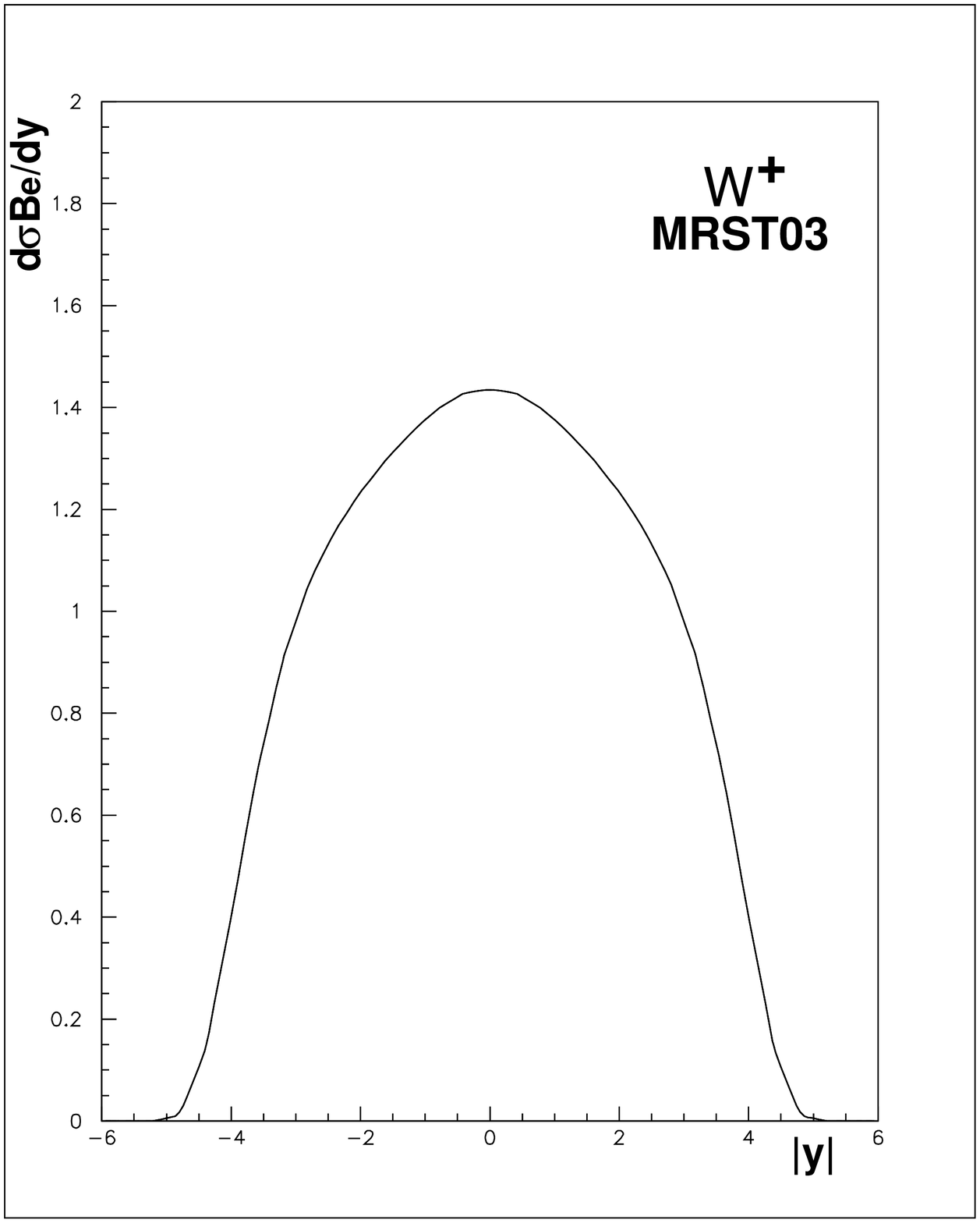,width=0.3\textwidth,height=4cm}
\epsfig{figure=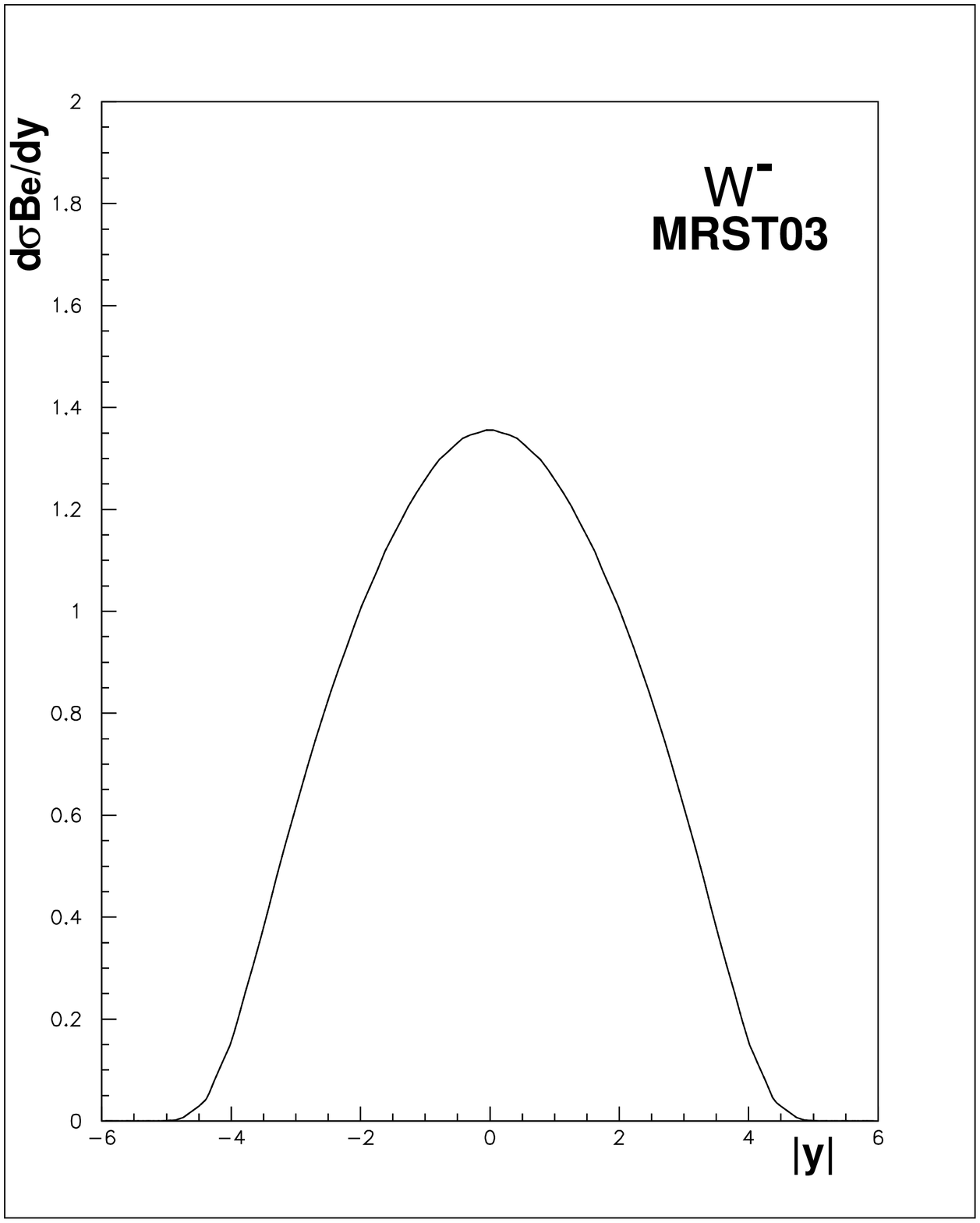,width=0.3\textwidth,height=4cm}
\epsfig{figure=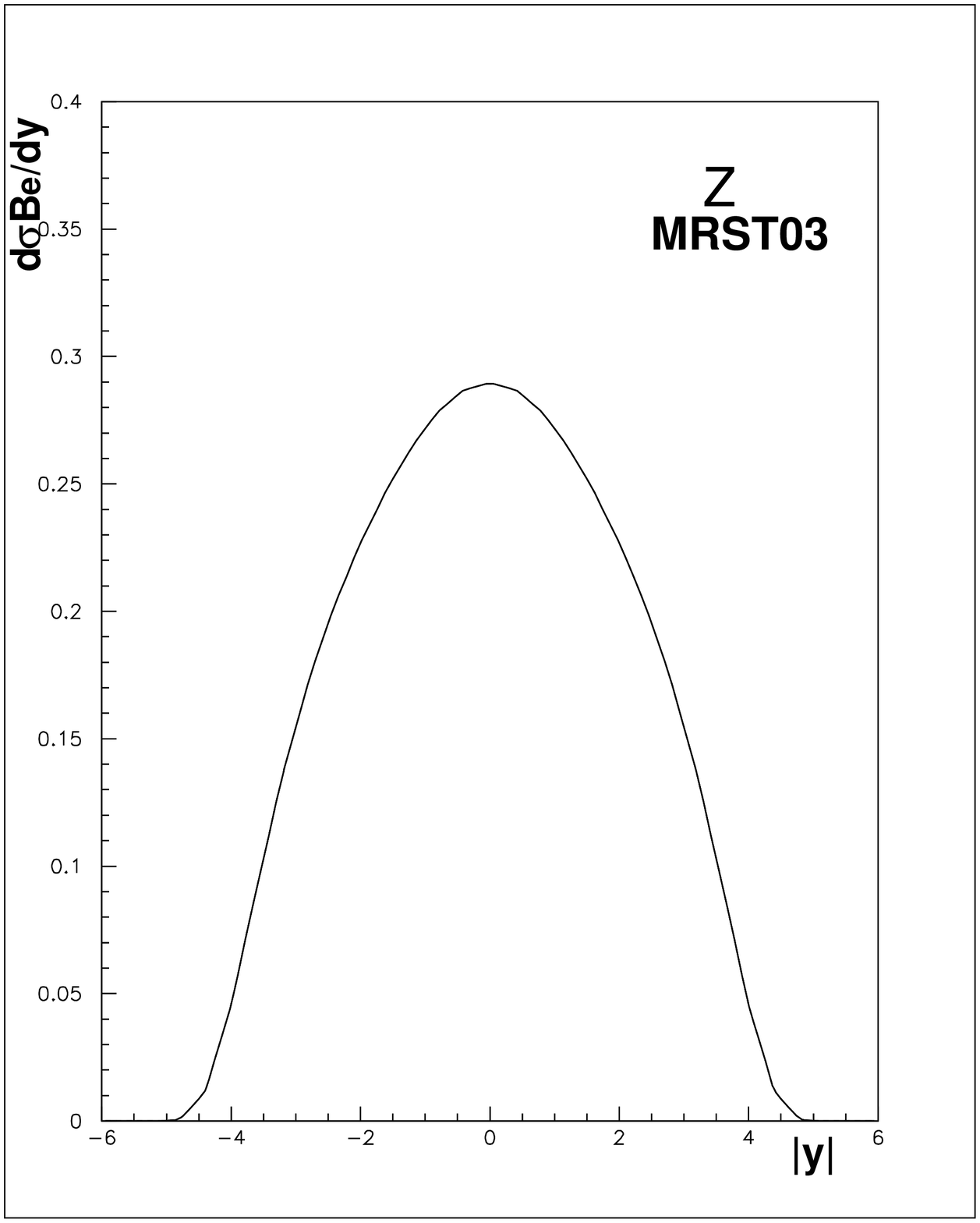,width=0.3\textwidth,height=4cm} 
}
\caption {LHC $W^+,W^-,Z$ rapidity distributions for the MRST03 PDFs: left plot $W^+$; middle plot $W^-$; 
right plot $Z$}
\label{fig:mrst03pred}
\end{figure}

The MRST group recently produced a PDF set, MRST03, which does not include any data for $x < 5\times 10^{-3}$.
The motivation behind this was as follows. In a global DGLAP fit to many data sets there is always a certain 
amount of tension between data sets. This may derive from the use of an inappropriate theoretical formalism
for the kinematic range of some of the data. Investigating the effect of kinematic cuts on the data, MRST 
found that a cut, $ x > 5\times 10^{-3}$, considerably reduced tension between the remaining data sets. An 
explanation may be the inappropriate use of the DGLAP formalism at small-$x$. 
The MRST03 PDF set is thus free of this bias 
BUT it is also only valid to use it for $x > 5\times 10^{-3}$. 
What is needed is an alternative theoretical formalism
for smaller $x$. However, the MRST03 PDF set may be used as a toy PDF set, to illustrate the effect of using 
very different PDF sets on our predictions. A comparison of Fig.~\ref{fig:mrst03pred} with 
Fig.~\ref{fig:WZrapFTZS13} or Fig.~\ref{fig:mrstcteq} shows how different the analytic 
predictions are from the conventional ones, and thus illustrates where we might  expect to see differences due 
to the need for an alternative formalism at small-$x$. 


\newpage


\subsection{$W$ and $Z$ production at the LHC\protect\footnote{Contributing author: Hasko Stenzel}} 
\label{sec:wjet}

The study of the production at the LHC of the electroweak bosons $W$ and $Z$ 
with subsequent decays in leptonic final states will provide several precision 
measurements of Standard Model parameters such as the mass of the $W$ boson 
or the weak mixing angle from the $Z$ boson forward-backward asymmetry. 
Given their large cross section and clean experimental signatures, the bosons will 
furthermore serve as calibration tool and luminosity monitor. More challenging, 
differential cross sections in rapidity or transverse momentum may be used 
to further constrain parton distribution functions. Eventually these measurements 
for single inclusive boson production may be applied to boson pair production in 
order to derive precision predictions for background estimates to discovery channels 
like $H\rightarrow W^+W^-$.

This contribution is devoted to the estimation of current uncertainties in the calculations 
for Standard Model cross sections involving $W$ and $Z$ bosons with particular emphasis 
on the PDF and perturbative uncertainties. All results are obtained at NLO 
with MCFM \cite{MCFM} version 4.0 interfaced to LHAPDF \cite{LHAPDF} for a convenient 
selection of various PDF families and evaluation of their intrinsic uncertainties. 
The cross sections are evaluated within a typical experimental acceptance and for 
momentum cuts summarised in Table~\ref{tab:expcuts}. The electromagnetic
 decays of 
$W$ and $Z$ are considered (massless leptons) and the missing transverse energy is 
assigned to the neutrino momentum sum (in case of $W$ decays). 
\begin{table}[htb]
\caption[cuts]{\label{tab:expcuts}{Experimental acceptance cuts used for the calculation 
of cross-sections.}}
\begin{center}
\begin{tabular}{ll}
&  \\
\hline
Observable & cut \\ \hline
$p_T^{lept}$ & $> 25$ GeV\\
$p_T^{jet}$ & $> 25$ GeV\\
$|\eta_{lept}|$ & $< 3.0$ \\
$|\eta_{jet}|$ & $< 4.0$ \\
$R(lepton-jet)$ & $> 0.8$ \\
$R(lepton-lepton)$ & $> 0.2$ \\
$E_T^{miss}$ & $> $25 GeV \\
\hline
\end{tabular}
\end{center}
\end{table}
Jets in the processes $W/Z+jets$ are produced in an inclusive mode with at least 
one jet in the event reconstructed with the $k_T$-algorithm. MCFM includes 
one- and two-jet processes at NLO and three-jet processes at LO. In the case of boson pair 
production the cuts of Table~\ref{tab:expcuts} can only be applied to 
the two leading leptons, hence a complete acceptance is assumed for additional leptons e.g. from $ZZ$ or $WZ$ decays.    
  
The calculations with MCFM are carried out for a given fixed set of electroweak 
input parameters using the effective field theory approach \cite{MCFM}. The  
PDF family CTEQ61 provided by the CTEQ collaboration \cite{Stump:2003yu} is 
taken as nominal PDF input while MRST2001E given by the MRST group \cite{Martin:2002aw} is considered for 
systematic purposes. The difference between CTEQ61 and MRST2001E alone can't be considered as 
systematic uncertainty but merely as cross-check. The systematic uncertainty is 
therefore estimated for each family separately with the family members, 40 for CTEQ61 and 
30 for MRST2001E, which are variants of the nominal PDF obtained with different assumptions  
while maintaining a reasonable fit of the input data. The value of $\alpha_s$ is not a free 
input parameter for the cross section calculation but taken from the corresponding value in the 
PDF. 

Important input parameters are renormalisation and factorisation scales. The central results 
are obtained with $\mu_R=\mu_F=M_V$, $V=W, Z$ for single boson production and $\mu_R=\mu_F=M_V+M_V^\prime$ 
for pair production ($V^\prime$ being the second boson in the event). Missing 
higher orders are estimated by a variation of the scales in the range 
$1/2 \leq x_{\mu R} \leq 2$ and independently  $1/2 \leq x_{\mu F} \leq 2$ where 
$\mu=x_\mu \cdot M_V$, following prescriptions applied to other processes \cite{Hasko}, keeping 
in mind that the range of variation of the scales is purely conventional.   

\subsubsection{Single $W$ and $Z$ cross sections}
Detailed studies of single $W$ and $Z$ production including detector simulation 
are presented elsewhere in these proceedings, here these channels are mainly studied 
for comparison with the associated production with explicitly reconstructed jets and with pair production. 
The selected process is inclusive in the sense that additional 
jets, present in the NLO calculation, are not explicitly reconstructed. The
experimentally required 
lepton isolation entailing a jet veto in a restricted region of phase space 
is disregarded at this stage.
   
\begin{figure}
\centering
\includegraphics[width=0.49\textwidth]{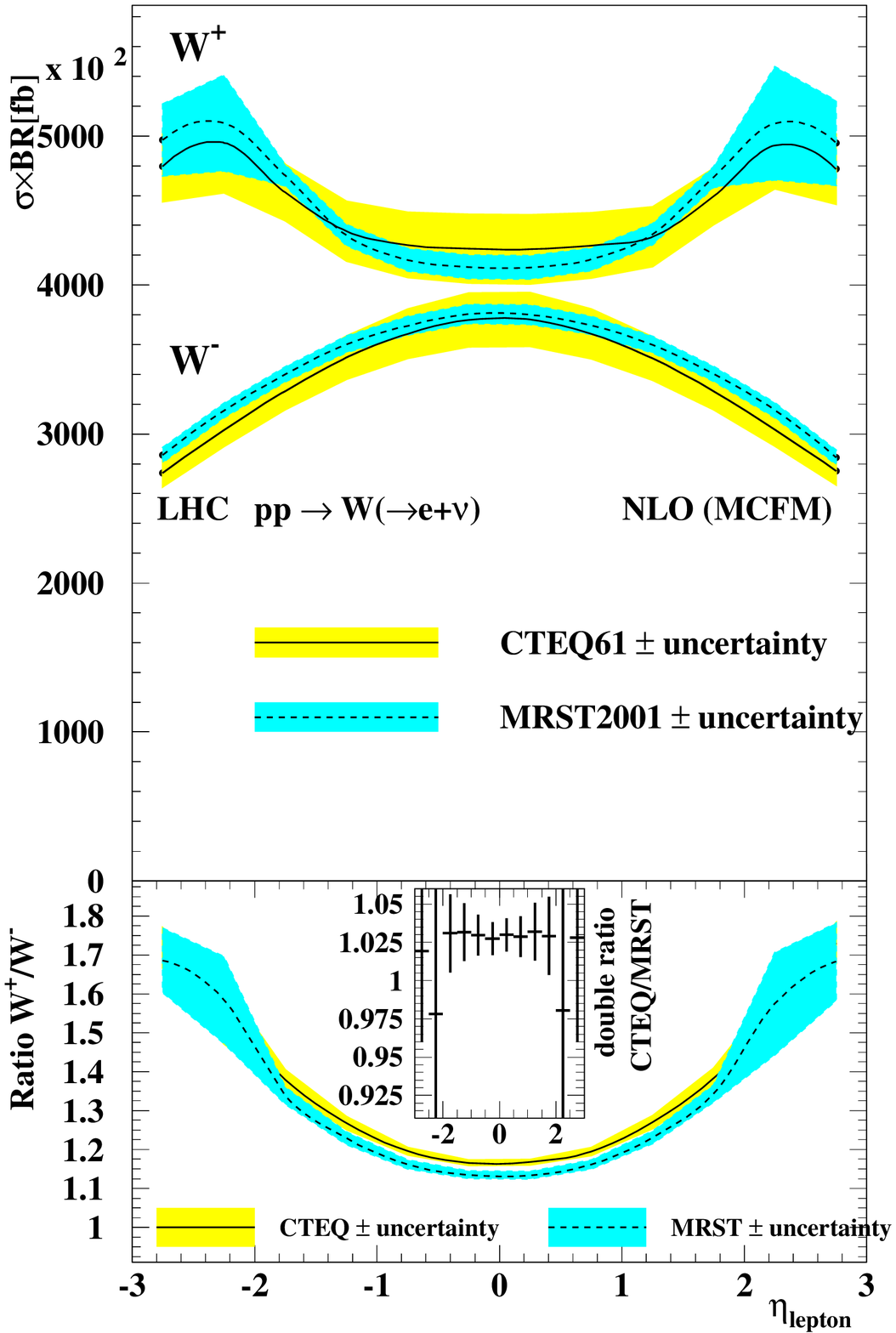}
\includegraphics[width=0.49\textwidth]{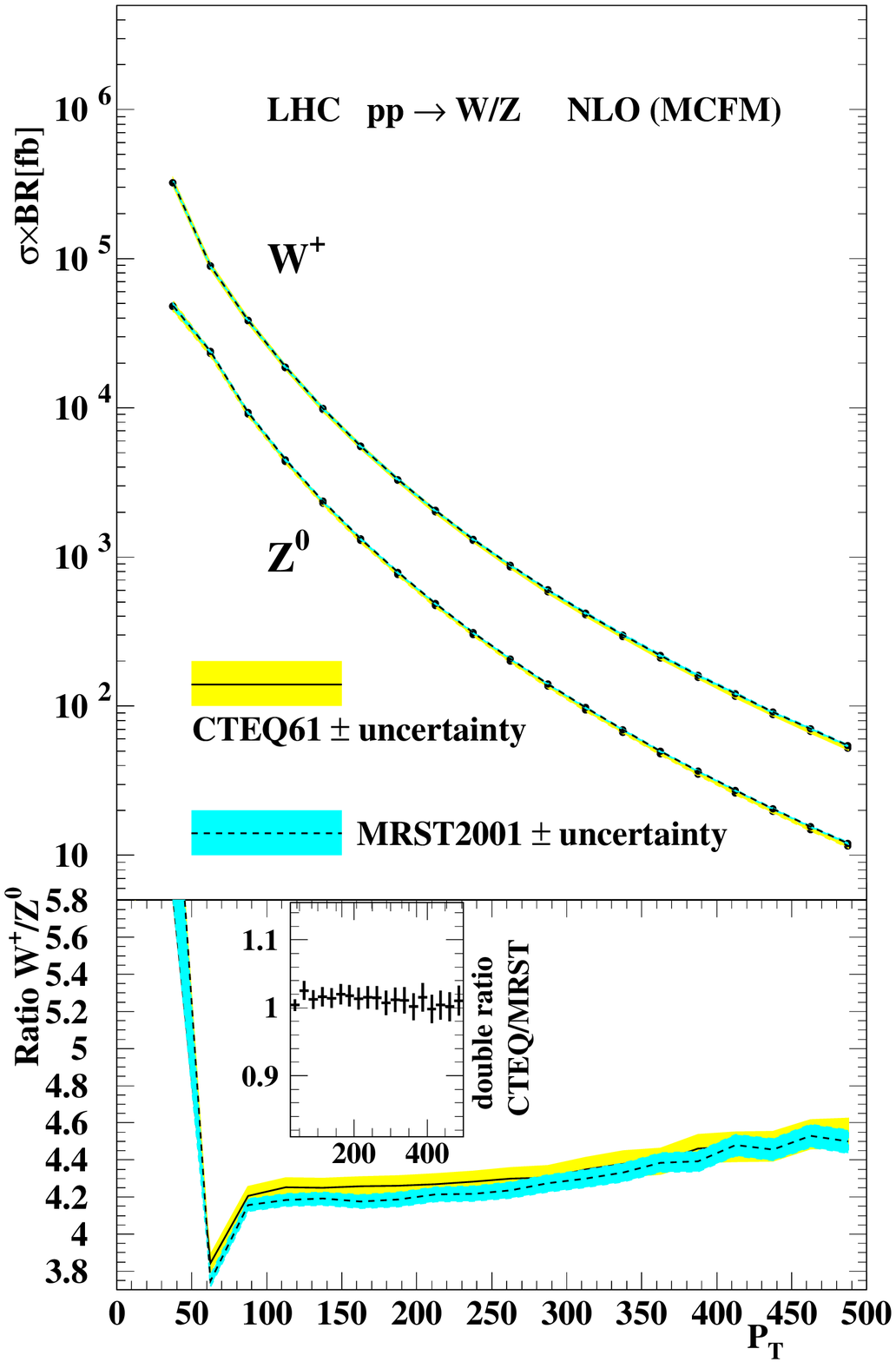}
\caption{Left: pseudo-rapidity distribution of the decay lepton from inclusive $W$ production and 
right: $p_T$ spectra of $W$ and $Z$. The bands represent the PDF-uncertainty.  
The lower inserts show on the left side the ratio $W^+/W^-$ resp. 
the double-ratio CTEQ/MRST 
and on the right side the ratios for $W^+/Z^0$. }
\label{fig:wz}       
\end{figure}

As an example the pseudo-rapidity distribution of the lepton from $W$ decays and the $p_T$ spectra  
for $Z$ and $W^+$ are shown in 
fig.~\ref{fig:wz}. 
The cross section for $W^+$ is larger than for $W^-$ as a direct consequence of the 
difference between up- and down-quark PDFs, and this difference survives in the 
pseudo-rapidity distribution of the decay lepton with a maximum around $|\eta|$=2.5. 
In the central part the PDF uncertainty, represented by the bands in fig.~\ref{fig:wz}, 
amounts to about 5$\%$ for CTEQ and 2$\%$ for MRST, and within the uncertainty 
CTEQ and MRST are fully consistent. Larger differences are visible in the peaks 
for the $W^+$, where at the same time the PDF uncertainty increases. In the ratio 
$W^+/W^-$ the PDF uncertainty is reduced to about 1-2$\%$ in the central region and 
a difference of about 3$\%$ is observed between CTEQ and MRST, as can be seen from the 
double-ratio CTEQ/MRST. The uncertainty of the double ratio is calculated from the CTEQ 
uncertainty band alone. 

In the case of $Z$ production the rapidity and $p_T$ spectra can be fully reconstructed 
from the $e^+e^-$ pair. A measurement of the $Z$ $p_T$ spectrum may be used to tune 
the Monte Carlo description of $W$ $p_T$, which is relevant for measurements of the $W$ mass.
The $p_T$ spectra are shown in the right part of fig.~\ref{fig:wz}. The total yield 
for $W^+$ is about six times larger than for $Z^0$ but for $p_T > 150$ GeV the ratio 
stabilises around 4.5. At small values of $p_T$ the fixed-order calculation becomes 
trustless and should be supplemented by resummed calculations. The PDF uncertainties 
for the $p_T$ spectra themselves are again about 5$\%$ and about 2$\%$ in the ratio, 
CTEQ and MRST being consistent over the entire $p_T$ range.  

\begin{figure}
\centering
\includegraphics[width=0.32\textwidth]{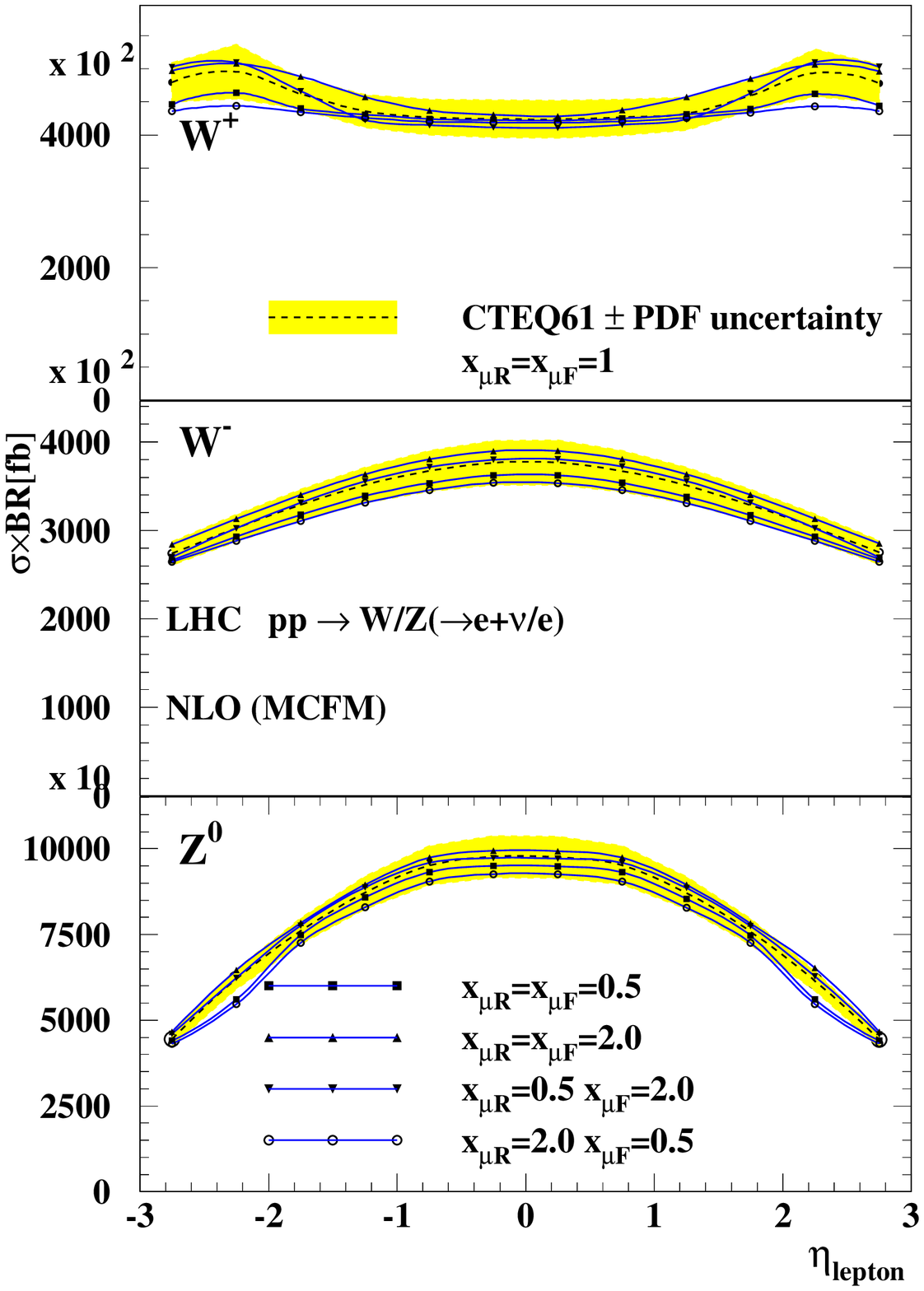}
\includegraphics[width=0.32\textwidth]{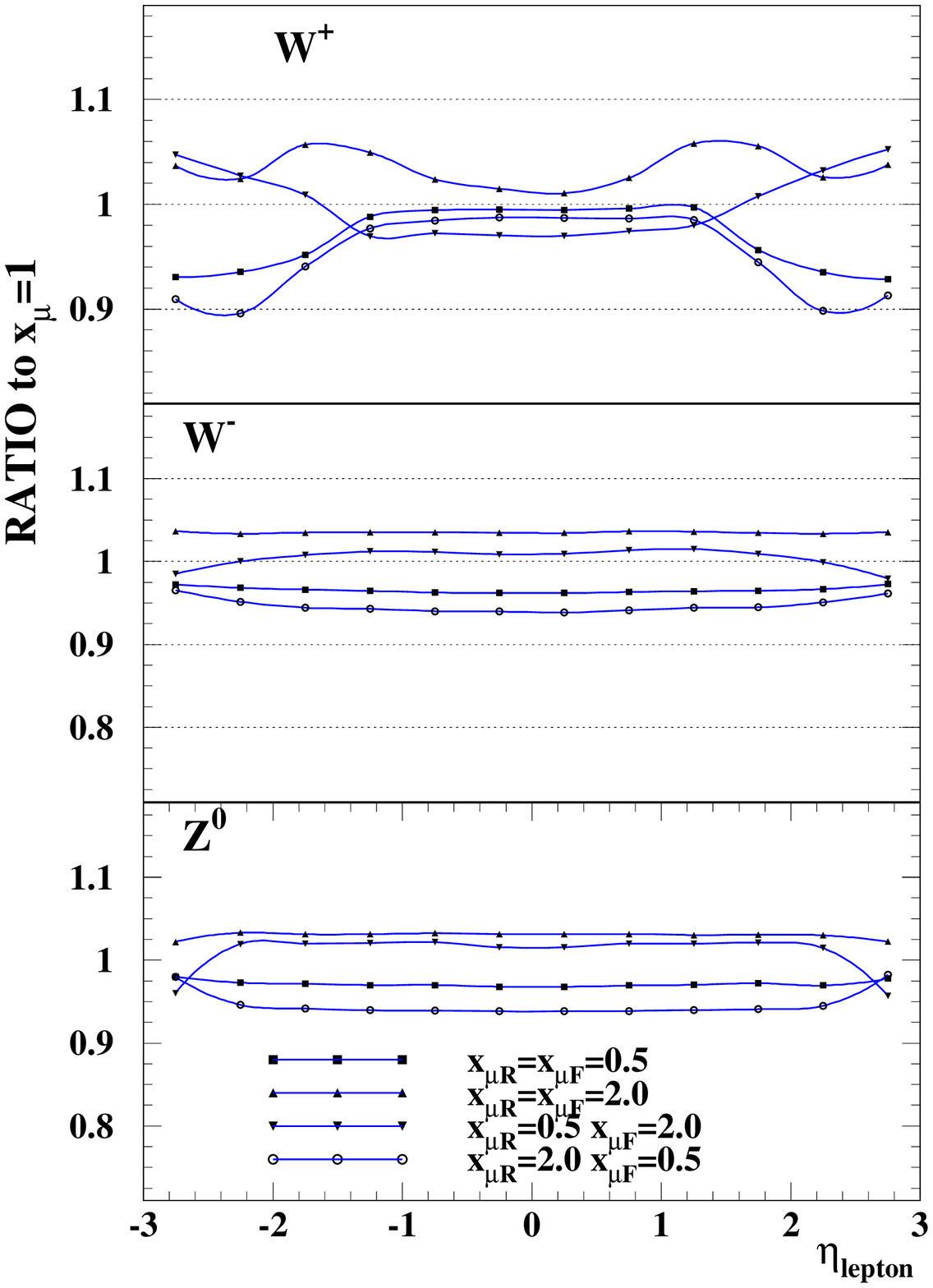}
\includegraphics[width=0.32\textwidth]{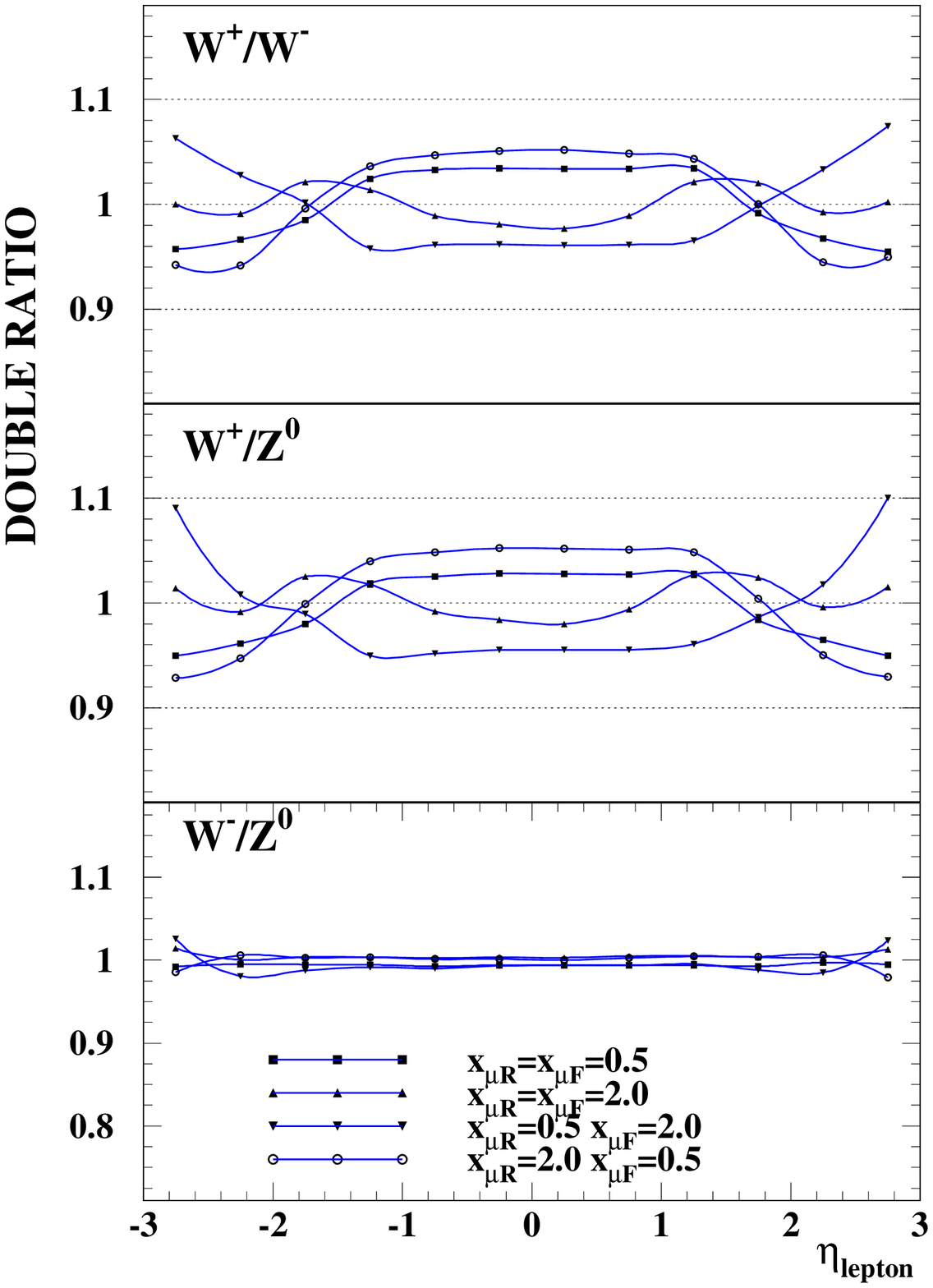}
\caption{Left: pseudo-rapidity distribution of the decay lepton from inclusive $W/Z$ production 
for different values of $x_{\mu R}$ and $x_{\mu F}=1$, 
centre: the ratio of predictions with respect to $x_\mu=1$ and 
right: double ratio $V/V^\prime$ of cross sections for actual scale settings normalised to the nominal scale. }
\label{fig:zwpert}       
\end{figure}

The perturbative uncertainties are estimated by variations of the renormalisation 
and factorisation scales in by a factor of two. The scale variation entails a global change 
in the total cross section of the order of 5$\%$. The $\eta$ distribution of leptons 
from $W/Z$ decays are shown in fig.~\ref{fig:zwpert}, comparing the nominal cross section 
with $x_{\mu R}=x_{\mu F}=1$, to alternative scale settings. The nominal cross section is 
drawn with its PDF uncertainty band, illustrating that the perturbative uncertainties are 
of the same size. For $W^-$ and $Z^0$ the shape of the 
distribution is essentially unaltered, but for $W^+$ the region around the maxima is 
changed more than the central part, leading to a shape deformation. The scale variation 
uncertainty is strongly correlated for $W^-$ and $Z^0$ and cancels in the ratio $W^-/Z^0$, 
but for $W^+$ it is almost anti-correlated with $W^-$ and $Z^0$ and partly enhanced 
in the ratio.    
    
Globally the perturbative uncertainty is dominated by the asymmetric scale setting 
$x_{\mu R} = 2, x_{\mu R} = 1/2$ for which a change of $-5\%$ is observed, 
the largest upward shift of $3.5\%$ is obtained for $x_{\mu R} = 2, x_{\mu R} = 2$, 
locally the uncertainty for $W^+$ can be much different.    
It can be expected that the perturbative uncertainties are reduced for NNLO 
calculations to the level of $1\%$.

The integrated cross sections and systematic uncertainties within the experimental acceptance 
are summarised in Table~\ref{tab:wz}. 
 
\begin{table}[htb]
\caption[single W/Z]{\label{tab:wz}{Total cross-sections and systematic uncertainties within the experimental acceptance.}}
\begin{center}
\begin{tabular}{|l|ccc|}
\hline
            & $W^+$ & $W^-$ & $Z^0$ \\ \hline
CTEQ61 [pb] & 5438 & 4002 & 923.9 \\
& & &\\
$\Delta_{PDF}^{CTEQ}$ [pb] & $\pm 282$ & $\pm 221$ & $\pm 49.1$ \\
& & & \\
$\Delta_{PDF}^{CTEQ}$ [$\%$] & $\pm 5.2$ & $\pm 5.5$ & $\pm 5.3$ \\
& & & \\
\hline
MRST [pb] & 5480 & 4110 & 951.1 \\
& & &\\
$\Delta_{PDF}^{MRST}$ [pb] & $\pm 103$ & $\pm 83.4$ & $\pm 17.4$ \\
& & & \\
$\Delta_{PDF}^{MRST}$ [$\%$] & $\pm 1.9$ & $\pm 2.1$ & $\pm 1.9$ \\
& & & \\
\hline
$\Delta_{pert}$ [$\%$] & $+ 3.5$ & $+ 3.5$ & $+ 3.1$ \\
                    & $- 5.2$ & $- 5.4$ & $- 5.5$ \\

\hline
\end{tabular}
\end{center}
\end{table}

\subsubsection{$W/Z+jet$ production}
In the inclusive production of $W/Z+jet$ at least one jet is requested to 
be reconstructed, isolated from any lepton by $R>0.8$. Additional jets 
are in case of overlap eventually merged at reconstruction level by the 
$k_T$-prescription. Given the presence of a relatively hard ($p_T>25$ GeV) jet, 
it can be expected that PDF- and perturbative uncertainties are different 
than for single boson production. The study of this process at the LHC, 
other than being a stringent test of perturbative QCD, may in addition 
contribute to a better understanding of the gluon PDF. 

\begin{figure}
\centering
\includegraphics[width=0.49\textwidth]{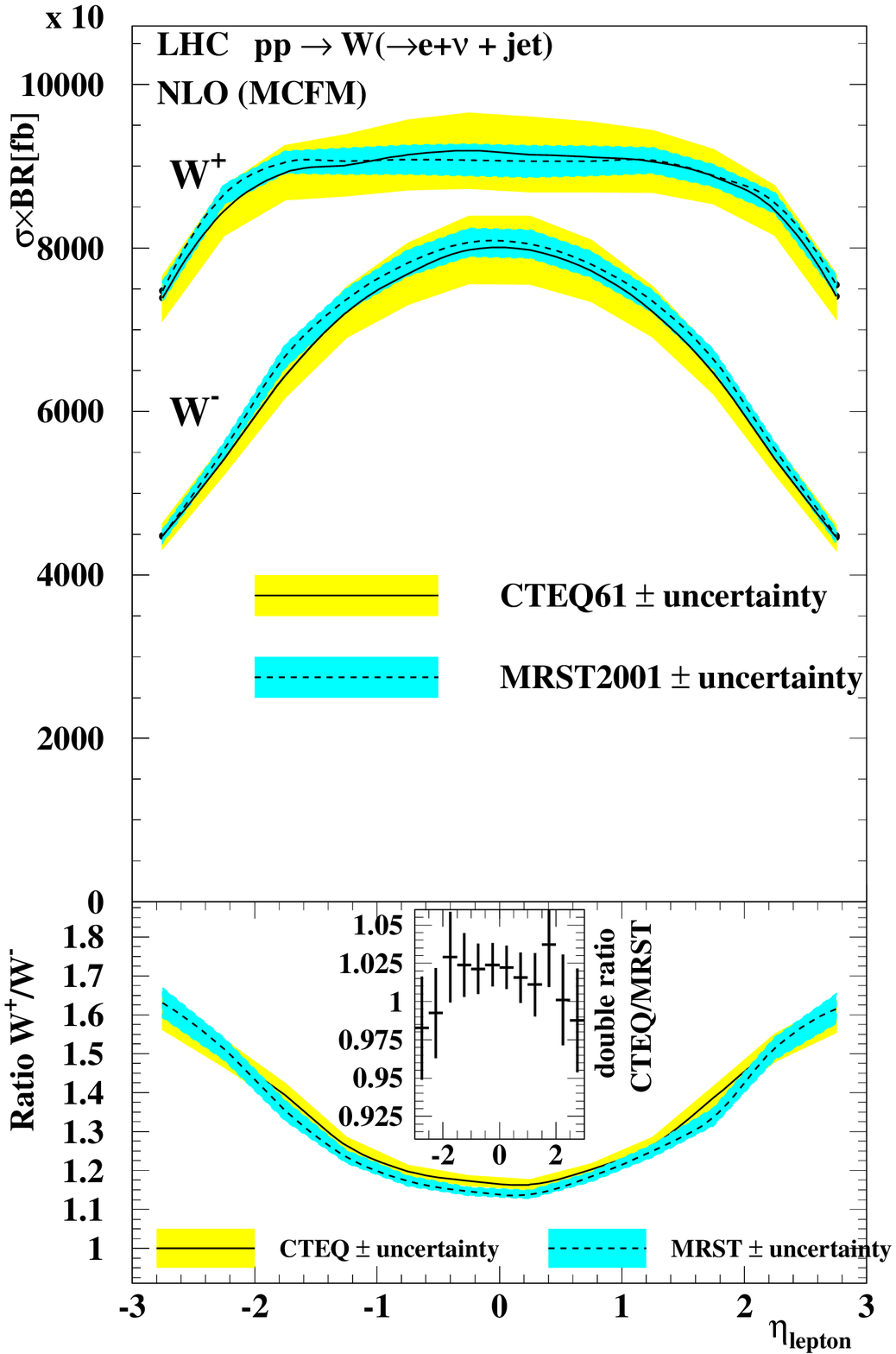}
\includegraphics[width=0.49\textwidth]{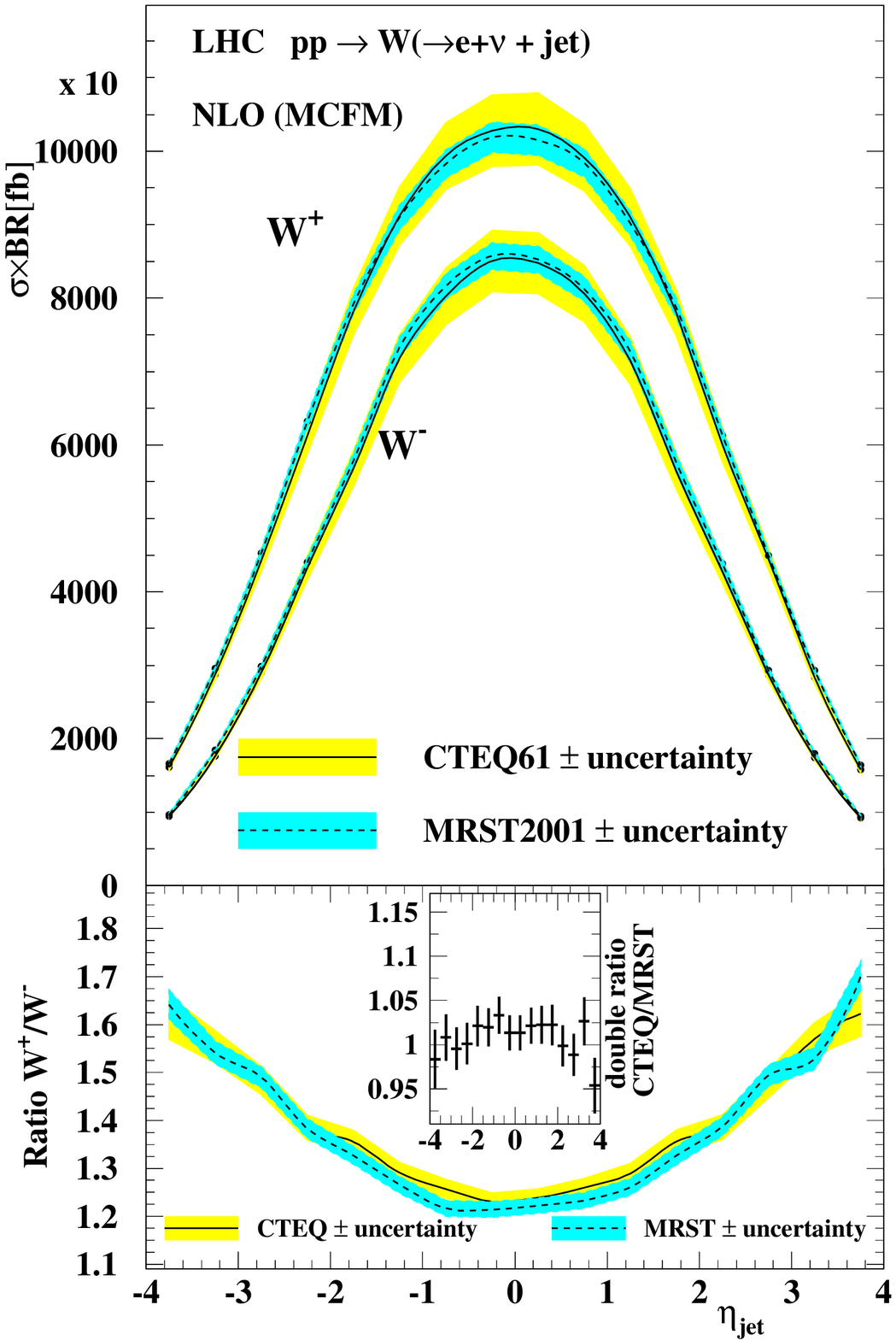}
\caption{Left: pseudo-rapidity distribution of the decay lepton from inclusive $W$+jet production and 
right: pseudo-rapidity of the associated leading jet. The bands represent the PDF-uncertainty.}  
\label{fig:wj}       
\end{figure}

The first difference with respect to single boson production appears 
in the lepton pseudo-rapidities, shown in fig.~\ref{fig:wj}. The peaks in 
the lepton spectrum from $W^+$ disappeared, the corresponding 
spectrum from $W^-$ is stronger peaked at central rapidity while 
the ratio $W^+/W^-$ with jets is essentially the same as without jets. 
The PDF uncertainties are slightly smaller (4.2-4.4$\%$) compared to single 
bosons. The jet pseudo-rapidities are shown in the right part of fig.~\ref{fig:wj}, 
they are much stronger peaked in the central region but the ratio $W^+/W^-$ 
for jets is similar to the lepton ratio.    
 
\begin{figure}
\centering
\includegraphics[width=0.6\textwidth]{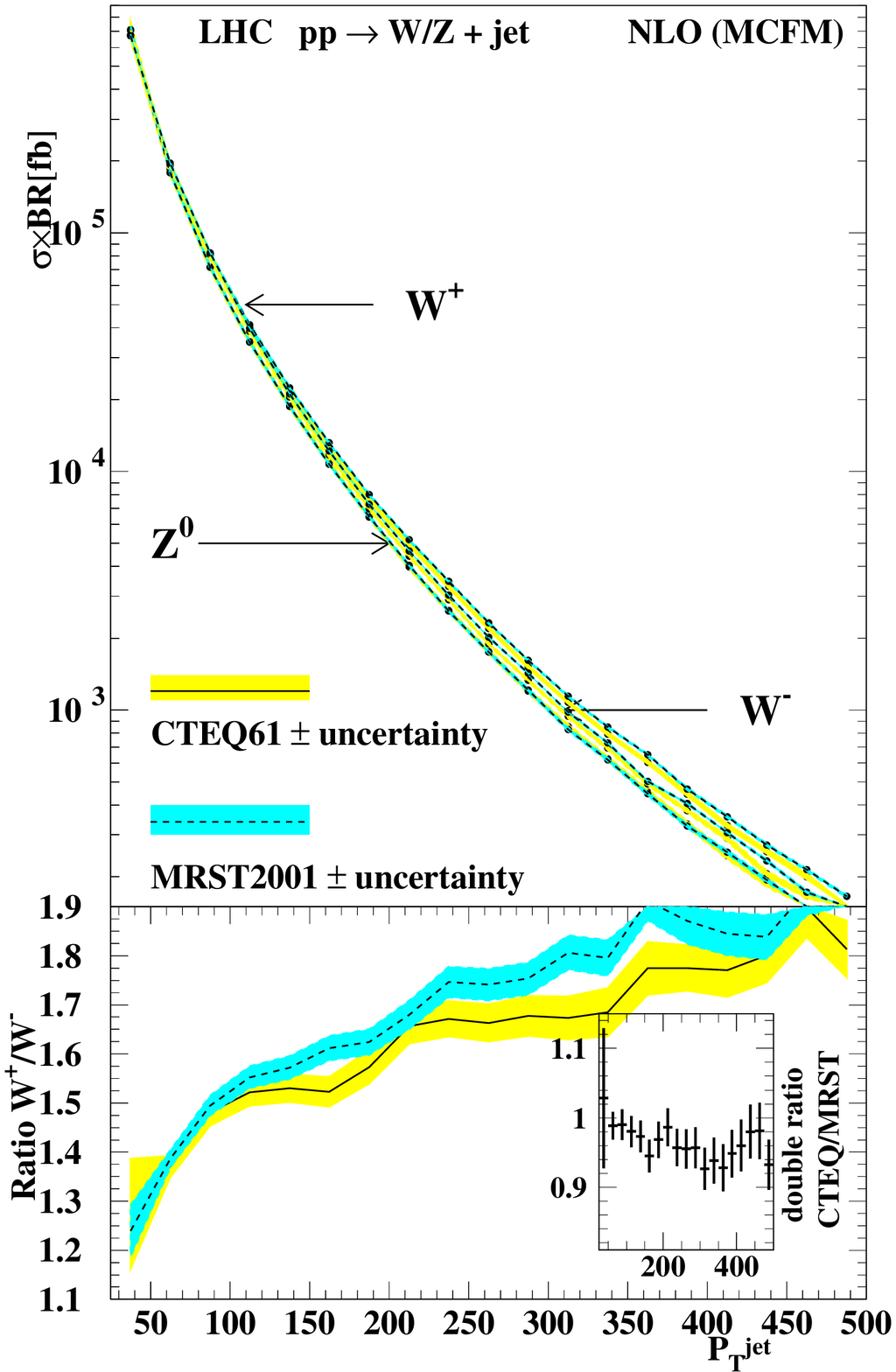}
\caption{Transverse momentum distribution of the jet from inclusive $W/Z+jet$ production}  
\label{fig:jetpt}       
\end{figure}

The transverse momenta of associated jets from $W/Z+jet$ production is shown in 
fig.~\ref{fig:jetpt}, the spectra are steeply falling and the ratio  $W^+/W^-$ is 
increasing from $1.3$ at low $p_T$ to almost 2 at 500 GeV $p_T$.  

\begin{figure}
\centering
\includegraphics[width=0.32\textwidth]{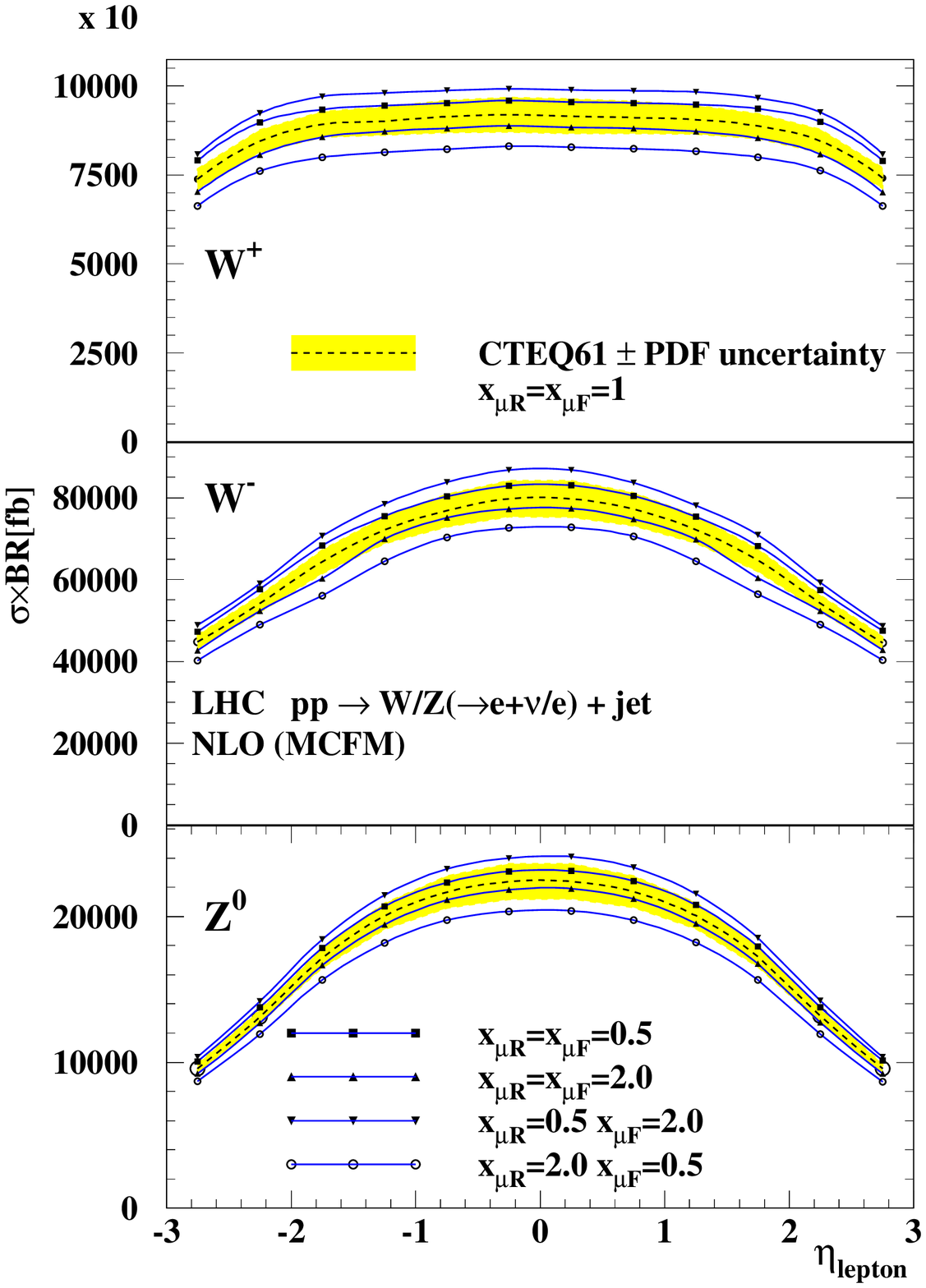}
\includegraphics[width=0.32\textwidth]{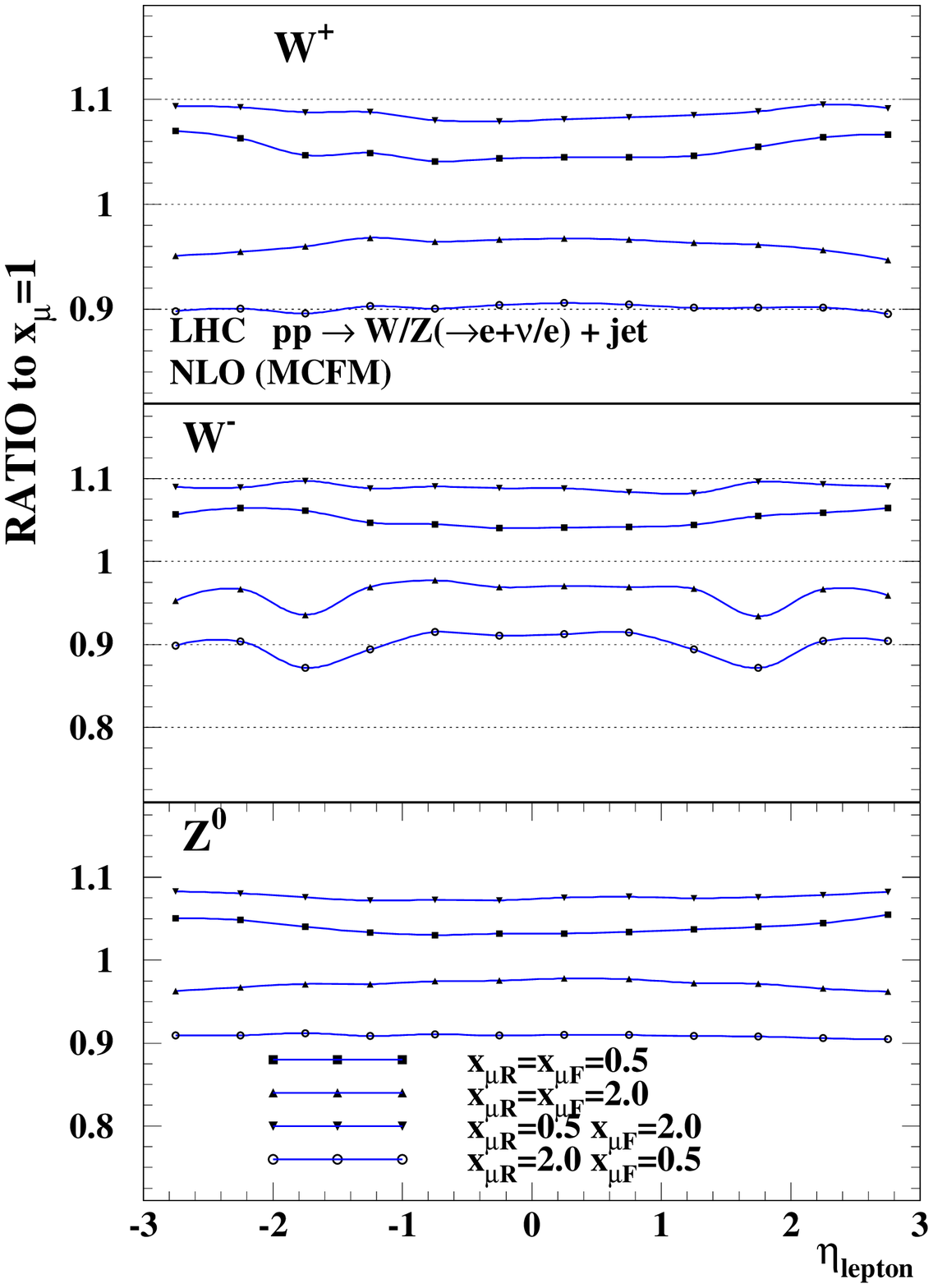}
\includegraphics[width=0.32\textwidth]{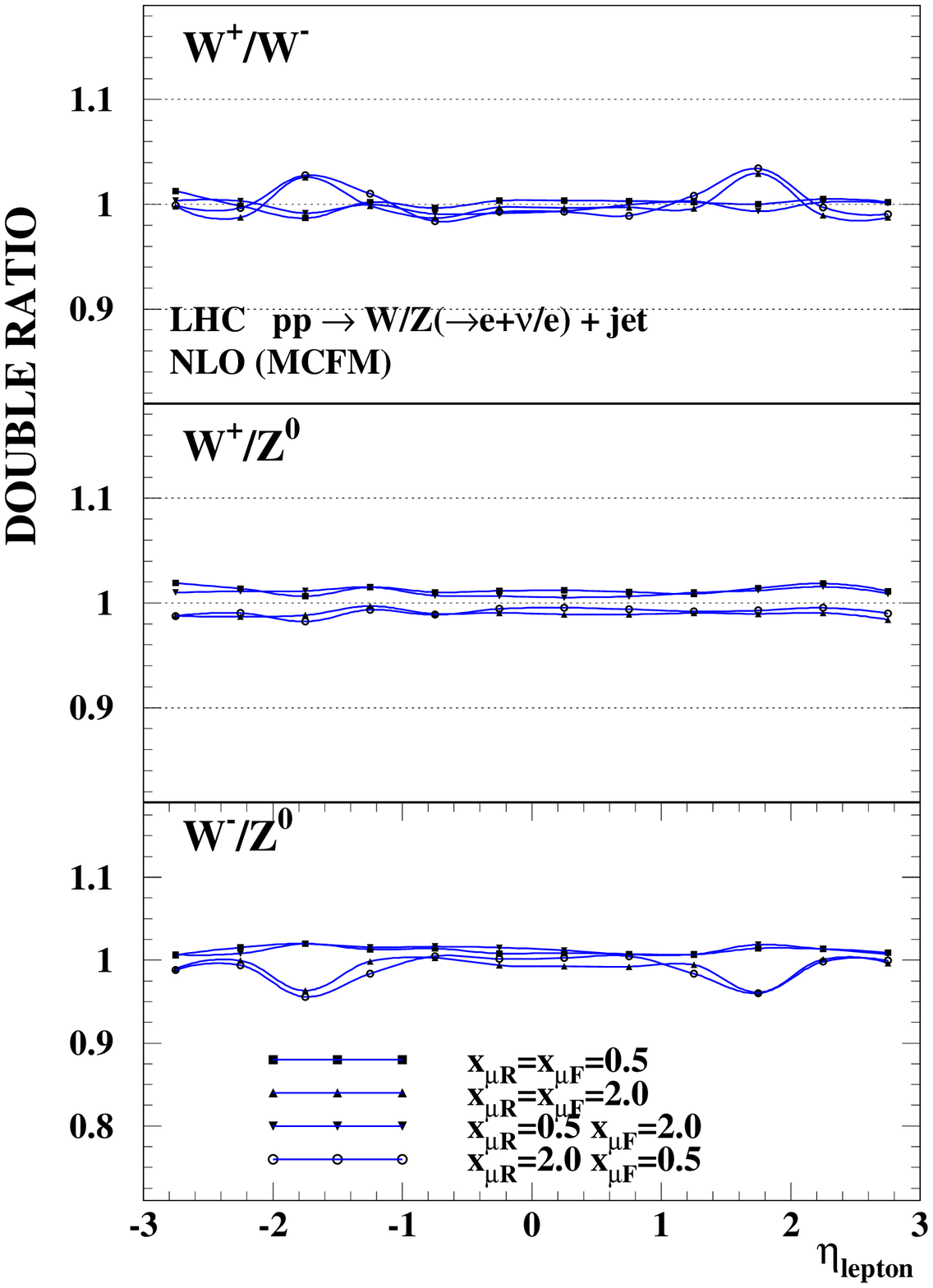}
\caption{Left: pseudo-rapidity distribution of the decay lepton from inclusive $W/Z+jet$ 
production 
for different values of $x_{\mu R}$ and $x_{\mu F}=1$, 
centre: the ratio of predictions with respect to $x_{\mu}=1$ and 
right: double ratio $V/V^\prime$ of cross sections for actual scale settings normalised to the nominal scale.}
\label{fig:zwjpert}       
\end{figure}

The perturbative uncertainties are investigated in the same way as 
for the single boson production and are shown in fig.~\ref{fig:zwjpert}.
The scale variation entails here a much larger uncertainty between 
8 and 10$\%$, almost twice as large as for single bosons. In contrast to the latter case, 
the scale variation is correlated for $W$ and $Z$ and cancels in the 
ratio $W^+/W^-$, with an exception for $W^-$ where a bump appears at $|\eta|=1.8$ 
for $x_{\mu R}=2$.  
   
The total cross sections and their systematic uncertainties are summarised in 
Table~\ref{tab:wzj}. 

\begin{table}
\caption[W/Z+jet]{\label{tab:wzj}{Total cross-sections and systematic uncertainties within the experimental acceptance for $W/Z+jet$ processes.}}
\begin{center}
\begin{tabular}{|l|ccc|}
\hline
            & $W^++jet$ & $W^-+jet$ & $Z^0+jet$ \\ \hline
CTEQ61 [pb] & 1041 & 784.5 & 208.1 \\
& & &\\
$\Delta_{PDF}^{CTEQ}$ [pb] & $\pm 44.1$ & $\pm 34.3$ & $\pm 9.01$ \\
& & & \\
$\Delta_{PDF}^{CTEQ}$ [$\%$] & $\pm 4.2$ & $\pm 4.4$ & $\pm 4.3$ \\
& & & \\
\hline
MRST [pb] & 1046 & 797.7 & 211.3 \\
& & &\\
$\Delta_{PDF}^{MRST}$ [pb] & $\pm 17.6$ & $\pm 14.8$ & $\pm 3.67$ \\
& & & \\
$\Delta_{PDF}^{MRST}$ [$\%$] & $\pm 1.7$ & $\pm 1.9$ & $\pm 1.8$ \\
& & & \\
\hline
$\Delta_{pert}$ [$\%$] & $+ 8.7$ & $+ 8.9$ & $+ 7.6$ \\
                    & $- 9.8$ & $- 10.0$ & $- 9.1$ \\

\hline
\end{tabular}
\end{center}
\end{table}
 
\clearpage
\subsubsection{Vector Boson pair production}
In the Standard Model the non-resonant production of vector bosons pairs in the 
continuum is suppressed by factors of $10^4$-$10^5$ with respect to single 
Boson production. The cross sections for $WW$, $WZ$ and $ZZ$ within the experimental
acceptance range from 500 fb ($WW$) to 10 fb ($ZZ$). Given the expected limited 
statistics for these processes, the main goal of their experimental study 
is to obtain the best estimate of the background they represent for searches 
of the Higgs boson or new physics yielding boson pairs.

The selection of boson pairs follows in extension the single boson selection cuts 
applied to 2, 3 or 4 isolated leptons. Again real gluon radiation and virtual loops have
 been taken into account at NLO but without applying lepton-jet isolation cuts. Lepton-lepton 
separation is considered only for the two leading leptons. 

\begin{figure}
\centering
\includegraphics[width=0.49\textwidth]{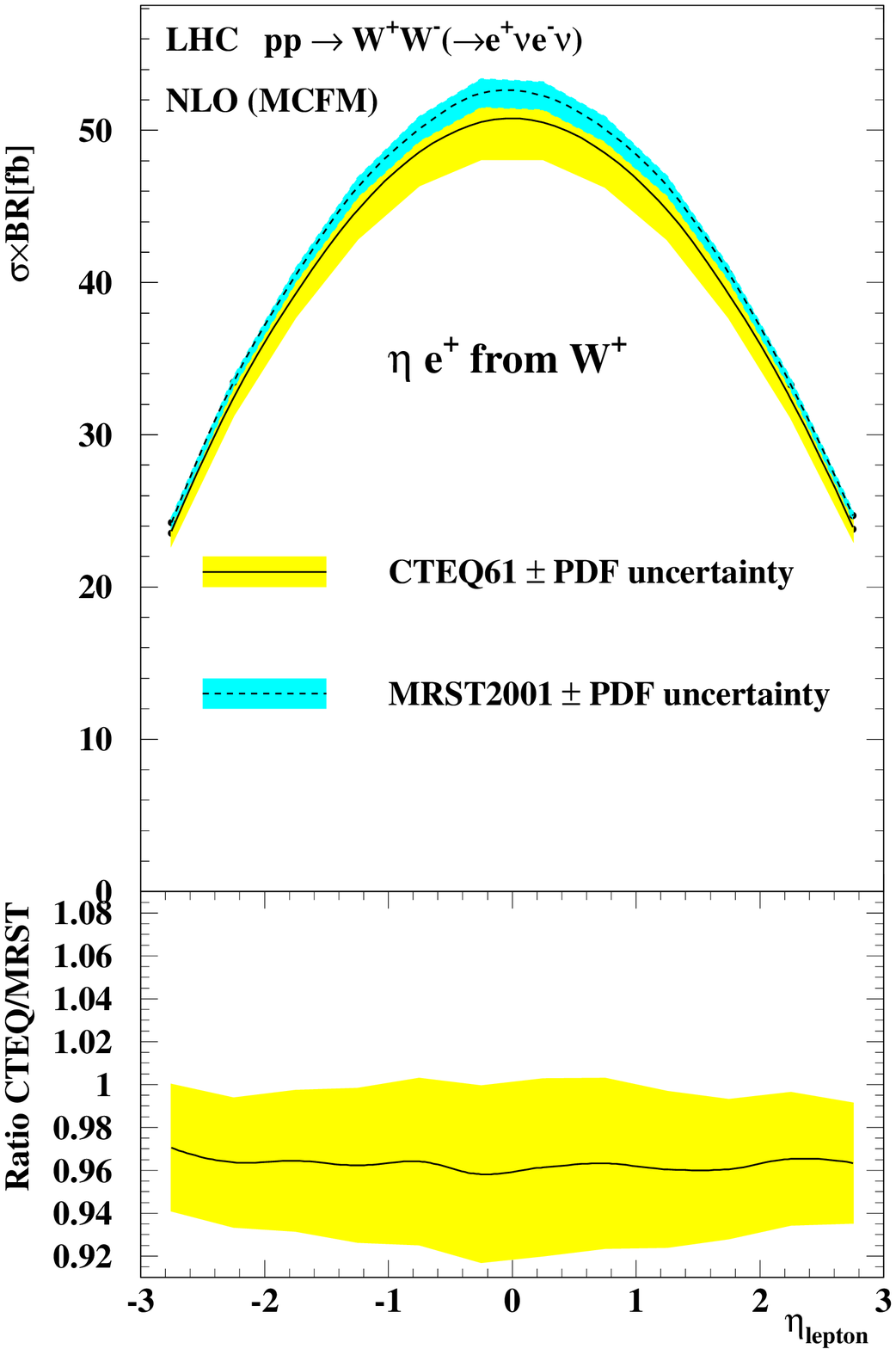}
\includegraphics[width=0.49\textwidth]{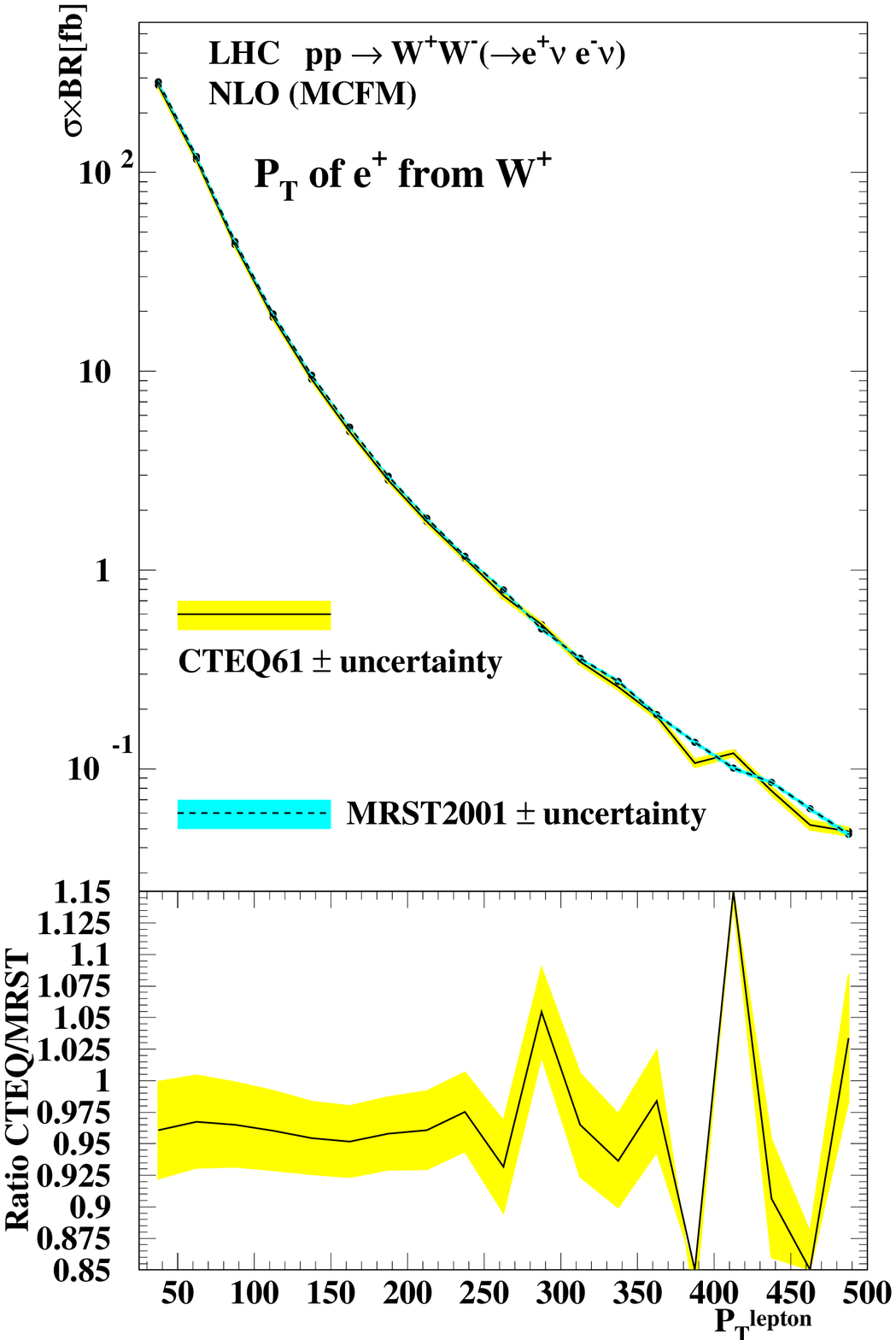}
\caption{Left: pseudo-rapidity distribution of the decay lepton from inclusive $WW$ production and 
right: transverse momentum of the decay lepton.}  
\label{fig:ww}       
\end{figure}

The pseudo-rapidity and transverse momentum distributions taking the $e^+$ from $W^+W^-$ 
production as example are shown in fig.\ref{fig:ww}. The pseudo-rapidity is 
strongly peaked and the cross section at $\eta=0$ twice as large as at $|\eta|=3$.
The PDF uncertainties are smaller than for single bosons, between 3.5 and 4 $\%$. 

\begin{figure}
\centering
\includegraphics[width=0.44\textwidth]{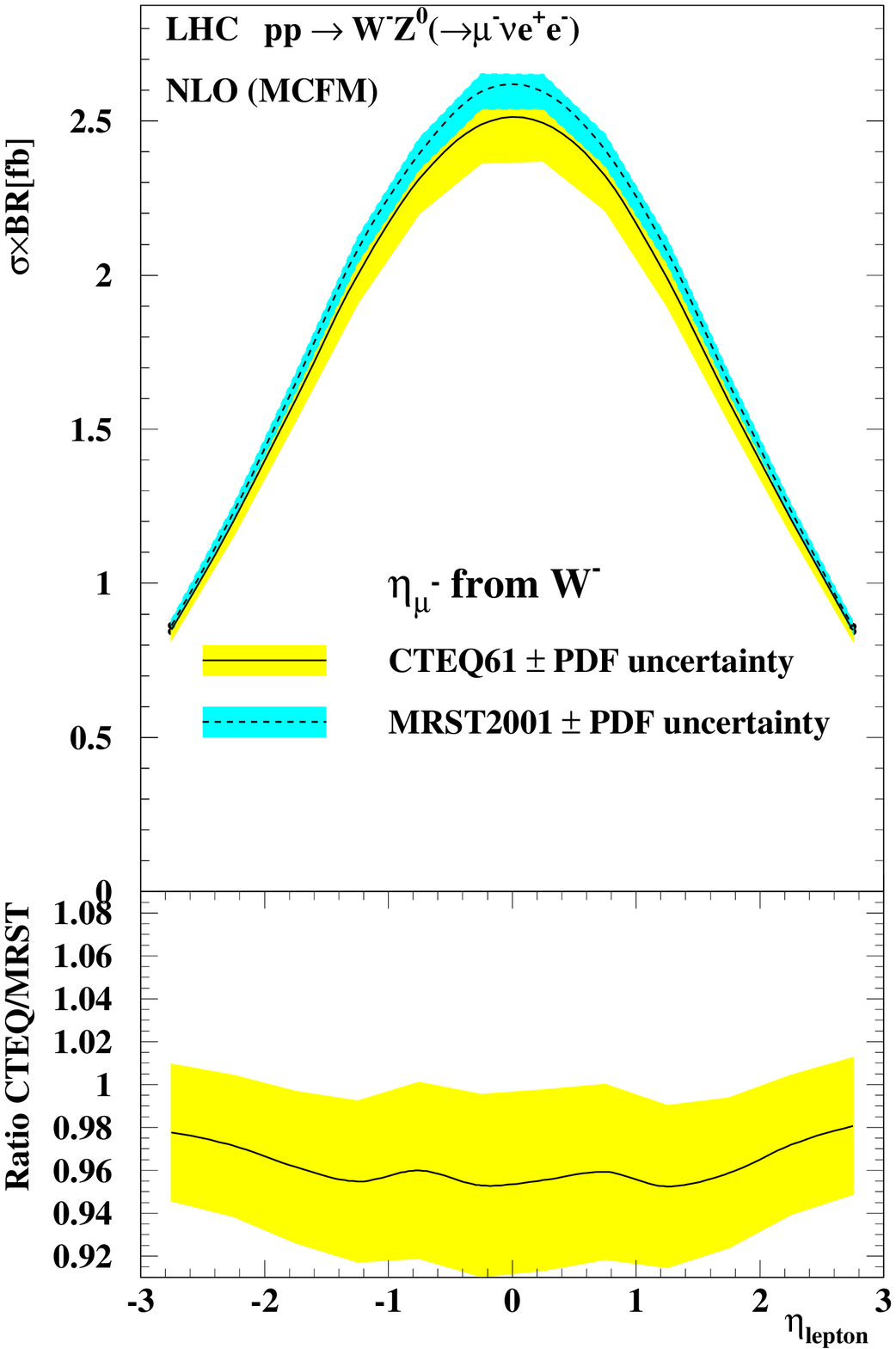}
\includegraphics[width=0.44\textwidth]{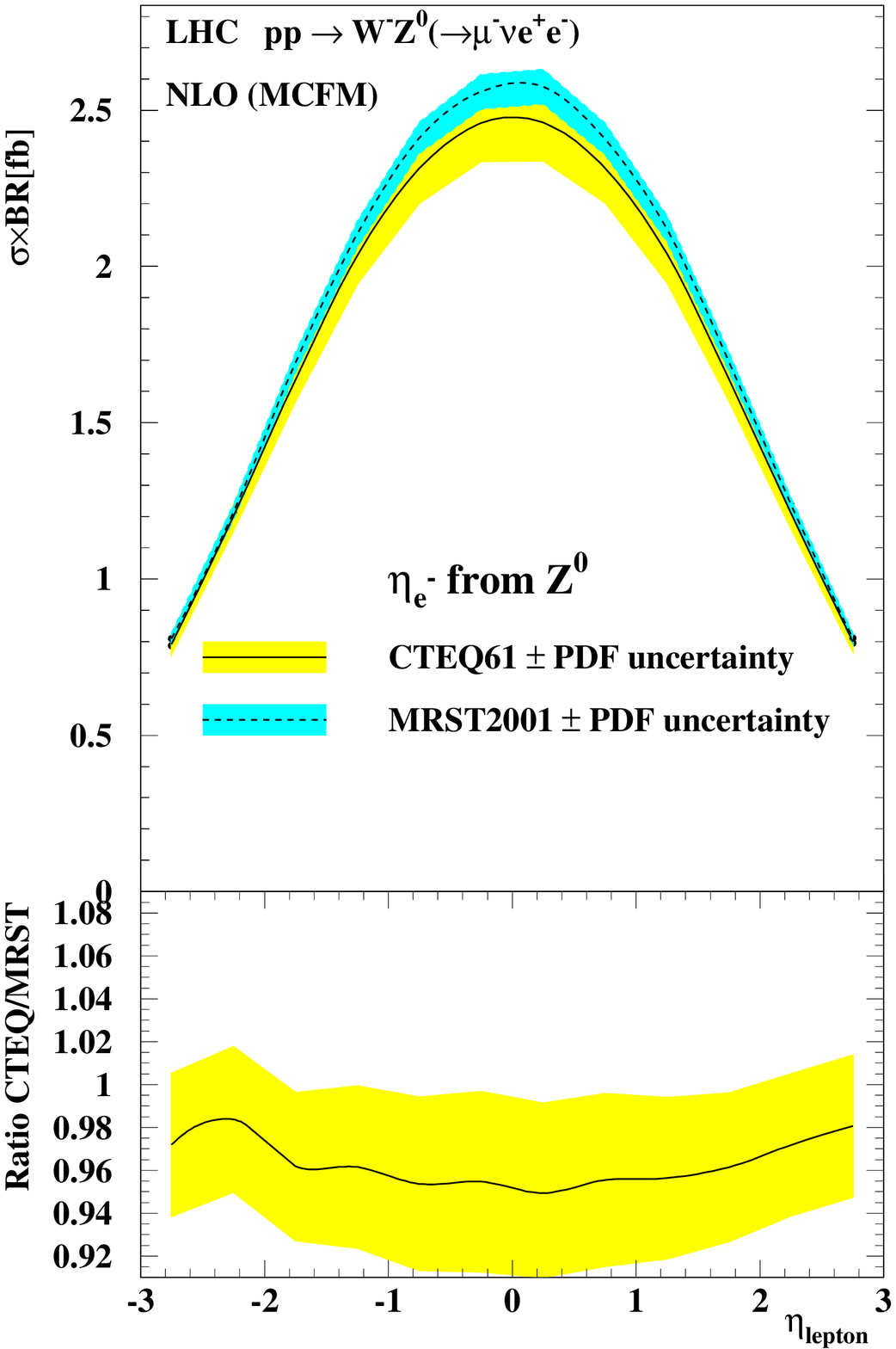}
\caption{Left: pseudo-rapidity distribution of the decay lepton of the $W^-$from inclusive $W^-Z^0$ production and 
right: pseudo-rapidity distribution of a decay lepton of the $Z^0$.}  
\label{fig:wmz}       
\end{figure}

The same shape of lepton distributions is also found for the other lepton and for
the other pair production 
processes, as shown for the $W^-Z^0$ case in fig.\ref{fig:wmz}.    
  
\begin{figure}
\centering
\includegraphics[width=0.44\textwidth]{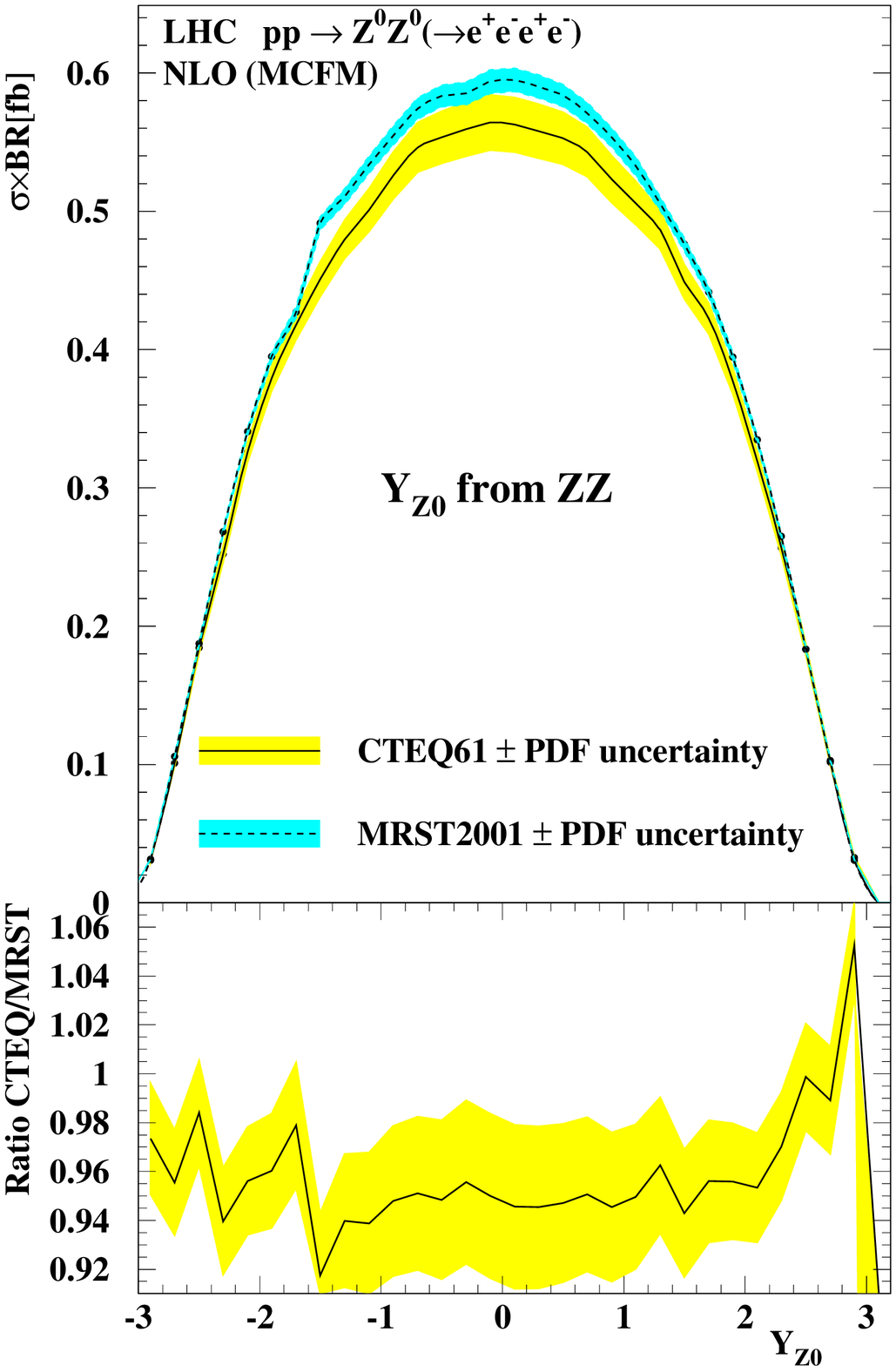}
\includegraphics[width=0.44\textwidth]{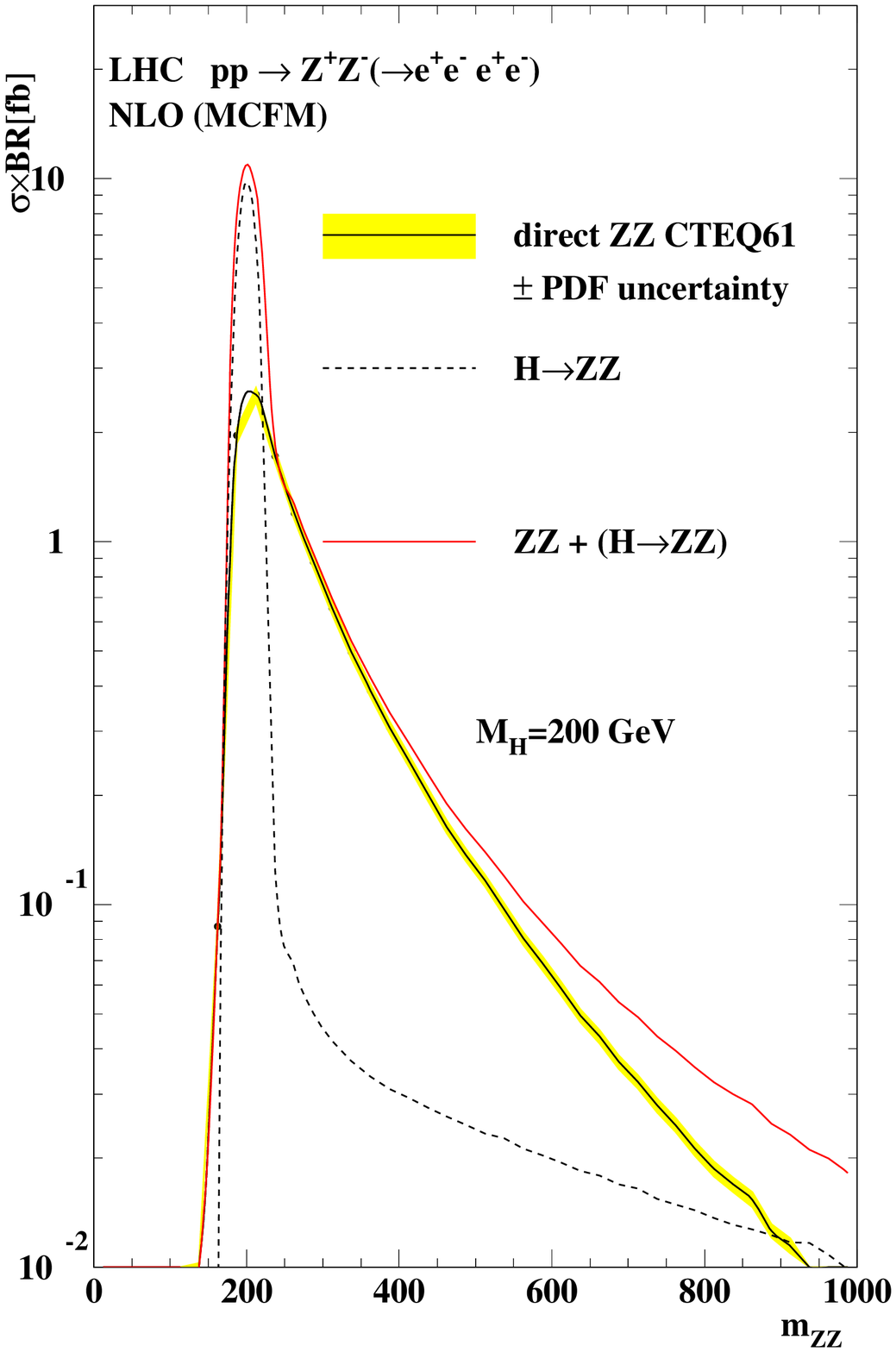}
\caption{Left: rapidity distribution of the leading $Z$ from inclusive $ZZ$ production and 
right: invariant mass of the $ZZ$ pair for non-resonant continuum production 
compared to resonant pair production via the SM Higgs decay.}  
\label{fig:zz}       
\end{figure}

The rapidity distribution of the leading $Z^0$ from $ZZ$ production is shown in the left part 
of fig.\ref{fig:zz}. With both $Z$'s being fully reconstructed, the invariant mass 
of the $ZZ$ system can be compared in the right part of fig.\ref{fig:zz} to the 
invariant mass spectrum of the Higgs decaying into the same final state for an  
intermediate mass of $m_H=200$ GeV. In this case a clear peak appears at low 
invariant masses above the continuum, and the mass spectrum is also harder at high 
masses in presence of the Higgs.

\begin{figure}
\centering
\includegraphics[width=0.44\textwidth]{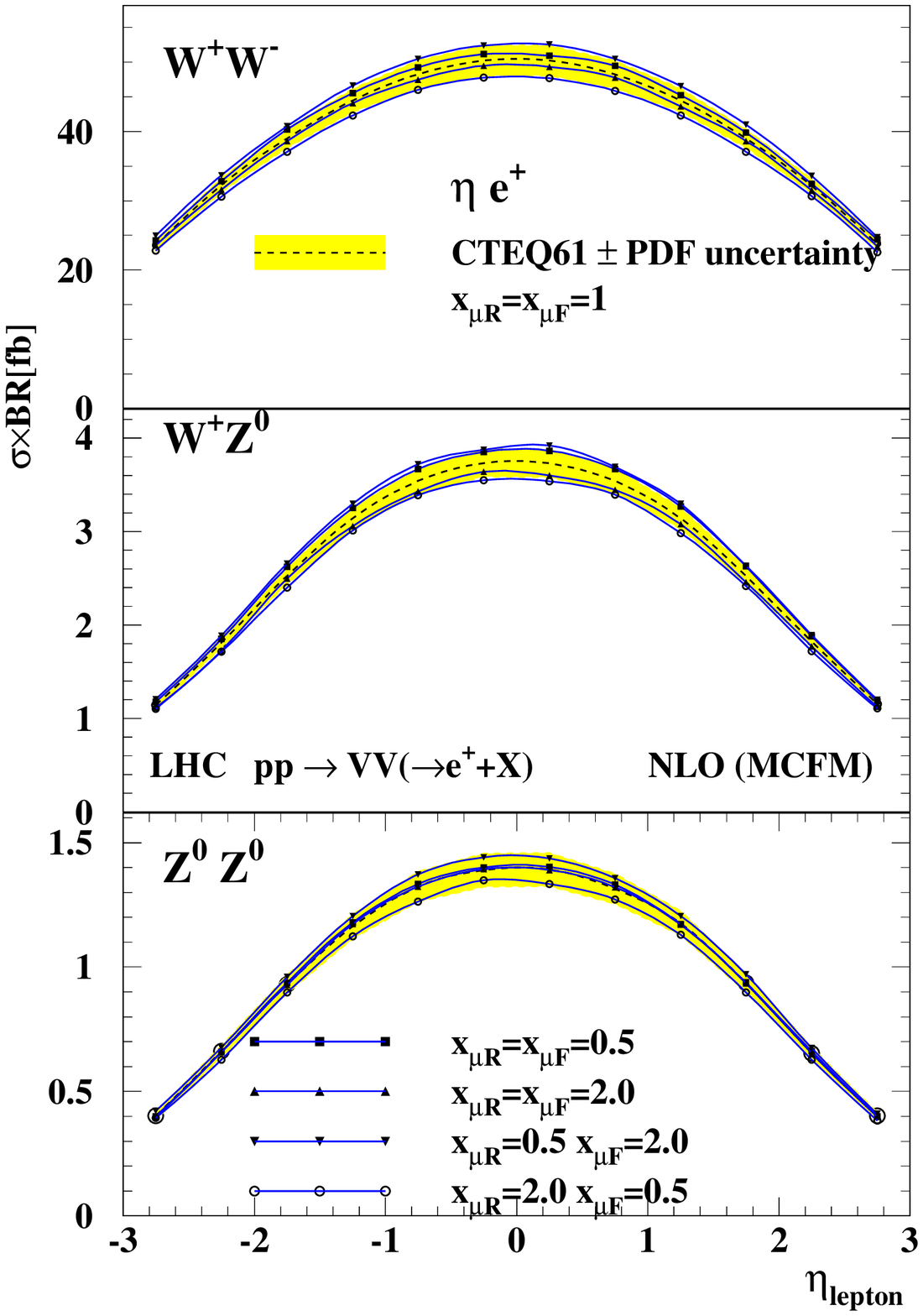}
\includegraphics[width=0.44\textwidth]{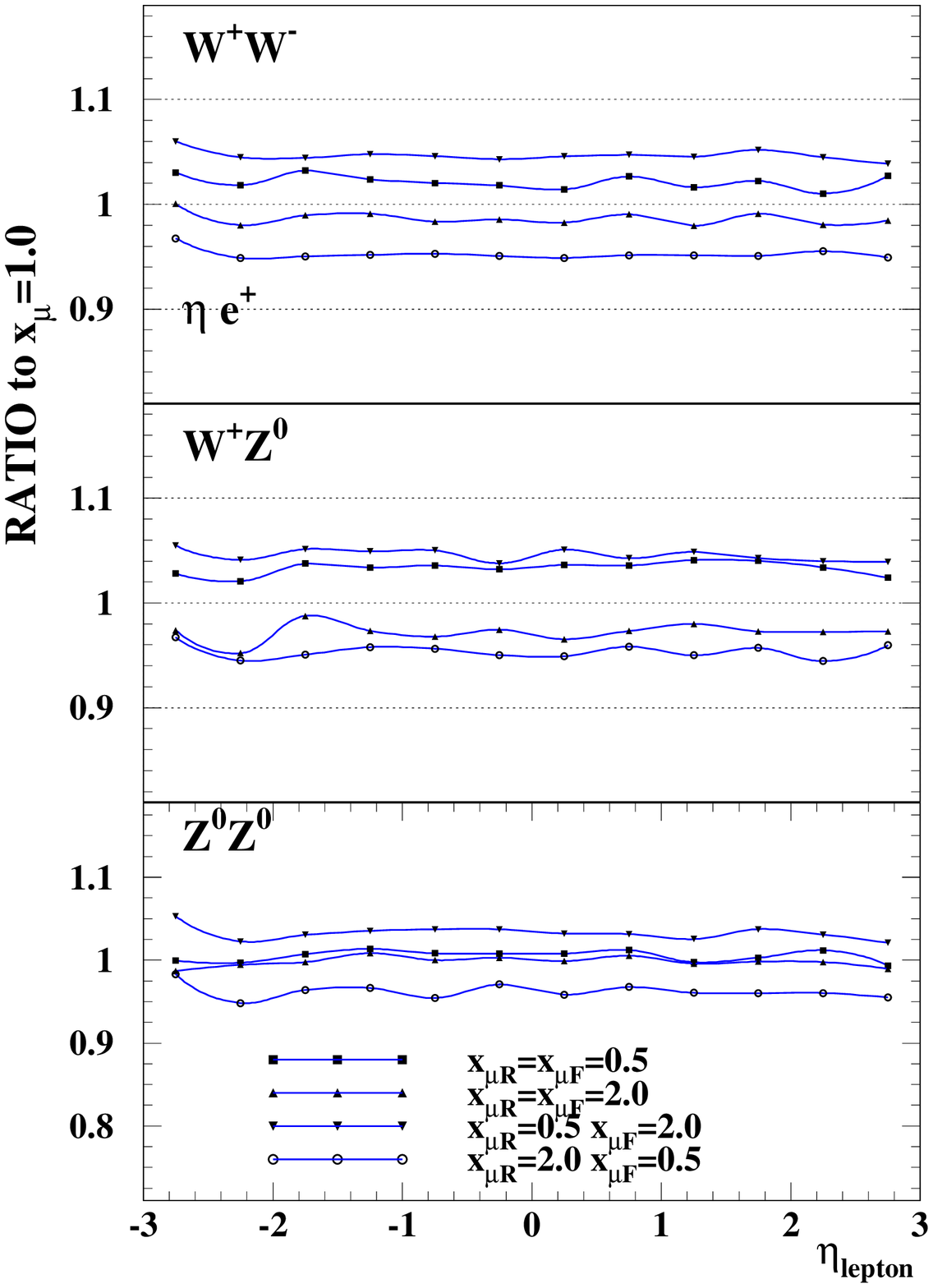}
\caption{Left: pseudo-rapidity distributions of leptons from various boson 
pair production processes and different scale settings and 
right: ratio of predictions relative to 
$x_\mu=1$.}  
\label{fig:vvscale}       
\end{figure}

The perturbative uncertainties, obtained as for the other processes, are shown 
in fig.\ref{fig:vvscale} for the lepton distributions. The systematic 
uncertainties range from 3.3 to 4.9 $\%$ and are slightly smaller than for 
single bosons, given the larger scale $\mu = 2M_V$ and better applicability of 
perturbative QCD. The perturbative uncertainty is essentially constant across 
the pseudo-rapidity and largely correlated between different pair production processes.  

\begin{figure}
\centering
\includegraphics[width=0.44\textwidth]{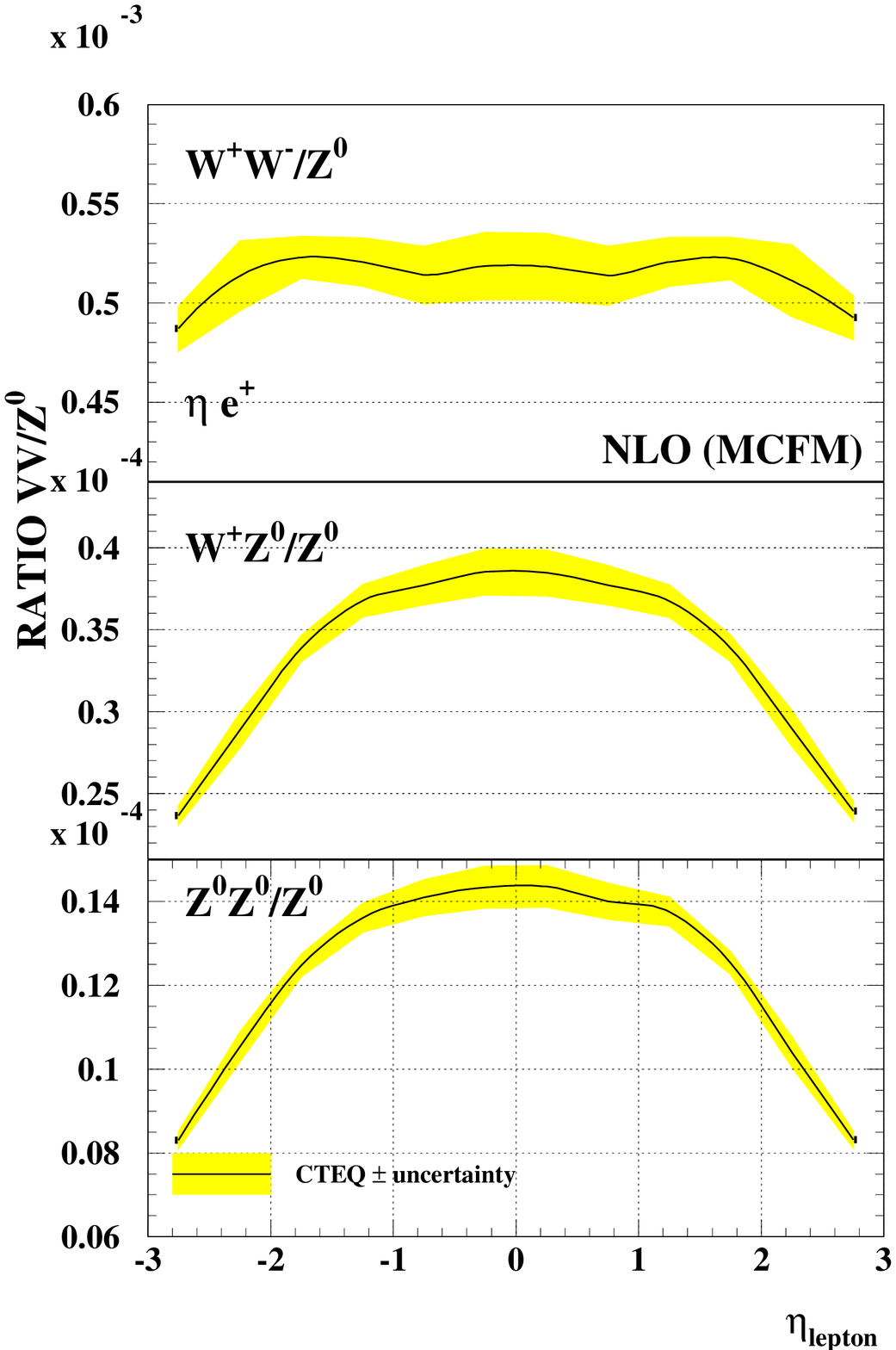}
\includegraphics[width=0.44\textwidth]{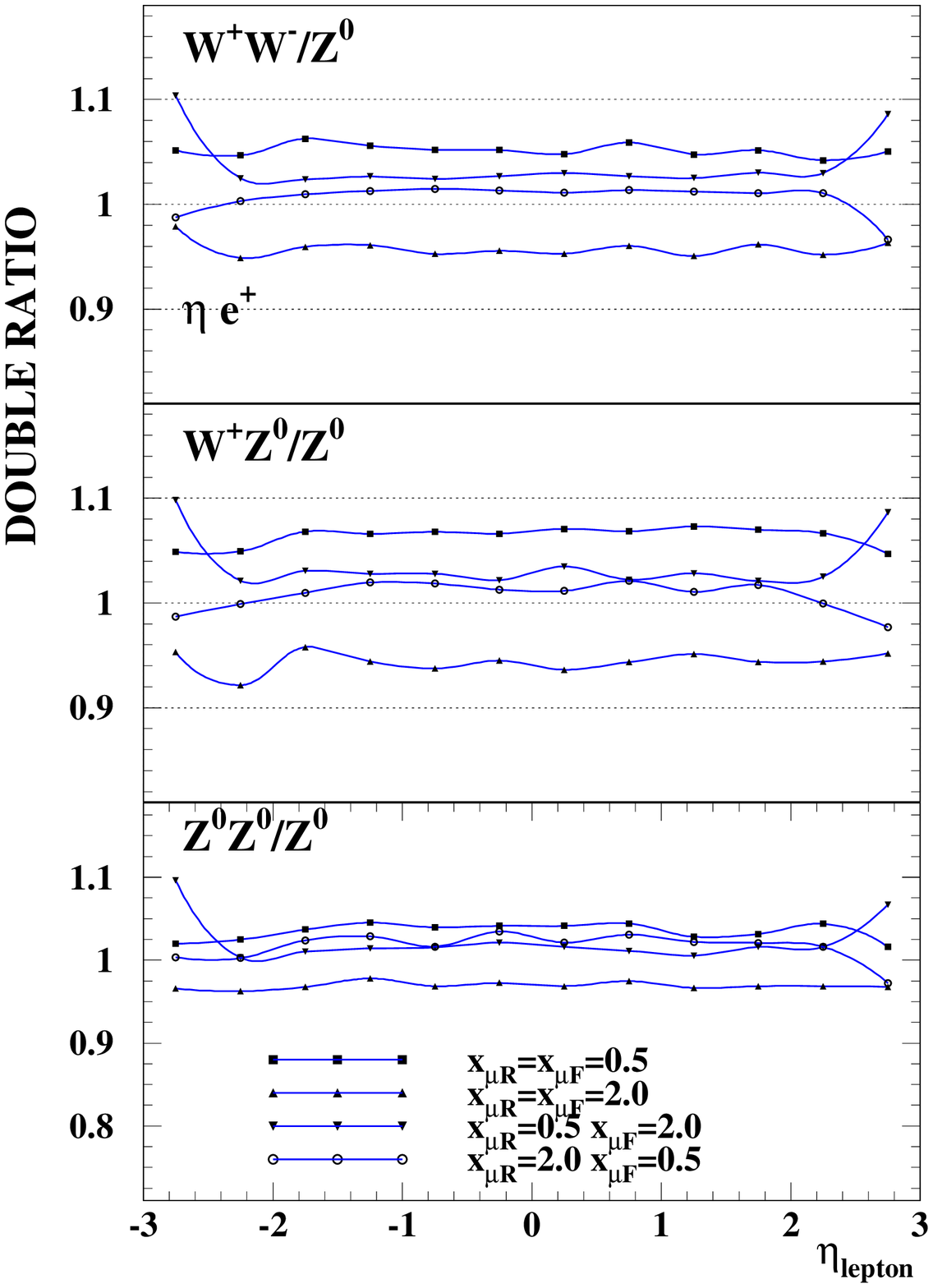}
\caption{Left: the ratio of pseudo-rapidity distributions of leptons from 
boson pair production processes normalised to single $Z$ production and 
right: the double ratio $VV/Z$ of predictions for different scales relative to 
$x_\mu=1$.}  
\label{fig:vvz}       
\end{figure}

The ratio of boson pair production to single $Z$ production is 
of particular interest, as similar quark configurations contribute 
to both process types, though evidently in a somewhat different 
$x,Q^2$ regime. This ratio is shown in fig.\ref{fig:vvz} for the 
lepton distribution, given the different shapes of pseudo-rapidity 
is not flat but its PDF uncertainty is reduced to the level of 2 $\%$. 
The perturbative uncertainties of the $VV/Z$ ratio, however, are 
only reduced for the $ZZ/Z$ case and even slightly larger 
for other ratios because the scale variations have partly an 
opposite effect on the cross sections for $Z$ and e.g. $WW$ production.       

The total cross sections and their systematic uncertainties are summarised in 
Table~\ref{tab:vv}. 

\begin{table}
\caption[VV]{\label{tab:vv}{Total cross-sections and systematic uncertainties within the experimental acceptance for pair production processes.}}
\begin{center}
\begin{tabular}{|l|cccc|}
\hline
            & $WW$ & $ZZ$ & $W^+Z^0$ & $W^-Z^0$  \\ \hline
CTEQ61 [fb] & 475.7 & 11.75 & 31.81 & 20.77 \\
& & & &\\
$\Delta_{PDF}^{CTEQ}$ [fb] & $\pm 17.0$ & $\pm 0.48$ & $\pm 1.12$ & $\pm 0.80$\\
& & & & \\
$\Delta_{PDF}^{CTEQ}$ [$\%$] & $\pm 3.6$ & $\pm 4.1$ & $\pm 3.5$  & $\pm 3.8$ \\
& & & & \\
\hline
MRST [fb] & 494.2 & 12.34 & 32.55 & 21.62 \\
& & & &\\
$\Delta_{PDF}^{MRST}$ [fb] & $\pm 6.3$ & $\pm 0.19$ & $\pm 0.49$ & $\pm 0.41$  \\
& & &  &\\
$\Delta_{PDF}^{MRST}$ [$\%$] & $\pm 1.3$ & $\pm 1.6$ & $\pm 1.5$ & $\pm 1.9$ \\
& & &  &\\
\hline
$\Delta_{pert}$ [$\%$] & $+ 4.6$ & $+ 3.3$ & $+ 4.6$ & $+ 4.8$ \\
                    & $- 4.9$ & $- 3.8$ & $- 4.7$ & $- 4.7$ \\

\hline
\end{tabular}
\end{center}
\end{table}

\newpage


\subsection{Study of next-to-next-to-leading order QCD 
predictions for W and Z production at LHC\protect\footnote{Contributing author:G\"unther Dissertori}}
\label{sec:gd_WZprodNNLO}
  
It has been in 2004 that the first differential 
next-to-next-to-leading order (NNLO) QCD calculation for vector boson production 
in hadron collisions was completed by Anastasiou \textit{et al.} \cite{Anastasiou2004}. This group has
calculated the rapidity dependence for W and Z production at NNLO. They have
shown that the perturbative expansion stabilizes at this order in perturbation theory
and that the renormalization and factorization scale uncertainties are  drastically 
reduced, down to the level of one per-cent. 
It is therefore interesting to perform a more detailed study of these
NNLO predictions for various observables which can be measured at LHC,
as well as to investigate their systematic uncertainties.
  
In the study presented here we have calculated both the differential (in rapidity)
and inclusive cross sections 
for W, Z and high-mass Drell-Yan (Z/$\gamma^*$) production. Here "inclusive"
refers to the results  obtained by integrating the differential cross sections over a
rapidity range similar to the experimentally accessible region, which
might be more relevant  than the complete cross section which also includes the large-rapidity tails. 
  
Such a prediction would then be compared to the
experimental measurements at LHC, which will allow for precise tests of the Standard Model
as well as to put strong constraints on the parton distribution functions (PDFs) of the
proton. It is clear that in the experiment only the rapidity and transverse momenta of the
leptons from the vector boson decays will be accessible, over a finite range in phase space.
In order to compute the rapidity of the vector boson by taking into account the
finite experimental lepton acceptance, Monte Carlo simulations have to be employed which 
model vector boson production at the best possible precision in QCD, as for example
the program \textsc{MC@NLO} \cite{MCatNLO}.  The so computed acceptance corrections will
include further systematic uncertainties, which are not discussed here.


\subsubsection{Parameters and analysis method}
 \label{subsec:gd_par}
 
 The NNLO predictions have been implemented in the computer code \textsc{VRAP}
 \cite{DixonPrivate}, which has been modified in order to include \textsc{ROOT} \cite{ROOT}
 support for producing ntuples, histograms and plots. The code allows to specify the
  collision energy (14 TeV in our case), the exchanged vector boson ($\gamma^*, 
  \mathrm{Z}$, $\mathrm{Z}/\gamma^*$,  $\mathrm{W}^+$,  $\mathrm{W}^-$), the scale $Q$ of the
  exchanged boson ($M_ \mathrm{Z}, M_ \mathrm{W}$ or off-shell, e.g.\  $Q=400\, \mathrm{GeV}$),
  the renormalization and factorization scales, the invariant mass of the di-lepton system 
  (fixed or integrated over a specified range), the value of the electro-magnetic coupling
  ($\alpha_\mathrm{QED} = 1/128$ or $\alpha_\mathrm{QED}(Q)$) and the number of light fermions
  considered. Regarding the choice of pdfs, the user can select a pdf set from the MRST2001
  fits \cite{Martin:2002dr} 
  or from the ALEKHIN fits \cite{Alekhin:2002fv}, 
  consistent at NNLO with variable flavour scheme. We have chosen
  the MRST2001 NNLO fit, mode 1 with $\alpha_s(M_\mathrm{Z}) = 0.1155$ 
  \cite{Martin:2002dr}, as reference set.
  
  The program is run to compute the differential cross section $d\sigma/dY$, $Y$ being the 
  boson rapidity, at a fixed number of points in $Y$. This result is then parametrized using a 
  spline interpolation, and the thus found function can be integrated over any desired rapidity range,
  such as $|Y| < 2, |Y| <2.5$ or $|Y| < 3$, as well as over finite bins in rapidity. For the study of
  on-shell production the integration range
  over the di-lepton invariant mass $M_{ll}$ was set to $M_V - 3 \Gamma_V < M_{ll}
  <  M_V + 3 \Gamma_V$, with $M_V$ and $\Gamma_V$ the vector boson mass and width. This
simulates an experimental selection over a finite signal range.
  
  The systematic uncertainties have been divided into several categories: The PDF uncertainty
  is estimated by taking the maximum deviation from the reference set when
   using different PDFs from within 
  the MRST2001 set or the ALEKHIN set. The latter difference is found to give the maximal
  variation in all of the investigated cases. The renormalization and factorization scales $\mu = 
  \mu_\mathrm{R}, \mu_\mathrm{F}$
  have been varied between $0.5 < \mu / Q < 2$, both simultaneously as well as fixing
  one to $\mu = Q$ and varying the other. The maximum deviation from the reference setting
  $\mu = Q$ is taken as uncertainty. The observed difference when using either a 
  fixed or a running electro-magnetic coupling constant is also studied as possible systematic 
  uncertainty due to higher-order QED effects.
  Since it is below the one per-cent level, it is not discussed further. Finally, in the case of Z
  production it has been checked that neglecting photon exchange and interference contributions
  is justified in view of the much larger PDF and scale uncertainties.


\subsubsection{Results for W and Z production}
 \label{subsec:gd_results_WZ}
 
 \begin{figure}[htb]
 \begin{center}
 \begin{tabular}{cc}
 \includegraphics[width=0.48\textwidth]{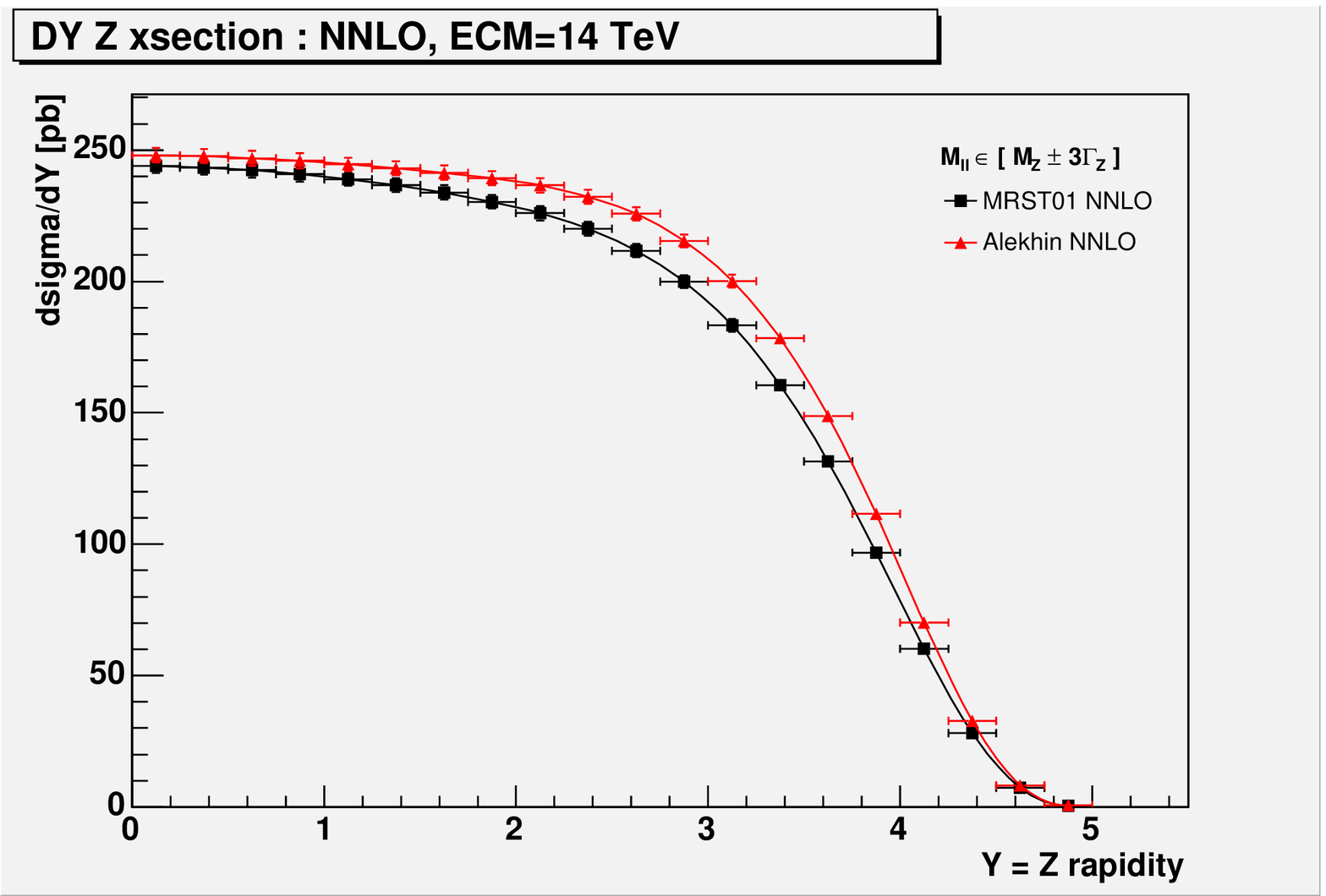} &\
  \includegraphics[width=0.48\textwidth]{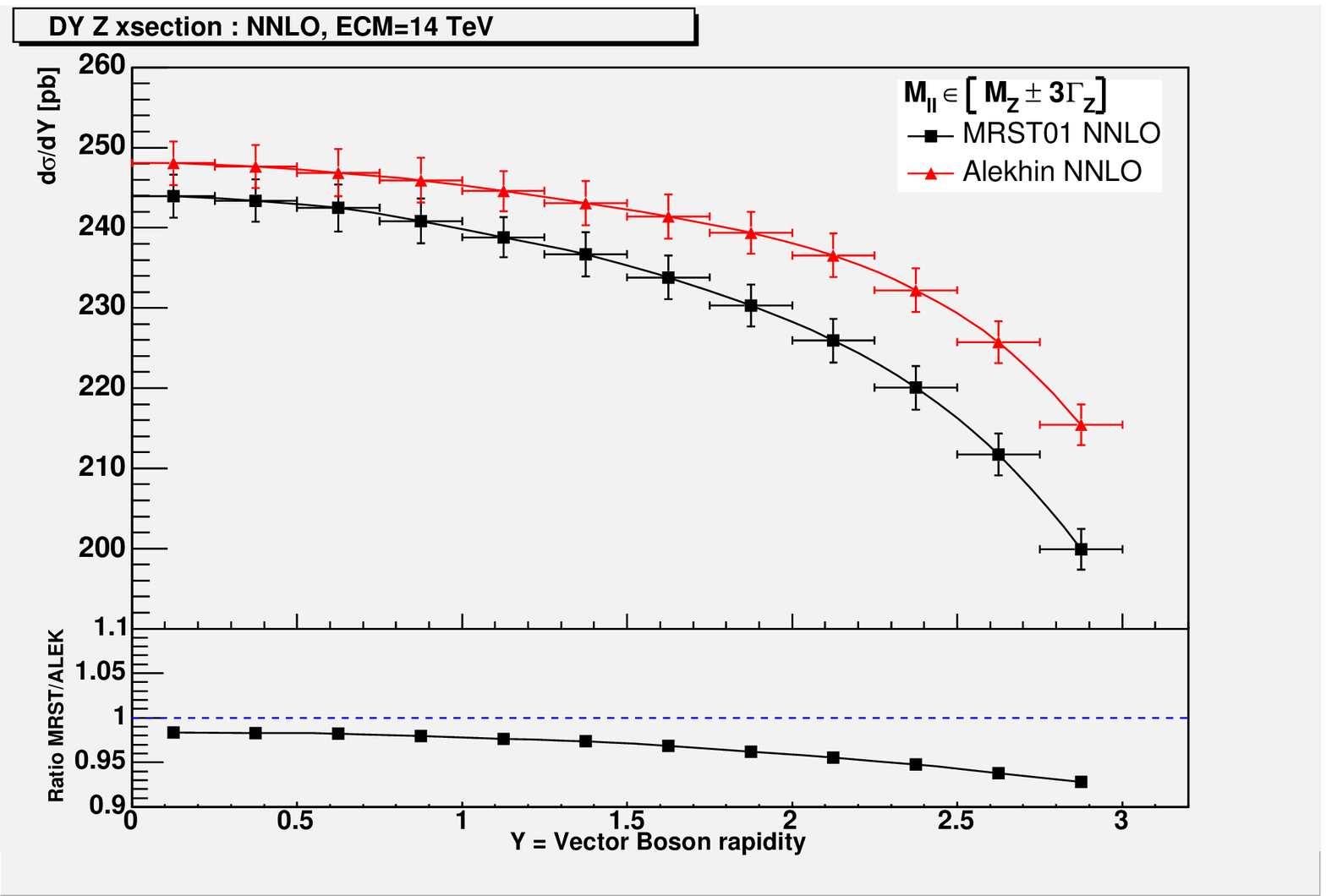}
 \end{tabular}
 \caption{\label{fig:gd_Zprod}  Left~: Drell-Yan Z production cross section ($\times$ BR) at LHC energies, 
 as a function of the Z rapidity, for two different PDF choices. Right~: Zoom into a restricted rapidity region, 
 with the ratio of the predictions for the two different PDF sets as lower inset. The error bars indicate the scale uncertainties.  }
 \end{center}
 \end{figure}
 
 In Figure \ref{fig:gd_Zprod} the results for  Z production at LHC are shown for two different 
 choices of PDF set, as a function of the boson rapidity. It can be seen that the predictions
  differ by about 2\% at central rapidity,
 and the difference increases to about 5\% at large rapidity. 
 A similar picture is obtained when integrating the
 differential cross section up to rapidities of 2, 2.5 and 3 (Table \ref{tab:gd_ZWprod}). 
 The more of the high-rapidity tail is included, the larger the uncertainty due to the PDF choice. 
 From Table 1 it can also be seen that the scale uncertainties are slightly below the one per-cent level. 
 It is worth noting that the choice of the integration range over the di-lepton invariant mass can have a
 sizeable impact on the cross section. For example, increasing the range from the standard value to
 $66\, \mathrm{GeV} < M_\mathrm{Z} < 116\, \mathrm{GeV}$ increases the cross section by 8\%.
 
 \begin{table}[htb]
 \begin{center}
 \begin{tabular}{|l|c|c|c||c|c|c|}
 \hline
  Channel & \multicolumn{3}{c||}{Z prod.} & \multicolumn{3}{c|}{W prod.} \\
  \hline
  range & $|Y| < 2$ & $|Y| < 2.5$ & $|Y| < 3$ & $|Y| < 2$ & $|Y| < 2.5$ & $|Y| < 3$ \\
  \hline
  cross section [nb] & 0.955 &  1.178 & 1.384 & 9.388 & 11.648 &  13.800 \\
  $\Delta$ PDF  [\%] &  2.44 & 2.95 & 3.57 & 5.13 & 5.47 & 5.90 \\
  $\Delta$ scale [\%] & 0.85 & 0.87 & 0.90 & 0.99 & 1.02 & 1.05 \\
  \hline  
 \end{tabular}
\caption{\label{tab:gd_ZWprod} NNLO QCD results for W and Z production at the LHC for the integration over different rapidity ranges. 
Also given are the relative uncertainties due to the choice of the PDFs and 
of the renormalization and factorization scale. The numbers include the branching ratio $Z (W) \rightarrow ee (e \nu)$.}
\end{center}
 \end{table}

The results for W production (Table \ref{tab:gd_ZWprod}) have been obtained by first 
calculating separately the cross sections
for $\mathrm{W}^+$ and $\mathrm{W}^-$ production, and then adding these up. Again we observe an
increase of the PDF uncertainty when going to larger rapidity ranges. Compared to the Z production,
here the PDF uncertainties are larger, between 5 and 6\%, whereas the scale uncertainties are of the
same level, $\approx 1\%$. It is interesting to note that the PDF uncertainty for $\mathrm{W}^-$ production
is about 10 - 20\% (relative) lower than that for $\mathrm{W}^+$.

\begin{figure}[htb]
 \begin{center}
 \begin{tabular}{cc}
 \includegraphics[width=0.48\textwidth]{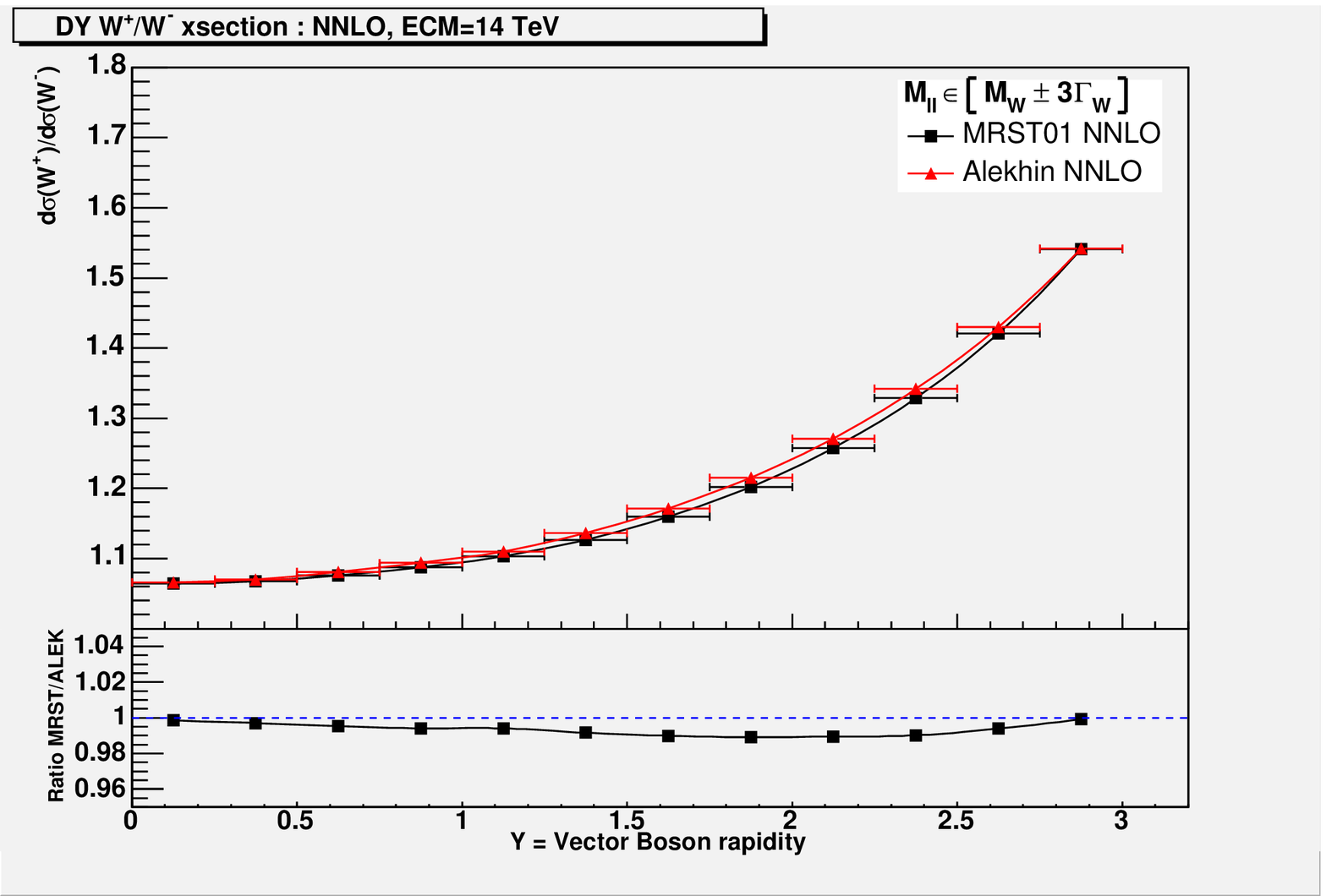} &\
  \includegraphics[width=0.48\textwidth]{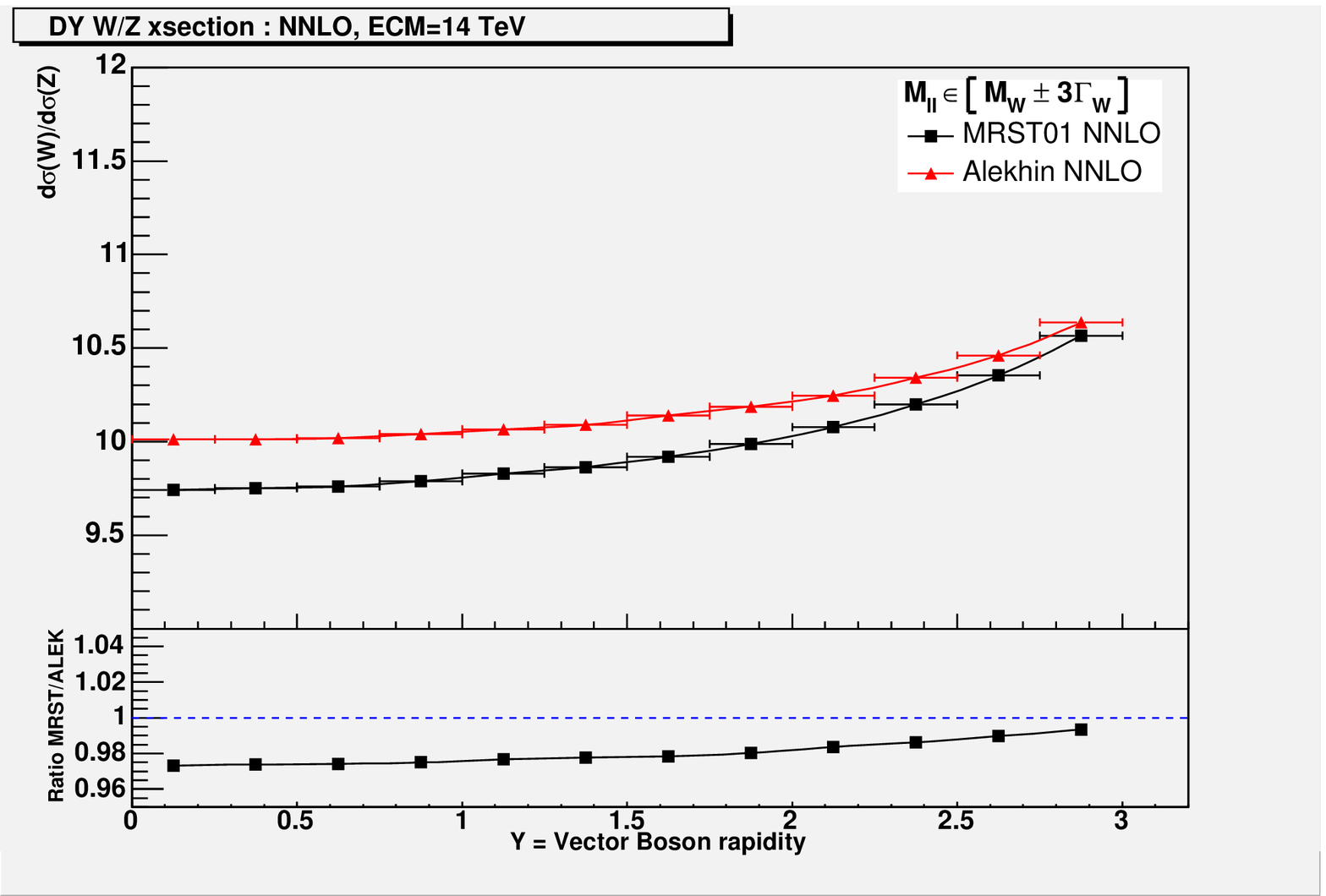}
 \end{tabular}
 \caption{\label{fig:gd_ratios}  Ratio of the production cross sections for  $\mathrm{W}^+$, 
 $\mathrm{W}^-$ (left), and W, Z (right), as a function of rapidity, for two different PDF sets. 
The inserts show the ratios of the results for the two PDF choices. }
 \end{center}
 \end{figure}

A considerable reduction in systematic uncertainty can be obtained by calculating cross section ratios.
Two options have been investigated, namely the ratios $\sigma(\mathrm{W}^+)/ \sigma(\mathrm{W}^-)$ 
and $\sigma(\mathrm{W})/ \sigma(\mathrm{Z})$. As can be seen from Figure \ref{fig:gd_ratios}, the PDF
uncertainties are reduced to the 0.7\% level in the former ratio, and to about 2\% in the latter. The scale
uncertainties are reduced to the 0.15\% level in both cases. Taking such ratios has also the potential
advantage of reduced experimental systematic uncertainties, such as those related to the 
acceptance corrections.

\subsubsection{Results for high-mass Drell-Yan processes}
 \label{subsec:gd_results_hmDY}
 
 Similarly to on-shell W and Z production we have also analyzed the high-mass
 Drell-Yan process,  namely $\mathrm{Z}/\gamma^*$ production at a scale of 
 $Q=400$ GeV. In this case the di-lepton invariant mass has been integrated
 over the range  $M_{ll} = 400 \pm 50$ GeV. Here the PDF uncertainties are found
 between 3.7\% and 5.1\% for the various integration ranges over rapidity, somewhat
 larger than for on-shell production. However, by normalizing the high-mass 
 production cross section to the on-shell case, the PDF uncertainties are considerably
 reduced, being 1.2 - 1.5\%. 
 
 The systematic  uncertainties related to the renormalization and factorization scale
 are reduced ($\Delta$ scale $\approx 0.2\%$)
 when going to the high-mass exchange, as expected from perturbative
 QCD with a decreasing strong coupling constant. In this case a normalization
 of the cross section to the on-shell case does not give an improvement. However,
 since the scale uncertainties are well below the PDF uncertainties, this is less of an issue
 for the moment.
  

\subsubsection{Summary}
 \label{subsec:gd_summary}
 
 We have studied NNLO QCD predictions for W and Z production at LHC energies.
 We have identified the choice of PDF set as the dominant systematic uncertainty,
 being between 3 and 6\%. The choice of the renormalization and factorization
 scale leads to much smaller uncertainties, at or below the 1\% level.  In particular
 we have shown that the systematic uncertainties can be sizeably reduced by taking
 ratios of cross sections, such as  $\sigma(\mathrm{W}^+)/ \sigma(\mathrm{W}^-)$,
 $\sigma(\mathrm{W})/ \sigma(\mathrm{Z})$ or 
 $\sigma(\mathrm{Z}/\gamma^*, Q=400\, \mathrm{GeV})$/$\sigma(\mathrm{Z}/\gamma^*, Q=M_\mathrm{Z})$. 
For such ratios it can be expected
 that also part of the experimental uncertainties cancel. With theoretical uncertainties
from QCD  at the few per-cent level the production of W and Z bosons will most likely be the 
best-known cross section at LHC. 

Concerning the next steps, it should be considered that 
at this level of precision it might become relevant
to include also higher-order electro-weak corrections.  In addition, since experimentally the boson rapidity
will be reconstructed from the measured lepton momenta,
a detailed study is needed to evaluate the precision at which the acceptance correction
factors for the leptons from the boson decays can be obtained. For this Monte Carlo programs
such as MC@NLO should be employed, which combine next-to-leading-order matrix elements
with parton showers and correctly take account of spin correlations. 
  

%% file: pdfexp.tex
\section{Experimental determination of   Parton Distributions
\protect\footnote{Subsection coordinators: A.~Glazov, S.~Moch}}
\label{section:exppdf}
\subsection{Introduction}

With HERA  currently in its second stage of operation,
it is possible to assess the potential precision limits of HERA data
and to estimate the potential impact of the measurements which are
expected at HERA-II, in particular with respect to the PDF uncertainties.

Precision limits of the structure function analyses at HERA are
examined in section~\ref{sec:prec}. Since large amounts of luminosity
are already collected, the systematic uncertainty becomes most important. 
A detailed study of error sources with particular emphasis on correlated errors 
for the upcoming precision analysis of the inclusive 
DIS cross section at low $Q^2$ using 2000 data taken by the H1 experiment 
is presented.
A new tool, based on the ratio of cross sections measured by different reconstruction
methods, is developed and its ability to qualify and unfold
various correlated error sources is demonstrated.

An important issue is the consistency of the HERA data. 
In section~\ref{sec:mandy}, the H1 and ZEUS published PDF analyses are compared, 
including a discussion of the different treatments of correlated systematic uncertainties. 
Differences in the data sets and the analyses are investigated by 
putting the H1 data set through both PDF analyses and by putting the ZEUS and H1 data sets 
through the same (ZEUS) analysis, separately. 
Also, the HERA averaged data set (section~\ref{sec:ave}) is put through the ZEUS PDF analysis 
and the result is compared to that obtained when putting the ZEUS and H1 data sets through this analysis 
together, using both the Offset and Hessian methods of treating correlated systematic uncertainties.

The HERA experimental data can not only be cross checked with respect to each
other but also combined into one common dataset, as discussed in section~\ref{sec:ave}.
In this respect, a method to combine measurements of the structure functions performed
by several experiments in a common kinematic domain is presented.
This method generalises the standard averaging procedure by taking into account
point-to-point correlations which are introduced by the systematic uncertainties of the measurements. 
The method is applied to the neutral and charged current DIS cross section data 
published by the H1 and ZEUS collaborations. 
The averaging improves in particular the accuracy due to the cross calibration 
of the H1 and ZEUS measurements.

The flavour decomposition of the light quark sea  is discussed in 
section~\ref{sec:mkbrdbarubar}. For low $x$ and thus low $Q^2$
domain at HERA only measurement of the photon exchange induced structure
functions $F_2$ and $F_L$ is possible, which is insufficient to disentangle
individual quark flavours. A general strategy in this case is to assume
flavour symmetry of the sea. Section~\ref{sec:mkbrdbarubar} considers 
PDF uncertainties if this assumption is released. These uncertainties
can be significantly reduced if  HERA would run in deuteron-electron collision
mode.

The impact of projected HERA-II data on PDFs is estimated in
section~\ref{sec:clair}. 
In particular, next-to-leading order (NLO) QCD predictions for inclusive jet cross sections 
at the LHC centre-of-mass energy are presented using the estimated PDFs.
A further
important measurement which could improve understanding of the gluon density at low $x$ 
and, at the same time, provide consistency checks of the low $Q^2$ QCD evolution is
the measurement of the longitudinal structure function $F_L$. 
Perspectives of this measurement are examined in section~\ref{sec:flmax}, 
while 
the impact of this measurement is also estimated in section~\ref{sec:clair}.

Further improvements for consistently including final-state observables 
in global QCD analyses are discussed in section~\ref{sec:jetsinfits}.
There, a method for ``a posteriori'' inclusion of PDFs, whereby the
Monte Carlo run calculates a grid (in $x$ and $Q$) of cross section
weights that can subsequently be combined with an arbitrary PDF.
The procedure is numerically equivalent to using an interpolated 
form of the PDF. The main novelty relative to prior work is the use
of higher-order interpolation, which substantially improves the 
tradeoff between accuracy and memory use.
An accuracy of about $0.01$\% has been reached for the single
inclusive cross-section in the central rapidity region $|y|<0.5$ for
jet transverse momenta from $100$ to $5000 \mathrm{GeV}$.
This method will make it possible to consistently include measurements done 
at HERA, Tevatron and LHC in global QCD analyses.

%% file: precisionlimits.tex
\subsection{Precision Limits for HERA DIS Cross Section Measurement
\protect\footnote{Contributing authors: G.~La\v{s}tovi\v{c}ka-Medin, A.~Glazov, T.~La\v{s}tovi\v{c}ka}}

\label{sec:prec}

The published precision low $Q^2$ cross section data
\cite{h1alphas} of the H1 experiment became an important data set
in various QCD fit analyses~\cite{h1alphas,cteq,mrst,Alekhin:2002fv}.
Following success of these data the H1 experiment plans to analyse
a large data sample, taken during 2000 running
period\footnote{Data statistics will be increased further by
adding data taken in year 1999.}, in order to reach precision
limits of low $Q^2$ inclusive cross sections measurements at HERA.
The precision is expected to approach 1\% level.

The aim of this contribution is to calculate realistic error
tables for 2000 H1 data and pursue paths how to reach such a high
precision. Correlated error sources are studied in particular and
a new tool, based on the ratio of cross sections measured by
different reconstruction methods, is developed. All errors,
including correlated errors, are treated in the same manner as
in~\cite{h1alphas}. Error tables are provided and used in QCD fit
analysis, see Sec~\ref{sec:clair}, in order to study the impact of the
new data on PDFs. The new data are expected to reach higher
precision level than \cite{h1alphas} due to the following reasons

\begin{itemize}
\item Larger data statistics - Statistical errors will decrease by
factor of $1.5-2$, compared to~\cite{h1alphas}, depending on the
kinematic region.

\item Very large Monte Carlo simulations (MC) - Due to a progress in
computing a number of simulated events can be significantly
increased in order to minimise statistical error of MC, 
to understand uncorrelated errors and to estimate
correlated errors more precisely.

\item During past years increasing knowledge, arriving from
various H1 analyses, enabled better understanding of the detector
and its components as well as improving quality of MC.

\item Data taking in 2000 was particularly smooth. Both HERA and
H1 were running at peak performance for HERA-I running period.

\end{itemize}

This contribution uses existing 2000 data and MC
ntuples along with the full analysis chain. It applies all
preliminary technical work done on these data, including
calibration, alignment, trigger studies etc. Quoted errors are
assumed to be achieved in the final version of analysis
yet the analysis has not been finalised, all the numbers
in the paper are preliminary and may change in the publication.

The uncertainties of the cross section measurement are divided
into a number of different types. Namely, these are {\it
statistical uncertainties} of the data, {\it uncorrelated
systematics} and {\it correlated systematics}. The term
'correlated' refers to the fact that cross section measurements in
kinematic bins are affected in a correlated way while different
correlated systematic error sources are considered uncorrelated
among each other. The classification of the systematic errors into
types is sometimes straightforward (MC statistics is uncorrelated error source)
but sometimes is rather arbitrary (radiative corrections are assumed
to be uncorrelated error source). The main goal of this classification
is to preserve correlation between data points while keeping the
treatement as simple as possible.

The cross section uncertainties depend on the method used to
reconstruct event kinematics. There are various methods existing,
involving a measurement of the scattered electron as well as of
the hadronic finale state. In the following two of them, so called
{\it electron method} and {\it sigma method}, are
employed~\cite{kinematics}. The electron method uses only the
measurement of the scattered electron, namely its energy and polar
angle, while the sigma method uses both the scattered electron and
the hadronic final state. An advantage of the sigma method is 
a proper treatment of QED radiation from the incoming beam
electron (ISR).

The {\it statistical uncertainty} of the data is typically
0.5-1\%, depending on the kinematic region analysed and the
definition of the kinematic bins. In the following we adapt the
bin definition used in~\cite{h1alphas}, apart from merging bins at
low~$y$ which was done in the published data in order to increase
statistics.

The {\it uncorrelated systematics} consists from various
contributions. A cross section uncertainty due to the Monte Carlo
statistics is the one with very good potential to be minimised. In
the following we assume 100 million simulated events to be used in
analysis of 2000 data. Estimates were calculated with available 12
million simulated events and corresponding statistical errors
scaled by a factor of $\sqrt{100/12}$. As a result the uncertainty
is very small and typically on the level of few permile.

Additional contributions to the uncorrelated systematics are
efficiencies. We assume  for trigger efficiency
0.3\% and backward tracker tracker efficiency 0.3\% uncertainty.
Radiative corrections are expected to affect the final cross
section by  0.4\%.

\begin{figure}
\centering a)\includegraphics[width=0.45\linewidth]{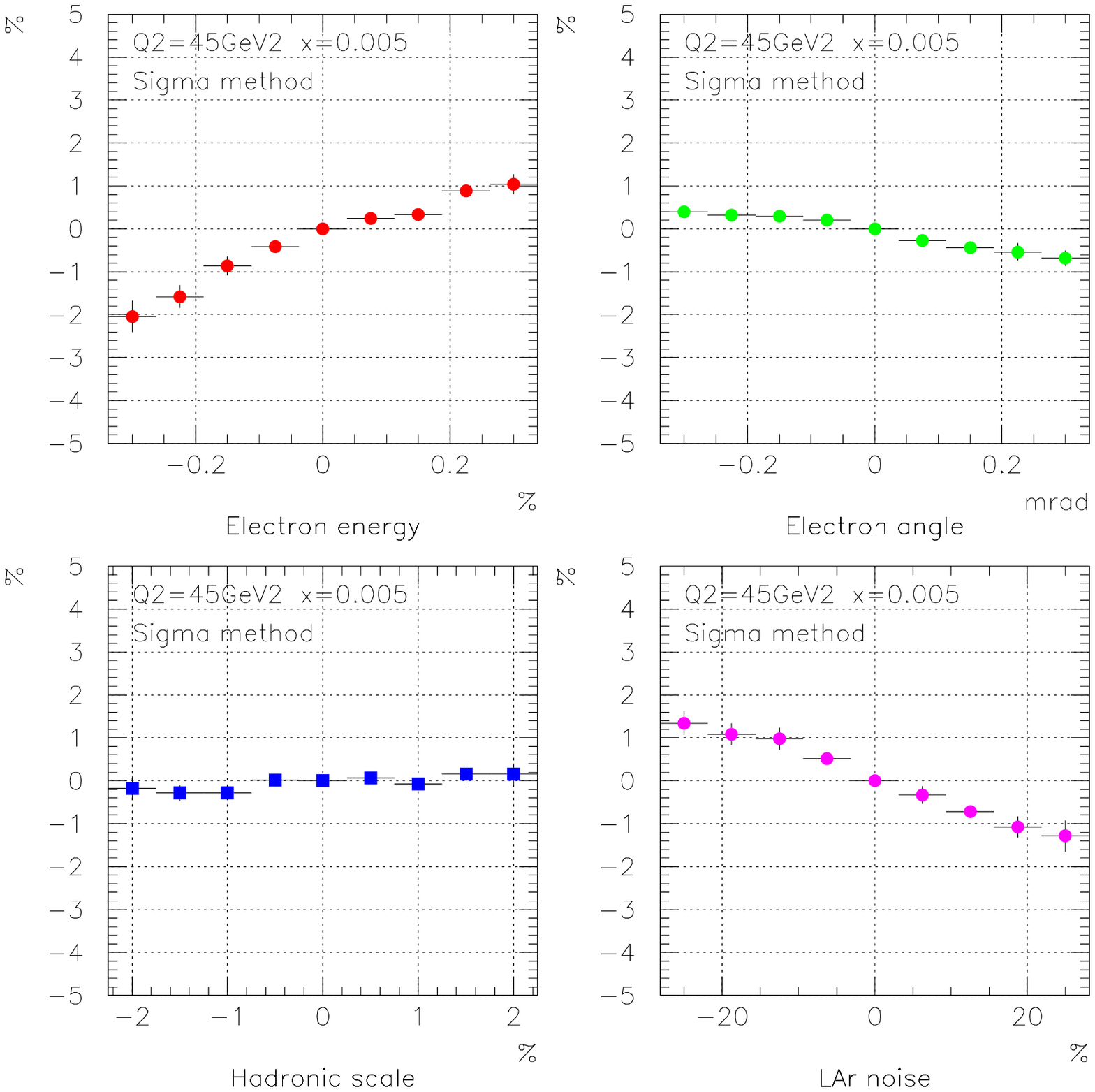}
~~~~~~b)\includegraphics[width=0.45\linewidth]{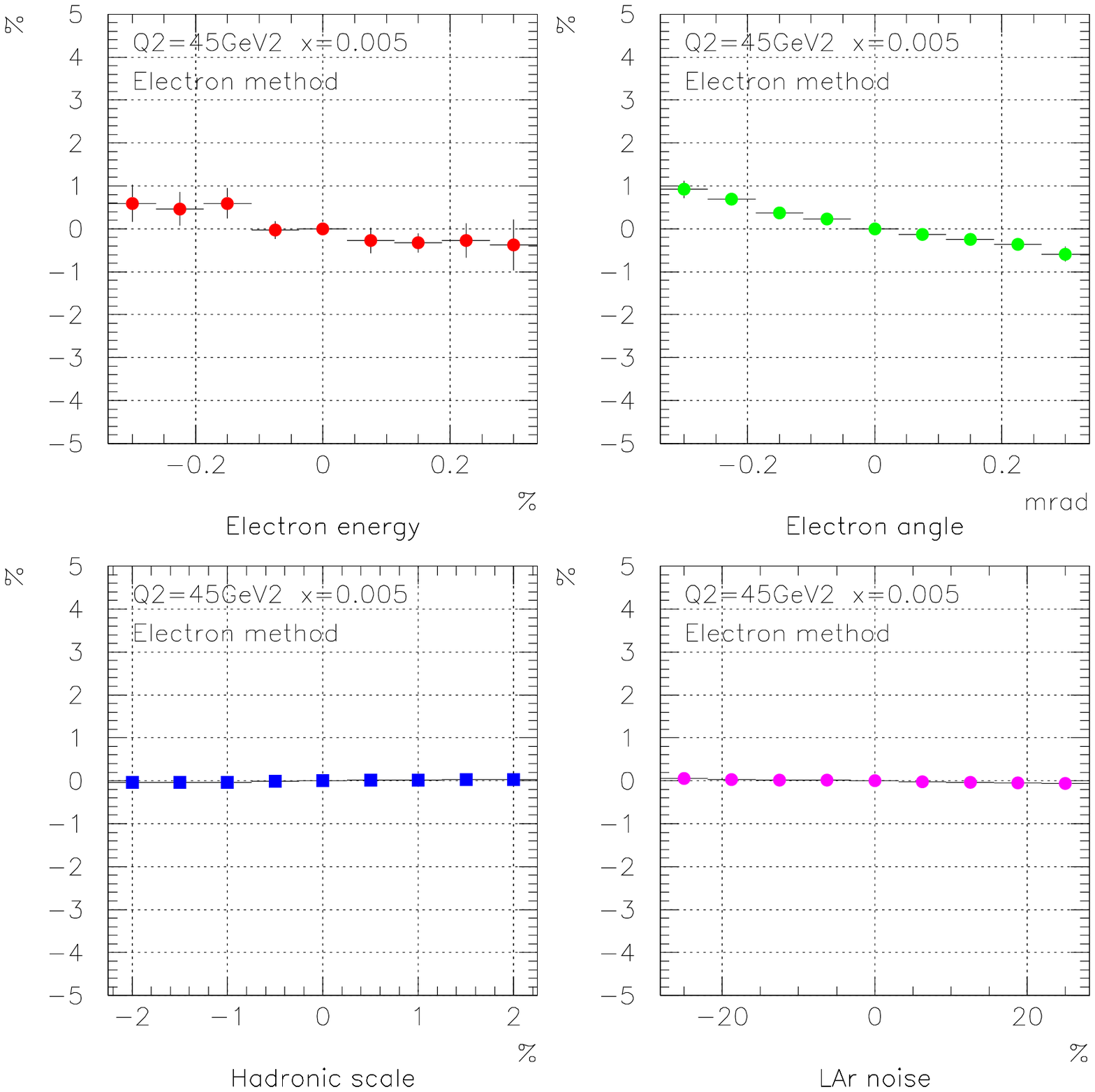}
\caption{A scan of the cross section measurement change in \%
depending on a variation of (from top-left) electron energy,
electron polar angle, hadronic final state calibration scale and
noise level in LAr calorimeter (bottom-right). The sigma method
(a) and the electron method (b) were used to reconstruct
kinematics of events.}\label{cross_scan}
\end{figure}

Effect of {\it correlated uncertainties} on the cross section
measurement is studied in the following manner. Particular source
of correlated uncertainty, for instance the scattered electron
energy measurement, is varied by assumed error and the change of
the measured cross section is quoted as the corresponding cross
section measurement error. An example of cross section change on
various correlated error source is shown in Fig.~\ref{cross_scan}
for bin of $Q^2=45\,{\rm GeV}^2$ and $x=0.005$. The kinematics of
events was reconstructed with the sigma method (a) and the
electron method (b). Errors are calculated as so called standard
errors of the mean in calculation of which the available Monte
Carlo sample was split into nine sub-samples. It is clearly seen
that the cross section measurement with the sigma method in this
kinematic bin is particularly sensitive to the electron energy
measurement (top-left) and to noise description in LAr calorimeter
(bottom-right). On the contrary, the electron polar angle
measurement and the calibration of the hadronic final state play a
little role. The electron method is mainly sensitive to the
electron energy measurement. The importance of the systematic
sources vary from bin to bin. 

There are five individual sources contributing to the correlated
cross section uncertainties:

\begin{itemize}
\item Uncertainties of 0.15\% at $E_e=27\,$GeV and 1\% at 7\,GeV
are assigned to the electron energy scale for the backward
calorimeter. The uncertainty is treated as a linear function of
$E_e$ interpolating between the results at 27\,GeV and 7\,GeV.

\item The uncertainty on the scattered electron polar angle
measurement is 0.3~mrad . The corresponding error on the cross
section measurement is typically well below 1\% but may be larger
at lowest values of $Q^2$.

\item The uncertainty on the hadronic energy scale comprises a
number of systematic error sources corresponding to the $E-p_z$
decomposition: an uncertainty of the  hadronic energy scale
calibration of 2\% for the central and forward calorimeter, 
an uncertainty of 3\% for the fraction carried
by tracks and a 5\% uncertainty of the  hadronic energy
scale measured in backward calorimeter.

\item The uncertainty on the hadronic energy scale is further
affected by the subtracted noise in the calorimetery. The noise
is described to the level of 10\% and the corresponding error is
propagated to the cross section uncertainty. The largest influence
is in the low $y$ region, which is measured with the sigma method.

\item The uncertainty due to the photoproduction background at
large $y$ is estimated from the normalisation error of the PHOJET
simulations to about 10\%. At low and medium values of $y \lesssim
0.5$ it is negligible.

\end{itemize}

The total systematic error is calculated from the quadratic
summation over all sources of the uncorrelated and correlated
systematic uncertainties. The total error of the DIS cross section
measurement is obtained from the statistical and systematical
errors added in quadrature.

\begin{table}
\centering
\includegraphics[angle=270,width=\linewidth]{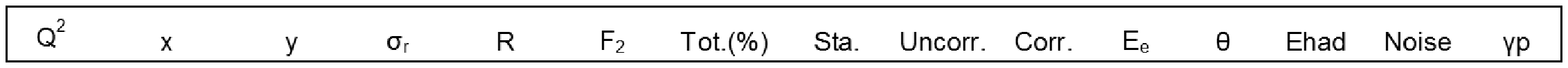}
\vspace{2mm}
\includegraphics[width=0.97\linewidth]{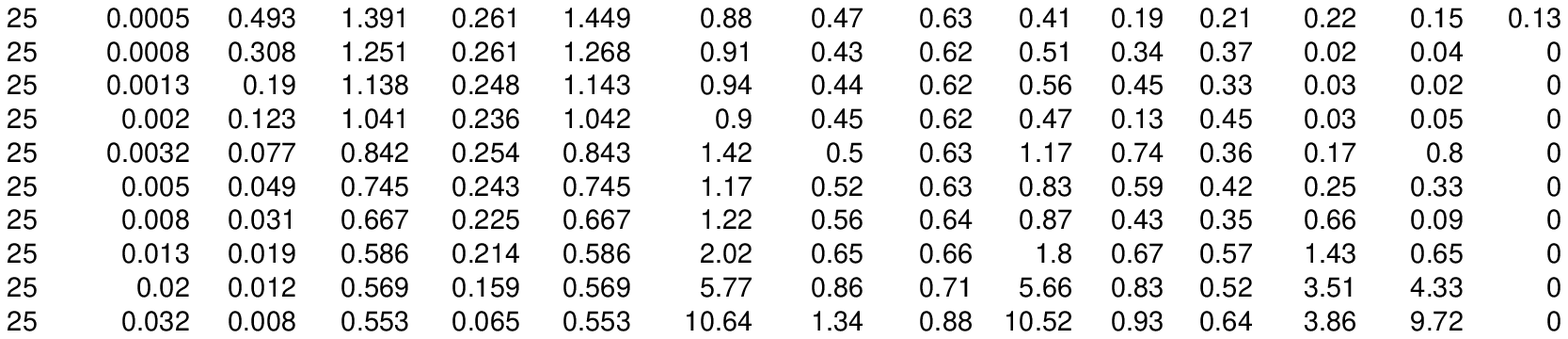}
\caption{An example of the error table for $Q^2 = 25\,{\rm GeV}^2$
for 2000 data, large Monte Carlo sample and suppressed systematic
errors compared to~[1], see text for details. Absolute errors are
shown. The table format is identical to the one published
in~[1].}\label{table_new}
\end{table}

\begin{table}
\centering
\includegraphics[angle=270,width=\linewidth]{header.eps}
\vspace{2mm}
\includegraphics[width=0.97\linewidth]{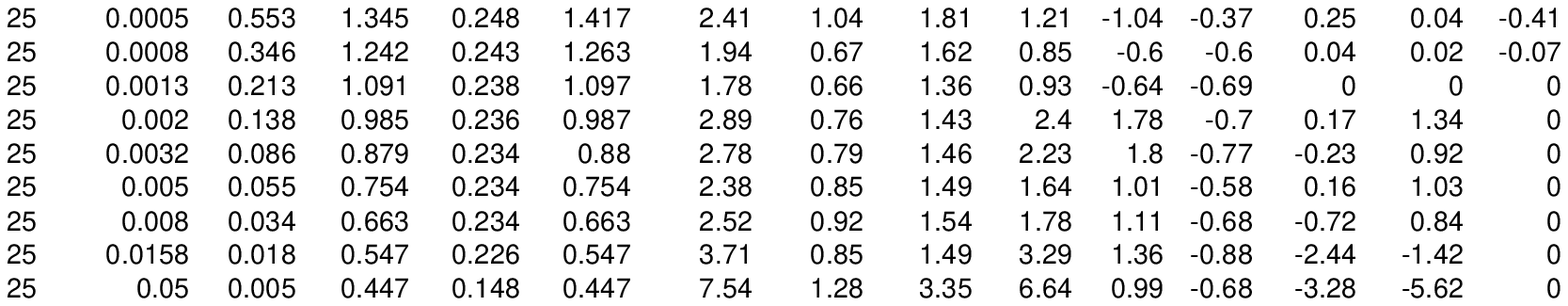}
\caption{An example of the full error table for $Q^2 = 25\,{\rm
GeV}^2$, published H1 data. The definition of kinematic bins is
not identical to that in Table~\ref{table_new}, some bins were
merged to enlarge statistics.}\label{table_published}
\end{table}

An example of the full error table for kinematic bin of $Q^2 =
25\,{\rm GeV}^2$ is shown in Table~\ref{table_new}. For a
comparison the corresponding part of the published data
from~\cite{h1alphas} is presented in Table~\ref{table_published}.
One can see that precision about 1\% can be reached especially in
four lowest $x$ bins, where the electron method was used to
reconstruct the event kinematics. The key contributions to the
seen improvement in the cross section measurement precision are
the electron energy measurement, very large Monte Carlo
statistics, well understood noise in LAr calorimeter and precisely
controlled efficiencies entering the analysis.

Full error table, covering the kinematic region of \mbox{$5 \le
Q^2 \le 150\,{\rm GeV}^2$} and \mbox{$0.01 \le y \le 0.6$} was
produced. The electron method was applied for kinematic bins at
$y>0.1$ while the sigma method otherwise. The measurement of the
proton structure function $F_2$ was simulated using fractal
parametrisation~\cite{tomas,*tomas2} for central values,
accounting for all sources of correlated and uncorrelated errors.
This table was used  to estimate effect of
precise low~$Q^2$ data on the determination of proton PDFs from
QCD fits.

\begin{figure}
\centering \includegraphics[width=\linewidth]{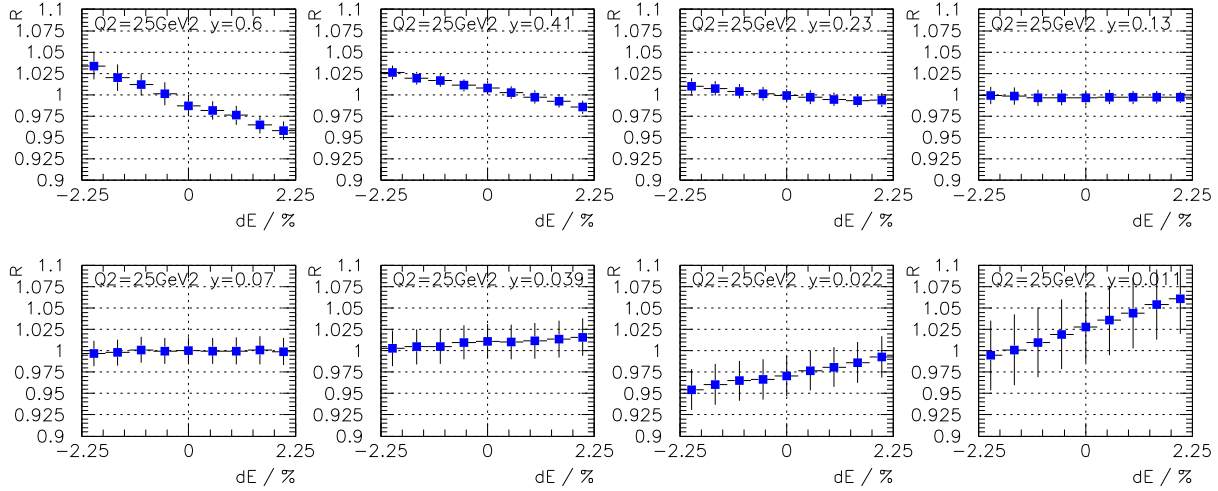}
\caption{A scan of the cross section ratio R in bins of $Q^2$ and
$y$ as a function of the hadronic final state calibration
variation.}\label{chi2_scan}
\end{figure}

The fact that different kinematics reconstruction methods 
are affected differently by
the correlated systematic uncertainties
may be employed as a  tool to estimate these uncertainties.
We define

\begin{equation}
R_i=\frac{\sigma_r^{el,i}}{\sigma_r^{\Sigma,i}}
\end{equation}

to be the cross section measurement ratio, where the reduced cross
section $\sigma_r^{el,i}$ and $\sigma_r^{\Sigma,i}$ is measured
using the electron method and the sigma method, respectively.
Kinematic bins, indexed by $i$, cover a region of the analysis
phase space where both reconstruction methods are applicable for
the measurement. The statistical error of $R_i$ measurement is
again evaluated by splitting the sample to a number of sub-samples
and calculating the standard error of the mean. An example of a
scan of the cross section ratio $R_i$ dependence on the hadronic
final state calibration variation in a bin of $Q^2 = 25\,{\rm
GeV}^2$ and various inelasticity $y$ is shown in
Fig.~\ref{chi2_scan}.

An error of a particular correlated uncertainty source $j$ can be
estimated by searching for lowest $\chi^2 = \sum_i (R_i(\alpha_j) -
1)^2/\sigma^2_i$, where summation runs over kinematic bins,
$\sigma_i$ is the error of $R_i$ measurement
and $\alpha_j$ is the variation of the source $j$. However, since there
is a number of correlated error sources the correct way to find
correlated uncertainties is account for all of them.

Unfolding of the correlated error sources can be linearised and
directly solved by minimising the following function:
\begin{equation}
{\cal L} = \sum_i \frac{1}{\sigma_i^2} ( R_i + \sum_j \alpha_j
\frac{\partial R_{i}}{\partial \alpha_j} - 1 )^2.
\end{equation}
The partial derivatives $\frac{\partial R_{i}}{\partial \alpha_j}$
for systematic source $\alpha_j$ are obtained from linear
fits to distributions as shown in Fig.~\ref{chi2_scan}. Parameters
$\alpha_j$ and their respective errors are obtained by matrix
inversion technique.

\begin{figure}
\centering
\includegraphics[width=0.7\linewidth]{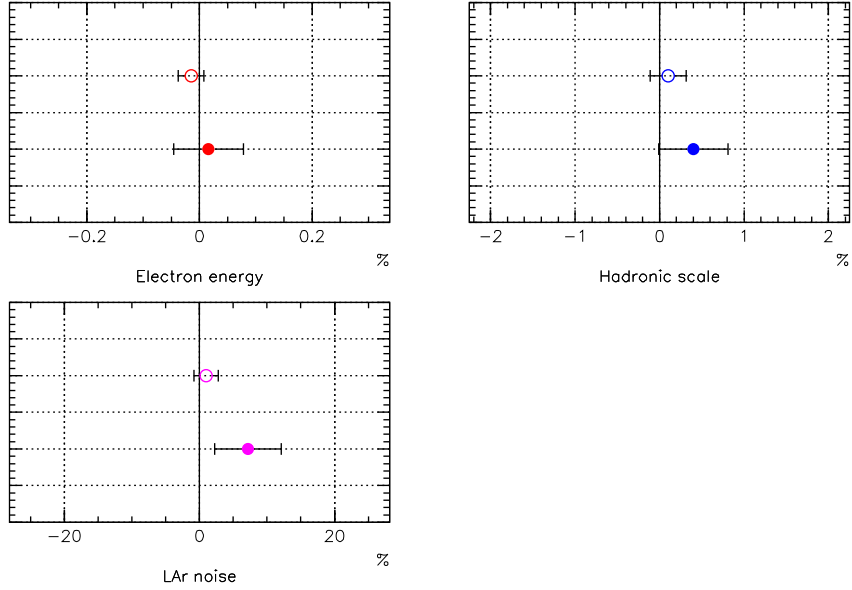}
\caption{Errors on the electron energy measurement (top-left),
hadronic scale calibration (top-right) and noise in LAr
calorimeter (bottom-left). Open points correspond to $\chi^2$ scan
in one correlated error source. Closed points show the result of
complete unfolding, taking into account
correlations.}\label{unfolding}
\end{figure}

The procedure was tested on available Monte Carlo sample for 2000
H1 data. Half of the sample, six million events, was used to
simulate data. Full analysis chain was applied to measure the
cross section and thus $R_i$. Kinematic bins were selected
according to \mbox{$15 \le Q^2 \le 60\,{\rm GeV}^2$} and
\mbox{$0.011 \le y \le 0.6$}, i.e. in the main region of the data.
The results are shown in Fig.~\ref{unfolding}. Closed points
correspond to unfolded errors of the electron energy measurement
(top-left), hadronic final state calibration and noise in the LAr
calorimeter (bottom-left). There is no sensitivity observed to the
electron polar angle measurement. All values are within
statistical errors compatible with zero, as expected. For the
final analysis the statistical errors are expected to be
approximately three times smaller due to the significantly larger
statistics than used for the presented study. This will enable the
method to gain sufficient control over systematic correlated
errors. Apart from being able to evaluate calibration of the
scattered electron and of the hadronic final state, it gives a
very good handle on the LAr calorimeter noise. 

For a comparison, open points in Fig.~\ref{unfolding} correspond
to a $\chi^2$ scan in one correlated error source. The statistical
errors are smaller, as expected, and compatible with zero.
However, the unfolding method is preferred since it takes into
account all correlated error sources correctly.

In summary, a study of the DIS cross section uncertainties
realistically achievable at HERA has been performed. For $x\in 0.001-0.01$ a 
precision of $1\%$ can be reached across for a wide range of $
Q^2 \in 5-150$~GeV$^2$, allowing improved estimate of $W,Z$ production cross section
in the central rapidity region of LHC. The accuracy of the DIS cross section
measurement can be verified using different kinematic reconstruction 
methods available at the HERA collider.

%% file: comparisonh1zeus.tex
\subsection{Comparison and combination of ZEUS and H1 PDF analyses
\protect\footnote{Contributing authors: A.~Cooper-Sarkar, C.~Gwenlan}}

\label{sec:mandy}

Parton Density Function (PDF) determinations are usually global 
fits~\cite{mrst,cteq,zeus}, which use fixed target 
DIS data as well as HERA data. In such analyses the high statistics HERA NC 
$e^+p$ data, which span the range $6.3 \times 10^{-5} < x < 0.65,
2.7 < Q^2 < 30,000$GeV$^2$, 
have determined the low-$x$ sea and 
gluon distributions, whereas the fixed target data have determined 
the valence distributions and the higher-$x$ sea distributions. 
The $\nu$-Fe fixed target data have been the most important input  
for determining the valence distributions, but these data suffer 
from uncertainties due to heavy target corrections. Such uncertainties 
are also present for deuterium fixed target data, 
which have been used to determine the shape of the high-$x$ $d$-valence quark.

HERA data on neutral and charged current (NC and CC) 
$e^+p$ and $e^-p$ inclusive double differential 
cross-sections are now available, 
and have been used by both the H1 and ZEUS collaborations~\cite{zeusj,h1a} 
in order to determine the parton distributions functions (PDFs) using data 
from within a single experiment. The HERA high $Q^2$ cross-section 
data can be used to determine the valence 
distributions, thus eliminating uncertainties from heavy target corrections. 
The PDFs are presented with full accounting for uncertainties from correlated 
systematic errors (as well as from statistical and uncorrelated sources).
Peforming an analysis within a single experiment has considerable advantages
in this respect, since the global fits 
have found significant tensions between 
different data sets, which make a rigorous statistical treatment of 
uncertainties difficult. 

\begin{figure}[tbp]
\centerline{
\epsfig{figure=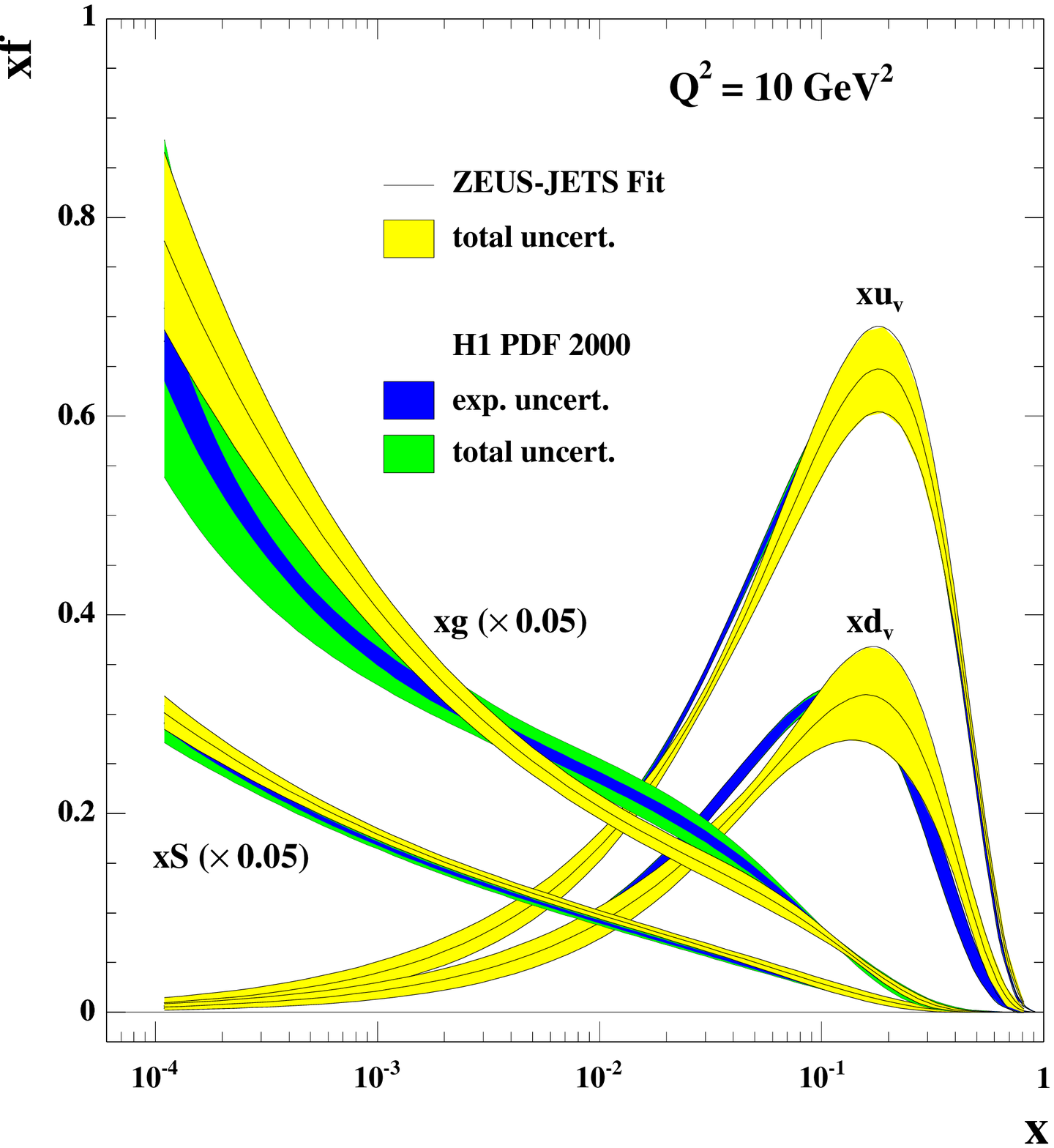,height=6cm}
\epsfig{figure=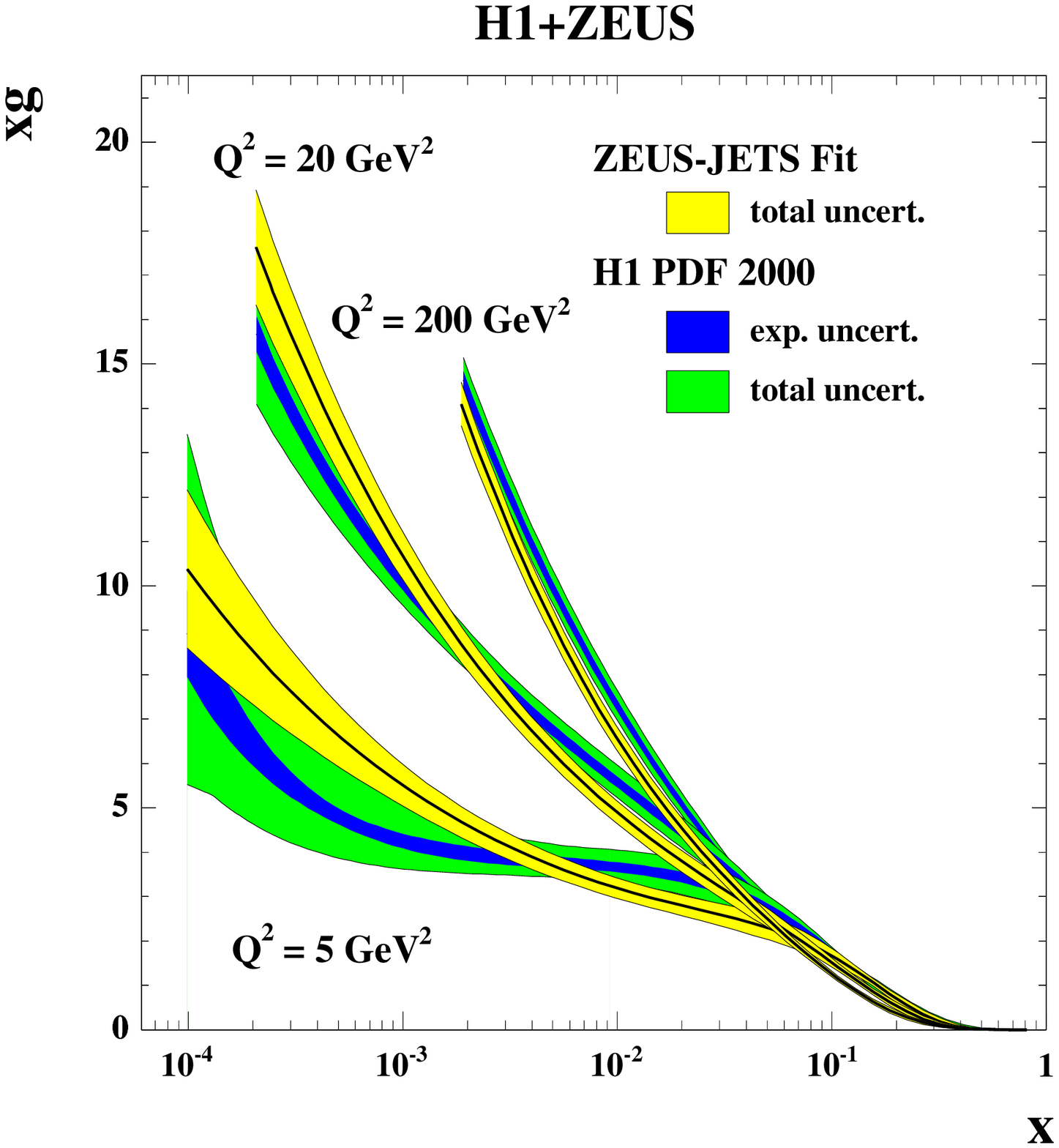,height=6cm}}
\caption {Left plot: Comparison of PDFs from ZEUS and H1 analyses at $Q^2=10$GeV$^2$.
Right plot: Comparison of gluon from ZEUS and H1 analyses, at various $Q^2$. 
Note that the ZEUS analysis total uncertainty includes both experimental and 
model uncertainties.}
\label{fig:h1zeus}
\end{figure}

Fig.~\ref{fig:h1zeus} compares the results of the H1 and ZEUS analyses. 
Whereas the extracted PDFs are broadly compatible within errors, there is a 
noticeable difference in the shape of the gluon PDFs.
Full details of the analyses are given in the relevant publications,
in this contribution we examine the differences in the two analyses, recapping 
only salient details.

\subsubsection{Comparing ZEUS and H1 published PDF analyses}
The kinematics
of lepton hadron scattering is described in terms of the variables $Q^2$, the
invariant mass of the exchanged vector boson, Bjorken $x$, the fraction
of the momentum of the incoming nucleon taken by the struck quark (in the 
quark-parton model), and $y$ which measures the energy transfer between the
lepton and hadron systems.
The differential cross-section for the NC process is given in terms of the
structure functions by
\begin{equation}
\frac {d^2\sigma(e^{\pm}p) } {dxdQ^2} =  \frac {2\pi\alpha^2} {Q^4 x}
\left[Y_+\,F_2(x,Q^2) - y^2 \,F_L(x,Q^2)
\mp Y_-\, xF_3(x,Q^2) \right],  \label{Eqn:NC}
\end{equation}
where $\displaystyle Y_\pm=1\pm(1-y)^2$. 
The structure functions $F_2$ and $xF_3$ are 
directly related to quark distributions, and their
$Q^2$ dependence, or scaling violation, 
is predicted by pQCD. At $Q^2 \leq 1000$~GeV$^2$ $F_2$ dominates the
charged lepton-hadron cross-section and for $x \leq 10^{-2}$, $F_2$ itself 
is sea quark dominated but its $Q^2$ evolution is controlled by
the gluon contribution, such that HERA data provide 
crucial information on low-$x$ sea-quark and gluon distributions.
At high $Q^2$, the structure function $xF_3$ becomes increasingly important, 
and
gives information on valence quark distributions. The CC interactions 
enable us to separate the flavour of the valence distributions 
at high-$x$, since their (LO) cross-sections are given by, 
\[
\frac {d^2\sigma(e^+ p) } {dxdQ^2} = \frac {G_F^2 M_W^4} {(Q^2 +M_W^2)^2 2\pi x}
x\left[(\bar{u}+\bar{c}) + (1 - y)^2 (d + s) \right],
\]
\[
\frac {d^2\sigma(e^- p) } {dxdQ^2} = \frac {G_F^2 M_W^4} {(Q^2 +M_W^2)^2 2\pi x}
x\left[(u + c) + (1 - y)^2 (\bar{d} + \bar{s}) \right].
\]
For both HERA analyses the QCD predictions for the structure functions 
are obtained by solving the DGLAP evolution equations~\cite{gl,ap,l,d} 
at NLO in the \msbar scheme with the
renormalisation and factorization scales chosen to be $Q^2$. 
These equations yield the PDFs
 at all values of $Q^2$ provided they
are input as functions of $x$ at some input scale $Q^2_0$. 
The resulting PDFs are then convoluted with coefficient functions, to give the
structure functions which enter into the expressions for the cross-sections.
For a full explanation of the relationships between DIS cross-sections, 
structure functions, PDFs and the QCD improved parton model see 
ref.~\cite{dcs}.

The HERA data are all in a kinematic region where there is no
sensitivity to target mass and higher 
twist contributions but a minimum $Q^2$ cut must be imposed 
to remain in the kinematic region where
perturbative QCD should be applicable. For ZEUS this is $Q^2 > 2.5$~GeV$^2$, 
and for H1 it is $Q^2 > 3.5$~GeV$^2$. Both collaborations have included the 
sensitivity to this cut as part of their model errors.

In the ZEUS analysis, the PDFs for $u$ valence, $xu_v(x)$,  $d$ valence, $xd_v(x)$, 
total sea, $xS(x)$, the 
gluon, $xg(x)$, and the difference between the $d$ and $u$
contributions to the sea, $x(\bar{d}-\bar{u})$, are each parametrized  
by the form 
\begin{equation}
  p_1 x^{p_2} (1-x)^{p_3} P(x),
\label{eqn:pdf}
\end{equation}
where $P(x) = 1 +p_4 x$, at $Q^2_0 = 7$GeV$^2$. The total sea 
$xS=2x(\bar{u} +\bar{d} +\bar{s}+ \bar{c} +\bar{b})$, where 
$\bar{q}=q_{sea}$ for each flavour, $u=u_v+u_{sea}, d=d_v+d_{sea}$ and 
$q=q_{sea}$ for all other flavours. 
The flavour structure of the light quark sea 
allows for the violation of the Gottfried sum rule. However, there is no 
information on the shape of the $\bar{d}-\bar{u}$ distribution in a fit 
to HERA data alone and so this distribution has its shape fixed consistent 
with the Drell-Yan data and its normalisation consistent 
with the size of the Gottfried sum-rule violation. 
A suppression of the strange sea with respect to the non-strange sea 
of a factor of 2 at $Q^2_0$, is also imposed
consistent with neutrino induced dimuon data from CCFR. 
Parameters are further restricted as follows.
The normalisation parameters, $p_1$, for the $d$ and $u$ valence and for the 
gluon are constrained to impose the number sum-rules and momentum sum-rule. 
The $p_2$ parameter which constrains the low-$x$ behaviour of the $u$ and $d$ 
valence distributions is set equal, 
since there is no information to constrain any difference. 
When fitting to HERA data alone it is also necessary to constrain 
the high-$x$ sea and gluon shapes, because HERA-I data do not have high 
statistics at large-$x$, in the region where these distributions are small.
The sea shape has been restricted by setting $p_4=0$ for the sea, 
but the gluon shape is constrained by including data on jet production in the 
PDF fit. Finally the ZEUS analysis has 11 free PDF parameters. 
ZEUS have included reasonable variations of 
these assumptions about the input parametrization 
in their analysis of model uncertainties. 
The strong coupling constant was fixed to $\asmz =  0.118$~\cite{Eidelman:2004wy}.
Full account has been taken of correlated experimental 
systematic errors by the Offset Method, 
as described in ref~\cite{zeus,Mandy}.

For the H1 analysis, the value of $Q^2_0 = 4$GeV$^2$, and 
the choice of quark distributions which are 
parametrized is different. The quarks are considered as $u$-type and $d$-type
with different parametrizations for, $xU= x(u_v+u_{sea} + c)$, 
$xD= x(d_v +d_{sea} + s)$, $x\bar{U}=x(\bar{u}+\bar{c})$ and 
$x\bar{D}=x(\bar{d}+\bar{s})$, with $q_{sea}=\bar{q}$, as 
usual, and the the form of the quark and gluon parametrizations
given by Eq.~\ref{eqn:pdf}. For $x\bar{D}$ and $x\bar{U}$ the polynomial, 
$P(x)=1.0$,
for the gluon and $xD$, $P(x)= (1+p_4 x)$, and for $xU$, 
$P(x)= (1 +p_4 x +p_5 x^3)$. The parametrization is then further restricted 
as follows.
Since the valence distributions must vanish as $x \to 0$, 
the low-$x$ parameters, $p_1$
 and $p_2$ are set equal for $xU$ and $x\bar{U}$, and for $xD$ and 
$x\bar{D}$. Since there is no information on the flavour structure of the sea 
it is 
also necessary to set $p_2$ equal for $x\bar{U}$ and $x\bar{D}$. 
The normalisation, $p_1$, of the gluon is determined from the momentum 
sum-rule and the $p_4$ parameters for $xU$ and $xD$ are determined from the 
valence number sum-rules.
Assuming that the strange and charm quark distributions can be expressed as 
$x$ independent fractions, $f_s$ and $f_c$, of the $d$ and $u$ type sea, 
gives the further constraint $p_1(\bar{U})=p_1(\bar{D}) (1-f_s)/(1-f_c)$. 
Finally there are 10 free parameters. H1 has also included reasonable 
variations of 
these assumptions in their analysis of model uncertainties. 
The strong coupling constant was fixed to $\asmz =  0.1185$ and this is 
sufficiently similar to the ZEUS choice that we can rule it out as a cause of
any significant difference. 
Full account has been taken of correlated experimental 
systematic errors by the Hessian Method, see ref.~\cite{Mandy}.
 
For the ZEUS analysis, the heavy quark production scheme used is the
general mass variable flavour number scheme of Roberts and Thorne~\cite{hq}.
For the H1 analysis, the zero mass variable flavour number scheme is used. 
It is well known that these choices have a small effect on the steepness of 
the gluon at very small-$x$, such that the zero-mass choice produces a 
slightly less steep gluon. However, there is no effect on the more striking 
differences in the gluon shapes at larger $x$.

There are two differences in 
the analyses which are worth further investigation. 
The different choices for the form of the PDF parametrization at 
$Q^2_0$ and the 
different treatment of the correlated experimental uncertainties.

\subsubsection{Comparing different PDF analyses of the same data set and 
comparing different data sets using the same PDF analysis.}

\begin{figure}[tbp]
\vspace{-2.0cm} 
\centerline{
\epsfig{figure=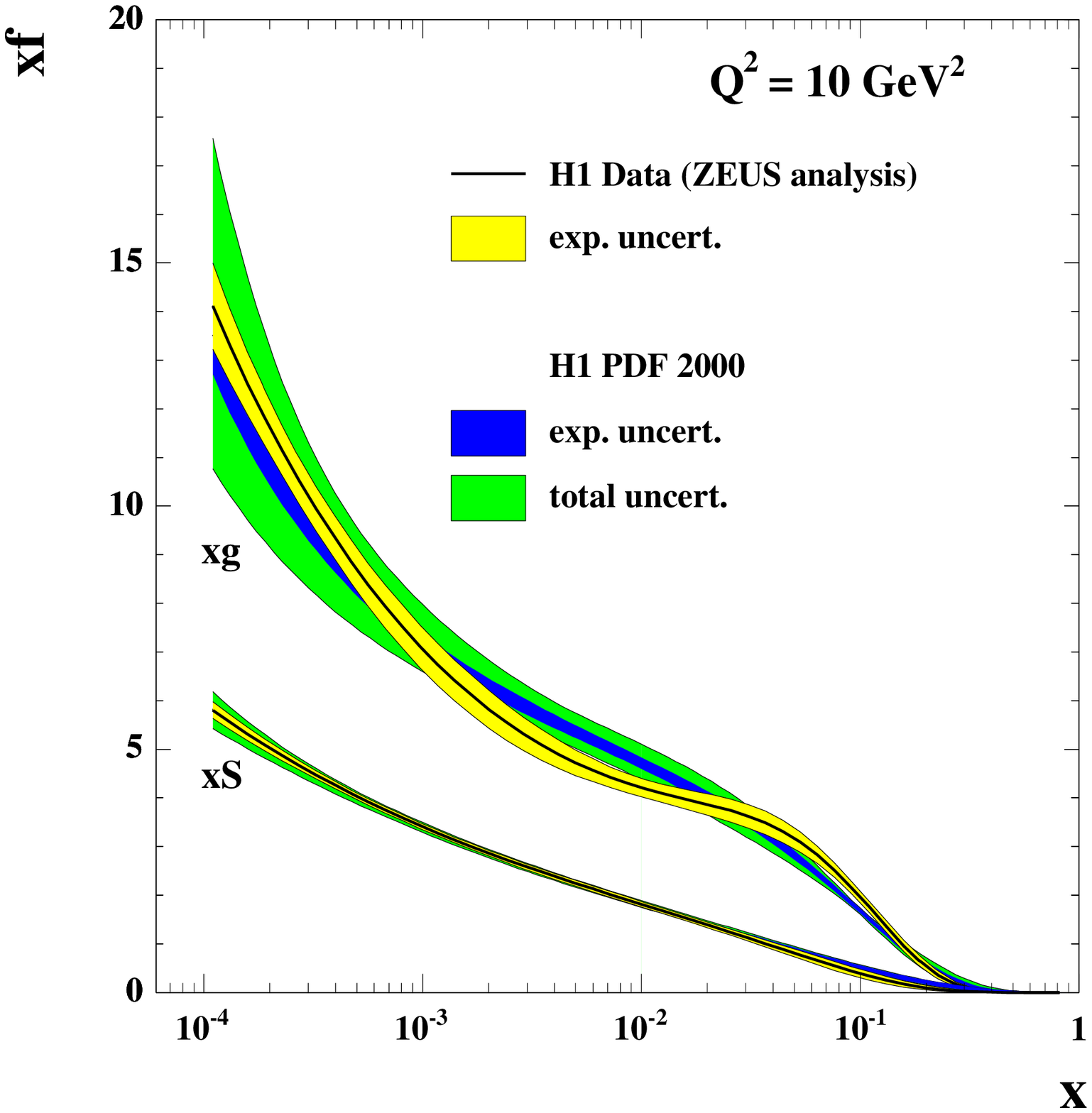,width=0.33\textwidth}
\epsfig{figure=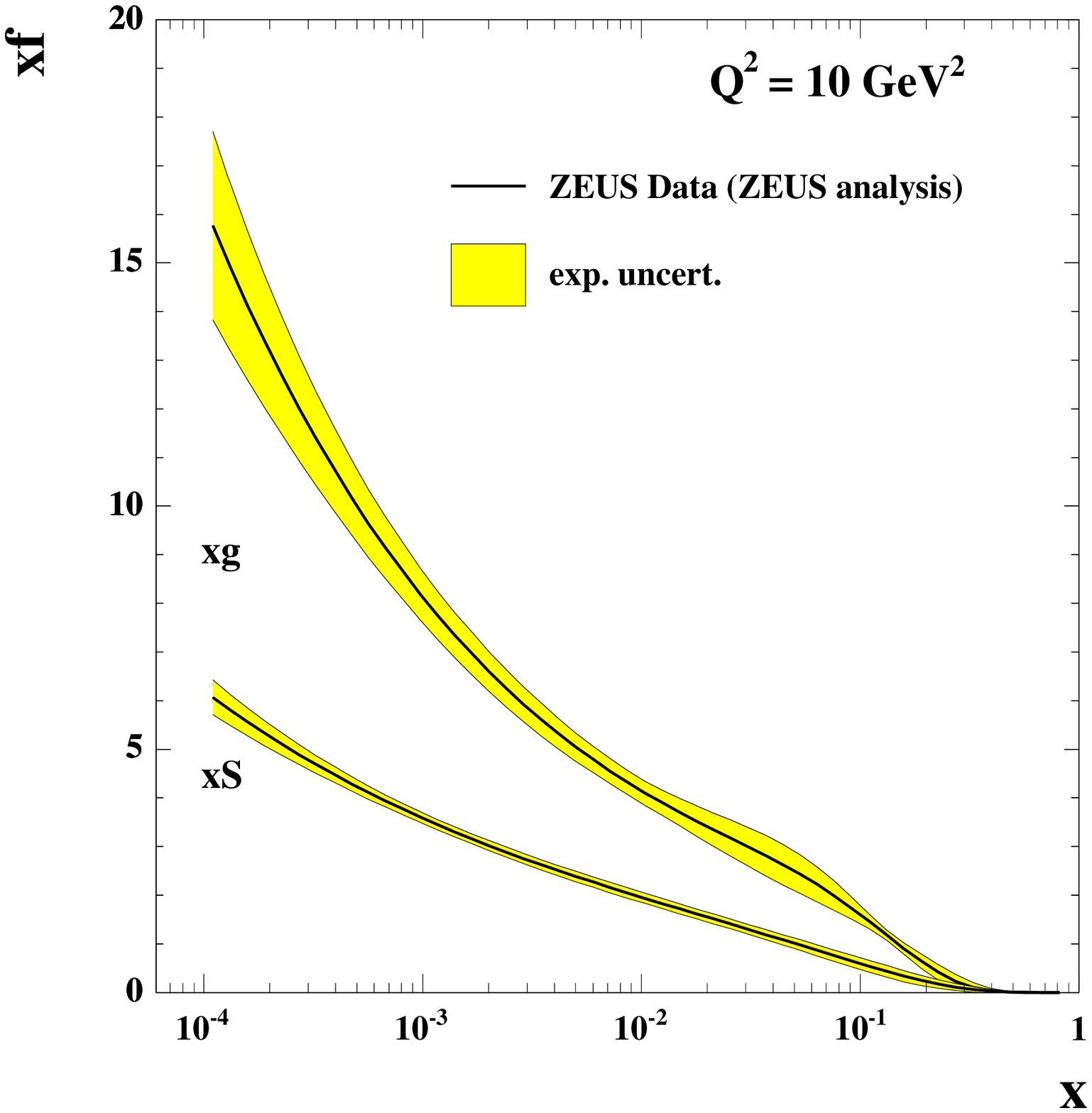,width=0.33\textwidth}
\epsfig{figure=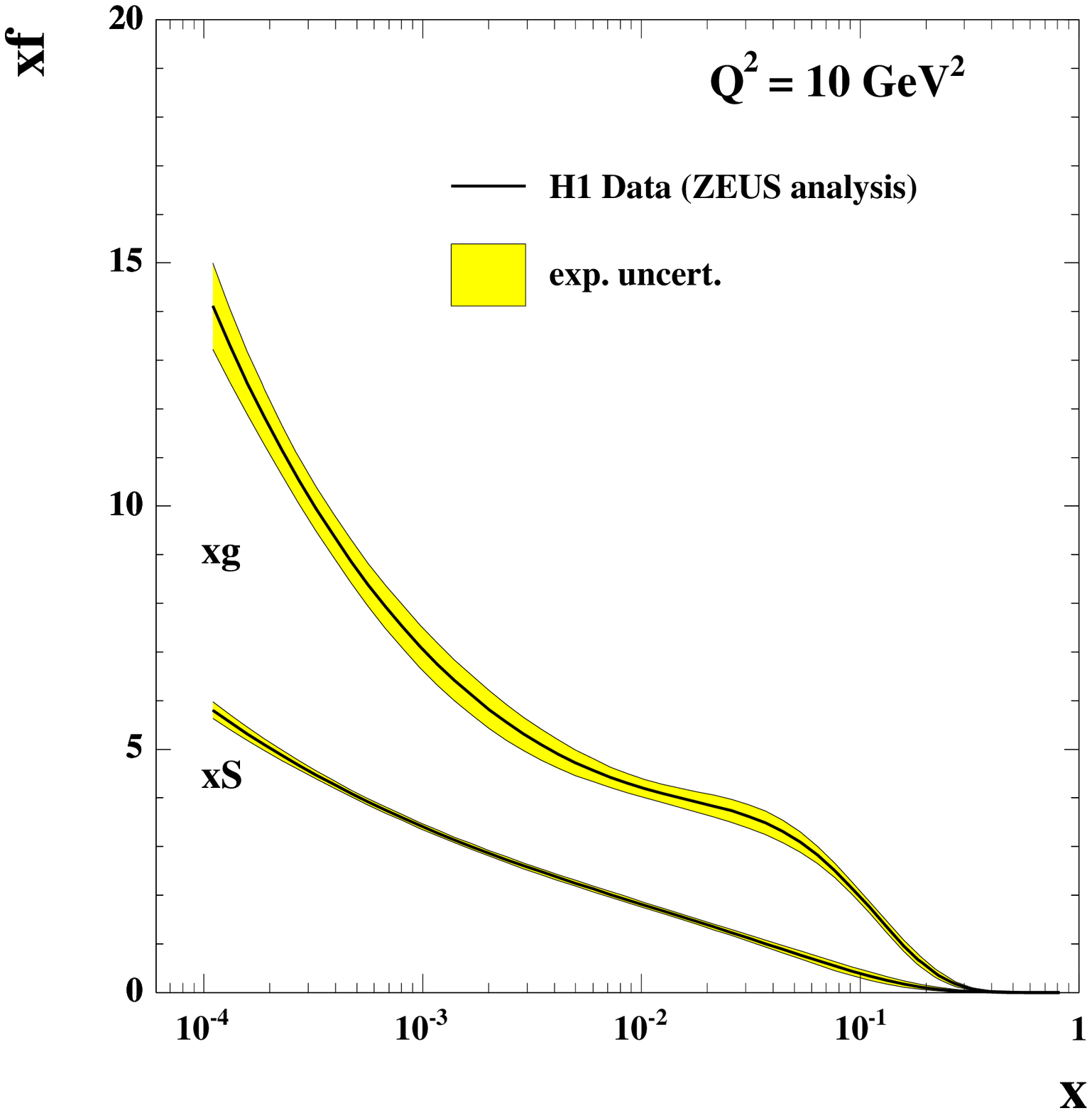,width=0.33\textwidth}
}
\caption {Sea and gluon distributions at $Q^2=10$GeV$^2$ extracted from 
different data sets and different analyses. 
Left plot: H1 data put through both ZEUS and H1 analyses.
Middle plot: ZEUS data put through ZEUS analysis. Right plot: H1 data put 
through ZEUS analysis.
}
\label{zazdzahd}
\end{figure}

So far we have compared the results of putting two different data sets into 
two different analyses. Because there are many differences in the assumptions 
going into these analyses
it is instructive to consider:(i) putting both data sets through the same analysis and (ii) putting one of the data sets through both analyses.
For these comparisons, the ZEUS analysis does NOT include the jet data, 
so that the data sets are more directly comparable, involving just the 
inclusive double differential cross-section data. 
Fig.~\ref{zazdzahd} compares the sea and gluon PDFs, 
at $Q^2=10$GeV$^2$, extracted from H1 data using the H1 PDF analysis 
with those extracted from H1 data using the ZEUS PDF analysis. 
These alternative analyses of the same data 
set give results which are compatible within the model dependence error 
bands. Fig.~\ref{zazdzahd} also compares the sea and gluon PDFs extracted from 
ZEUS data using the ZEUS analysis with those extracted from H1 data using 
the ZEUS analysis. From 
this comparison we can see that the different data sets lead to somewhat 
different gluon shapes even when put through exactly the same analysis. 
Hence the most of the difference in shape of the ZEUS and H1 PDF analyses can 
be traced back to a difference at the level of the data sets.

\subsubsection{Comparing the Offset and Hessian method of assessing correlated experimental uncertainties}

Before going further it is useful to discuss the treatment of correlated 
systematic errors in the ZEUS and H1 analyses. A full discussion of the 
treatment of correlated systematic errors in PDF 
analyses is given in ref~\cite{dcs}, only salient details are recapped here. 
Traditionally, experimental collaborations have evaluated an overall systematic
uncertainty on each data point and these have been treated as uncorrelated, 
such that they are simply added to the statistical uncertainties in quadrature
when evaluating $\chi^2$. However, modern deep inelastic scattering experiments
have very small statistical uncertainties, so that the contribution of 
systematic uncertainties becomes dominant and consideration of 
point to point correlations between systematic uncertainties is essential.

For both ZEUS and H1 analyses
the formulation of the $\chi^2$ including correlated systematic uncertainties
 is constructed as follows. The correlated uncertainties
are included in the theoretical prediction, $F_i(p,s)$, such that
\[ 
F_i(p,s) = F_i^{\rm NLOQCD}(p) + 
\sum_{\lambda} s_{\lambda} \Delta^{\rm sys}_{i\lambda}
\]
where, $F_i^{\rm NLOQCD}(p)$, represents the prediction 
from NLO QCD in terms of the theoretical parameters $p$,
and the parameters $s_\lambda$ represent independent variables 
for each source of
 systematic uncertainty. They have zero mean and unit variance by construction.
The symbol 
$\Delta^{\rm sys}_{i\lambda}$ represents the one standard deviation correlated 
systematic error on data point $i$ due to correlated error 
source $\lambda$.
The $\chi^2$ is then formulated as 
\begin{equation}
\chi^2 = \sum_i \frac{\left[ F_i(p,s)-F_i(\rm meas) \right]^2}{\sigma_i^2} + \sum_\lambda s^2_\lambda 
\label{eq:chi2}
\end{equation}
where, $F_i(\rm meas)$, represents a measured data point and the symbol 
$\sigma_i$ represents the one standard deviation uncorrelated 
error on data point $i$, from both statistical and systematic sources. 
The experiments use this $\chi^2$ in different ways. ZEUS uses the Offset 
method and H1 uses the Hessian method.
 
Traditionally, experimentalists have used `Offset' methods to account for
correlated systematic errors. The $\chi^2$ is formluated without any terms
due to correlated systematic errors ($s_\lambda=0$ in Eq.~\ref{eq:chi2}) for
evaluation of the central values of the fit parameters. 
However, the data points are then offset to account for each 
source of systematic error in turn 
(i.e. set $s_\lambda = + 1$ and then $s_\lambda = -1$ for each source 
$\lambda$) 
and a new fit is performed for each of these
variations. The resulting deviations of the theoretical parameters 
from their  central  values are added in 
quadrature. (Positive and  negative deviations are added 
in quadrature separately.) This method does not assume that the systematic
uncertainties are Gaussian distributed.
An equivalent (and much more efficient) procedure to perform the
Offset method has been given by 
Pascaud and Zomer~\cite{pascaud}, and this is what is actually used.
The Offset method is a conservative method of error estimation  
as compared to the Hessian method. 
It gives fitted theoretical predictions which are as close as 
possible to the central values of the published data. It does not use the full 
statistical power of the fit to improve the estimates of $s_\lambda$, 
since it choses to mistrust the systematic error estimates,
but it is correspondingly more robust.

The Hessian method is an alternative procedure in which the systematic
uncertainty parameters $s_\lambda$ are allowed to vary in the main fit 
when determining the values of the theoretical parameters. 
Effectively, the theoretical prediction is not fitted
to the central values of the published experimental data, but  
these data points are allowed to move
collectively, according to their correlated systematic uncertainties.
 The theoretical prediction determines the 
optimal settings for correlated systematic shifts of experimental data points 
such that the most consistent fit to all data sets is obtained. Thus, 
in a global fit, systematic shifts in 
one experiment are correlated to those in another experiment by the fit.
In essence one is allowing the theory to calibrate the detectors. This requires
great confidence in the theory, but more significantly, it requires confidence
in the many model choices which go into setting the boundary conditions for 
the theory (such as the parametrization at $Q^2_0$). 

\begin{figure}[tbp]
\vspace{-2.0cm} 
\centerline{
\epsfig{figure=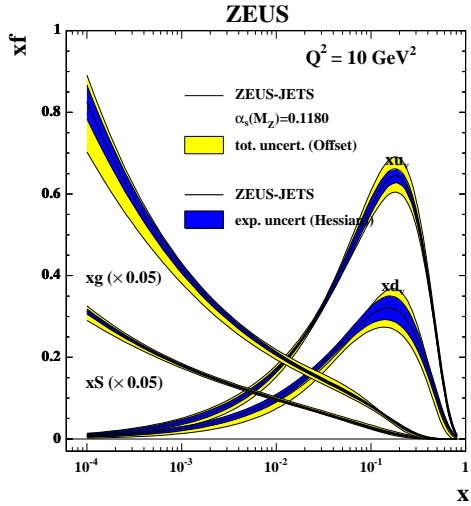,height=7cm}
}
\caption {PDFs at $Q^2=10$GeV$^2$, for the ZEUS analysis of ZEUS data 
performed by the Offset and the Hessian methods.
}
\label{fig:offhess}
\end{figure}

The ZEUS analysis can be performed using the Hessian method as well as the 
Offset method and
Fig.~\ref{fig:offhess} compares the PDFs, and their uncertainties, 
extracted from ZEUS data using these two methods. 
The central values of the different methods are in good agreement but 
the use of the Hessian method results in smaller uncertainties, for a 
the standard set of model assumptions, since the input data can be shifted 
within their correlated systematic uncertainties to suit the theory better. 
However, model uncertainties are more significant for the Hessian method than 
for the Offset method. The experimental 
uncertainty band for any one set of model choices 
is set by the usual $\chi^2$ tolerance, $\Delta \chi^2=1$, but the 
acceptability of a different set of choices is judged by the hypothesis 
testing criterion, such that
the $\chi^2$ should be approximately in the range $N \pm \surd (2N)$,
where $N$ is the number of degrees of freedom. The PDF
parameters obtained for the different model choices can differ 
by much more than their experimental uncertainties, because each model choice
can result in somewhat 
different values of the systematic uncertainty parameters, $s_\lambda$, and 
thus a different estimate of the shifted positions of the data points. This 
results in a larger spread of model uncertainty than in the Offset method, 
for which the data points cannot move. Fig~\ref{fig:h1zeus} illustrates the 
comparability of the ZEUS (Offset) total uncertainty  
estimate to the H1 (Hessian) experimental 
plus model uncertainty estimate. 

Another issue which arises in relation to the Hessian method is that
the data points should not be shifted far outside their 
one standard deviation systematic uncertainties. This can indicate 
inconsistencies between data sets, or parts of data sets, with respect to 
the rest of the data. The CTEQ collaboration have considered data
inconsistencies in their most recent global fit~\cite{cteq}.
They use the Hessian method but they increase the resulting uncertainty 
estimates, by 
increasing the $\chi^2$ tolerance to $\Delta \chi^2 = 100$, to allow for both 
model uncertainties and data inconsistencies. 
 In setting this tolerance they have considered the distances
from the $\chi^2$-minima of individual data sets 
to the global minimum for all data sets. These distances
 by far exceed the range allowed by the $\Delta \chi^2 =1$ criterion.
Strictly speaking such variations can indicate that data sets are inconsistent
but the CTEQ collaboration take the view 
that all of the current 
world data sets must be considered acceptable and compatible at some level,
even if strict statistical criteria are not met, since the conditions
for the application of strict criteria, namely Gaussian error distributions, 
are also not met. It is not possible to simply drop ``inconsistent'' data 
sets, as then the partons in some regions would lose important constraints.
On the other hand the level of ``inconsistency'' should 
be reflected in the uncertainties of the PDFs.
This is achieved by raising the $\chi^2$ tolerance. This 
results in uncertainty estimates which are comparable to 
those achieved by using the Offset method~\cite{Mandy}.

\subsubsection{Using both H1 and ZEUS data in the same PDF analysis}

\begin{figure}[tbp]
\centerline{
\epsfig{figure=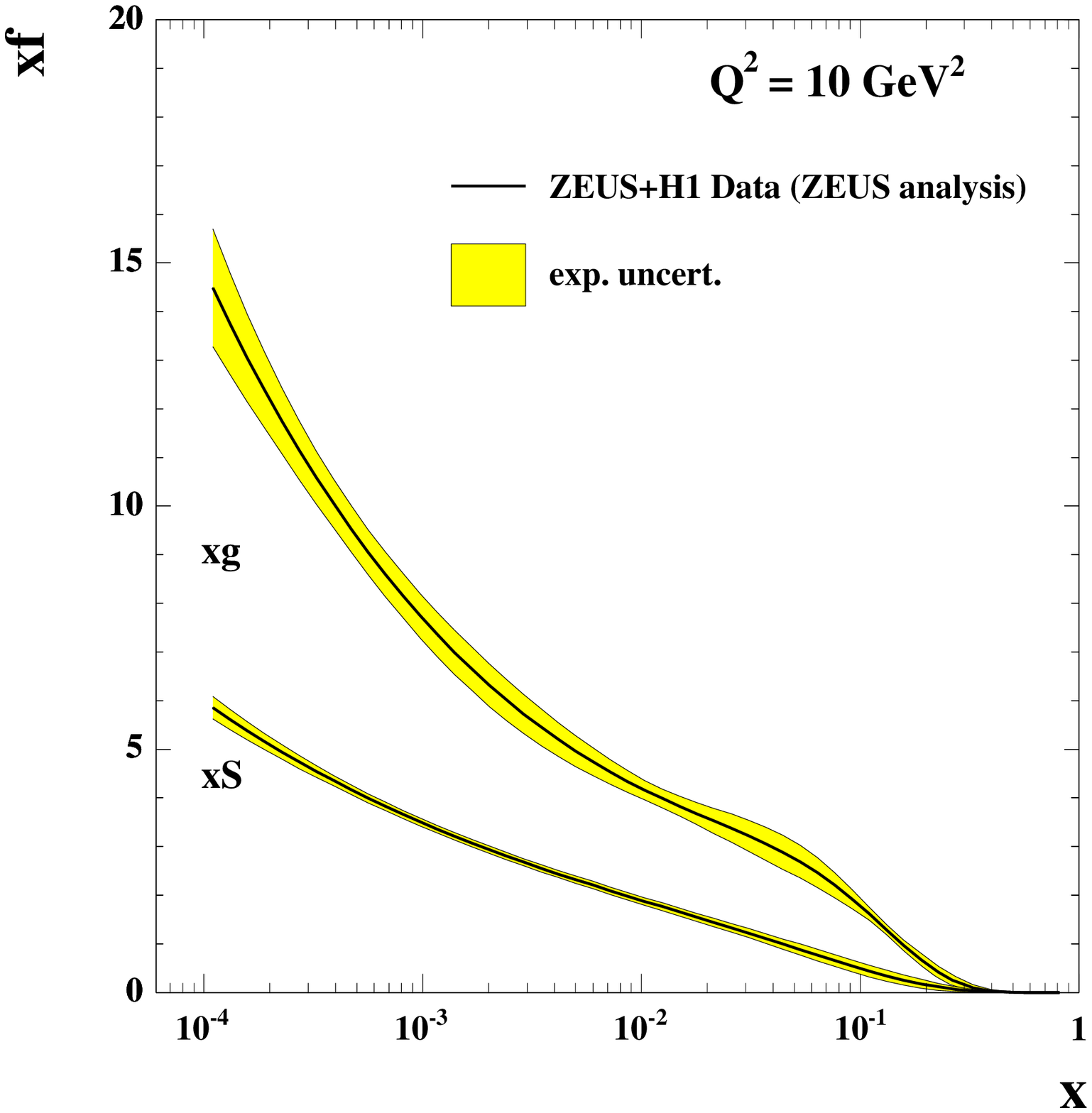,height=5cm}
\epsfig{figure=SEAGLU_ZEUSData_ZEUSANAL.eps,height=5cm}}
\centerline{
\epsfig{figure=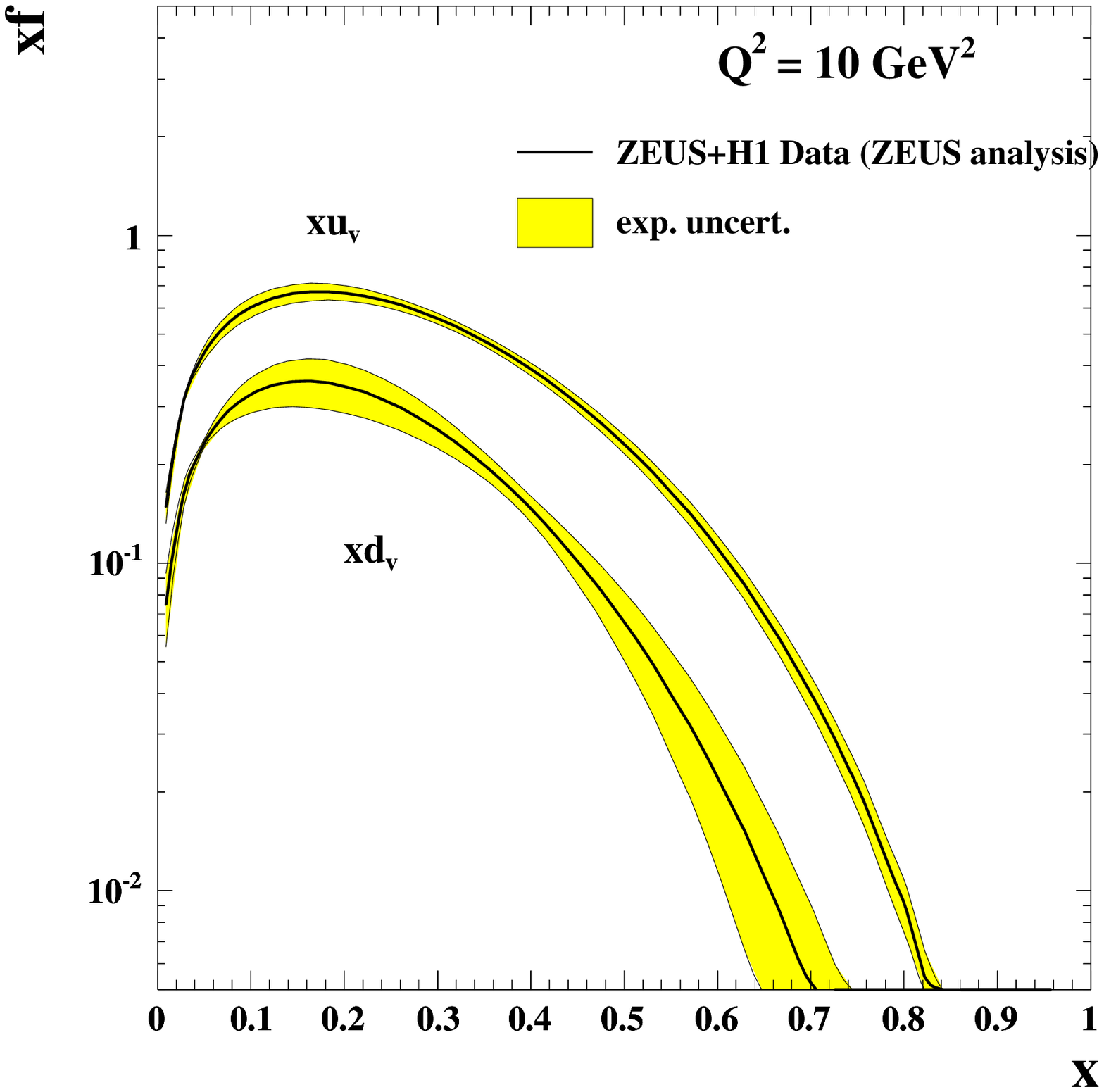,height=5cm}
\epsfig{figure=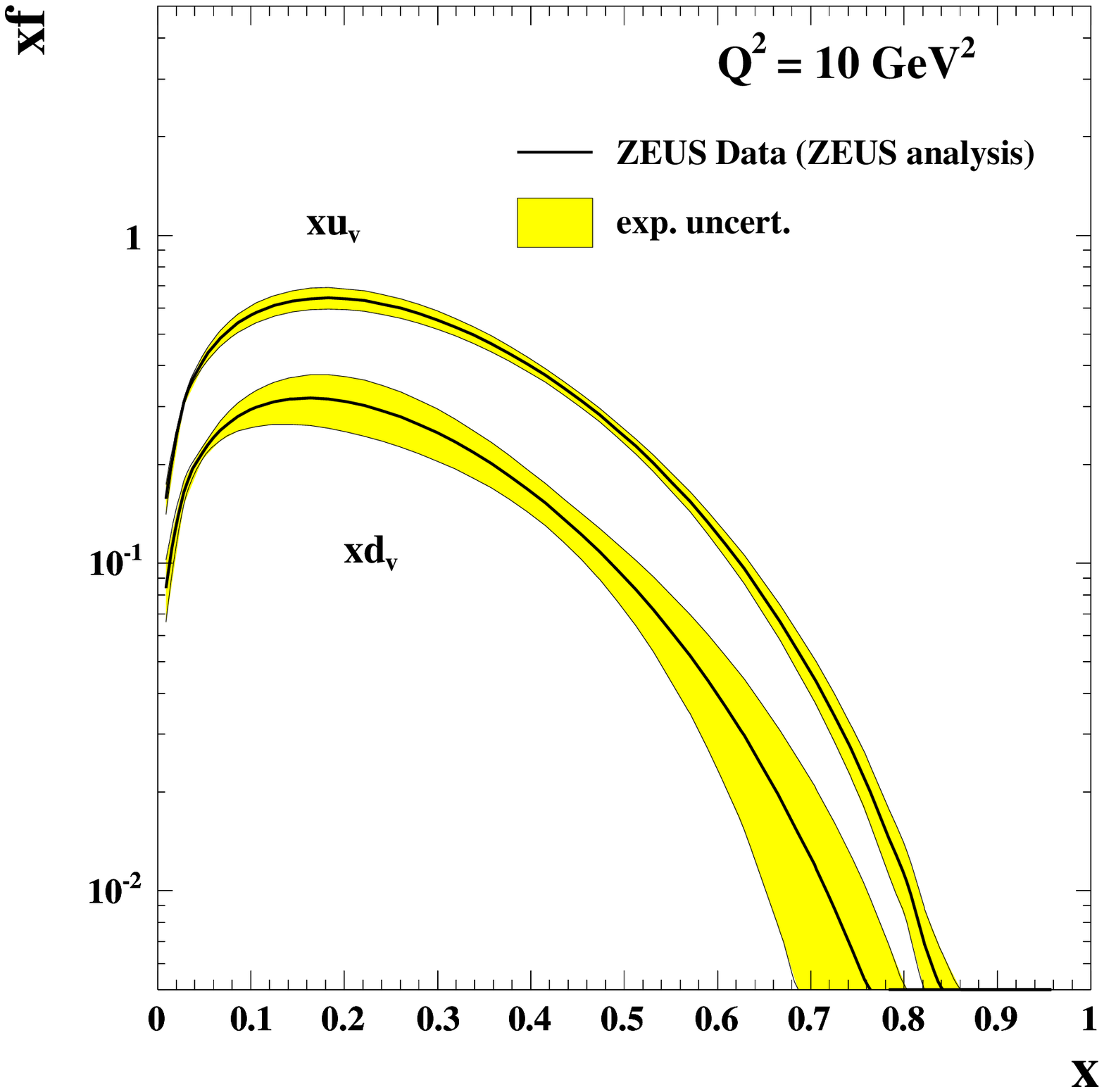,height=5cm}
}
\caption {Top plots: Sea and gluon distributions at $Q^2=10$GeV$^2$ extracted 
from H1 and ZEUS data using the ZEUS analysis (left) compared to those 
extracted from ZEUS data alone using the ZEUS analysis (right).
Bottom Plots: Valence distributions at $Q^2=10$GeV$^2$, extracted from 
H1 and ZEUS data using the ZEUS analysis (left) compared to those extracted 
from ZEUS data alone using the ZEUS analysis (right). 
}
\label{fig:zh1tog}
\end{figure}

Using data from a single experiment avoids questions of data consistency, 
but to
get the most information from HERA
it is necessary to put ZEUS and H1 data sets into the same 
analysis together, and then questions of consistency arise. 
Fig~\ref{fig:zh1tog} compares the sea and gluon PDFs and the 
$u$ and $d$ valence PDFs extracted from the ZEUS PDF 
analysis of ZEUS data alone,
to those extracted from the ZEUS PDF analysis of both H1 and ZEUS data.
It is noticeable that, for the low-$x$ sea and gluon PDFs, combining the 
data sets does not bring a reduction in uncertainty
equivalent to doubling the statistics. This is because the data which 
determine these PDFs are systematics limited. In fact there is some degree of 
tension between the ZEUS and the H1 data sets, such that the $\chi^2$ per 
degree of freedom rises for both data sets when they are fitted together. 
The Offset method of treating the systematic errors 
reflects this tension such that the overall uncertainty is not much improved 
when H1 data are added to ZEUS data. However, the uncertainty on the high-$x$
valence distributions is reduced by the input of 
H1 data, since the data are still statistics limited at high $x$.

\subsubsection{Combining the H1 and ZEUS data sets before PDF analysis} 

\begin{figure}[tbp]
\centerline{
\epsfig{figure=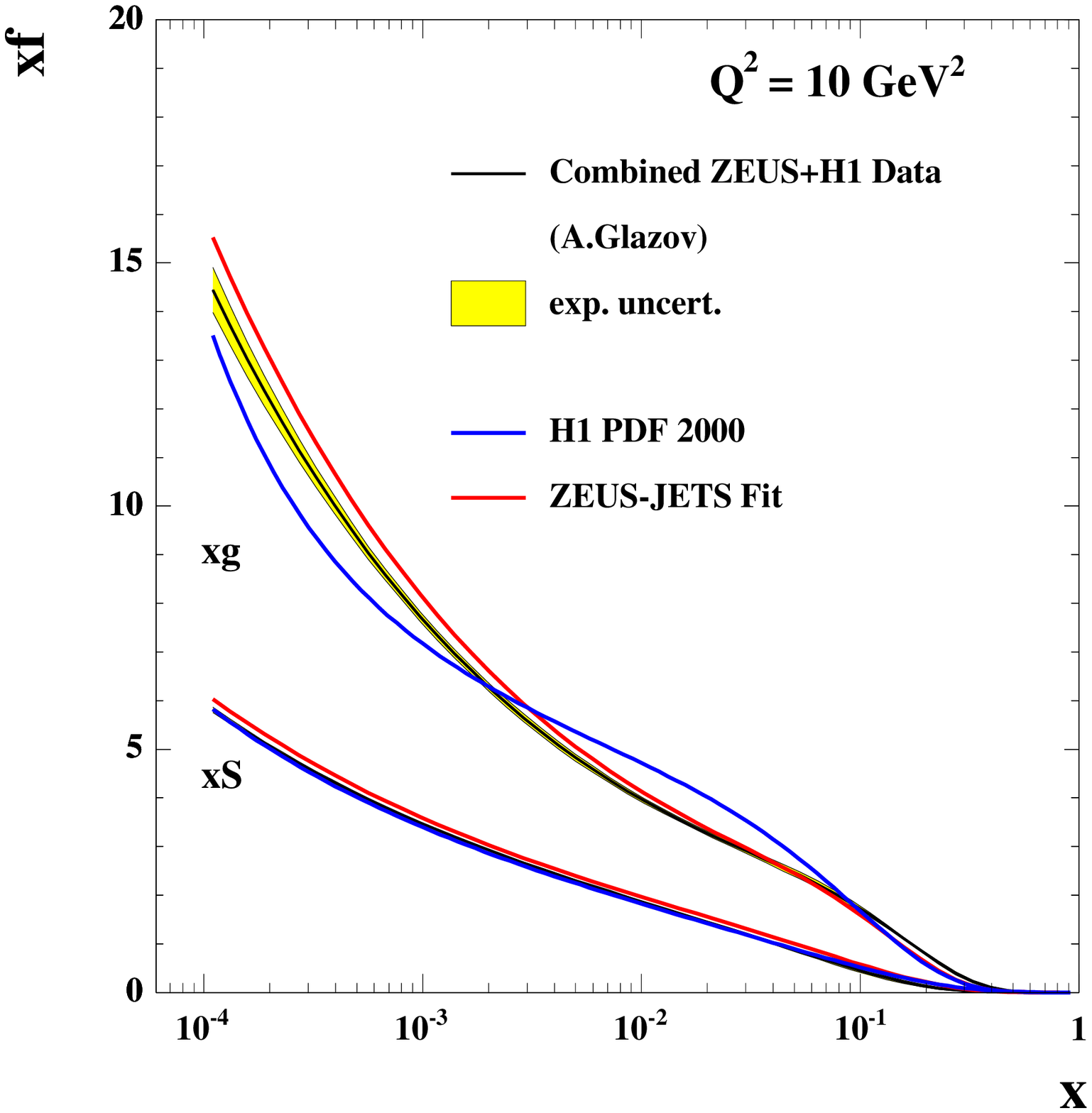,height=5cm}
\epsfig{figure=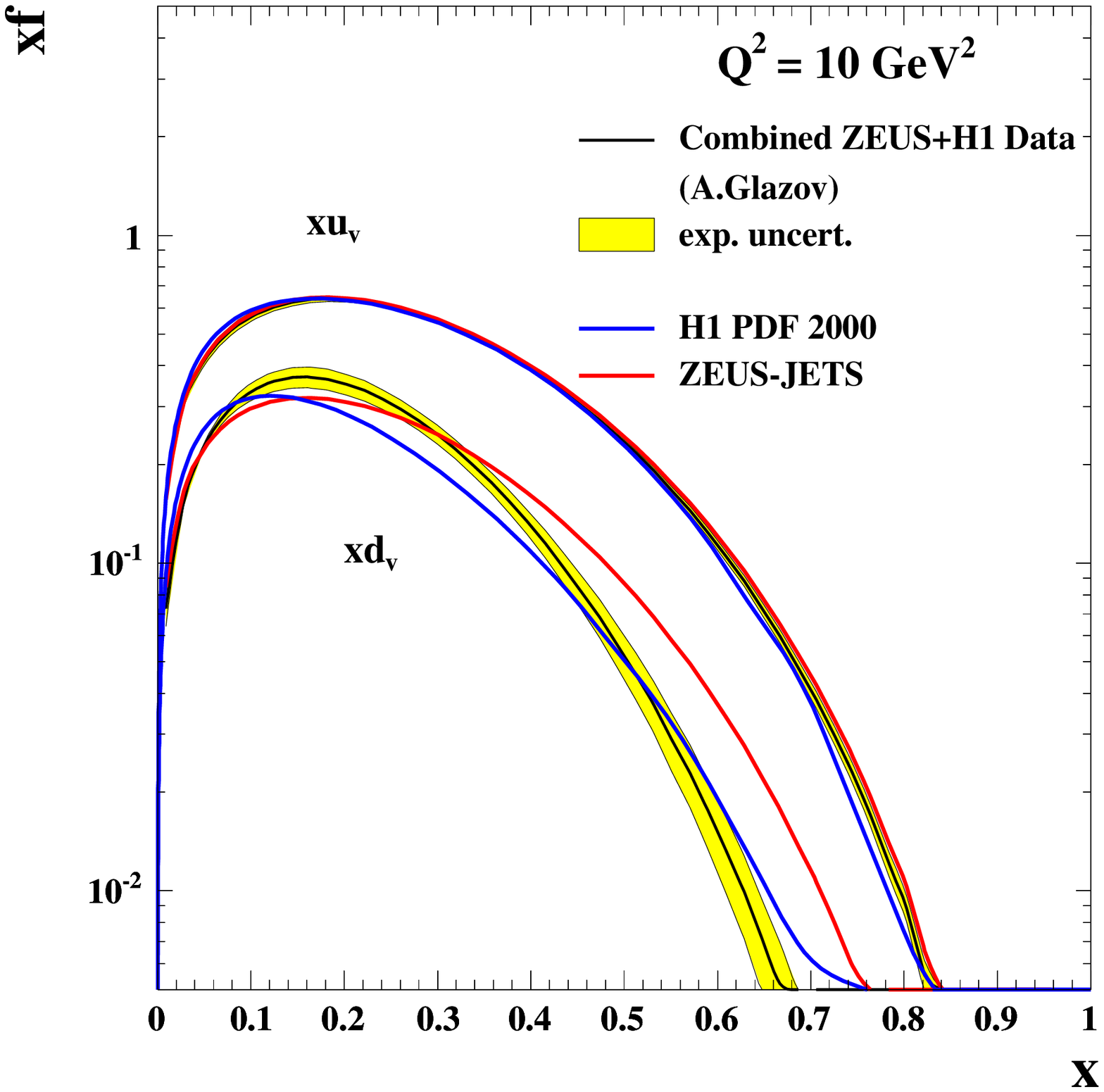,height=5cm}}

\caption {Left plot: Sea and gluon distributions at $Q^2=10$GeV$^2$, extracted 
from the combined H1 and ZEUS data set using the ZEUS analysis.
Right plot: Valence distributions at $Q^2=10$GeV$^2$, extracted from 
the combined H1 and ZEUS data set using the ZEUS analysis.
}
\label{fig:glazov}
\end{figure}

Thus there could be an advantage in combining ZEUS and H1  
data in a PDF fit if the tension between the data 
sets could be resolved. It is in this context the question of combining these 
data into a single data set arises. The procedure for combination is detailed 
in the contribution of S. Glazov to these proceedings (section~\ref{sec:ave}).
Essentially, since ZEUS and H1 are measuring the same physics in the same 
kinematic region, one can try to combine them using a 'theory-free' 
Hessian fit in which the only assumption is that there is a true 
value of the cross-section, for each process, at each $x,Q^2$ point. 
The systematic uncertainty parameters, $s_\lambda$, of each experiment 
are fitted to determine the best fit to this assumption. 
Thus each experiment is calibrated to the other. This works well because the 
sources of systematic uncertainty in each experiment are rather different. 
Once the procedure has been performed the resulting systematic uncertainties 
on each of the combined data points are significantly smaller than the 
statistical errors. Thus one can legitimately make a fit to the combined data 
set in which these statistical and systematic uncertainties are simply 
combined in quadrature. The result of making such a fit, using the ZEUS 
analysis, is shown in Fig.~\ref{fig:glazov}.
The central values of the ZEUS and H1 published 
analyses are also shown for comparison.
Looking back to Fig.~\ref{fig:zh1tog} one can see that there has been a 
dramatic reduction in the level of uncertainty 
compared to the ZEUS Offset method fit to the separate ZEUS and H1 data sets.
This result is very promising. A preliminary study of model dependence, varying
the form of the polynomial, $P(x)$, used in the PDF paremtrizations at $Q^2_0$,
also indicates that model dependence is relatively small. 

\begin{figure}[tbp]
\centerline{
\epsfig{figure=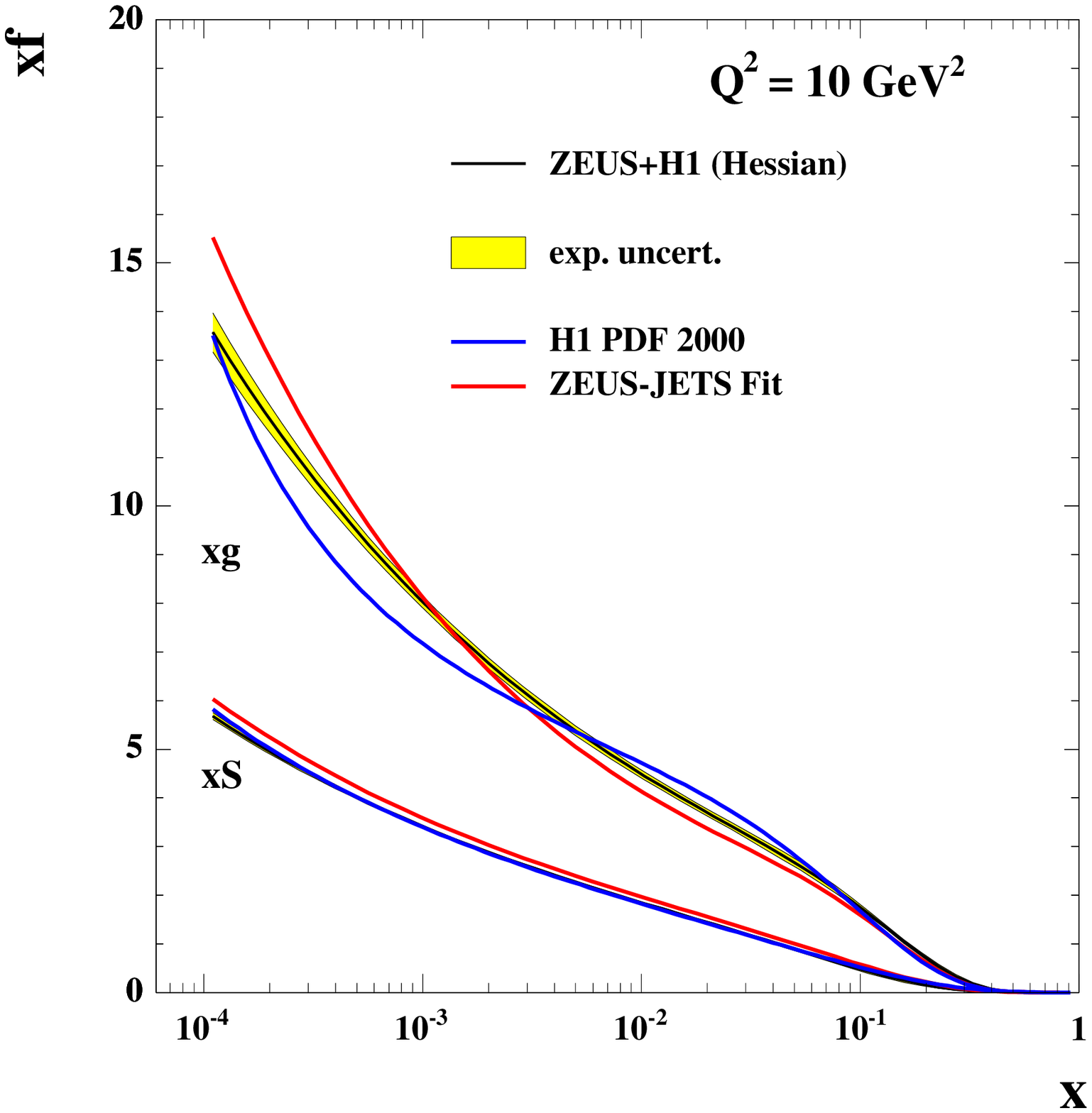,height=5cm}
\epsfig{figure=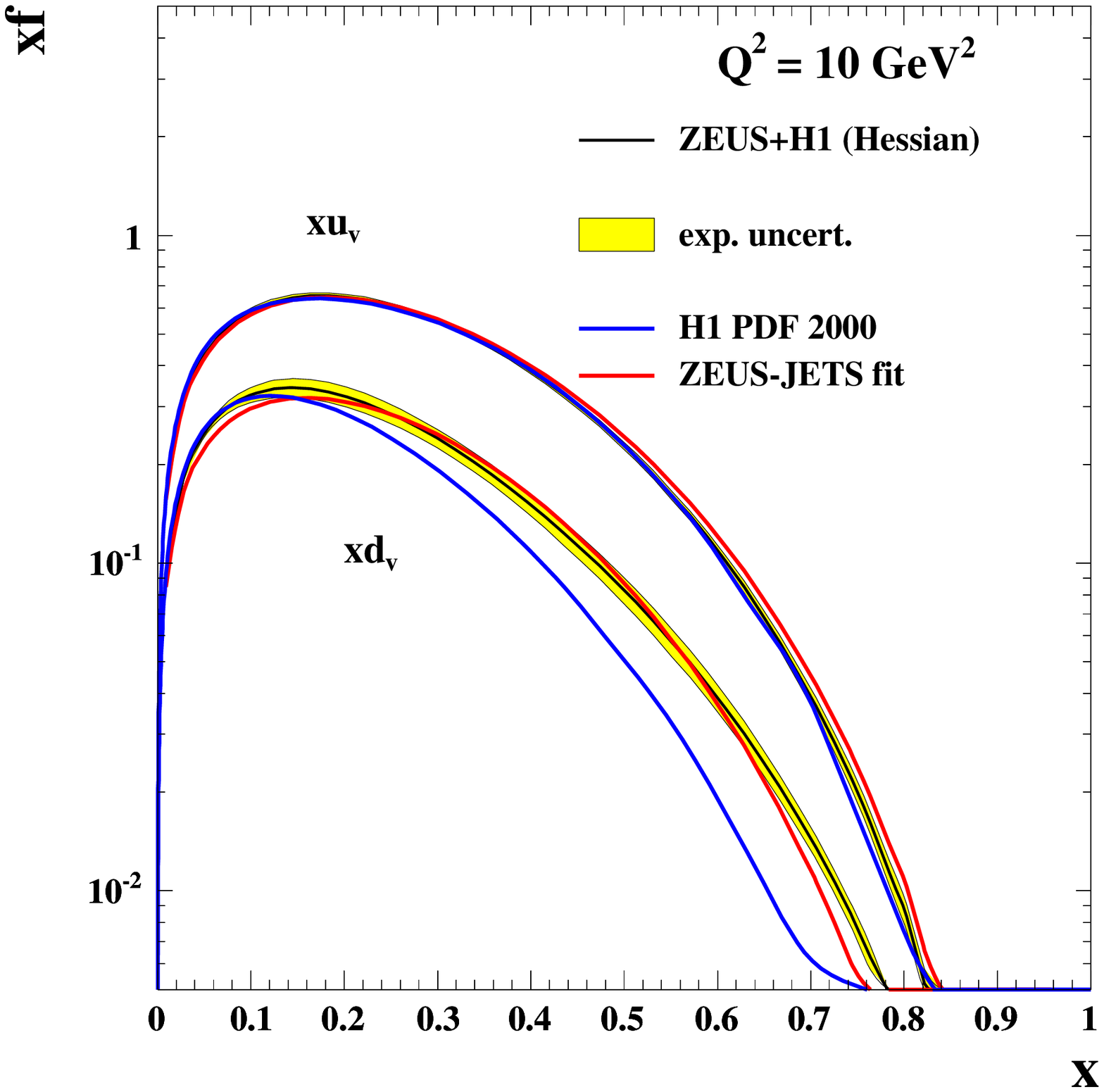,height=5cm}}

\caption {Left plot: Sea and gluon distributions at $Q^2=10$GeV$^2$, extracted
from the H1 and ZEUS data sets using the ZEUS analysis done by Hessian
method.
Right plot: Valence distributions at $Q^2=10$GeV$^2$, extracted
from the H1 and ZEUS data sets using the ZEUS analysis done by Hessian
method. 
}
\label{fig:zh1hess}
\end{figure}

The tension between ZEUS and H1 data could have been resolved by
putting them both into a PDF fit using the Hessian method to shift the data 
points. That is, rather than calibrating the two experiments to each other in 
the 'theory-free' fit, we could have used the theory of 
pQCD to calibrate each experiment. Fig.~\ref{fig:zh1hess} shows the PDFs 
extracted when the ZEUS and H1 data sets are put through the ZEUS PDF analysis
procedure using the Hessian method. 
The uncertainties on the resulting PDFs are comparable to those found for the 
fit to the combined data set, see Fig.~\ref{fig:glazov}. 
However, the central values of the resulting 
PDFs are rather different- particularly for the less well known gluon and
$d$ valence PDFs. For both of the fits shown in 
Figs.~\ref{fig:glazov},~\ref{fig:zh1hess} the values of the systematic error
 parameters, $s_\lambda$, for each experiment have been allowed to float so 
that the data points are shifted to give a better fit to our assumptions, but 
the values of the systematic error parameters chosen 
by the 'theory-free' fit and by the PDF fit are rather different. A 
representaive sample of these values is given in 
Table~\ref{tab:sl}. These discrepancies might be somewhat alleviated by a full 
consideration of model errors in the PDF fit, or of appropriate $\chi^2$ 
tolerance when combining the ZEUS and H1
experiments in a PDF fit, but these differences should make us wary about the 
uncritical use of the Hessian method.
\begin{table}[t]
\begin{tabular}{ccc}\\
 \hline
Syatematic uncertainty $s_\lambda$  &  in PDF fit&  in Theory-free fit \\
 \hline
 ZEUS electron efficiency  & 1.68 & 0.31 \\
 ZEUS electron angle  & -1.26 & -0.11 \\
 ZEUS electron energy scale  & -1.04 & 0.97 \\
 ZEUS hadron calorimeter energy scale  & 1.05 & -0.58 \\
 H1 electron energy scale  & -0.51 & 0.61 \\
 H1 hadron energy scale  & -0.26 & -0.98 \\
 H1 calorimeter noise  & 1.00 & -0.63 \\
 H1 photoproduction background  & -0.36 & 0.97 \\

 \hline\\
\end{tabular}
\caption{Systematic shifts for ZEUS and H1 data as determine by a joint pQCD 
PDF fit, and as determined by the theory-free data combination fit}
\label{tab:sl}
\end{table}

%% file: averaging.tex
\subsection{Averaging of DIS Cross Section Data
\protect\footnote{Contributing author: A.~Glazov}}

\label{sec:ave}

The QCD fit procedures (Alekhin~\cite{Alekhin:2002fv}, CTEQ~\cite{cteq}, MRST~\cite{mrst},
H1~\cite{h1a}, ZEUS~\cite{zeus}) use data from a number of individual experiments directly 
to extract the parton distribution functions (PDF).
All  programs 
use both the central values of measured cross section data
as well as information about the correlations
among the experimental data points. 

The direct extraction procedure has several shortcomings.
The number of input datasets is large containing several individual publications.
The data points are correlated because of
common systematic uncertainties, 
within and also  across the publications.
Handling of the experimental data 
without additional expert knowledge becomes difficult.
Additionally, as it is discussed in Sec.~\ref{sec:mandy},
the treatment of the correlations produced by the systematic 
errors is not  unique.
In the Lagrange 
Multiplier method~\cite{pascaud} 
each systematic error  is treated as a parameter 
and thus fitted to QCD. Error propogation is then used to
estimate resulting uncertainties on PDFs.
In the so-called ``offset'' method 
(see e.g.~\cite{zeus}) the datasets are shifted in turn by each
systematic error before fitting. The resulting 
fits are used to form an envelope function to estimate the PDF
uncertainty.
Each method  has its own advantages  and shortcomings, and 
it is difficult to
select the standard one. 
Finally, some global QCD analyses use non-statistical 
criteria to estimate the PDF uncertainties
 ($\Delta \chi^2 \gg 1$). This is driven by the 
apparent discrepancy between
different experiments which is often difficult to quantify. 
Without a model independent consistency
check of the data it might be the only safe procedure.

These drawbacks can be significantly reduced by  averaging 
of the input structure function data in a model independent way before 
performing
a QCD analysis of that data.
One combined dataset of deep inelastic scattering (DIS) cross section measurements 
is much easier to handle compared to a scattered set of individual experimental
measurements, while retaining the full correlations between data points.
The averaging method proposed here  is unique and removes
the drawback of the offset method, which fixes the size of the systematic
uncertainties. In the averaging procedure the correlated 
systematic uncertainties are floated coherently allowing in some cases
reduction of the uncertainty.
In addition, study of a global
 $\chi^2/dof$ of the average and distribution of the pulls
allows a model independent consistency check between the experiments.  
In case of discrepancy between the input datasets, localised enlargement of the uncertainties for the average can be performed.

A standard way to represent a cross section measurement of a single experiment
is given in the case of the $F_2$ structure function by:
\begin{equation}
\begin{array}{lcl}
 \chi^2_{exp}(\left\{F^{i,true}_{2}\right\},\left\{\alpha_j\right\}) 
&=& \sum_i
 \frac{\textstyle\left[F_{2}^{i,true}-\left(F^i_{2} 
+ \sum_j \frac{\partial F^i_2}{\partial \alpha_j} \alpha_j\right)\right]^2}
{\textstyle \sigma^2_{i}} 
 + \sum_j \frac{\textstyle\alpha^2_j} {\textstyle \sigma^2_{\alpha_j}}. \label{b}
\end{array}
\end{equation}
Here $F^i_{2}$ ($\sigma^2_{i}$) are the measured 
central values (statistical and uncorrelated systematic uncertainties) 
of the $F_2$ structure function\footnote{The structure
function is measured for different $Q^2$ (four momentum transfer squared)
and Bjorken-$x$ values which are omitted here for simplicity.
},  
$\alpha_j$ are the correlated systematic uncertainty sources and
$\partial F^i_2/\partial \alpha_j$ are the sensitivities of the
measurements to these systematic sources. 
Eq.~\ref{b}
corresponds to the correlated probability distribution functions for the
structure function $F^{i,true}_{2}$ and for the systematic uncertainties
$\alpha_j$. Eq.~\ref{b} resembles Eq.~\ref{eq:chi2} where the theoretical 
predictions for $F_2$ are substituted by $F^{i,true}_{2}$.

The $\chi^2$ function Eq.~\ref{b} by construction has a  minimum $\chi^2=0$ for 
$F^{i,true}_{2}=F^{i}_{2}$ and $\alpha_j = 0 $. 
One can show  
that the total uncertainty for $F^{i,true}_2$ determined from the
formal minimisation of Eq.~\ref{b} is equal to the sum in quadrature
of the statistical and systematic uncertainties.
The reduced covariance matrix  $cov(F^{i,true}_2,F^{j,true}_2)$ 
quantifies the correlation between experimental points.

In the analysis of  data from more than one experiment, the
$\chi^2_{tot}$ function is taken as a sum of the $\chi^2$ functions Eq.~\ref{b} for each
experiment. 
 The QCD fit is then performed in terms of parton density functions
which are used to calculate predictions for $F^{i,true}_{2}$.

Before performing the QCD fit, the $\chi^2_{tot}$ function can be
minimised with respect to $F^{i,true}_{2}$ and $\alpha_j$. If none of
correlated sources is present, this minimisation is equivalent to taking
an average of the structure function measurements. If the systematic 
sources are included, the minimisation corresponds to a generalisation of
the averaging procedure which contains correlations among the measurements.

Being a sum of positive definite quadratic functions, $\chi^2_{tot}$ is also
a positive definite quadratic and thus has a unique minimum which can be
found as a solution of a system of  linear equations. Although this system
of the equations has a large dimension 
it has a simple structure allowing fast and precise solution.

A dedicated program has been developed to perform this averaging of the 
DIS cross section data ({\tt http://www.desy.de/\~{}glazov/f2av.tar.gz}).
This program can calculate the simultaneous averages for neutral current
(NC) and charged current (CC) electron- 
and positron-proton scattering cross section data including correlated 
systematic sources. 
The output of the program includes the central values and uncorrelated
uncertainties of the average  cross section data. The correlated
systematic uncertainties can be represented in terms of (i) covariance
matrix, (ii) dependence of the average cross section on the original systematic
sources together with the correlation matrix for the systematic sources, (iii) and
finally the correlation matrix of the systematic sources can be diagonalised,
in this case the form of $\chi^2$ for the average data is identical to Eq.~\ref{b} but the
original systematic sources are not preserved.

\begin{figure}
\includegraphics[height=0.3\textheight]{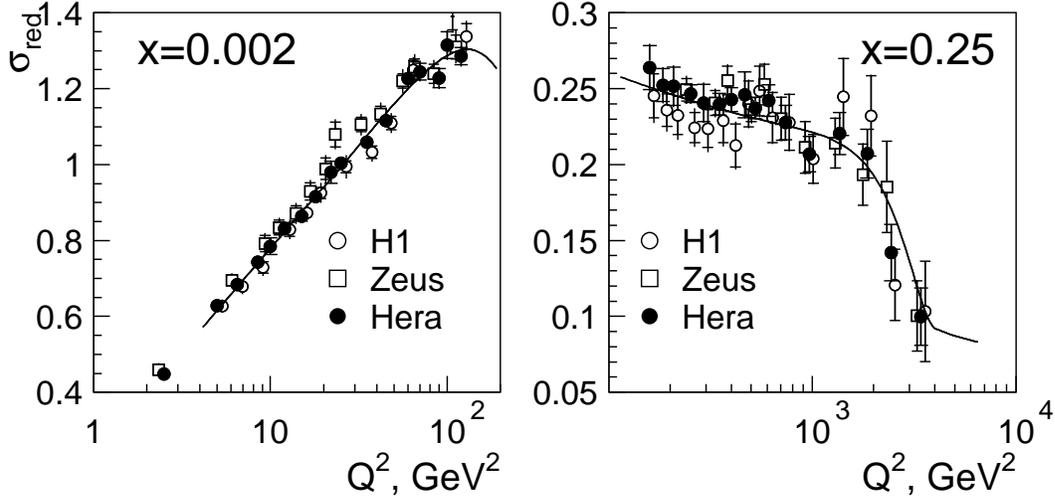}
  \caption{\label{fig}$Q^2$ dependence of the NC reduced cross section for $x=0.002$ and
 $x=0.25$ bins. H1 data is shown as open circles, ZEUS data is shown
as open squares and the average of H1 and ZEUS data is shown as filled circles.
The line represents the expectation from the H1 PDF 2000 QCD fit. }
\end{figure}

The first application of the averaging program has been a determination of
the average of the published H1 and ZEUS
data~\cite{h1a,h1alphas,h1ep300,h1em320,z1a,z2a,z3a,z4a,Chekanov:2003yv,z6a}. 
Nine individual
NC and CC cross section measurements are included from H1 and seven are
included from ZEUS. Several sources of systematic uncertainties are
correlated between datasets, the correlations among H1 and ZEUS
datasets are taken from~\cite{h1a} and \cite{zeusj}, respectively. 
No correlations are assumed between H1 and ZEUS systematic uncertainties
apart from a  common $0.5\%$ luminosity measurement uncertainty.
The total number of 
data points is 1153 (552 unique points)
 and the number of  correlated systematic sources, including
 normalisation uncertainties, is 43.

The averaging can take place
only if  most of the
data from the experiments are quoted at the same $Q^2$ and $x$ values.
Therefore, before the averaging the  data points are interpolated
to a common $Q^2,x$ grid. This interpolation is based on the  H1 PDF 2000 
QCD fit~\cite{h1a}. 
The interpolation of data points in principle introduces a
model dependency. For H1 and ZEUS structure function data both
experiments employ rather similar $Q^2,x$ grids. About $20\%$ of the
input points are interpolated, for most of the cases the correction
factors are small (few percent) and stable if different QCD fit
parametrizations~\cite{cteq,mrst} are used.

The cross section data have also been corrected to a fixed center of mass energy
squared $S=101570$~GeV$^2$. This has introduced a small correction
for the data taken at $S=90530$~GeV$^2$. The correction is based on H1-2000 PDFs,
it is only significant for high inelasticity $y>0.6$ and does 
not exceed $6\%$.

The HERA data sets agree very well: $\chi^2/dof$ for the average is 
$521/601$. The distribution of pulls does not show any significant 
tensions across the kinematic plane. Some systematic trends can be
observed at low $Q^2<50$~GeV$^2$, where ZEUS NC data lie systematically
higher than the  H1 data, although this difference is within  the 
normalisation uncertainty. 
An example of the resulting average DIS cross section is shown
in Fig.~\ref{fig}, where the data points are displaced in $Q^2$ for clarity.

A remarkable side feature of the averaging is a significant reduction of the correlated
systematic uncertainties. For example the uncertainty on the scattered electron energy 
measurement in the H1 backward calorimeter is reduced by a 
factor of three. The reduction
of the correlated systematic uncertainties thus
leads to a significant reduction of the
total errors, especially for low $Q^2<100$~GeV$^2$, where systematic 
uncertainties
limit the measurement accuracy. For this domain the 
total errors are often reduced by a factor two compared
to the total errors of the individual H1 and ZEUS measurements. 

The reduction of the correlated systematic uncertainties is achieved since 
the dependence of the measured cross section on the systematic sources 
is significantly different between H1 and ZEUS experiments. 
This difference
is due mostly to
the difference in the kinematic reconstruction methods used 
by the two collaborations, and
to a lesser extent to the individual features of 
the H1 and ZEUS detectors.
For example, the cross section dependence on the scattered electron energy 
scale has a very  particular behaviour for H1 data which relies on kinematic
reconstruction using only the scattered electron in one region of phase
space. ZEUS uses the double angle reconstruction method where the
pattern of this dependence is completely different leading to a
measurement constraint.

In summary,
a generalised averaging procedure to include point-to-point correlations
caused by the systematic uncertainties has been developed. This averaging procedure
has been applied to H1 and ZEUS DIS cross section data. The data show good consistency.
The averaging of H1 and ZEUS data leads to a significant reduction of the correlated
systematic uncertainties and thus a  large improvement in precision
for low $Q^2$ measurements. The goal of the averaging procedure is to obtain 
HERA DIS cross section set which takes into account all correlations among the experiments.

%% file: longitudinal.tex
\subsection{The longitudinal structure function $F_L$
\protect\footnote{Contributing authors: J.~Feltesse, M.~Klein}}
\label{sec:flmax}
\subsubsection{Introduction}
\begin{figure}[ht]
   \centering
   \epsfig{file=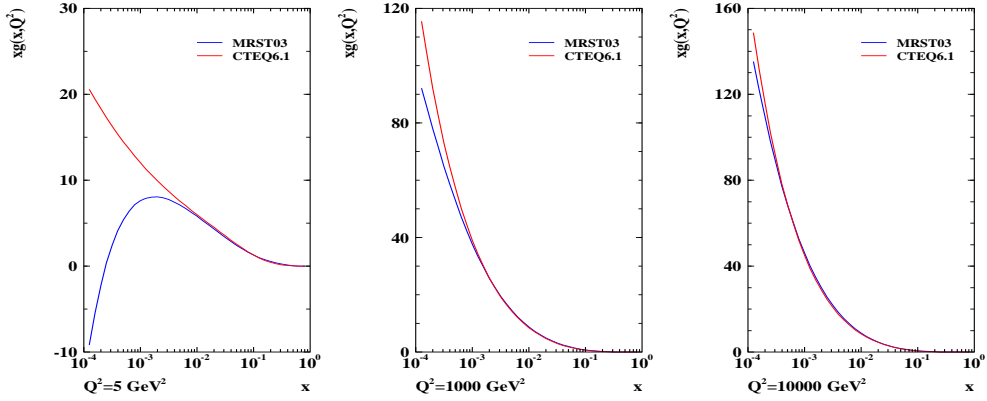,width=5.2cm,height=13cm,angle=90.} 
   \caption{Gluon momentum distributions determined by MRST and CTEQ in NLO QCD,
                   as a function of $x$ for $Q^2= 5$\,GeV$^2$,
                   close to the initial scale of the fits, and at higher $Q^2$ as the result of
                   the DGLAP evolution.}
   \label{fig:evol}
\end{figure}

\begin{figure}[htbp]
  \begin{picture}(200,200) 
  \put(60,0){
   \epsfig{file=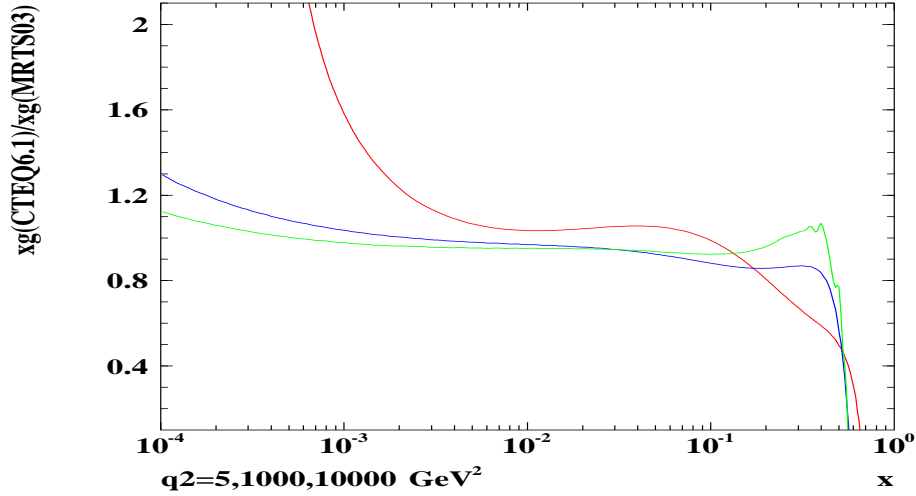,angle=90.,width=12cm,height=6.5cm} 
   }
   \end{picture}
   \caption{Ratio of the gluon distributions of CTEQ to MRST as a function of $x$
                   for low and large $Q^2$.}
   \label{fig:gluerat}
\end{figure}

\begin{figure}[htbp]
   \centering
   \epsfig{file=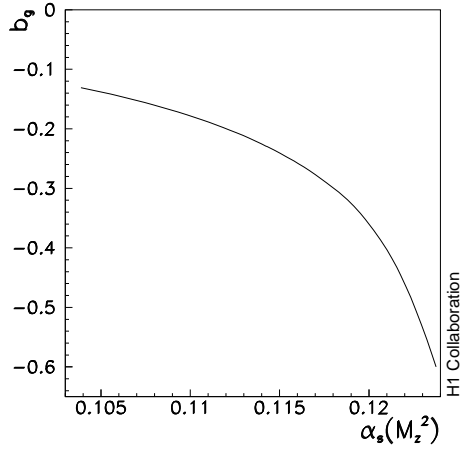,width=6cm} 
   \caption{Correlation of the low $x$ behaviour of the gluon distribution,
                   characterised by the power $x^{-b_g}$, with the strong coupling
                   constant $\alpha_s$ as obtained in the H1 NLO QCD fit to H1 and BCDMS data.}
   \label{fig:bgalf}
\end{figure}

At low $x$ the sea quarks are determined by the accurate data on \F.
The charm contribution to \Fc is directly measured while there is no
separation of up and down quarks at low $x$ which are assumed
to have the same momentum distribution, see
Sect.~\ref{sec:mkbrdbarubar}. 
Within this assumption,
and setting the strange sea to be a fraction of the up/down sea,  the proton quark
content at low $x$ is determined. The gluon distribution \xg, however,
is determined only by the derivative \pdff which is not well measured \cite{h1alphas}.
It is thus not surprising that rather different gluon distributions 
are obtained in global NLO analyses, as is illustrated in Figure \ref{fig:evol}.
The figure displays the result of recent fits by MRST and CTEQ on
the gluon distribution at low and high $Q^2$. It can be seen that there are
striking differences at the initial scale, $Q^2 = 5$\,GeV$^2$, which at high $Q^2$
get much reduced due to the evolution mechanism. The  ratio of these distributions,
however, exhibits differences at lower $x$ at the level of 10\% even in the LHC
Higgs and $W$ production
kinematic range, see Figure \ref{fig:gluerat}. One also observes a striking problem at large $x$
which is beyond the scope of this note, however. In a recent QCD analysis
it was observed \cite{h1alphas} that the dependence of the gluon distribution
at low $x$, $xg \propto x^{b_G}$, is correlated to the value of \amz,
see Figure \ref{fig:bgalf}. 

In the Quark-Parton Model the longitudinal structure function \FL is
zero \cite{Callan:1969uq}.
In DGLAP QCD, to lowest order, \FLc is given by \cite{Altarelli:1978tq}
\begin{equation}
        F_L(x,Q^2)= \frac{\alpha_s}{4 \pi} x^2
        \int_x^1 \frac{dz}{z^3} \cdot \left[ \frac{16}{3}F_2(z,Q^2) + 8
        \sum e_q^2 \left(1-\frac{x}{z} \right) zg(z,Q^2) \right]
\label{altmar}
\end{equation}
with contributions from quarks and from gluons.
Approximately this equation can be solved
\cite{Cooper-Sarkar:1987ds} and the gluon distribution appears as a measurable
quantity,
\begin{equation}
 xg(x) = 1.8[ \frac{3 \pi}{2 \alpha_s} F_L(0.4 x) - 
             F_2(0.8 x] \simeq \frac{8.3}{\alpha_s} F_L,
\end{equation}
determined by measurements
of \Fc and \FLc. Since \FLc, at low $x$, is not much smaller than \Fc, to a 
good approximation \FLc is a direct measure for the  gluon distribution at low $x$.

Apart from providing a very useful constraint to the determination of 
the gluon distribution, see also Sect.~\ref{sec:clair}, a measurement of \FL is of
principal theoretical interest. It provides a crucial test of QCD to high orders.
A significant departure
of an \FLc measurement from the prediction which is based on the measurement
of \F and \pdff only, would require theory to be modified. There are
known reasons as to why the theoretical description of gluon radiation at
low $x$ may differ from conventional DGLAP evolution: the neglect of
$\ln (1/x)$, in contrast to BFKL evolution, or the importance of NLL resummation
effects on the gluon splitting function (see
Sect.~\ref{sec:pdf,res,sx}). 
Furthermore recent
calculations of deep inelastic scattering to NNLO predict very large effects
from the highest order on \FLc contrary to \Fc \cite{Moch:2004xu}.

Within the framework of the colour dipole model there exists a testable
prediction  for \FL, and the longitudinal structure function, unlike \Fc,
may be subject to large higher twist  effects \cite{Bartels:2000hv}. 
\subsubsection{Indirect Determinations of $F_L$  at Low $x$}
So far first estimates on \FL at low $x$ have been obtained by the H1
Collaboration. These result from data on the inclusive $ep \rightarrow eX$
scattering cross section 
\begin{equation}
  \frac{Q^4 x}{2\pi \alpha^2 Y_+} \cdot \frac{d^2\sigma}{dxdQ^2} =    
    [ F_2(x,Q^2) - f(y) \cdot F_L(x,Q^2)] = \sigma_r
       \label{sig}
  \end{equation}  
obtained at fixed, large energy, $s=4 E_e E_p$. The cross section
is defined by the two proton structure functions, \Fc and \FLc,
with $Y_+ = 1+ (1-y)^2$ and $f(y)=y^2/Y_+$. At
fixed $s$ the inelasticity $y$ is fixed by $x$ and $Q^2$ as
$y=Q^2/sx$.  Thus one can only measure a combination $F_2 -f(y) F_L$.
Since HERA accesses a large range of $y$, and $f(y)$ is large only at
large $y > 0.4$, assumptions
have been made on \FLc to extract \Fc at larger $y$. Since the cross
section measurement accuracy has reached the few per cent level \cite{h1alphas},
the effect of the \FLc assumption on \Fc at lowest $x$  has been non-negligible.
The determination of \F has thus been restricted to a region in which $y < 0.6$.
The proton structure function \F is known over a few orders of magnitude in $x$
rather well, from HERA and at largest $x$ from fixed target data.
Thus H1 did interpret
the cross section at higher $y$ as a determination of \FL imposing
assumptions about the behaviour of \F at lowest $x$. These
were derived from QCD fits to the H1 data \cite{Adloff:1996yz} or at lower $Q^2$,
where QCD could not be trusted, from the derivative of \Fc \cite{flshape}.
Recently, with the established $x$ behaviour \cite{Adloff:2001rw} of 
$F_2(x,Q^2) = c(Q^2) x^{-\lambda(Q^2)}$, a new method \cite{flshape} has been used
to determine \FLc. This ``shape method" is based on the observation that the shape
of $\sigma_r$, Eq.\,\ref{sig}, at high $y$ is   driven by  $f \propto y^2$
and sensitivity to \FLc is restricted to a very narrow range of $x$ corresponding
to $y=0.3-0.9$. Assuming that \FL in this range, for each bin in $Q^2$, does
not depend on $x$, one obtains a simple relation, $\sigma_r = c x^{-\lambda} - f F_L$.
which  has been used to determine \FL. 
\begin{figure}[htbp]
   \centering
   \epsfig{file=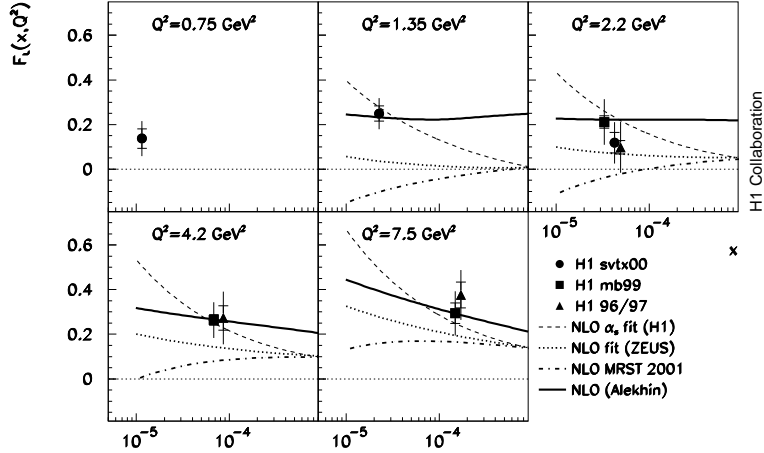,width=10.cm} 
   \caption{Data on the longitudinal structure function obtained
                  using assumptions on the behaviour of the other structure
                  function $F_2$ in comparison with NLO QCD fit predictions.
                   The data labeled svtx00 and  mb99  data are preliminary. }
   \label{fig:flshape}
\end{figure}
Figure \ref{fig:flshape} shows the existing, preliminary data on \FL at low $Q^2$ from the H1
Collaboration in comparison with predictions from NLO DGLAP
QCD fits to HERA and further cross section data. One can see that the accuracy
and the $x$ range  of these \FL determinations are rather limited although
the data have some discriminative power already.
\subsubsection{Backgrounds and Accuracy}
The longitudinal structure function contribution to $\sigma_r$ represents
a small correction of the cross section in a small part of the kinematic range
only. The demands for the \FLc measurement are extremely high: the
cross section needs to be measured at the per cent level and the scattered electron
be uniquely identified up to high $y $. The method of unfolding \Fc and \FLc
consists in  a measurement of $\sigma_r$ at fixed $x$ and $Q^2$ with varying $s$.
This allows both structure functions to be determined from a straight line
variation of  $\sigma_r$ as a function of $f(y)$, see \cite{Klein:2004zq}.

 At large $y$, corrresponding to low $x$,
and low $Q^2$ the scattering kinematics at HERA resembles that of a
fixed target scattering experiment: the electron scattered off quarks
at very low $x$ (``at rest'') is going in the backward detector region,
i.e. in the direction of the electron beam. The scattered electron is
accompanied by part of the hadronic final state which is related to
the struck quark.  High inelasticities  $y \simeq 1 -E_e'/E_e$ demand to identify
scattered electrons down to a few GeV of energy $E_e'$. Thus  a
considerable background is to be isolated and removed which stems from
hadrons or photons, from the $\pi_0 \rightarrow \gamma \gamma$ decay.
These particles may originate both from a genuine DIS event but to a
larger extent stem from photoproduction processes, in which the scattered
electron escapes mostly non recognised in electron beam direction. Removal of
this background in H1 is possible by requiring a track associated to the
Spacal cluster, which rejects photons, and by measuring its charge
which on a statistical basis removes the remaining part of the
background as was demonstrated before \cite{h1alphas,flshape}.

The scattered electron kinematics, $E_e'$ and $\theta_e$, can be
accurately reconstructed using the high resolution Spacal calorimeter energy
determination and the track measurements in the Backward
Silicon Tracker (BST) and the Central Jet Drift Chamber (CJC). Reconstruction
of the hadronic final state allows the energy momentum constraint to
be imposed, using the ``$E-p_z$" cut, which removes radiative
corrections, and the Spacal energy scale to be calibrated at large
$E_e'$ using the double angle method. At low energies $E_e'$ the
Spacal energy scale can be calibrated to a few \% using 
the  $\pi_0$ mass constraint and be cross checked with the BST
momentum measurement and with QED Compton
events. The luminosity is measured to 1-2\%. Any common normalisation
uncertainty may be removed, or further constrained, by comparing cross
section data at very low $y$ where the contribution of \FLc is
negligible.

Subsequently two case studies are presented which illustrate
the potential of  measuring \FLc  directly in unfolding it from
the large \Fc contribution to the cross section, a study using
a set of 3 low proton beam energies and a simulation for
just one low $Ep$ data set combined with standard 920 GeV
data. Both studies use essentially the same correlated systematic 
errors and differ slightly in the assumptions on the background
and efficiency uncertainties which regard the errors on cross section ratios.
The following assumptions on the
correlated systematics are used: $\delta E_e'/E_e' = 0.003$ at large
$E_e$ linearly rising to $0.03$ at 3\,GeV; $\delta \theta_e
=0.2$\,mrad in the BST acceptance region and 1\,mrad at larger angles;
$\delta E_h/E_h =0.02$. These and further assumed systematic
uncertainties represent about the state of analysis reached so far in
inclusive low $Q^2$ cross section measurements of H1.  
\subsubsection{Simulation Results}
\begin{figure}[htbp]
\vspace*{-1.0cm}
\begin{center}
\epsfig{file=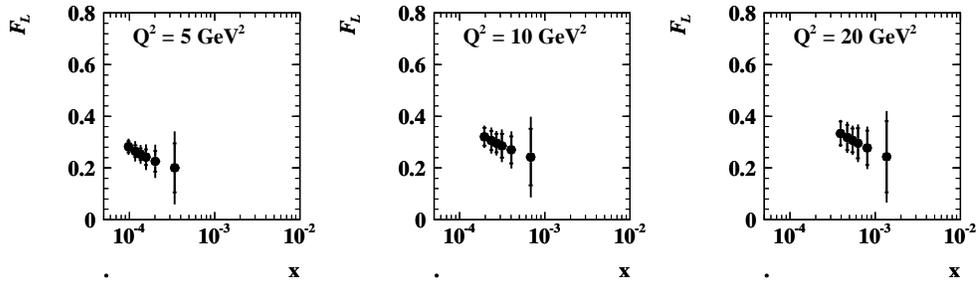,height=13.9cm,angle=90.,
bbllx=0pt,bblly=0pt,bburx=557pt,bbury=792pt}
\vspace*{-5.0cm}
\caption[*]{Simulated measurement of the longitudinal structure function $F_L(x,Q^2)$
  using the H1 backward apparatus to reconstruct the scattered
  electron up to maximum inelasticities of $y=0.9$ corresponding to a
  mimimum electron energy of $E_e'$ of about 3~GeV. The inner error
  bar is the statistical error.  The full error bar denotes the
  statistical and systematic uncertainty added in quadrature.  }
\label{ffl}
\end{center}
\end{figure}

A simulation has been performed for $E_e = 27.6$\,GeV and for four
different proton beam energies, $E_p=920, \,575, \,465$ and $400\,$GeV
assuming luminosities of 10, 5, 3 and 2\,\ipb, respectively.  The beam energies
are chosen such that the cross section data are equidistant in $f(y)$.
If the luminosity scales as expected  as $E_p^2$, the low $E_p$ luminosities are
equivalent to 35\,\ipb at standard HERA settings.
Further systematic errors regard the residual radiative corrections,
assumed to be 0.5\%, and the
photoproduction background, 1-2\% depending on $y$. This assumption
on the background demands an
improvement by a factor of about two at high $y$ which can be expected
from a high statistics subtraction of background using the charge
assignment of the electron scattering candidate.  An extra
uncorrelated efficiency correction is assumed of 0.5\%. The resulting
cross section measurements are accurate to 1-2\%.  For each $Q^2$ and $x$ point
this choice provides up to four cross section measurements. The
two structure functions are then obtained from a fit to $\sigma_r=F_2 +
f(y) F_L$ taking into account the correlated systematics. This
separation provides also accurate data of $F_2$, independently of \FLc. The
simulated data on \FLc span nearly one order of magnitude in $x$ and
are shown in Figure\,\ref{ffl}.  For the chosen luminosity the statistical
and systematic errors on \FLc are of similar size.  The overall
accuracy on \FL, which may be obtained according to the assumed
experimental uncertainties, is thus estimated to be of the order of 10-20\%. 

Based on recent information  about aspects of the machine conditions in a low
proton beam energy mode, a further case study was performed \cite{joelring} 
for only one reduced proton beam energy.  In this simulation, for the
standard electron  beam
energy of $E_e=27.6$\,GeV, proton beam energies of  $E_p=920$ and 460 GeV
were chosen with luminosities of 30 and 10\,$pb^{-1}$,  respectively.
According to \cite{ferdi} it  would  take  about three  weeks  to
change the configuration of the machine and to tune the luminosity plus 
10 weeks to record 10 $pb^{-1}$ of  good  data  with  High Voltage of trackers on.
Uncertainties besides the correlated errors specified above are assumed for
photo-production background subtraction varying from 0\%  at y=0.65 to 4\% at y = 0.9,
and of 0.5\% for the residual radiative corrections. An overall uncertainty of 1\%  is assumed
on the  measurement of the cross section at  low  beam  energy  settings, which  covers  relative
uncertainties on electron identification, trigger efficiency, vertex efficiency, and relative luminosity.

\begin{figure}[htbp]
\vspace*{0.1cm}
\begin{center}
\epsfig{file=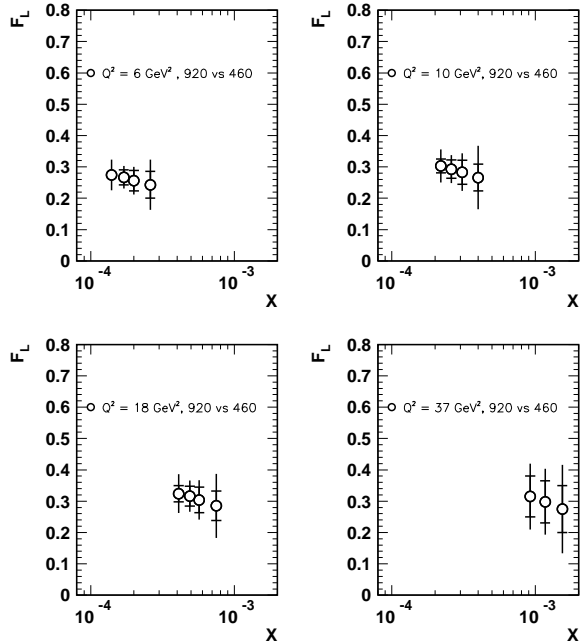,height=8.5cm}
\caption[*]{Simulated measurement of the longitudinal structure function $F_L(x,Q^2)$
  for data at 920 GeV (30 $pb^{-1}$) and 460 GeV (10 $pb^{-1}$). The inner error
  bar is the statistical error.  The full error bar denotes the
  statistical and systematic uncertainty added in quadrature.  }
\label{figjoel}
\end{center}
\end{figure}

To evaluate the errors two independent methods have been considered an
analytic calculation and a fast  Monte-Carlo simulation technique.   The  two
methods provide statistical and systematic errors which are in excellent agreement.
The overall result of this simulation of \FLc  is  displayed  in  Figure \ref{figjoel}.
In many bins the overall precision on \FL is around or below 20\%. It is remarkable
that the overall precision would stay below 25\% even if the statistical error
or the larger source of systematic uncertainty would turn out to be twice larger
than assumed to be in this study.

\subsubsection{Summary}

It has been demonstrated with two detailed studies   that  a  direct  measurement  of  the
longitudinal structure function \FL  may be performed at HERA
at the five sigma level of accuracy, in the $x$  range from  $10^{-4}$ 
 to  $10^{-3}$  in  four bins of $Q^2$. This 
 measurement  requires about three months of running  
and  tuning  time  at  reduced  proton  beam
energy.   In addition it would provide the first measurement of the 
diffractive longitudinal  structure function  at  the  three  sigma
level (see the contribution of P.~Newman in the summary of Working
Group 4).
The exact choice of the parameters of such a measurement are
subject to further studies. 
In conclusion an accurate measurement of \FL is feasible,
it requires efficient detectors, dedicated beam time and analysis skills.
It would be the right sign of completion to have measured \Fc first,
in 1992 and onwards, and to conclude the HERA data taking with a dedicated
measurement of the second important structure function \FL, which 
is related to the gluon density in the low $x$ range of the LHC.

%% file: mkbrsea.tex

\subsection{Determination of the Light Quark Momentum Distributions at Low $x$ at HERA
\protect\footnote{Contributing authors: M.~Klein, B.~Reisert}}
\label{sec:mkbrdbarubar}
Based on the data taken in the first phase of HERA's operation (1993-2000),
the HERA collider experiments have measured a complete set of neutral
(NC) and charged (CC) current 
double differential $e^{\pm}p$ inclusive scattering cross sections, based 
on about 120\,pb$^{-1}$ of positron-proton and 15\,pb$^{-1}$
of electron-proton data. The NC and CC deep inelastic scattering (DIS) cross sections for 
unpolarised $e^{\pm}p$ scattering are
determined by structure functions and quark momentum
distributions in the proton as follows:
\begin{eqnarray}
 \sigma_{NC}^{\pm} \sim Y_+ F_2 \mp Y_-  xF_3, ~~ \\  
 F_2 \simeq e_u^2  x(U + \bU) + e_d^2  x (D+\bD), ~~ \\
 xF_3 \simeq 2x [a_u e_u (U-\bU) + a_d e_d (D-\bD)], ~~\\
 \sigma_{CC}^+ \sim x\bU+ (1-y)^2xD, ~~\\
 \sigma_{CC}^- \sim xU +(1-y)^2 x\bD . ~~
\end{eqnarray}
Here $y=Q^2/sx$ is the inelasticity, $s=4E_eE_p$ and $Y_{\pm} = 1 \pm (1-y)^2$.
The parton distribution $U=u+c+b$ is the sum of the momentum distributions of the up-type
quarks with charge $e_u=2/3$ and axial vector coupling $a_u=1/2$, while
$D=d+s$ is the sum of the momentum distributions of the down type
quarks with charge $e_d=-1/3$, $a_d =-1/2$. Similar relationships hold
for the anti-quark distributions $\bU$ and $\bD$.  

\begin{figure}
   \centering
    \epsfig{file=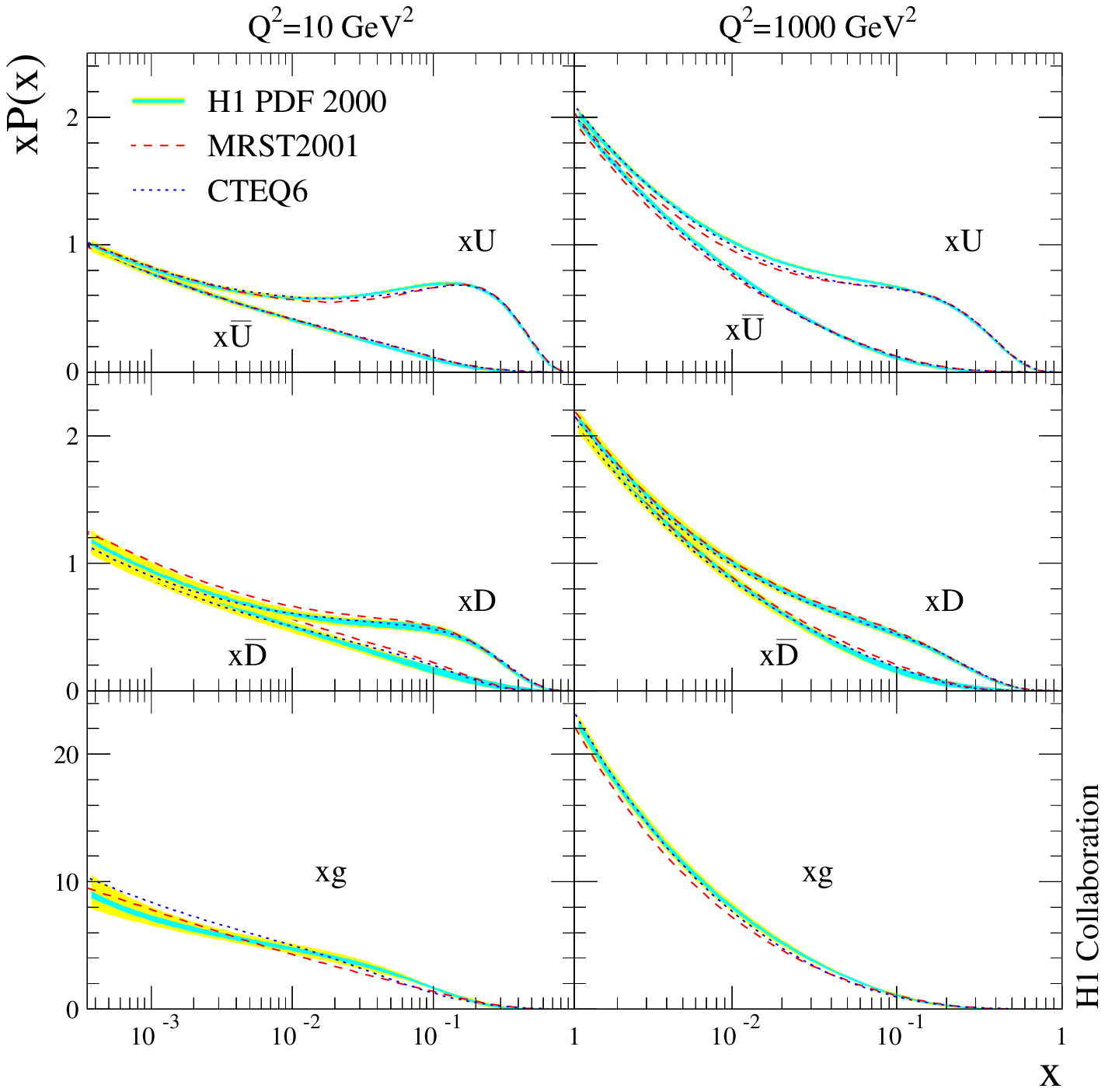,width=7cm}
      \epsfig{file=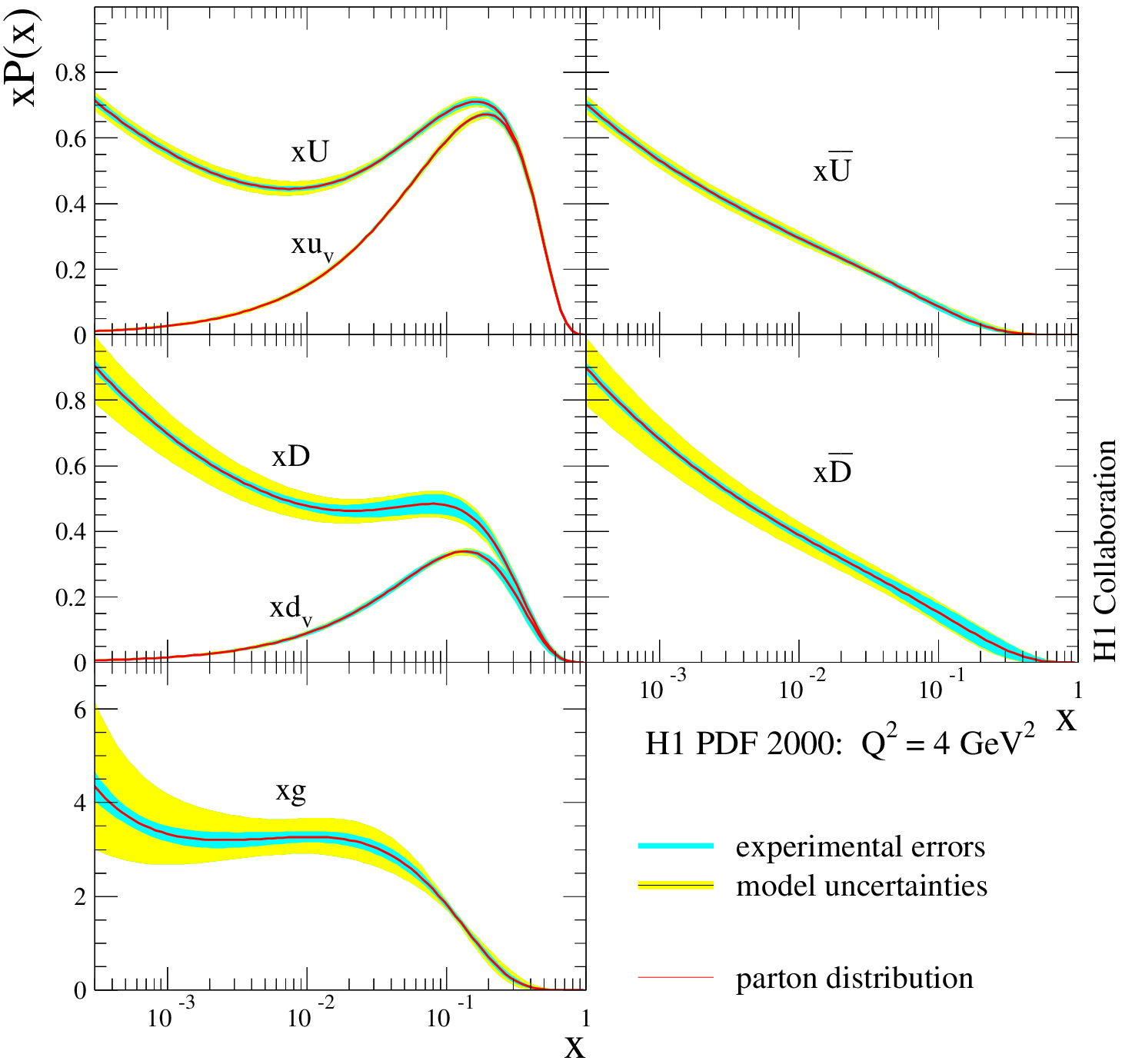,width=7cm} 
   \caption{Determination of the sum of up, anti-up, down and anti-downquark
    distributions and of the gluon distribution in the proton based on the
    H1 neutral and charged current cross section data. Left: for $Q^2$ of
    10  and 1000 GeV$^2$ compared with results from MRST and
    CTEQ; Right: the parton distributions with their experimental and model
    uncertainties as determined by H1 at the starting scale $Q^2_0=4$\,GeV$^2$.}\label{fig:UD}
\end{figure}
As is illustrated 
in  Fig.\,\ref{fig:UD}  the H1 experiment \cite{h1a} has determined
all four quark distributions and the gluon distribution $xg$.
The accuracy achieved so far by H1, for $x=0.01, 0.4$ and $0.65$, is
$1\%,\,3\%, 7\%$ for the sum of up quark distributions and
$2\%,\,10\%, 30\%$ for the sum of down quark distributions, respectively.
The extracted parton distributions are in reasonable agreement
with the results obtained in global fits by the MRST \cite{mrst} and CTEQ \cite{cteq}
collaborations. The H1 result is also consistent 
with the pdfs determined by  the ZEUS Collaboration \cite{zeusj}
which uses jet data to improve the accuracy for the gluon
distribution and imports a $\bd-\bu$ asymmetry fit result from
MRST. New data which are being taken (HERA II) will improve
the accuracy of these determinations further. At the time this is
written, the available data per experiment have been grown to roughly 
150\,pb$^{-1}$ for both $e^+p$ and $e^-p$ scattering, and
more is still to come.  These data will be particularly important
to improve the accuracy at large $x$, which at HERA is related to high
$Q^2$.

\begin{figure}
   \centering
  \epsfig{file=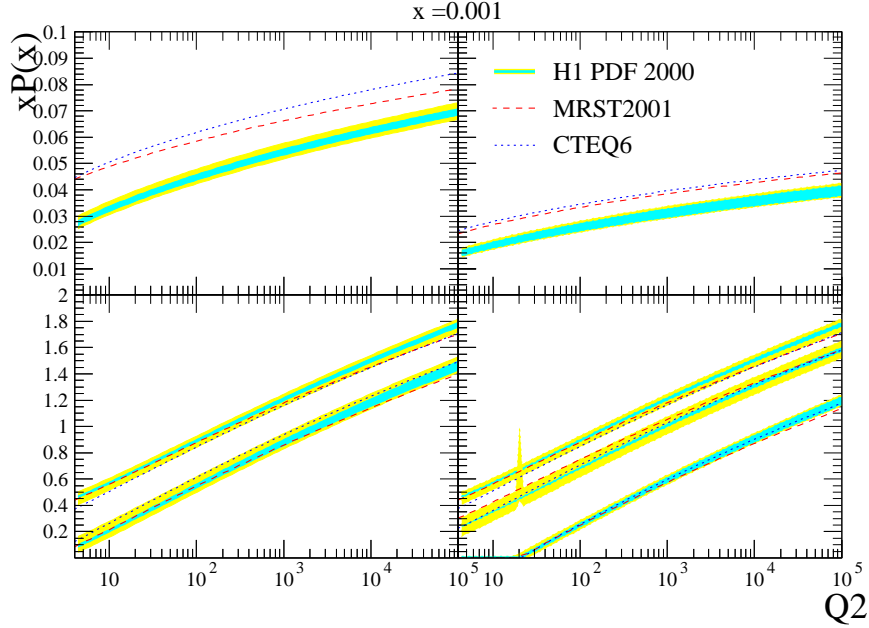,width=12cm} 
   \caption{Parton distributions and their uncertainties as determined by H1
   extrapolated to the region of the LHC, for $x=0.001$ near to the rapidity
   plateau. Top left: $u$ valence; top right: $d$ valence; bottom left: $\bu$ and
   below $c$; bottom right, in decreasing order: $\bd$, $s$, $b$.
   The results are compared with recent fits to global data by MRST
   and CTEQ. Note that at such small $x$ the valence quark distributions
   are very small. With increasing $Q^2$ the relative importance of the heavy
   quarks compared to the light quarks increases while the absolute difference
   of the quark distributions is observed to be rather independent of $Q^2$.
   The beauty contribution to the cross section thus becomes sizeable,
   amounting to about 5\% for $pp \rightarrow HW$.
   }
   \label{fig:extraq}
\end{figure}
As is clear from the above equations, the NC and CC cross section data are sensitive
directly  to only these four quark distribution combinations. Disentangling
the individual quark flavours (up, down, strange, charm and beauty) requires
additional assumptions.  While informations on the $c$ and $b$ densities
are being obtained from measurements of $F_2^{c\overline{c}}$ and
$F_2^{b\overline{b}}$ of improving accuracy, the determination
of the strange quark density at HERA is less straightforward and may
rest on $s W^+ \rightarrow c$ and strange ($\Phi$) particle production \cite{zeusphi}.
The relative contributions from the heavy quarks become increasingly
important with $Q^2$, as is illustrated in Fig.\,\ref{fig:extraq}.

The larger $x$ domain is dominated by the valence quarks. At HERA
the valence quark distributions are not directly determined but extracted
from the differences $u_v=U-\overline{U}$ and $d_v=D-\overline{D}$. Note 
that this implies the assumption that sea and anti-quarks are equal
which in non-perturbative QCD models may not hold.
A perhaps more striking assumption is inherent in these fits and
regards  the sea quark asymmetries at low $x$ which is the main subject
of the subsequent discussion.

\begin{figure}
   \centering
  \epsfig{file=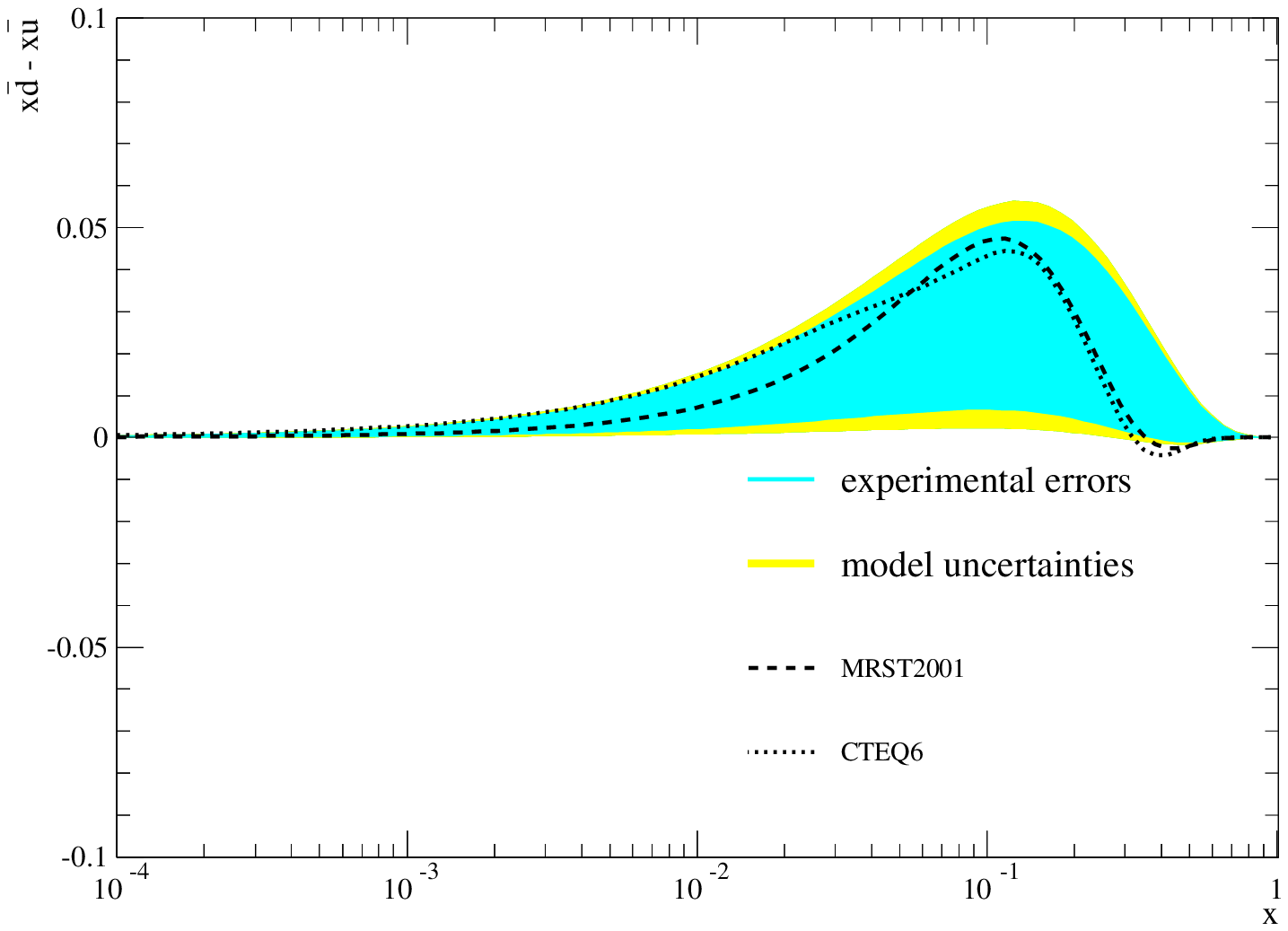,width=7.5cm} 
  \epsfig{file=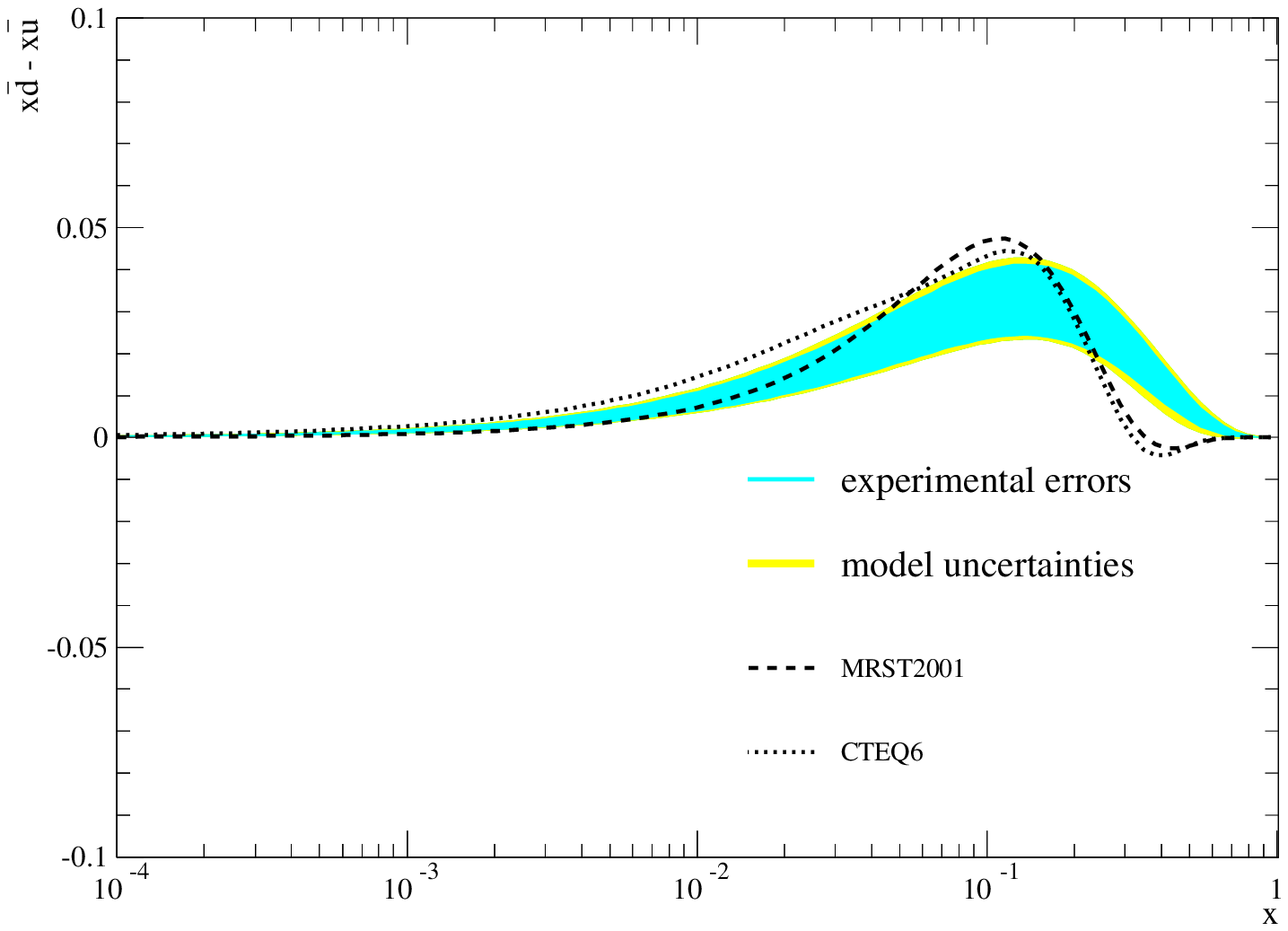,width=7.5cm} 
   \caption{Determination of the difference $x(\bd -\bu)$ in the H1 PDF 2000
                   fit to the H1 data (left) and the H1 and the BCDMS $\mu p$ and $\mu D$
                   data (right). The sea quark difference is enforced to tend to zero at low $x$.
                   The global fit results of MRST and CTEQ include Drell Yan data
                   which suggest a sea quark asymmetry at $x \sim 0.1$.}
   \label{fig:dubarh1}
\end{figure}
Fig.\,\ref{fig:dubarh1} shows the difference $x\bd - x\bu$ as determined in 
the H1 PDF 2000 fit based on the H1 data alone (left) and using in addition the
BCDMS proton and deuteron data (right). One observes a trend of these fits
to reproduce the asymmetry near $x \sim 0.1$ which in the MRST and CTEQ
fits, shown in Fig.\,\ref{fig:dubarh1}, is due to fitting the Drell Yan data
from the E866/NuSea experiment \cite{Towell:2001nh}. While this enhancement is not very
stable in the H1 fit \cite{benjamin} and not significant either, with the BCDMS
data an asymmetry is observed which reflects the violation of the Gottfried
sum rule.

\begin{figure}[htb]
   \centering
    \epsfig{file=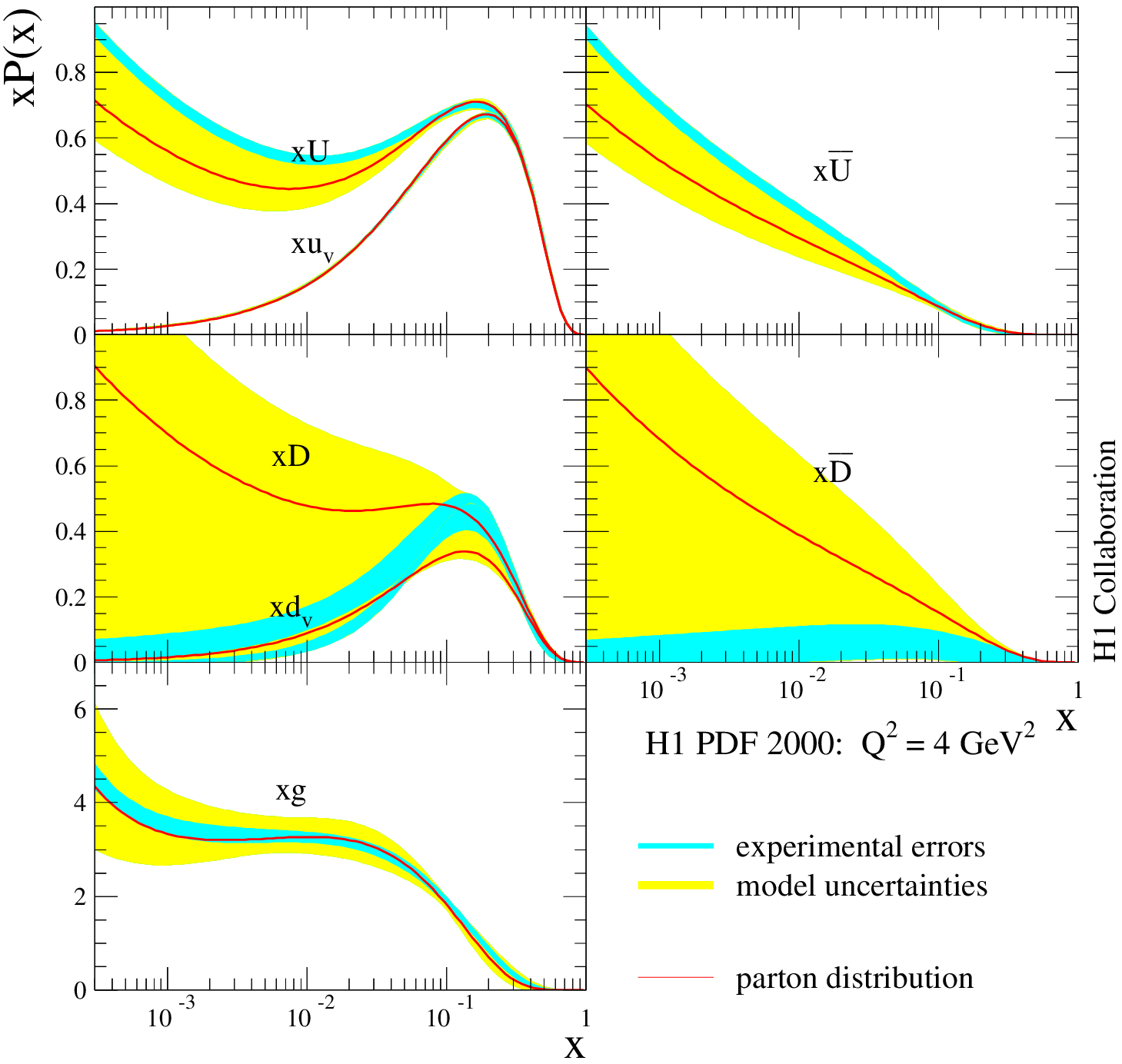,width=7cm} 
    \epsfig{file=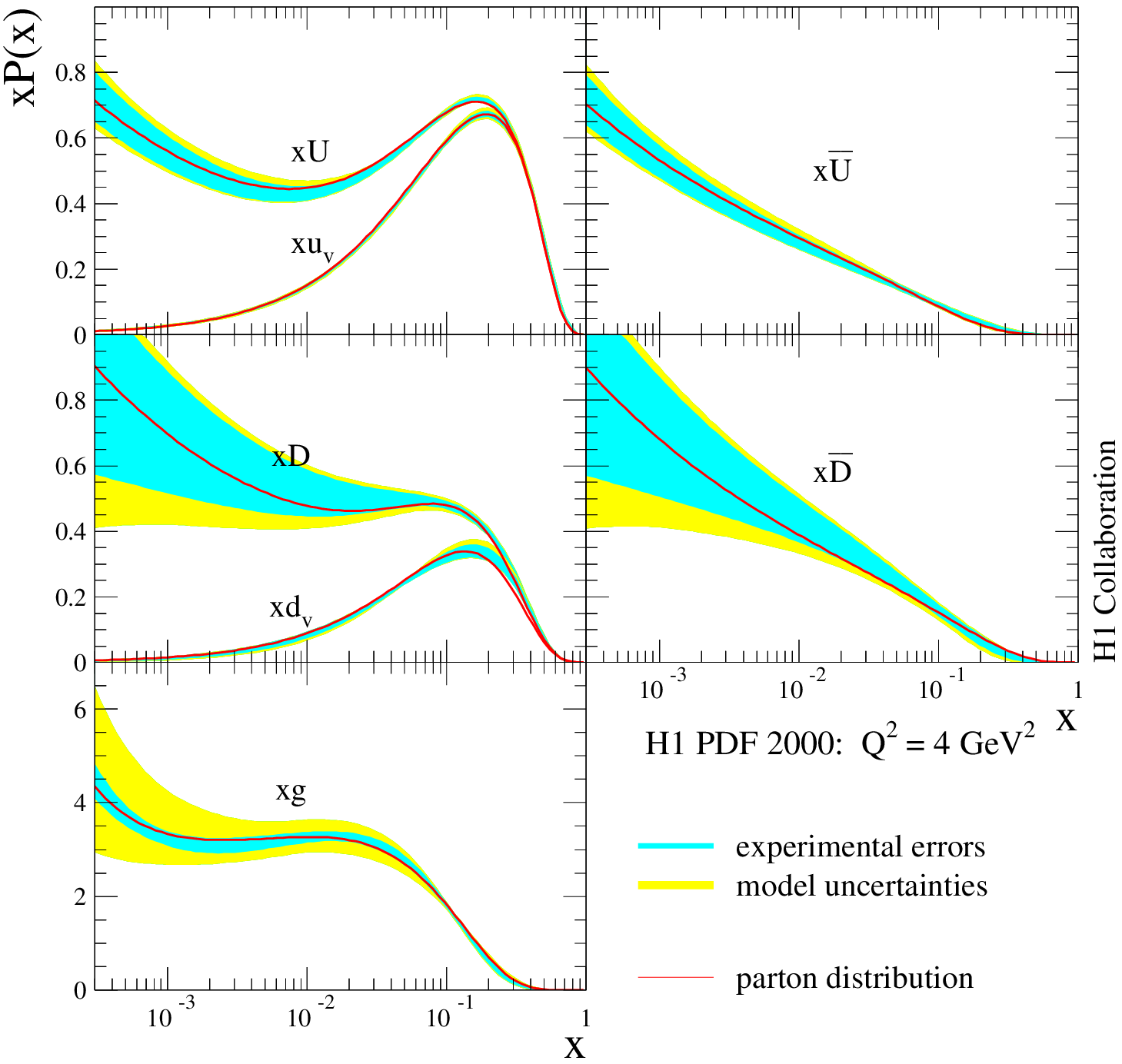,width=7cm} 
   \caption{Determinations of the quark and gluon momentum distributions 
    releasing the constraint $x\bd = x\bu$ at low $x$, from the H1 NC and CC
    data alone (left) and from the H1 $ep$ and the BCDMS $\mu p$ and $\mu D$
    data (right). Since at low $x < 0.01$ there is no further constraint than
    that given from $F_2$ the uncertainties of $\bU$ and in particular of $\bD$
    become sizeable.}
   \label{fig:oweh}
\end{figure}
In the H1 fit \cite{h1a} the parton distributions  at the initial
scale $Q^2=4$\,GeV$^2$  are parameterised
as $xP =A_p x^{B_P} (1-x)^{C_P} \cdot f_P(x)$. The function $f_P$ is a polynomial
in $x$ which is determined by requiring ``$\chi^2$ saturation" of the fits,
i.e. starting from $f_P=1$ additional terms $D_Px$, $E_Px^2$ etc. are added
and only considered if they cause a significant improvement in $\chi^2$,
half integer powers were considered in \cite{benjamin}. 
The result for fitting the H1 data has been as follows: $f_g=(1+D_gx)$, 
$f_U=(1+D_Ux+F_Ux^3)$, $f_D=(1+D_Dx)$ and $f_{\bU}=f_{\bD}=1$.
The parton distributions at low $x$ are thus parameterised as
$xP \rightarrow A_P x^{B_P}$.
The strange (charm) anti-quark distribution is coupled to the total amount of
down (up) anti-quarks as $\bs=f_c\bD$ ($\bcc=f_c\bU$). Two assumptions
have been made on the behaviour of the quark and anti-quark distributions
at low $x$. It has been assumed that quark and anti-quark distributions
are equal and, moreover, that the sea is flavour symmetric.
This implies that the slopes $B$ of all four quark distributions
are set equal $B_U=B_D=B_{\bU}=B_{\bD}$. Moreover, the nomalisations of
up and down quarks are the same, i.e. $A_{\bU}(1-f_c)=A_{\bD}(1-f_s)$,
which ensures that   $\bd/\bu \rightarrow 1$ as $x$ tends to zero.
The consequence of this assumption is illustrated in Fig.\,\ref{fig:dubarh1}.
While the DIS data suggest some asymmetry at larger $x$, the up-down quark
asymmetry is enforced to vanish at lower $x$. This results in a rather fake high
accuracy in the determination of the four quark distributions at low $x$,
despite the fact that at low $x$ there is only one combination of them measured,
which is $F_2 = x[4(U+\bU) + (D+\bD)]/9$.  If one relaxes both the conditions
on the slopes and normalisations, the fit to the H1 data decides to completely remove the
down quark contributions as is seen in Fig.\,\ref{fig:oweh} (left plot).

\begin{figure}[htb]
   \centering
    \epsfig{file=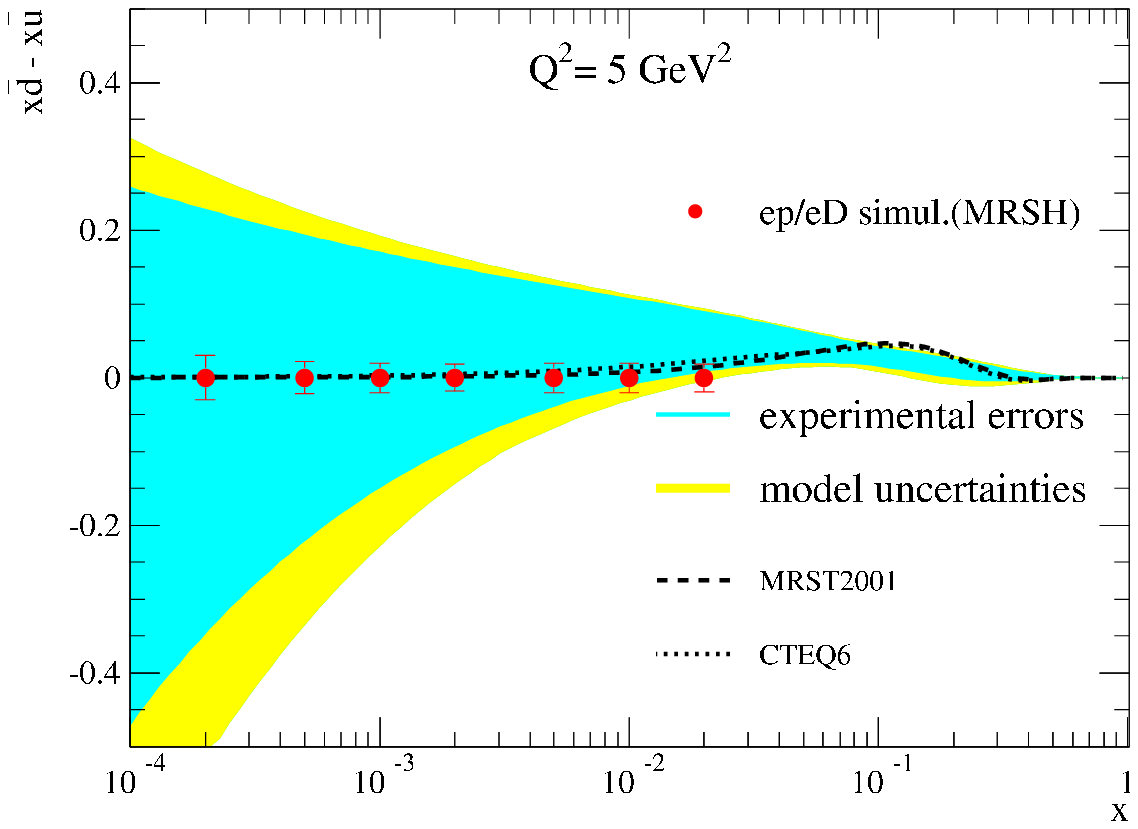,width=8cm}
   \caption{Simulation of the difference of sea quark distributions, here assumed to be zero,
at low $x$ based on additional 20\,pb$^{-1}$ of electron-deuteron data at HERA.
The error band represents the uncertainty of the H1 NLO QCD fit to the
H1 $ep$ and the BCDMS $\mu p$ and $\mu d$ data without the constraint
$\overline{d} = \overline{u}$ at low $x$. The dashed curves represent calculations
using recent global fits by MRST and by CTEQ.}
  \label{fig:better}
\end{figure}

%
%
%
In DIS the up and down quark asymmetry can be constrained using 
deuteron data because the nucleon structure function 
determines a different linear combination according to
$F_2^N=5x (U+\bU+D+\bD)/18 +x(c+\bcc-s-\bs)/6$ with $N=(p+n)/2$. Unfortunately,
there are only data at rather large $x$ available. The effect of including
the BCDMS data on the low $x$ behaviour of the parton distributions
is illustrated in Fig.\,\ref{fig:oweh} (right plot). It restores some amount of
down quarks at low $x$ , the errors, however, in particular
of the down quarks, are still very large. The result is a 
large sea quark asymmetry uncertainty, which  is shown in Fig.\,\ref{fig:better}.
At HERA a proposal had been made \cite{eda,*edb,*edc} to operate the machine in electron-deuteron mode.
Measuring the behaviour at low $x$ would not require high luminosity. Such data
would constrain \footnote{
Constraints on the sea quark distributions may also be obtained from $W^+/W^-$ 
production at the TeVatron.  However,  the sensitivity  is
limited to larger $x \ge 0.1$ \cite{wpwma,*wpwmb}  since  $W's$ produced in collisions
involving sea quarks of smaller $x$ will be boosted so strongly, 
that their decay products are not within the acceptance of the collider detectors.
$W^+$ and $W^-$ production at the LHC has been discussed in \cite{james}. }
 a possible sea quark asymmetry with very high accuracy, 
as is also shown in Fig.\,\ref{fig:better}.

Deuterons at HERA would require a new source and modest
modifications to the preaccelerators. The H1 apparatus could be used in its 
standard mode with a forward proton detector added to take data
at half the beam energy. Tagging the spectator protons with high accuracy
at HERA, for the first time in DIS, one could reconstruct the electron-neutron
scattering kinematics essentially free of nuclear corrections \cite{eda}. Since the
forward scattering amplitude is related to diffraction
one would also be able to constrain shadowing to the per cent level  \cite{mark}.
The low $x$ measurements would require small luminosity amounts,
of less than 50\,pb$^{-1}$. Long awaited constraints of the $d/u$ ratio
at large $x$ and $Q^2$ would require extended running, as would CC data.
Besides determining the parton distributions better, the measurement
of the singlet $F_2^N$ structure function would give important
constraints on the evolution and theory at low $x$ \cite{stefano}.
It  would  also result in an estimated factor of two improvement on the
measurement of $\alpha_s$ at HERA \cite{bkp}.  
For the development of QCD, of low $x$ physics in
particular, but as well for understanding physics at the LHC and also for 
superhigh energy neutrino astrophysics, HERA  $eD$ data remain to
be important.

%% file: futureimpact.tex
\subsection{Impact of future HERA data on the determination of proton PDFs using the ZEUS NLO QCD fit
\protect\footnote{Contributing authors: C.~Gwenlan, A.~Cooper-Sarkar, C.~Targett-Adams.}}

\label{sec:clair}

\subsubsection{PDF fits to HERA data}

Recently, the ZEUS Collaboration have performed a 
combined NLO QCD fit to inclusive neutral and charged current 
DIS data~\cite{z1a,z3a,Chekanov:2003yv,z2a,z4a,z6a} 
as well as high precision jet data in DIS~\cite{pl:b547:164} and $\gamma p$ scattering~\cite{epj:c23:615}.
This is called the ZEUS-JETS PDF fit~\cite{zeusj}. 
The use of only HERA data eliminates the uncertainties from heavy-target corrections and 
removes the need for isospin symmetry assumptions. It also avoids the difficulties that can 
sometimes arise from combining data-sets from several different experiments, thereby 
allowing a rigorous statistical treatment of the PDF uncertainties. Furthermore, PDF 
uncertainties from current global fits are, in general, limited by (irreducible) 
experimental systematics. In contrast, those from fits to HERA data alone, are largely limited by 
the statistical precision of existing measurements. Therefore, the impact of future data from HERA 
is likely to be most significant in fits to only HERA data. 

\begin{table}
\begin{tabular}{llcc}
\hline
   &   &{\bf HERA-I}  &{\bf HERA-II}  \\
{\bf data sample}     & {\bf kinematic coverage}  &{\bf $\mathcal{L}$ (${\rm pb}^{-1}$)}  &{\bf $\mathcal{L}$ (${\rm pb}^{-1}$)}  \\
   &   &        &{\bf (assumed)}  \\
\hline
96-97 NC $e^+p$~\cite{z1a} &  $2.7<Q^2<30000$ ${\rm GeV}^2$; $6.3 \cdot 10^{-5}<x<0.65$    &30  &30  \\
94-97 CC $e^+p$~\cite{z2a} &  $280<Q^2<17000$ ${\rm GeV}^2$; $6.3 \cdot 10^{-5}<x<0.65$  & 48 &48\\
98-99 NC $e^-p$~\cite{z3a} &  $200<Q^2<30000$ ${\rm GeV}^2$; $0.005<x<0.65$  &  16 &350 \\
98-99 CC $e^-p$~\cite{z4a} &  $280<Q^2<17000$ ${\rm GeV}^2$; $0.015<x<0.42$  &  16 &350 \\
99-00 NC $e^+p$~\cite{Chekanov:2003yv} & $200<Q^2<30000$ ${\rm GeV}^2$; $0.005<x<0.65$ &  63 &350\\
99-00 CC $e^+p$~\cite{z6a} &  $280<Q^2<17000$ ${\rm GeV}^2$; $0.008<x<0.42$    &  61 &350 \\
96-97 inc. DIS jets~\cite{pl:b547:164}  &  $125<Q^2<30000$ ${\rm GeV}^2$; $E_{\rm T}^{Breit}>8$ GeV   &  37  &500 \\
96-97 dijets in $\gamma p$~\cite{epj:c23:615} &   $Q^2 \lesssim 1$ ${\rm GeV}^2$; $E_{\rm T}^{jet1,2}>14,11$ ${\rm GeV}$   &  37  &500 \\ \hline
optimised jets~\cite{chris} &   $Q^2 \lesssim 1$ ${\rm GeV}^2$; $E_{\rm T}^{jet1,2}>20,15$ ${\rm GeV}$    &  -  &500 \\
\hline
\end{tabular}
\caption{The data-sets included in the ZEUS-JETS and HERA-II projected PDF fits. 
The first column lists the type of data and the second gives the kinematic coverage. 
The third column gives the integrated luminosities of the HERA-I measurements included in the ZEUS-JETS fit.
The fourth column gives the luminosities assumed in the HERA-II projection. 
Note that the 96-97 NC and the 94-97 CC measurements have not had their luminosity scaled for the HERA-II projection.}
\label{tab:HERAIIASSUMPTIONS}
\end{table}

\begin{figure}[Htp]
{\includegraphics[width=15cm,height=9cm]{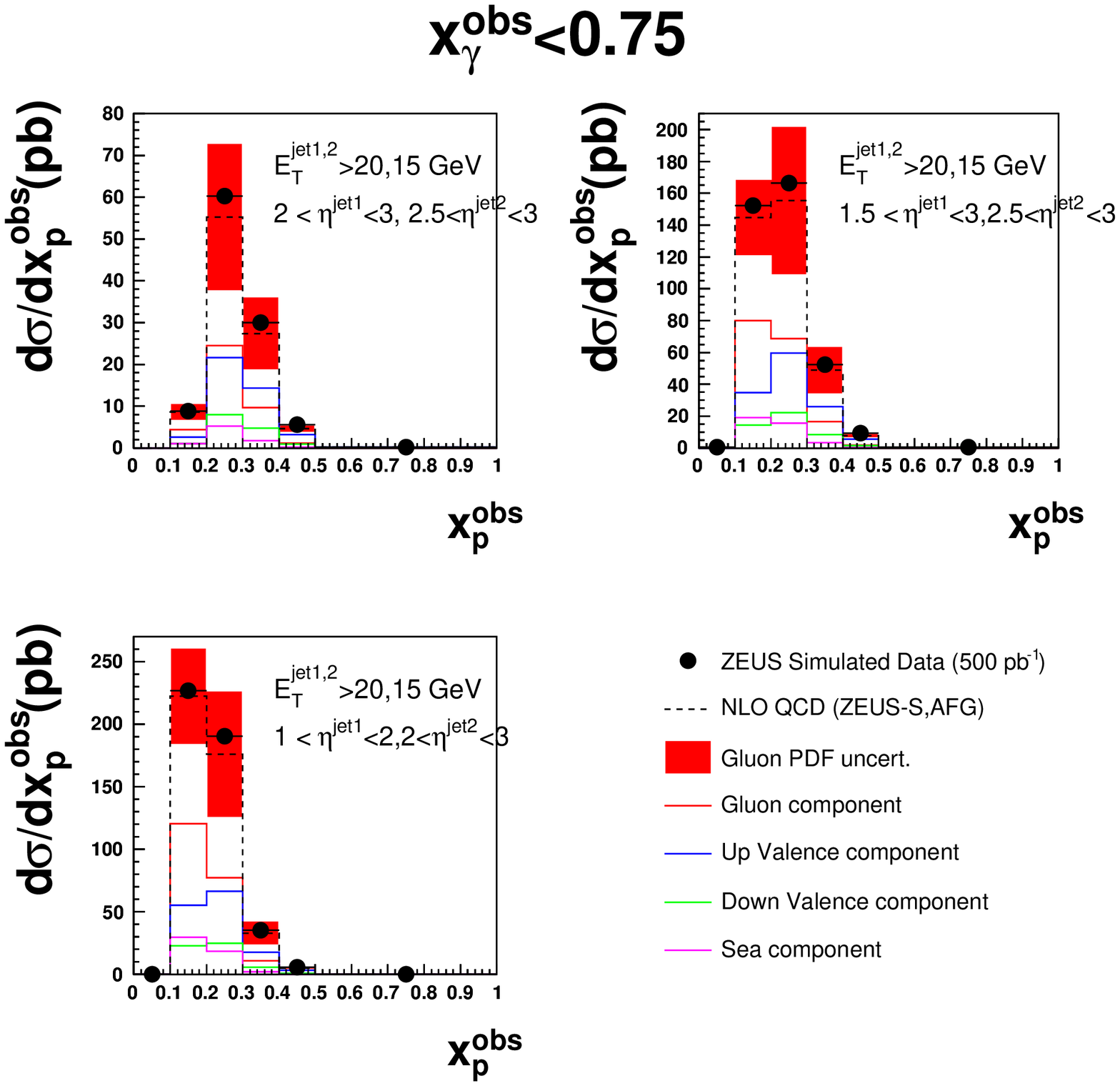}
\includegraphics[width=15cm,height=9cm]{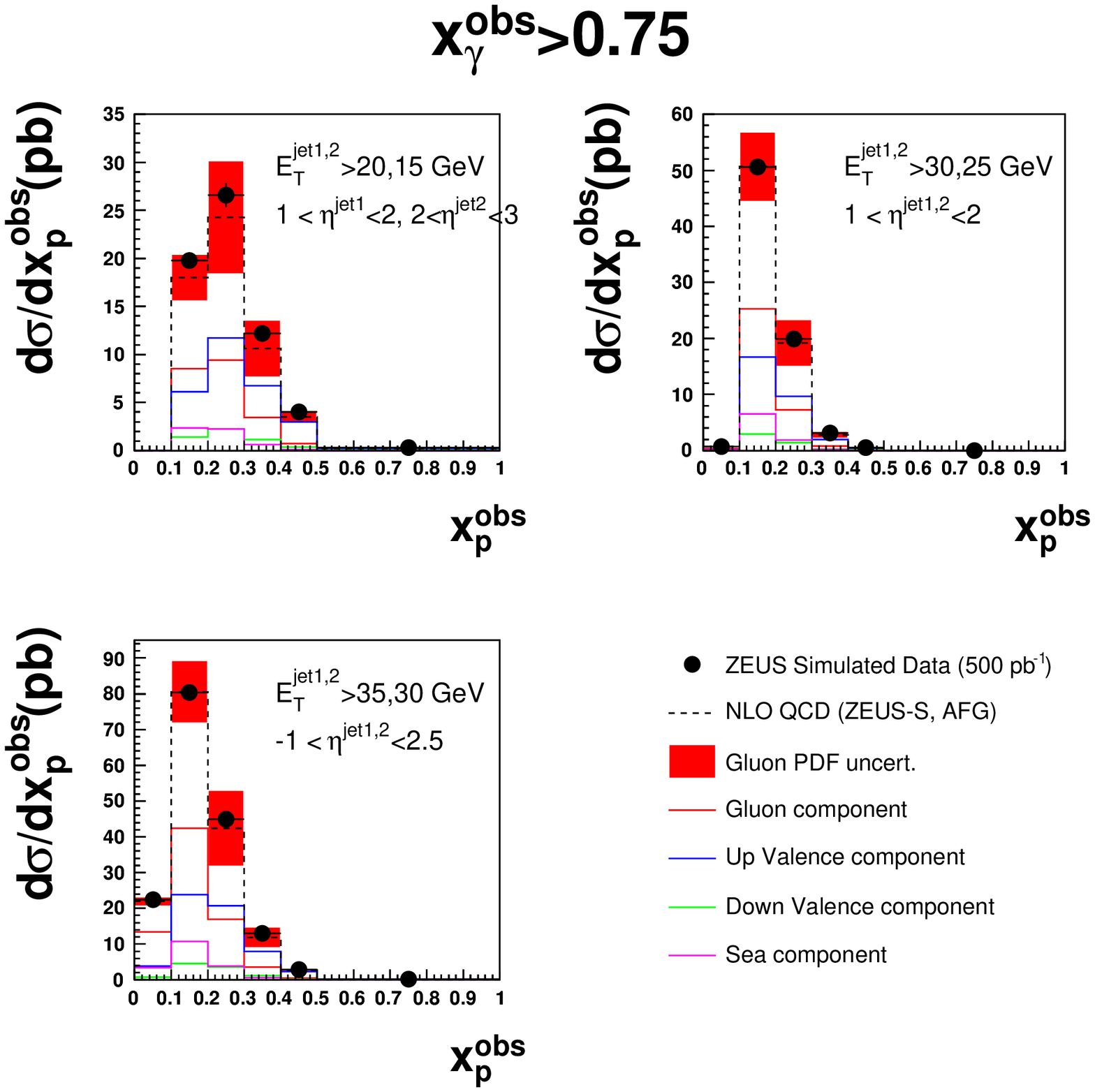}}
\caption{The optimised jet cross sections included in the HERA-II projected fit. 
The solid points show the simulated data generated using the NLO QCD programme of Frixione-Ridolfi, 
using the CTEQ5M1 proton and the AFG photon PDFs. The error bars show the statistical uncertainties, 
which correspond to $500$ ${\rm pb}^{-1}$ of HERA data. Systematic uncertainties have been neglected. 
The dashed line shows the NLO QCD prediction using the ZEUS-S proton and AFG photon PDFs. The shaded band shows the 
contribution to the cross section uncertainty arising from the uncertainty in the gluon distribution in the proton.}
\label{Fig:OptJet}
\end{figure}
\begin{figure}[Ht]
\includegraphics[width=15cm,height=9cm]{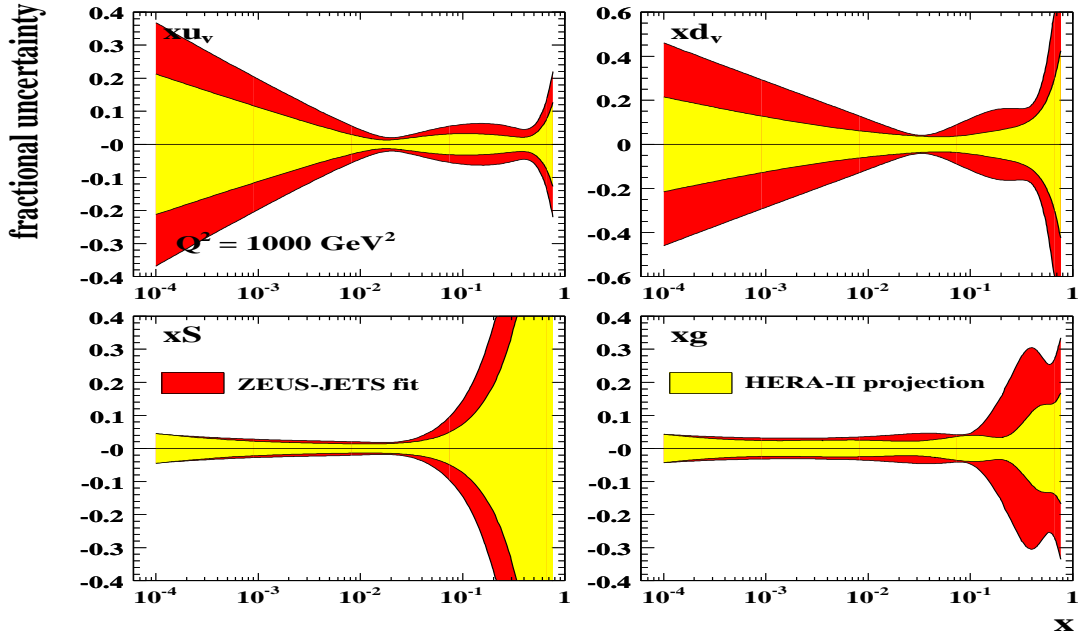}
\vspace{-0.5cm}
\caption{The fractional PDF uncertainties, as a function of $x$, for the $u$-valence, 
$d$-valence, sea-quark and gluon distributions at $Q^2=$ 1000 ${\rm GeV}^2$. 
The red shaded bands show the results of the ZEUS-JETS fit and the yellow shaded bands show the 
results of the HERA-II projected fit.}
\label{Fig:PDFS1}
\end{figure}
\begin{figure}[Ht]
\includegraphics[width=15cm,height=10.5cm]{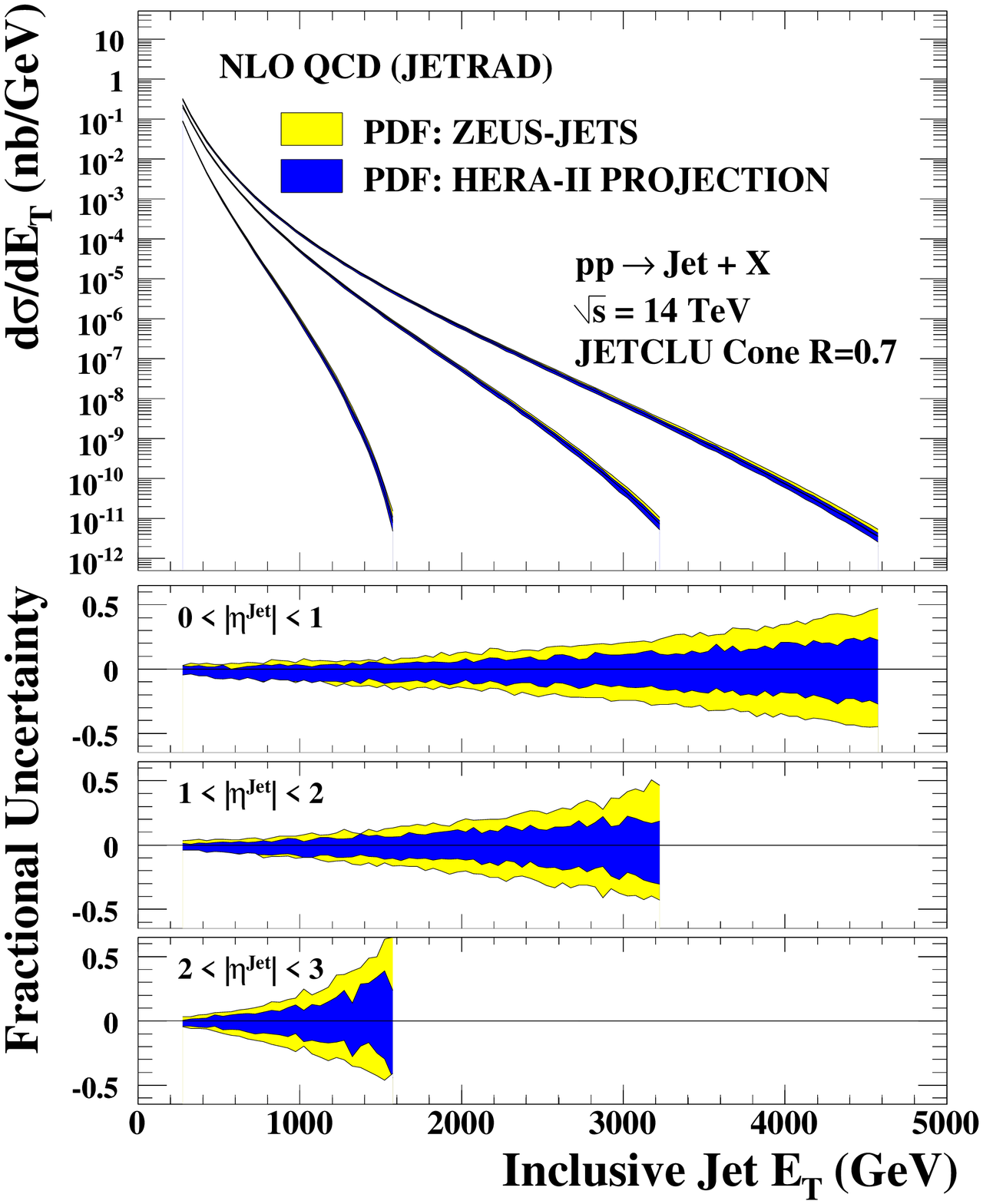}
\vspace{-0.2cm}
\caption{NLO QCD inclusive jet cross section predictions at $\sqrt{s}$=14 TeV in three 
regions of pseudo-rapidity. The yellow and blue bands show the PDF uncertainties from the ZEUS-JETS 
and HERA-II projected fits, respectively.}
\label{Fig:JETS}
\end{figure}

\subsubsection{The ZEUS NLO QCD fit}
The ZEUS-JETS PDF fit has been used as the basis for all results shown in this contribution. 
The most important details of the fit are summarised here. 
A full description may be found elsewhere~\cite{zeusj}. 
The fit includes the full set of 
ZEUS inclusive neutral and charged current $e^{\pm}p$ data from HERA-I (1994-2000), 
as well as two sets of high precision jet 
data in $e^+p$ DIS ($Q^2 >> 1$ ${\rm GeV}^2$)
and $\gamma p$ ($Q^2 \sim 0$) scattering.
The inclusive data used in the fit, span the kinematic 
range $6.3 \times 10^{-5} < x < 0.65$ and $2.7 < Q^2 < 30 000$ ${\rm GeV}^2$. 

The PDFs are obtained by solving the NLO DGLAP equations within the $\overline{{\rm MS}}$ scheme. 
These equations yield the PDFs at all values of $Q^2$ provided they are input as functions of $x$ 
at some starting scale $Q_0^2$. The resulting PDFs are convoluted with coefficient functions to 
give predictions for structure functions and, hence, cross sections. In the ZEUS fit, the  $xu_v(x)$ ($u$-valence), 
$xd_v(x)$ ($d$-valence), $xS(x)$ (total sea-quark), $xg(x)$ (gluon) and $x(\bar{d}(x)-\bar{u}(x))$ 
PDFs are parameterised at a starting scale of $Q_0^2 = 7$ ${\rm GeV}^2$ by the form,
\begin{eqnarray}
\label{PDFparm}
xf(x) = p_{1} x^{p_2} (1-x)^{p_{3}} P(x),
\end{eqnarray}
where $P(x)=(1+p_4 x)$. No advantage in the $\chi^2$ results from using more complex polynomial forms. 
The normalisation parameters, $p_1(u_v)$ and $p_1(d_v)$, 
are constrained by quark number sum rules while $p_1(g)$ is constrained by the momentum sum rule. 
Since there is no information to constrain any difference in the low-$x$ behaviour of the $u$- and $d$-valence 
quarks, $p_{2}(u_v)$ has been set equal to $p_2(d_v)$. The data from HERA are currently less precise than the 
fixed target data in the high-$x$ regime. Therefore, the high-$x$ sea and gluon distributions 
are not well constrained in current fits to HERA data alone. 
To account for this, the sea shape has been restricted by setting $p_4(S)=0$. The high-$x$ gluon shape is constrained by 
the inclusion of HERA jet data. 
In fits to only HERA data, there is no information on the shape of $\bar{d}-\bar{u}$. Therefore, 
this distribution has its shape fixed consistent with Drell-Yan data and its normalisation 
set consistent with the size of the Gottfried sum rule violation. A suppression of the strange 
sea with respect to the non-strange sea of a factor of 2 at $Q_0^2$ is also imposed, consistent 
with neutrino induced dimuon data from CCFR. The value of the strong coupling has been fixed 
to $\alpha_s(M_{\rm Z})=0.1180$. 
After all constraints, the ZEUS-JETS fit has 11 free parameters.
Heavy quarks were treated in the variable flavour number scheme of Thorne \& Roberts~\cite{hq}. 
Full account was taken of correlated experimental systematic uncertainties, using the Offset Method~\cite{zeus,Mandy}.

The results of two separate studies are presented. The first study 
provides an estimate of how well the PDF uncertainties may be known by the end of HERA-II, 
within the currently planned running scenario, while the second study investigates the impact of a future HERA 
measurement of ${ F_L}$ on the gluon distribution. All results presented, are based on the recent 
ZEUS-JETS PDF analysis~\cite{zeusj}.

\subsubsection{PDF uncertainty estimates for the end of HERA running}
The data from HERA-I are already very precise and cover a wide kinematic region.
However, HERA-II is now running efficiently and is expected to provide a substantial increase in luminosity. 
Current estimates suggest that, by the end of HERA running (in mid-2007), an integrated luminosity of $700$ ${\rm pb}^{-1}$ should be achievable. 
This will allow more precise measurements of cross sections 
that are curently statistically limited: in particular, the 
high-$Q^2$ NC and CC data, as well as high-$Q^2$ and/or high-$E_{\rm T}$ jet data. In addition to the simple increase in luminosity, 
recent studies~\cite{chris} have shown that future jet cross section measurements, in kinematic regions optimised for sensitivity to PDFs, should 
have a significant impact on the gluon uncertainties.
In this contribution, the effect on the PDF uncertainties, 
of both the higher precision expected from HERA-II and the possibility of optimised jet cross section measurements,
has been estimated in a new QCD fit. This fit will be referred to as the ``HERA-II projection''. 

In the HERA-II projected fit, the statistical uncertainties on the currently available HERA-I data have been reduced.
For the high-$Q^2$ inclusive data, a total integrated luminosity of $700$ ${\rm pb}^{-1}$ was assumed, equally divided between $e^+$ and $e^-$. 
For the jet data, an integrated luminosity of $500$ ${\rm pb}^{-1}$ was assumed. The central values and systematic uncertainties were 
taken from the published data in each case. In addition to the assumed increase in precision of the measurements, 
a set of optimised jet cross sections were also included, for forward dijets in $\gamma p$ collisions, as defined in a 
recent study~\cite{chris}. Since no real 
data are yet available, simulated points were generated using the NLO QCD program of Frixione-Ridolfi~\cite{np:b507}, 
using the CTEQ5M1~\cite{cteq} proton and AFG~\cite{afg} photon PDFs. 
The statistical uncertainties were taken to correspond to $500$ ${\rm pb}^{-1}$. For this study, systematic uncertainties on 
the optimised jet cross sections were ignored. The simulated optimised jet cross section points, compared to the predictions of NLO QCD using the ZEUS-S proton PDF~\cite{zeus}, are shown in Fig.~\ref{Fig:OptJet}. 

Table~\ref{tab:HERAIIASSUMPTIONS} lists the data-sets included in 
the ZEUS-JETS and HERA-II projected fits. The luminosities of the 
(real) HERA-I measurements and those assumed for the HERA-II projection are also given.

The results are summarised in Fig.~\ref{Fig:PDFS1}, which shows the 
fractional PDF uncertainties, for the $u$- and $d$-valence, 
sea-quark and gluon distributions, at $Q^2 = 1000$ ${\rm GeV}^2$. 
The yellow bands show the results of the ZEUS-JETS fit 
while the red bands show those for the HERA-II projection. 
Note that the same general features are observed for all values of $Q^2$. 
In fits to only HERA data, the information on the valence quarks comes from the high-$Q^2$ NC and CC cross sections. 
The increased statistical precision of the high-$Q^2$ data, as assumed in the HERA-II projected fit, 
gives a significant improvement in the valence uncertainties over the whole range of $x$. For the sea 
quarks, a significant improvement in the uncertainties at high-$x$ is also observed. In contrast, the low-$x$ 
uncertainties are not visibly reduced. This is due to the fact that the data constraining the low-$x$ region 
tends to be at lower-$Q^2$, which are already systematically limited. This is also the reason why the low-$x$ gluon 
uncertainties are not significantly reduced. However, the mid-to-high-$x$ gluon, which is constrained 
by the jet data, is much improved in the HERA-II projected fit. Note that about half 
of the observed reduction in the gluon uncertainties is due to the inclusion of the simulated optimised jet cross sections.

\vspace{-0.2cm}
\paragraph{Inclusive jet cross sections at the LHC}
The improvement to the high-$x$ partons, observed in the HERA-II projection compared to the ZEUS-JETS fit, will be particularly 
relevant for high-scale physics at the LHC.
This is illustrated in Fig.~\ref{Fig:JETS}, which shows NLO QCD predictions from the JETRAD~\cite{np:b403:633} programme for 
inclusive jet production at $\sqrt{s}=14$ TeV. The results are shown for both 
the ZEUS-JETS and the HERA-II projected PDFs. The uncertainties on the cross sections, resulting from the PDFs, have been 
calculated using the LHAPDF interface~\cite{Whalley:2005nh}. For the ZEUS-JETS PDF, the uncertainty reaches $\sim 50\%$ at 
central pseudo-rapidities, for the highest jet transverse energies shown. 
The prediction using the HERA-II projected PDF shows a 
marked improvement at high jet tranverse energy.

\subsubsection{Impact of a future HERA measurement of $F_{ L}$ on the gluon PDF}

The longitudinal structure function, ${ F_L}$, is directly related to the gluon density in 
the proton.
In principle, ${ F}_{ L}$ can be extracted by measuring the NC DIS cross section at fixed $x$ and $Q^2$, 
for different values of $y$ (see Eqn.~\ref{Eqn:NC}). A precision measurement could be achieved by varying the centre-of-mass 
energy, since $s=Q^2/xy\approx4E_eE_p$, where $E_e$ and $E_p$ are the electron and proton beam energies, 
respectively. Studies~\cite{Klein:2004zq}
(Sec.~\ref{sec:flmax}) 
have shown that this would be most efficiently achieved by 
changing the proton beam energy. However, such a measurement has not yet been performed at HERA.

There are several reasons why a measurement of ${ F_L}$ at low-$x$ could be important. 
The gluon density is not well known at low-$x$ and so 
different PDF parameterisations can give quite different predictions for ${ F_L}$ 
at low-$x$. Therefore, a precise measurement of the longitudinal sturcture function could both pin down the 
gluon PDF and reduce its uncertainties. Furthermore, predictions of ${ F_L}$ also depend upon the 
nature of the underlying theory (e.g. order in QCD, resummed calculation etc). Therefore, a measurement of 
${ F_L}$ could also help to discriminate between different theoretical models. 

\paragraph{Impact on the gluon PDF uncertainties}
\begin{figure}[Ht]
\includegraphics[width=15cm,height=9cm]{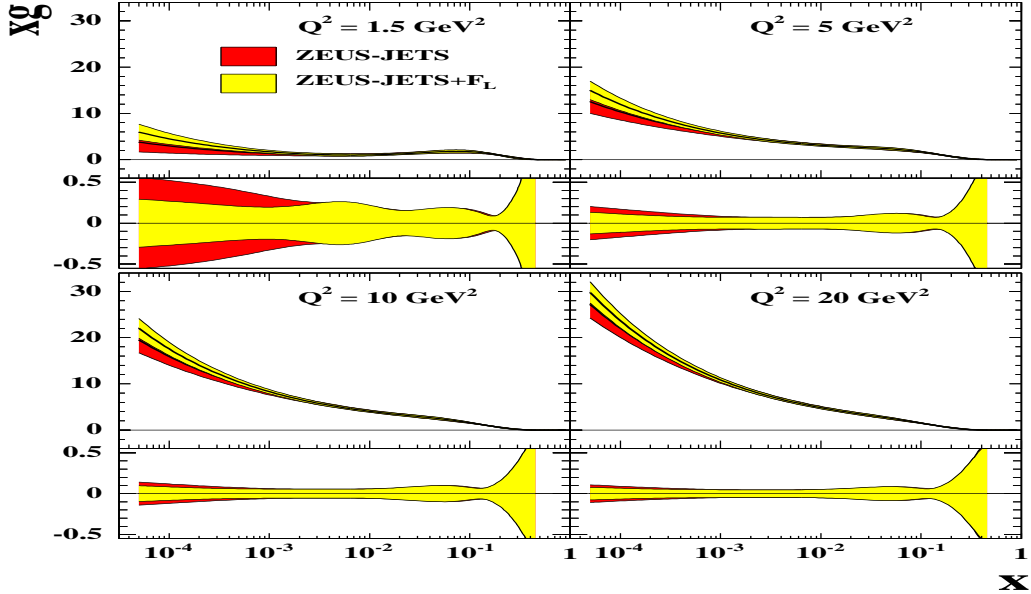}
\vspace{-0.5cm}
\caption{The gluon PDFs, showing also the fractional uncertainty, for fits with and 
without inclusion of the simulated ${ F_L}$ data, for $Q^2 =$ 1.5, 5, 10 and 20 
${\rm GeV}^2$. The red shaded bands show the results of the ZEUS-JETS fit and the 
yellow shaded band show the results of the ZEUS-JETS+${ F_L}$ fit.}
\label{Fig:FL}
\end{figure}
The impact of a possible future HERA measurement of ${ F_L}$ on the gluon PDF uncertainties 
has been investigated, using a set of simulated ${ F_L}$ data-points~\cite{Klein:2004zq}.
(see Sec.~\ref{sec:flmax}). 
The simulation was performed using the GRV94~\cite{zphys:c67:433} proton PDF for the central values, and 
assuming $E_e = 27.6$ GeV and $E_p = 920, 575, 465$ and $400$ GeV, with luminosities of 10, 5, 3 and 2 ${\rm pb}^{-1}$, respectively. 
Assuming that the luminosity scales simply as $E_p^2$, this scenario would nominally 
cost $35$ ${\rm pb}^{-1}$ of luminosity under standard HERA conditions. However, this estimate takes no account of time taken 
for optimisation of the machine with each change in $E_p$, which could be considerable. The systematic uncertainties 
on the simulated data-points were calculated assuming a $\sim 2\%$ precision on the inclusive NC 
cross section measurement. 
A more comprehensive description of the simulated data is given 
in contribution for this proceedings, see Sec.~\ref{sec:flmax}.

\begin{figure}[Hct]
\includegraphics[width=15cm,height=9cm]{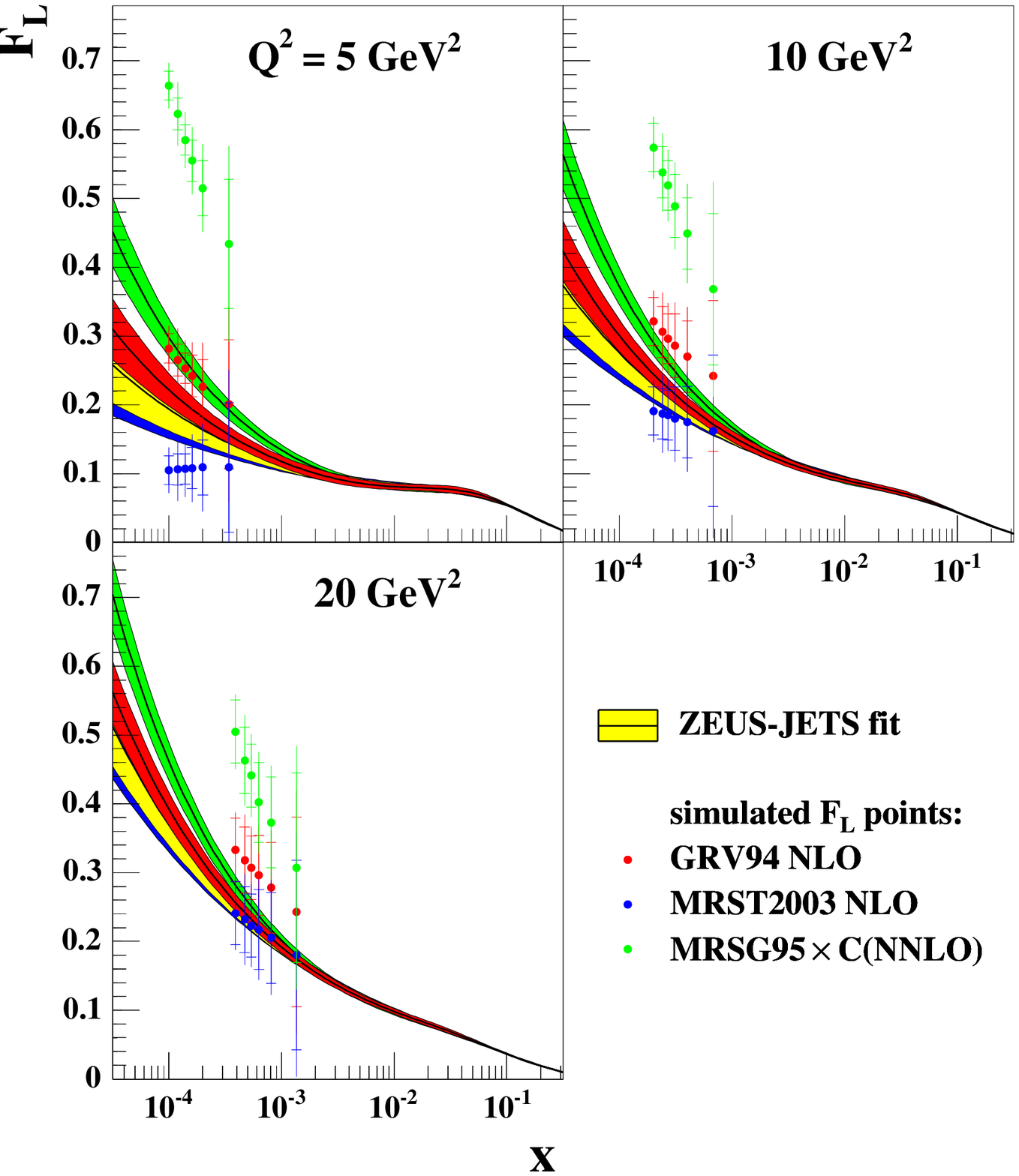}
\vspace{-0.5cm}
\caption{The distribution of the longitudinal structure function ${ F_L}$ at 
$Q^2=$5, 10 and 20 ${\rm GeV}^2$. 
The blue, red and green points show the simulated ${ F_L}$ data-points, respectively 
labelled maximum, middle and minimum in Table~\ref{tab:FLEXTREMES}. The blue, red and green shaded bands 
show the NLO QCD predictions, in the case where the data-points of the corresponding colour 
have been included in the fit. For comparison, the yellow shaded band shows 
the prediction of the ZEUS-JETS fit.}
\label{Fig:EXTREMEFL}
\end{figure}
The simulated data were included in the ZEUS-JETS fit. 
Figure~\ref{Fig:FL} shows the gluon distribution and fractional uncertainties for fits with and without 
inclusion of the simulated ${ F_L}$ data. The 
results indicate that the gluon uncertainties are reduced at low-$x$, but the improvement is only 
significant at relatively low $Q^2 \lesssim 20$ ${\rm GeV}^2$. 

\paragraph{Discrimination between theoretical models}
In order to assess whether a HERA measurement of ${ F_L}$ could discriminate between theoretical models, 
two more sets of ${ F_L}$ data-points have been simulated~\cite{robert}, using different theoretical 
assumptions. The first of the two sets was generated using the MRSG95~\cite{pl:b354:155} proton PDF, which 
has a large gluon density. 
The PDFs were then convoluted with the NNLO order coefficient 
functions, which are large and positive. This gives the ``maximum'' 
set of ${ F_L}$ data-points. In contrast, the second set has been generated using the MRST2003~\cite{epj:c35:325} 
proton PDF, which has a negative gluon at low-$x$ and low-$Q^2$, thus providing a ``minimum'' set of 
${ F_L}$ data. The original set of ${ F_L}$ points described in the previous subsection lies 
between these two extremes. The details of all three sets are summarised in Tab.~\ref{tab:FLEXTREMES}.

Figure~\ref{Fig:EXTREMEFL} shows the results of including, individually, each set of simulated ${ F_L}$ 
data into the ZEUS NLO QCD fit. The results show that the NLO fit is relatively stable to the inclusion of 
the extreme sets of data. 
This indicates that a measurement of ${ F_L}$ could discriminate between certain theoretical models. However, it should 
be noted that the maximum and minimum models studied here were chosen specifically to give the widest possible 
variation in ${ F_L}$. There are many other alternatives that would lie between these extremes and the 
ability of an ${ F_L}$ measurement to discriminate 
between them would depend both on the experimental precision of the measurement itself, as well as the 
theoretical uncertainties on the models being tested.

\begin{table}
\begin{tabular}{lll}
\hline
  & \tablehead{1}{}{}{PDF}
  & \tablehead{1}{}{}{QCD order of coefficient functions}\\
\hline
Maximum ${ F_{L}}$ & MRSG95   & NNLO \\
Middle ${ F_{L}}$ & GRV94    & NLO\\
Minimum ${ F_{L}}$ & MRST2003 & NLO\\
\hline
\end{tabular}
\caption{Summary of the PDFs used to generate the simulated ${ F_L}$ data-points.
The extreme maximum $F_L$ points were generated using the MRSG95 PDF, and 
convoluted with NNLO coefficient functions. The middle points were generated using the GRV94 PDF, 
and the extreme minimum points were generated using the MRST2003 PDF, which has a negative gluon at low-$x$.}
\label{tab:FLEXTREMES}
\end{table}

%% file: jetsinfits.tex
\subsection{A Method to Include Final State Cross-sections Measured in
Proton-Proton Collisions to Global NLO QCD Analysis
\protect\footnote{Contributing authors: T.~Carli, G.~Salam, F.~Siegert.}}
\label{sec:jetsinfits}

The Large Hadron Collider (LHC), currently under construction at CERN,
will collide protons on protons with an energy of $7$~{\rm TeV}.
Together with its high collision rate the high available centre-of-mass
energy 
will make it possible to test new interactions at very short
distances that might be revealed
in the production cross-sections of Standard Model (SM) particles at very
high transverse momentum ($P_T$)
as deviation from the SM theory.

The sensitivity to new physics crucially depends on experimental uncertainties in the measurements
and on theoretical uncertainties in the SM predictions. It is therefore important to work out
a strategy to minimize both the experimental and theoretical uncertainties from LHC data.
For instance, one could use single inclusive jet  or Drell-Yan
cross-sections at low $P_T$ to constrain the PDF uncertainties at high $P_T$.
Typical residual renormalisation and factorisation scale uncertainties
in next-to-leading order (NLO) calculations for single inclusive
jet-cross-section are about $5-10\%$ and should hopefully be reduced
as NNLO calculations become available. The impact of PDF uncertainties on the other
hand can be substantially larger in some regions, especially at large $P_T$,
and for example at $P_T = 2000$~{\rm GeV} dominate the overall
uncertainty of $20\%$.
If a suitable combination of data measured
at the Tevatron and LHC can be included in global NLO QCD analyses,
the PDF uncertainties can be constrained.

The aim of this contribution is to propose a method for consistently
including final-state observables in global QCD analyses.

For inclusive data like the proton structure function $F_2$ in deep-inelastic
scattering (DIS) the perturbative coefficients
are known analytically. During the fit the cross-section
can therefore be quickly calculated from the strong coupling ($\as$) and the PDFs
and can be compared to the measurements.
However, final state observables, where detector acceptances
or jet algorithms are involved in the definition of the perturbative coefficients
(called ``weights'' in the following),
have to be calculated using NLO Monte Carlo programs. Typically such programs
need about one day of CPU time to calculate accurately the cross-section.
It is therefore necessary to find a way to calculate the perturbative
coefficients with high precision in a long run and to include
 $\as$ and the PDFs  ``a posteriori''.

To solve this problem
many methods have been proposed in the past
\cite{Graudenz:1995sk,Kosower:1997vj,Stratmann:2001pb,h1alphas,wobisch,zeusj}.  In principle the
highest efficiencies can be obtained by taking moments with respect to
Bjorken-$x$ \cite{Graudenz:1995sk,Kosower:1997vj}, because this
converts convolutions into multiplications. This can have notable
advantages with respect to memory consumption, especially in cases
with two incoming hadrons.  On the other hand, there are complications
such as the need for PDFs in moment space and the
associated inverse Mellin transforms.

Methods in $x$-space have traditionally been somewhat less efficient,
both in terms of speed (in the `a posteriori' steps --- not a major
issue here) and in terms
of memory consumption. They are, however, somewhat more transparent
since they provide direct information on the $x$ values of relevance.
Furthermore they can be used with any PDF.
The use of $x$-space methods can be further improved by using
methods developed originally for PDF evolution \cite{Ratcliffe:2000kp,Dasgupta:2001eq}.
\subsubsection{PDF-independent representation of cross-sections}
\paragraph{Representing the PDF on a grid}
We make the assumption that PDFs can be accurately represented by storing their values on a
two-dimensional grid of points and using $n^{\mathrm{th}}$-order interpolations between those points.
Instead of using the parton momentum fraction $x$ and the factorisation scale $Q^2$, we use a
variable transformation that provides good coverage of the full $x$
and $Q^2$ range
with uniformly spaced grid points:%
\footnote{An alternative for the $x$ grid is to use $y = \ln 1/x +
  a(1-x)$ with $a$ a parameter that serves to increase the density of
  points in the large $x$ region.}
\begin{equation}
\label{eq:ytau}
y(x) = \ln \frac{1}{x} \; \; \; {\rm and} \; \; \;
\tau(Q^2) = \ln \ln \frac{Q^2}{\Lambda^2}.
\end{equation}
The parameter $\Lambda$ is to be chosen of the order of $\Lambda_{\mathrm{QCD}}$, but not necessarily identical.
The PDF $q(x,Q^2)$ is then represented by its values $q_{i_y,i_\tau}$ at the 2-dimensional
grid point $(i_y \, \delta y, i_\tau \, \delta \tau)$, where $\delta
y$ and $\delta \tau$ denote the grid spacings,
and obtained elsewhere by interpolation:
\begin{equation}
\label{eq:interp}
q(x,Q^2) = \sum_{i=0}^n \sum_{\iota=0}^{n'} q_{k+i,\kappa+\iota} \,\,
I_i^{(n)}\left(  \frac{y(x)}{\delta y}  - k \right)\,
I_\iota^{(n')}\left(  \frac{\tau(Q^2)}{\delta\tau}-\kappa  \right),
\end{equation}
where $n$, $n'$ are the interpolation orders.
The interpolation function $I_i^{(n)}(u)$ is 1 for $u=i$ and otherwise is given by:
\begin{equation}
\label{eq:Ii}
I_i^{(n)}(u) = \frac{(-1)^{n-i}}{i!(n-i)!} \frac{u (u-1) \ldots (u-n)}{u-i}\,.
\end{equation}
Defining $\mathrm{int} (u)$ to be the largest integer such that $\mathrm{int}(u) \le u$,
$k$ and $\kappa$ are defined as:
\begin{eqnarray}
\label{eq:kchoice}
k(x) =& \mathrm{int} \left( \frac{y(x)}{\delta y} - \frac{n-1}{2} \right), &
\kappa(x) = \mathrm{int} \left( \frac{\tau(Q^2)}{\delta \tau} - \frac{n'-1}{2} \right).
\end{eqnarray}
Given finite grids whose vertex indices range from $0\ldots N_y-1$ for
the $y$ grid and $0\ldots N_\tau-1$ for the $\tau$ grid, one should
additionally require that eq.~(\ref{eq:interp}) only uses available
grid points. This can be achieved by remapping $k \to
\max(0,\min(N_y-1-n,k))$ and $\kappa \to
\max(0,\min(N_\tau-1-n',\kappa))$.

\paragraph{Representing the final state cross-section weights on a
  grid (DIS case)}
Suppose that we have an NLO Monte Carlo program that produces events $m=1\dots N$.
Each event $m$ has an $x$ value, $x_m$, a $Q^2$ value, $Q^2_m$, as well as a weight, $w_m$,
and a corresponding order in $\as$, $p_m$.
Normally one would obtain the final result $W$ of the Monte Carlo integration from:\footnote{Here, and in the following,
renormalisation and factorisation scales have been set equal for simplicity.}
\begin{equation}
\label{eq:normalint}
 W = \sum_{m=1}^N \,w_m \, \left( \frac{\alpha_s(Q_m^2)} {2\pi}\right)^{p_m}  \, q(x_m,Q^2_m).
\end{equation}

Instead one introduces a weight grid $W_{i_y,i_\tau}^{(p)}$ and then for each event updates
a portion of the grid with:\\
$i = 0\dots n,\; \iota = 0\dots n':$
\begin{eqnarray}
\label{eq:weight2evolve}
W_{k+i,\kappa + \iota}^{(p_m)} \to W_{k+i,\kappa + \iota}^{(p_m)} + w_m\,
  I_i^{(n)} \left(\frac{y(x_m)}{\delta y} - k\right)
  I_{\iota}^{(n')}\left(\frac{\tau(Q^2_m)}{\delta \tau} - \kappa \right), \\
 \;\;\; {\rm where} \;\;\;
  k \equiv k(x_m),\; \kappa \equiv \kappa(Q^2_m). \nonumber
  \end{eqnarray}
The final result for $W$, for an arbitrary PDF, can then be obtained \emph{subsequent}
to the Monte Carlo run:
\begin{equation}
\label{eq:WfinalxQ}
W = \sum_p \sum_{i_y} \sum_{i_\tau}
W_{i_y,i_\tau}^{(p)} \, \left( \frac{\alpha_s\left({Q^2}^{(i_\tau)}\right)}{2\pi}\right)^{p} q \!\left(x^{(i_y)}, {Q^2}^{(i_\tau)} \right)\,,
\end{equation}
where the sums index with $i_y$ and $i_\tau$ run over the number of grid points and
we have have explicitly introduced $x^{(i_y)}$ and ${Q^2}^{(i_\tau)}$ such that:
\begin{equation}
\label{eq:xQdefs}
 y(x^{(i_y)}) = i_y \, \delta y \quad {\rm and} \quad
\tau\left({Q^2}^{(i_\tau)}\right) =  i_\tau \, \delta \tau.
\end{equation}

\paragraph{Including renormalisation and factorisation scale dependence }
If one has the weight matrix $W_{i_y,i_\tau}^{(p)}$ determined separately order by order in
$\as$, it is straightforward to vary the renormalisation $\mu_R$  and
factorisation $\mu_F$ scales a posteriori (we assume that they were kept equal
in the original calculation).

It is helpful to introduce some notation relating to the DGLAP evolution 
equation:
\begin{equation}
  \label{eq:DGLAP}
  \frac{d q(x,Q^2)}{d \ln Q^2} = \frac{\alpha_s(Q^2)}{2\pi} (P_0 \otimes q)(x,Q^2)
                + \left(\frac{\alpha_s(Q^2)}{2\pi}\right)^2 (P_1
                \otimes q)(x,Q^2) + \ldots,
\end{equation}
where the $P_0$ and $P_1$ are the LO and NLO matrices of DGLAP
splitting functions that operate on vectors (in flavour space) $q$ of
PDFs.
Let us now restrict our attention to the NLO case where we have just
two values of $p$, $p_{\mathrm{LO}}$ and $p_{\mathrm{NLO}}$.
Introducing $\xi_R$ and $\xi_F$ corresponding to the factors by which one varies $\mu_R$ and $\mu_F$ respectively,
for arbitrary $\xi_R$ and $\xi_F$ we may then write:
\begin{eqnarray}
  \label{eq:Wfinalxi}
  W(\xi_R, \xi_F) = \sum_{i_y} \sum_{i_\tau}
    \left(\frac{\alpha_s\left(\xi_R^2 {Q^2}^{(i_\tau)}\right)\,
               }{2\pi}
     \right)^{p_{\mathrm{LO}}}
  W_{i_y,i_\tau}^{(p_{\mathrm{LO}})}
   q \!\left(x^{(i_y)}, \xi_F^2  {Q^2}^{(i_\tau)} \right) +
   \nonumber \\
  \left(\frac{\alpha_s\left(\xi_R^2 {Q^2}^{(i_\tau)} \right)\,
             }{2\pi}
   \right)^{p_{\mathrm{NLO}}}
  \left[
    \left( W_{i_y,i_\tau}^{(p_{\mathrm{NLO}})} + 2\pi  \beta_0 p_{\mathrm{LO}} \ln \xi_R^2
         \,W_{i_y,i_\tau}^{(p_{\mathrm{LO}})}
    \right)  q \!\left(x^{(i_y)}, \xi_F^2 {Q^2}^{(i_\tau)} \right)
    \right. \\\left.
     - \ln \xi_F^2 \,W_{i_y,i_\tau}^{(p_{\mathrm{LO}})}
     (P_0\otimes q) \!\left(x^{(i_y)}, \xi_F^2 {Q^2}^{(i_\tau)} \right)
  \right] \,, \nonumber
\end{eqnarray}
where $\beta_0 = (11 N_c - 2n_f)/(12\pi)$ and $N_c$ ($n_f$) is the number of colours (flavours).
Though this formula is
given for $x$-space based approach, a similar formula applies for
moment-space approaches. Furthermore it is straightforward to extend
it to higher perturbative orders.

 \paragraph{Representing the weights in the case of two incoming hadrons}
In hadron-hadron scattering one can use analogous procedures with one more dimension.
Besides $Q^2$, the weight grid depends on the momentum fraction of the first ($x_1$) and
second ($x_2$) hadron.

In the case of jet production in proton-proton collisions
the weights generated by the Monte Carlo program as well as the PDFs
can be organised in seven possible initial state combinations of partons:
\begin{eqnarray}
\mathrm{gg}: \;\; F^{(0)}(x_{1}, x_{2}; Q^{2}) &=& G_{1}(x_{1})G_{2}(x_{2})\\
\mathrm{qg}: \;\; F^{(1)}(x_{1}, x_{2}; Q^{2}) &=& \left(Q_{1}(x_{1})+
                  \overline Q_{1}(x_{1})\right) G_{2}(x_{2})\\
\mathrm{gq}: \;\; F^{(2)}(x_{1}, x_{2}; Q^{2}) &=&  G_{1}(x_{1})\left(Q_{2}(x_{2})+
                  \overline Q_{2}(x_{2})\right)\\
\mathrm{qr}: \;\; F^{(3)}(x_{1}, x_{2}; Q^{2}) &=&  Q_{1}(x_{1}) Q_{2}(x_{2})
                                        + \overline Q_{1}(x_{1}) \overline Q_{2}(x_{2}) -D(x_{1}, x_{2})\\
\mathrm{qq}: \;\; F^{(4)}(x_{1}, x_{2}; Q^{2}) &=& D(x_{1}, x_{2})\\
\mathrm{q\bar q}: \;\; F^{(5)}(x_{1}, x_{2}; Q^{2}) &=& \overline D(x_{1}, x_{2})\\
\mathrm{q\bar r}: \;\; F^{(6)}(x_{1}, x_{2}; Q^{2}) &=& Q_{1}(x_{1}) \overline Q_{2}(x_{2})
                   + \overline Q_{1}(x_{1}) Q_{2}(x_{2})       -\overline D(x_{1}, x_{2}),
\end{eqnarray}
where $g$ denotes gluons, $q$ quarks and $r$ quarks of different flavour $q \neq r$
and we have used the generalized PDFs defined as:
\begin{eqnarray}
 G_{H}(x) = f_{0/H}(x,Q^{2}), &&
 Q_{H}(x) = \sum_{i = 1}^{6} f_{i/H}(x,Q^{2}), \;\;
 \overline Q_{H}(x) = \sum_{i = -6}^{-1} f_{i/H}(x,Q^{2}), \nonumber \\
 D(x_{1}, x_{2}) &=& \mathop{\sum_{i = -6}^{6}}_{i\neq0} f_{i/H_1}(x_{1},Q^2) f_{i/H_2}(x_{2},Q^{2}), \\
 \overline D(x_{1}, x_{2}, \mu^{2}_{F}) &=&
  \mathop{\sum_{i = -6}^{6}}_{i\neq0} f_{i/H_1}(x_{1},Q^{2}) f_{-i/H_2}(x_{2},Q^{2}), \nonumber \;\;
\end{eqnarray}
where $f_{i/H}$ is the PDF of flavour $i=-6 \dots 6$ for hadron $H$
and $H_1$ ($H_2$) denotes the first or second hadron\footnote{
In the above equation we follow the standard PDG Monte Carlo numbering
scheme \cite{Eidelman:2004wy} where gluons
are denoted as $0$, quarks have values from $1$-$6$ and anti-quarks have the corresponding
negative values.}.

The analogue of eq.~\ref{eq:WfinalxQ} is then given by:
\begin{equation}
\label{eq:WfinalxQ_twohadrons}
W = \sum_p \sum_{l=0}^{6} \sum_{i_{y_1}} \sum_{i_{y_2}} \sum_{i_\tau}
W_{i_{y_1},i_{y_2},i_\tau}^{(p)(l)} \, \left( \frac{\alpha_s\left({Q^2}^{(i_\tau)}\right)}{2\pi}\right)^{p}
F^{(l)}\left(x_1^{(i_{y_1})}, x_2^{(i_{y_1})},  {Q^2}^{(i_\tau)}\right).
\end{equation}

 \paragraph{Including scale depedence in the case of two incoming hadrons}
It is again possible to choose arbitrary renormalisation and
factorisation scales, specifically for NLO accuracy:
\begin{eqnarray}
  \label{eq:Wfinalxi_twohadrons}
  W(\xi_R, \xi_F) = \sum_{l=0}^{6} \sum_{i_{y_1}} \sum_{i_{y_2}} \sum_{i_\tau}
    \left(\frac{\alpha_s\left(\xi_R^2 {Q^2}^{(i_\tau)}\right)\,
               }{2\pi}
     \right)^{p_{\mathrm{LO}}}
     W_{i_{y_1},i_{y_2},i_\tau}^{(p_{\mathrm{LO}})(l)}
  F^{(l)}\left(x_1^{(i_{y_1})}, x_2^{(i_{y_1})}, \xi_F^2{Q^2}^{(i_\tau)}\right)
    +
   \nonumber \\
  \left(\frac{\alpha_s\left(\xi_R^2 {Q^2}^{(i_\tau)} \right)\,
             }{2\pi}
   \right)^{p_{\mathrm{NLO}}}
  \left[
    \left(
      W_{i_{y_1},i_{y_2},i_\tau}^{(p_{\mathrm{NLO}})(l)}
      + 2\pi  \beta_0 p_{\mathrm{LO}} \ln \xi_R^2
         \,
         W_{i_{y_1},i_{y_2},i_\tau}^{(p_{\mathrm{LO}})(l)}
    \right)
    F^{(l)}\left(x_1^{(i_{y_1})}, x_2^{(i_{y_1})}, \xi_F^2{Q^2}^{(i_\tau)}\right)
    \right. \\\left.
     - \ln \xi_F^2 \,
     W_{i_{y_1},i_{y_2},i_\tau}^{(p_{\mathrm{LO}})(l)}
     \left(
     F^{(l)}_{q_1 \to P_0\otimes q_1}\left(x_1^{(i_{y_1})}, x_2^{(i_{y_1})},
       \xi_F^2{Q^2}^{(i_\tau)}\right) +
     F^{(l)}_{q_2 \to P_0\otimes q_2}\left(x_1^{(i_{y_1})}, x_2^{(i_{y_1})},
       \xi_F^2{Q^2}^{(i_\tau)}\right)
     \right)
  \right] \,, \nonumber
\end{eqnarray}
where $F^{(l)}_{q_1 \to P_0\otimes q_1}$ is calculated as $F^{(l)}$,
but with $q_1$ replaced wtih $P_0 \otimes q_1$, and analogously for
$F^{(l)}_{q_2 \to P_0\otimes q_2}$.

\subsubsection{Technical implementation}

To test the scheme discussed above we use the NLO Monte Carlo program
NLOJET++ \cite{Nagy:2003tz,*Nagy:2001fj,*Nagy:2001xb} and the CTEQ6 PDFs \cite{cteq}.
The grid $W_{i_{y_1},i_{y_2},i_\tau}^{(p)(l)}$ of eq.~\ref{eq:WfinalxQ_twohadrons}
is filled in a NLOJET++ user module. This module has access to the event
weight and parton momenta and it is here that one specifies and calculates the physical
observables that are being studied (e.g. jet algorithm).

Having filled the grid we construct the cross-section
in a small standalone program which reads the weights from the grid
and multiplies them with an arbitrary  $\as$ and
PDF  according to eq.~\ref{eq:WfinalxQ_twohadrons}.
This program  runs very fast (in the order of seconds) and can be called
in a PDF fit.

The connection between these two programs is accomplished via a C++ class, which provides methods e.g.
for creating and optimising the grid, filling weight events and saving it to disk.
The classes are general enough to be extendable for the use with other NLO calculations.

The complete code for the NLOJET++ module, the C++ class and the standalone job is
available from the authors.
It is still in a development, testing and tuning stage, but help and more ideas are welcome.

\paragraph{The C++ class}
\label{sec:class}
The main data members of this class are the grids implemented as
arrays of three-dimensional ROOT histograms, with each grid point at the bin centers\footnote{
ROOT histograms are easy to implement, to represent and to manipulate.
They are therefore ideal in an early development phase.
An additional
advantage is the automatic file compression to save space.
The overhead of storing some empty bins is largely reduced by optimizing the
$x_1$, $x_2$ and $Q^2$ grid boundaries using the NLOJET++ program before final filling.
To avoid this residual overhead and to exploit certain symmetries in the grid, a special data
class (e.g. a sparse matrix) might be constructed in the future.}:
\begin{equation}
 {\rm TH3D[p][l][iobs](x_1,x_2,Q^2)},
\end{equation}
where the $l$ and $p$ are explained  in eq.~\ref{eq:WfinalxQ_twohadrons}
and $iobs$ denotes the observable bin, e.g. a given $P_T$ range\footnote{
For the moment we construct a grid for each initial state parton configuration. It will be easy
to merge the $qg$ and the $gq$ initial state parton configurations in one grid.
In addition, the weights for some of the initial state parton configurations
are symmetric in $x_1$ and $x_2$.
This could be exploited in future applications to further reduce the grid size.
}.

The  C++ class initialises, stores and fills the grid using the following main
methods:
\begin{itemize}
\item \emph{Default constructor:} Given the pre-defined kinematic regions of interest, it initializes the grid.
\item \emph{Optimizing method:} Since in some bins the weights will be zero over a large
                                     kinematic region in $x_1, x_2, Q^2$,
                                     the optimising method implements an automated procedure to adapt the
                                     grid boundaries for each observable bin.
                                     These boundaries are calculated in a first (short) run.
                                     In the present implementation, the optimised grid has a fixed
                                     number of grid points. Other choices, like
                                     a fixed grid spacing, might be implemented in the future.
\item \emph{Loading method:} Reads the saved weight grid from a ROOT file
\item \emph{Saving method:} Saves the complete grid to a ROOT file, which will be automatically compressed.
\end{itemize}

 \paragraph{The user module for NLOJET++}
The user module has to be adapted specifically to the exact definition of the cross-section calculation.
If a grid file already exists in the directory where NLOJET++ is started,
the grid is not started with the default constructor, but with the optimizing method (see \ref{sec:class}).
In this way the grid boundaries are optimised for each observable bin.
This is necessary to get very fine grid spacings without exceeding the computer memory.
The grid is filled at the same place where the standard NLOJET++ histograms are filled.
After a certain number of events, the grid is saved in a root-file and the calculation
is continued.

\paragraph{The standalone program for constructing the cross-section}
The standalone program calculates the cross-section in the following way:
\begin{enumerate}
\item Load the weight grid from the ROOT file
\item Initialize the PDF interface\footnote{We use the C++ wrapper
of the LHAPDF interface \cite{Whalley:2005nh}.}, load $q(x,Q^2)$ on a
helper PDF-grid (to increase the performance)
\item For each observable bin, loop over $i_{y_1},i_{y_2}, i_\tau, l, p$
      and calculate $F^{l}(x_1, x_2,Q^2)$ from the appropriate PDFs $q(x,Q^2)$,
      multiply  $\as$
      and the weights from the grid
      and sum over the initial state parton configuration $l$,
      according to eq.~\ref{eq:WfinalxQ_twohadrons}.
\end{enumerate}

\subsubsection{Results}
\label{sec:results}
We calculate the single inclusive jet cross-section as a function of the jet transverse
momentum ($P_T$) for jets within a rapidity of $|y|<0.5$. To define the jets we use
the seedless cone jet algorithm as implemented in NLOJET++ using the four-vector recombination
scheme and the midpoint algorithm. The cone radius has been put to $R=0.7$, the overlap fraction
was set to $f=0.5$. We set the renormalisation and factorization scale to $Q^2=P_{T,max}^2$,
where $P_{T,max}$ is the $P_T$ of the highest $P_T$ jet in the required rapidity region\footnote{
Note that beyond LO the $P_{T,max}$ will in general differ from the $P_T$
of the other jets, so when binning an inclusive jet cross section, the
$P_T$ of a given jet may not correspond to the renormalisation scale
chosen for the event as a whole. For this reason we shall need separate
grid dimensions for the jet $P_T$ and for the renormalisation scale. Only
in certain moment-space approaches \cite{Kosower:1997vj} has this
requirement so far been efficiently circumvented.}.

In our test runs, to be independent from statistical fluctuations
(which can be large in particular in the NLO case),
we fill in addition to the grid a reference histogram in the standard way
according to eq.~\ref{eq:normalint}.

The choice of the grid architecture depends on the required accuracy, on the exact cross-section definition
and on the available computer resources. Here, we will just sketch the influence of the grid architecture
and the interpolation method on the final result. We will investigate an example where we
calculate the inclusive jet cross-section in $N_{\mathrm{obs}} = 100$ bins in the kinematic range
$100\, \leq P_T \leq 5000\,\mathrm{GeV}$.
In future applications this can serve as guideline for a user to adapt the grid method to his/her specific problem.
We believe that the code is transparent and flexible enough to adapt to many applications.

As reference for comparisons of different grid architectures and interpolation methods
we use the following:
\begin{itemize}
\item \emph{Grid spacing in $y(x)$:} $10^{-5} \leq x_1, x_2 \leq 1.0$ with $N_y=30$
\item \emph{Grid spacing in $\tau(Q^2)$:} $100\,\mathrm{GeV} \leq Q \leq 5000\,\mathrm{GeV}$ with $N_\tau=30$
\item \emph{Order of interpolation:} $n_y=3,\, n_\tau=3$
\end{itemize}
The grid boundaries correspond to the user setting for the first run which determines
the grid boundaries for each observable bin.
In the following we call this grid architecture $30^2$x$30$x$100 (3,3)$.
Such a grid takes about $300$~{\rm Mbyte}
of computer memory. The root-file where the grid is stored has about $50$~{\rm Mbyte}.

\begin{figure}[htbp]
\centering
\includegraphics[width=0.49\textwidth]{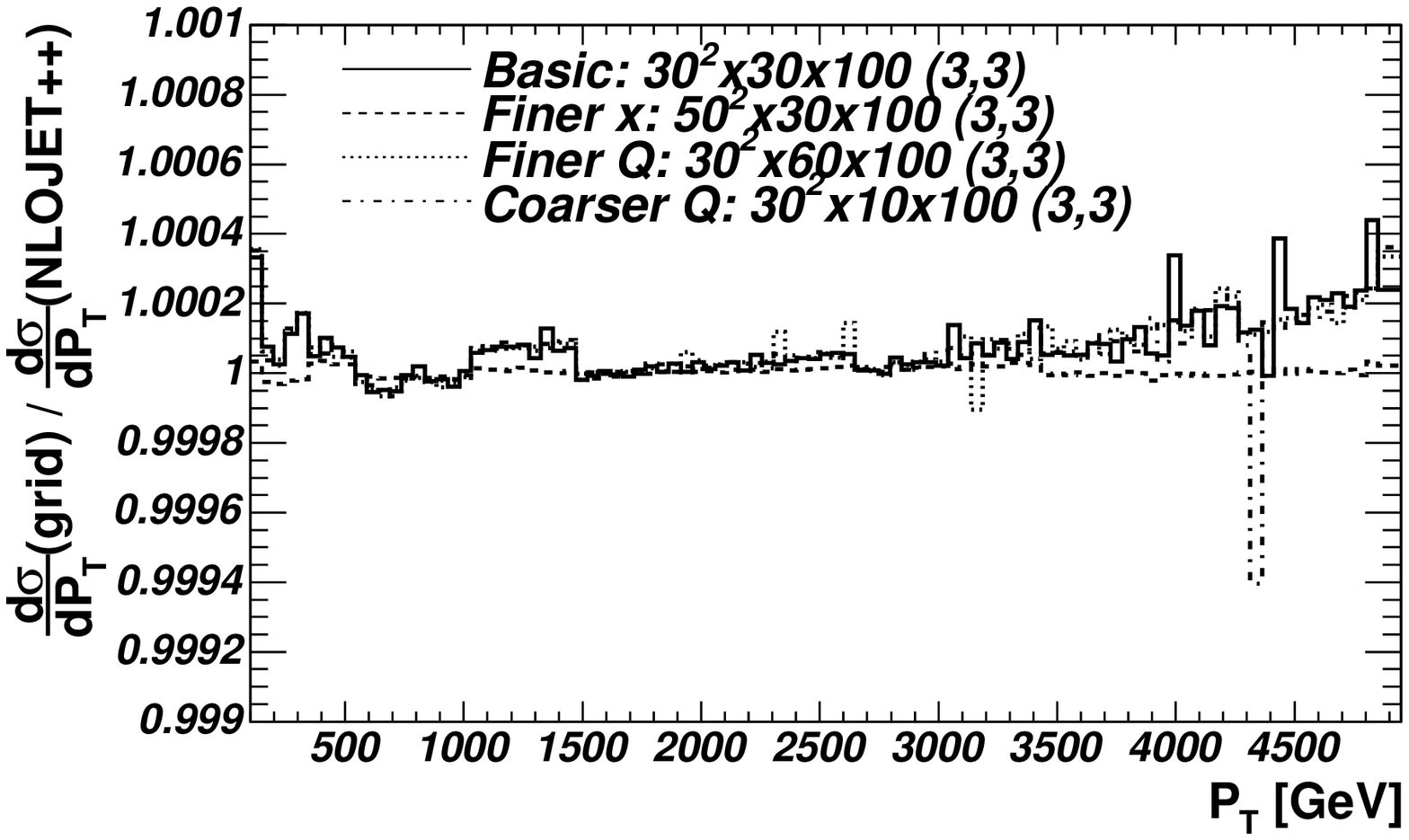}
\includegraphics[width=0.49\textwidth]{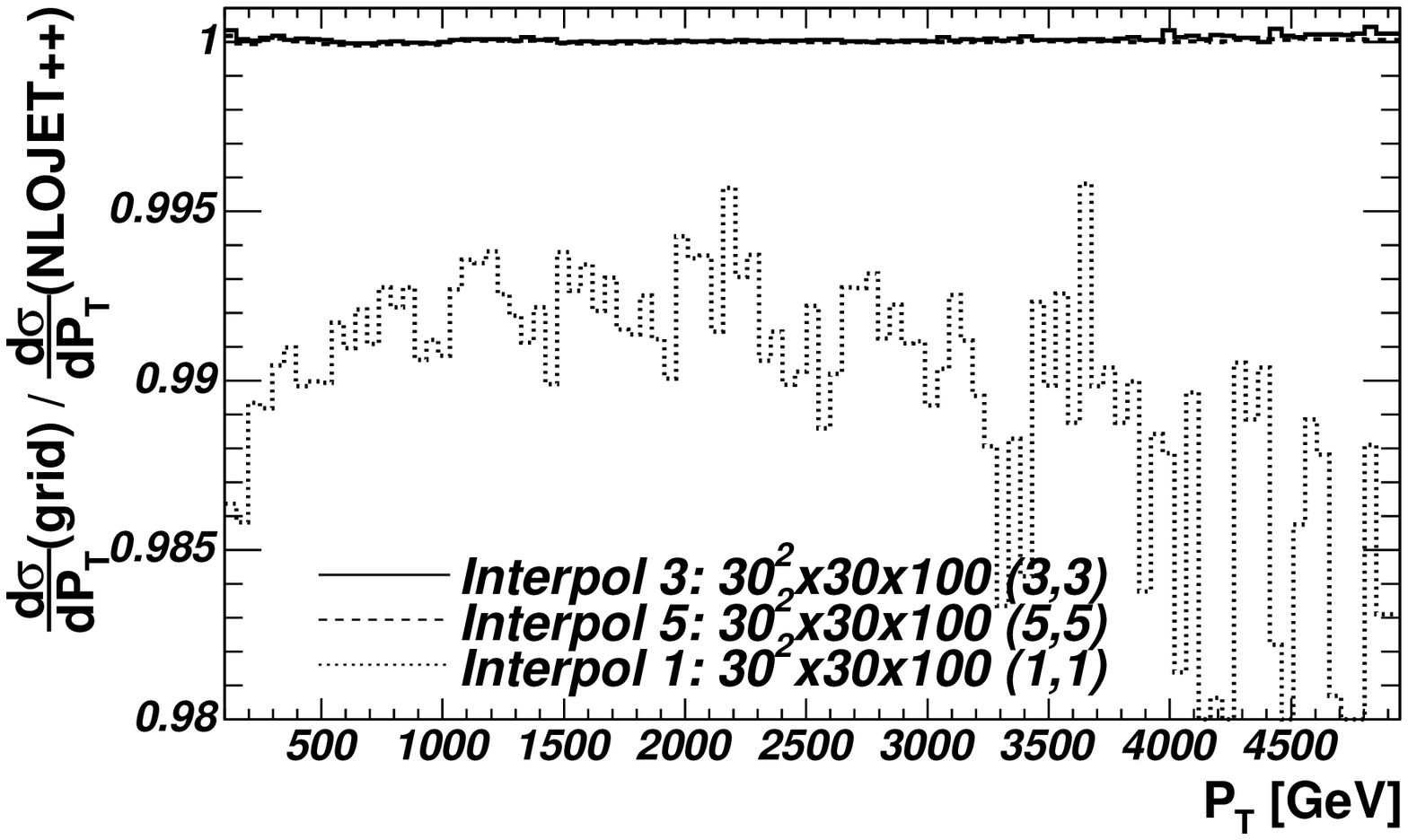}
\includegraphics[width=0.49\textwidth]{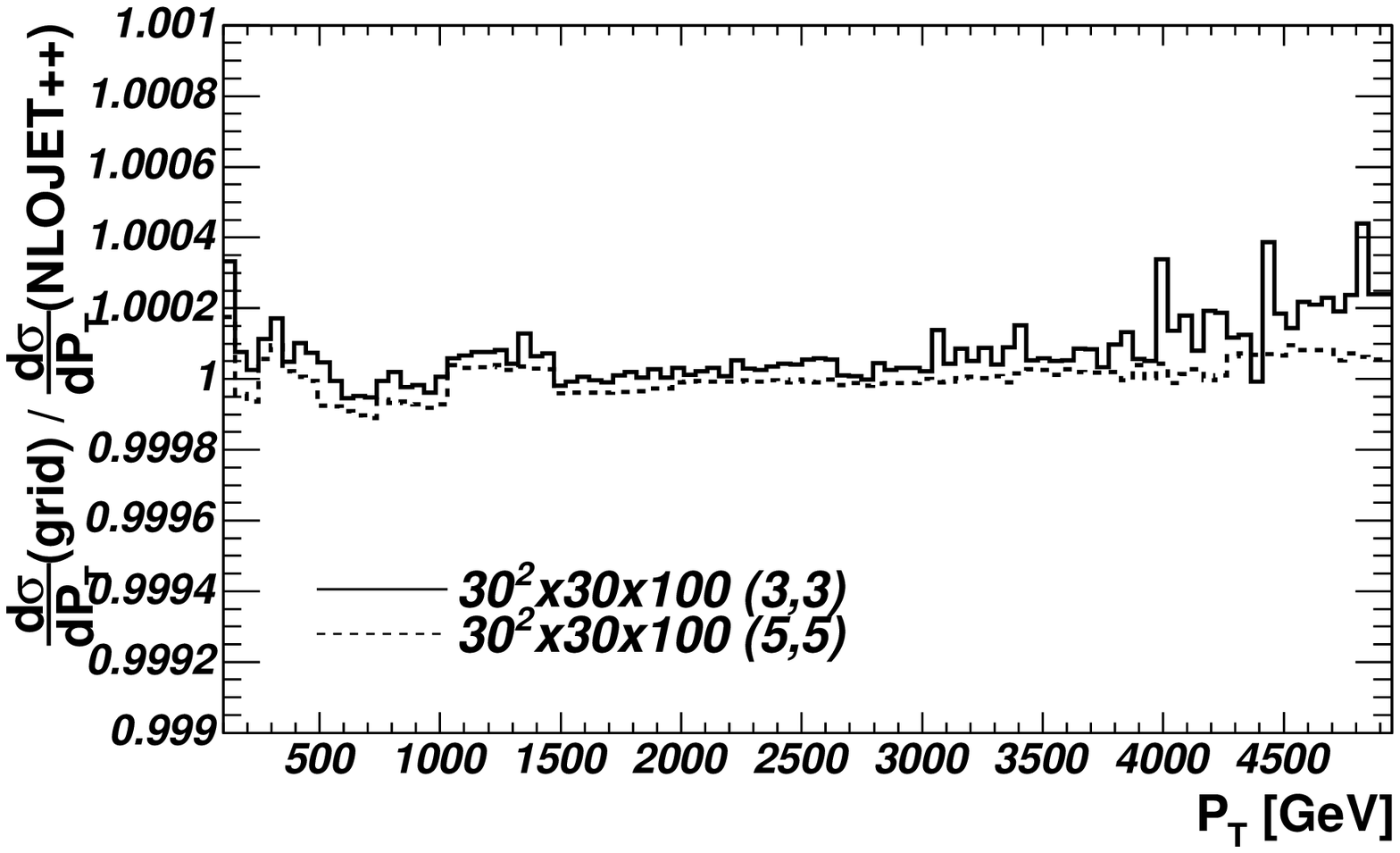}
\includegraphics[width=0.49\textwidth]{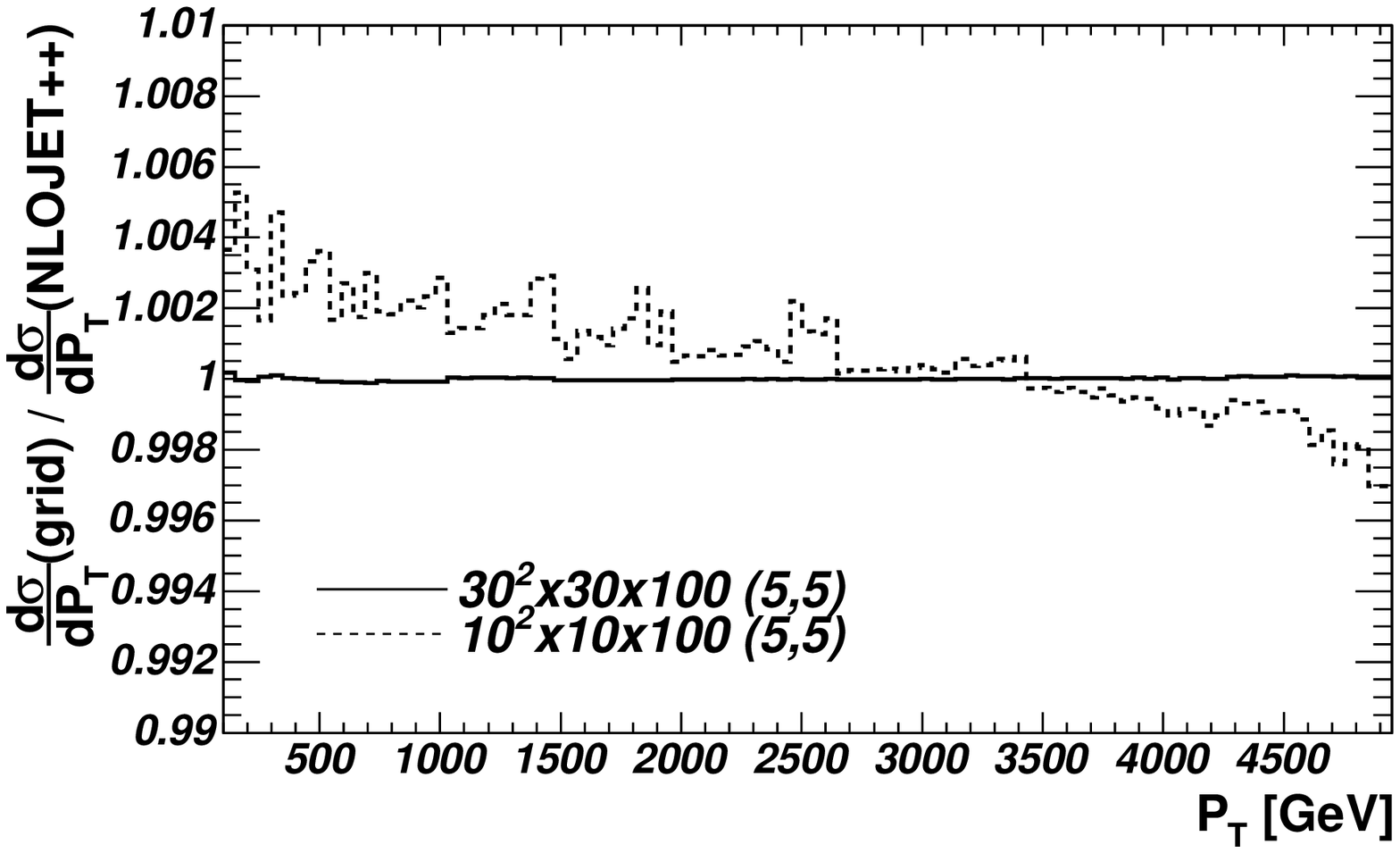}
\begin{picture}(0,0)
\put( -450,0){c)}
\put( -210,0){d)}
\put( -450,135){a)}
\put( -210,135){b)}
\end{picture}
\caption{Ratio between the single inclusive jet cross-section with $100$ $P_T$ bins calculated
with the grid technique and the reference cross-section calculated in the standard way.
Shown are the standard
grid, grids with finer $x$ and $Q^2$ sampling (a) with interpolation
of order $1$, $3$ and $5$ (b) (and on a finer scale in c)) and
a small grid (d).
}
\label{fig:101obsbins}
\end{figure}

The result is shown in Fig.~\ref{fig:101obsbins}a).
The reference cross-section is reproduced everywhere to within $0.05\%$. The typical precision is about $0.01\%$.
At low and high $P_T$ there is a positive bias of about $0.04\%$.
Also shown in Fig.~\ref{fig:101obsbins}a) are the results obtained with different grid architectures.
For a finer  $x$ grid ($50^2$x$30$x$100 (3,3)$) the accuracy is further improved (within $0.005\%$)
and there is no bias.
A finer ($30^2$x$60$x$100 (3,3)$) as well as a coarser  ($30^2$x$10$x$100 (3,3)$) binning in $Q^2$
does not improve the precision.

Fig.~\ref{fig:101obsbins}b) and Fig.~\ref{fig:101obsbins}c)
show for the grid ($30^2$x$30$x$100$) different interpolation methods.
With an interpolation of order $n=5$ the precision is $0.01\%$ and the bias at low and high $P_T$
observed for the $n=3$ interpolation disappears. The result is similar to the one
obtained with finer $x$-points. Thus by increasing the interpolation order
the grid can be kept smaller.
An order $n=1$ interpolation gives a systematic negative bias of about $1\%$
becoming even larger towards high $P_T$.

Depending on the available computer resources and the specific problem, the user
will have to choose a proper grid architecture.
In this context, it is interesting that a very small grid
$10^2$x$10$x$100 (5,5)$ that takes only about  $10$~{\rm Mbyte} computer memory
reaches still a precision of $0.5\%$, if an interpolation of order $n=5$ is used
(see Fig.~\ref{fig:101obsbins}d)).

We have developed a technique to store the perturbative coefficients calculated by an NLO
Monte Carlo program on a grid allowing for  a-posteriori inclusion of an arbitrary parton
density function (PDF) set. We extended a technique that was already successfully used to analyse
HERA data to the more demanding case of proton-proton collisions at LHC energies.

The technique can be used to constrain PDF uncertainties, e.g. at high momentum transfers,
from data that will be measured at LHC and allows the consistent
inclusion of final state observables
in global QCD analyses. This will help increase the sensitivity of LHC
to find new physics as deviations from the Standard Model predictions.

Even for the large kinematic range for the parton momentum fractions $x_1$ and $x_2$ and
of the squared momentum transfer $Q^2$ accessible at LHC, grids of moderate size
seem to be sufficient. The single inclusive jet cross-section in the central region $|y|<0.5$
can be calculated with a precision of $0.01\%$ in a realistic example with $100$ bins
in the transverse jet energy range $100\, \leq P_T \leq 5000\,\mathrm{GeV}$.
In this example, the grid  occupies about $300$~{\rm Mbyte} computer memory.
With smaller grids of order $10$~{\rm Mbyte}
the reachable accuracy is still $0.5\%$. This is probably sufficient for all practical
applications.

%% file: evpfits.tex
\section{GLAP evolution and parton fits
\protect\footnote{Subsection coordinators: A.~Glazov, S.~Moch}}
\label{section:evpfits}
\subsection{Introduction}
The high-precision data from HERA and the anticipated data from LHC 
open the possibility for a precise determination of parton
distributions. This, however, requires
an improvement in the theoretical description of 
DIS and hard hadronic
scattering processes, as well as an 
improvement of the  techniques used to
extract parton distributions from the data.

The determination of perturbative QCD corrections has undergone
substantial progress recently.
The key ingredient of a complete next-to-next-to-leading order (NNLO) prediction in perturbative QCD 
are the recently calculated three-loop splitting functions which govern the 
scale dependence of PDFs. 
Extensions in the accuracy of the perturbative predictions yet beyond NNLO are 
given by the three-loop coefficient functions for $F_2$, while the coefficient functions 
for $F_L$ at this order are actually required to complete the NNLO predictions.
Section~\ref{sec:nnloprecision} briefly discusses the recent results 
and their phenomenological implications.
Certain mathematical aspects, which are important in the calculation of 
higher order corrections in massless QCD are presented in section~\ref{sec:mellinmath1}.
In particular, algebraic relations in Mellin-$N$ space are pointed out, which are 
of importance for harmonic sums, harmonic polylogarithms and multiple
$\zeta$-values. 

These  calculation of the PDF evolution to NNLO 
in perturbative QCD are used in 
section~\ref{sec:pdfevolution} to provide an update and extension 
of a set of benchmark tables for the evolution of parton distributions of hadrons.
These benchmark  tables were first presented in the report of the QCD/SM working group at the 
2001 Les Houches workshop, but based on approximate 
NNLO splitting functions, which are superseded by the exact results
which are now available.
In addition, section~\ref{sec:pdfevolution} now includes also reference tables for 
the case of polarized PDF evolution.

Whereas in principle the  $x$-shapes of PDFs at low scales can be
determined from first principles using non-perturbative methods, in
practice at present this is only possible using models (briefly touched in 
in section~\ref{sec:gi-pdf-xshape}). Therefore, an accurate determination 
of PDFs requires a global QCD fit to the data, which is the subject of 
sections~\ref{sec:alek}--\ref{sec:cteq}.

Section~\ref{sec:alek} discusses in particular the impact on parton fits
of NNLO corrections on the one hand, and of the inclusion of Drell-Yan data 
and future LHC data on the other hand.
It then presents values for a benchmark fit together with a table of
correlation coefficients for 
the parameter obtained in the fit. 
This benchmark fit is then re-examined in sec.~\ref{sec:thorne}, 
along with a comparison between PDFs 
and the associated uncertainty obtained using the approaches of Alekhin and the 
MRST group. The differences between these benchmark partons and the actual
global fit partons are also discussed, and used to explore
complications inherent in extracting PDFs with uncertainties.
Finally, in section~\ref{sec:cteq} the stability of PDF determinations in NLO global 
analyses is re-investigated and the results of the CTEQ PDF group on this
issue are summarized.

An alternative approach to a completely bias-free parameterization of PDFs is presented 
in section~\ref{sec:nnpdf}. 
There, a neural network approach to global fits
of parton distribution functions is introduced and 
work on unbiased parameterizations of deep-inelastic 
structure functions with 
faithful estimation of their uncertainties is reviewed together with a summary of 
the current status of neural network parton distribution fits.

%% file: nnloprecision.tex
\subsection{Precision Predictions for Deep-Inelastic Scattering
\protect\footnote{Contributing authors: S.~Moch, J.A.M.~Vermaseren, A.~Vogt}}
\label{sec:nnloprecision}

With high-precision data from HERA and in view of the 
outstanding importance of hard scattering processes at the LHC, 
a quantitative understanding of deep-inelastic processes is indispensable,
necessitating calculations beyond the standard next-to-leading order of 
perturbative QCD. 

In this contribution we review recent results for the complete 
next-to-next-to-leading order (NNLO, N$^2$LO) approximation 
of massless perturbative QCD for the structure functions 
$F_{\,1}$, $F_{\,2}$, $F_{\,3}$ and $F_L$ in DIS. 
These are based on the second-order coefficient functions~\cite{%
vanNeerven:1991nn,Zijlstra:1991qc,Zijlstra:1992kj,Zijlstra:1992qd,Moch:1999eb},
the three-loop splitting functions which govern the evolution 
of unpolarized parton distributions of hadrons~\cite{Moch:2004pa,Vogt:2004mw} 
and the three-loop coefficient functions for  
$F_L = F_{\,2} - 2x F_1$ in electromagnetic (photon-exchange) DIS~\cite{%
Moch:2004xu,Vermaseren:2005qc}. 
Moreover we discuss partial N$^3$LO results for $F_2$, based on the 
corresponding three-loop coefficient functions also presented in 
Ref.~\cite{Vermaseren:2005qc}.
For the splitting functions $P$ and coefficient functions $C$ 
we employ the convention 
\begin{eqnarray}
  \label{eq:exp-convention}
  P(\as) \, = \, \sum_{n=0}\, \left(\frac{\as}{4\pi}\right)^{n+1} P^{(n)} \, ,
\quad\quad\quad
  C(\as) \, = \, \sum_{n=0}\, \left(\frac{\as}{4\pi}\right)^{n} C^{(n)} 
\end{eqnarray}
for the expansion in the running coupling constant $\as$. 
For the longitudinal structure function $F_L$ the third-order 
corrections are required to complete the NNLO predictions, 
since the leading contribution to the coefficient function $C_{L}$ is of 
first order in the strong coupling constant $\as$. 

In the following we briefly display selected results to demonstrate the 
quality of precision predictions for DIS and their effect on the evolution. 
The exact (analytical) results to third order for the quantities in Eq.~(\ref{eq:exp-convention}) 
are too lengthy, about ${\cal O}(100)$ pages in normalsize fonts and will not be reproduced here.
Also the method of calculation is well documented in the literature~\cite{%
Moch:1999eb,Moch:2002sn,Moch:2004pa,Vogt:2004mw,Vermaseren:2005qc,Moch:2004sf}.
In particular, it proceeds via the Mellin transforms of the functions 
of the Bjorken variable $x$,
\begin{eqnarray}
\label{eq:mellintrafo}
  A(N) &=& \int\limits_0^1 \! dx \: x^{N-1} A(x) \, .
\end{eqnarray}
Selected mathematical aspects of Mellin transforms are discussed in 
section~\ref{sec:mellinmath1}.

\subsubsection{Parton evolution}

The well-known $2n_f-1$ scalar non-singlet and $2 \times 2$ 
singlet evolution equations for $n_f$ flavors read
\begin{equation}
  \label{eq:evolns}
  \frac{d}{d \ln \mu_f^{\,2}} \: q_{\rm ns}^{\: i} \: =\: 
  P_{\rm ns}^{\,i} \,\otimes\, q_{\rm ns}^{\, i}\, ,
\,\,\,\,\,\,\,\,\,\,\,\,\,\,\,\,\, i=\pm,\mbox{v} \: ,
\end{equation}
for the quark flavor asymmetries $q_{\rm ns}^{\pm}$ and 
the valence distribution $q_{\rm ns}^{\rm v}$, and 
\begin{equation}
  \label{eq:evols}
  \frac{d}{d \ln\mu_f^{\,2}}
  \left( \begin{array}{c} \! q_{\rm s}^{} \! \\ g  \end{array} \right) 
  \: = \: \left( \begin{array}{cc} P_{\rm qq} & P_{\rm qg} \\ 
  P_{\rm gq} & P_{\rm gg} \end{array} \right) \otimes 
  \left( \begin{array}{c} \!q_{\rm s}^{}\! \\ g  \end{array} \right) 
\end{equation}
for the singlet quark distribution $q_{\rm s}^{}$ and the gluon distribution
$g$, respectively. 
Eqs.~(\ref{eq:evolns}) and~(\ref{eq:evols}) 
are governed by three independent types of non-singlet splitting functions, 
and by the $2 \times 2$ matrix of singlet splitting functions.
Here $\otimes$ stands for the Mellin convolution.
We note that benchmark numerical solutions to NNLO accuracy 
of Eqs.~(\ref{eq:evolns}) and~(\ref{eq:evols}) for a specific set of 
input distributions are given in section~\ref{sec:pdfevolution}. 
Phenomenological QCD fits of parton distributions in data analyses 
are extensively discussed in sections~\ref{sec:alek}--\ref{sec:cteq}. 
An approach based on neural networks is described in section~\ref{sec:nnpdf}.

\begin{figure}[ht]
\begin{center}
\includegraphics[width=10.5cm,angle=0]{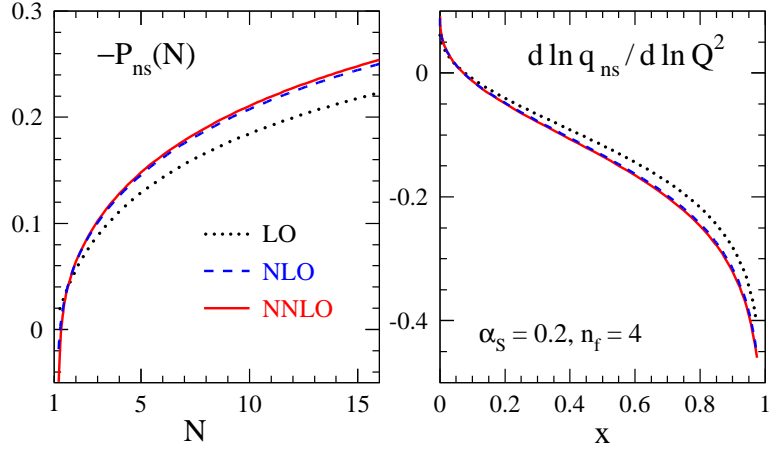}
\end{center}
\vspace*{-5mm}
\caption{On the left we show the perturbative expansion of 
 $P_{\,\rm ns}^{\, \rm v}(N)$, and 
 on the right the resulting perturbative expansion of the logarithmic scale derivative 
 $\,d \ln q_{\,\rm ns}/ d\ln \mu_f^{\,2}\,$ is displayed for a model input. 
 See the text for 
 details.}
\label{fig:gamdqns}
\end{figure}

Let us start the illustration of the precision predictions 
by looking at the parton evolution and at large Mellin-$N$ 
(large Bjorken-$x$) behavior.
Fig.~\ref{fig:gamdqns} shows the stability of the 
perturbative expansion which is very benign and indicates, for $\as \lsim 0.2$,
corrections of less than 1\% beyond NNLO.
On the left we show the results for the perturbative expansion 
of $P_{\rm ns}$ in Mellin space, 
cf. Eqs.~(\ref{eq:exp-convention}), (\ref{eq:mellintrafo}).
We employ four active flavors, $n_f =4$, and 
an order-independent value for the strong coupling constant,
\begin{equation}
  \label{eq:asvalue}
  \as (\mu_{0}^{\,2}) \: = \: 0.2 \:\: ,
\end{equation}
which corresponds to $\mu_{0}^{\,2} \,\simeq\, 25\ldots 50$ 
GeV$^2$ for $\as (M_Z^{\, 2}) = 0.114 \ldots 0.120$ beyond the leading order. 
On the right of Fig.~\ref{fig:gamdqns} 
the perturbative expansion of the logarithmic derivative, 
cf. Eqs.~(\ref{eq:exp-convention}), (\ref{eq:evolns}), is illustrated
at the standard choice $\mu_r = \mu_f$ of the renormalization scale. 
We use the schematic, but characteristic model distribution, 
\begin{eqnarray}
  \label{eq:shapens}
  x q_{\,\rm ns}(x,\mu_{0}^{\,2}) & = & x^{\, 0.5} (1-x)^3 \, .
\end{eqnarray}
The normalization of $q_{\,\rm ns}$
is irrelevant at this point, as we consider the logarithmic scale derivative only.

\begin{figure}[ht]
\begin{center}
\includegraphics[width=10.5cm,angle=0]{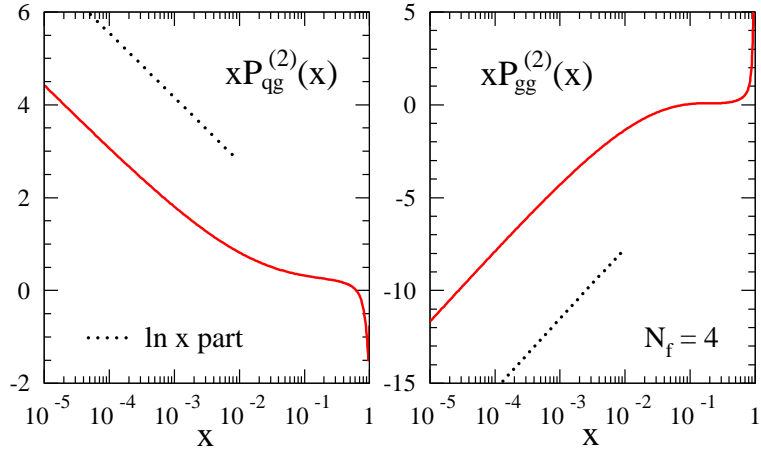}
\end{center}
\vspace*{-5mm}
\caption{The three-loop gluon-quark (left) and gluon-gluon (right) 
splitting functions together with the leading small-$x$ contribution (dotted line).
}
\label{fig:p2ignf4s}
\end{figure}

Next, let us focus on the three-loop splitting functions at 
small momentum fractions $x$, where the splitting functions 
$P_{\,\rm ig}$ in the lower row of the $2 \times 2$ matrix 
in Eq.~(\ref{eq:evols}), representing $\rm g \!\rightarrow\! i$ splittings,
 are most important.
In Fig.~\ref{fig:p2ignf4s} we show, again for $n_f =4$, the three-loop splitting functions 
$P^{(2)}_{\,\rm qg}$ and  $P^{(2)}_{\,\rm gg}$ together 
with the leading small-$x$ term indicated separately for $x<0.01$.
In the present singlet case the 
leading logarithmic small-$x$ limits $\sim x^{\,-1} \ln x$ of 
Refs.~\cite{Catani:1994sq,Fadin:1998py} are confirmed 
together with the general structure of the BFKL limit~\cite{%
  Kuraev:1977fs,Balitsky:1978ic,Jaroszewicz:1982gr}.
The same holds for the leading small-$x$ terms $\ln^4 x$ 
in the non-singlet sector~\cite{Kirschner:1983di,Blumlein:1996jp}, 
with the qualification that 
a new, unpredicted leading logarithmic contribution is found for the 
color factor $d^{abc}d_{abc}$ entering at three loops for the first time.

It is obvious  from Fig.~\ref{fig:p2ignf4s} 
(see also Refs.~\cite{Moch:1999eb,Moch:2004pa,Vogt:2004mw,Moch:2004sf}) 
that the leading $\:x\! \rightarrow 0$-terms alone are insufficient 
for collider phenomenology at HERA or the LHC as they 
do not provide good approximations of the full results 
at experimentally relevant small values of~$x$.
Resummation of the small-$x$ terms 
and various phenomenological improvements 
are discussed in detail in section~\ref{sec:pdf,res}.

\begin{figure}[ht]
\begin{center}
\includegraphics[width=10.5cm,angle=0]{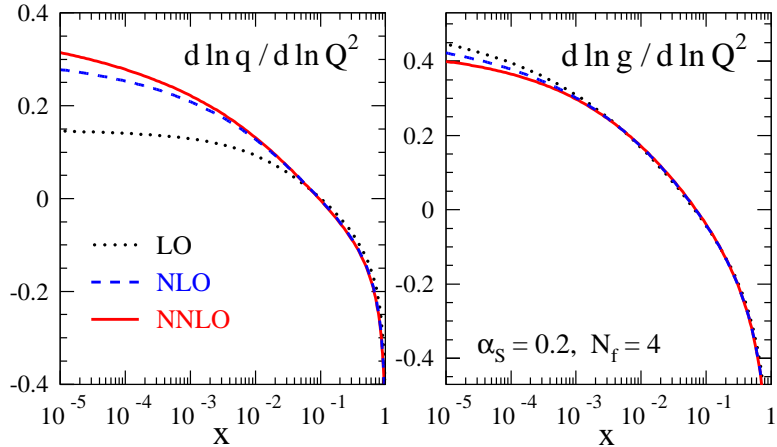}
\end{center}
\vspace*{-5mm}
\caption{The perturbative expansion of the scale derivatives~(\ref{eq:evols}) 
of the singlet distributions~(\ref{eq:shapes}).
}
\label{fig:dqsgledn}
\end{figure}

In the same limit of small $x$, it is instructive to look 
at the evolution of parton distributions.
Again, we choose the reference scale of Eq.~(\ref{eq:asvalue}), 
$n_f=4$ and the sufficiently realistic model distributions
\begin{eqnarray}
  \label{eq:shapes}
  xq_{\rm s}(x,\mu_{0}^{\,2}) &\! = \! &
  0.6\: x^{\, -0.3} (1-x)^{3.5}\, (1 + 5.0\: x^{\, 0.8\,})  \nonumber \\
  xg (x,\mu_{0}^{\,2})\:\: &\! = \! &
  1.6\: x^{\, -0.3} (1-x)^{4.5}\, (1 - 0.6\: x^{\, 0.3\,})
\end{eqnarray}
irrespective of the order of the expansion to facilitate 
direct comparisons of the various contributions.
Of course, this order-independence 
does not hold for actual data-fitted parton distributions like those 
in sections~\ref{sec:alek}--\ref{sec:cteq}.  
In Fig.~\ref{fig:dqsgledn} we display 
the perturbative expansion of the scale derivative 
for the singlet quark and gluon densities 
at $\,\mu_f^{\,2} \: =\: \mu_0^{\,2\,}$ for the 
initial conditions specified in Eqs.~(\ref{eq:asvalue}) and (\ref{eq:shapes}).
For the singlet quark distribution the total NNLO corrections, 
while reaching 10\% at $x = 10^{\,-4}$, remain smaller than the NLO results by a 
factor of eight or more over the full $x$-range.
For the gluon distribution already the NLO corrections are small and the 
NNLO contribution amounts to only 3\% for $x$ as low as $10^{\,-4}$.
Thus, we see in Fig.~\ref{fig:dqsgledn} that the perturbative 
expansion is very stable. 
It appears to converge rapidly at $x >~ 10^{-3}$, while relatively 
large third-order corrections are found for very small momenta $\:x \lsim 10^{-4}$.

\subsubsection{Coefficient functions}

While the previous considerations were addressing the evolution 
of parton distributions, we now turn to the further 
improvements of precision predictions due to the full third-order coefficient 
functions for the structure functions $F_2$ and $F_L$ in electromagnetic 
DIS~\cite{Moch:2004xu,Vermaseren:2005qc}. The results
for $F_L$ complete the NNLO description of 
unpolarized electromagnetic DIS, and 
the third-order coefficient functions for $F_{\,2}$ form, at not too 
small values of the Bjorken variable $x$, the dominant part of the 
next-to-next-to-next-to-leading order (N$^3$LO) corrections.
Thus, they facilitate improved determinations of the strong coupling $\as$ 
from scaling violations.

\begin{figure}[ht]
\begin{center}
\includegraphics[width=10.5cm,angle=0]{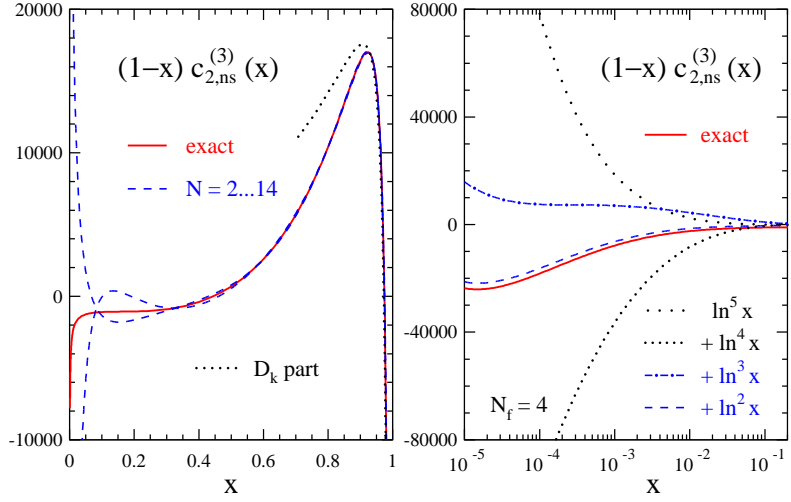}
\end{center}
\vspace*{-5mm}
\caption{
 The three-loop non-singlet coefficient function $c_{\,2,\rm ns}^{\,(3)}(x)$ 
 in the large-$x$ (left) and the small-$x$ (right) region, multiplied by 
 $(1\!-\!x)$ for display purposes.
 }
\label{fig:c2ns3f4c}
\end{figure}

Let us start with the three-loop coefficient functions for $F_2$ in the 
non-singlet case.
In Fig.~\ref{fig:c2ns3f4c} we display the three-loop non-singlet coefficient function 
$c_{\,2,\rm ns}^{\,(3)}(x)$ for $n_f=4$ flavors.
We also show the soft-gluon enhanced terms ${\cal D}_k$ dominating the
large-$x$ limit, 
\begin{eqnarray}
  \label{eq:softgluonDk}
  {\cal D}_k &=& {\ln^{\,2k-1}(1-x) \over (1-x)_+} \, ,
\end{eqnarray}
and the small-$x$ approximations obtained by successively including enhanced 
logarithms $\ln^k x $. 
However the latter are insufficient for an accurate description of the exact result.
The dashed band in Fig.~\ref{fig:c2ns3f4c} shows the uncertainty of previous estimates~\cite{%
vanNeerven:2001pe} mainly based on the calculation of fixed Mellin 
moments~\cite{Larin:1994vu,Larin:1997wd,Retey:2000nq}.
For a detailed discussion of the soft-gluon resummation of the
the ${\cal D}_k$ terms, we refer to section~\ref{sec:pdf,res}. 

\begin{figure}[ht]
\begin{center}
\includegraphics[width=4.5cm,angle=0]{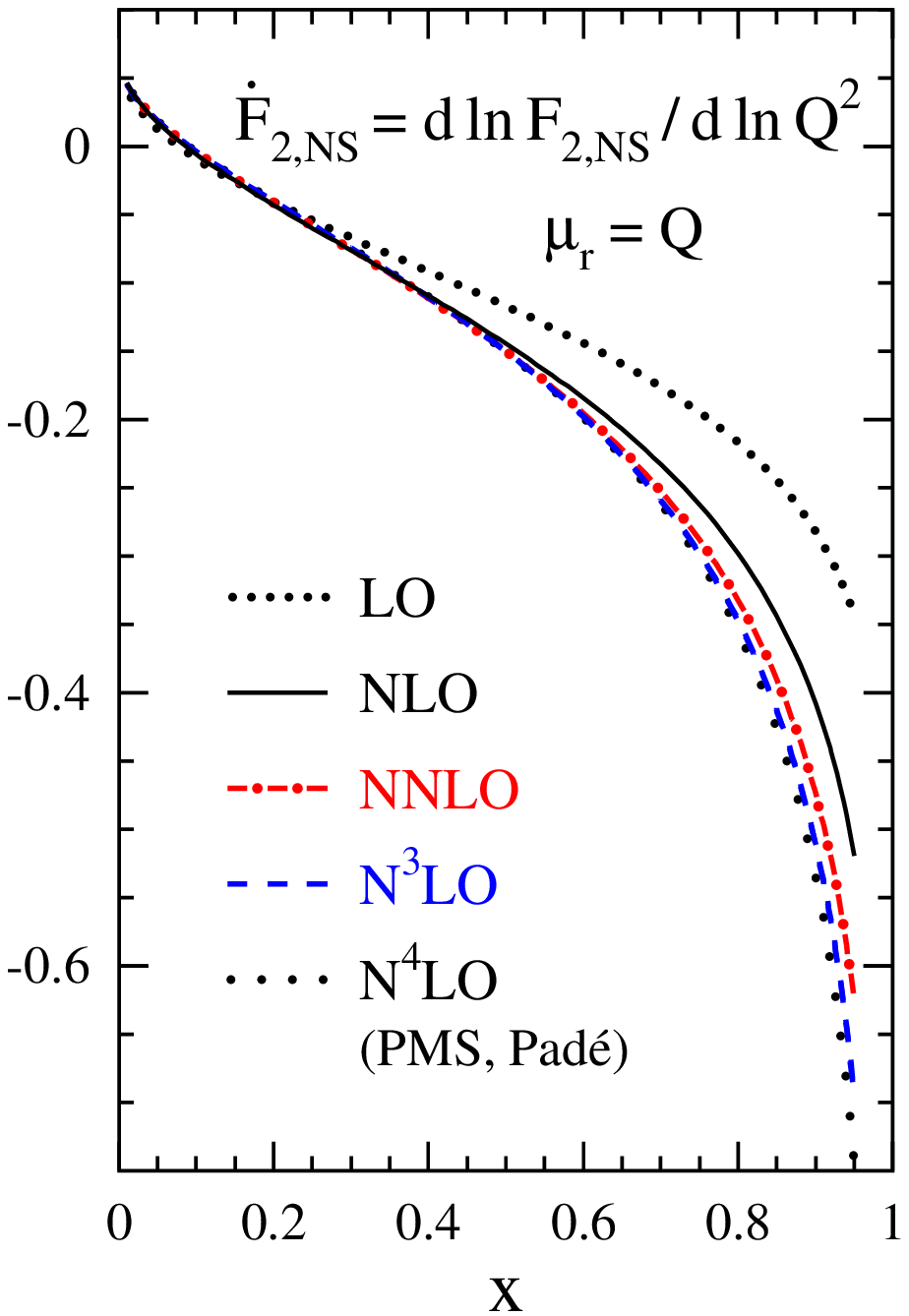}
\includegraphics[width=4.5cm,angle=0]{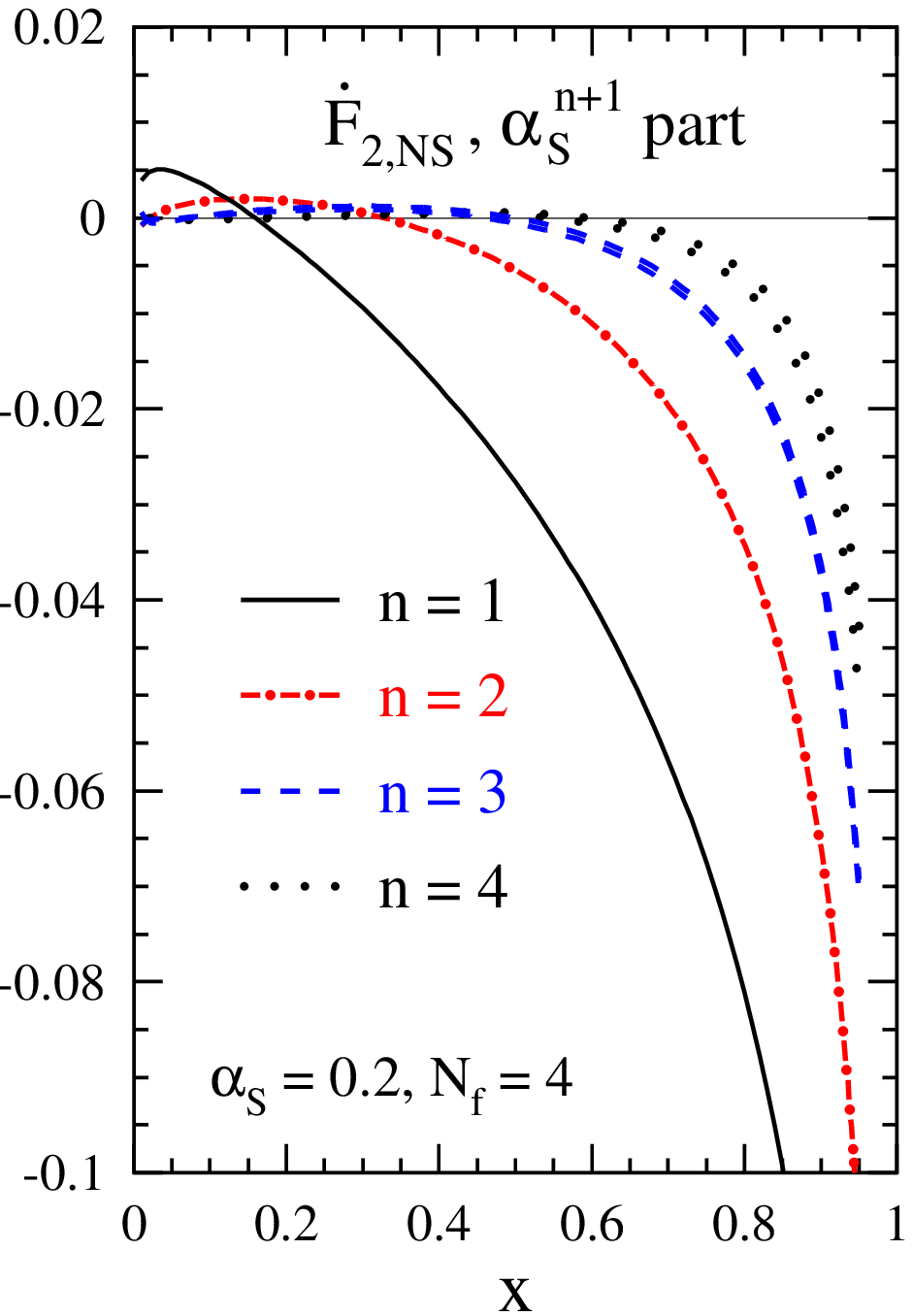}
\includegraphics[width=4.5cm,angle=0]{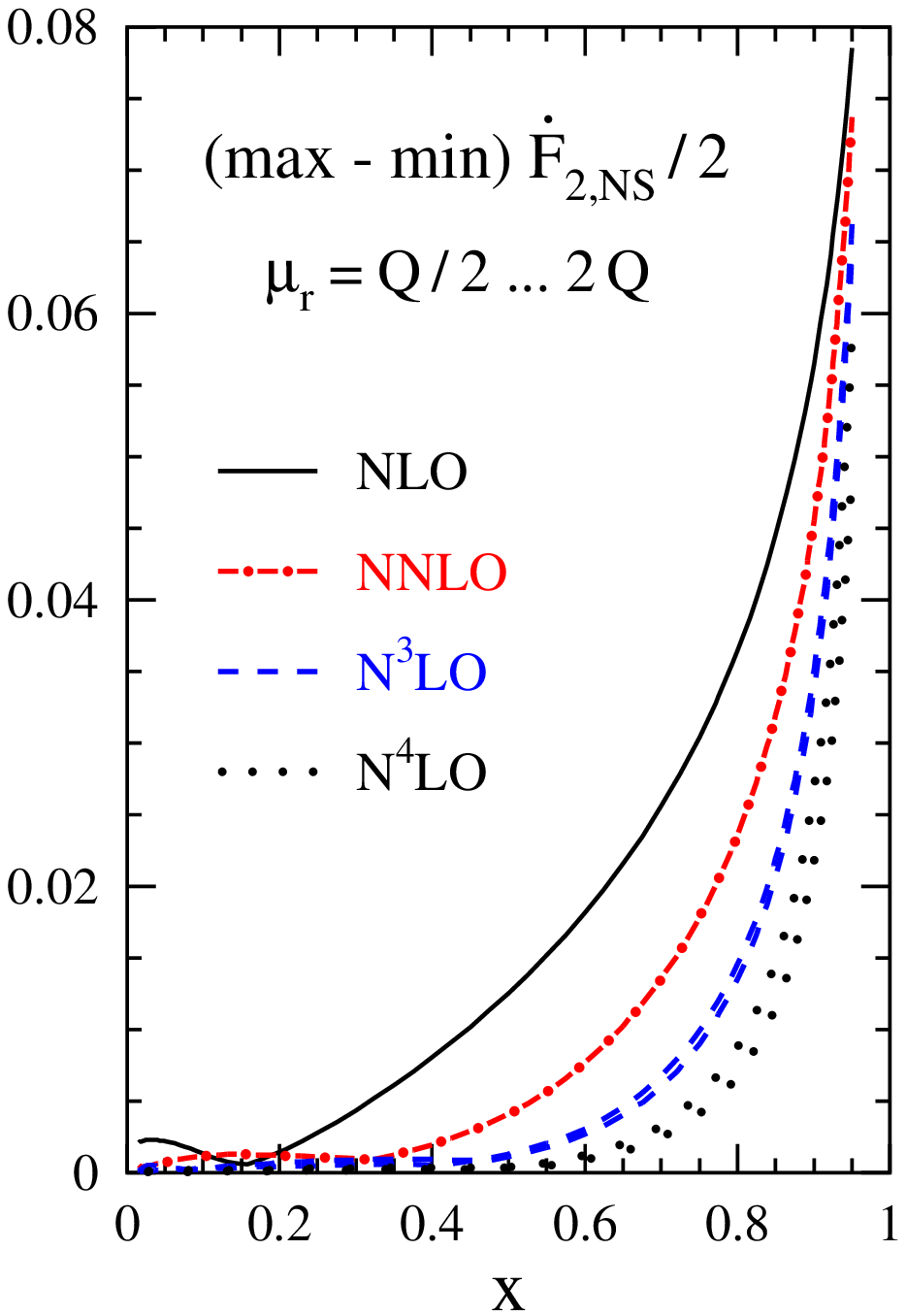}
\end{center}
\vspace*{-5mm}
\caption{
 The perturbative expansion of the logarithmic scale derivative 
 of the non-singlet structure function $F_{2,\rm ns}$. The results up to NNLO
 are exact, while those at N$^3$LO are very good approximations. The N$^4$LO 
 corrections have been estimated by various methods.
 }
\label{fig:f2plots}
\end{figure}

Building on the coefficient functions, it is interesting to study 
the perturbative expansion of the logarithmic scale derivative for the 
non-singlet structure function $F_{2,\rm ns}$.
To that end we use in Fig.~\ref{fig:f2plots} again the input shape 
Eq.~(\ref{eq:shapens}) (this time for $F_{2,\rm ns}$ itself) irrespective of the order of the expansion, 
$n_f=4$ flavors and the reference scale of Eq.~(\ref{eq:asvalue}). 
The N$^4$LO approximation based on Pad\'e summations of the perturbation series 
can be expected to correctly indicate at least the rough size of the four-loop corrections, 
see Ref.~\cite{Vermaseren:2005qc} for details.
From Fig.~\ref{fig:f2plots} we see that the three-loop results for $F_{\,2}$ 
can be employed to effectively extend the main part of DIS analyses to the 
N$^3$LO at $x > 10^{-2}$ where the effect of the unknown fourth-order splitting functions 
is expected to be very small. 
This has, for example, the potential for a `gold-plated' determination of $\as(M_Z)$ 
with an error of less than 1\% from the truncation of the perturbation series.
On the right hand side of Fig.~\ref{fig:f2plots}
the scale uncertainty which is conventionally estimated by
\begin{eqnarray}
  \label{eq:screl}
  \Delta \dot{f} & \equiv &
  {1 \over 2}\, 
  \left( \max\, [ \dot{f}(x,\mu_r^2 )] - \min\, [\dot{f}(x,\mu_r^2)] \right)\, ,
\end{eqnarray}
is plotted, where the scale varies $\mu_r \in [Q/2, 2Q]$.

\begin{figure}[ht]
\begin{center}
\includegraphics[width=10.5cm,angle=0]{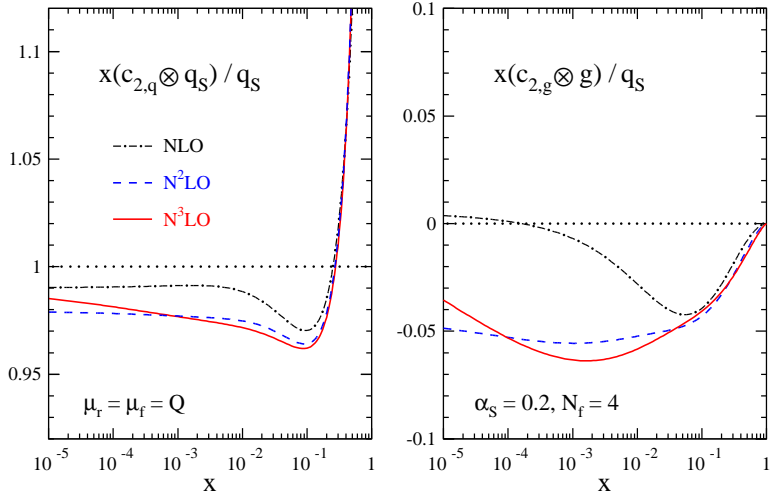}
\end{center}
\vspace*{-5mm}
\caption{The perturbative expansion up to three loops (N$^3$LO) of the quark 
 (left) and gluon (right) contributions to singlet structure function
 $F_{2}$. }
\label{fig:c2sgcnv1}
\end{figure}

In the singlet case, we can study the quark and gluon contributions to the structure function 
$F_2$. In Fig.~\ref{fig:c2sgcnv1} we plot 
the perturbative expansion up to N$^3$LO of the quark 
and gluon contributions to structure function 
$F_{2,\rm s}$ at the scale (\ref{eq:asvalue}) using the 
distributions~(\ref{eq:shapes}).
All curves have been normalized to the leading-order result 
$\,F_{2,\rm s}^{\,\rm LO} = \langle e^2 \rangle \: q_{\rm s}\,$.
Fig.~\ref{fig:c2sgcnv1} nicely illustrates the perturbative stability of 
the structure function $F_2$.

\begin{figure}[ht]
\begin{center}
\includegraphics[width=7.5cm,angle=0]{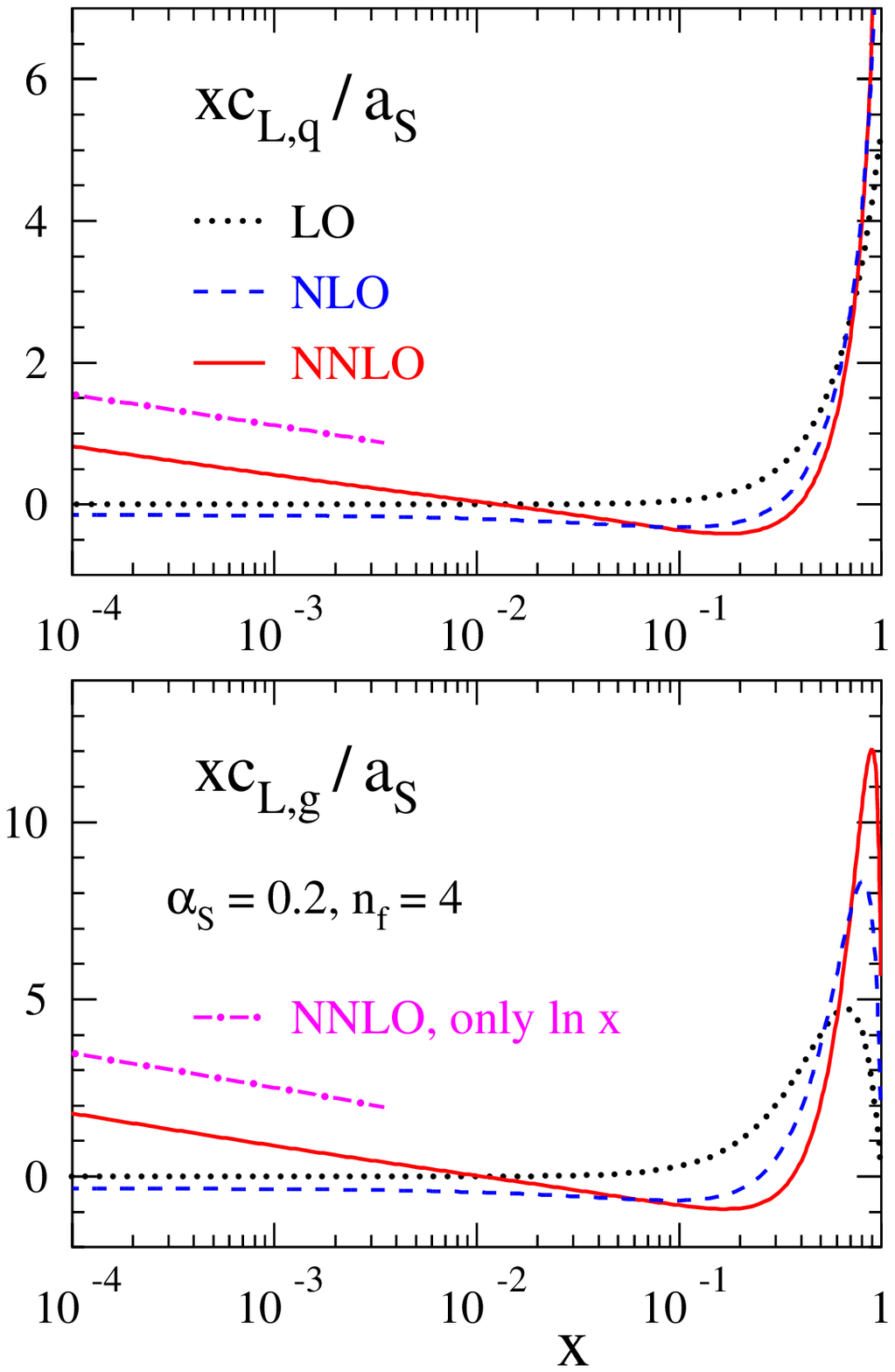}
\includegraphics[width=7.5cm,angle=0]{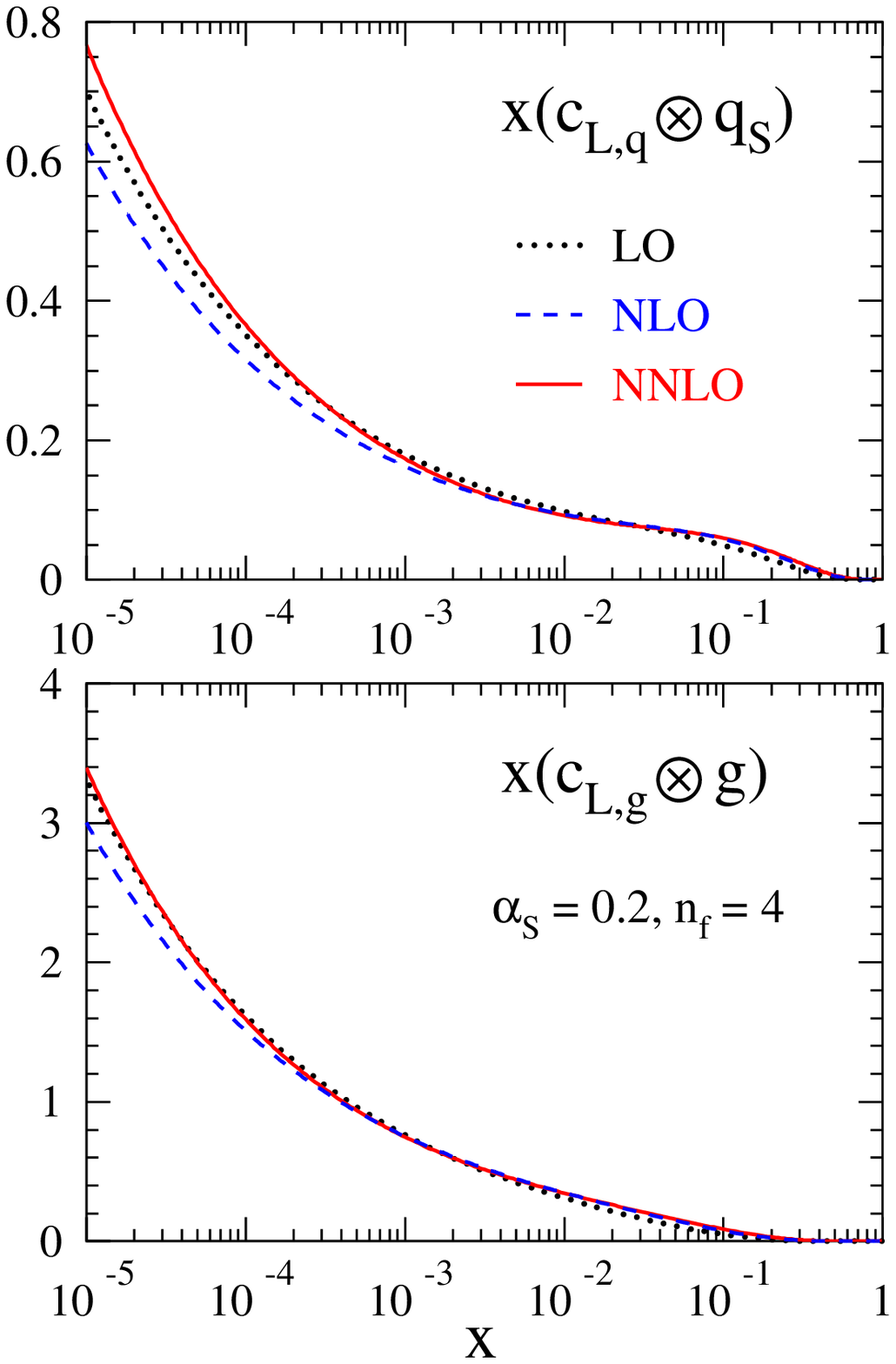}
\end{center}
\vspace*{-5mm}
\caption{The perturbative expansion to N$^2$LO of the longitudinal 
singlet-quark and gluon coefficient functions to third order multiplied by 
$x$ for display purposes (left) and 
of the quark and gluon contributions to singlet structure function $F_{L}$ (right).
}
\label{fig:clnum4}
\end{figure}

Finally, we address the longitudinal structure function $F_L$
at three loops. 
In the left part of Fig.~\ref{fig:clnum4} 
we plot the singlet-quark and gluon coefficient functions 
$c_{L,\rm q}$ and $c_{L,\rm g}$ for $F_L$ up to the third order for four flavors and the 
$\as$-value of Eq.~(\ref{eq:asvalue}).
The curves have been divided by $a_{\rm s} = \alpha_s/(4\pi)$ to account for the 
leading contribution being actually of first order in the strong coupling constant $\as$. 
Both the second-order and the third-order contributions are rather large 
over almost the whole $x$-range. 
Most striking, however, is the behavior at very small values of $x$, where 
the anomalously small one-loop parts are negligible against the 
(negative) constant two-loop terms, which in turn are completely 
overwhelmed by the (positive) new three-loop corrections 
$xc_{L,a}^{(3)}\sim \ln x\, +\, ${\it const\ }, 
which we have indicated in Fig.~\ref{fig:clnum4}.

To assess the effect for longitudinal structure function $F_L$, we convolute 
in Fig.~\ref{fig:clnum4} on the right the coefficient functions with the input shapes 
Eq.~(\ref{eq:shapes}) for $n_f=4$ flavors and the reference scale 
of Eq.~(\ref{eq:asvalue}). 
A comparison of the left and right plots in Fig.~\ref{fig:clnum4} clearly reveals 
the smoothening effect of the Mellin convolutions. 
For the chosen  input conditions, the (mostly positive) NNLO corrections 
to the flavor-singlet $F_{L}$ amount to less than 20\% 
for $5\cdot 10^{-5} < x < 0.3$.
In data fits we expect that the parton distributions, in particular the gluon 
distribution, will further stabilize the overall NNLO/NLO ratio. 
Thus, at not too small scales, $F_L$ is a quantity of good perturbative stability, 
for the $x$-values accessible at HERA, see Ref.~\cite{Moch:2004xu} for more details.


%% file: mellinmath.tex
\subsection{Mathematical Structure of Higher Order Corrections
\protect\footnote{Contributing authors: 
J.~Bl\"umlein, H.~B\"ottcher, A.~Guffanti, V.~Ravindran}}
\label{sec:mellinmath1}

The QCD anomalous dimensions and Wilson coefficients for structure functions are single scale quantities
and may be expressed in simple form in Mellin space in terms of polynomials of harmonic sums and ration 
functions of the Mellin variable. Unlike the case in various calculations using  representations in 
momentum-fraction ($z$-) space the use of multiple nested harmonic sums leads to a synchronization in language.
Furthermore, significant simplifications w.r.t. the number of functions needed can be achieved. This is due to 
algebraic \cite{Blumlein:1998if,Blumlein:2003gb} relations between these quantities, which 
in a similar way are also present between harmonic polylogarithms \cite{Remiddi:1999ew} 
and multiple $\zeta$-values \cite{Borwein:1999js}. These relations result from 
the the specific index
pattern of the objects considered and their multiplication relation and do not refer to further more specific 
properties. In Table~\ref{tb:table1jb} we illustrate the level of complexity which one meets in case of harmonic sums.
To three--loop order weight {\sf w=6} harmonic sums occur. The algebraic relations for the whole class of harmonic sums 
lead to a reduction by a factor of $\sim 4$ (column 3). As it turns out, 
physical pseudo-observables, as anomalous 
dimensions and 
Wilson-coefficients in the $\overline{\rm MS}$ scheme, to 2-- resp. 3--loop order depend on harmonic sums only, in 
which the index $\{-1\}$ never occurs. The algebraic reduction for this class is illustrated in column 5. We also 
compare the complexity of only non-alternating harmonic sums and their algebraic reduction, which is much lower. 
This class of sums is, however, not wide enough to describe the above physical quantities.
In addition to the algebraic relations of harmonic sums structural relations exist, which reduces the basis further
\cite{JB}.

\begin{table}[htp]
\begin{center}
  \mbox{\epsfig{file=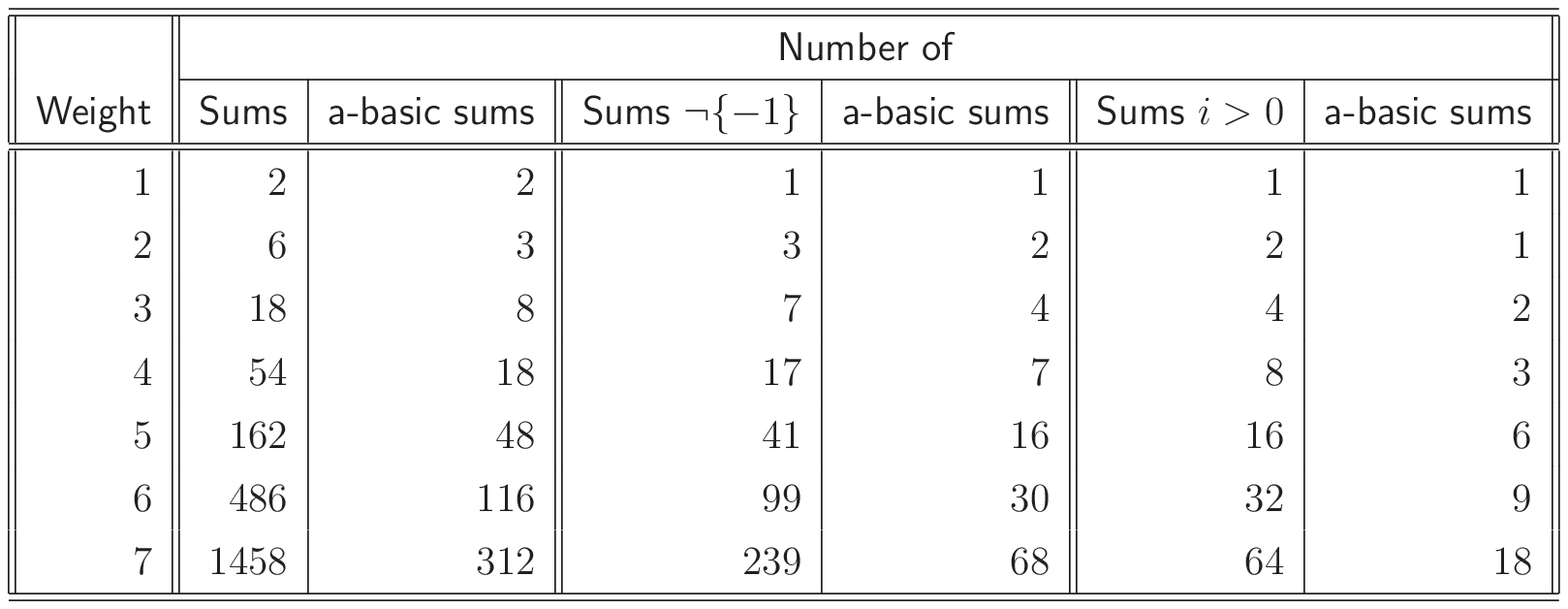,width=15cm}}
  \vspace{3mm}
  \caption{Number of alternating 
    and non-alternating harmonic sums in dependence of their weight, 
    \protect\cite{JB}.}
  \label{tb:table1jb}
\end{center}
\end{table}

\noindent
Using all these relations one finds that 5 basic functions are sufficient to describe all 2--loop Wilson coefficients
for deep--inelastic scattering \cite{JBSM} and further 8 \cite{Blumlein:2004bb} 
for the 3--loop anomalous dimensions. 
Their analytic continuations to complex values of the Mellin variable are given in 
\cite{Blumlein:2000hw,Blumlein:2005jg}.
These functions are the (regularized) Mellin transforms of~:
\begin{eqnarray}
\label{eqBAS}
& & 
\frac{\ln(1+x)}{1+x}, \hspace{1cm} 
\frac{\Li_2(x)}{1 \pm x}, \hspace{1cm} \frac{S_{1,2}(x)}{1\pm x},
\hspace{1cm} \frac{\Li_4(x)}{x \pm
1}, \nonumber\\ & &\frac{S_{1,3}(x)}{1+x}, \hspace{1cm}
\frac{S_{2,2}(x)}{x \pm 1}, \hspace{1.4cm} \frac{\Li_2^2(x)}{1+x}, \hspace{1cm} \frac{S_{2,2}(-x)
- \Li_2^2(-x)/2}{x \pm 1}~.
\end{eqnarray}
It is remarkable, that the numerator--functions in (\ref{eqBAS}) are Nielsen integrals
\cite{Nielsen} and polynomials thereof, although one might expect harmonic polylogarithms \cite{Remiddi:1999ew} 
outside this class
in general. The representation of the Wilson coefficients and anomalous dimensions in the way described 
allows for compact expressions and very fast and precise numerical evaluation well suited for fitting procedures
to experimental data.

\subsubsection{Two--loop Processes at LHC in Mellin Space}
\label{sec:mellinmath2}

\newcommand{\myappendix}{\setcounter{equation}{0}\appendix}
\newcommand{\mysection}{\setcounter{equation}{0}\section}
\renewcommand{\theequation}{\thesection.\arabic{equation}}
\newcommand{\Slash} {\slash \!\!\!}
\renewcommand{\thefootnote}{\arabic{footnote}}
\newcommand{\xo}{\mbox{$x_1^0$}}
\newcommand{\xt}{\mbox{$x_2^0$}}

Similar to the case of the Wilson coefficients in section~\ref{sec:mellinmath1} one may consider
the Wilson coefficients for inclusive hard processes at hadron colliders, as the Drell--Yan process
to $O(\alpha_s^2)$ \cite{%
Matsuura:1989sm,Hamberg:1991np,Ravindran:2003gi}, scalar or pseudoscalar Higgs--boson production 
to $O(\alpha_s^3)$ in the heavy--mass 
limit \cite{%
Catani:2001ic,Harlander:2001is,Harlander:2002wh,Harlander:2002vv,Anastasiou:2002yz,Ravindran:2003um}, 
and the 2--loop time--like Wilson coefficients for fragmentation \cite{%
Rijken:1996vr,Rijken:1996ns,Rijken:1996np}.
These quantities have been analyzed in \cite{Blumlein:2004bc,Blumlein:2005im} w.r.t. their general structure 
in Mellin 
space. The cross section for the Drell--Yan process and Higgs production 
is given by
\begin{eqnarray}
\sigma\left(\frac{\hat{s}}{s}, Q^2\right) 
= \int_x^1 \frac{d x_1}{x_1} \int_{x/x_1}^1 \frac{d x_2}{x_2} 
f_a(x_1,\mu^2) f_b(x_2, \mu^2) 
\hat{\sigma}\left( \frac{x}{x_1 x_2}, \frac{Q^2}{\mu^2}\right)~, 
\end{eqnarray}
with $x = \hat{s}/s$. Here, $f_c(x,\mu^2)$ are the initial state parton densities and $\mu^2$ 
denotes the factorization scale. The Wilson coefficient of the process
is $\hat{\sigma}$ and $Q^2$ is the time--like virtuality of the 
$s$--channel boson.
Likewise, for the fragmentation process of final state partons into 
hadrons in $pp$--scattering one considers the double differential final 
state distribution 
\begin{eqnarray}
\frac{d^2 \sigma^H}{d x d \cos \theta} =
\frac{3}{8} (1 + \cos^2 \theta)
\frac{d \sigma_T^H}{d x} 
+ \frac{3}{4} \sin^2 \theta
  \frac{d \sigma_L^H}{d x}~. 
\end{eqnarray}
Here,
\begin{eqnarray}
\frac{d \sigma^H_k}{d x} &=& \int_x^1 \frac{dz}{z} \Biggl[\sigma_{\rm 
tot}^{(0)} \left\{ 
D_S^H \left(\frac{x}{z}, M^2\right) 
C_{k,q}^{\rm S}(z,Q^2/M^2)
+ D_g^H \left(\frac{x}{z}, M^2\right) 
C_{k,q}^{\rm S}(z,Q^2/M^2)\right\}\nonumber\\
& & + \sum_{p=1}^{N_f}\sigma_{p}^{(0)}
D_{{\rm NS},p}^H \left(\frac{x}{z}, M^2\right)
C_{k,q}^{\rm NS}(z,Q^2/M^2) \Biggr]~.
\end{eqnarray}
In the subsystem cross-sections $\sigma$ the initial state parton 
distributions are included.
$D_{k}^H$ denote the non--perturbative fragmentation functions and 
$C_{k,i}^{\rm S,NS}(z,Q^2/M^2)$ the respective time--like Wilson 
coefficients describing the fragmentaion process for a parton $i$ into the 
hadron $H$.

Although these Wilson coefficients  are not directly related to the 
2--loop Wilson coefficients for deeply inelastic scattering, 
one finds for these functions at most the same set of basic functions 
as given above. Again one obtains very fast and concise numerical 
programs 
also for these processes working in Mellin space, which will be well 
suited for inclusive analyses of experimental collider data at LHC in the 
future. 
\subsubsection{\boldmath Non--Singlet Parton Densities at $O(\alpha_s^3)$} 
\label{sec:mellinmath3} 

\begin{figure}
\includegraphics[width=6.5cm]{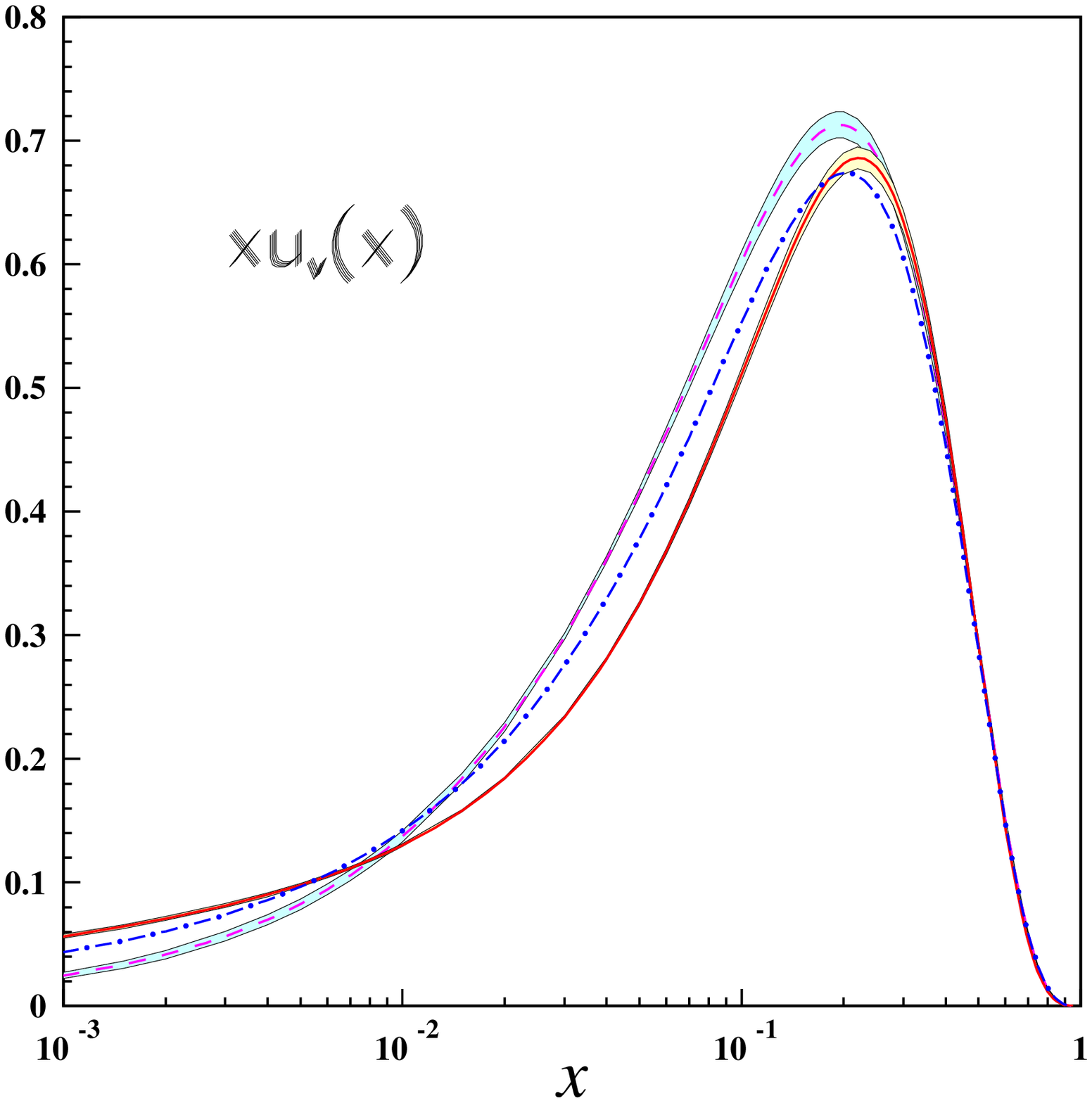} \quad\quad
\includegraphics[width=6.5cm]{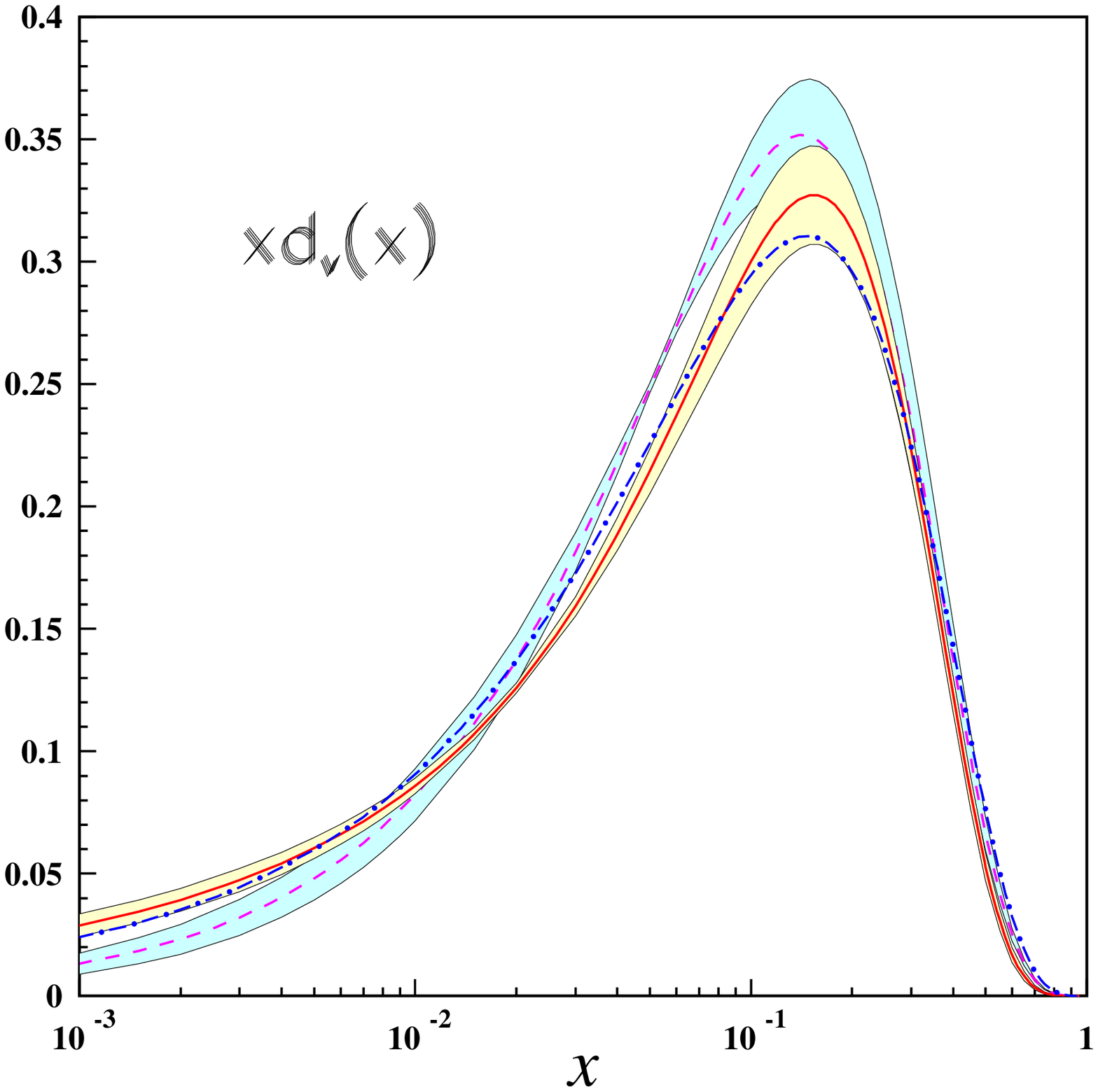}
\caption{
$xu_v$ and $xd_v$ at $Q_0^2 = 4 \GeV^2$ (full lines) \protect\cite{Blumlein:2004ip};
dashed lines \protect\cite{Alekhin:2000ch}; dash-dotted lines \protect\cite{Martin:2002dr}.
}
\label{fig:pdf1}
\end{figure}

The precision determination of the QCD--scale $\Lambda_{\rm QCD}$ and of 
the idividual parton densities is an important issue for the whole physics 
programme at LHC since all measurements rely on the detailed knwoledge of 
this parameter and distribution functions. In Ref.~\cite{Blumlein:2004ip} 
first results were reported of a world data analysis for charged 
lepton-$p(d)$ scattering  w.r.t. the flavor non--singlet sector at 
$O(\alpha_s^3)$ accuracy. The flavor non--singlet distributions 
$xu_v(x,Q^2)$ and $xd_v(x,Q^2)$ were determined along with fully 
correlated error bands giving parameterizations both for the values and 
errors of these distributions for a wide range in $x$ and $Q^2$.
In Figure~\ref{fig:pdf1} these distributions including their error are 
shown.
The value of the strong coupling constant $\alpha_s(M_Z^2)$ was determined 
as $0.1135 + 0.0023 - 0.0026~(\rm exp.)$ The full analysis is given in 
\cite{BBG}, including the determination of higher twist contributions in 
the large $x$ region both for $F_2^p(x,Q^2)$ and $F_2^d(x,Q^2)$. 

\subsubsection{Scheme-invariant evolution for unpolarzed DIS structure
functions}

The final HERA-II data on unpolarized DIS structure functions, combined
with the present world data from other experiments, will allow to reduce 
the experimental
error on the strong coupling constant, $\alpha_s(M_Z^2)$, to the level of 
$1\%$ \cite{Botje:1996hy}.
On the theoretical side the NLO analyzes have intrinsic limitations 
which allow no better than $5$\% accuracy in the determination of 
$\alpha_s$ \cite{Blumlein:1996gv}.
In order to match the expected experimental accuracy, analyzes of DIS 
structure functions need then to be carried out at the  NNLO--level.
To perform a full NNLO analysis the knowledge of the 3-loop $\beta$-function 
coefficient, $\beta_2$, the 2-- resp. 3--loop Wilson 
coefficients and the 3-loop anomalous dimensions is required. 
With the calculation of the latter \cite{Moch:2004pa,Vogt:2004mw},
the whole scheme--independent set of quantities is known, thus allowing a 
complete NNLO study of DIS structure functions.

Besides the standard approach solving the QCD evolution equations for 
parton densities in the $\overline{\rm MS}$ scheme it appears appealing 
to study scheme--invariant evolution equations \cite{Blumlein:2004xs}.
Within this approach the input distributions at a scale $Q^2_0$ are {\sf
measured} experimentally. The only parameter to be determined by a fit to 
data is the QCD--scale $\Lambda_{\rm QCD}$. To perform an analysis in the 
whole kinematic region the non--singlet \cite{Blumlein:2004ip} 
contribution has to be separated from the singlet terms of two measured 
observables. In practice these can be chosen to be $F_2(x,Q^2)$ and
$\partial F_2(x,Q^2)/\partial \ln(Q^2)$ or  $F_2(x,Q^2)$ and
$F_L(x,Q^2)$ if the latter structure function is measured well enough.
Either $\partial F_2(x,Q^2)/\partial \ln(Q^2)$ or  $F_L(x,Q^2)$ play a 
role synonymous to the gluon distribution while  $F_2(x,Q^2)$ takes the 
role of the singlet--quark distribution compared to the standard analysis.
These equations do no longer describe the evolution of {\sf universal} 
quantities depending on the choice of a scheme but of process--dependent
quantities which are observables and thus factorization  
scheme--indedependent. Since the respective evolution kernels are 
calculated in perturbation theory the dependence on the renormalization 
scale remains and becomes smaller with the order in the coupling constant 
included. 

Physical evolution kernels have been studied before in 
\cite{Furmanski:1981cw,Catani:1996sc,Blumlein:2000wh}. The 3--loop 
scheme--invariant evolution equations were solved in the massless case in 
\cite{Blumlein:2004xs}. This analysis is extended including the heavy 
flavor contributions at present \cite{BBG}. The large complexity of the 
evolution kernels can only be handeled in Mellin space since in $z$--space
various inverse and direct Mellin convolutions would be required 
numerically, causing significant accuracy and run--time problems. The 
inclusion of the heavy flavor contributions is possible using the 
parameterizations \cite{Alekhin:2003ev}. 

In Fig.~\ref{fig:pdf1} we present the scheme invariant evolution for the structure 
functions $F_2$ and $\partial F_2/\partial t$ to NNLO with $t= - 2/\beta_0 \ln(\alpha_s(Q^2)/\alpha_s(Q_0^2))$.
The input distribution at the reference scale are not extracted from data, 
but rather built up as a convolution of Wilson coefficients and PDFs,
the latter being parametrised according to \cite{Giele:2002hx}.

\begin{figure}
\includegraphics[width=6.5cm]{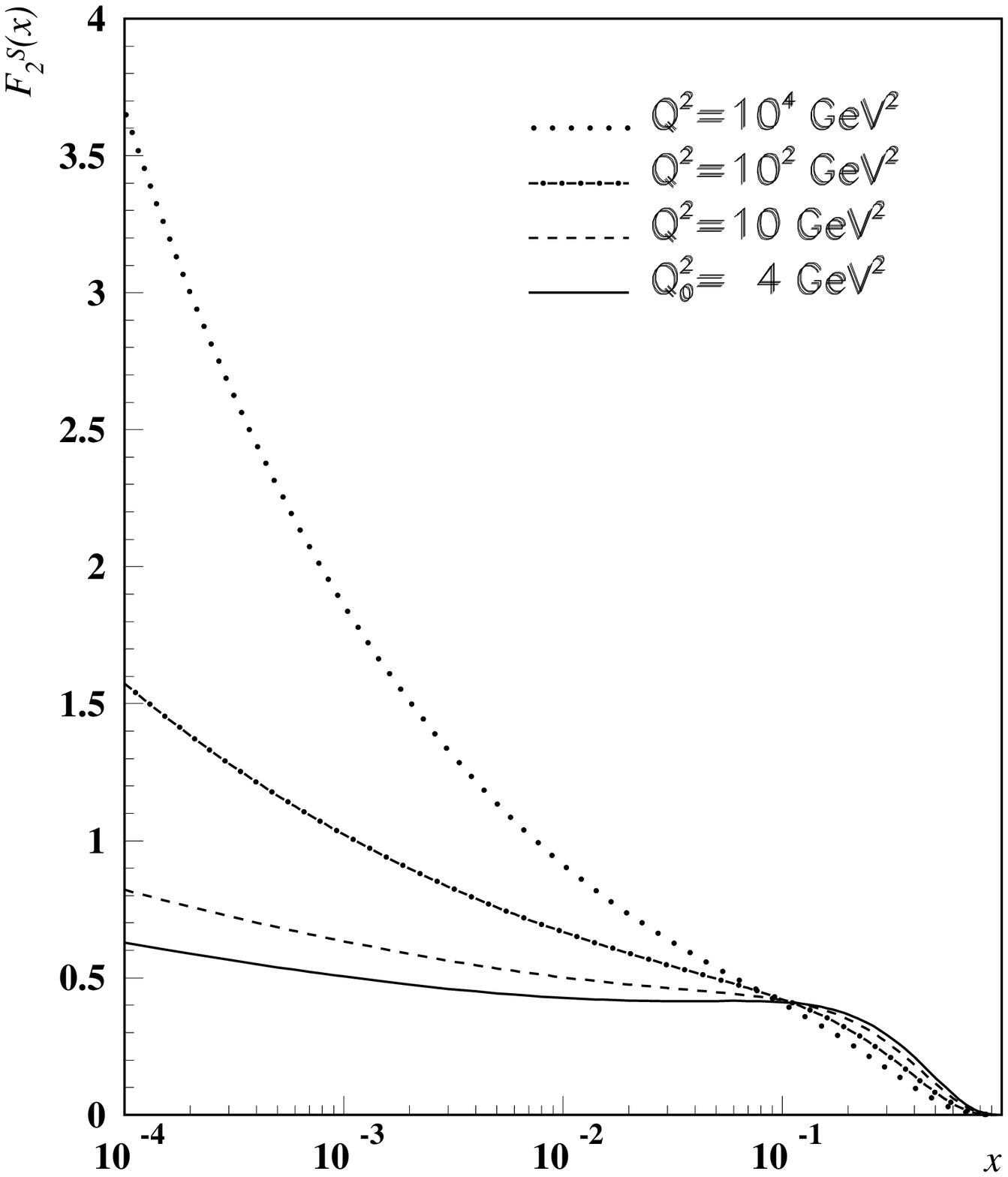}\qquad\quad
\includegraphics[width=6.5cm]{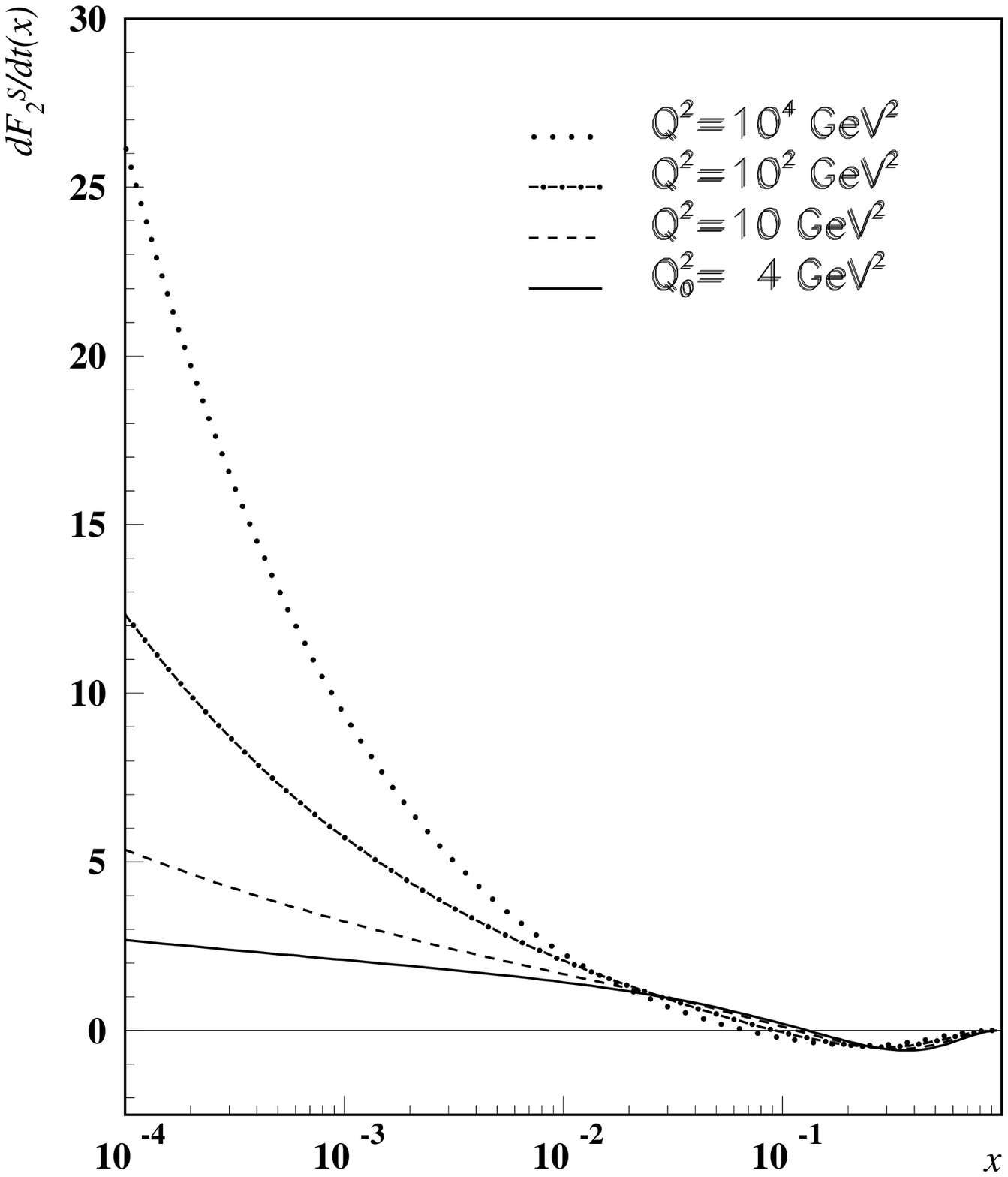}
\caption{NNLO scheme invariant evolution for the singlet part of the 
         structure function $F_2$ and its slope $\partial F_2/\partial t$
	 for four massless flavours, \protect\cite{Blumlein:2004xs}.
}
\end{figure}

Scheme--invariant evolution equations allow a widely un--biased approach 
to determine the initial conditions for QCD evolution, which in general is 
a source of systematic effects which are difficult to control. On the 
other hand, their use requires to consider all correlations of the input 
measurements in a detailed manner experimentally. At any scale $Q^2$ 
mappings are available to project the observables evolved onto the 
quark--singlet and the gluon density in whatever scheme. In this way the 
question whether sign changes in the unpolarized gluon distribution 
in the $\overline{\rm MS}$ scheme do occur or do not occur in the small 
$x$ region can be  
answered uniquely. As in foregoing analyses 
\cite{Blumlein:2002be,Blumlein:2004ip}
correlated error propagation throughout the evolution is being performed.

%% file: evolution.tex
\subsection{Updated reference results for the evolution of parton 
distributions$\,$\protect\footnote{Contributing authors: G.P.~Salam, A.~Vogt}}
\label{sec:pdfevolution}

In this contribution we update and extend our benchmark tables, first presented
in the report of the QCD/SM working group at the 2001 Les Houches workshop
\cite{Giele:2002hx}, for the evolution of parton distributions of hadrons in 
perturbative QCD. Since then the complete next-to-next-to-leading order (NNLO) 
splitting functions have been computed \cite{Moch:2004pa,Vogt:2004mw}, 
see also section~\ref{sec:nnloprecision}.
Thus we can now replace the NNLO results of 2001 which were based on the 
approximate splitting functions of Ref.~\cite{vanNeerven:2000wp}.
Furthermore we now include reference tables for the polarized case treated in 
neither Ref.~\cite{Giele:2002hx} nor the earlier study during the 1995/6 HERA 
workshop \cite{Blumlein:1996rp}. Since the spin-dependent NNLO splitting 
functions are still unknown, we have to restrict ourselves to the polarized
leading-order (LO) and next-to-leading-order (NLO) evolution.

As in Ref.~\cite{Giele:2002hx}, we employ two entirely independent and 
conceptually different {\sc Fortran} programs. At this point, the $x$-space 
code of G.S. is available from the author upon request, while the Mellin-space 
program of A.V. has been published in Ref.~\cite{Vogt:2004ns}.
The results presented below correspond to a direct iterative solution of the 
N$^{\rm m}$LO evolution equations for the parton distributions $f_p (x,
\mu_{\rm f}^2) \equiv p(x,\mu_{\rm f}^2) $, where $p = q_i, \,\bar{q}_i\, ,g$ 
with $ i = 1, \,\ldots , N_{\rm f}$,
\begin{equation}
\label{gsav-eq1}
 \frac {d\:\! f_p(x,\mu_{\rm f}^2)} {d \ln \mu_{\rm f}^2} \: = \: \sum_{l=0}^m 
 \, a_{\rm s}^{\,l+1}(\mu_{\rm r}^2) \int_x^1 \! \frac{dy}{y}\, \sum_{p'} \:\!
 P_{pp'}^{(l)}\bigg(\frac{x}{y},\frac{\mu_{\rm f}^2}{\mu_{\rm r}^2}\bigg)
 \, f_{p'}(y,\mu_{\rm f}^2) 
\end{equation}
with the strong coupling, normalized as $a_{\rm s} \equiv \alpha_{\rm s}/ 
(4\pi)$, given in terms of 
\begin{equation}
\label{gsav-eq2}
  \frac{d\, a_{\rm s}}{d \ln \mu_r^2} \: = \: \beta^{}_{\rm N^mLO}(a_{\rm s})
  \: = \: - \sum_{l=0}^m \, a_{\rm s}^{\,l+2} \,\beta_{\:\! l} 
\end{equation}
with $\,\beta_0 = 11 - 2/3\, N_{\rm f}\,$ etc.
$\mu_{\rm r}$ and $\mu^{}_{\rm f}$ represent the renormalization and mass-%
factorization scales in the $\overline{\mbox{MS}}$ scheme. The reader is 
referred to Refs.~\cite{Giele:2002hx,Vogt:2004ns} for the scale dependence of 
the splitting functions $P^{(l)}$ and a further discussion of our solutions of 
Eqs.~(\ref{gsav-eq1}) and (\ref{gsav-eq2}).
 
For the unpolarized case we retain the initial conditions as set up at the Les 
Houches meeting: The evolution is started at 
\begin{equation}
\label{gsav-eq3}
  \mu_{\rm f,0}^2 \: = \: 2 \mbox{ GeV}^2 \:\: .
\end{equation}
Roughly along the lines of the CTEQ5M parametrization \cite{Lai:1999wy}, the 
input distributions are chosen as
\begin{eqnarray}
\label{gsav-eq4}
  xu_{\rm v}(x,\mu_{\rm f,0}^2) &\! =\! & 5.107200\: x^{0.8}\: (1-x)^3  
    \nonumber \\
  xd_{\rm v}(x,\mu_{\rm f,0}^2) &\! =\! & 3.064320\: x^{0.8}\: (1-x)^4  
    \nonumber \\
  xg\,(x,\mu_{\rm f,0}^2)       &\! =\! & 1.700000\, x^{-0.1} (1-x)^5 
    \\
  x\bar{d}\,(x,\mu_{\rm f,0}^2) &\! =\! & .1939875\, x^{-0.1} (1-x)^6
    \nonumber\\
  x\bar{u}\,(x,\mu_{\rm f,0}^2) &\! =\! & (1-x)\: x\bar{d}\,(x,\mu_{\rm f,0}^2)
    \nonumber\\
  xs\,(x,\mu_{\rm f,0}^2)       &\! =\! & x\bar{s}\,(x,\mu_{\rm f,0}^2) 
    \: = \: 0.2\, x(\bar{u}+\bar{d}\,)(x,\mu_{\rm f,0}^2) 
    \nonumber  
\end{eqnarray}
where, as usual, $q_{i,\rm v} \equiv q_i - \bar{q}_i$. 
The running couplings are specified by Eq.~(\ref{gsav-eq2}) and
\begin{equation}
\label{gsav-eq5}
  \alpha_{\rm s}(\mu_{\rm r}^2\! =\! 2\mbox{ GeV}^2) \: = \: 0.35 \:\: .
\end{equation}
For simplicity initial conditions (\ref{gsav-eq4}) and (\ref{gsav-eq5}) are 
employed regardless of the order of the evolution and the (fixed) ratio of the 
renormalization and factorization scales.
 
For the evolution with a fixed number $N_{\rm f} > 3$ of quark flavours the 
quark distributions not specified in Eq.~(\ref{gsav-eq4}) are assumed to vanish 
at $\mu_{\rm f,0}^2$, and Eq.~(\ref{gsav-eq5}) is understood to refer to the 
chosen value of $N_{\rm f}$. For the evolution with a variable $ N_{\rm f} = 3 
\ldots 6 $, Eqs.~(\ref{gsav-eq3}) and (\ref{gsav-eq4}) always refer to three 
flavours. $N_{\rm f}$ is then increased by one unit at the heavy-quark pole 
masses taken as
\begin{equation}
\label{gsav-eq6}
  m_{\rm c} \: = \: \mu_{\rm f,0}      \: , \quad 
  m_{\rm b} \: = \: 4.5 \mbox{ GeV}^2  \: , \quad
  m_{\rm t} \: = \: 175 \mbox{ GeV}^2  \:\: ,
\end{equation}
i.e., Eqs.~(\ref{gsav-eq1}) and (\ref{gsav-eq2}) are solved for a fixed number
of flavours between these thresholds, and the respective matching conditions 
are invoked at $\mu_{\rm f}^2 = m_{\rm h\,}^2$, $\,h = c,\, b,\, t$. 
The matching conditions for the unpolarized parton distributions have been 
derived at NNLO in Ref.~\cite{Buza:1996wv}, and were first implemented in an 
evolution program in
Ref.~\cite{Chuvakin:2001ge}. Note that, while the parton distributions are 
continuous up to NLO due to our choice of the matching scales, $\alpha_{\rm s}$ 
is discontinuous at these flavour thresholds already at this order for $\mu_{\rm r} 
\neq \mu_{\rm f}^{}$, see Refs.~\cite{Larin:1994va,Chetyrkin:1997sg}.
Again the reader is referred to Refs.~\cite{Giele:2002hx,Vogt:2004ns} for more
details.

\begin{figure}[p]
\begin{center}
\mbox{\includegraphics[height=10.8cm]{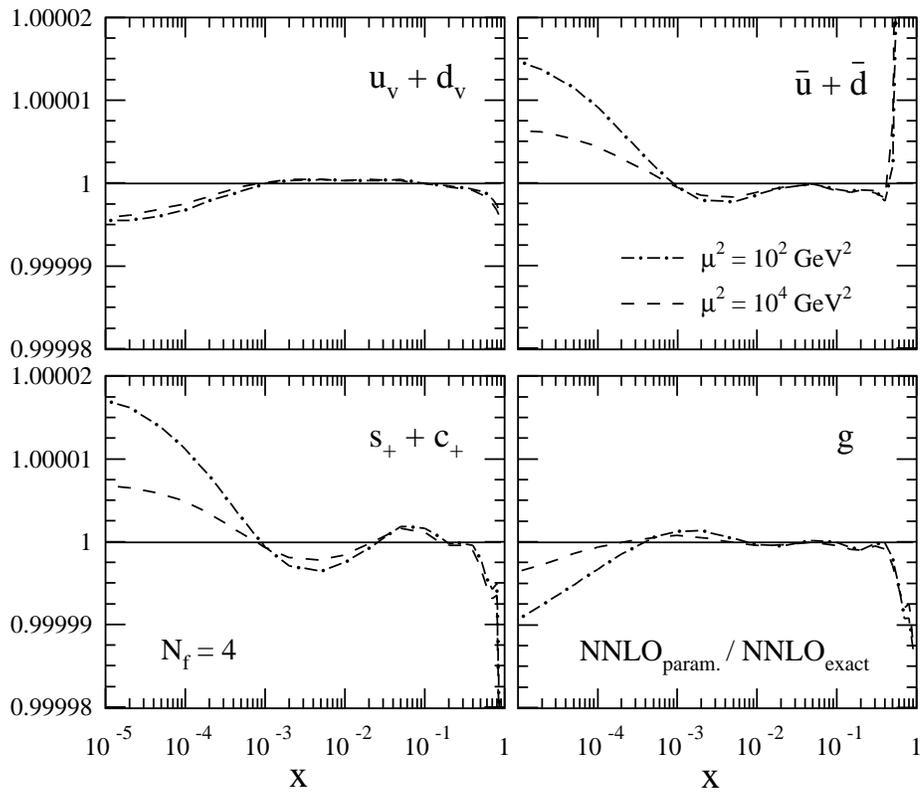}}
\end{center}
\vspace{-10mm}
\caption{
 \label{gsav-f1}
 Relative effects of using the parametrized three-loop splitting functions of 
 Refs.~\protect\cite{Moch:2004pa,Vogt:2004mw}, instead of the exact expressions
 from the same source, on the NNLO evolution for the input (\ref{gsav-eq3}) -- 
 (\ref{gsav-eq5}) at two representative values of $\mu = \mu_{\rm r} = 
 \mu_{\rm f}^{}$.}
\end{figure}

Since the exact NNLO splitting functions $P^{(2)}$ are rather lengthy and not
directly suitable for use in a Mellin-space program (see, however, Ref.~\cite
{Blumlein:2005jg}), the reference tables shown below have been computed using
the parametrizations (4.22) -- (4.24) of Ref.~\cite{Moch:2004pa} and (4.32) --
(4.35) of Ref.~\cite{Vogt:2004mw}. Likewise, the operator matrix element 
$\widetilde{A}_{\rm hg}^{\rm S,2}$ entering the NNLO flavour matching is taken 
from Eq.~(3.5) of Ref.~\cite{Vogt:2004ns}. The relative error made by using 
the parametrized splitting functions is illustrated in Fig.~\ref{gsav-f1}. It 
is generally well below $10^{-4}$, except for the very small sea quark 
distributions at very large $x$.
\begin{figure}[p]
\begin{center}
\mbox{\includegraphics[height=10.8cm]{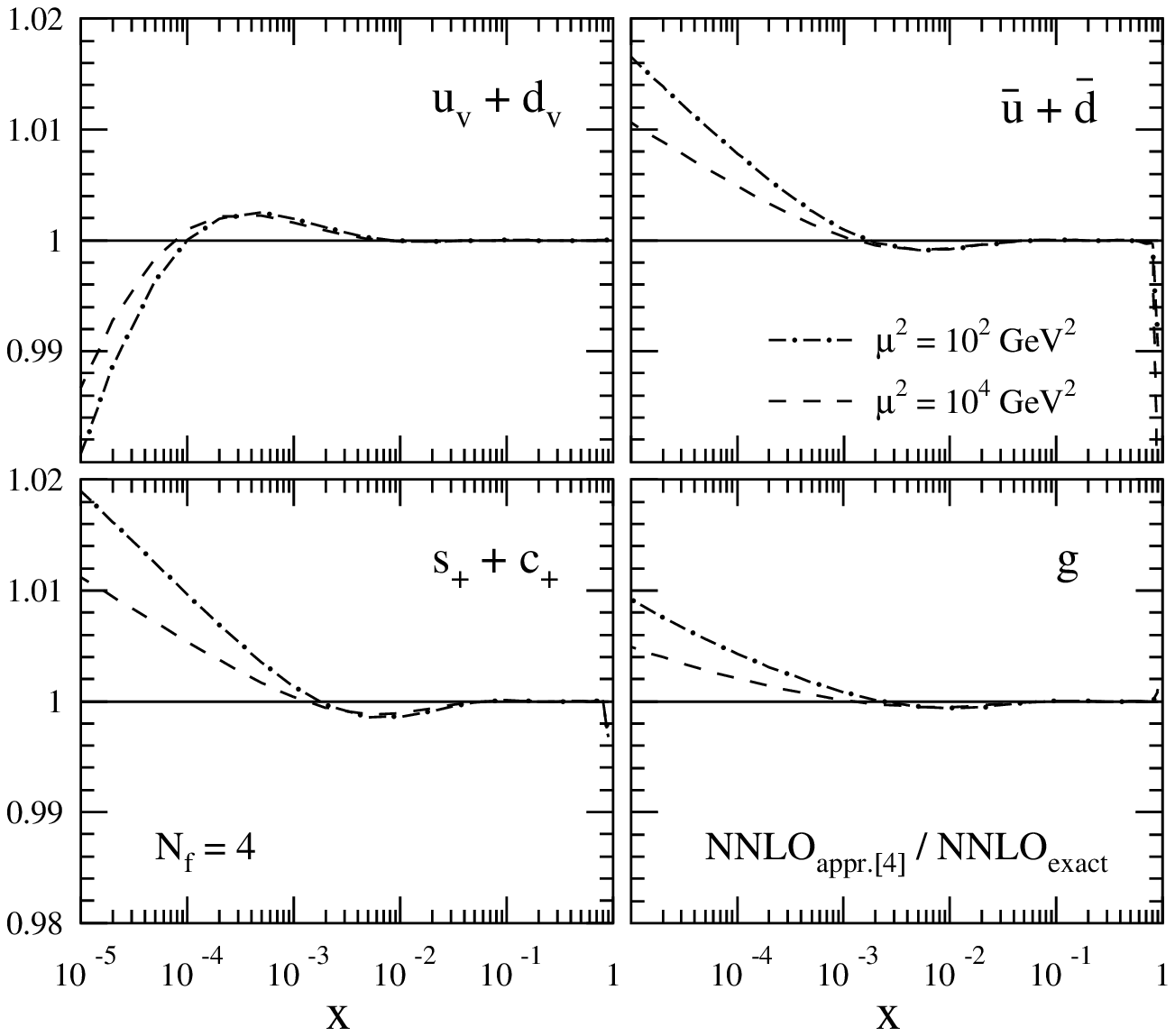}}
\end{center}
\vspace{-10mm}
\caption
{\label{gsav-f2}
 Relative errors made by using the previous average approximations 
\protect\cite{vanNeerven:2000wp} for the three-loop splitting functions 
(used, e.g., 
 in Refs.~\protect\cite{Martin:2002dr,Alekhin:2002fv})
 instead of the full 
 results~\protect\cite{Moch:2004pa,Vogt:2004mw}, on the NNLO evolution of the input 
 (\ref{gsav-eq3}) -- (\ref{gsav-eq5}) at  $\,\mu_{\rm r}=\mu_{\rm f}^{}$.}

\end{figure}

Eqs.~(\ref{gsav-eq3}), (\ref{gsav-eq5}) and (\ref{gsav-eq6}) are used for the 
(longitudinally) polarized case as well, where Eq.~(\ref{gsav-eq4}) replaced by 
the sufficiently realistic toy input~\cite{Vogt:2004ns}
\begin{eqnarray}
\label{gsav-eq7}
  xu_{\rm v} &\! =\! & +1.3\: x^{0.7}\: (1-x)^3\: (1+3x) \nonumber\\
  xd_{\rm v} &\! =\! & -0.5\: x^{0.7}\: (1-x)^4\: (1+4x) \nonumber\\
  xg\:       &\! =\! & +1.5\: x^{0.5}\: (1-x)^5  \nonumber \\
  x\bar{d}\: &\! =\! & x\bar{u}
             \: = \: -0.05\: x^{0.3}\: (1-x)^7  \nonumber\\
  xs\:       &\! =\! & x\bar{s}
             \: = \: +0.5\: x \bar{d} \:\: .
\end{eqnarray}
As Eq.~(\ref{gsav-eq4}) in the unpolarized case, this input is employed 
regardless of the order of the evolution.

As in Ref.~\cite{Giele:2002hx}, we have compared the results of our two 
evolution programs, under the conditions specified above, at 500 
$x$-$\mu_{\rm f}^2$ points covering the range $10^{-8} \leq x \leq 0.9$ and 
$2 \mbox{ GeV}^2 \leq \mu_{\rm f}^2 \leq 10^6 \mbox{ GeV}^2 $. 
A representative subset of our results at $\mu_{\rm f}^2 = 10^4 \mbox{ GeV}^4$, a scale relevant to high-$E_T$ jets and close to $m_{\rm W}^2$, $m_{\rm Z}^2$ 
and, possibly, $m_{\rm Higgs}^2$, is presented in Tables \ref{gsav-t4}$\,$--%
$\,$\ref{gsav-t3}.
These results are given in terms of the valence distributions, defined below 
Eq.~(\ref{gsav-eq4}), $L_{\pm} \equiv \bar{d} \pm \bar{u}$, and the quark-%
antiquark sums $\,q_+ \! \equiv q \! -\!\bar{q}\,$ for $\,q = s,\, c$ and, for 
the variable-$N_{\rm f}$ case, $b$.

For compactness an abbreviated notation is employed throughout the tables, 
i.e., all numbers $a\cdot 10^b$ are written as $a^b$. In the vast majority of 
the $x$-$\mu_{\rm f}^2$ points our results are found to agree to all five 
figures displayed, except for the tiny NLO and NNLO sea-quark distributions at 
$x = 0.9$, in the tables. Entries where the residual offsets between our 
programs lead to a different fifth digit after rounding are indicated by the 
subscript `$*$'. In these cases the number with the smaller modulus is given in
the tables.

The approximate splitting functions \cite{vanNeerven:2000wp}, as mentioned 
above employed in the previous version \cite{Giele:2002hx} of our reference 
tables, have been used in (global) NNLO fits of the unpolarized parton
distributions \cite{Martin:2002dr,Alekhin:2002fv}, which in turn have been
widely employed for obtaining NNLO cross sections, in particular for $W$ and 
Higgs production. The effect of replacing the approximate results by the full 
splitting functions \cite{Moch:2004pa,Vogt:2004mw} is illustrated in Figure 
\ref{gsav-f2}. Especially at scales relevant to the above-mentioned processes,
the previous approximations introduce an error of less than 0.2\% for $x\!\gsim
\! 10^{-3}$, and less than 1\% even down to $\,x \simeq 10^{-5}$. Consequently 
the splitting-function approximations used for the evolution the parton 
distributions of Refs.~\cite{Martin:2002dr,Alekhin:2002fv} are confirmed to a 
sufficient accuracy for high-scale processes at the LHC. 

The unchanged unpolarized LO and NLO reference tables of Ref.~\cite
{Giele:2002hx} are not repeated here. Note that the one digit of the first 
(FFN) $\alpha_{\rm s}$ value was mistyped in the header of Table 1 in that 
report$\,$\footnote{We thank H. B\"ottcher and J. Bl\"umlein for pointing this 
out to us.}, the correct value can be found in Table 3 below.

\begin{table}[htp]
\caption{Reference results for the $N_{\rm f}\!=\!4$ next-next-to-leading-order
 evolution for the initial conditions (\ref{gsav-eq3}) -- (\ref{gsav-eq5}).
 The corresponding value of the strong coupling is $\alpha_{\rm s}
 (\mu_{\rm r}^2 \! =\! 10^4 \mbox{ GeV}^2) = 0.110141$. The valence 
 distributions $s_v$ and $c_v$ are equal for the input (\ref{gsav-eq4}). The 
 notation is explained below Eq.~(\ref{gsav-eq4}) and in the paragraph below
 Eq.~(\ref{gsav-eq7}).}
\label{gsav-t4}
\vspace{2mm}
\begin{center}
\begin{tabular}{||c||r|r|r|r|r|r|r|r||}
\hline \hline
\multicolumn{9}{||c||}{} \\[-3mm]
\multicolumn{9}{||c||}{NNLO, $\, N_{\rm f} = 4\,$, 
                       $\,\mu_{\rm f}^2 = 10^4 \mbox{ GeV}^2$} \\
\multicolumn{9}{||c||}{} \\[-0.3cm] \hline \hline
 & & & & & & & \\[-0.3cm]
\multicolumn{1}{||c||}{$x$} &
\multicolumn{1}{c|} {$xu_v$} &
\multicolumn{1}{c|} {$xd_v$} &
\multicolumn{1}{c|} {$xL_-$} &
\multicolumn{1}{c|} {$2xL_+$} &
\multicolumn{1}{c|} {$xs_v$} &
\multicolumn{1}{c|} {$xs_+$} &
\multicolumn{1}{c|} {$xc_+$} &
\multicolumn{1}{c||}{$xg$} \\[0.5mm] \hline \hline
\multicolumn{9}{||c||}{} \\[-3mm]
\multicolumn{9}{||c||}{$\mu_{\rm r}^2 = \mu_{\rm f}^2$} \\
\multicolumn{9}{||c||}{} \\[-0.3cm] \hline \hline
 & & & & & & & \\[-0.3cm]
$\! 10^{-7}\!$ &
$\! 1.5287^{-4}\!$ &$\! 1.0244^{-4}\!$ &$\! 5.7018^{-6}\!$ &$\! 1.3190^{+2}\!$ &
$\! 3.1437^{-5}\!$ &$\! 6.4877^{+1}\!$ &$\! 6.4161^{+1}\!$ &$\! 9.9763^{+2}\!$\\
$\! 10^{-6}\!$ &
$\! 6.9176^{-4}\!$ &$\! 4.4284^{-4}\!$ &$\! 2.5410^{-5}\!$ &$\! 6.8499^{+1}\!$ &
$\! 9.4279^{-5}\!$ &$\! 3.3397^{+1}\!$ &$\! 3.2828^{+1}\!$ &$\! 4.9124^{+2}\!$\\
$\! 10^{-5}\!$ &
$\! 3.0981^{-3}\!$ &$\! 1.8974^{-3}\!$ &$\! 1.0719^{-4}\!$ &$\! 3.3471^{+1}\!$ &
$\! 2.2790^{-4}\!$ &$\! 1.6059^{+1}\!$ &$\! 1.5607^{+1}\!$ &$\! 2.2297^{+2}\!$\\
$\! 10^{-4}\!$ &
$\! 1.3722^{-2}\!$ &$\! 8.1019^{-3}\!$ &$\! 4.2558^{-4}\!$ &$\! 1.5204^{+1}\!$ &
$\! 3.6644^{-4}\!$ &$\! 7.0670^{+0}\!$ &$\! 6.7097^{+0}\!$ &$\! 9.0668^{+1}\!$\\
$\! 10^{-3}\!$ &
$\! 5.9160^{-2}\!$ &$\! 3.4050^{-2}\!$ &$\! 1.6008^{-3}\!$ &$\! 6.3230^{+0}\!$ &
$\! 1.4479^{-4}\!$ &$\! 2.7474^{+0}\!$ &$\! 2.4704^{+0}\!$ &$\! 3.1349^{+1}\!$\\
$\! 10^{-2}\!$ &
$\! 2.3078^{-1}\!$ &$\! 1.2919^{-1}\!$ &$\! 5.5688^{-3}\!$ &$\! 2.2752^{+0}\!$ &
$\!-5.7311^{-4}\!$ &$\! 8.5502^{-1}\!$ &$\! 6.6623^{-1}\!$ &$\! 8.1381^{+0}\!$\\
$\! 0.1    \!$ &
$\! 5.5177^{-1}\!$ &$\! 2.7165^{-1}\!$ &$\! 1.0023^{-2}\!$ &$\! 3.9019^{-1}\!$ &
$\!-3.0627^{-4}\!$ &$\! 1.1386^{-1}\!$ &$\! 5.9773^{-2}\!$ &$\! 9.0563^{-1}\!$\\
$\! 0.3    \!$ &
$\! 3.5071^{-1}\!$ &$\! 1.3025^{-1}\!$ &$\! 3.0098^{-3}\!$ &$\! 3.5358^{-2}\!$ &
$\!-3.1891^{-5}\!$ &$\! 9.0480^{-3}\!$ &$\! 3.3061^{-3}\!$ &$\! 8.4186^{-2}\!$\\
$\! 0.5    \!$ &
$\! 1.2117^{-1}\!$ &$\! 3.1528^{-2}\!$ &$\! 3.7742^{-4}\!$ &$\! 2.3867^{-3}\!$ &
$\!-2.7215^{-6}\!$ &$\! 5.7965^{-4}\!$ &$\! 1.7170^{-4}\!$ &$\! 8.1126^{-3}\!$\\
$\! 0.7    \!$ &
$\! 2.0077^{-2}\!$ &$\! 3.0886^{-3}\!$ &$\! 1.3434^{-5}\!$ &$\! 5.4244^{-5}\!$ &
$\!-1.0106^{-7}\!$ &$\! 1.2936^{-5}\!$ &$\! 3.5304^{-6}\!$ &$\! 3.8948^{-4}\!$\\
$\! 0.9    \!$ &
$\! 3.5111^{-4}\!$ &$\! 1.7783^{-5}\!$ &$\! 8.651 ^{-9}\!$ &$\! 2.695 ^{-8}\!$ &
$\!-1.476^{-10}\!$ &$\! 7.132 ^{-9}\!$ &$\! 2.990 ^{-9}\!$ &$\! 1.2136^{-6}\!$\\
\hline \hline
\multicolumn{9}{||c||}{} \\[-3mm]
\multicolumn{9}{||c||}{$\mu_{\rm r}^2 = 2\, \mu_{\rm f}^2$} \\
\multicolumn{9}{||c||}{} \\[-0.3cm] \hline \hline
 & & & & & & & \\[-0.3cm]
$\! 10^{-7}\!$ &
$\! 1.3416^{-4}\!$ &$\! 8.7497^{-5}\!$ &$\! 4.9751^{-6}\!$ &$\! 1.3020^{+2}\!$ &
$\! 2.1524^{-5}\!$ &$\! 6.4025^{+1}\!$ &$\! 6.3308^{+1}\!$ &$\! 1.0210^{+3}\!$\\
$\! 10^{-6}\!$ &
$\! 6.2804^{-4}\!$ &$\! 3.9406^{-4}\!$ &$\! 2.2443^{-5}\!$ &$\! 6.6914^{+1}\!$ &
$\! 6.5149^{-5}\!$ &$\! 3.2602^{+1}\!$ &$\! 3.2032^{+1}\!$ &$\! 4.9626^{+2}\!$\\
$\! 10^{-5}\!$ &
$\! 2.9032^{-3}\!$ &$\! 1.7575^{-3}\!$ &$\! 9.6205^{-5}\!$ &$\! 3.2497^{+1}\!$ &
$\! 1.5858^{-4}\!$ &$\! 1.5570^{+1}\!$ &$\! 1.5118^{+1}\!$ &$\! 2.2307^{+2}\!$\\
$\! 10^{-4}\!$ &
$\! 1.3206^{-2}\!$ &$\! 7.7673^{-3}\!$ &$\! 3.9093^{-4}\!$ &$\! 1.4751^{+1}\!$ &
$\! 2.5665^{-4}\!$ &$\! 6.8388^{+0}\!$ &$\! 6.4807^{+0}\!$ &$\! 9.0162^{+1}\!$\\
$\! 10^{-3}\!$ &
$\! 5.8047^{-2}\!$ &$\! 3.3434^{-2}\!$ &$\! 1.5180^{-3}\!$ &$\! 6.1703^{+0}\!$ &
$\! 1.0388^{-4}\!$ &$\! 2.6695^{+0}\!$ &$\! 2.3917^{+0}\!$ &$\! 3.1114^{+1}\!$\\
$\! 10^{-2}\!$ &
$\! 2.2930^{-1}\!$ &$\! 1.2857^{-1}\!$ &$\! 5.4626^{-3}\!$ &$\! 2.2492^{+0}\!$ &
$\!-3.9979^{-4}\!$ &$\! 8.4058^{-1}\!$ &$\! 6.5087^{-1}\!$ &$\! 8.0993^{+0}\!$\\
$\! 0.1    \!$ &
$\! 5.5428^{-1}\!$ &$\! 2.7326^{-1}\!$ &$\! 1.0072^{-2}\!$ &$\! 3.9297^{-1}\!$ &
$\!-2.1594^{-4}\!$ &$\! 1.1439^{-1}\!$ &$\! 5.9713^{-2}\!$ &$\! 9.0851^{-1}\!$\\
$\! 0.3    \!$ &
$\! 3.5501^{-1}\!$ &$\! 1.3205^{-1}\!$ &$\! 3.0557^{-3}\!$ &$\! 3.6008^{-2}\!$ &
$\!-2.2632^{-5}\!$ &$\! 9.2227^{-3}\!$ &$\! 3.3771^{-3}\!$ &$\! 8.5022^{-2}\!$\\
$\! 0.5    \!$ &
$\! 1.2340^{-1}\!$ &$\! 3.2166^{-2}\!$ &$\! 3.8590^{-4}\!$ &$\! 2.4459^{-3}\!$ &
$\!-1.9420^{-6}\!$ &$\! 5.9487^{-4}\!$ &$\! 1.7699^{-4}\!$ &$\! 8.2293^{-3}\!$\\
$\! 0.7    \!$ &
$\! 2.0597^{-2}\!$ &$\! 3.1751^{-3}\!$ &$\! 1.3849^{-5}\!$ &$\! 5.5722^{-5}\!$ &
$\!-7.2616^{-8}\!$ &$\! 1.3244^{-5}\!$ &$\! 3.5361^{-6}\!$ &$\! 3.9687^{-4}\!$\\
$\! 0.9    \!$ &
$\! 3.6527^{-4}\!$ &$\! 1.8544^{-5}\!$ &$\! 9.050 ^{-9}\!$ &$\! 2.663 ^{-8}\!$ &
$\!-1.075^{-10}\!$ &$\! 6.713 ^{-9}\!$ &$\! 2.377 ^{-9}\!$ &$\! 1.2489^{-6}\!$\\
\hline \hline
\multicolumn{9}{||c||}{} \\[-3mm]
\multicolumn{9}{||c||}{$\mu_{\rm r}^2 = 1/2\, \mu_{\rm f}^2$} \\
\multicolumn{9}{||c||}{} \\[-0.3cm] \hline \hline
 & & & & & & & \\[-0.3cm]
$\! 10^{-7}\!$ &
$\! 1.7912^{-4}\!$ &$\! 1.2521^{-4}\!$ &$\!6.4933_*^{-6}\!$&$\! 1.2714^{+2}\!$ &
$\! 4.9649^{-5}\!$ &$\! 6.2498^{+1}\!$ &$\! 6.1784^{+1}\!$ &$\! 9.2473^{+2}\!$\\
$\! 10^{-6}\!$ &
$\! 7.7377^{-4}\!$ &$\! 5.1222^{-4}\!$ &$\! 2.8719^{-5}\!$ &$\! 6.7701^{+1}\!$ &
$\! 1.4743^{-4}\!$ &$\! 3.2999^{+1}\!$ &$\! 3.2432^{+1}\!$ &$\! 4.6863^{+2}\!$\\
$\! 10^{-5}\!$ &
$\! 3.3184^{-3}\!$ &$\! 2.0760^{-3}\!$ &$\! 1.1977^{-4}\!$ &$\! 3.3644^{+1}\!$ &
$\! 3.5445^{-4}\!$ &$\! 1.6147^{+1}\!$ &$\! 1.5696^{+1}\!$ &$\! 2.1747^{+2}\!$\\
$\! 10^{-4}\!$ &
$\! 1.4184^{-2}\!$ &$\! 8.4455^{-3}\!$ &$\! 4.6630^{-4}\!$ &$\! 1.5408^{+1}\!$ &
$\! 5.6829^{-4}\!$ &$\! 7.1705^{+0}\!$ &$\! 6.8139^{+0}\!$&$\!8.9820_*^{+1}\!$\\
$\! 10^{-3}\!$ &
$\! 5.9793^{-2}\!$ &$\! 3.4418^{-2}\!$ &$\! 1.6996^{-3}\!$ &$\! 6.4042^{+0}\!$ &
$\! 2.2278^{-4}\!$ &$\! 2.7892^{+0}\!$ &$\! 2.5128^{+0}\!$ &$\! 3.1336^{+1}\!$\\
$\! 10^{-2}\!$ &
$\! 2.3106^{-1}\!$ &$\! 1.2914^{-1}\!$ &$\! 5.7016^{-3}\!$ &$\! 2.2876^{+0}\!$ &
$\!-8.9125^{-4}\!$ &$\! 8.6205^{-1}\!$ &$\! 6.7377^{-1}\!$ &$\! 8.1589^{+0}\!$\\
$\! 0.1    \!$ &
$\! 5.5039^{-1}\!$ &$\! 2.7075^{-1}\!$ &$\! 1.0031^{-2}\!$ &$\! 3.8850^{-1}\!$ &
$\!-4.7466^{-4}\!$ &$\! 1.1332^{-1}\!$ &$\! 5.9489^{-2}\!$ &$\! 9.0795^{-1}\!$\\
$\! 0.3    \!$ &
$\! 3.4890^{-1}\!$ &$\! 1.2949^{-1}\!$ &$\! 2.9943^{-3}\!$ &$\! 3.5090^{-2}\!$ &
$\!-4.9304^{-5}\!$ &$\! 8.9667^{-3}\!$ &$\! 3.2670^{-3}\!$ &$\! 8.4309^{-2}\!$\\
$\! 0.5    \!$ &
$\! 1.2026^{-1}\!$ &$\! 3.1269^{-2}\!$ &$\! 3.7428^{-4}\!$ &$\! 2.3729^{-3}\!$ &
$\!-4.1981^{-6}\!$ &$\! 5.7783^{-4}\!$ &$\! 1.7390^{-4}\!$&$\!8.1099_*^{-3}\!$\\
$\! 0.7    \!$ &
$\! 1.9867^{-2}\!$ &$\! 3.0534^{-3}\!$ &$\! 1.3273^{-5}\!$ &$\! 5.4635^{-5}\!$ &
$\!-1.5541^{-7}\!$ &$\! 1.3275^{-5}\!$ &$\! 3.9930^{-6}\!$ &$\! 3.8824^{-4}\!$\\
$\! 0.9    \!$ &
$\! 3.4524^{-4}\!$ &$\! 1.7466^{-5}\!$ &$\! 8.489 ^{-9}\!$ &$\! 3.030 ^{-8}\!$ &
$\!-2.255^{-10}\!$ &$\! 8.863 ^{-9}\!$ &$\! 4.803 ^{-9}\!$ &$\! 1.2026^{-6}\!$\\
\hline \hline
\end{tabular}
\end{center}
\end{table}

\begin{table}[htp]
\caption{As Table \ref{gsav-t4}, but for the variable-$N_{\rm f}$ evolution
 using the flavour matching conditions of Ref.~\protect\cite
 {Buza:1996wv,Larin:1994va,Chetyrkin:1997sg}.
 The corresponding values for the strong coupling $\alpha_{\rm s}(\mu_{\rm r}^2
 \! =\! 10^4 \mbox{ GeV}^2)$ are given by $0.115818$, $0.115605$ and $0.115410$
 for $\mu_{\rm r}^2 / \mu_{\rm f}^2 = 0.5$, $1$ and $2$, respectively. For
 brevity the small, but non-vanishing valence distributions $s_v$, $c_v$ and 
 $b_v$ are not displayed.}
\label{gsav-t5}
\vspace{2mm}
\begin{center}
\begin{tabular}{||c||r|r|r|r|r|r|r|r||}
\hline \hline
\multicolumn{9}{||c||}{} \\[-3mm]
\multicolumn{9}{||c||}{NNLO, $\, N_{\rm f} = 3\,\ldots 5\,$,
                       $\,\mu_{\rm f}^2 = 10^4 \mbox{ GeV}^2$} \\
\multicolumn{9}{||c||}{} \\[-0.3cm] \hline \hline
 & & & & & & & \\[-0.3cm]
\multicolumn{1}{||c||}{$x$} &
\multicolumn{1}{c|} {$xu_v$} &
\multicolumn{1}{c|} {$xd_v$} &
\multicolumn{1}{c|} {$xL_-$} &
\multicolumn{1}{c|} {$2xL_+$} &
\multicolumn{1}{c|} {$xs_+$} &
\multicolumn{1}{c|} {$xc_+$} &
\multicolumn{1}{c|} {$xb_+$} &
\multicolumn{1}{c||}{$xg$} \\[0.5mm] \hline \hline
\multicolumn{9}{||c||}{} \\[-3mm]
\multicolumn{9}{||c||}{$\mu_{\rm r}^2 = \mu_{\rm f}^2$} \\
\multicolumn{9}{||c||}{} \\[-0.3cm] \hline \hline
 & & & & & & & \\[-0.3cm]
$\! 10^{-7}\!$ &
$\! 1.5978^{-4}\!$ &$\! 1.0699^{-5}\!$ &$\! 6.0090^{-6}\!$ &$\! 1.3916^{+2}\!$ &
$\! 6.8509^{+1}\!$ &$\! 6.6929^{+1}\!$ &$\! 5.7438^{+1}\!$ &$\! 9.9694^{+3}\!$\\
$\! 10^{-6}\!$ &
$\! 7.1787^{-4}\!$ &$\! 4.5929^{-4}\!$ &$\! 2.6569^{-5}\!$ &$\! 7.1710^{+1}\!$ &
$\! 3.5003^{+1}\!$ &$\! 3.3849^{+1}\!$ &$\! 2.8332^{+1}\!$ &$\! 4.8817^{+2}\!$\\
$\! 10^{-5}\!$ &
$\! 3.1907^{-3}\!$ &$\! 1.9532^{-3}\!$ &$\! 1.1116^{-4}\!$ &$\! 3.4732^{+1}\!$ &
$\! 1.6690^{+1}\!$ &$\! 1.5875^{+1}\!$ &$\! 1.2896^{+1}\!$ &$\! 2.2012^{+2}\!$\\
$\! 10^{-4}\!$ &
$\! 1.4023^{-2}\!$ &$\! 8.2749^{-3}\!$ &$\! 4.3744^{-4}\!$ &$\! 1.5617^{+1}\!$ &
$\! 7.2747^{+0}\!$ &$\! 6.7244^{+0}\!$ &$\! 5.2597^{+0}\!$ &$\! 8.8804^{+1}\!$\\
$\! 10^{-3}\!$ &
$\! 6.0019^{-2}\!$ &$\! 3.4519^{-2}\!$ &$\! 1.6296^{-3}\!$ &$\! 6.4173^{+0}\!$ &
$\! 2.7954^{+0}\!$ &$\! 2.4494^{+0}\!$ &$\! 1.8139^{+0}\!$ &$\! 3.0404^{+1}\!$\\
$\! 10^{-2}\!$ &
$\! 2.3244^{-1}\!$ &$\! 1.3000^{-1}\!$ &$\! 5.6100^{-3}\!$ &$\! 2.2778^{+0}\!$ &
$\! 8.5749^{-1}\!$ &$\! 6.6746^{-1}\!$ &$\! 4.5073^{-1}\!$ &$\! 7.7912^{+0}\!$\\
$\! 0.1    \!$ &
$\! 5.4993^{-1}\!$ &$\! 2.7035^{-1}\!$ &$\! 9.9596^{-3}\!$ &$\! 3.8526^{-1}\!$ &
$\! 1.1230^{-1}\!$ &$\! 6.4466^{-2}\!$ &$\! 3.7280^{-2}\!$ &$\! 8.5266^{-1}\!$\\
$\! 0.3    \!$ &
$\! 3.4622^{-1}\!$ &$\! 1.2833^{-1}\!$ &$\! 2.9572^{-3}\!$ &$\! 3.4600^{-2}\!$ &
$\! 8.8410^{-3}\!$ &$\! 4.0134^{-3}\!$ &$\! 2.1047^{-3}\!$ &$\! 7.8898^{-2}\!$\\
$\! 0.5    \!$ &
$\! 1.1868^{-1}\!$ &$\! 3.0811^{-2}\!$ &$\! 3.6760^{-4}\!$ &$\! 2.3198^{-3}\!$ &
$\! 5.6309^{-4}\!$ &$\! 2.3752^{-4}\!$ &$\! 1.2004^{-4}\!$ &$\! 7.6398^{-3}\!$\\
$\! 0.7    \!$ &
$\! 1.9486^{-2}\!$ &$\! 2.9901^{-3}\!$ &$\! 1.2957^{-5}\!$ &$\! 5.2352^{-5}\!$ &
$\! 1.2504^{-5}\!$ &$\! 5.6038^{-6}\!$ &$\! 2.8888^{-6}\!$ &$\! 3.7080^{-4}\!$\\
$\! 0.9    \!$ &
$\! 3.3522^{-4}\!$ &$\! 1.6933^{-5}\!$ &$\! 8.209 ^{-9}\!$ &$\! 2.574 ^{-8}\!$ &
$\! 6.856 ^{-9}\!$ &$\! 4.337 ^{-9}\!$ &$\! 2.679 ^{-9}\!$ &$\! 1.1721^{-6}\!$\\
\hline \hline
\multicolumn{9}{||c||}{} \\[-3mm]
\multicolumn{9}{||c||}{$\mu_{\rm r}^2 = 2\, \mu_{\rm f}^2$} \\
\multicolumn{9}{||c||}{} \\[-0.3cm] \hline \hline
 & & & & & & & \\[-0.3cm]
$\! 10^{-7}\!$ &
$\! 1.3950^{-4}\!$ &$\! 9.0954^{-5}\!$ &$\! 5.2113^{-6}\!$ &$\! 1.3549^{+2}\!$ &
$\! 6.6672^{+1}\!$ &$\! 6.5348^{+1}\!$ &$\! 5.6851^{+1}\!$ &$\! 1.0084^{+3}\!$\\
$\! 10^{-6}\!$ &
$\! 6.4865^{-4}\!$ &$\! 4.0691^{-4}\!$ &$\! 2.3344^{-5}\!$ &$\! 6.9214^{+1}\!$ &
$\! 3.3753^{+1}\!$ &$\! 3.2772^{+1}\!$ &$\! 2.7818^{+1}\!$ &$\! 4.8816^{+2}\!$\\
$\! 10^{-5}\!$ &
$\! 2.9777^{-3}\!$ &$\! 1.8020^{-3}\!$ &$\! 9.9329^{-5}\!$ &$\! 3.3385^{+1}\!$ &
$\! 1.6015^{+1}\!$ &$\! 1.5306^{+1}\!$ &$\! 1.2601^{+1}\!$ &$\! 2.1838^{+2}\!$\\
$\! 10^{-4}\!$ &
$\! 1.3452^{-2}\!$ &$\! 7.9078^{-3}\!$ &$\! 4.0036^{-4}\!$ &$\! 1.5035^{+1}\!$ &
$\! 6.9818^{+0}\!$ &$\! 6.4880^{+0}\!$ &$\! 5.1327^{+0}\!$ &$\! 8.7550^{+1}\!$\\
$\! 10^{-3}\!$ &
$\! 5.8746^{-2}\!$ &$\! 3.3815^{-2}\!$ &$\! 1.5411^{-3}\!$ &$\! 6.2321^{+0}\!$ &
$\! 2.7012^{+0}\!$ &$\! 2.3747^{+0}\!$ &$\! 1.7742^{+0}\!$ &$\! 3.0060^{+1}\!$\\
$\! 10^{-2}\!$ &
$\! 2.3063^{-1}\!$ &$\! 1.2923^{-1}\!$ &$\! 5.4954^{-3}\!$ &$\! 2.2490^{+0}\!$ &
$\! 8.4141^{-1}\!$ &$\! 6.5083^{-1}\!$ &$\! 4.4354^{-1}\!$ &$\! 7.7495^{+0}\!$\\
$\! 0.1    \!$ &
$\! 5.5279^{-1}\!$ &$\! 2.7222^{-1}\!$ &$\! 1.0021^{-2}\!$ &$\! 3.8897^{-1}\!$ &
$\! 1.1312^{-1}\!$ &$\! 6.2917^{-2}\!$ &$\! 3.7048^{-2}\!$ &$\! 8.5897^{-1}\!$\\
$\! 0.3    \!$ &
$\! 3.5141^{-1}\!$ &$\! 1.3051^{-1}\!$ &$\! 3.0134^{-3}\!$ &$\! 3.5398^{-2}\!$ &
$\! 9.0559^{-3}\!$ &$\! 3.8727^{-3}\!$ &$\! 2.0993^{-3}\!$ &$\! 8.0226^{-2}\!$\\
$\! 0.5    \!$ &
$\! 1.2140^{-1}\!$ &$\! 3.1590^{-2}\!$ &$\! 3.7799^{-4}\!$ &$\! 2.3919^{-3}\!$ &
$\! 5.8148^{-4}\!$ &$\! 2.2376^{-4}\!$ &$\! 1.1918^{-4}\!$ &$\! 7.8098^{-3}\!$\\
$\! 0.7    \!$ &
$\! 2.0120^{-2}\!$ &$\! 3.0955^{-3}\!$ &$\! 1.3462^{-5}\!$ &$\! 5.4194^{-5}\!$ &
$\! 1.2896^{-5}\!$ &$\! 5.0329^{-6}\!$ &$\! 2.8153^{-6}\!$ &$\! 3.8099^{-4}\!$\\
$\! 0.9    \!$ &
$\! 3.5230^{-4}\!$ &$\! 1.7849^{-5}\!$ &$\! 8.687 ^{-9}\!$ &$\! 2.568 ^{-8}\!$ &
$\! 6.513 ^{-9}\!$ &$\! 3.390 ^{-9}\!$ &$\! 2.407 ^{-9}\!$ &$\! 1.2188^{-6}\!$\\
\hline \hline
\multicolumn{9}{||c||}{} \\[-3mm]
\multicolumn{9}{||c||}{$\mu_{\rm r}^2 = 1/2\, \mu_{\rm f}^2$} \\
\multicolumn{9}{||c||}{} \\[-0.3cm] \hline \hline
 & & & & & & & \\[-0.3cm]
$\! 10^{-7}\!$ &
$\! 1.8906^{-4}\!$ &$\! 1.3200^{-4}\!$ &$\! 6.9268^{-6}\!$ &$\! 1.3739^{+2}\!$ &
$\! 6.7627^{+1}\!$ &$\! 6.5548^{+1}\!$ &$\! 5.5295^{+1}\!$ &$\! 9.4403^{+2}\!$\\
$\! 10^{-6}\!$ &
$\! 8.1001^{-4}\!$ &$\! 5.3574^{-4}\!$ &$\! 3.0345^{-5}\!$ &$\! 7.2374^{+1}\!$ &
$\! 3.5337^{+1}\!$ &$\! 3.3846^{+1}\!$ &$\! 2.7870^{+1}\!$ &$\! 4.7444^{+2}\!$\\
$\! 10^{-5}\!$ &
$\! 3.4428^{-3}\!$ &$\! 2.1524^{-3}\!$ &$\! 1.2531^{-4}\!$ &$\! 3.5529^{+1}\!$ &
$\! 1.7091^{+1}\!$ &$\! 1.6065^{+1}\!$ &$\! 1.2883^{+1}\!$ &$\! 2.1802^{+2}\!$\\
$\! 10^{-4}\!$ &
$\! 1.4580^{-2}\!$ &$\! 8.6744^{-3}\!$ &$\! 4.8276^{-4}\!$ &$\! 1.6042^{+1}\!$ &
$\! 7.4886^{+0}\!$ &$\! 6.8276^{+0}\!$ &$\! 5.3044^{+0}\!$ &$\! 8.9013^{+1}\!$\\
$\! 10^{-3}\!$ &
$\! 6.0912^{-2}\!$ &$\! 3.5030^{-2}\!$ &$\! 1.7393^{-3}\!$ &$\! 6.5544^{+0}\!$ &
$\! 2.8656^{+0}\!$ &$\! 2.4802^{+0}\!$ &$\! 1.8362^{+0}\!$ &$\! 3.0617^{+1}\!$\\
$\! 10^{-2}\!$ &
$\! 2.3327^{-1}\!$ &$\! 1.3022^{-1}\!$ &$\! 5.7588^{-3}\!$ &$\! 2.2949^{+0}\!$ &
$\! 8.6723^{-1}\!$ &$\! 6.7688^{-1}\!$ &$\! 4.5597^{-1}\!$&$\!7.8243_*^{+0}\!$\\
$\! 0.1    \!$ &
$\! 5.4798^{-1}\!$ &$\! 2.6905^{-1}\!$ &$\! 9.9470^{-3}\!$ &$\! 3.8192^{-1}\!$ &
$\! 1.1124^{-1}\!$ &$\! 6.7091^{-2}\!$ &$\! 3.7698^{-2}\!$ &$\! 8.4908^{-1}\!$\\
$\! 0.3    \!$ &
$\! 3.4291^{-1}\!$ &$\! 1.2693^{-1}\!$ &$\! 2.9239^{-3}\!$ &$\! 3.4069^{-2}\!$ &
$\! 8.6867^{-3}\!$ &$\! 4.3924^{-3}\!$ &$\! 2.1435^{-3}\!$ &$\! 7.8109^{-2}\!$\\
$\! 0.5    \!$ &
$\! 1.1694^{-1}\!$ &$\! 3.0310^{-2}\!$ &$\! 3.6112^{-4}\!$ &$\! 2.2828^{-3}\!$ &
$\! 5.5537^{-4}\!$ &$\! 2.7744^{-4}\!$ &$\! 1.2416^{-4}\!$ &$\! 7.5371^{-3}\!$\\
$\! 0.7    \!$ &
$\! 1.9076^{-2}\!$ &$\! 2.9217^{-3}\!$ &$\! 1.2635^{-5}\!$ &$\! 5.2061^{-5}\!$ &
$\! 1.2677^{-5}\!$ &$\! 7.2083^{-6}\!$ &$\! 3.0908^{-6}\!$ &$\! 3.6441^{-4}\!$\\
$\! 0.9    \!$ &
$\! 3.2404^{-4}\!$ &$\! 1.6333^{-5}\!$ &$\! 7.900 ^{-9}\!$ &$\! 2.850 ^{-8}\!$ &
$\! 8.407 ^{-9}\!$ &$\! 6.795 ^{-9}\!$ &$\! 3.205 ^{-9}\!$ & $\!1.1411^{-6}\!$\\
\hline \hline
\end{tabular}
\end{center}
\end{table}

\begin{table}[htp]
\caption{Reference results for the $N_{\rm f}=4$ (FFN) and the variable-$N_{\rm 
 f}$ (VFN) polarized leading-order evolution of the initial distributions
 (\ref{gsav-eq7}), shown together with these boundary conditions. The 
 respective values for $\alpha_{\rm s} (\mu_{\rm r}^2\! =\! \mu_{\rm f}^2\! 
 =\! 10^4 \mbox{ GeV}^2)$ read $0.117574$ (FFN) and $0.122306$ (VFN). 
 The notation is the same as for the unpolarized case.}
\label{gsav-t1}
\begin{center}
\vspace{-2mm}
\begin{tabular}{||c||r|r|r|r|r|r|r|r||}
\hline \hline
\multicolumn{9}{||c||}{} \\[-3mm]
\multicolumn{1}{||c||}{$x$} &
\multicolumn{1}{c|} {$xu_v$} &
\multicolumn{1}{c|} {$-xd_v$} &
\multicolumn{1}{c|} {$-xL_-$} &
\multicolumn{1}{c|} {$-2xL_+$} &
\multicolumn{1}{c|} {$xs_+$} &
\multicolumn{1}{c|} {$xc_+$} &
\multicolumn{1}{c|} {$xb_+$} &
\multicolumn{1}{c||}{$xg$} \\[0.5mm] \hline \hline
\multicolumn{9}{||c||}{} \\[-3mm]
\multicolumn{9}{||c||}{Pol.~input, $\,\mu_{\rm f}^2 = 2 \mbox{ GeV}^2$} \\
\multicolumn{9}{||c||}{} \\[-0.3cm] \hline \hline
 & & & & & & & \\[-0.3cm]
$\! 10^{-7}\!\!$ &
$\! 1.6366^{-5}\!$ &$\! 6.2946^{-6}\!$ &$\! 7.9433^{-5}\!$ &$\! 1.5887^{-3}\!$ &
$\!-3.9716^{-4}\!$ &$\! 0.0   ^{+0}\!$ &$\! 0.0   ^{+0}\!$ &$\! 4.7434^{-4}\!$\\
$\! 10^{-6}\!\!$ &
$\! 8.2024^{-5}\!$ &$\! 3.1548^{-5}\!$ &$\! 1.5849^{-4}\!$ &$\! 3.1698^{-3}\!$ &
$\!-7.9244^{-4}\!$ &$\! 0.0   ^{+0}\!$ &$\! 0.0   ^{+0}\!$ &$\! 1.5000^{-3}\!$\\
$\! 10^{-5}\!\!$ &
$\! 4.1110^{-4}\!$ &$\! 1.5811^{-4}\!$ &$\! 3.1621^{-4}\!$ &$\! 6.3241^{-3}\!$ &
$\!-1.5810^{-3}\!$ &$\! 0.0   ^{+0}\!$ &$\! 0.0   ^{+0}\!$ &$\! 4.7432^{-3}\!$\\
$\! 10^{-4}\!\!$ &
$\! 2.0604^{-3}\!$ &$\! 7.9245^{-4}\!$ &$\! 6.3052^{-4}\!$ &$\! 1.2610^{-2}\!$ &
$\!-3.1526^{-3}\!$ &$\! 0.0   ^{+0}\!$ &$\! 0.0   ^{+0}\!$ &$\! 1.4993^{-2}\!$\\
$\! 10^{-3}\!\!$ &
$\! 1.0326^{-2}\!$ &$\! 3.9716^{-3}\!$ &$\! 1.2501^{-3}\!$ &$\! 2.5003^{-2}\!$ &
$\!-6.2507^{-3}\!$ &$\! 0.0   ^{+0}\!$ &$\! 0.0   ^{+0}\!$ &$\! 4.7197^{-2}\!$\\
$\! 10^{-2}\!\!$ &
$\! 5.1723^{-2}\!$ &$\! 1.9886^{-2}\!$ &$\! 2.3412^{-3}\!$ &$\! 4.6825^{-2}\!$ &
$\!-1.1706^{-2}\!$ &$\! 0.0   ^{+0}\!$ &$\! 0.0   ^{+0}\!$ &$\! 1.4265^{-1}\!$\\
$\! 0.1    \!\!$ &
$\! 2.4582^{-1}\!$ &$\! 9.1636^{-2}\!$ &$\! 2.3972^{-3}\!$ &$\! 4.7943^{-2}\!$ &
$\!-1.1986^{-2}\!$ &$\! 0.0   ^{+0}\!$ &$\! 0.0   ^{+0}\!$ &$\! 2.8009^{-1}\!$\\
$\! 0.3    \!\!$ &
$\! 3.6473^{-1}\!$ &$\! 1.1370^{-1}\!$ &$\! 5.7388^{-4}\!$ &$\! 1.1478^{-2}\!$ &
$\!-2.8694^{-3}\!$ &$\! 0.0   ^{+0}\!$ &$\! 0.0   ^{+0}\!$ &$\! 1.3808^{-1}\!$\\
$\! 0.5    \!\!$ &
$\! 2.5008^{-1}\!$ &$\! 5.7710^{-2}\!$ &$\! 6.3457^{-5}\!$ &$\! 1.2691^{-3}\!$ &
$\!-3.1729^{-4}\!$ &$\! 0.0   ^{+0}\!$ &$\! 0.0   ^{+0}\!$ &$\! 3.3146^{-2}\!$\\
$\! 0.7    \!\!$ &
$\! 8.4769^{-2}\!$ &$\! 1.1990^{-2}\!$ &$\! 1.9651^{-6}\!$ &$\! 3.9301^{-5}\!$ &
$\!-9.8254^{-6}\!$ &$\! 0.0   ^{+0}\!$ &$\! 0.0   ^{+0}\!$ &$\! 3.0496^{-3}\!$\\
$\! 0.9    \!$ &
$\! 4.4680^{-3}\!$ &$\! 2.1365^{-4}\!$ &$\! 9.689^{-10}\!$ &$\! 1.9378^{-8}\!$ &
$\!-4.8444^{-9}\!$ &$\! 0.0   ^{+0}\!$ &$\! 0.0   ^{+0}\!$ &$\! 1.4230^{-5}\!$\\
\hline \hline
\multicolumn{9}{||c||}{} \\[-3mm]
\multicolumn{9}{||c||}{LO, $\, N_{\rm f} = 4\,$, 
                       $\,\mu_{\rm f}^2 = 10^4 \mbox{ GeV}^2$} \\
\multicolumn{9}{||c||}{} \\[-0.3cm] \hline \hline
 & & & & & & & \\[-0.3cm]
$\! 10^{-7}\!\!$ &
$\! 4.8350_*^{-5}\!$ &$\! 1.8556^{-5}\!$ &$\! 1.0385^{-4}\!$ &$\! 3.5124^{-3}\!$ &
$\!\!-1.2370^{-3}\!$ &$\!\!-7.1774^{-4}\!$ &$\! 0.0^{+0}\!$ &$\! 1.4116^{-2}\!$\\
$\! 10^{-6}\!\!$ &
$\! 2.3504^{-4}\!$ &$\! 9.0090^{-5}\!$ &$\! 2.0700^{-4}\!$ &$\! 7.7716^{-3}\!$ &
$\!\!-2.8508^{-3}\!$ &$\!\!-1.8158^{-3}\!$ &$\! 0.0^{+0}\!$ &$\! 4.2163^{-2}\!$\\
$\! 10^{-5}\!\!$ &
$\! 1.1220^{-3}\!$ &$\! 4.2916^{-4}\!$ &$\! 4.1147^{-4}\!$ &$\! 1.6007^{-2}\!$ &
$\!\!-5.9463^{-3}\!$ &$\!\!-3.8889^{-3}\!$ &$\! 0.0^{+0}\!$ &$\! 1.0922^{-1}\!$\\
$\! 10^{-4}\!\!$ &
$\! 5.1990^{-3}\!$ &$\! 1.9818^{-3}\!$ &$\! 8.0948^{-4}\!$ &$\! 2.8757^{-2}\!$ &
$\!\!-1.0331^{-2}\!$ &$\!\!-6.2836^{-3}\!$ &$\! 0.0^{+0}\!$ &$\! 2.4069^{-1}\!$\\
$\! 10^{-3}\!\!$ &
$\! 2.2900^{-2}\!$ &$\! 8.6763^{-3}\!$ &$\! 1.5309^{-3}\!$ &$\! 4.0166^{-2}\!$ &
$\!\!-1.2428^{-2}\!$ &$\!\!-4.7739^{-3}\!$ &$\! 0.0^{+0}\!$ &$\! 4.2181^{-1}\!$\\
$\! 10^{-2}\!\!$ &
$\! 9.1489^{-2}\!$ &$\! 3.4200^{-2}\!$ &$\! 2.4502^{-3}\!$ &$\! 3.3928^{-2}\!$ &
$\!\!-4.7126^{-3}\!$ &$\! 7.5385^{-3}\!$ &$\! 0.0^{+0}\!$ &$\! 4.9485^{-1}\!$\\
$\! 0.1    \!\!$ &
$\! 2.6494^{-1}\!$ &$\! 9.1898^{-2}\!$ &$\! 1.5309^{-3}\!$ &$\! 8.5427^{-3}\!$ &
$\! 3.3830^{-3}\!$ &$\! 1.1037^{-2}\!$ &$\! 0.0^{+0}\!$ &$\! 2.0503^{-1}\!$\\
$\! 0.3    \!\!$ &
$\! 2.2668^{-1}\!$ &$\! 6.2946^{-2}\!$ &$\! 2.1104^{-4}\!$ &$\! 6.6698^{-4}\!$ &
$\! 7.2173^{-4}\!$ &$\! 1.7769^{-3}\!$ &$\! 0.0^{+0}\!$ &$\! 3.3980^{-2}\!$\\
$\! 0.5    \!\!$ &
$\! 9.7647^{-2}\!$ &$\! 1.9652^{-2}\!$ &$\! 1.4789^{-5}\!$ &$\!\!-1.8850^{-5}\!$ &
$\! 8.3371^{-5}\!$ &$\! 1.5732^{-4}\!$ &$\! 0.0^{+0}\!$ &$\! 4.3802^{-3}\!$\\
$\! 0.7    \!\!$ &
$\! 1.9545^{-2}\!$ &$\! 2.3809^{-3}\!$ &$\! 2.7279^{-7}\!$ &$\!\!-4.1807^{-6}\!$ &
$\! 3.4543^{-6}\!$ &$\! 4.8183^{-6}\!$ &$\! 0.0^{+0}\!$ &$\! 2.6355^{-4}\!$\\
$\! 0.9    \!\!$ &
$\! 4.1768^{-4}\!$ &$\! 1.7059^{-5}\!$ &$\! 5.494 ^{-11}\!$&$\!\!-7.6712^{-9}\!$ &
$\! 4.1103^{-9}\!$ &$\! 4.3850^{-9}\!$ &$\! 0.0^{+0}\!$ &$\! 9.8421^{-7}\!$\\
\hline \hline
\multicolumn{9}{||c||}{} \\[-3mm]
\multicolumn{9}{||c||}{LO, $\, N_{\rm f} = 3\ldots 5\,$, 
                        $\,\mu_{\rm f}^2 = 10^4 \mbox{ GeV}^2$} \\
\multicolumn{9}{||c||}{} \\[-0.3cm] \hline \hline
 & & & & & & & \\[-0.3cm]
$\! 10^{-7}\!\!$ &
$\! 4.9026^{-5}\!$ &$\! 1.8815^{-5}\!$ &$\! 1.0422^{-4}\!$ &$\! 3.5315^{-3}\!$ &
$\!\!-1.2447^{-3}\!$ &$\!\!-7.2356^{-4}\!$ &$\!-6.2276^{-4}\!$ &$\! 1.3726^{-2}\!$\\
$\! 10^{-6}\!\!$ &
$\! 2.3818^{-4}\!$ &$\! 9.1286^{-5}\!$ &$\! 2.0774^{-4}\!$ &$\! 7.8108^{-3}\!$ &
$\!\!-2.8667^{-3}\!$ &$\!\!-1.8280^{-3}\!$ &$\!-1.5301^{-3}\!$ &$\! 4.1011^{-2}\!$\\
$\! 10^{-5}\!\!$ &
$\! 1.1359^{-3}\!$ &$\! 4.3445^{-4}\!$ &$\! 4.1289^{-4}\!$ &$\! 1.6070^{-2}\!$ &
$\!\!-5.9705^{-3}\!$ &$\!\!-3.9060^{-3}\!$ &$\!-3.1196^{-3}\!$ &$\! 1.0615^{-1}\!$\\
$\! 10^{-4}\!\!$ &
$\! 5.2567^{-3}\!$ &$\! 2.0035^{-3}\!$ &$\! 8.1206^{-4}\!$ &$\! 2.8811^{-2}\!$ &
$\!\!-1.0345^{-2}\!$ &$\!\!-6.2849^{-3}\!$ &$\!-4.5871^{-3}\!$ &$\! 2.3343^{-1}\!$\\
$\! 10^{-3}\!\!$ &
$\! 2.3109^{-2}\!$ &$\! 8.7537^{-3}\!$ &$\! 1.5345^{-3}\!$ &$\! 4.0125^{-2}\!$ &
$\!\!-1.2390^{-2}\!$ &$\!\!-4.7174^{-3}\!$ &$\!-2.4822^{-3}\!$ &$\! 4.0743^{-1}\!$\\
$\! 10^{-2}\!\!$ &
$\! 9.2035^{-2}\!$ &$\! 3.4391^{-2}\!$ &$\! 2.4501^{-3}\!$ &$\! 3.3804^{-2}\!$ &
$\!\!-4.6512^{-3}\!$ &$\!\! 7.5994^{-3}\!$ &$\! 6.4665^{-3}\!$ &$\! 4.7445^{-1}\!$\\
$\! 0.1    \!\!$ &
$\! 2.6478^{-1}\!$ &$\! 9.1762^{-2}\!$ &$\! 1.5206^{-3}\!$ &$\! 8.5181^{-3}\!$ &
$\! 3.3438^{-3}\!$ &$\! 1.0947^{-2}\!$ &$\! 6.5223^{-3}\!$ &$\! 1.9402^{-1}\!$\\
$\! 0.3    \!\!$ &
$\! 2.2495^{-1}\!$ &$\! 6.2376^{-2}\!$ &$\! 2.0811^{-4}\!$ &$\! 6.6195^{-4}\!$ &
$\! 7.0957^{-4}\!$ &$\! 1.7501^{-3}\!$ &$\! 9.2045^{-4}\!$ &$\! 3.1960^{-2}\!$\\
$\! 0.5    \!\!$ &
$\! 9.6318^{-2}\!$ &$\! 1.9353^{-2}\!$ &$\! 1.4496^{-5}\!$ &$\!\!-1.8549^{-5}\!$ &
$\! 8.1756^{-5}\!$ &$\! 1.5424^{-4}\!$ &$\! 7.8577^{-5}\!$ &$\! 4.1226^{-3}\!$\\
$\! 0.7    \!\!$ &
$\! 1.9147^{-2}\!$ &$\! 2.3281^{-3}\!$ &$\! 2.6556^{-7}\!$ &$\!\!-4.0936^{-6}\!$ &
$\! 3.3746^{-6}\!$ &$\! 4.7024^{-6}\!$ &$\! 2.4901^{-6}\!$ &$\! 2.4888^{-4}\!$\\
$\! 0.9    \!\!$ &
$\! 4.0430^{-4}\!$ &$\! 1.6480^{-5}\!$ &$\! 5.285^{-11}\!$ &$\!\!-7.4351^{-9}\!$ &
$\! 3.9818^{-9}\!$ &$\! 4.2460^{-9}\!$ &$\! 2.6319^{-9}\!$ &$\! 9.2939^{-7}\!$\\
\hline \hline
\end{tabular}
\end{center}
\end{table}

\begin{table}[htp]
\caption{Reference results for the polarized next-to-leading-order polarized 
 evolution of the initial distributions (\ref{gsav-eq7}) with $N_{\rm f}\!=\!4$
 quark flavours. 
 The corresponding value of the strong coupling is $\alpha_{\rm s}
 (\mu_{\rm r}^2 \! =\! 10^4 \mbox{ GeV}^2) = 0.110902$. As in the leading-order 
 case, the valence distributions $s_v$ and $c_v$ vanish for the input 
 (\ref{gsav-eq7}).} 
\label{gsav-t2}
\vspace{2mm}
\begin{center}
\begin{tabular}{||c||r|r|r|r|r|r|r||}
\hline \hline
\multicolumn{8}{||c||}{} \\[-3mm]
\multicolumn{8}{||c||}{Pol.~NLO, $\, N_{\rm f} = 4$, 
                       $\, \mu_{\rm f}^2 = 10^4 \mbox{ GeV}^2$} \\
\multicolumn{8}{||c||}{} \\[-0.35cm] \hline \hline
 & & & & & & & \\[-0.3cm]
\multicolumn{1}{||c||}{$x$} &
\multicolumn{1}{c|} {$xu_v$} &
\multicolumn{1}{c|} {$xd_v$} &
\multicolumn{1}{c|} {$xL_-$} &
\multicolumn{1}{c|} {$2xL_+$} &
\multicolumn{1}{c|} {$xs_+$} &
\multicolumn{1}{c|} {$xc_+$} &
\multicolumn{1}{c||}{$xg$} \\ \hline \hline
\multicolumn{8}{||c||}{} \\[-3mm]
\multicolumn{8}{||c||}{$\mu_{\rm r}^2 = \mu_{\rm f}^2$} \\
\multicolumn{8}{||c||}{} \\[-0.3cm] \hline \hline
 & & & & & & & \\[-0.3cm]
$\! 10^{-7}\!$ &
$\! 6.7336^{-5}\!$ &$\!-2.5747^{-5}\!$ &$\!-1.1434^{-4}\!$ &$\!-5.2002^{-3}\!$ &
$\!-2.0528^{-3}\!$ &$\!-1.5034^{-3}\!$ &$\! 2.6955^{-2}\!$\\
$\! 10^{-6}\!$ &
$\! 3.1280^{-4}\!$ &$\!-1.1938^{-4}\!$ &$\!-2.3497^{-4}\!$ &$\!-1.0725^{-2}\!$ &
$\!-4.2774^{-3}\!$ &$\!-3.1845^{-3}\!$ &$\! 6.5928^{-2}\!$\\
$\! 10^{-5}\!$ &
$\! 1.4180^{-3}\!$ &$\!-5.3982^{-4}\!$ &$\!-4.8579^{-4}\!$ &$\!-1.9994^{-2}\!$ &
$\!-7.8594^{-3}\!$ &$\!-5.6970^{-3}\!$ &$\! 1.4414^{-1}\!$\\
$\! 10^{-4}\!$ &
$\! 6.2085^{-3}\!$ &$\!-2.3546^{-3}\!$ &$\!-9.8473^{-4}\!$ &$\!-3.1788^{-2}\!$ &
$\!-1.1749^{-2}\!$ &$\!-7.5376^{-3}\!$ &$\! 2.7537^{-1}\!$\\
$\! 10^{-3}\!$ &
$\! 2.5741^{-2}\!$ &$\!-9.7004^{-3}\!$ &$\!-1.8276^{-3}\!$ &$\!-3.8222^{-2}\!$ &
$\!-1.1427^{-2}\!$ &$\!-3.6138^{-3}\!$ &$\! 4.3388^{-1}\!$\\
$\! 10^{-2}\!$ &
$\! 9.6288^{-2}\!$ &$\!-3.5778^{-2}\!$ &$\!-2.6427^{-3}\!$ &$\!-2.6437^{-2}\!$ &
$\!-1.2328^{-3}\!$ &$\! 1.0869^{-2}\!$ &$\! 4.8281^{-1}\!$\\
$\! 0.1    \!$ &
$\! 2.5843^{-1}\!$ &$\!-8.9093^{-2}\!$ &$\!-1.4593^{-3}\!$ &$\!-7.5546^{-3}\!$ &
$\! 3.4258^{-3}\!$ &$\! 1.0639^{-2}\!$ &$\! 2.0096^{-1}\!$\\
$\! 0.3    \!$ &
$\! 2.1248^{-1}\!$ &$\!-5.8641^{-2}\!$ &$\!-1.9269^{-4}\!$ &$\!-1.2210^{-3}\!$ &
$\! 3.5155^{-4}\!$ &$\! 1.3138^{-3}\!$ &$\! 3.4126^{-2}\!$\\
$\! 0.5    \!$ &
$\! 8.9180^{-2}\!$ &$\!-1.7817^{-2}\!$ &$\!-1.3125^{-5}\!$ &$\!-9.1573^{-5}\!$ &
$\! 1.9823^{-5}\!$ &$\! 8.5435^{-5}\!$ &$\! 4.5803^{-3}\!$\\
$\! 0.7    \!$ &
$\! 1.7300^{-2}\!$ &$\!-2.0885^{-3}\!$ &$\!-2.3388^{-7}\!$ &$\!-1.9691^{-6}\!$ &
$\! 1.8480^{-7}\!$ &$\! 1.3541^{-6}\!$ &$\! 2.9526^{-4}\!$\\
$\! 0.9    \!$ &
$\! 3.4726^{-4}\!$ &$\!-1.4028^{-5}\!$ &$\!-4.407^{-11}\!$ &$\!-4.247 ^{-9}\!$ &
$\!-1.903 ^{-9}\!$ &$\!-1.683 ^{-9}\!$ &$\! 1.2520^{-6}\!$\\
\hline \hline
\multicolumn{8}{||c||}{} \\[-3mm]
\multicolumn{8}{||c||}{$\mu_{\rm r}^2 = 2\, \mu_{\rm f}^2$} \\
\multicolumn{8}{||c||}{} \\[-0.3cm] \hline \hline
 & & & & & & & \\[-0.3cm]
$\! 10^{-7}\!$ &
$\! 6.1781^{-5}\!$ &$\!-2.3641^{-5}\!$ &$\!-1.1137^{-4}\!$ &$\!-4.6947^{-3}\!$ &
$\!-1.8092^{-3}\!$ &$\!-1.2695^{-3}\!$ &$\! 2.2530^{-2}\!$\\
$\! 10^{-6}\!$ &
$\! 2.8974^{-4}\!$ &$\!-1.1068^{-4}\!$ &$\!-2.2755^{-4}\!$ &$\!-9.8528^{-3}\!$ &
$\!-3.8580^{-3}\!$ &$\!-2.7838^{-3}\!$ &$\! 5.7272_*^{-2}\!$\\
$\! 10^{-5}\!$ &
$\! 1.3281^{-3}\!$ &$\!-5.0612^{-4}\!$ &$\!-4.6740^{-4}\!$ &$\!-1.8799^{-2}\!$ &
$\!-7.2908^{-3}\!$ &$\!-5.1629^{-3}\!$ &$\! 1.2975^{-1}\!$\\
$\! 10^{-4}\!$ &
$\! 5.8891^{-3}\!$ &$\!-2.2361^{-3}\!$ &$\!-9.4412^{-4}\!$ &$\!-3.0787^{-2}\!$ &
$\!-1.1292^{-2}\!$ &$\!-7.1363^{-3}\!$ &$\! 2.5644^{-1}\!$\\
$\! 10^{-3}\!$ &
$\! 2.4777^{-2}\!$ &$\!-9.3502^{-3}\!$ &$\!-1.7632^{-3}\!$ &$\!-3.8610^{-2}\!$ &
$\!-1.1658^{-2}\!$ &$\!-3.9083^{-3}\!$ &$\! 4.1725^{-1}\!$\\
$\! 10^{-2}\!$ &
$\! 9.4371^{-2}\!$ &$\!-3.5129^{-2}\!$ &$\!-2.6087^{-3}\!$ &$\!-2.8767^{-2}\!$ &
$\!-2.3430_*^{-3}\!$ &$\! 9.7922_*^{-3}\!$ &$\! 4.7804^{-1}\!$\\
$\! 0.1    \!$ &
$\! 2.6008^{-1}\!$ &$\!-8.9915^{-2}\!$ &$\!-1.4923^{-3}\!$ &$\!-8.3806^{-3}\!$ &
$\! 3.1932^{-3}\!$ &$\! 1.0585^{-2}\!$ &$\! 2.0495^{-1}\!$\\
$\! 0.3    \!$ &
$\! 2.1837^{-1}\!$ &$\!-6.0497^{-2}\!$ &$\!-2.0143^{-4}\!$ &$\!-1.2157^{-3}\!$ &
$\! 3.9810^{-4}\!$ &$\! 1.4042^{-3}\!$ &$\! 3.5366^{-2}\!$\\
$\! 0.5    \!$ &
$\! 9.3169^{-2}\!$ &$\!-1.8699^{-2}\!$ &$\!-1.3954^{-5}\!$ &$\!-7.9331^{-5}\!$ &
$\! 3.0091^{-5}\!$ &$\! 9.9849^{-5}\!$ &$\! 4.7690^{-3}\!$\\
$\! 0.7    \!$ &
$\! 1.8423^{-2}\!$ &$\!-2.2357^{-3}\!$ &$\!-2.5360^{-7}\!$ &$\!-1.0062^{-6}\!$ &
$\! 7.6483^{-7}\!$ &$\! 2.0328^{-6}\!$ &$\! 3.0796^{-4}\!$\\
$\! 0.9    \!$ &
$\! 3.8293^{-4}\!$ &$\!-1.5559^{-5}\!$ &$\!-4.952^{-11}\!$ &$\!-1.955 ^{-9}\!$ &
$\!-7.298^{-10}\!$ &$\!-4.822^{-10}\!$ &$\! 1.3247^{-6}\!$\\
\hline \hline
\multicolumn{8}{||c||}{} \\[-3mm]
\multicolumn{8}{||c||}{$\mu_{\rm r}^2 = 1/2\, \mu_{\rm f}^2$} \\
\multicolumn{8}{||c||}{} \\[-0.3cm] \hline \hline
 & & & & & & & \\[-0.3cm]
$\! 10^{-7}\!$ &
$\! 7.4443^{-5}\!$ &$\!-2.8435^{-5}\!$ &$\!-1.1815^{-4}\!$ &$\!-5.7829^{-3}\!$ &
$\!-2.3341^{-3}\!$ &$\!-1.7739^{-3}\!$ &$\! 3.2071^{-2}\!$\\
$\! 10^{-6}\!$ &
$\! 3.4143^{-4}\!$ &$\!-1.3016^{-4}\!$ &$\!-2.4482^{-4}\!$ &$\!-1.1668^{-2}\!$ &
$\!-4.7305^{-3}\!$ &$\!-3.6168^{-3}\!$ &$\! 7.5123^{-2}\!$\\
$\! 10^{-5}\!$ &
$\! 1.5256^{-3}\!$ &$\!-5.8002^{-4}\!$ &$\!-5.1085^{-4}\!$ &$\!-2.1193^{-2}\!$ &
$\!-8.4295^{-3}\!$ &$\!-6.2295_*^{-3}\!$ &$\! 1.5788^{-1}\!$\\
$\! 10^{-4}\!$ &
$\! 6.5726^{-3}\!$ &$\!-2.4891^{-3}\!$ &$\!-1.0409^{-3}\!$ &$\!-3.2697^{-2}\!$ &
$\!-1.2166^{-2}\!$ &$\!-7.8952^{-3}\!$ &$\! 2.9079^{-1}\!$\\
$\! 10^{-3}\!$ &
$\! 2.6766^{-2}\!$ &$\!-1.0070^{-2}\!$ &$\!-1.9171^{-3}\!$ &$\!-3.7730^{-2}\!$ &
$\!-1.1160^{-2}\!$ &$\!-3.2890_*^{-3}\!$ &$\! 4.4380^{-1}\!$\\
$\! 10^{-2}\!$ &
$\! 9.8073^{-2}\!$ &$\!-3.6370^{-2}\!$ &$\!-2.6942^{-3}\!$ &$\!-2.4056^{-2}\!$ &
$\!-1.2354_*^{-4}\!$ &$\! 1.1929^{-2}\!$ &$\! 4.8272^{-1}\!$\\
$\! 0.1    \!$ &
$\! 2.5628^{-1}\!$ &$\!-8.8133^{-2}\!$ &$\!-1.4304^{-3}\!$ &$\!-6.9572^{-3}\!$ &
$\! 3.5561^{-3}\!$ &$\! 1.0604^{-2}\!$ &$\! 1.9831^{-1}\!$\\
$\! 0.3    \!$ &
$\! 2.0709^{-1}\!$ &$\!-5.6988^{-2}\!$ &$\!-1.8541^{-4}\!$ &$\!-1.3308^{-3}\!$ &
$\! 2.5993^{-4}\!$ &$\! 1.1855^{-3}\!$ &$\! 3.3524^{-2}\!$\\
$\! 0.5    \!$ &
$\! 8.5835^{-2}\!$ &$\!-1.7089^{-2}\!$ &$\!-1.2463^{-5}\!$ &$\!-1.1920^{-4}\!$ &
$\! 2.6972_*^{-6}\!$ &$\! 6.4995^{-5}\!$ &$\! 4.5044^{-3}\!$\\
$\! 0.7    \!$ &
$\! 1.6405^{-2}\!$ &$\!-1.9723^{-3}\!$&$\!-2.1859_*^{-7}\!$&$\!-3.6817^{-6}\!$ &
$\!-7.4795_*^{-7}\!$ &$\! 3.4496^{-7}\!$ &$\! 2.9100^{-4}\!$\\
$\! 0.9    \!$ &
$\! 3.2011^{-4}\!$ &$\!-1.2870^{-5}\!$ &$\!-4.000^{-11}\!$ &$\!-8.173^{-9}\!$ &
$\!-3.886 ^{-9}\!$ &$\!-3.686 ^{-9}\!$ &$\! 1.2230^{-6}\!$\\
\hline \hline
\end{tabular}
\end{center}
\end{table}

\begin{table}[htp]
\caption{As Table \ref{gsav-t2}, but for the variable-$N_{\rm f}$ evolution 
 using Eqs.~(\ref{gsav-eq3}), (\ref{gsav-eq4}) and (\ref{gsav-eq7}).
 The corresponding values for the strong coupling $\alpha_{\rm s}(\mu_{\rm r}^2 
 \! =\! 10^4 \mbox{ GeV}^2)$ are given by $0.116461$, $0.116032$ and $0.115663$
 for $\mu_{\rm r}^2 / \mu_{\rm f}^2 = 0.5$, $1$ and $2$, respectively.} 
\label{gsav-t3}
\vspace{-2mm}
\begin{center}
\begin{tabular}{||c||r|r|r|r|r|r|r|r||}
\hline \hline
\multicolumn{9}{||c||}{} \\[-3mm]
\multicolumn{9}{||c||}{Pol.~NLO, $\, N_{\rm f} = 3\,\ldots 5\,$,
                       $\,\mu_{\rm f}^2 = 10^4 \mbox{ GeV}^2$} \\
\multicolumn{9}{||c||}{} \\[-0.3cm] \hline \hline
 & & & & & & & \\[-0.3cm]
\multicolumn{1}{||c||}{$x$} &
\multicolumn{1}{c|} {$xu_v$} &
\multicolumn{1}{c|} {$-xd_v$} &
\multicolumn{1}{c|} {$-xL_-$} &
\multicolumn{1}{c|} {$-2xL_+$} &
\multicolumn{1}{c|} {$xs_+$} &
\multicolumn{1}{c|} {$xc_+$} &
\multicolumn{1}{c|} {$xb_+$} &
\multicolumn{1}{c||}{$xg$} \\[0.5mm] \hline \hline
\multicolumn{9}{||c||}{} \\[-3mm]
\multicolumn{9}{||c||}{$\mu_{\rm r}^2 = \mu_{\rm f}^2$} \\
\multicolumn{9}{||c||}{} \\[-0.3cm] \hline \hline
 & & & & & & & \\[-0.3cm]
$\! 10^{-7}\!\!$ &
$\!6.8787_*^{-5}\!$&$\! 2.6297^{-5}\!$ &$\! 1.1496^{-4}\!$ &$\! 5.2176^{-3}\!$ &
$\!-2.0592^{-3}\!$ &$\!-1.5076^{-3}\!$ &$\!-1.2411^{-3}\!$ &$\! 2.5681^{-2}\!$\\
$\! 10^{-6}\!\!$ &
$\! 3.1881^{-4}\!$ &$\! 1.2165^{-4}\!$ &$\! 2.3638^{-4}\!$ &$\! 1.0770^{-2}\!$ &
$\!-4.2953^{-3}\!$ &$\!-3.1979^{-3}\!$ &$\!-2.4951^{-3}\!$ &$\! 6.3021^{-2}\!$\\
$\! 10^{-5}\!\!$ &
$\! 1.4413^{-3}\!$ &$\! 5.4856^{-4}\!$ &$\! 4.8893^{-4}\!$ &$\! 2.0077^{-2}\!$ &
$\!-7.8934^{-3}\!$ &$\!-5.7228^{-3}\!$ &$\!-4.1488^{-3}\!$ &$\! 1.3809^{-1}\!$\\
$\! 10^{-4}\!\!$ &
$\! 6.2902^{-3}\!$ &$\! 2.3849^{-3}\!$ &$\! 9.9100^{-4}\!$ &$\! 3.1883^{-2}\!$ &
$\!-1.1785^{-2}\!$ &$\!-7.5596^{-3}\!$ &$\!-4.8420^{-3}\!$ &$\! 2.6411^{-1}\!$\\
$\! 10^{-3}\!\!$ &
$\! 2.5980^{-2}\!$ &$\! 9.7872^{-3}\!$ &$\! 1.8364^{-3}\!$ &$\! 3.8224^{-2}\!$ &
$\!-1.1416^{-2}\!$ &$\!-3.5879^{-3}\!$ &$\!-1.1723^{-3}\!$ &$\! 4.1601^{-1}\!$\\
$\! 10^{-2}\!\!$ &
$\! 9.6750^{-2}\!$ &$\! 3.5935^{-2}\!$ &$\! 2.6452^{-3}\!$ &$\! 2.6306^{-2}\!$ &
$\!-1.1774^{-3}\!$ &$\! 1.0917^{-2}\!$ &$\! 8.1196^{-3}\!$ &$\! 4.6178^{-1}\!$\\
$\! 0.1    \!$ &
$\! 2.5807^{-1}\!$ &$\! 8.8905^{-2}\!$ &$\! 1.4509^{-3}\!$ &$\! 7.4778^{-3}\!$ &
$\! 3.4207^{-3}\!$ &$\! 1.0591^{-2}\!$ &$\! 6.1480^{-3}\!$ &$\! 1.9143^{-1}\!$\\
$\! 0.3    \!$ &
$\! 2.1104^{-1}\!$ &$\! 5.8186^{-2}\!$ &$\! 1.9054^{-4}\!$ &$\! 1.2026^{-3}\!$ &
$\! 3.4999^{-4}\!$ &$\! 1.3015^{-3}\!$ &$\! 7.2795^{-4}\!$ &$\! 3.2621^{-2}\!$\\
$\! 0.5    \!$ &
$\! 8.8199^{-2}\!$ &$\! 1.7601^{-2}\!$ &$\! 1.2924^{-5}\!$ &$\! 8.9668^{-5}\!$ &
$\! 1.9771^{-5}\!$ &$\! 8.4378^{-5}\!$ &$\! 5.2125^{-5}\!$ &$\! 4.4207^{-3}\!$\\
$\! 0.7    \!$ &
$\! 1.7027^{-2}\!$ &$\! 2.0531^{-3}\!$ &$\! 2.2921^{-7}\!$ &$\! 1.9243^{-6}\!$ &
$\! 1.8384^{-7}\!$ &$\! 1.3298^{-6}\!$ &$\! 1.2157^{-6}\!$ &$\! 2.8887^{-4}\!$\\
$\! 0.9    \!$ &
$\! 3.3898^{-4}\!$ &$\! 1.3676^{-5}\!$ &$\! 4.284 ^{-11}\!$ &$\! 4.260 ^{-9}\!$ &
$\!-1.916 ^{-9}\!$ &$\!-1.701 ^{-9}\!$ &$\!-7.492 ^{-11}\!$ &$\! 1.2435^{-6}\!$\\
\hline \hline
\multicolumn{9}{||c||}{} \\[-3mm]
\multicolumn{9}{||c||}{$\mu_{\rm r}^2 = 2\, \mu_{\rm f}^2$} \\
\multicolumn{9}{||c||}{} \\[-0.3cm] \hline \hline
 & & & & & & & \\[-0.3cm]
$\! 10^{-7}\!\!$ &
$\!6.2819_*^{-5}\!$&$\! 2.4035^{-5}\!$ &$\! 1.1180^{-4}\!$ &$\! 4.6896^{-3}\!$ &
$\!-1.8050^{-3}\!$ &$\!-1.2637^{-3}\!$ &$\!-1.0544^{-3}\!$ &$\! 2.1305^{-2}\!$\\
$\! 10^{-6}\!\!$ &
$\! 2.9408^{-4}\!$ &$\! 1.1232^{-4}\!$ &$\! 2.2855^{-4}\!$ &$\! 9.8538^{-3}\!$ &
$\!-3.8554^{-3}\!$ &$\!-2.7780^{-3}\!$ &$\!-2.2077^{-3}\!$ &$\! 5.4411^{-2}\!$\\
$\! 10^{-5}\!\!$ &
$\! 1.3450^{-3}\!$ &$\! 5.1245^{-4}\!$ &$\! 4.6965^{-4}\!$ &$\! 1.8815^{-2}\!$ &
$\!-7.2936^{-3}\!$ &$\!-5.1597^{-3}\!$ &$\!-3.8359^{-3}\!$ &$\! 1.2368^{-1}\!$\\
$\! 10^{-4}\!\!$ &
$\! 5.9485^{-3}\!$ &$\! 2.2582^{-3}\!$ &$\! 9.4866^{-4}\!$ &$\! 3.0816^{-2}\!$ &
$\!-1.1297^{-2}\!$ &$\!-7.1323^{-3}\!$ &$\!-4.7404^{-3}\!$ &$\! 2.4503^{-1}\!$\\
$\! 10^{-3}\!\!$ &
$\! 2.4951^{-2}\!$ &$\! 9.4134^{-3}\!$ &$\! 1.7698^{-3}\!$ &$\! 3.8618^{-2}\!$ &
$\!-1.1654^{-2}\!$ &$\!-3.8925^{-3}\!$ &$\!-1.5608^{-3}\!$ &$\! 3.9912^{-1}\!$\\
$\! 10^{-2}\!\!$ &
$\! 9.4706^{-2}\!$ &$\! 3.5243^{-2}\!$ &$\! 2.6108^{-3}\!$ &$\! 2.8761^{-2}\!$ &
$\!-2.3471^{-3}\!$ &$\! 9.7827^{-3}\!$ &$\! 7.5188^{-3}\!$ &$\! 4.5698^{-1}\!$\\
$\! 0.1    \!$ &
$\! 2.5982^{-1}\!$ &$\! 8.9780^{-2}\!$ &$\! 1.4862^{-3}\!$ &$\! 8.3807^{-3}\!$ &
$\! 3.1615^{-3}\!$ &$\! 1.0522^{-2}\!$ &$\! 6.1973^{-3}\!$ &$\! 1.9561^{-1}\!$\\
$\! 0.3    \!$ &
$\! 2.1732^{-1}\!$ &$\! 6.0165^{-2}\!$ &$\! 1.9984^{-4}\!$ &$\! 1.2086^{-3}\!$ &
$\! 3.9371^{-4}\!$ &$\! 1.3919^{-3}\!$ &$\! 7.6929^{-4}\!$ &$\! 3.3906^{-2}\!$\\
$\! 0.5    \!$ &
$\! 9.2445^{-2}\!$ &$\! 1.8539^{-2}\!$ &$\! 1.3804^{-5}\!$ &$\! 7.8411^{-5}\!$ &
$\! 2.9799^{-5}\!$ &$\! 9.8805^{-5}\!$ &$\! 5.7333^{-5}\!$ &$\! 4.6166^{-3}\!$\\
$\! 0.7    \!$ &
$\!1.8219^{-2}\!$ &$\! 2.2090^{-3}\!$&$\! 2.5004_*^{-7}\!$&$\! 9.8927_*^{-7}\!$ &
$\! 7.5552^{-7}\!$ &$\! 2.0057^{-6}\!$ &$\! 1.4438^{-6}\!$ &$\! 3.0231^{-4}\!$\\
$\! 0.9    \!$ &
$\! 3.7653^{-4}\!$ &$\! 1.5285^{-5}\!$ &$\! 4.855^{-11}\!$ &$\! 2.005 ^{-9}\!$ &
$\!-7.599^{-10}\!$ &$\!-5.171^{-10}\!$ &$\! 3.809^{-10}\!$ &$\! 1.3232^{-6}\!$\\
\hline \hline
\multicolumn{9}{||c||}{} \\[-3mm]
\multicolumn{9}{||c||}{$\mu_{\rm r}^2 = 1/2\, \mu_{\rm f}^2$} \\
\multicolumn{9}{||c||}{} \\[-0.3cm] \hline \hline
 & & & & & & & \\[-0.3cm]
$\! 10^{-7}\!\!$ &
$\! 7.6699^{-5}\!$ &$\! 2.9289^{-5}\!$ &$\! 1.1912^{-4}\!$ &$\! 5.8548^{-3}\!$ &
$\!-2.3667^{-3}\!$ &$\!-1.8030^{-3}\!$ &$\!-1.4521^{-3}\!$ &$\! 3.1009^{-2}\!$\\
$\! 10^{-6}\!\!$ &
$\! 3.5067^{-4}\!$ &$\! 1.3364^{-4}\!$ &$\! 2.4707^{-4}\!$ &$\! 1.1806^{-2}\!$ &
$\!-4.7934^{-3}\!$ &$\!-3.6731^{-3}\!$ &$\!-2.7846^{-3}\!$ &$\! 7.2690^{-2}\!$\\
$\! 10^{-5}\!\!$ &
$\! 1.5611^{-3}\!$ &$\! 5.9329^{-4}\!$ &$\! 5.1593^{-4}\!$ &$\! 2.1406^{-2}\!$ &
$\!-8.5248^{-3}\!$ &$\!-6.3125^{-3}\!$ &$\!-4.4072^{-3}\!$ &$\! 1.5274^{-1}\!$\\
$\! 10^{-4}\!\!$ &
$\! 6.6957^{-3}\!$ &$\! 2.5346^{-3}\!$ &$\! 1.0509^{-3}\!$ &$\! 3.2903^{-2}\!$ &
$\!-1.2252^{-2}\!$ &$\!-7.9608^{-3}\!$ &$\!-4.8402^{-3}\!$ &$\! 2.8097^{-1}\!$\\
$\! 10^{-3}\!\!$ &
$\! 2.7125^{-2}\!$ &$\! 1.0200^{-2}\!$ &$\! 1.9310^{-3}\!$ &$\! 3.7698^{-2}\!$ &
$\!-1.1127^{-2}\!$ &$\!-3.2334^{-3}\!$ &$\!-7.5827^{-4}\!$ &$\! 4.2756^{-1}\!$\\
$\! 10^{-2}\!\!$ &
$\! 9.8758^{-2}\!$ &$\! 3.6602^{-2}\!$ &$\! 2.6980^{-3}\!$ &$\! 2.3675^{-2}\!$ &
$\! 5.1386^{-5}\!$ &$\! 1.2092^{-2}\!$ &$\! 8.6053^{-3}\!$ &$\! 4.6241^{-1}\!$\\
$\! 0.1    \!$ &
$\! 2.5572^{-1}\!$ &$\! 8.7847^{-2}\!$ &$\! 1.4179^{-3}\!$ &$\! 6.7523^{-3}\!$ &
$\! 3.5944^{-3}\!$ &$\! 1.0578^{-2}\!$ &$\! 6.0904^{-3}\!$ &$\! 1.8838^{-1}\!$\\
$\! 0.3    \!$ &
$\! 2.0497^{-1}\!$ &$\! 5.6318^{-2}\!$ &$\! 1.8228^{-4}\!$ &$\! 1.2965^{-3}\!$ &
$\! 2.6142^{-4}\!$ &$\! 1.1713^{-3}\!$ &$\! 6.8941^{-4}\!$ &$\! 3.1884^{-2}\!$\\
$\! 0.5    \!$ &
$\! 8.4404^{-2}\!$ &$\! 1.6775^{-2}\!$ &$\! 1.2174^{-5}\!$ &$\! 1.1604^{-4}\!$ &
$\! 2.8309^{-6}\!$ &$\! 6.3682^{-5}\!$ &$\! 4.7009^{-5}\!$ &$\! 4.3221^{-3}\!$\\
$\! 0.7    \!$ &
$\! 1.6013^{-2}\!$ &$\! 1.9215^{-3}\!$&$\! 2.1196_*^{-7}\!$&$\! 3.6047^{-6}\!$ &
$\!-7.4260^{-7}\!$ &$\! 3.1714^{-7}\!$ &$\! 9.6419^{-7}\!$ &$\! 2.8268^{-4}\!$\\
$\! 0.9    \!$ &
$\! 3.0848^{-4}\!$ &$\! 1.2377^{-5}\!$ &$\! 3.829 ^{-11}\!$ &$\! 8.129 ^{-9}\!$ &
$\!-3.873 ^{-9}\!$ &$\!-3.681 ^{-9}\!$ &$\!-6.816 ^{-10}\!$ &$\! 1.2009^{-6}\!$\\
\hline \hline
\end{tabular}
\end{center}
\end{table}

%% file: nonperturbativeshape.tex
\subsection{Non-perturbative $x$-shape of PDFs
\protect\footnote{Contributing author: G.~Ingelman}}

\label{sec:gi-pdf-xshape}
The $x$-shape of parton density functions at a low scale $Q_0^2$ is due to the
dynamics of the bound state proton and is hence an unsolved problem of
non-perturbative QCD. Usually this is described by parameterizations of data using more or less arbitrary functional forms. More understanding can be obtained by a recently developed physical model \cite{Alwall-Ingelman,*Alwall-Ingelmanb}, which is phenomenologically successful in describing data. 

\begin{figure}
\begin{center}
\includegraphics*[width=0.2\columnwidth]{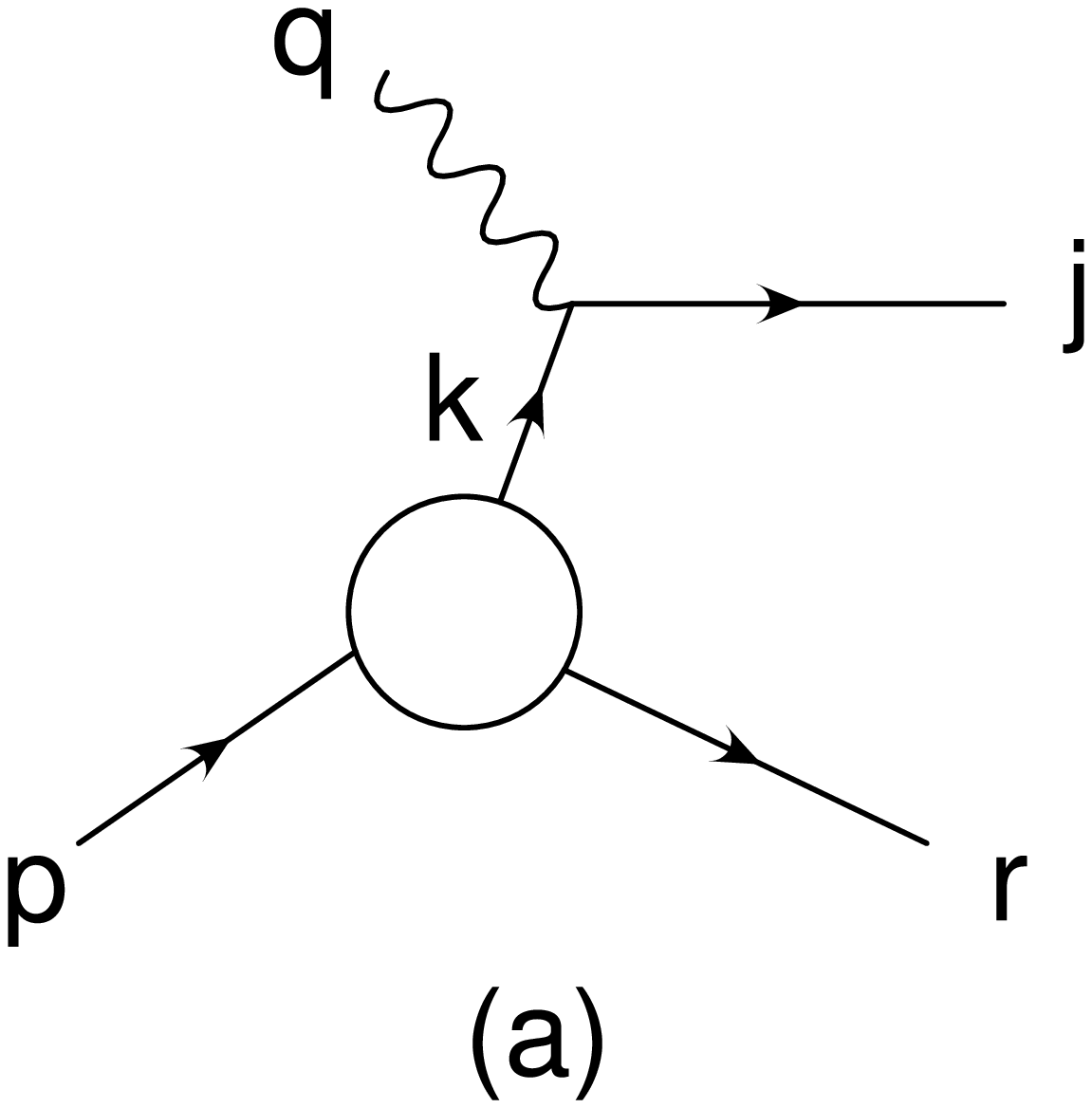} \hspace*{10mm}
\includegraphics*[width=0.2\columnwidth]{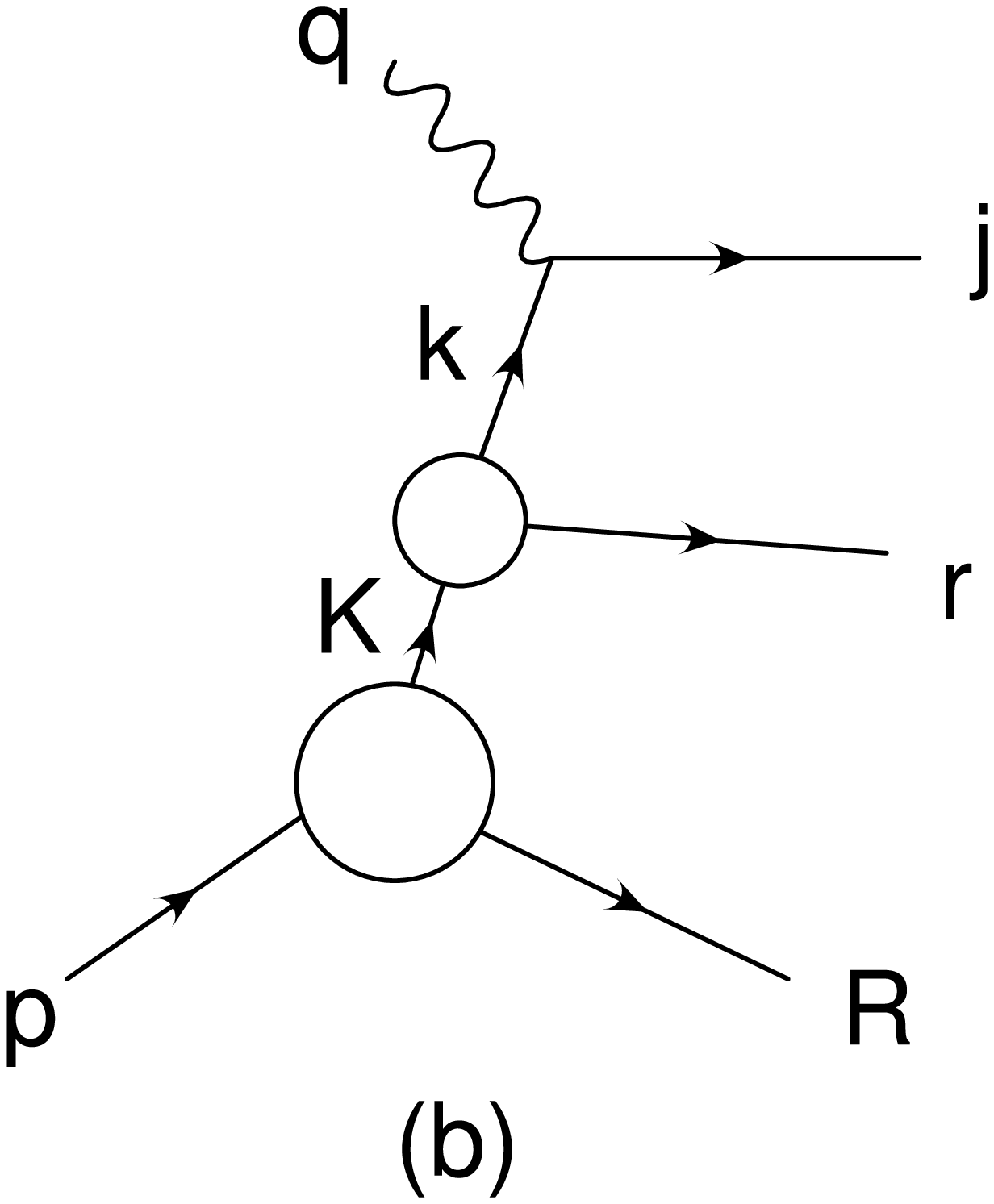} \hspace*{10mm}
\includegraphics*[width=0.3\columnwidth]{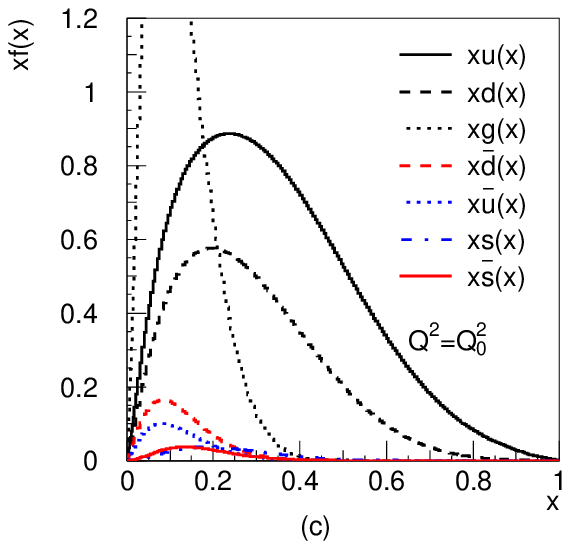}
\caption{\label{fig:gi-fluct} Probing (a) a valence parton in the proton and (b) a sea parton in a hadronic fluctuation (letters are four-momenta) resulting in (c) parton distributions at the starting scale $Q_0^2$.}
\end{center}
\end{figure}

The model gives the four-momentum $k$ of a single probed valence parton (Fig.~\ref{fig:gi-fluct}a) by assuming that, in the nucleon rest frame, the shape of the momentum distribution for a parton of type $i$ and mass $m_i$ can be taken as a Gaussian $f_i(k) = N(\sigma_i,m_i) \exp\left\{-\left[(k_0-m_i)^2+
k_x^2+k_y^2+k_z^2\right]/2\sigma_i^2\right\}$, which may be motivated as a result of the many interactions binding the parton in the nucleon. The width of the distribution should be of order hundred MeV from the Heisenberg uncertainty relation applied to the nucleon size, {\em i.e.}\ $\sigma_i=1/d_N$. The momentum fraction $x$ of the parton is then defined as the light-cone fraction $x=k_+/p_+$ and is therefore invariant under longitudinal boosts (e.g.\ to the infinite momentum frame). Constraints are imposed on the final-state momenta to obtain a kinematically allowed final state, which also ensures that $0<x<1$ and $f_i(x)\to0$ for $x\to 1$. 

The sea partons are obtained using a hadronic basis for the non-perturbative dynamics of the bound state proton and considering hadronic fluctuations
\begin{equation} \label{eq:gi-hadronfluctuation}
|p\rangle = \alpha_0|p_0\rangle + \alpha_{p\pi^0}|p\pi^0\rangle +
\alpha_{n\pi^+}|n\pi^+\rangle + \ldots + \alpha_{\Lambda K}|\Lambda
K^+\rangle + \ldots
\end{equation}
Probing a parton $i$ in a hadron $H$ of a baryon-meson fluctuation
$|BM\rangle$ (Fig.~\ref{fig:gi-fluct}b) gives a sea parton with
light-cone fraction $x=x_H\, x_i$ of the target proton. The momentum
of the probed hadron is given by a similar Gaussian, but with a
separate width parameter $\sigma_H$. Also here, kinematic constraints
ensure physically allowed final states.

\begin{figure}
\begin{center}
\includegraphics*[width=0.9\columnwidth]{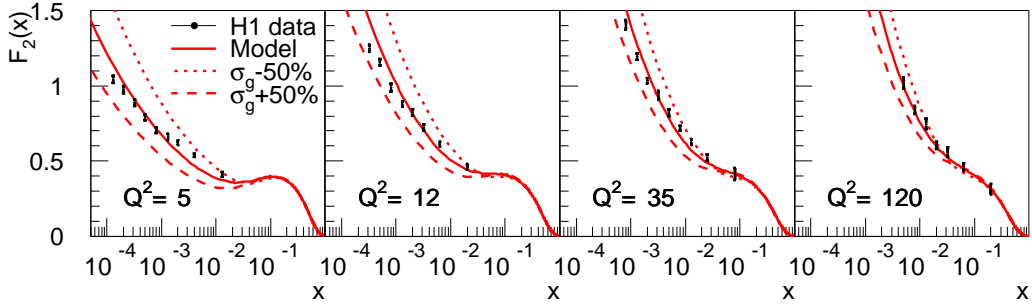}
\caption{$F_2(x,Q^2)$ from H1 
compared to the model with  $\pm50\%$ variation of the width parameter $\sigma_g$ of the gluon distribution. \label{fig:gi-F2} }
\end{center}
\end{figure}

Using a Monte Carlo method the resulting valence and sea parton $x$-distributions are obtained without approximations. These apply at a low scale $Q_0^2$ and the distributions at higher $Q^2$ are obtained using perturbative QCD evolution at next-to-leading order. To describe all parton distributions (Fig.~\ref{fig:gi-fluct}c), the model has only four shape parameters and three normalization parameters, plus the starting scale: 
\\
\begin{equation}
\label{eq:gi-params}
\begin{array}{c}
\sigma_u=230\;\mathrm{MeV} \quad \sigma_d=170\;\mathrm{MeV} \quad 
\sigma_g=77\;\mathrm{MeV} \quad \sigma_H=100\;\mathrm{MeV}\\
\alpha_\mathrm{p\pi^0}^2=0.45 \quad \alpha_{n\pi^+}^2=0.14 \quad 
\alpha_{\Lambda K}^2=0.05 \quad Q_0=0.75\;\mathrm{GeV}
\end{array}
\end{equation}
These are determined from fits to data as detailed in \cite{Alwall-Ingelman} and illustrated in Fig.~\ref{fig:gi-F2}. The model reproduces the inclusive proton structure function and gives a natural explanation of observed quark asymmetries, such as the difference between the up and down valence distributions and between the anti-up and anti-down sea quark distributions. Moreover, its asymmetry in the momentum distribution of strange and anti-strange quarks in the nucleon is large enough to reduce the NuTeV anomaly to a level which does not give a significant indication of physics beyong the Standard Model. 

Recent fits of PDF's at very low $x$ and $Q^2$ have revealed problems with the gluon density, which in some cases even becomes negative. The reason for this is that the DGLAP evolution, driven primarily by the gluon at small $x$, otherwise gives too large parton densities and thereby a poor fit to $F_2$ in the genuine DIS region at larger $Q^2$. It has been argued \cite{Alwall:2004wk} that the root of the problem is the application of the formalism for DIS also in the low-$Q^2$ region, where the momentum transfer is not large enough that the parton structure of the proton is clearly resolved. The smallest distance that can be resolved is basically given by the momentum transfer of the exchanged photon through $d=0.2/\sqrt{Q^2}$, where $d$ is in Fermi if $Q^2$ is in GeV$^2$. This indicates that partons are resolved only for $Q^2\gsim 1\, \rm{GeV}^2$. For $Q^2\lsim 1\, \rm{GeV}^2$, there is no hard scale involved and a parton basis for the description is not justified. Instead, the interaction is here of a soft kind between the nearly on-shell photon and the proton. The cross section is then dominated by the process where the photon fluctuates into a virtual vector meson state which then interacts with the proton in a strong interaction. 
The quantum state of the photon can be expressed as $|\gamma\rangle = C_0|\gamma_0\rangle + \sum_V \frac{e}{f_V}|V\rangle + \int_{m_0}dm (\cdots)$.
The sum is over $V = \rho^0, \omega, \phi \ldots$ as in the original vector meson dominance model (VDM), whereas the generalised vector meson dominance model (GVDM) also includes the integral over a continuous mass spectrum (not written out explicitly here). 

\begin{figure}
\begin{center}
\includegraphics*[width=0.9\columnwidth]{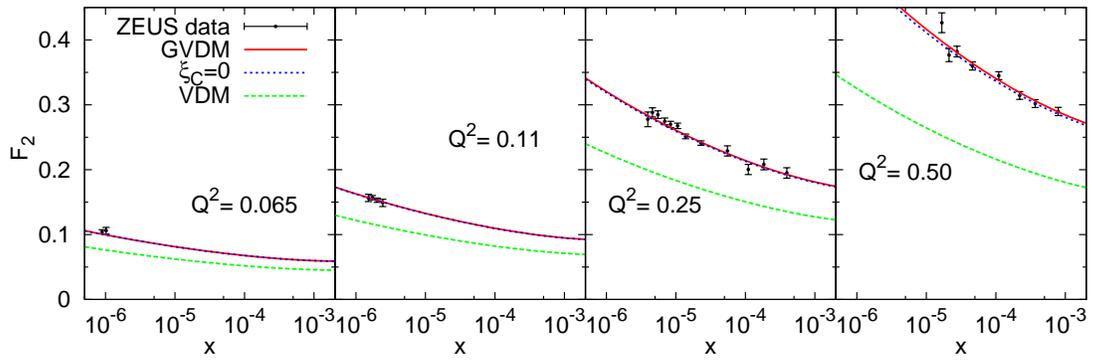}
\caption{$F_2$ data at low $Q^2$ from ZEUS 
compared to the full GVDM in eq.~(\ref{eq:gi-F2-GVDM}) (full curves), when excluding the longitudinal contribution of the continuum ($\xi_C=0$) and excluding the continuous contribution altogether (setting $r_C=0$) giving VDM. \label{fig:gi-F2-GVDM}} 
\end{center}
\end{figure}

Applied to $ep$ at low $Q^2$ this leads to the expression \cite{Alwall:2004wk}
\begin{eqnarray}\label{eq:gi-F2-GVDM}
F_2(x,Q^2) & = & \frac{(1-x)Q^2}{4\pi^2\alpha} 
      \left\{  \sum_{V=\rho, \omega, \phi} r_V \left(\frac{m_V^2}{Q^2 + m_V^2}\right)^2
	\left(1 + \xi_V\frac{Q^2}{m_V^2}\right) \right.
\nonumber \\
 & & \left. +\; r_C\left[ (1-\xi_C)\frac{m_0^2}{Q^2 + m_0^2} + 
        \xi_C \frac{m_0^2}{Q^2}\ln{(1 + \frac{Q^2}{m_0^2})}
	\right] \right\}
	A_\gamma \frac{Q^{2\epsilon}}{x^\epsilon}
\end{eqnarray}
where the hadronic cross-section $\sigma(ip\rightarrow X) = A_i s^\epsilon + B_i s^{-\eta}\approx A_i s^\epsilon \approx A_i (Q^2/x)^\epsilon$ has been used for the small-$x$ region of interest. The parameters involved are all essentially known from GVDM phenomenology. With $\epsilon = 0.091$, $\xi = 0.34$, $m_0=1.5$~GeV and $A_\gamma = 71\, \mu\rm{b}$, this GVDM model gives a good fit ($\chi^2/\rm{d.o.f.} = 87/66 = 1.3$) as illustrated in Fig.~\ref{fig:gi-F2-GVDM}. Using this model at very low $Q^2$ in combination with the normal parton density approach at larger $Q^2$ it is possible to obtain a good description of data over the full $Q^2$ range \cite{Alwall:2004wk}. This involves, however, a phenomenological matching of these two approaches, since a theoretically well justified combination is an unsolved problem. 

Neglecting the GVDM component when fitting PDF's to data at small $Q^2$ may thus lead to an improper gluon distribution, which is not fully universal and therefore may give incorrect results when used for cross section calculations at LHC.

%% file: alekhin.tex
\subsection{Towards precise determination of the nucleon PDFs
\protect\footnote{Contributing author: S.~I.~Alekhin}}

\label{sec:alek}

The nucleon parton distribution functions (PDFs) available to the moment 
are extracted from the rather limited set of experimental distributions
(the deep-inelastic scattering (DIS) structure functions, the Drell-Yan (DY)
and jet production cross sections).
Other high-energy processes potentially could provide 
additional constraints on PDFs, however insufficient theoretical understanding 
does not allow to use those data without risk of having uncontrolled 
theoretical inaccuracies. Even for the case of the existing global fits of 
the PDFs performed by the MRST and CTEQ groups missing next-to-next-to-leading 
(NNLO) order QCD corrections to the Drell-Yan and jet production cross 
sections are not small as compared to the accuracy of the corresponding data 
used and therefore might give non-negligible effect. In this section  
we outline progress in the QCD fits with consistent account 
of the NNLO corrections.

\subsubsection{Impact of the NNLO evolution kernel fixation on PDFs}

In order to allow account of the NNLO corrections in the fit of 
PDFs one needs analytical expressions for
the 3-loop corrections to the QCD evolution kernel. Until recent times 
these expressions were known only in the approximate form of 
Ref.~\cite{vanNeerven:2000wp} derived from the partial information  
about the kernel, including the set of its Mellin moments and 
the low-$x$ asymptotics~\cite{Larin:1997wd,Retey:2000nq,Catani:1994sq}
However with the refined calculations of Ref.~\cite{Moch:2004pa,Vogt:2004mw}
the exact expression for the NNLO kernel has been available. 
These improvement is of particular importance for analysis of the 
low-$x$ data including the HERA ones due to general rise of the high-order QCD 
correction at low $x$. 
We illustrate impact of the NNLO evolution kernel validation on PDFs using  
the case of fit to the global DIS data~\cite{%
Whitlow:1991uw,Benvenuti:1989rh,Benvenuti:1989fm,Arneodo:1996qe,h1alphas,z1a}.
The exact NNLO corrections to the DIS coefficient functions are 
know~\cite{Kazakov:1992xj,Zijlstra:1992qd}
that allowed to perform approximate NNLO fit of PDFs to these 
data~\cite{Alekhin:2002fv} using the approximate NNLO 
corrections to the evolution kernel of Ref.~\cite{vanNeerven:2000wp}. 
Taking into account exact NNLO evolution kernel
the analysis of Ref.~\cite{Alekhin:2002fv} was updated recently to the exact 
NNLO case~\cite{Alekhin:2005gq}. 

\begin{figure}
\centerline{\epsfig{file=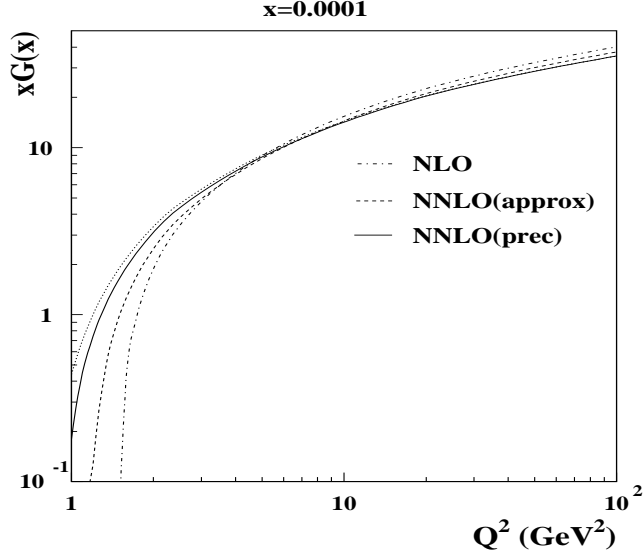,width=10cm,height=8cm}}
\caption{The gluon distributions obtained in the different variants of PDFs 
fit to the DIS data
(solid: the fit with exact NNLO evolution; dashes: 
the fit with approximate NNLO evolution; dots: the approximate NNLO 
gluons evolved with the exact NNLO kernel; dashed-dots: the NLO fit).} 
\label{fig:gluon}
\end{figure}

The gluon distributions at small $x$ 
obtained in these two variants of the fit are compared in Fig.\ref{fig:gluon}.
With the exact NNLO corrections the QCD evolution of gluon
distribution at small $x$ gets weaker and as a result
at small $x/Q$ the gluon distribution obtained using 
the precise NNLO kernel is quite different 
from the approximate one. In particular, the 
approximate NNLO gluon distribution is negative at 
$Q^2\lesssim1.3~{\rm GeV}^2$, while the  
precise one remains positive even below $Q^2=1~{\rm GeV}^2$. 
For the NLO case the positivity of gluons 
at small $x/Q$ is even worse than for the 
approximate NNLO case due to the approximate 
NNLO corrections dampen the gluon evolution at small $x$ too, therefore 
account of the NNLO corrections is crucial in this respect.
(cf. discussion of Ref.~\cite{Huston:2005jm}).
Positivity of the PDFs is not mandatory beyond the QCD leading order, 
however it 
allows probabilistic interpretation of the parton model and facilitates   
modeling of the soft processes, such as underlying events in the 
hadron-hadron collisions at LHC. 
The change of gluon distribution at small $x/Q$ 
as compared to the fit with approximate NNLO evolution  
is rather due the change in evolution kernel than due to 
shift in the fitted parameters of PDFs. 
This is clear from comparison of the exact NNLO gluon distribution
to one obtained from the approximate NNLO fit and evolved to low $Q$ 
using the exact NNLO kernel (see Fig.\ref{fig:gluon}).
In the vicinity of crossover in the gluon distribution 
to the negative values   
its relative change due to variation of the evolution kernel 
is quite big and therefore further fixation of the kernel at small $x$
discussed in Ref.~\cite{Altarelli:2003hk} 
might be substantial for validation of the PDFs at low $x/Q$. 
For the higher-mass kinematics at LHC 
numerical impact of the NNLO kernel update is not dramatic.
Change in the Higgs and $W/Z$ bosons  
production cross sections due to more precise definition of the NNLO 
PDFs is comparable to the errors coming from the PDFs uncertainties, i.e. at 
the level of several percent.

\subsubsection{NNLO fit of PDFs to the combined DIS and Drell-Yan data}  

\begin{figure}
\centerline{\epsfig{file=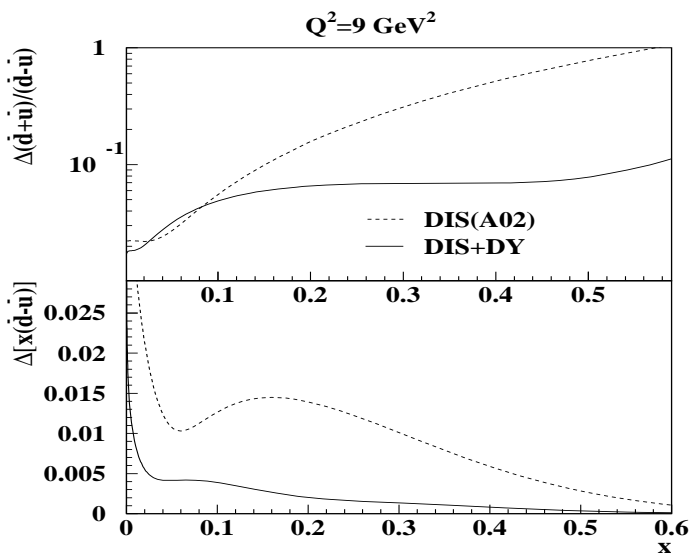,width=10cm,height=8cm}}
\caption{Uncertainties in the non-strange sea distributions
obtained from NNLO QCD fit to the DIS data combined with the 
fixed target Drell-Yan data (solid curves). The same 
uncertainties obtained in fit to the DIS data 
only~[8] are given for comparison by dashes.}
\label{fig:dy}
\end{figure}

The DIS process provide very clean source of information about PDFs
both from experimental and theoretical side, however very poorly 
constrains the gluon and sea distributions at $x\gtrsim 0.3$.
The well known way to improve precision of the sea distributions 
is to combine DIS data with the Drell-Yan ones. 
The cross section of process $N N \rightarrow l^+ l^-$ reads 
$$
\sigma_{DY}\propto \sum_i \left[ q_i(x_1)\bar{q}_i(x_2)
+q_i(x_2)\bar{q}_i(x_1) \right]+{\rm higher{^{_{\_}}}order~terms}, 
$$
where $q(\bar{q})_i$ are the quarks(antiquarks) distribution and 
$x_{1,2}$ give the momentum fractions carried by each of 
the colliding partons. The quark distributions are determined  
by the DIS data with the precision of several 
percent in the wide region of $x$ and therefore 
precision of the sea distribution extracted from the 
combined fit to the DIS and DY data is basically determined by the 
latter. The Fermilab fixed-target experiments
provide measurements of the DY cross sections for the isoscalar 
target~\cite{Moreno:1990sf}
and the ratio of cross sections for the deuteron and proton 
targets~\cite{Towell:2001nh}
with the accuracy better than 20\% at $x \lesssim 0.6$.
Fitting PDFs to these data combined with the global 
DIS data of Ref.~\cite{%
Whitlow:1991uw,Benvenuti:1989rh,Benvenuti:1989fm,Arneodo:1996qe,h1alphas,z1a}
we can achieve comparable precision in the sea distributions.
Recent calculations of Ref.~\cite{Anastasiou:2003yy}
allow to perform this fit with full account of the NNLO 
correction. Using these calculations the DY data of 
Refs.~\cite{Moreno:1990sf,Towell:2001nh} were included into 
the NNLO fit of Ref.~\cite{Alekhin:2005gq} that 
leads to significant improvement in the precision of 
sea distributions (see Fig.~\ref{fig:dy}). Due to the DY data on the 
deuteron/proton ratio the isospin asymmetry of sea is also 
improved. It is worth to note that  
the precision achieved for the total sea distribution is in
good agreement to the rough estimates given above.  
The value of $\chi^2/{\rm NDP}$ obtained in the fit is 1.1 and the spread of 
$\chi^2/{\rm NDP}$ over separate experiments used in the fit is not dramatic, 
its biggest value is 1.4. We rescaled the  
errors in data for experiments with $\chi^2/{\rm NDP}>1$ 
in order to bring $\chi^2/{\rm NDP}$ for this 
experiments to 1 and found that overall impact of this rescaling 
on the PDFs errors is marginal. 
This proofs sufficient statistical consistency of the data sets used in the fit
and disfavors huge increase in the value of $\Delta\chi^2$ criterion
suggested by the CTEQ collaboration for estimation of errors 
in the global fit of PDFs.
A particular feature of the PDFs obtained is good stability with respect to 
the choice of factorization/renormalization scale in the DY cross section:
Variation of this scale from $M_{\mu^+\mu^-}/2$ to $2M_{\mu^+\mu^-}$ leads 
to variation of PDFs comparable to their uncertainties due to errors in data. 

\subsubsection{LHC data and flavor separation of the sea at small $x$}

Combination of the existing DIS and fixed-target
DY data provide good constraint on the total sea 
quarks distribution and allows separation of the $\bar{u}$- and 
$\bar{d}$-quark
distributions up to the values of $x$ sufficient for most practical 
applications at the LHC. 
At small $x$ the total sea is also well constrained by the
precise HERA data on the inclusive structure functions, however 
$\bar{u}/\bar{d}$ separation is poor in this region due to 
lack of the deuteron target 
data at HERA. The problem of the sea flavor separation 
is regularly masked due to additional constraints imposed on PDFs. 
In particular, most often the Regge-like behavior of the sea isospin 
asymmetry
$x(\bar{d}-\bar{u})\propto x^{a_{ud}}$ is assumed 
with $a_{ud}$ selected around value of 0.5 motivated by the intercept of 
the meson trajectories. This assumption automatically provides constraint 
$\bar{d}=\bar{u}$ at $x\rightarrow 0$ and therefore leads to suppression of 
the uncertainties both in $\bar{u}$ and $\bar{d}$ at small $x$.
If we do not assume the Regge-like behavior of $x(\bar{d}-\bar{u})$ its 
precision determined from the NNLO fit to the combined DIS and DY data of 
Section 1.2
is about 0.04 at $x=10^{-4}$ furthermore this constraint is defined rather by
assumption about the shape of PDFs at small $x$ than by data used in 
the fit. The strange sea distribution 
is known much worse than the non-strange ones.
It is essentially defined only by the CCFR 
experiment from the cross section of dimuon production in the
neutrino nucleus collisions~\cite{Bazarko:1994tt}.
In this experiment the strange sea distribution was probed 
at $x=0.01\div 0.2$ and the shape obtained is similar to one of  
the non-strange sea with the strangeness suppression factor about 0.5.  
This is in clear disagreement with the Regge-like constraint 
on $x(\bar{d}-\bar{s})$ or $x(\bar{u}-\bar{s})$ and therefore 
we cannot use even this assumption to predict the strange sea at small $x$. 

\begin{figure}
\centerline{\epsfig{file=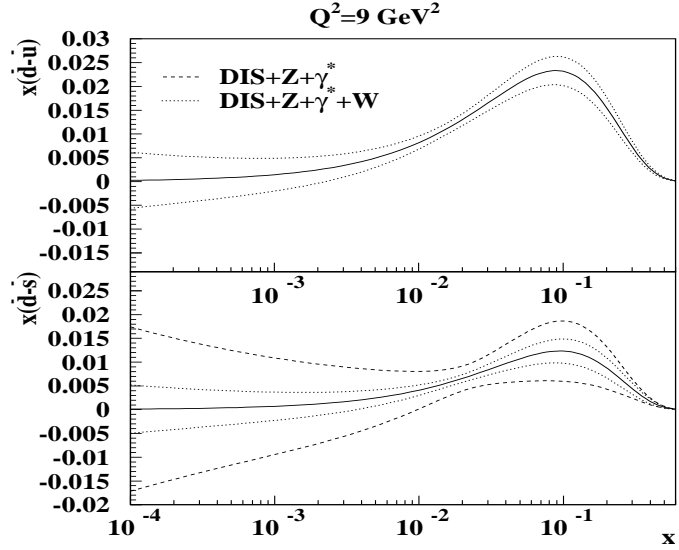,width=10cm,height=8cm}}
\caption{The $1\sigma$ error band for $x(\bar{d}-\bar{u})$ (upper panel)
and $x(\bar{d}-\bar{s})$ (lower panel) expected for the fit of PDFs to the 
LHC data combined with the global DIS ones. 
Dashed curves correspond to the case of $Z/\gamma*$-production, dots are 
for the combination $Z/\gamma*$- with the $W^+/W^-$-production.  
Solid curves are for the central values obtained from the reference fit to the 
global DIS data}.
\label{fig:sea}
\end{figure}

The LHC data on $\mu^+ \mu^-$ 
production cross section can be used for further 
validation of the sea distributions at
small $x$. Study of this process at the lepton pair masses down to 15 GeV 
will allow to probe PDFs at $x$ down to $10^{-4}$, while  
with both leptons detected full kinematics can be reliably reconstructed.
In order to check impact of the foreseen LHC data on the sea flavor 
separation we generated 
sample of pseudo-data for the process $p p \rightarrow \mu^+ \mu^- X$
at $\sqrt{s}=14~{\rm TeV}$ with integral luminosity of 10 1/fb
corresponding to the first stage of the LHC operation. 
In order to meet typical limitations of the LHC detectors
only events with the lepton pair absolute rapidity 
less than 2.5 were accepted;
other detector effects were not taken into account.
For generation of these pseudo-data we used 
PDFs obtained in the dedicated version of fit~\cite{Alekhin:2005gq}
with the sea distributions parameterized as 
$xS_{u,d,s}=\eta_{u,d,s} x^a(1-x)^{b_{u,d,s}}$
with the constraints $\eta_u=\eta_d=\eta_s$ and $b_s=(b_u+b_d)/2$ imposed. 
These constraints are necessary for stability of the fit in view of 
limited impact of the DIS data on the flavor separation and, besides, 
the former one guarantees
SU(3) symmetry in the sea distributions at small $x$. 
The generated pseudo-data were added to the basic DIS data sample and 
the errors in PDFs parameters were re-estimated 
with no constraints on the sea distributions imposed at this stage. 
Since dimuon data give extra information about the PDFs products
they allow to disentangle the strange distribution,
if an additional constraint on the non-strange sea distributions is set. 
The dashed curves in the lower panel of Fig.\ref{fig:sea}
give the $1\sigma$ bands for $x(\bar{d}-\bar{s})$ as they are 
defined by the LHC simulated data combined with the global DIS ones
given $(\bar{d}-\bar{u})$ is fixed. One can see that $\bar{d}/\bar{s}$
(and $\bar{u}/\bar{s}$)
separation at the level of several percents
would be feasible down to x=$10^{-4}$ in this case. 
The supplementary  constraint on $(\bar{d}-\bar{u})$ can be obtained 
from study of the $W$-boson charge asymmetry. 
To estimate impact of this process
we simulated the single $W^+$- and $W^-$-production data similarly to the 
case of the $\mu^+\mu^-$-production and took into account this sample too.  
In this case one can achieve
separation of all three flavors with the precision better than 0.01
(see Fig.\ref{fig:sea}).
Note that strange sea separation is also improved due to certain sensitivity 
of the $W$-production cross section to the strange sea contribution.
The estimates obtained refer to the ideal case of full kinematical 
reconstruction of the $W$-bosons events. For the case of using the   
charge asymmetry of muons produced from the $W$-decays the 
precision of the PDFs would be 
worse. Account of the backgrounds and the detector effects would also 
deteriorate it, however these losses can be at least partially compensated by 
rise of the LHC luminosity at the second stage of operation. 

\begin{table}
\begin{center}
\begin{tabular}{ccc}   
Valence    &$a_u$&$0.718\pm0.085$\\
       &$b_u$&$3.81\pm0.16$\\
       &$\epsilon_u$&$-1.56\pm0.46$\\
       &$\gamma_u$&$3.30\pm0.49$\\
       &$a_d$&$1.71\pm0.20$\\
       &$b_d$&$10.00\pm0.97$\\
       &$\epsilon_d$&$-3.83\pm0.23$\\
       &$\gamma_d$&$4.64\pm0.41$\\ \hline
Sea       &$A_S$&$0.211\pm0.016$\\
       &$a_{s}$&$-0.048\pm0.039$\\
       &$b_{s}$& $2.20\pm0.20$\\ \hline
Glue     &$a_G$&$0.356\pm0.095$\\
       &$b_G$&$10.9\pm1.4$\\ \hline
       &$\alpha_{\rm s}(M_{\rm Z})$&$0.1132\pm0.0015$\\ 
\end{tabular}
\end{center}
\caption{Values of the parameters obtained in the benchmark fit.}
\label{tab:pars}
\end{table}

\subsubsection{Benchmarking of the PDFs fit}

For the available nucleon PDFs the
accuracy at percent level is reached in some kinematical 
regions. For this reason benchmarking of the codes used in these PDFs
fits is becoming important issue. 
A tool for calibration of the QCD evolution codes was provided by Les Houches 
workshop \cite{Giele:2002hx}.
To allow benchmarking of the PDFs errors calculation we performed a test fit   
suggested in Les Houches workshop too. 
This fit reproduces basic features of the existing global fits of PDFs, 
but is simplified a lot to facilitate its reproduction. 
We use for the analysis data on the proton DIS structure functions $F_2$ 
obtained by the BCDMS, NM, H1, and ZEUS collaborations and 
ratio of the deuteron and proton structure functions $F_2$ obtained by 
the NMC. The data tables with full description of experimental errors 
taken into account are available 
online\footnote{https://mail.ihep.ru/$\tilde{~}$alekhin/benchmark/TABLE}. 
Cuts for the momentum transferred $Q^2>9~{\rm GeV}^2$
and for invariant mass of the hadronic system $W^2>15~{\rm GeV}^2$
are imposed in order to avoid influence of the power corrections
and simplify calculations. The contribution of the $Z$-boson exchange 
at large $Q$ is not taken into account for the same purpose.
The PDFs are parameterized in the form
$$
xp_i(x,1~{GeV})=N_i x^{a_i} (1-x)^{b_i} (1 + \epsilon_i \sqrt{x} + \gamma_i x),
$$
to meet choice common for many popular global fits of PDFs.
Some of the parameters $\epsilon_i$ and $\gamma_i$
are set to zero since they were found to be consistent to zero  
within the errors. We assume isotopic symmetry for sea distribution
and the strange sea is 
the same as the non-strange ones suppressed by factor of 0.5.
Evolution of the PDFs is performed 
in the NLO QCD approximation within the $\overline{\rm MS}$ scheme. 
The heavy quarks  contribution is accounted in the massless scheme with 
the variable number of flavors (the thresholds for $c$- and $b$-quarks 
are 1.5 GeV and 4.5 GeV correspondingly).
All experimental errors including correlated ones are taken into account 
for calculation of the errors in PDFs using the covariance matrix 
approach~\cite{Alekhin:2000es} and assuming linear propagation of errors. 
The results of the benchmark fit obtained with the code used in 
analysis of Refs.\cite{Alekhin:2002fv,Alekhin:2005gq} are given in 
Tables~\ref{tab:pars} and~\ref{tab:cor}. The total number of the fitted PDF parameters left is 14.
The normalization parameters $N_i$ for the gluon and 
valence quark distributions are calculated from the momentum and 
fermion number conservation. The remaining normalization parameter 
$A_S$ gives the total momentum carried by the sea distributions. 
Important note is that in view of many model assumptions made in the fit 
these results can be used mainly for the purposes of benchmarking rather 
for the phenomenological studies. 

\begin{sidewaystable}
\begin{center}
\begin{tabular}{ccccccccccccccc}   
&$a_u$& $b_u$& $\epsilon_u$& $\gamma_u$& $a_d$& $b_d$& $\epsilon_d$& 
$\gamma_d$& $A_S$& $a_{s}$& $b_{s}$& $a_G$& $b_G$& $\alpha_{\rm s}(M_{\rm Z})$ \\
$a_u$&   1.000 & 0.728 &-0.754 &-0.708 & 0.763 & 0.696 &-0.444 & 0.215 &-0.216 &-0.473 &-0.686 & 0.593 & 0.777 &-0.006 \\
$b_u$&    0.728 & 1.000 &-0.956 &-0.088 & 0.377 & 0.620 &-0.420 & 0.387 & 0.175 &-0.182 &-0.713 & 0.067 & 0.505 &-0.337 \\
$\epsilon_u$&  -0.754 &-0.956 & 1.000 & 0.105 &-0.388 &-0.662 & 0.503 &-0.485 &-0.229 & 0.059 & 0.600 &-0.047 &-0.503 & 0.276 \\
$\gamma_u$&  -0.708 &-0.088 & 0.105 & 1.000 &-0.741 &-0.390 & 0.219 & 0.107 & 0.597 & 0.591 & 0.310 &-0.716 &-0.675 &-0.088 \\
$a_d$&    0.763 & 0.377 &-0.388 &-0.741 & 1.000 & 0.805 &-0.622 & 0.248 &-0.367 &-0.509 &-0.528 & 0.652 & 0.664 & 0.101 \\
$b_d$&   0.696 & 0.620 &-0.662 &-0.390 & 0.805 & 1.000 &-0.904 & 0.728 & 0.017 &-0.193 &-0.512 & 0.272 & 0.576 &-0.136 \\
$\epsilon_d$&   -0.444 &-0.420 & 0.503 & 0.219 &-0.622 &-0.904 & 1.000 &-0.896 &-0.132 &-0.019 & 0.245 &-0.038 &-0.362 & 0.173 \\
$\gamma_d$&    0.215 & 0.387 &-0.485 & 0.107 & 0.248 & 0.728 &-0.896 & 1.000 & 0.346 & 0.240 &-0.107 &-0.241 & 0.120 &-0.228 \\
$A_S$&  -0.216 & 0.175 &-0.229 & 0.597 &-0.367 & 0.017 &-0.132 & 0.346 & 1.000 & 0.708 & 0.127 &-0.375 &-0.026 & 0.047 \\
$a_{s}$&  -0.473 &-0.182 & 0.059 & 0.591 &-0.509 &-0.193 &-0.019 & 0.240 & 0.708 & 1.000 & 0.589 &-0.595 &-0.241 &-0.011\\
$b_{s}$&  -0.686 &-0.713 & 0.600 & 0.310 &-0.528 &-0.512 & 0.245 &-0.107 & 0.127 & 0.589 & 1.000 &-0.508 &-0.402 &-0.109\\
$a_G$&    0.593 & 0.067 &-0.047 &-0.716 & 0.652 & 0.272 &-0.038 &-0.241 &-0.375 &-0.595 &-0.508 & 1.000 & 0.565 & 0.587\\
$b_G$&   0.777 & 0.505 &-0.503 &-0.675 & 0.664 & 0.576 &-0.362 & 0.120 &-0.026 &-0.241 &-0.402 & 0.565 & 1.000 &-0.138\\
$\alpha_{\rm s}(M_{\rm Z})$&  -0.006 &-0.337 & 0.276 &-0.088 & 0.101 &-0.136 & 0.173 &-0.228 & 0.047 &-0.011 &-0.109 & 0.587 &-0.138 & 1.000\\
\end{tabular}
\end{center}
\caption{Correlation coefficients for the parameters
obtained in the benchmark fit.}
\label{tab:cor}
\end{sidewaystable}

%% file: thorne.tex
\subsection{Benchmark Partons from DIS data and a Comparison with Global Fit Partons
\protect\footnote{Contributing author: R.S.~Thorne.}} 
\label{sec:thorne}

In this article I consider the uncertainties on partons arising from the
errors on the experimental data that are used in a parton analysis.
Various groups \cite{Botje},\cite{Gielea,*Gieleb},\cite{Alekhin:2002fv},\cite{cteq},\cite{h1alphas},
\cite{Martin:2002aw},\cite{zeus} have 
concentrated on the experimental errors and have obtained
estimates of the uncertainties on parton distributions within a
NLO QCD framework, using a variety of competing procedures. Here the two 
analyses, performed by myself and S. Alekhin (see Sec.~\ref{sec:alek})
minimise the differences one obtains for the central values of the partons
and the size of the uncertainties 
by fitting to exactly the same data sets with the same cuts, and 
using the same theoretical prescription. 
In order to be conservative we use only DIS data -- BCDMS proton 
\cite{Benvenuti:1989rh} and deuterium \cite{Benvenuti:1989fm} fixed target data, 
NMC data on proton DIS and on the 
ratio $F_2^n(x,Q^2)/F_2^p(x,Q^2)$ \cite{Arneodo:1996qe}, and H1 \cite{h1alphas}
and ZEUS \cite{z1a} DIS data. We also apply cuts of $Q^2=9\GeV^2$ and 
$W^2=15\GeV^2$ in order to avoid the influence of higher twist.
We each use NLO perturbative QCD in the ${\overline{\rm MS}}$ renormalization
and factorization scheme, with the zero-mass variable flavour number scheme 
and quark masses of $m_c=1.5\GeV$ and $m_b=4.5\GeV$. There is a very minor 
difference between $\alpha_S(\mu^2)$ used in 
the two fitting programs due to the different methods of 
implementing heavy quark thresholds (the differences being formally of higher 
order), as observed in the study by M. Whalley for this workshop 
\cite{Whalley}. If the couplings in the two approaches have the same value at
$\mu^2=M_Z^2$, then the MRST value is $\sim 1\%$ higher for 
$Q^2\sim 20\GeV^2$.   

We each input our
parton distributions at $Q_0^2=1\GeV^2$ with a parameterization of the form 
\begin{equation}
xf_i(x,Q_0^2)=A_i(1-x)^{b_i}(1+\epsilon_i x^{0.5}+\gamma_i x)x^{a_i}.
\label{eq:param}
\end{equation}      
The input sea is constrained to be $40\%$ up and anti-up quarks, $40\%$ 
down and anti-down quarks, and $20\%$ strange and antistrange. No difference
between $\bar u$ and $\bar d$ is input.  
There is no
negative term for the gluon, as introduced in \cite{Martin:2002aw}, since this 
restricted form of data shows no strong requirement for it in order to obtain 
the best fit. Similarly we are able to set $\epsilon_g$, $\gamma_g$, 
$\epsilon_S$ and 
$\gamma_S$ all equal to zero. $A_g$ is set by the momentum sum rule and 
$A_{u_V}$ and $A_{d_V}$ are set by valence 
quark number. Hence, there are nominally
13 free parton parameters. However, the MRST fitting program exhibited 
instability in the error matrix due to a very high correlation between 
$u_V$ parameters, so $\epsilon_u$ was set at its best fit value of 
$\epsilon_u=-1.56$, while 12 parameters were free to vary. The 
coupling was also allowed to vary in order to obtain the best fit. The 
treatment of the errors on the data was exactly as for the published partons 
with uncertainties for each group, i.e. as in \cite{Alekhin:2002fv} and 
\cite{mrst}. This means that all detail on correlations between errors is
included for the Alekhin fit (see Sec.~\ref{sec:alek}), 
assuming that these errors are distributed
in the Gaussian manner. The errors in the MRST fit are treated as
explained in the appendix of \cite{mrst}, and the correlated errors are
not allowed to move the central values of the data to as great an extent
for the HERA data, and cannot do so at all for the fixed target data,
where the data used are averaged over the different beam 
energies. The Alekhin approach is more statistically rigorous. The 
MRST approach is more pragmatic, reducing the ability of the data to move
relative to the theory comparison by use of correlated errors 
(other than normalization), and is in some ways similar to the offset method 
\cite{zeus}. The danger of this movement of data relative to theory 
has been suggested by the 
joint analysis of H1 and ZEUS data at this workshop (see Sec.~\ref{sec:mandy}), where  
letting the joint data sets determine the movement due to
correlated errors gives different results from when the data sets are 
compared to theoretical results. 

\subsubsection{Comparison Between the Benchmark Parton Distributions.}
\label{sec:benchcomp}  

\begin{figure}[tbp] 
\centerline{\hspace{-2cm}
\epsfig{figure=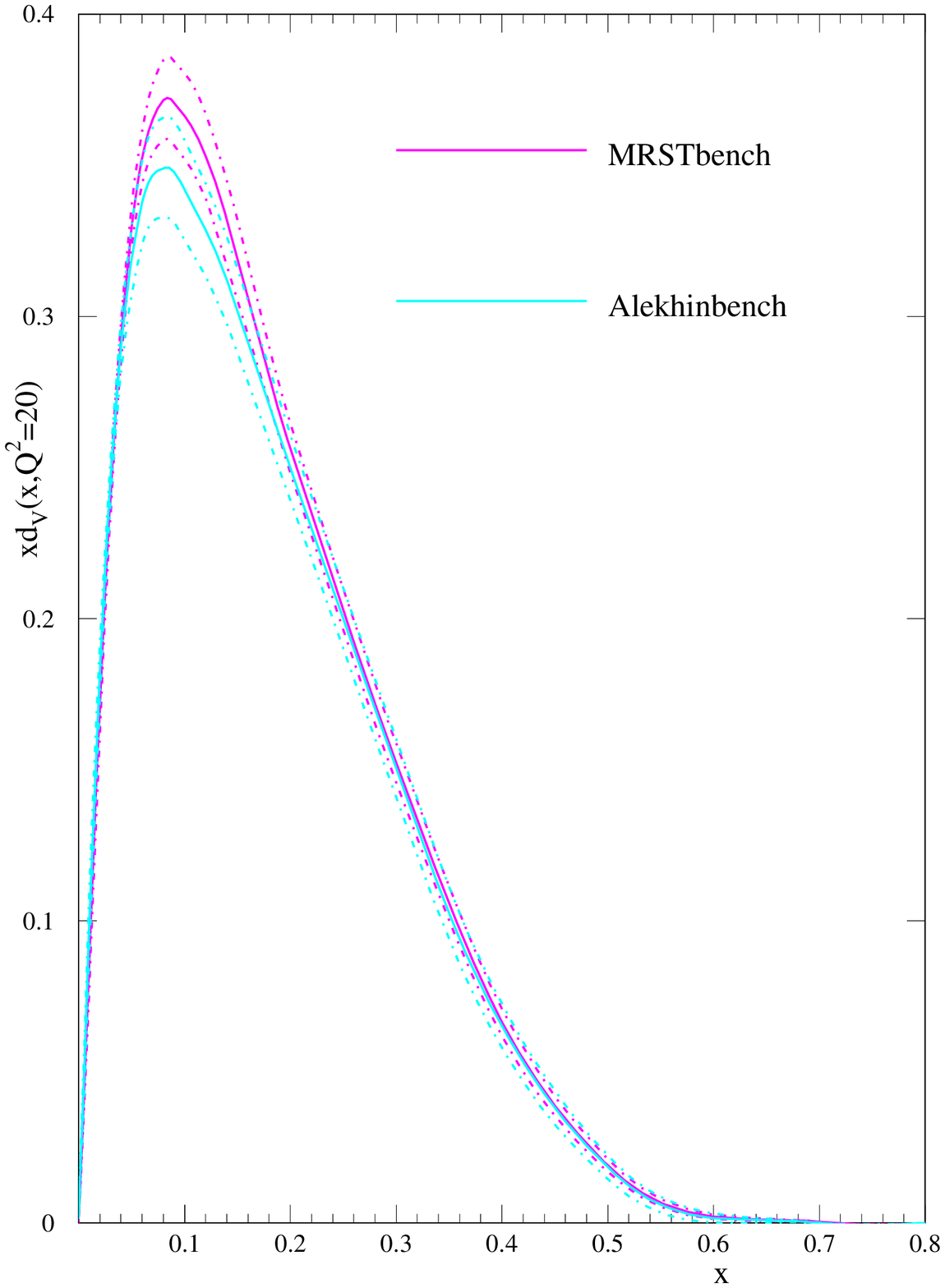,width=0.5\textwidth}
\epsfig{figure=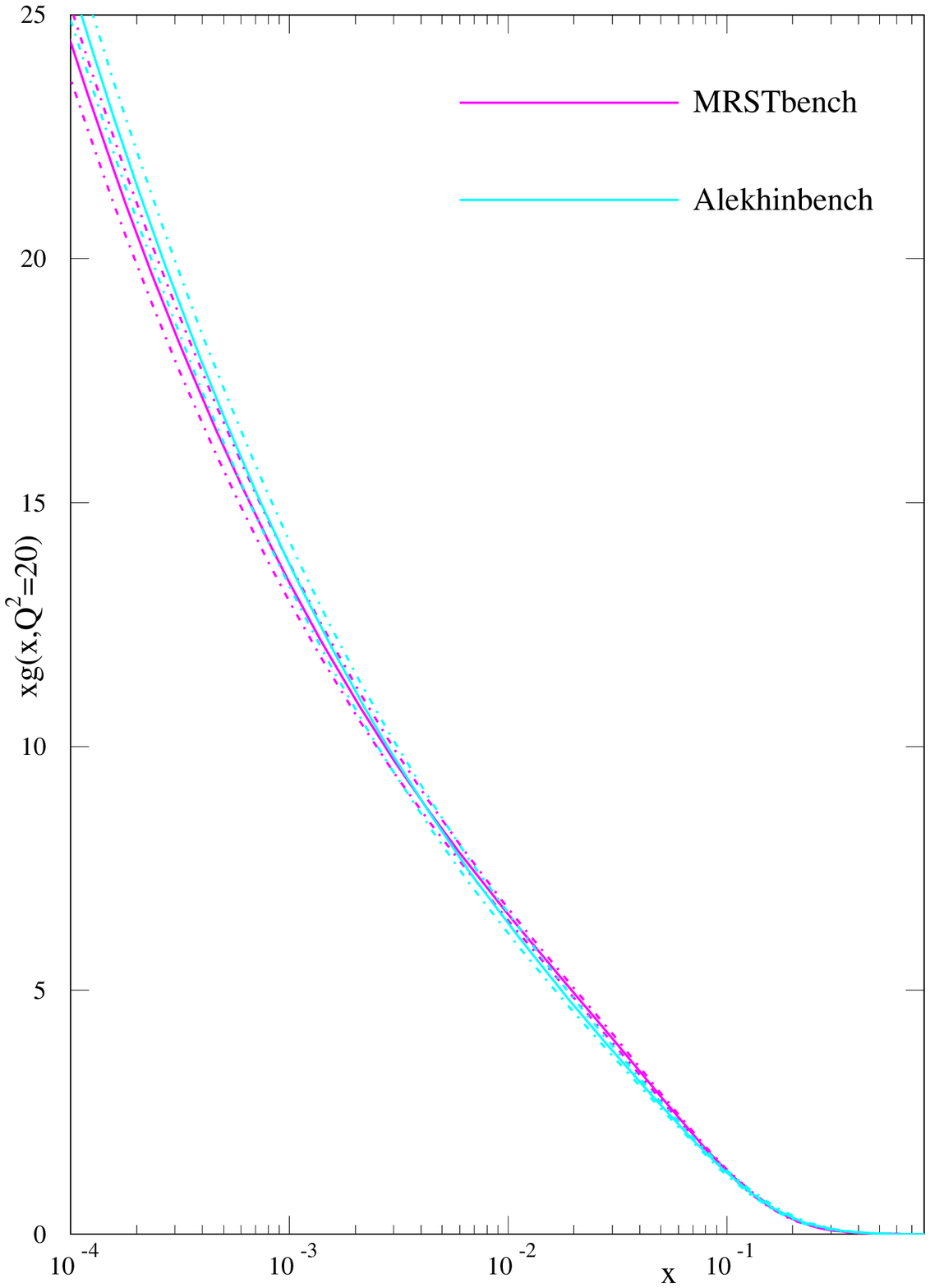,width=0.5\textwidth}}
\caption {Left plot: $xd_V(x,20)$ from the  
MRST benchmark partons compared to that from the Alekhin benchmark
partons.
Right plot: $xg(x,20)$ from the  
MRST benchmark partons compared to that from the Alekhin benchmark
partons.}
\label{fig:comppartonsa}
\end{figure}

I compare the results of the two approaches to fitting the restricted
data chosen for the benchmarking. The input parameters for the Alekhin 
fit are presented in Sec.~\ref{sec:alek}. Those for the MRST type fit are 
similar, but there are some differences which are 
best illustrated by comparing the 
partons at a typical $Q^2$ for the data, e.g. $Q^2=20\GeV^2$. 
A comparison is shown for the $d_V$ quarks and the gluon in Fig. 
\ref{fig:comppartonsa}. 
 
From the plots it is  clear that 
there is generally good agreement between the parton 
distributions. The central values are usually very close, and nearly
always within the uncertainties. The difference in the central values is 
mainly due to the different treatment of correlated errors,
and partially due to the difference in the coupling definition. 
The uncertainties are similar in the two sets, but 
are generally about $1.2-1.5$ times larger for the Alekhin 
partons, due to the increased freedom in the use of the 
correlated experimental errors. The values of $\alpha_S(M_Z^2)$ are quite 
different, $\alpha_S(M_Z^2)=0.1132\pm 0.0015$ compared to $0.1110 \pm 0.0012$.
However, as mentioned earlier, one expects a $1\%$ difference due to 
the different threshold prescriptions -- the MRST $\alpha_S$ would be larger
at $Q^2\sim 20\GeV^2$, where the data are concentrated, so correspondingly
to fit the data it receives a $1\%$ shift downwards for $Q^2=M_Z^2$. Once this
systematic effect is taken into account, the values of $\alpha_S(M_Z^2)$ are
very compatible. Hence, there is no surprising inconsistency between the two 
sets of parton distributions.

\subsubsection{Comparison of the Benchmark Parton Distributions and Global Fit 
Partons.}
\label{sec:comppart}

\begin{figure}[tbp] 
\centerline{\hspace{-2cm}
\epsfig{figure=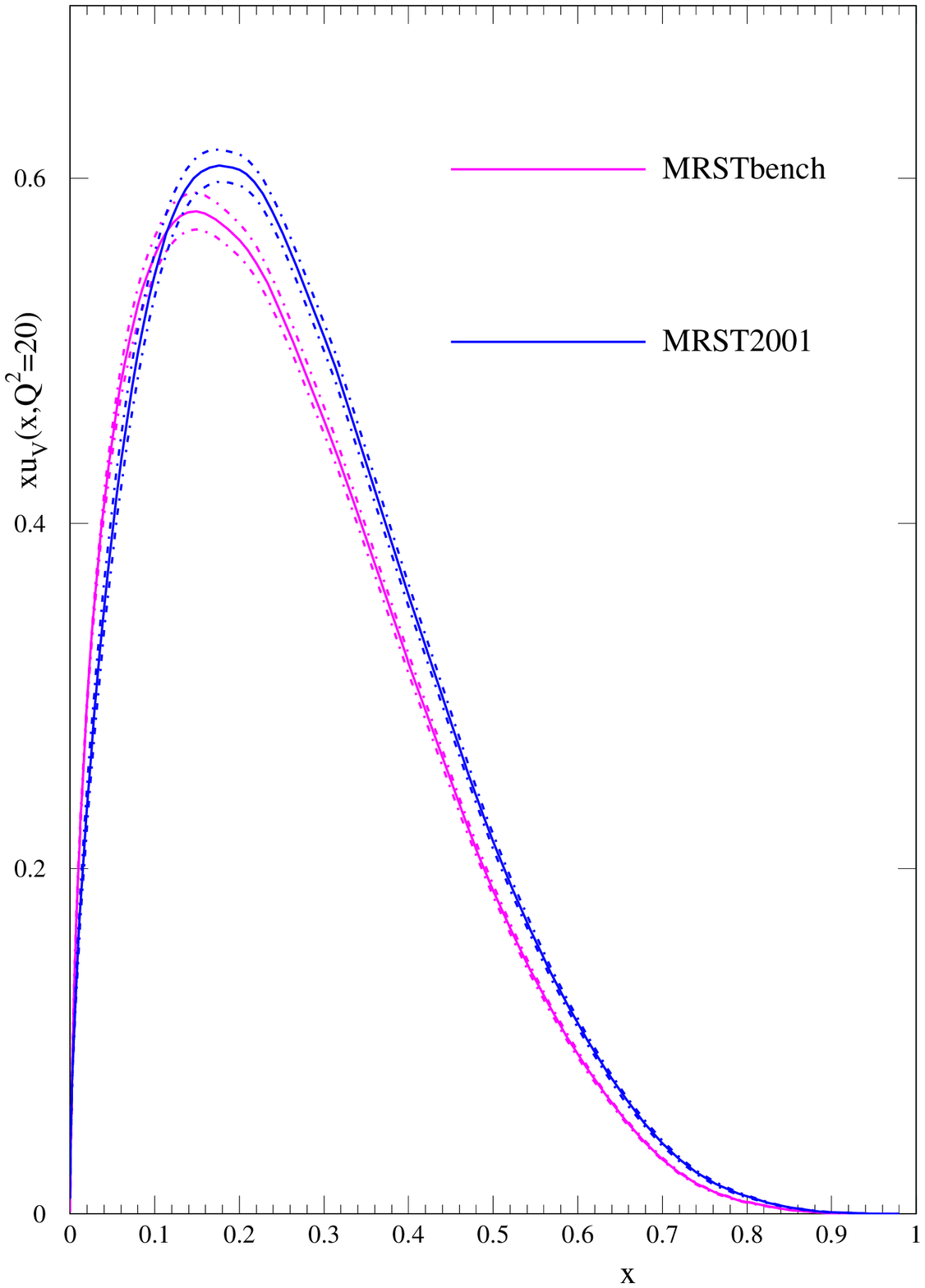,width=0.5\textwidth}
\epsfig{figure=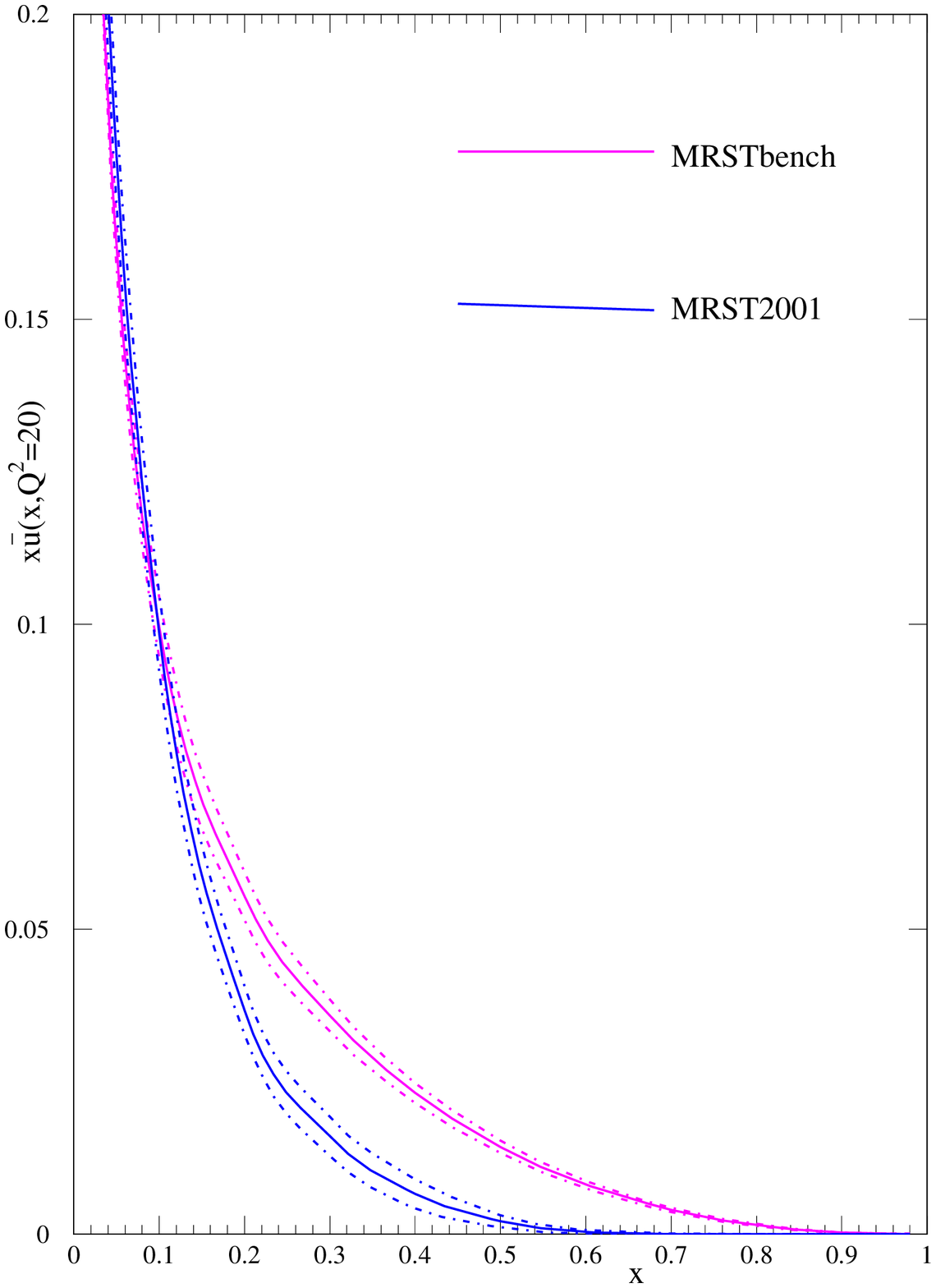,width=0.5\textwidth}}
\caption {Left plot: $xu_V(x,20)$ from the  
MRST benchmark partons compared to that from the MRST01
partons.
Right plot: $x\bar u(x,20)$ from the  
MRST benchmark partons compared to that from the MRST01
partons with emphasis on large $x$.}
\label{fig:comppartonsu}
\end{figure}

It is also illuminating to show the comparison between the benchmark partons 
and the published partons from a global fit. This is done below for the 
MRST01 partons. For example, $u_V(x,Q^2)$ and $\bar u(x,Q^2)$ are shown in 
Fig. \ref{fig:comppartonsu}. It is striking that the uncertainties in the two 
sets are rather similar. This is despite the fact that the uncertainty on 
the benchmark partons 
is obtained from allowing $\Delta \chi^2 =1$ in the fit while that for the 
MRST01 partons is obtained from $\Delta \chi^2=50$.\footnote{Though it is 
meant to be interpreted as a one sigma error in the former case and a 
$90\%$ confidence limit in the latter.} This illustrates the 
great improvement in precision which is obtained due to 
the increase in data from the relaxation of the cuts  and the inclusion of 
types of data other than DIS. For the $u_V$ partons, which are those most 
directly constrained by the DIS data in the benchmark fit, the comparison 
between the two sets of partons is reasonable, but hardly  
perfect -- the central values differing by a few standard deviations. 
This is particularly important given that in this comparison the 
treatment of the data in the fit has been exactly the same in both cases.
There is a minor difference in theoretical approach because of the 
simplistic treatment of heavy
flavours in the benchmark fit. However, this would influence the gluon and 
sea quarks rather than valence quarks. Moreover, the region sensitive to this
simplification would be  $Q^2\sim m_c^2$ (the lower charge weighting for 
bottom quarks greatly
reducing the effect near $Q^2=m_b^2$) which is removed by the 
$Q^2$ cut of $9 \GeV^2$. 
Indeed, introducing the variable flavour number scheme 
usually used for the MRST 
partons modifies the benchmark partons only very minimally.
Hence, if the statistical analysis is correct,
the benchmark partons should agree with the global partons
within their uncertainties (or at most 1.5 times their uncertainties, allowing
for the effect of the correlated errors), which they do not. 
For the $\bar u$ partons the comparison is far worse, the benchmark partons 
being far larger at high $x$. 

\begin{figure}[tbp] 
\centerline{\hspace{-2cm}
\epsfig{figure=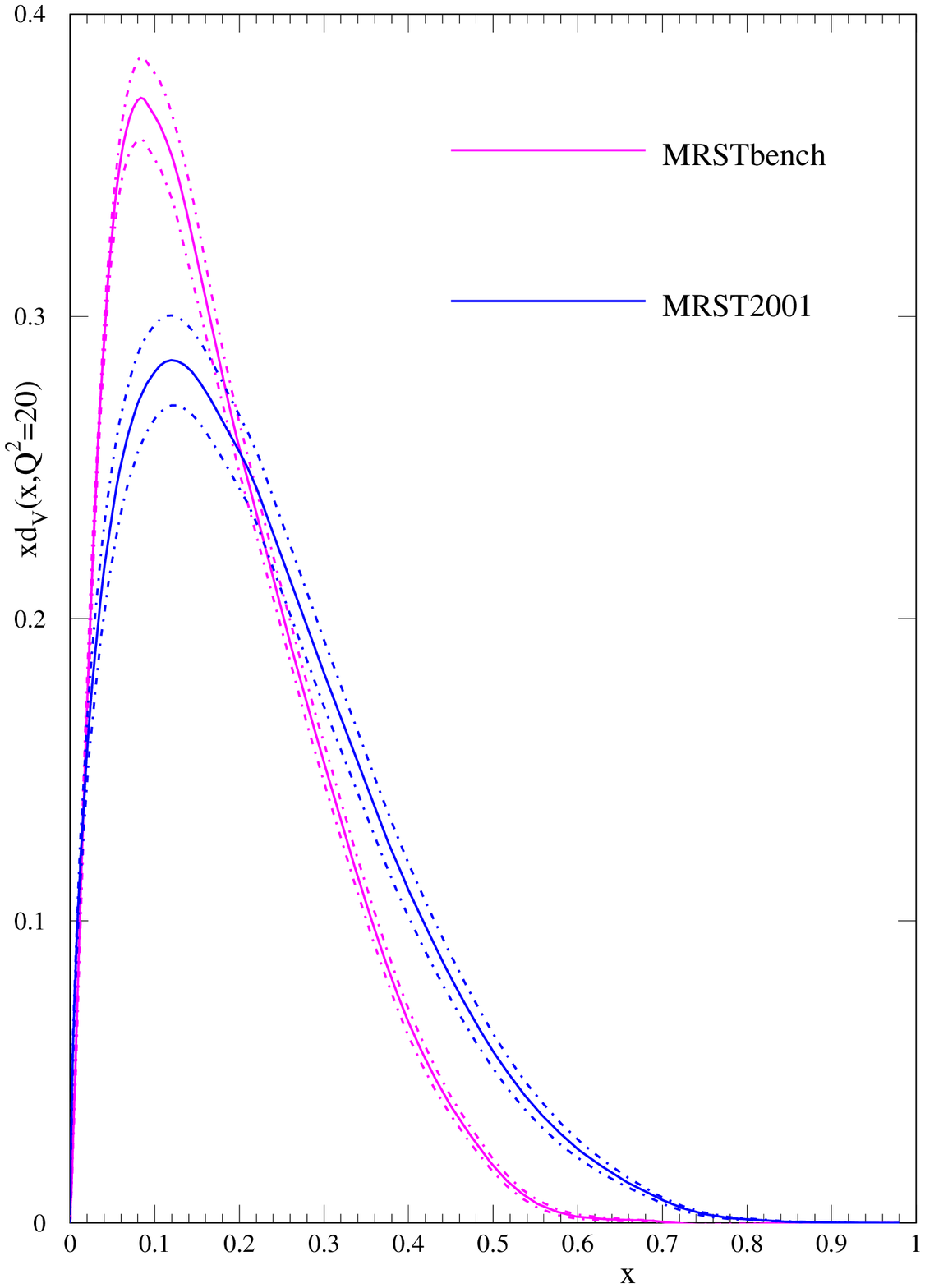,width=0.5\textwidth}
\epsfig{figure=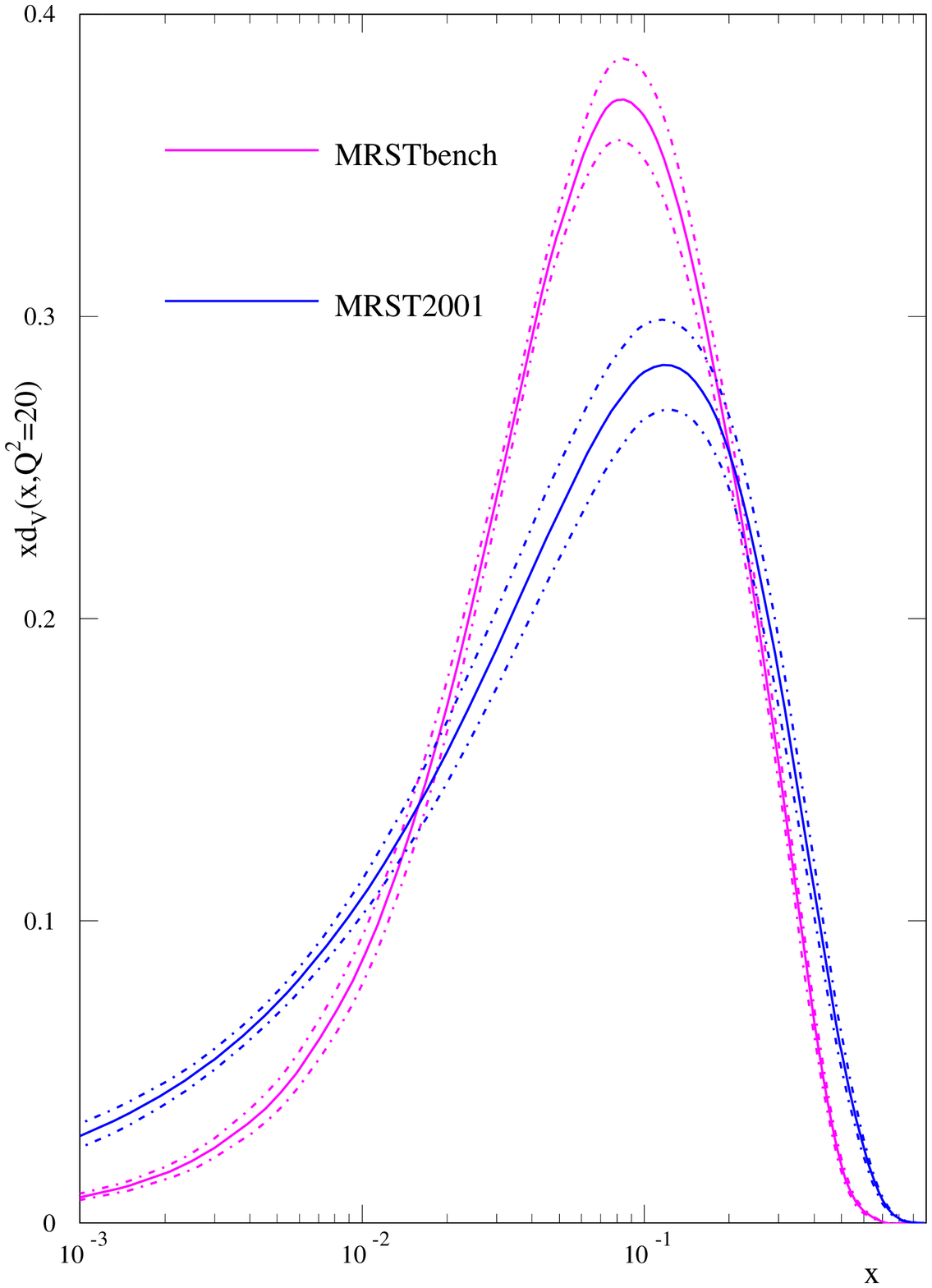,width=0.5\textwidth}}
\caption {Left plot: $xd_V(x,20)$ from the  
MRST benchmark partons compared to that from the MRST01
partons.
Right plot: $xd_V(x,20)$ from the  
MRST benchmark partons compared to that from the MRST01
partons with emphasis on small $x$.}
\label{fig:comppartonsdv}
\end{figure}

This disagreement in the high-$x$ $\bar u$ partons can be understood better 
if one also looks at the high-$x$ $d_V$ distribution shown in Fig.
\ref{fig:comppartonsdv}.
Here the benchmark distribution is very much smaller than for MRST01. However,
the increase in the sea distribution, which is common to protons and neutrons, 
at high-$x$ has allowed a good fit to the high-$x$ BCDMS deuterium data even 
with the very small high-$x$ $d_V$ distribution. 
In fact it is a better fit than 
in \cite{mrst}. However, the fit can be shown to break down with the 
additional inclusion of high-$x$ SLAC data \cite{Whitlow:1991uw}
on the deuterium structure 
function. More dramatically, the shape of the $\bar u$ is 
also completely
incompatible with the Drell-Yan data usually included in the global fit, e.g.
\cite{Moreno:1990sf,E866}. Also in Fig. \ref{fig:comppartonsdv} we see that the 
$d_V$ distributions are very different at smaller $x$. The benchmark 
set is markedly 
inconsistent with NMC data on $F_2^n(x,Q^2)/F_2^p(x,Q^2)$ which is at 
small $x$, but below the cut of $Q^2=9\GeV^2$. 

\begin{figure}[tbp] 
\centerline{\hspace{-2cm}
\epsfig{figure=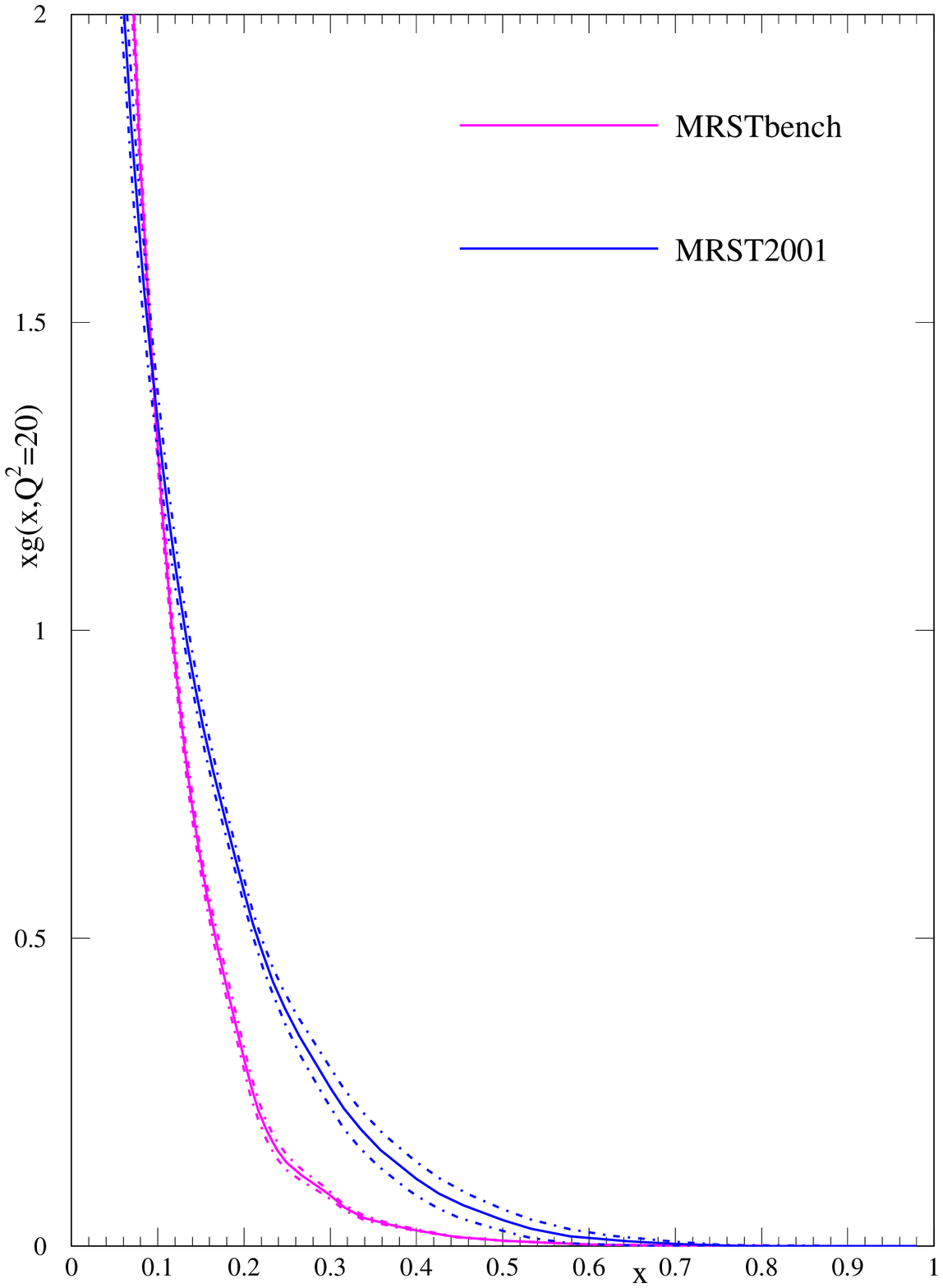,width=0.5\textwidth}
\epsfig{figure=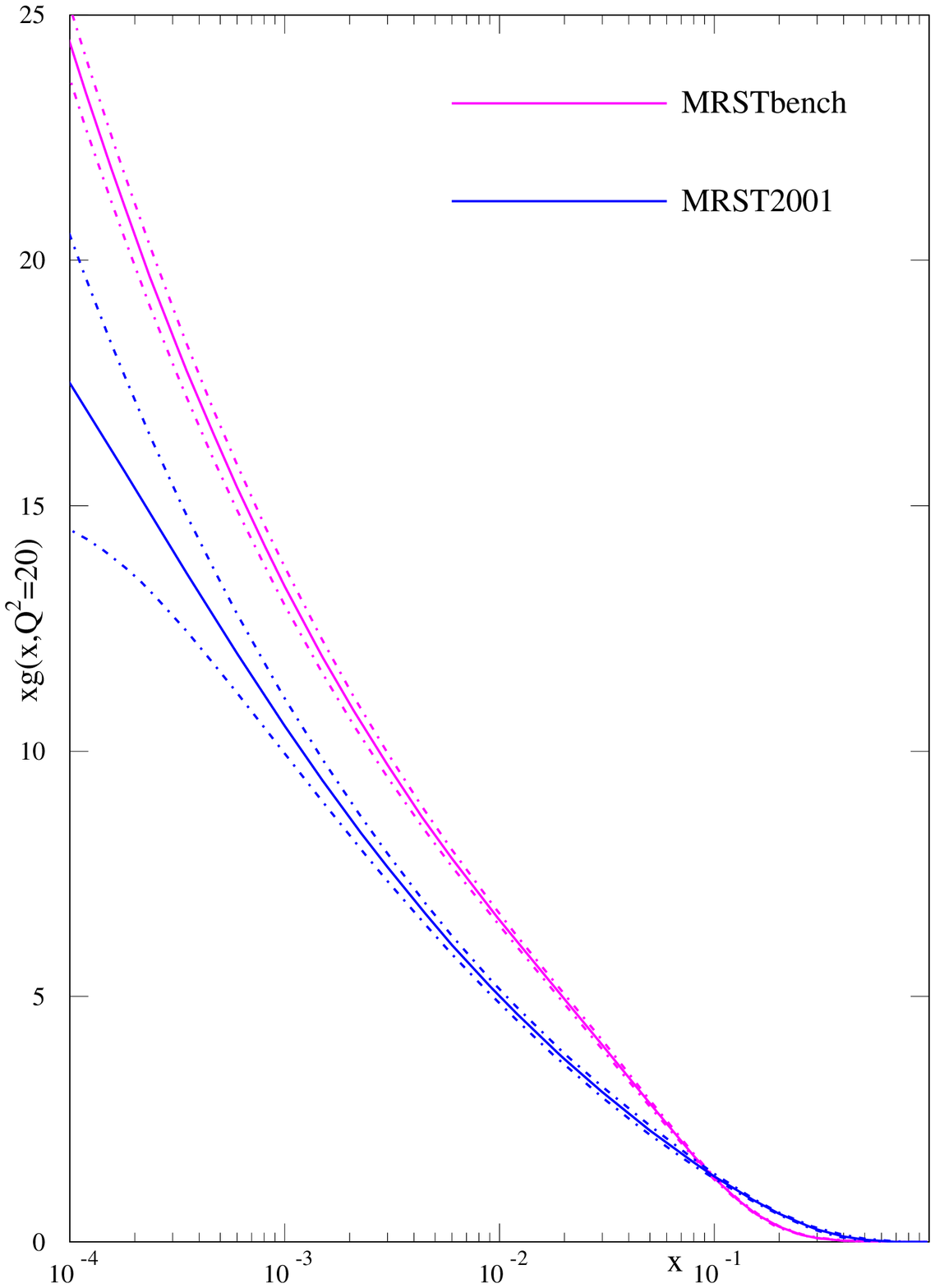,width=0.5\textwidth}}
\caption {Left plot: $xg(x,20)$ from the  
MRST benchmark partons compared to that from the MRST2001
partons.
Right plot: $xg(x,20)$ from the  
MRST benchmark partons compared to that from the MRST2001
partons with emphasis on small $x$.}
\label{fig:comppartonsg}
\end{figure}

The gluon from the benchmark set is also compared to the MRST01 gluon in Fig.
\ref{fig:comppartonsg}. Again there is an enormous difference at high 
$x$. Nominally the benchmark gluon has little to constrain it at high 
$x$. However, the momentum sum rule determines it to be very small in this 
region in order to 
get the best fit to HERA data, similar to the gluon from 
\cite{h1alphas}. As such, the gluon
has a small uncertainty and is many standard deviations from the 
MRST01 gluon. Indeed, the input gluon at high $x$ is so small that its value 
at higher $Q^2$ is dominated by the evolution of $u_V$ quarks to gluons,
rather than by the input gluon. Hence, the uncertainty is dominated by the 
quark 
parton input uncertainty rather than its own, and since the up quark is well
determined the uncertainty on the high-$x$ gluon is small for the 
benchmark partons.  The smallness of the high-$x$ gluon 
results in the benchmark partons producing a very poor
prediction indeed for the Tevatron jet data \cite{D0,CDF}, which are the usual 
data that constrain the high-$x$ gluon in global fits.  

It is also illustrative 
to look at small $x$. Here the benchmark gluon is only a couple of standard
deviations
from the MRST01 gluon, suggesting that its size is not completely 
incompatible with a good fit to the HERA small-$x$ data at $Q^2$ below the 
benchmark cut. However, the uncertainty in the benchmark gluon is much 
smaller than in the MRST01 gluon, despite the much smaller amount of 
low-$x$ data in the fit for the benchmark partons. 
This comes about as a result of the artificial choice made in the gluon
input at $Q_0^2$. Since it does not have the term introduced in 
\cite{mrst}, allowing the freedom for the input gluon to be negative at 
very small $x$, the gluon
is required by the fit to be valence-like. Hence, at 
input it is simply very small at small $x$. At higher $Q^2$ it becomes much 
larger, but in a manner driven entirely by evolution, i.e. it is determined
by the input gluon at moderate $x$, 
which is well constrained. In this framework
the small-$x$ gluon does not have any intrinsic uncertainty -- its uncertainty
is a reflection of moderate $x$. This is a feature of e.g. the CTEQ6 gluon 
uncertainty \cite{cteq}, where the input gluon is valence-like. In this case 
the percentage gluon uncertainty does not get any larger once $x$ 
reaches about $0.001$. The alternative treatment in \cite{mrst}
gives the expected increase in the gluon uncertainty as $x\to 0$,
since in this case the uncertainty is determined 
largely by that in the input gluon at small $x$. 
The valence-like input form for a gluon is an example of fine-tuning, the form
being unstable to evolution in either direction. The artificial 
limit on the small-$x$ uncertainty is a consequence of this.

\subsubsection{Conclusions.}
\label{sec:conc}

I have demonstrated that different approaches to fitting 
parton distributions that use exactly the same data and theoretical 
framework produce partons that are very similar and have comparable 
uncertainties. There are certainly some differences due to the alternative 
approaches to dealing with experimental errors, but these are relatively 
small. However, the partons extracted using a
very limited data set are completely incompatible,
even allowing for the uncertainties,
with those obtained from a global fit with an identical treatment of errors 
and a minor difference in theoretical procedure.  
This implies that the inclusion of more data from a variety of 
different experiments moves the central values of the partons in a 
manner indicating either that the different experimental data are 
inconsistent with each other, 
or that the theoretical framework is inadequate for correctly describing the 
full range of data. To a certain extent both explanations are probably
true. Some 
data sets are not entirely consistent with each other (even if they
are seemingly equally reliable). Also, there are a wide variety of reasons 
why NLO perturbative QCD might require modification for some data sets, or in 
some kinematic regions \cite{epj:c35:325}. Whatever the reason for the 
inconsistency between the MRST benchmark partons and the MRST01 partons, the 
comparison exhibits the dangers in extracting partons from a very limited 
set of data and taking them seriously. It also clearly
illustrates the problems in determining the 
true uncertainty on parton distributions.

%% file: cteq.tex
\subsection{Stability of PDF fits \protect 
\footnote{Contributing authors: J.~Huston, J.~Pumplin.}
\label{sec:cteq}}

One of the issues raised at the workshop is the reliability of determinations of 
parton distribution functions (PDFs), which might be compromised for example by 
the neglect of NNLO effects or non-DGLAP evolution in the standard analysis, or 
hidden assumptions made in parameterizing the PDFs at nonperturbative scales.  
We summarize the results of the CTEQ PDF group on this issue. 
For the full story see \cite{Huston:2005jm}.

\subsubsection{Stability of PDF determinations}
The stability of NLO global analysis was seriously challenged 
by an analysis \cite{epj:c35:325} which found a 20\% variation 
in the cross section predicted for $W$ production at the LHC -- a critical 
``standard candle'' process for hadron colliders -- when certain cuts on input 
data are varied. If this instability were confirmed, it would significantly 
impact the phenomenology of a wide range of physical processes for the
Tevatron Run II and the LHC. The CTEQ PDF group therefore performed an 
independent study of this issue within their global analysis framework.  
In addition, to explore the dependence of the results on assumptions about the 
parameterization of PDFs at the starting scale $Q_{0} \! = \! 1.3\, \mathrm{GeV}$, 
we also studied the effect of allowing a negative gluon distribution at small $x$ -- 
a possibility that is favored by the MRST NLO analysis, and that is closely tied 
to the W cross section controversy. 

The stability of the global analysis was investigated by varying the inherent 
choices that must be made to perform the analysis. These choices include the 
selection of experimental data points based on kinematic cuts, the functional 
forms used to parameterize the initial nonperturbative parton distribution functions, 
and the treatment of $\alpha_s$. 

The stability of the results is most 
conveniently measured by differences in the global $\chi^2$ for the relevant fits. 
To quantitatively define a change of $\chi^{2}$ that characterizes a significant 
change in the quality of the PDF fit is a difficult issue in global QCD analysis. 
In the context of the current analysis, we have argued that an increase by 
$\Delta\chi^{2}\sim 100$ (for $\sim$ \! 2000 data points) represents roughly 
a 90\% confidence level uncertainty on PDFs due to the uncertainties of the 
current input experimental data~\cite{%
Pumplin:2000vx,Stump:2001gu,Pumplin:2001ct,cteq}. 
In other words, PDFs with $\chi^2 - \chi^2_{\mathrm{Best Fit}} > 100$ are 
regarded as not tolerated by current data. 

The CTEQ6 and previous CTEQ global fits imposed ``standard'' cuts 
$Q > 2 \,\mathrm{GeV}$ and $W > 3.5 \, \mathrm{GeV}$ on the input data set, 
in order to suppress higher-order terms in the perturbative expansion and the 
effects of resummation and power-law (``higher twist'') corrections. 
We examined the effect of stronger cuts on $Q$ to see if the fits are stable. 
We also examined the effect of imposing cuts on $x$, which should serve 
to suppress any errors due to deviations from DGLAP evolution, such as those 
predicted by BFKL. The idea is that any inconsistency in the global fit due to 
data points near the boundary of the accepted region will be revealed by an 
improvement in the fit to the data that remain after those near-boundary 
points have been removed. 
In other words, the decrease in $\chi^2$ for the subset of data that is retained, 
when the PDF shape parameters are refitted to that subset alone, measures the 
degree to which the fit to that subset was distorted in the original fit by 
compromises imposed by the data at low $x$ and/or low $Q$. 

The main results of this study are presented in Table~\ref{tab:tableI}. 
Three fits are shown, from three choices of the cuts on input data as specified 
in the table. They are labeled `standard', `intermediate' and `strong'. 
$N_{\rm pts}$ is the number of data points that pass the cuts in each case, 
and $\chi^2_{N_{\rm pts}}$ is the $\chi^2$ value for that subset of data. 
The fact that the changes in $\chi^2$ in each column are insignificant compared 
to the uncertainty tolerance is strong evidence that our NLO global fit results 
are very stable with respect to choices of kinematic cuts.
\begin{table}[t]
\begin{center}
\begin{tabular}{||r|r|r|r||c|c|c|c||c||}
\hline \hline  
Cuts   & $Q_{\mathrm{min}}$  &   $x_{\mathrm{min}}$  &  $N_{\mathrm{pts}}$ &
$\chi^2_{1926}$ & $\chi^2_{1770}$ & $\chi^2_{1588}$ &
$\sigma_W^{\mathrm{LHC}} \! \times \! B_{\ell \nu} \, [\mathrm{nb}]$ 
\\
\hline 
standard     & $2     \, \mathrm{GeV}$ & $0$     & $1926$ &  $2023$ & 
$1850$ & $1583$ & $20.02$ 
\\ 
intermediate & $2.5   \, \mathrm{GeV}$ & $0.001$ & $1770$ &    --   & 
$1849$ & $1579$ & $20.10$ 
\\ strong       & $3.162 \, \mathrm{GeV}$ & $0.005$ & $1588$ &    --   &   
--   & $1573$ & $20.34$ 
\\
\hline \hline
\end{tabular}
\caption{Comparisons of three fits with different choices of the cuts on input 
data at the $Q$ and $x$ values indicated. In these fits, a conventional 
positive-definite gluon parameterization was used. 
\label{tab:tableI}}
\end{center}
\end{table}
\begin{table}[ht]
\begin{center}
\begin{tabular}{||r|r|r|r||c|c|c|c||c||}
\hline \hline  
Cuts  & $Q_{\mathrm{min}}$  &   $x_{\mathrm{min}}$  &  $N_{\mathrm{pts}}$ &
$\chi^2_{1926}$ & $\chi^2_{1770}$ & $\chi^2_{1588}$ &
$\sigma_W^{\mathrm{LHC}} \! \times \! B_{\ell \nu} \, [\mathrm{nb}]$ 
\\
\hline 
standard     & $2     \, \mathrm{GeV}$ & $0$     & $1926$ & $2011$ & $1845$ &
$1579$ & $19.94$ 
\\ 
intermediate & $2.5   \, \mathrm{GeV}$ & $0.001$ & $1770$ &   --   & $1838$ &
$1574$ & $19.80$ 
\\ 
strong       & $3.162 \, \mathrm{GeV}$ & $0.005$ & $1588$ &   --   &   --   &
$1570$ & $19.15$ 
\\
\hline \hline
\end{tabular}
\caption{Same as Table~\ref{tab:tableI} except that the gluon parameterization 
is extended to allow negative values. \label{tab:tableII} }
\end{center}
\end{table}

We extended the analysis to a series of fits in which the gluon distribution 
$g(x)$ is allowed to be negative at small $x$, at the scale 
$Q_0 \! = \! 1.3 \, \mathrm{GeV}$ where we begin the DGLAP evolution. 
The purpose of this additional study is to determine whether the feature 
of a negative gluon PDF is a key element in the stability puzzle, 
as suggested by the findings of~\cite{epj:c35:325}. 
The results are presented in Table~\ref{tab:tableII}. 
Even in this extended case, we find no evidence of instability. 
For example, $\chi^2$ for the subset of 1588 points that pass 
the \textit{strong} cuts increases only from 1570 to 1579 when the fit is 
extended to include the full standard data set. 

Comparing the elements of Table~\ref{tab:tableI} and Table~\ref{tab:tableII} 
shows that our fits with $g(x) < 0$ have slightly smaller values of$\chi^2$: 
e.g., $2011$ versus $2023$ for the standard cuts. 
However, the difference $\Delta \chi^2 \! = \! 12$ between these values is again 
not significant according to our tolerance criterion.

\subsubsection{W cross sections at the LHC}
\begin{figure}[ht]\centering 
\resizebox*{0.6\textwidth}{!} {\includegraphics[clip=true]{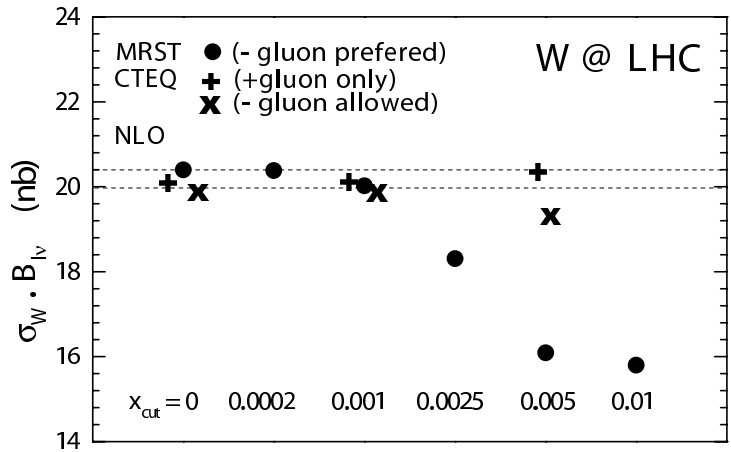} }
\caption{ Predicted total cross section of $W^+ + W^-$ production at the LHC 
for the fits obtained in our stability study, compared to the NLO results of 
Ref.~\protect\cite{epj:c35:325}. 
The $Q$-cut values associated with the CTEQ points are given in the two tables. 
The overall PDF uncertainty of the prediction is $\sim 5\%$. } 
\label{fig:WtotXs}
\end{figure}

The last columns of Tables~\ref{tab:tableI} and \ref{tab:tableII} show the 
predicted cross section for $W^+ + W^-$ production at the LHC. This prediction 
is also very stable: it changes by only $1.6 \%$ for the positive-definite 
gluon parameterization, which is substantially less than the overall PDF uncertainty 
of $\sigma_W$ estimated previously with the standard cuts. 
For the negative gluon parameterization, the change is $4 \%$--larger, 
but still less than the overall PDF uncertainty. 
These results are explicitly displayed, and compared to the MRST results 
in Fig.~\ref{fig:WtotXs}.
We see that this physical prediction is indeed insensitive to the kinematic cuts 
used for the fits, and to the assumption on the positive definiteness of the gluon 
distribution. 

We also studied the stability of the prediction for $\sigma_{W}$ 
using the Lagrange Multiplier (LM) method of Refs.~\cite{%
Pumplin:2000vx,Stump:2001gu,Pumplin:2001ct}. 
Specifically, we performed a series of fits to the global data set that are 
constrained to specific values of $\sigma_W$ close to the best-fit prediction. 
The resulting variation of $\chi^{2}$ versus $\sigma_{W}$ measures the uncertainty 
of the prediction. We repeated the constrained fits for each case of fitting 
choices (parameterization and kinematic cuts). 
In this way we gain an understanding of the stability of the uncertainty, 
in addition to the stability of the central prediction.  

\begin{figure}[htb]\centering 
\resizebox*{0.4\textwidth}{!}{\includegraphics[clip=true]{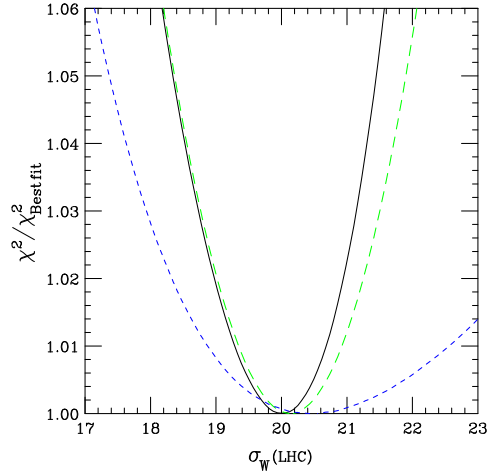}}
\caption{ Lagrange multiplier results for the $W$ cross section (in $\mathrm{nb}$) 
at the LHC using a positive-definite gluon. The three curves, in order of 
decreasing steepness, correspond to the three sets of kinematic cuts labeled 
standard/intermediate/strong in Table~\ref{tab:tableI}. 
\label{fig:ChiVsSigCutsPG}}
\end{figure}

Figure~\ref{fig:ChiVsSigCutsPG} shows the results of the LM study for the three 
sets of kinematic cuts described in Table~\ref{tab:tableI}, all of which have 
a positive-definite gluon distribution.
The $\chi^2$ shown along the vertical axis is normalized to its value for the best 
fit in each series. In all three series, $\chi^2$ depends almost quadratically on 
$\sigma_W$. We observe several features:
\begin{itemize}
\item 
  The location of the minimum of each curve represents the best-fit prediction 
  for $\sigma_W^{\mathrm{LHC}}$ for the corresponding choice of cuts. The fact 
  that the three minima are close together displays the stability of the predicted 
  cross section already seen in Table~\ref{tab:tableI}.
\item
  Although more restrictive cuts make the global fit less sensitive to possible 
  contributions from resummation, power-law and other nonperturbative effects,
  the loss of constraints caused by the removal of precision HERA data points at 
  small $x$ and low $Q$ results directly in increased uncertainties on the PDF 
  parameters and their physical predictions. This is shown in 
  Fig.~\ref{fig:ChiVsSigCutsPG} by the increase of the width of the curves with 
  stronger cuts. The uncertainty of the predicted $\sigma_W$ increases by more 
  than a factor of 2 in going from the standard cuts to the strong cuts.
\end{itemize}

\begin{figure}[htb]\centering 
\resizebox*{0.4\textwidth}{!}{\includegraphics[clip=true]{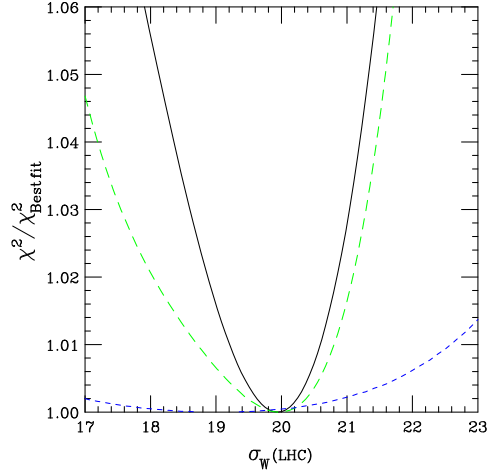}}
\caption{ Lagrange multiplier results for the $W$ cross section (in $\mathrm{nb}$) 
at the LHC using a functional form where the gluon is not required to be 
positive-definite. The three curves, in order of decreasing steepness, 
correspond to the three sets of kinematic cuts labeled standard/intermediate/strong 
in Table~\ref{tab:tableII}. 
\label{fig:ChiVsSigCutsNG}}
\end{figure}

Figure~\ref{fig:ChiVsSigCutsNG} shows the results of the LM study for the three 
sets of kinematic cuts described in Table~\ref{tab:tableII}, all of which have 
a gluon distribution which is allowed to go negative.

We observe:
\begin{itemize}
\item
  Removing the positive definiteness condition necessarily lowers the value of
  $\chi^2$, because more possibilities are opened up in the $\chi^2$ minimization 
  procedure.  But the decrease is insignificant compared to other sources of 
  uncertainty. Thus, a negative gluon PDF is allowed, but not required.
\item 
  The minima of the two curves occur at approximately the same $\sigma_{W}$. 
  Allowing a negative gluon makes no significant change in the central prediction 
  -- merely a decrease of about $1\,\%$, which is small compared to the overall 
  PDF uncertainty. 
\item
  For the standard set of cuts, allowing a negative gluon PDF would expand the 
  uncertainty range only slightly.  For the intermediate and strong cuts, allowing 
  a negative gluon PDF would significantly expand the uncertainty range. 
\end{itemize}

\begin{figure}[hbt]  
\mbox{        
\resizebox{0.48\textwidth}{!}{        
\includegraphics[clip=true,height=.25\textheight]{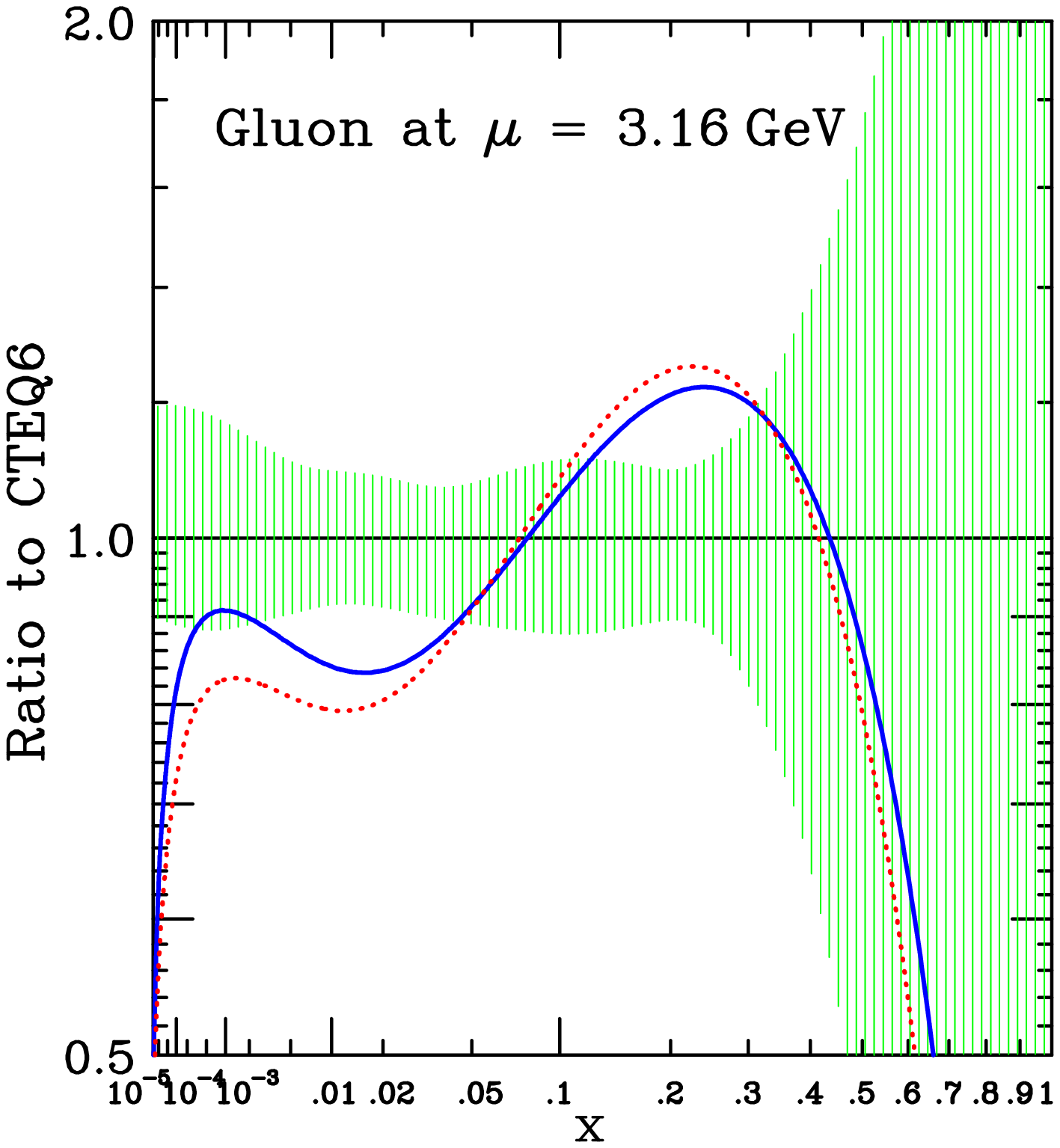}} 
\hfill        
\resizebox{0.48\textwidth}{!}{        
\includegraphics[clip=true,height=.25\textheight]{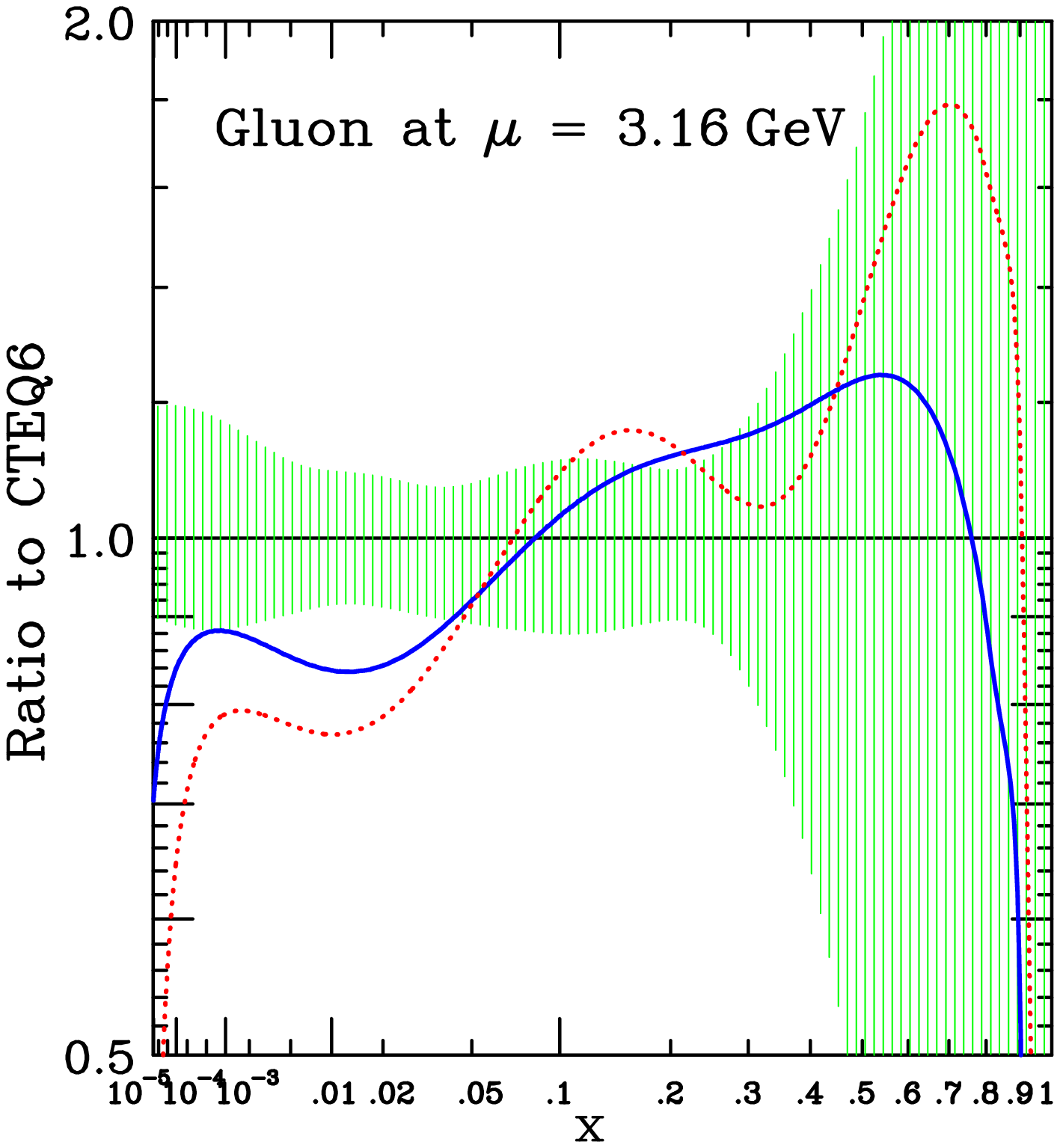}}        
}  
\caption{Left: mrst2002 NLO (solid) and NNLO (dotted);
Right: mrst2004 NLO (solid) and NNLO (dotted);
Shaded region is uncertainty according to the 40 eigenvector sets of CTEQ6.1.   
\label{fig:GluUncQ3s35}}
\end{figure}
We examined a number of aspects of our analysis that might account for the 
difference in conclusions between our stability study and that 
of~\cite{epj:c35:325}. 
A likely candidate seems to be that in order to obtain stability, it is necessary 
to allow a rather free parametrization of the input gluon distribution. 
This suspicion is seconded by recent work by MRST \cite{Martin:2004ir}, in which 
a different gluon parametrization appears to lead to a best-fit gluon distribution 
that is close to that of CTEQ6. In summary, we found that the NLO PDFs and their 
physical predictions at the Tevatron and LHC are quite stable with respect to 
variations of the kinematic cuts and the PDF parametrization after all. 

\subsubsection{NLO and NNLO}
In recent years, some preliminary next-to-next-leading-order (NNLO) analyses 
for PDFs have been carried out either for DIS alone \cite{Alekhin:2003yh}, 
or in a global analysis context \cite{Martin:2002dr} -- even if all the necessary 
hard cross sections, such as inclusive jet production, are not yet available 
at this order. 
Determining the parton distributions at NNLO is obviously desirable on theoretical 
grounds, and it is reasonable to plan for having a full set of tools for a true 
NNLO global analysis in place by the time LHC data taking begins.  
At the moment, however, NNLO fitting is not a matter of pressing necessity, 
since the difference between NLO and NNLO appears to be very small compared 
to the other uncertainties in the PDF analysis. 
This is demonstrated in Fig.~\ref{fig:GluUncQ3s35}, which shows the NLO and NNLO 
gluon distributions extracted by the MRST group. 
The difference between the two curves is much smaller than the other uncertainties 
measured by the 40 eigenvector uncertainty sets of CTEQ6.1, which is shown by 
the shaded region.  The difference is also much smaller than the difference 
between CTEQ and MRST best fits.  
Similar conclusions \cite{Pumplin:2005yf} can be found using the NLO and NNLO 
fits by Alekhin.

%% file: nnpdf.tex
\subsection{The neural network approach to parton distributions 
\protect\footnote{Contributing authors: L.~Del Debbio, S.~Forte, 
  J.~I.~Latorre, A.~Piccione, J.~Rojo}}
\label{sec:nnpdf}

The requirements of precision physics at
 hadron colliders, as has been emphasized through
this workshop, have recently led
to a rapid improvement in the techniques for the determination of 
parton distribution functions (pdfs) of the nucleon.
Specifically it is now mandatory to 
determine accurately the uncertainty on these quantities, and
the different collaborations performing global pdf 
analysis\cite{Martin:2002dr,cteq61,Alekhin:2002fv} have performed 
estimations of these uncertainties using a variety
of techniques. The
main difficulty 
is that one is trying to determine the uncertainty on a function,
that is, a probability measure in a space of functions, and to extract it from
a finite set of experimental data, 
a problem which is mathematically ill-posed. 
It is also known that the standard approach to global parton fits 
have several shortcomings: 
the bias introduced by choosing fixed functional forms to
parametrize the parton distributions
(also known as {\it model dependence}), the problems to assess
faithfully the pdf uncertainties, the
combination of inconsistent experiments, and the lack
of general, process-independent error propagation techniques.
Although the problem of quantifying the uncertainties in pdfs has seen
a huge progress since its paramount importance was raised some
years ago, until now no unambiguous conclusions have been obtained.

In this contribution we present a novel strategy to address the problem
of constructing unbiased parametrizations of parton distributions
with a faithful estimation of their uncertainties, based on 
a combination of two techniques: Monte Carlo methods and neural networks.
This strategy, introduced in \cite{f2nn,nnpdf},
has been first implemented to address the marginally simpler problem
of parametrizing deep-inelastic structure functions $F(x,Q^2)$, which
we briefly summarize now. 
In a first step we construct a Monte Carlo sampling of the experimental data 
(generating artificial data replicas), and then
we train neural networks to each data replica, to
construct a probability measure in the space of structure functions
$\mathcal{P}\lc F(x,Q^2)\rc$. The probability measure constructed
in this way
contains all information from experimental data, including correlations,
with the only assumption of smoothness. Expectation values and moments over
this probability measure are then evaluated as averages over
the trained network sample,
\be
\label{probmeas}
\la \mathcal{F}\lc F(x,Q^2)\rc\ra=\int\mathcal{D}F
\mathcal{P}\lc F(x,Q^2)\rc
\mathcal{F}\lc F(x,Q^2)\rc=\frac{1}{N_{\rep}}
\sum_{k=1}^{N_{\rep}}\mathcal{F}\lp F^{(\net)(k)}(x,Q^2)\rp \ .
\ee
where $\mathcal{F}\lc F\rc$ is an arbitrary function of $F(x,Q^2)$.

The first step is the Monte Carlo sampling of experimental data, 
generating $N_{\rep}$ replicas of the original $N_{\dat}$ experimental data,
\be
F_i^{(\art)(k)} =\lp 1+r_N^{(k)}\sigma_N\rp\lc F_i^{(\rmexp)}+r_i^{s,(k)}
\sigma^{stat}_i+\sum_{l=1}^{N_{sys}}r^{l,(k)}\sigma_i^{sys,l} \rc, 
\qquad i=1,\ldots,N_{\dat} \ ,
\ee
where $r$ are gaussian random numbers with the same correlation
as the respective uncertainties, and $\sigma^{stat},\sigma^{sys},
\sigma_{N}$ are the statistical, systematic and normalization
errors.
The number of replicas $N_{\rep}$ has to be large enough so that the
replica sample
 reproduces central values, errors and correlations
of the experimental data.

The second step consists on training a neural network\footnote{For
a more throughly description of neural network, see \cite{f2nn}
and references therein}  on each of the data
replicas. Neural networks
are specially suitable to parametrize parton distributions since
they are unbiased, robust approximants and interpolate between
data points with the only assumption of smoothness. The neural network 
training consist on the minimization for each replica of the
$\chi^2$ defined with the inverse of the 
experimental covariance matrix,
\be
{\chi^2}^{(k)}=\frac{1}{N_{\dat}}\sum_{i,j=1}^{N_{\dat}}\lp
F_i^{(\art)(k)}-F_i^{(\net)(k)}\rp\mathrm{cov}^{-1}_{ij}
\lp F_j^{(\art)(k)}-F_j^{(\net)(k)}\rp \ .
\ee
Our minimization strategy is based on
Genetic Algorithms (introduced in \cite{rojo04}), which are specially suited
for finding global minima in highly nonlinear 
minimization problems.

The set of trained nets, once is validated through suitable statistical
estimators, becomes the sought-for probability measure 
$\mathcal{P}\lc F(x,Q^2)\rc$ in the space of structure functions. 
Now observables with
errors and correlations can be computed from averages over this
probability measure, using eq. (\ref{probmeas}). 
For example, the average and error of a
structure function $F(x,Q^2)$ at arbitrary $(x,Q^2)$ can be
computed as
\be
\la F(x,Q^2) \ra =\frac{1}{N_{\rep}}\sum_{k=1}^{N_{\rep}}
F^{(\net)(k)}(x,Q^2), \quad
\sigma(x,Q^2)=\sqrt{\la  F(x,Q^2)^2 \ra-\la F(x,Q^2) \ra^2} \ .
\ee
A more detailed account of the application
of the neural network approach to structure
functions can be found in \cite{nnpdf}, which describes
the most recent NNPDF parametrization of the proton structure 
function\footnote{
The source code, driver program and graphical web interface for
our structure function fits is available at {
\tt http://sophia.ecm.ub.es/f2neural}.}.

Hence this strategy can be used also to parametrize
parton distributions, provided one now takes into 
account perturbative QCD evolution. Therefore we need to define
a suitable evolution formalism. Since complex
neural networks are not allowed, 
we must use the convolution theorem to evolve
parton distributions in $x-$space using the inverse $\Gamma(x)$
of the Mellin space evolution factor $\Gamma(N)$, defined as
\be
q(N,Q^2)=q(N,Q_0^2) 
\Gamma\lp N,\aq,\aqq\rp \ ,
\ee
The only subtlety is that the x-space evolution
factor $\Gamma(x)$ is a distribution,
which must therefore be
regulated at $x=1$, yielding the final evolution equation,
\be
q(x,Q^2)= q(x,Q_0^2)\int_x^1 dy~\Gamma(y) +\int_x^1\frac{dy}{y}
\Gamma(y)\lp q\lp\frac{x}{y},Q_0^2\rp -yq(x,Q_0^2)\rp \ ,
\ee
where in the above equation $q(x,Q_0^2)$ is parametrized
using a neural network.
At higher orders in perturbation theory coefficient functions $C(N)$
are introduced through a modified evolution factor, $\tilde{\Gamma}(N)\equiv
 \Gamma(N) C(N)$.
We have benchmarked our evolution code with the
Les Houches benchmark tables \cite{Giele:2002hx} at NNLO up to an
accuracy of $10^{-5}$.
The evolution factor $\Gamma(x)$ and its integral are
computed and interpolated before
the neural network training in order to have a faster fitting
procedure.

\begin{figure}
\includegraphics[scale=0.4]{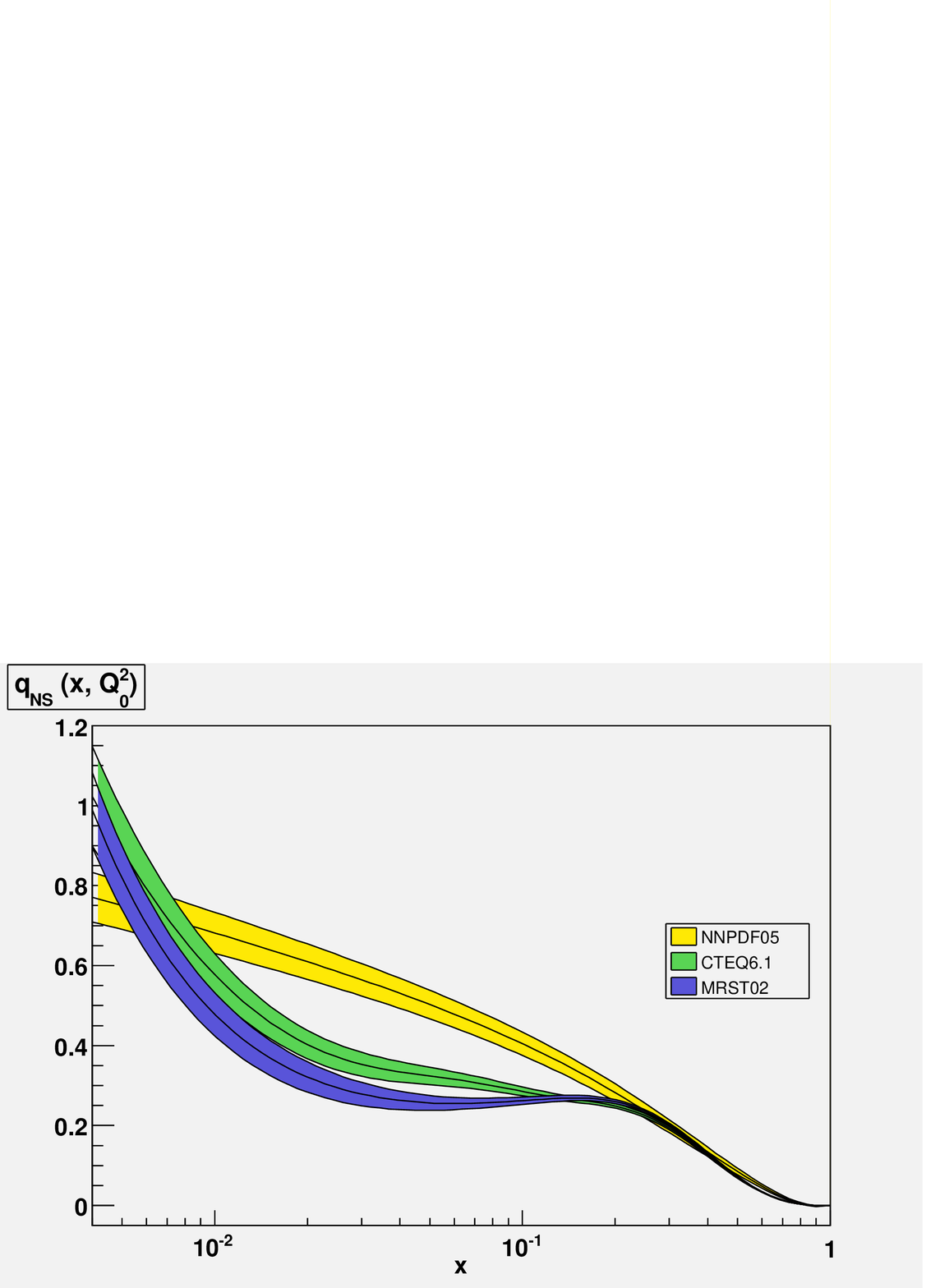}
\includegraphics[scale=0.4]{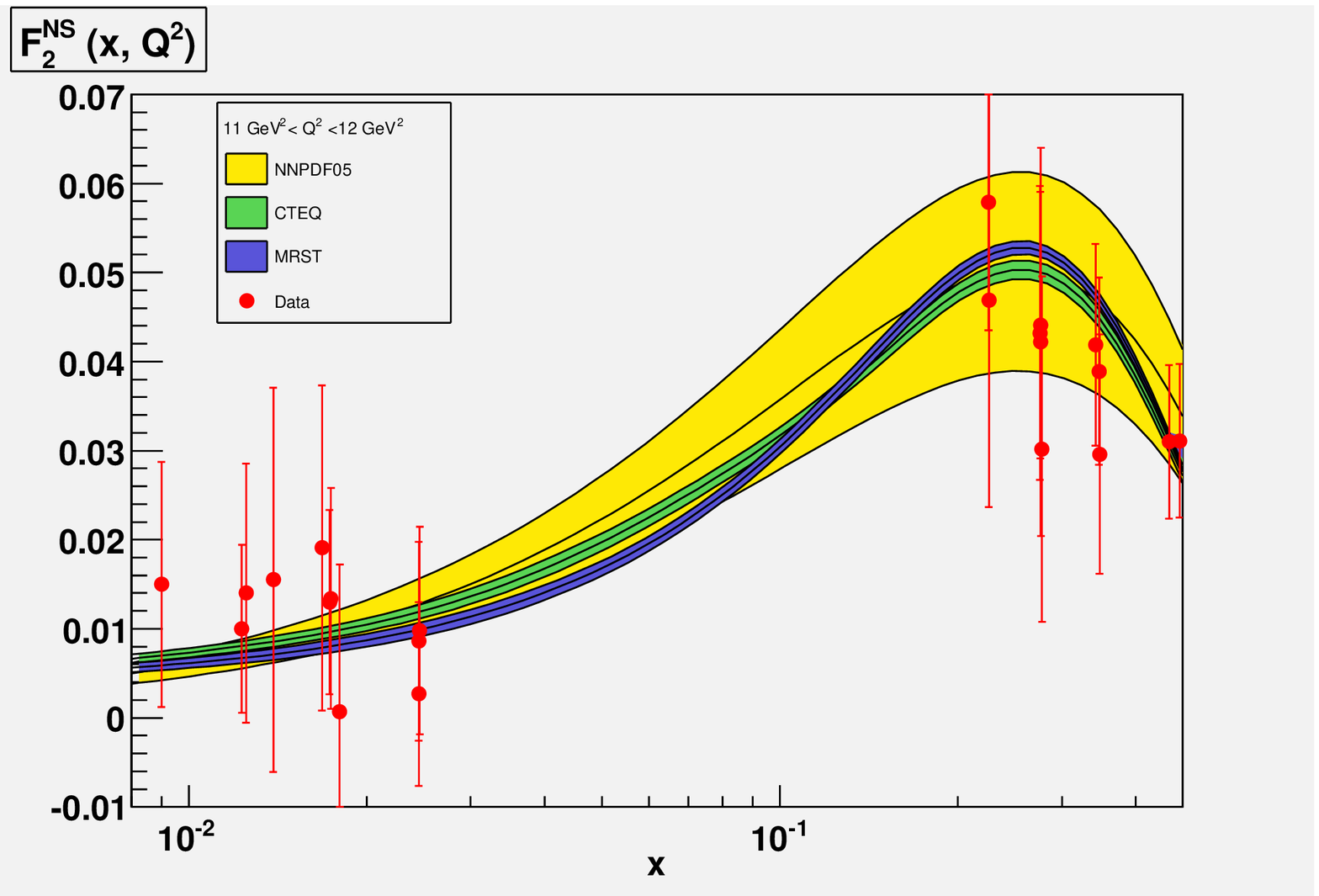}
\caption{Preliminary results for the 
NNPDF $q_{NS}$ fit at $Q_0^2=2\,\mathrm{GeV}^2$, and
the prediction for $F_2^{NS}(x,Q^2)$ compared with the CTEQ and MRST
results.}
\label{f2nn}
\end{figure}
As a first application of our method, we have extracted the nonsinglet
parton distribution $q_{NS}(x,Q^2_0)=\frac{1}{6}\lp
u+\bar{u}-d-\bar{d}\rp(x,Q^2_0)$ from the nonsinglet structure
function $F_2^{NS}(x,Q^2)$ as measured by the NMC \cite{Arneodo:1996qe} and BCDMS
\cite{Benvenuti:1989rh,Benvenuti:1989fm} collaborations. The preliminary results of a NLO
fit with fully correlated uncertainties \cite{qns} can be seen in
fig. \ref{f2nn} compared to other pdfs sets.
Our preliminary results appear to point
in the direction that 
the  uncertainties at small $x$ do not allow,
provided the current experimental data, to determine if
$q_{NS}(x,Q^2)$ grows at small $x$, as supported
by different theoretical arguments as well as 
by other global parton fits. However, more work is still needed
to confirm these results.
Only additional nonsinglet structure function data at small $x$
could settle in a definitive way this issue\footnote{
Like the experimental low $x$ deuteron structure function which
would be measured in an hypothetical electron-deuteron run
at HERA II, as it was pointed out during the workshop
by M.~Klein (section~\ref{sec:mkbrdbarubar}) and C.~Gwenlan}.

Summarizing, we have described a general technique to parametrize
experimental data in an bias-free way with a faithful
estimation of their uncertainties, which has been successfully applied to
structure functions and that now is being implemented in the
context of  parton distribution. The next step will be to construct a full
set of parton distributions from all available hard-scattering data
using the strategy described  in this contribution.

%% file: resumsec.tex
\section{Resummation\protect\footnote{Subsection
coordinator: S.~Forte}$^{,}$~\protect\footnote{Contributing authors:
    G.~Altarelli, J.~Andersen, R.~D.~Ball,
M.~Ciafaloni, D.~Colferai, G.~Corcella, S.~Forte,
    L.~Magnea, A.~Sabio Vera, G.~P.~Salam, A.~Sta\'sto}}
\label{sec:pdf,res}

\subsection{Introduction}
\label{sec:pdf,res,intro}

An accurate perturbative
determination of the hard partonic cross-sections (coefficient
functions) and of the anomalous dimensions which govern parton
evolution is necessary for the precise extraction of parton densities.
Recent progress in the determination of  higher order contributions to
these  quantities has been
reviewed in Sec.~\ref{sec:nnloprecision}.
 As is well known, such high-order
perturbative calculations display classes of terms containing large
logarithms, which ultimately signal the breakdown of perturbation
theory. Because these terms are scale--dependent and in general non
universal, lack of their inclusion can lead to significant
distortion of the parton densities in some kinematical regions, thereby
leading to loss of accuracy if parton distributions extracted from
deep-inelastic scattering (DIS)
or the Drell-Yan (DY) processes are used at the LHC. 

Logarithimic enhancement of higher order perturbative contribution
may take place when more than one large scale ratio is present. In DIS and DY this
happen in the two opposite limits when the center-of-mass energy of
the partonic collision is much higher than the characteristic scale of the
process, or
close to the threshold for the production of the 
final state. These correspond respectively to the small $x$ and large $x$
kinematical regions, where $0\le x\le 1$ is defined in terms of
the invariant mass $M^2$ of the non-leptonic final
state  as $M^2 =\frac{ (1 - x) Q^2}{x}$. The corresponding
perturbative contributions are respectively enhanced by powers of $
\ln\frac{1}{x}$ and $\ln (1-x)$, or, equivalently, in the space
of Mellin moments, by powers of $\frac{1}{N}$ and $\ln N$, where
$N\to0 $ moments dominate as $x\to 0$ while $N\to\infty$ moments
dominate as $x\to 1$. 

The theoretical status of small $x$ and large $x$ resummation is somewhat
different. 
Large $x$ logs are well
understood and the corresponding perturbative corrections have been
determined to all orders with very high accuracy. Indeed,  the coefficients
that determine their resummation  can be extracted from fixed-order
perturbative computations. Their resummation for DY and DIS
was originally derived in~\cite{Sterman:1986aj,Catani:1989ne} and
extended on very general grounds in~\cite{Forte:2002ni}.
The coefficients of the resulting exponentiation have now been
determined so that resummation can now be performed exactly at ${\rm
  N}^2{\rm LL}$ \cite{Vogt:2000ci,Moch:2004pa}, and to a very good
approximation at ${\rm N}^3{\rm LL}$
\cite{Moch:2005ba,Moch:2005ky,Laenen:2005uz}, including even some
non-logarithmic terms~\cite{Eynck:2003fn}. 
On the other hand, small $x$ logs are due to the fact that   at high energies, due to the
opening of phase space, both collinear~
\cite{gl,ap,d}
and
high-energy~%
\cite{Lipatov:1976zz,Kuraev:1977fs,Balitsky:1978ic,Lipatov:1985uk}
logarithms contribute, and thus the coefficients required for their resummation
can only be extracted from a simultaneous resolution of the DGLAP
equation, which  resums collinear logarithms, and the BFKL equation,
which resum the high-energy logarithms. Although the determination of
the kernels of these two equations has dramatically progressed in the
last several years, thanks to the computation of the N$^2$LO DGLAP
kernel~\cite{Moch:2004pa,Vogt:2004mw} and of the NLO BFKL
kernel~\cite{Lipatov:1976zz,Kuraev:1977fs,Balitsky:1978ic,Lipatov:1985uk,Camici:1997ij,Ciafaloni:1998gs},
the formalism which is needed to combine these two equations, as
required for sucessful phenomenology, has only  recently progressed to
the point of being usable for
realistic applications~\cite{Ciafaloni:1999yw,Ciafaloni:2003ek,Ciafaloni:2003rd,Altarelli:1999vw,Altarelli:2000mh,Altarelli:2001ji,Altarelli:2003hk,Altarelli:2003kt, Altarelli:2004dq,Altarelli:2005xx}.

In practice, however, neither small $x$ nor large $x$ resummation is
systematically incorporated in current parton fits, so
data points for which such effects may be important must be 
discarded. This is especially unsatisfactory
in the case of large $x$ resummation, where  resummed results
(albeit with a varying degree of logarithmic accuracy)
are available for essentially all processes of interest for a global parton fit,
in particular, besides DIS and DY, 
prompt photon
production~\cite{Catani:1999hs,Laenen:2000de}, jet
production~\cite{Kidonakis:1997gm,Kidonakis:2000gi} and heavy
quark electroproduction~\cite{Laenen:1998kp,Corcella:2003ib}.
Even if one were to conclude that resummation is  not
needed,  either because (at small $x$) it is affected by theoretical
uncertainties or because (at large $x$) its effects are small, this
conclusion 
could only be arrived at  after a careful study of  the impact of resummation
on the determination of parton
distributions, which is not available so far.

The purpose of this section is to provide a first assessment of the
potential impact of the inclusion of small $x$ and large $x$
resummation on the determination of parton distributions. In the case
of large $x$, this will be done by determining resummation effects on
parton distributions extracted from structure functions within a simplified
parton fit. In the case of small $x$, this will be done
through a study of the impact of small $x$ resummation on 
splitting functions, as well as the theoretical uncertainty involved
in the resummation process,
 in particular by comparing the results obtained
within the approach of
ref.~\cite{Ciafaloni:1999yw,Ciafaloni:2003ek,Ciafaloni:2003rd}
and that of
ref.~\cite{Altarelli:1999vw,Altarelli:2000mh,Altarelli:2001ji,Altarelli:2003hk,Altarelli:2003kt, Altarelli:2004dq,Altarelli:2005xx}.
We will also discuss numerical approaches to the solution of the 
small-$x$ (BFKL) evolution  equation.

\subsection{Soft gluons}
\label{sec:pdf,res,soft}

With the current level of theoretical control of soft gluon
resummations, available calculations for DIS or DY should be fully reliable over
most of the available phase space.  Specifically,
one expects current (resummed) predictions for DIS structure functions
to apply so long as the leading power correction can be neglected,
{\it i.e.} so long as $W^2 \sim (1 - x) Q^2 >> \Lambda^2$, with $x =
x_{Bj}$. Similarly, for the inclusive DY cross section, one
would expect the same to be true so long as $(1 - z)^2 Q^2 >>
\Lambda^2$, where now $z = Q^2/\hat{s}$, with $\hat{s} = x_1 x_2 S$
the partonic center of mass energy squared.
Indeed, as already mentioned, a consistent inclusion of resummation
effects in parton fits is feasible with present knowledge: 
on the one hand, recent fits show
that consistent parton sets can be obtained by making use of data from
a single process (DIS) (see Sec.~\ref{sec:mandy},\ref{sec:alek} and Ref.~\cite{Alekhin:2005gq}), 
on the other hand, even if one adopts the philosophy of global fits, resummed
calculations 
are available for all processes of interest. 

In practice, however, currently available global
parton fits are based on ${\rm NLO}$, or ${\rm N}^2{\rm LO}$
fixed-order perturbative calculations, so
data points which would lie
within the expected reach of resummed  calculations cannot be
fit consistently and must be discarded.  The effect is that large-$x$
quark distributions become less constrained, which has consequences on
the gluon distribution, as well as on medium-$x$ quark distributions,
through sum rules and evolution. The pool of untapped information is
growing, as more data at large values of $x$ have become available
from, say, the NuTeV collaboration at
Fermilab~\cite{Tzanov:2003gq,Naples:2003ne}. A related issue is the
fact that a growing number of QCD predictions for various processes of
interest at the LHC are now computed including resummation effects in
the hard partonic cross sections, which must be convoluted with parton
densities in order to make predictions at hadron level. Such
predictions are not fully consistent, since higher order effects are
taken into account at parton level, but disregarded in defining the
parton content of the colliding hadrons.

It is therefore worthwile to provide an assessment of the potential
impact of resummation on parton distributions. Here, we will do this by
computing resummation effects on quark distributions in the context of
a simplified parton fit.

\subsubsection{General Formalism in DIS}

Deep Inelastic Scattering structure functions $F_i (x, Q^2)$ are given
by the convolution of perturbative coefficient functions, typically
given in the $\overline{\mathrm{MS}}$ factorization scheme, and parton
densities. The coefficient functions $C_i^q$ for quark-initiated DIS
present terms that become large when the Bjorken variable $x$ for the
partonic process is close to $x = 1$, which forces gluon radiation
from the incoming quark to be soft or collinear.  At ${\cal
O}(\alpha_s)$, for example, the coefficient functions can be written
in the form
\beq 
 C_i^q \left(x,\frac{Q^2}{\mu_F^2},\alpha_s(\mu^2) \right) =
 \delta(1 - x) + {\frac{\alpha_s(\mu^2)}{2\pi}} H_i^q
 \left(x,\frac{Q^2}{\mu_F^2} \right) + {\cal O} \left( \alpha_s^2
 \right) \, .
\label{corma:coeff}
\eeq
Treating all quarks as massless, the part of $H_i^q$ which
contains terms that are logarithmically enhanced as $x \to 1$ reads
\beq
 H_{i, {\mathrm{soft}}}^q \left(x,\frac{Q^2}{\mu_F^2} \right)
 = 2 C_F \left\{ \left[ {\frac{\ln(1 - x)}{1 - x}} \right]_+ +
 {\frac{1} {(1 - x)_+}} \left( \frac{\ln Q^2}{\mu_F^2} - 
 \frac{3}{4} \right) \right\} \, .
\label{corma:largex}
\eeq
In moment space, where soft resummation is naturally performed, the
contributions proportional to $\alpha_s [\ln(1 - x)/(1 - x)]_+$ and to
$\alpha_s [1/(1 - x)]_+$ correspond to double $(\alpha_s \ln^2 N)$ and
single $(\alpha_s \ln N)$ logarithms of the Mellin variable $N$.
The Mellin transform of Eq.~(\ref{corma:largex}) in fact reads, at
large $N$,
\beq
 \hat{H}_{i, {\mathrm{soft}}}^q \left(N, \frac{Q^2}{\mu_F^2} 
 \right) = 2 C_F \left\{ {\frac{1}{2}} \ln^2 N + \left[ \gamma_E + 
 {\frac{3}{4}} - \frac{\ln{Q^2}}{\mu_F^2} \right] \ln N \right\} \, .
\label{corma:largen}
\eeq
All terms growing logarithmically with $N$, as well as all
$N$-independent terms corresponding to contributions proportional to
$\delta (1 - x)$ in $x$-space, have been shown to exponentiate.  In
particular, the pattern of exponentiation of logarithmic singularities
is nontrivial: one finds that the coefficient functions can be written
as
\beq
 \hat{C}_i^q \left(N,\frac{Q^2}{\mu_F^2},\alpha_s(\mu^2) \right) =
 {\cal R} \left(N,\frac{Q^2}{\mu_F^2},\alpha_s(\mu^2) \right) 
 \Delta \left(N,\frac{Q^2}{\mu_F^2},\alpha_s(\mu^2) \right) \, ,
\label{corma:expon}
\eeq
where ${\cal R} (N, Q^2/\mu_F^2, \alpha_s(\mu^2) )$ is a finite
remainder, nonsingular as $N \to \infty$, while~\cite{Forte:2002ni}
\beq
 \ln \Delta \left(N,\frac{Q^2}{\mu_F^2},\alpha_s(\mu^2) \right) = 
 \int_0^1 {dx {\frac{{x^{N - 1} - 1}}{1 - x}}}
 \left\{ \int_{\mu_F^2}^{(1 - x) Q^2} {\frac{d k^2}  {k^2}}
 A \left[ \alpha_s(k^2) \right] + B \left[ \alpha_s \left(Q^2 (1 - x) 
 \right) \right] \right\} \, .
\label{corma:delta}
\eeq
In \Eq{corma:delta} the leading logarithms (LL), of the form
$\alpha^n_s \ln^{n + 1} N$, are generated at each order by the
function $A$.  Next-to-leading logarithms (NLL), on the other hand, of
the form $\alpha_s^n \ln^n N$, require the knowledge of the function
$B$.  In general, resumming ${\rm N}^k{\rm LL}$ to all orders requires
the knowledge of the function $A$ to $k+1$ loops, and of the function
$B$ to $k$ loops. In the following, we will adopt the common standard
of NLL resummation, therefore we need the expansions
\beq
 A(\alpha_s) = \sum_{n = 1}^{\infty} \left({\frac{\alpha_s}  {\pi}} \right)^n A^{(n)}\ \ ;\ \ 
 B(\alpha_s) = \sum_{n = 1}^{\infty} \left({\frac{\alpha_s} {\pi}} \right)^n B^{(n)}
\label{corma:expan}
\eeq
to second order for $A$ and to first order for $B$. The relevant
coefficients are
\beqa
 A^{(1)} & = & C_F \, , \nonumber \\
 A^{(2)} & = & {\frac{1 }{ 2}} C_F \left[ C_A \left(
 {\frac{67}{18}} - {\frac{\pi^2}{6}} \right) - {\frac{5}{9}} n_f 
 \right] \, , \\
 B^{(1)} & = & - {\frac{3}{4}} C_F \, . \nonumber
\label{corma:resco}
\eeqa
Notice that in \Eq{corma:delta} the term $\sim A(\alpha_s(k^2))/k^2$
resums the contributions of gluons that are both soft and collinear,
and in fact the anomalous dimension $A$ can be extracted order by
order from the residue of the singularity of the nonsinglet splitting
function as $x \to 1$.  The function $B$, on the other hand, is
related to collinear emission from the final state current jet.

In \cite{Laenen:1998kp, Corcella:2003ib} soft resummation was extended
to the case of heavy quark production in DIS. In the case of heavy
quarks, the function $B(\alpha_s)$ needs to be replaced by a different
function, called $S(\alpha_s)$ in \cite{Corcella:2003ib}, which is
characteristic of processes with massive quarks, and includes effects
of large-angle soft radiation.  In the following, we shall consider
values of $Q^2$ much larger than the quark masses and employ the
resummation results in the massless approximation, as given in
\Eq{corma:delta}.

\subsubsection{Simplified parton fit}
\label{sec:pdf,res,soft,pfit}

We would like to use large-$x$ resummation in the DIS coefficient
functions to extract resummed parton densities from DIS structure
function data. Large-$x$ data typically come from fixed-target
experiments: in the following, we shall consider recent
charged-current (CC) data from neutrino-iron scattering, collected by
the NuTeV collaboration \cite{Tzanov:2003gq,Naples:2003ne}, and
neutral-current (NC) data from the NMC \cite {Arneodo:1996qe} and
BCDMS \cite{Benvenuti:1989rh,Benvenuti:1989fm} collaborations.

Our strategy will be to make use of data at different, fixed values of
$Q^2$.  We will extract from these data moments of the corresponding
structure functions, with errors; since such moments factor into a
product of moments of parton densities times moments of coefficient
functions, computing parton moments with errors is straightforward.
We then compare NLO to resummed partons in Mellin space, and
subsequently provide a translation back to $x$-space by means of
simple parametrization.  Clearly, given the limited data set we are
working with, our results will be affected by comparatively large
errors, and we will have to make simplifying assumptions in order to
isolate specific quark densities. Resummation effects are, however,
clearly visible, and we believe that our fit provides a rough
quantitative estimate of their size. A more precise quantitative
analysis would have to be performed in the context of a global fit.

The first step is to construct a parametrization of the chosen
data. An efficient and faithful parametrization of the NMC and BCDMS
neutral-current structure functions was provided in
\cite{f2nn,nnpdf}, where a large sample of Monte
Carlo copies of the original data was generated, taking properly into
account errors and correlations, and a neural network was trained on
each copy of the data. One can then use the ensemble of networks as a
faithful and unbiased representation of the probability distribution
in the space of structure functions. We shall make use of the
nonsinglet structure function $F_2^{\mathrm{ns}}(x,Q^2)$ extracted
from these data, as it is unaffected by gluon contributions, and
provides a combination of up and down quark densities which is
independent of the ones we extract from charged current data
(specifically, $F_2^{\mathrm{ns}}(x,Q^2)$ gives $u-d$).

\begin{figure}
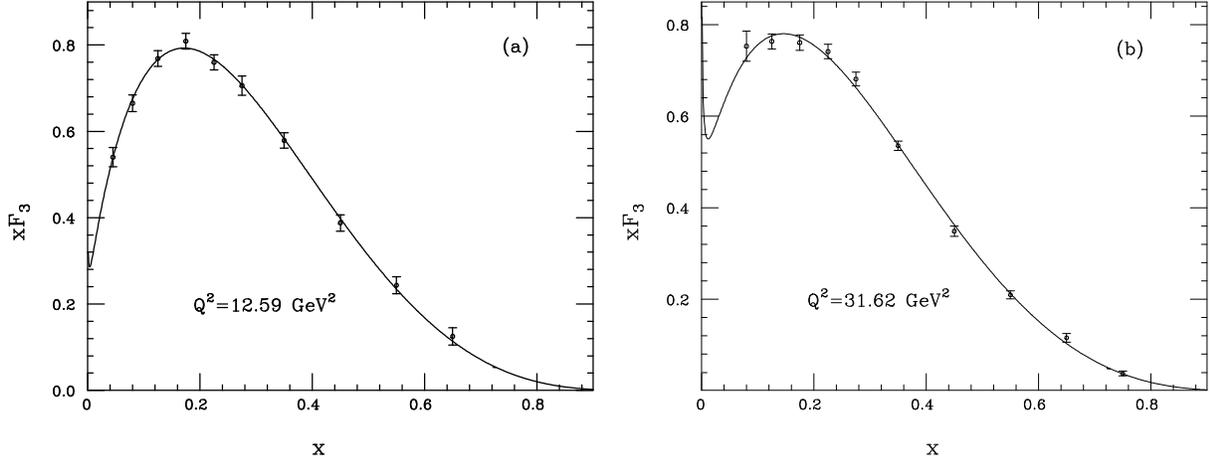

\centerline{\resizebox{0.49\textwidth}{!}{\includegraphics{f3_1259.ps}}
\hfill
\resizebox{0.49\textwidth}{!}{\includegraphics{f3_3162.ps}}}
\caption{NuTeV data on the structure function $x F_3$, at $Q^2 = 12.59
  ~{\rm GeV}^2$ (a) and at $Q^2 = 31.62 ~{\rm GeV}^2$ (b), along with
  the best fit curve parametrized by \Eq{corma:fitf3}.}
\label{corma:figf3}
\end{figure}

As far as the NuTeV data are concerned, we shall consider the data on
the CC structure functions $F_2$ and $F_3$.  The structure function
$F_3$ can be written as a convolution of the coefficient function
$C_3^q$ with quark and antiquark distributions, with no gluon
contribution, as
\beq
 x F_3 = \frac{1}{2} \left( x F_3^\nu + x F_3^{\bar \nu} \right) =
 x \left[\sum_{q,q'} |V_{qq'}|^2 \left( q - \bar q \right) \otimes 
 C_3^q \right].
\label{corma:f3}
\eeq
We consider data for $F_3$ at $Q^2 = 12.59$ and $31.62$ GeV$^2$, and, in
order to compute moments, we fit them using the functional form
\beq
x F_3(x) = C x^{-\rho} ( 1 - x )^\sigma ( 1 + k x )~.
\label{corma:fitf3}
\eeq
The best-fit values of $C$, $\rho$ and $\delta$, along with the
$\chi^2$ per degree of freedom, are given in
\cite{Corcella:2005us}. Here we show the relevant NuTeV data on
$x F_3$, along with our best-fit curves, in Fig.~\ref{corma:figf3}.
\par
\begin{figure}
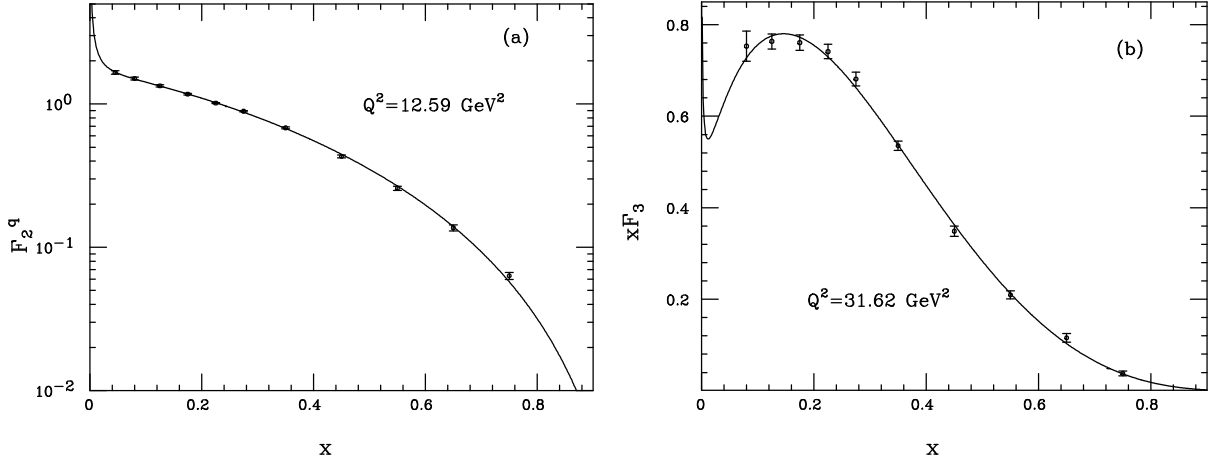

\centerline{\resizebox{0.49\textwidth}{!}{\includegraphics{f2_1259.ps}}%
\hfill%
\resizebox{0.49\textwidth}{!}{\includegraphics{f2_3162.ps}}}
\caption{NuTeV data on the quark-initiated contribution $F_2^q$ to the
structure function $F_2$, for $Q^2 = 12.59 ~{\rm GeV}^2$ (a), and $Q^2
= 31.62 ~{\rm GeV}^2$ (b). The solid lines are the best-fit
predictions.}
\label{corma:figF2}
\end{figure}

The analysis of NuTeV data on $F_2$ is slightly complicated by the
fact that gluon-initiated DIS gives a contribution, which, however, is
not enhanced but suppressed at large $x$. We proceed therefore by
taking the gluon density from a global fit, such as the NLO set CTEQ6M
\cite{cteq}, and subtract from $F_2$ the gluon contribution
point by point.  We then write $F_2$ as
\beq
 F_2 \equiv \frac{1}{2} \left( F_2^\nu + F_2^{\bar \nu} \right) = 
 x \sum_{q,q'} |V_{qq'}|^2 \left[(q + \bar q) \otimes C_2^q
 + g \otimes C_2^g \right] \equiv F_2^q + F_2^g \, ,
\label{corma:fitf2}
\eeq
and fit only the quark-initiated part $F_2^q$, using the same
parametrization as in \Eq{corma:fitf3}. Fig.~\ref{corma:figF2} shows
the data on $F_2^q$ and the best fit curves, as determined in
Ref.~\cite{Corcella:2005us}.
After the subtraction of the gluon contribution from $F_2$, the
structure functions we are considering ($F_2^q$, $x F_3$ and
$F_2^{\mathrm{ns}}$) are all given in factorized form as
\beq
 F_i (x, Q^2) = x \int_x^1 \frac{d \xi}{\xi} \, q_i \left(\xi, \mu_F^2 
 \right) C_i^q \left( \frac{x}{\xi}, \frac{Q^2}{\mu_F^2},
 \alpha_s (\mu^2) \right) \,,
\label{corma:factF}
\eeq
where $C_i^q$ is the relevant coefficient function and $q_i$ is a
combination of quark and antiquark distributions only.  Hereafter, we
shall take $\mu = \mu_F = Q$ for the factorization and renormalization
scales. At this point, to identify individual quark distributions from
this limited set of data, we need to make some simplifying
assumptions.  Following \cite{Corcella:2005us}, we assume isospin
symmetry of the sea, $\bar u = \bar d$, $s = \bar s$ and we further
impose a simple proportionality relation expressing the antistrange
density in terms of the other antiquarks, $\bar s = \kappa \bar u$. As
in \cite{Corcella:2005us}, we shall present results for $\kappa = \frac{1}{2}$.
With these assumptions, we can explicit solve for the remaining three
independent quark densities (up, down, and, say, strange), using the
three data sets we are considering.

Taking the Mellin moments of \Eq{corma:factF}, the convolution becomes
an ordinary product and we can extract NLO or NLL-resummed parton
densities, according to whether we use NLO or NLL coefficient
functions.  More precisely,
\beq
 \hat{q}_i^{\mathrm{NLO}}(N, Q^2) = \frac{\hat{F}_i(N -
 1, Q^2)}{\hat{C}_i^{\mathrm{NLO}}
 \left(N, 1, \alpha_s(Q^2) \right)}
 \, \, ; \, \, \, \, 
 \hat{q}_i^{\mathrm{res}}(N, Q^2) = \frac{\hat{F}_i(N - 
 1, Q^2)}{\hat{C}_i^{\mathrm{res}}
 \left(N, 1, \alpha_s(Q^2) \right)}~.
\label{ratio}
\eeq
After extracting the combinations $q_i$, one can derive the individual
quark densities, at NLO and including NLL large-$x$ resummation. We
concentrate our analysis on the up quark distribution, since
experimental errors on the structure functions are too large to see an
effect of the resummation on the other quark densities, such as $d$ or
$s$, with the limited data set we are using.

\subsubsection{Impact of the resummation}
\label{sec:pdf,res,soft,imp}

\begin{figure}[ht!]
\centerline{\resizebox{0.65\textwidth}{!}{\includegraphics{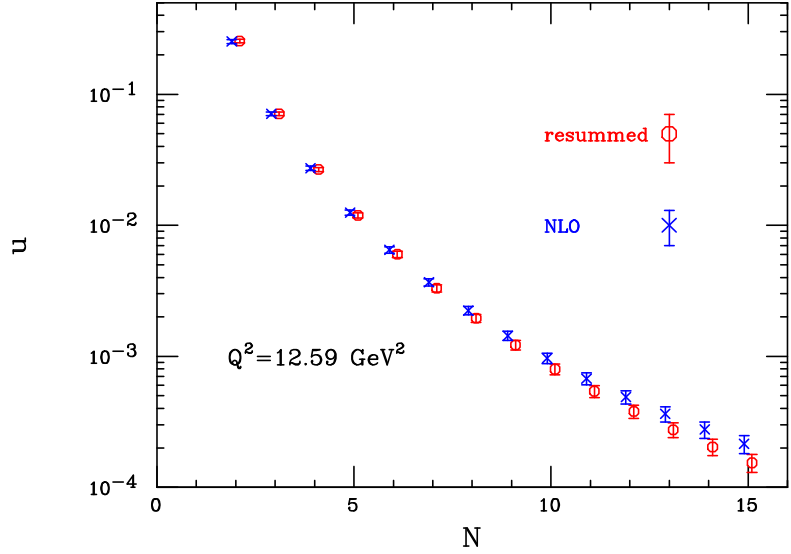}}}
\caption{NLO and resummed moments of the up quark distribution at $Q^2
= 12.59$~GeV$^2$}
\label{corma:n1}
\end{figure}

\begin{figure}[ht!]
\centerline{\resizebox{0.65\textwidth}{!}{\includegraphics{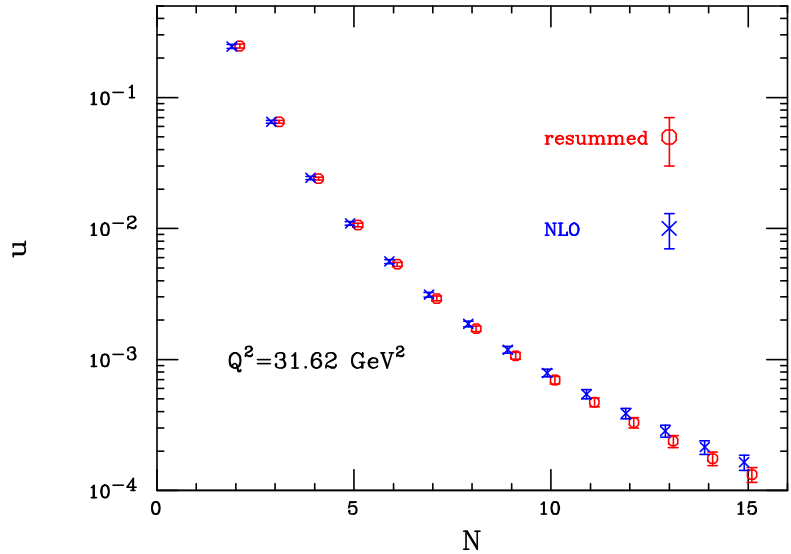}}}
\caption{As in Fig.~\ref{corma:n1}, but at $Q^2=31.62$~GeV$^2$.}
\label{corma:n2}
\end{figure}

We present results for moments of the up quark distribution in
Figs.~\ref{corma:n1} and \ref{corma:n2}.
Resummation effects become statistically significant around $N \sim
6-7$ at both values of $Q^2$. Notice that high moments of the resummed
up density are {\it suppressed} with respect to the NLO density, as a
consequence of the fact that resummation in the $\msb$ scheme enhances
high moments of the coefficient functions.

\begin{figure}[ht!]
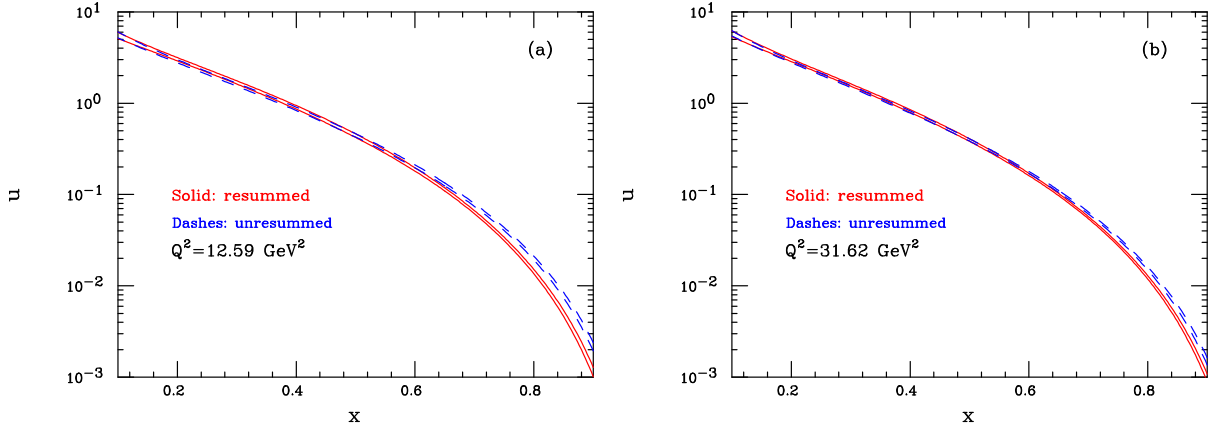

\centerline{\resizebox{0.49\textwidth}{!}{\includegraphics{ux2.ps}}%
\hfill%
\resizebox{0.49\textwidth}{!}{\includegraphics{ux3.ps}}}
\caption{NLO and resummed up quark distribution at $Q^2 =
  12.59$~GeV$^2$ (a) and at $Q^2 = 31.62$ GeV$^2$, using the
  parametrization given in the text. The band corresponds to one
  standard deviation in parameter space.}
\label{corma:x}
\end{figure}

In order to illustrate the effect in the more conventional setting of
$x$-space distributions, we fit our results for the moments to a
simple parametrization of the form $u(x) = D x^{-\gamma}(1 -
x)^\delta$. Our best fit values for the parameters, with statistical
errors, are given in Table~(\ref{corma:tab}), and the resulting
distributions are displayed in Fig.~\ref{corma:x}, with one standard
deviation uncertainty bands.
\begin{table}[t]
\centerline{
\begin{tabular}{llllccccc}\\
\hline
$\hspace{3mm} Q^2$ & PDF & $ \hspace{1cm} D$ & $ \hspace{1cm} \gamma$ & 
$\hspace{2mm} \delta$ \\ 
\hline
12.59 & NLO & $3.025 \pm 0.534$ & $0.418 \pm 0.101$ & $3.162 \pm 0.116$ \\
      & RES & $4.647 \pm 0.881$ & $0.247 \pm 0.109$ & $3.614 \pm 0.128$ \\
31.62 & NLO & $2.865 \pm 0.420$ & $0.463 \pm 0.086$ & $3.301 \pm 0.098$ \\
      & RES & $3.794 \pm 0.583$ & $0.351 \pm 0.090$ & $3.598 \pm 0.104$ \\
\hline \\
\end{tabular}}
\caption{Best fit values and errors for the up-quark $x$-space parametrization,
at the chosen values of $Q^2$.}
\label{corma:tab}
\end{table}
Once again, the effect of soft resummation is clearly visible at large
$x$: it suppresses the quark densities extracted from the given
structure function data with respect to the NLO prediction.
\begin{figure}[ht!]
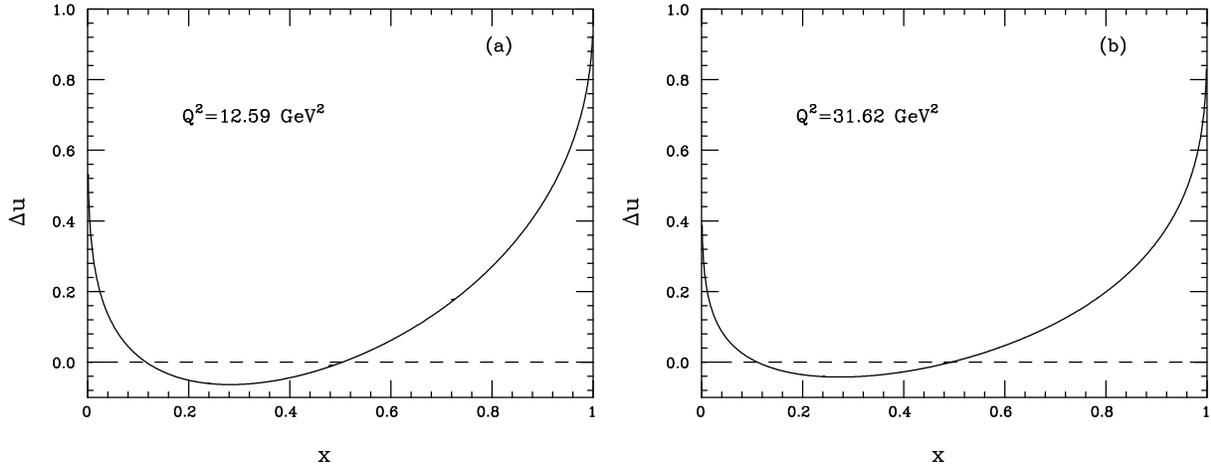

\centerline{\resizebox{0.49\textwidth}{!}{\includegraphics{deltau1.ps}}%
\hfill%
\resizebox{0.49\textwidth}{!}{\includegraphics{deltau.ps}}}
\caption{Central value for the relative change in the up quark
distribution, $\Delta u (x) \equiv \left(u_{\rm NLO} (x) - u_{\rm res}
(x) \right)/u_{\rm NLO} (x)$, at $Q^2=12.59$ (a) and 31.62~GeV$^2$
(b).}
\label{corma:del}
\end{figure}\par
In order to present the effect more clearly, we show in
Fig.~\ref{corma:del} the normalized deviation of the NLL-resummed
prediction from the NLO one, i.e.  $\Delta u(x) = \left(u_{\rm NLO}
(x) - u_{\rm res} (x) \right)/u_{\rm NLO} (x)$, at the two chosen
values of $Q^2$ and for the central values of the best-fit
parameters. We note a change in the sign of $\Delta u$ in the
neighborhhod of the point $x = 1/2$: although our errors are too large
for the effect to be statistically significant, it is natural that the
suppression of the quark distribution at large $x$ be compensated by
an enhancement at smaller $x$. In fact, the first moment of the
coefficient function is unaffected by the resummation: thus $C_i^q$,
being larger at large $x$, must become smaller at small $x$. The
further sign change at $x \sim 0.1$, on the other hand, should not be
taken too seriously, since our sample includes essentially no data at
smaller $x$, and of course we are using an $x$-space parametrization
of limited flexibility.

\begin{figure}[ht!]
\centerline{\resizebox{0.65\textwidth}{!}{\includegraphics{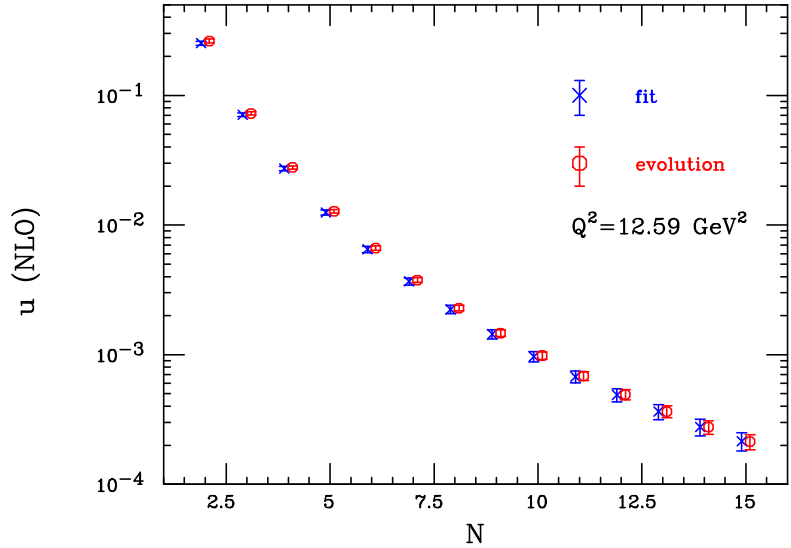}}}
\caption{Comparison of fitted moments of the NLO 
up quark distribution, at $Q^2 =
12.59$~GeV$^2$, with moments obtained via NLO evolution from 
$Q^2 = 31.62$~GeV$^2$.}
\label{corma:figev1}
\end{figure}

\begin{figure}[ht!]
\centerline{\resizebox{0.65\textwidth}{!}{\includegraphics{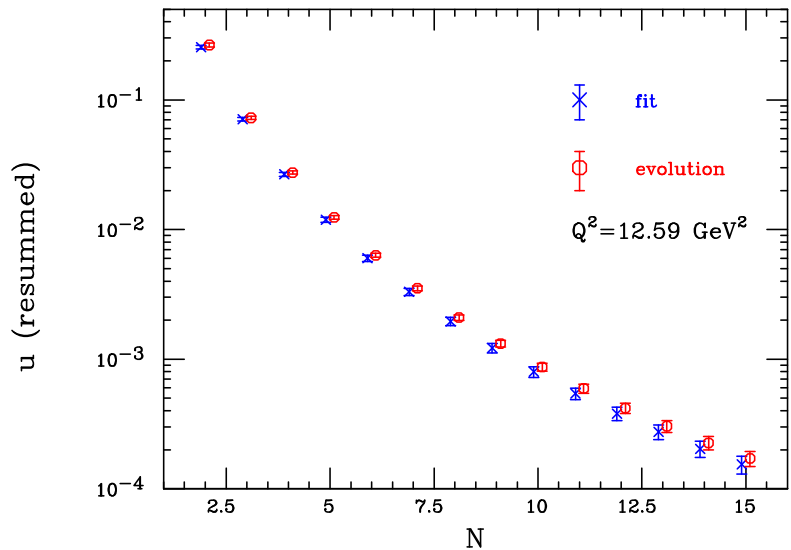}}}
\caption{As in Fig.~\ref{corma:figev1}, but comparing NLL-resummed moments
of the up quark density.}
\label{corma:figev2}
\end{figure}

Finally, we wish to verify that the up-quark distributions extracted by
our fits at $Q^2 = 12.59$ and $31.62$ GeV$^2$ are consistent with
perturbative evolution. To achieve this goal, we evolve our $N$-space
results at $Q^2 = 31.62$ GeV$^2$ down to 12.59 GeV$^2$, using NLO
Altarelli--Parisi anomalous dimensions, and compare the evolved moments
with the direct fit at 12.59~GeV$^2$.  Figures~\ref{corma:figev1}
and \ref{corma:figev2} show
that the results of our fits at 12.59 GeV$^2$ are compatible with the
NLO evolution within the confidence level of one standard deviation.
Note however that the evolution of resummed moments appears to give
less consistent results, albeit within error bands: this can probably 
be ascribed to a contamination between pertubative resummation and power 
corrections, which we have not disentangled in our analysis.
\par
Qualitatively, the observed effect on the up quark distribution is
easily described, at least within the limits of a simple
parametrization like the one we are employing: resummation increases
the exponent $\delta$, responsible for the power-law decay of the
distribution at large $x$, by about $10\%$ to $15\%$ at moderate
$Q^2$. The exponent $\gamma$, governing the small-$x$ behavior, and
the normalization $D$, are then tuned so that the first finite moment
(the momentum sum rule) may remain essentially unaffected.

In conclusion, our results indicate that quark distributions are suppressed at
large $x$ by soft gluon effects. Quantitatively, we observe an effect
ranging between $10\%$ and $20\%$ when $0.6 < x < 0.8$ at moderate
$Q^2$, where we expect power corrections not to play a significant
role. Clearly, a more detailed quantitative understanding of the
effect can be achieved only in the context of a broader and fully
consistent fit. We would like however to notice two things: first, the
effect of resummations propagates to smaller values of $x$, through
the fact that the momentum sum rule is essentially unaffected by the
resummation; similarly, evolution to larger values of $Q^2$ will shift
the Sudakov suppression to smaller $x$. A second point is that, in a
fully consistent treatment of hadronic cross section, there might be a
partial compensation between the typical Sudakov enhancement of the
partonic process and the Sudakov suppression of the quark
distribution: the compensation would, however, be channel-dependent,
since gluon-initiated partonic processes would be unaffected. We
believe it would be interesting, and phenomenologically relevant, to
investigate these issues in the context of a more comprehensive parton
fit.

\subsection{Small $x$}
\label{sec:pdf,res,sx}

Small $x$ structure functions are dominated by the flavour singlet
contribution, whose coefficient functions and anomalous dimensions
receive logarithmic enhancements, which make
perturbation theory converge more slowly. In the small $x$, i.e. high
energy limit, the cross section is quasi-constant and
characterised by the effective expansion parameter 
$\langle \alpha_s(\boldsymbol{k}^2) \rangle
\log\frac1{x}\log\frac{\boldsymbol{k}_{\max}^2}{\boldsymbol{k}_{\min}^2}$, where $x =
Q^2/s$, $\boldsymbol{k}^2 \lesssim Q^2$ is the transverse momentum of the
exchanged gluon, $s$ is the photon-proton centre of mass energy
squared and $Q^2$ is the hard scale. Such expansion parameter can be
large, due to both the double-logs and to the fact that $\langle \boldsymbol{k}^2
\rangle$ may drift towards the non-perturbative region. Even assuming
that truly non-perturbative effects are factored out --- as is the
case for structure functions --- the problem remains of resumming the
perturbative series with both kinds of
logarithms~\cite{gl,ap,d,Lipatov:1976zz,Kuraev:1977fs,Balitsky:1978ic,Lipatov:1985uk} 

In the BFKL approach one tries to resum the high-energy logarithms
first, by an evolution equation in $\log 1/x$, whose $\boldsymbol{k}$-dependent
evolution kernel is calculated perturbatively in $\alpha_s$. However, the
leading kernel~\cite{Lipatov:1976zz,Kuraev:1977fs,Balitsky:1978ic,Lipatov:1985uk} overestimates the hard cross-section, and
subleading
ones~%
\cite{Fadin:1998py,Camici:1997ij,Ciafaloni:1998gs}
turn out to be large and of alternating sign, pointing towards an
instability of the leading-$\log x$ (L$x$) hierarchy. The problem is
that, for any given value of the hard scales $Q, Q_0 \ll \sqrt{s}$ ---
think, for definiteness, of $\gamma^*(Q)$-$\gamma^*(Q_0)$ collisions ---, the
contributing kernels contain collinear enhancements in all
$\boldsymbol{k}$-orderings of the exchanged gluons of type $\sqrt{s} \gg \cdots
\boldsymbol{k}_1 \gg \boldsymbol{k}_2 \cdots$, or $\sqrt{s} \gg \cdots \boldsymbol{k}_2 \gg \boldsymbol{k}_1
\cdots$ and so on, to all orders in $\alpha_s$. Such enhancements are only
partly taken into account by any given truncation of the L$x$
hierarchy, and they make it unstable.
In the DGLAP evolution equation one resums collinear logarithms first,
but fixed order splitting functions do contain~\cite{Moch:2004pa,Vogt:2004mw}
high-energy logarithms also, and a further resummation is needed.

Two approaches to the simultaneous resummation of these two classes of
logs have recently reached the stage where their phenomenological
application can be envisaged.
The 
renormalisation group improved (CCSS)
approach~\cite{Ciafaloni:1999yw,Ciafaloni:2003ek,Ciafaloni:2003rd,Ciafaloni:2005cg}
is built up within the BFKL 
framework, by improving the whole hierarchy of subleading kernels in
the collinear region, so as to take into account all the
$\boldsymbol{k}$-orderings mentioned before, consistently with the RG.
In the duality (ABF)
approach~\cite{Ball:1999sh,Altarelli:1999vw,Altarelli:2000mh,Altarelli:2001ji,Altarelli:2003hk,Altarelli:2003kt,
  Altarelli:2004dq,Altarelli:2005xx} 
one concentrates on the problem of obtaining an improved anomalous
dimension (splitting function) 
for DIS which reduces to the ordinary perturbative result at large 
$N$ (large $x$), thereby automatically satisfying renormalization
group constraints, 
while including resummed BFKL corrections at small $N$ (small $x$),
determined through the renormalization-group improved (i.e. running
coupling) version of the BFKL kernel. 

We will briefly review the theoretical underpinnings of these two
approaches in turn, and then compare phenomenological  results
obtained in both approaches. 
Note that we shall use the notation of the CCSS or ABF papers in the
corresponding sections, in order to enable  a simpler connection with
the original literature, 
at the price of some notational discontinuity.
 In particular, $\ln\frac{1}{x}$ is called $Y$ by CCSS and $\xi$ by
 ABF; the  Mellin variable conjugate to  $\ln\frac{1}{x}$ is called $\omega$
 by CCSS and $N$ by ABF; and  the Mellin variable conjugated to
 $\ln \frac{Q^2}{{\boldsymbol{k}}^2}$ is called $\gamma$  by CCSS and $M$ by ABF.

\subsubsection{The renormalisation group improved approach}
\label{sec:pdf,res,sx,rgi}

The basic problem which is tackled in
the CCSS approach~\cite{Ciafaloni:1999yw,Ciafaloni:2003ek,Ciafaloni:2003rd,Ciafaloni:2005cg} is the calculation
of the (azimuthally averaged) gluon Green function ${\cal G}(Y;k,k_0)$ as a
function of the magnitudes of the external gluon transverse momenta $k
\equiv |\boldsymbol{k}|,\;k_0 \equiv |\boldsymbol{k}_0| $ and of the rapidity $Y\equiv \log
\frac{s}{k k_0}$. This is not yet a hard cross section, because one 
needs to incorporate the impact factors of the
probes~\cite{Catani:1990xk,Catani:1990eg,Collins:1991ty,Bartels:2000gt,Bartels:2001mv,Bartels:2002uz,Bartels:2001ge,Bartels:2002yj}.  Nevertheless, the Green
function exhibits most of the physical features of the hard process,
if we think of $k^2,\;k_0^2$ as external (hard) scales. The limits
$k^2\gg k_0^2$ ($k_0^2 \gg k^2$) correspond conventionally to the
ordered (anti-ordered) collinear limit. By definition, in the
$\omega$-space conjugate to $Y$ (so that $\hat{\omega} = \partial_Y$) one sets
\begin{align}
 \label{defGGF}
 {\cal G}_{\omega}(\boldsymbol{k},\boldsymbol{k}_0) &\equiv [\omega - {\cal K}_{\omega}]^{-1} (\boldsymbol{k},\boldsymbol{k}_0) \,,  \\
 \label{eqGGF}
 \omega {\cal G}_{\omega}(\boldsymbol{k},\boldsymbol{k}_0) &= \delta^2(\boldsymbol{k}-\boldsymbol{k}_0) +
 \int d^2 \boldsymbol{k}' \; {\cal K}_{\omega}(\boldsymbol{k},\boldsymbol{k}') {\cal G}_{\omega}(\boldsymbol{k}',\boldsymbol{k}_0) \;,
\end{align}
where ${\cal K}_{\omega}(\boldsymbol{k},\boldsymbol{k}')$ is a kernel to be defined, whose $\omega = 0$
limit is related to the BFKL $Y$-evolution kernel discussed before.

In order to understand the RG constraints, it is useful to switch from
$\boldsymbol{k}$-space to $\gamma$-space, where the variable $\gamma$ is
conjugated to $t\equiv \log \boldsymbol{k}^2 / \boldsymbol{k}_0^2$ at fixed $Y$, and to make
the following kinematical remark: the ordered (anti-ordered) region
builds up scaling violations in the Bjorken variable $x=\boldsymbol{k}^2/s$
($x_0=\boldsymbol{k}_0^2/s$) and, if $x$ ($x_0$) is fixed instead of $k k_0/s =
e^{-Y}$, the variable conjugated to $t$ is shifted~\cite{Salam:1998tj} by an
$\omega$-dependent amount, and becomes $\gamma+{\textstyle\frac{\omega}{2}} \sim
\partial_{\log \boldsymbol{k}^2}$~ ~($1-\gamma+{\textstyle\frac{\omega}{2}} \sim \partial_{\log
  \boldsymbol{k}_0^2}$). Therefore, the characteristic function
$\chi_{\omega}(\gamma)$ of ${\cal K}_{\omega}$ (with a factor
$\alpha_s$ factored out)
must be singular when either one of
the variables is small, as shown (in the frozen $\alpha_s$ limit) by
\begin{equation}\label{chicoll}
 \frac1{\omega}\chi_{\omega}(\gamma) \to \left[ \frac1{\gamma+{\textstyle\frac{\omega}{2}}} +
 \frac1{1-\gamma+{\textstyle\frac{\omega}{2}}} + \cdots \right] \left[ \gamma_{gg}^{(1)}(\alpha_s,\omega)
 + \cdots \right] \;,
\end{equation}
where $\gamma_{gg}^{(1)}$ is the one-loop gluon anomalous dimension,
and further orders may be added. Eq.~(\ref{chicoll}) ensures the
correct DGLAP evolution in either one of the collinear limits
(because, e.g., $\gamma+{\textstyle\frac{\omega}{2}}\sim\partial_{\log\boldsymbol{k}^2}$) and is
$\omega$-dependent, because of the shifts. Since higher powers of $\omega$
are related to higher subleading powers of $\alpha_s$~\cite{Ciafaloni:1998iv}, this
$\omega$-dependence of the constraint~(\ref{chicoll}) means that the
whole hierarchy of subleading kernels is affected.

To sum up, the kernel ${\cal K}_{\omega}$ is constructed so as to satisfy the
RG constraint~(\ref{chicoll}) and to reduce to the exact L$x$ + NL$x$
BFKL kernels in the $\omega\to0$ limit; it is otherwise interpolated on
the basis of various criteria (e.g., momentum conservation), which
involve a ``scheme'' choice.

The resulting integral equation has been solved in~\cite{Ciafaloni:1999yw,Ciafaloni:2003ek,Ciafaloni:2003rd} by
numerical matrix evolution methods in $\boldsymbol{k}$- and $x$-space.
Furthermore, introducing the integrated gluon density $g$, the
resummed splitting function $P_{\mathrm{eff}}(x,Q^2)$ is defined by the
evolution equation
\begin{equation}\label{evol}
 \frac{\partial g(x,Q^2)}{\partial\log Q^2} = \int\frac{\mathrm{d} z}{z} \;
 P_{\mathrm{eff}}\big(z,\alpha_s(Q^2)\big) g\Big(\frac{x}{z}, Q^2 \Big) \;,
\end{equation}
and has been extracted~\cite{Ciafaloni:1999yw,Ciafaloni:2003ek,Ciafaloni:2003rd} by a numerical deconvolution
method~\cite{Ciafaloni:2000cb}.
Note that in the RGI approach the running of the coupling is treated by
adopting in~(\ref{eqGGF}) the off-shell dependence of $\alpha_s$ suggested by
the BFKL and DGLAP limits, and then solving the ensuing
integral equation numerically.

It should be noted that the RGI
approach has the somewhat wider goal of calculating the off-shell
gluon density~(\ref{defGGF}), not only its splitting function.
Therefore, a comparison with the ABF approach, to be discussed below, is possible in the
``on-shell'' limit, in which the homogeneous (eigenvalue) equation of
RGI holds. In the frozen coupling limit we have simply
\begin{equation}\label{implicit}
 \chi_{\omega}(\alpha_s,\gamma-{\textstyle\frac{\omega}{2}}) = \omega \;, \qquad
 (\text{$\chi_{\omega}$ is at scale $k k_0$}) \;.
\end{equation}
Solving Eq.~(\ref{implicit}) for either $\omega$ or $\gamma$, we are able
to identify the effective characteristic function and its dual
anomalous dimension
\begin{equation}\label{dual}
 \omega = \chi_{\mathrm{eff}}(\alpha_s,\gamma) \;; \quad \gamma = \gamma_{\mathrm{eff}}(\alpha_s,\omega) \;,
\end{equation}
in the same spirit as the ABF
approach~\cite{Ball:1999sh,Altarelli:1999vw,Altarelli:2000mh,Altarelli:2001ji,Altarelli:2003hk,Altarelli:2003kt, Altarelli:2004dq,Altarelli:2005xx}.

\subsubsection{The duality approach}
\label{sec:pdf,res,sx,dual}

As already mentioned, in  the ABF approach one
constructs  an improved anomalous
dimension (splitting function) for DIS which reduces to the ordinary
perturbative result at large $N$ (large $x$) given by: 
\begin{equation}
\gamma(N,\alpha_s)=\alpha_s \gamma_0(N)~+~\alpha_s^2
\gamma_1(N)~+~\alpha_s^3
\gamma_2(N)~~\dots .
\label{gammadef}
\end{equation}
while including resummed BFKL corrections at small $N$ (small $x$)
which are determined  by the aforementioned BFKL
kernel $\chi(M,\alpha_s)$:
\begin{equation}
\chi(M,\alpha_s)=\alpha_s \chi_0(M)~+~\alpha_s^2 \chi_1(M)~+~\dots ,
\label{chidef}
\end{equation} 
which is the Mellin transform of the $\omega\to0$, angular averaged 
kernel ${\cal K}$ eq.~\ref{eqGGF}
with respect to $t=\ln \frac{k^2}{k_0^2}$.
The main theoretical tool which enables this construction is 
the duality relation between the kernels $\chi$ and $\gamma$
\begin{equation}
\chi(\gamma(N,\alpha_s),\alpha_s)=N,
\label{dualdef}
\end{equation}
[compare eq.~(\ref{dual})]
which is a consequence of the fact that the solutions of the BFKL and DGLAP
equations coincide at leading twist~\cite{Ball:1997vf,Ball:1999sh,Altarelli:1999vw}.
Further improvements are obtained exploiting
the symmetry
under gluon interchange of the BFKL gluon-gluon kernel and through the
inclusion of running coupling effecs. 

By using duality, one can
construct a more balanced expansion for both $\gamma$ and $\chi$,
the "double leading" (DL) expansion, where the
information from $\chi$ is used to include in $\gamma$ all powers of
$\alpha_s/N$ and, conversely $\gamma$ is used to improve $\chi$ by all
powers of $\alpha_s/M$. A great advantage of the DL expansion is that
it resums the collinear poles of $\chi$ at $M=0$, enabling the
imposition of the physical requirement of momentum conservation
$\gamma(1,\alpha_s)=0$, whence, by duality: 
\begin{equation}
\chi(0,\alpha_s)=1.
\label{mom}
\end{equation}
This procedure eliminates in a model independent way the alternating
sign poles $+1/M, -1/M^2,.....$ that appear in $\chi_0$,
$\chi_1$,\dots. These poles make the perturbative expansion of $\chi$
unreliable even in the central region of $M$: e.g., 
$\alpha_s \chi_0$ has a minimum at $M=1/2$, while, at realistic values
of $\alpha_s$,  $\alpha_s \chi_0+\alpha_s^2 \chi_1$ has a maximum. 

At
this stage, while the poles at $M=0$ are eliminated, those at $M=1$
remain, so that the DL expansion is still not finite near
$M=1$. The resummation of the $M=1$ poles can be accomplished by
exploiting the collinear-anticollinear symmetry,
as 
suggested in the CCSS approach discussed above. 
In Mellin space, this symmetry
implies that at the fixed-coupling level the kernel $\chi$
for evolution in  $\ln \frac{s}{k k_0}$
 must satisfy
$\chi(M)=\chi(1-M)$. This symmetry is however broken by the DIS choice
 of variables $\ln\frac{1}{ x}=\ln\frac {s}{Q^2}$ and by the running of the
 coupling. In the fixed coupling limit 
the kernel $\chi_{\rm DIS}$, dual to the DIS anomalous
dimension, is related to the symmetric one
$\chi_{\sigma}$ through  the implicit equation~\cite{Fadin:1998py} 
\begin{equation}
\chi_{\rm DIS}(M+1/2\chi_{\rm \sigma}(M))=\chi_{\sigma}(M),
\label{symm}
\end{equation} 
to be compared to eq.~(\ref{implicit}) of the CCSS approach.

Hence, the $M=1$ poles can be resummed  by performing the double-leading
resummation of $M=0$ poles of $\chi_{\rm DIS}$, determining the
associated $\chi_\sigma$ through eq. (\ref{symm}), then symmetrizing
it, 
and finally going back to DIS
variables by using eq. (\ref{symm}) again in reverse.
Using the momentum
conservation  eq. (\ref{mom}) and eq. (\ref{symm}), 
it is easy to show that $\chi_\sigma(M)$ is an entire
function of M, with 
$\chi_\sigma(-1/2)=\chi_\sigma(3/2)=1$ and has a minimum at $M=1/2$. 
Through this procedure one obtains order
by order from the DL expansion a symmetrized DL
kernel $\chi_{\rm DIS}$, and its corresponding dual anomalous dimension $\gamma$. The
kernel $\chi_{\rm DIS}$ has to all orders a minimum and satisfies a
momentum conservation constraint $\chi_{\rm DIS}(0)=\chi_{\rm DIS}(2)=1$.

The final ingredient of the ABF approach is a treatment of the running
coupling corrections to the resummed terms. Indeed, their inclusion
in the resummed anomalous dimension greatly softens
the asymptotic behavior near $x=0$.
Hence, the 
dramatic rise of structure functions at small $x$, 
which characterized resummations
based on leading--order BFKL evolution, and is
ruled out phenomenologically, is replaced by a much milder rise. This
requires a running coupling generalization of the duality
equation~(\ref{dualdef}), which is possible noting that in $M$ space the
running coupling $\alpha_s(t)$ becomes a differential operator,  
since $t \rightarrow d/dM$. Hence, 
 the BFKL evolution equation for
double moments $G(N,M)$, which is an algebraic equation at fixed
coupling, becomes a differential equation in $M$ for running
coupling. In the ABF approach, one solves this differential equation
analytically when the kernel is replaced by its quadratic
approximation near the minimum. The solution is expressed in terms of an
Airy  function if the kernel is linear in $\alpha_s$, for example in
the case of $\alpha_s \chi_0$, or of a Bateman function in the more
general case of a non linear dependence on $\alpha_s$ as is the case
for the DL kernels. The final result for the improved anomalous
dimension is given in terms of the DL expansion plus the ``Airy'' or
``Bateman'' anomalous dimension, with the terms already included in the DL
expansion subtracted away.  

For example, at leading DL order, i.e. only using $\gamma_0(N)$ and
$\chi_0(M)$, 
the improved anomalous dimension is
\begin{equation}
\gamma_I^{NL}(\alpha_s, N) = \Big[\alpha_s\gamma_0(N)+ \alpha_s^2 \gamma_1(N) 
+\gamma_s(\frac{\alpha_s}{N}) -\frac{n_c\alpha_s}{\pi N}\Big]
+\gamma_A(c_0,\alpha_s,N)-\frac{1}{2} +
\sqrt{\frac{2}{\kappa_0\alpha_s}[N-\alpha_s
c_0]} .\label{gamimp}
\end{equation}
The terms within square brackets
give the LO DL approximation, i.e. they contain the
fixed--coupling information from
$\gamma_0$ and (through $\gamma_s$) from $\chi_0$. The ``Airy'' anomalous dimension
$\gamma_A(c_0,\alpha_s,N)$ contains the running coupling resummation,
i.e. it is the exact solution of the 
running coupling BFKL equation which corresponds to a quadratic 
approximation to $\chi_0$ near $M=1/2$.
The last two terms
subtract the contributions to $\gamma_A(c_0,\alpha_s,N)$
which are already included in
$\gamma_s$ and
$\gamma_0$.
In the limit $\alpha_s
\rightarrow 0$ with $N$ fixed, $\gamma_I(\alpha_s,N)$ reduces to
$\alpha_s\gamma_0(N)+O(\alpha_s^2)$. For  $\alpha_s
\rightarrow 0$ with $\alpha_s/N$ fixed, $\gamma_I(\alpha_s,N)$  reduces to
$\gamma_s(\frac{\alpha_s}{N})+O(\alpha_s^2/N)$, i.e. the leading term of the
small $x$ expansion. Thus the Airy term is subleading in both
limits. However, if $N\to 0$ at fixed $\alpha_s$, the Airy term
replaces the leading singularity of the DL anomalous dimension,
which is a square root branch cut, with a simple pole, located on the
real axis at rather smaller $N$, thereby softening the small $x$ behaviour.
The quadratic approximation is sufficient to
give the correct asymptotic behaviour up to terms which
are of subleading order in comparison to those included in the DL
expression in eq.~(\ref{gamimp}). 
\begin{figure}[htbp]
  \includegraphics[width=0.75\textwidth]{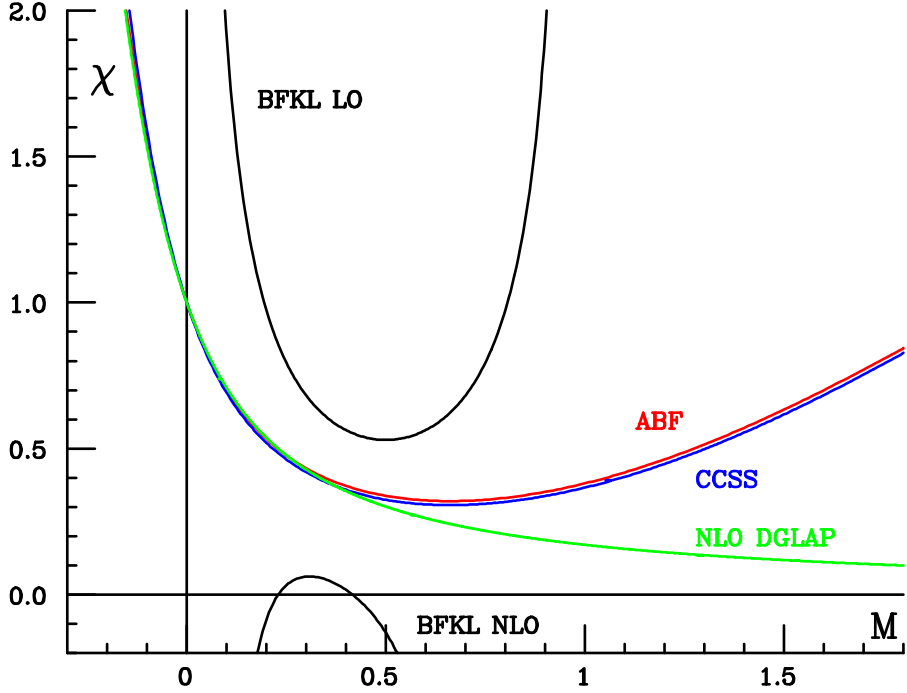}\centering
  \caption{The kernel $\chi$ (BFKL characteristic function) for fixed coupling
    ($\beta_0=0$) $\alpha_s=0.2$ and
    ${n_{\! f}}=0$. The BFKL curves are the LO and NLO truncations of
    eq.~(\ref{chidef}), the DGLAP curve is the dual
    eq.~(\ref{dualdef}) of the NLO anomalous dimension
    eq.~(\ref{gammadef}), while the CCSS and ABF curves are
    respectively the solution $\omega$ of
    eq.~(\ref{implicit}) and the solution $\chi_{\rm DIS}$ of eq.~(\ref{symm}).}
  \label{fig:chi}
\end{figure}

The running coupling resummation 
procedure can be applied to a symmetrized kernel, which possesses
a minimum to all orders, and then extended to next-to-leading
order~\cite{Altarelli:2004dq,Altarelli:2005xx}. This entails
various
technical complications, specifically related to the nonlinear
dependence of the symmetrized kernel on $\alpha_s$, to the need to
include interference between running coupling effects and 
the small $x$ resummation, and to the consistent treatment
of next-to-leading log $Q^2$ terms, in particular those related to the
running of the coupling. It should be noted that  even though the ABF
appraoch is limited to the description of
leading-twist evolution at zero-momentum transfer, it leads to a pair
of systematic dual perturbative expansions for the $\chi$ and $\gamma$
kernels. Hence, comparison with the CCSS approach is possible for
instance by
comparing the NLO ABF kernel to the RG improved L$x$+NL$x$ CCSS kernel.

\subsubsection{Comparison of results}
\label{sec:pdf,res,sx,imp}

Even though the basic underlying physical principles of the CCSS and
ABF approaches are close, there are technical differences in the
construction of the resummed  RG-improved (CCSS) or symmetrized DL (ABF)
kernel,
in the derivation from it of an anomalous dimension and associated
splitting function, and in the inclusion of running coupling effects.
Therefore,  we will compare results for the resummed
fixed-coupling $\chi$ kernel (BFKL characteristic function), then the
corresponding fixed-coupling splitting functions, and finally the
running coupling splitting functions which provide the final result
in both approaches. In order to assess the phenomenological impact on
parton evolution we will finally compare the convolution of the
splitting function with a ``typical'' gluon distribution.

\begin{figure}[htbp]
\includegraphics[width=0.85\textwidth]{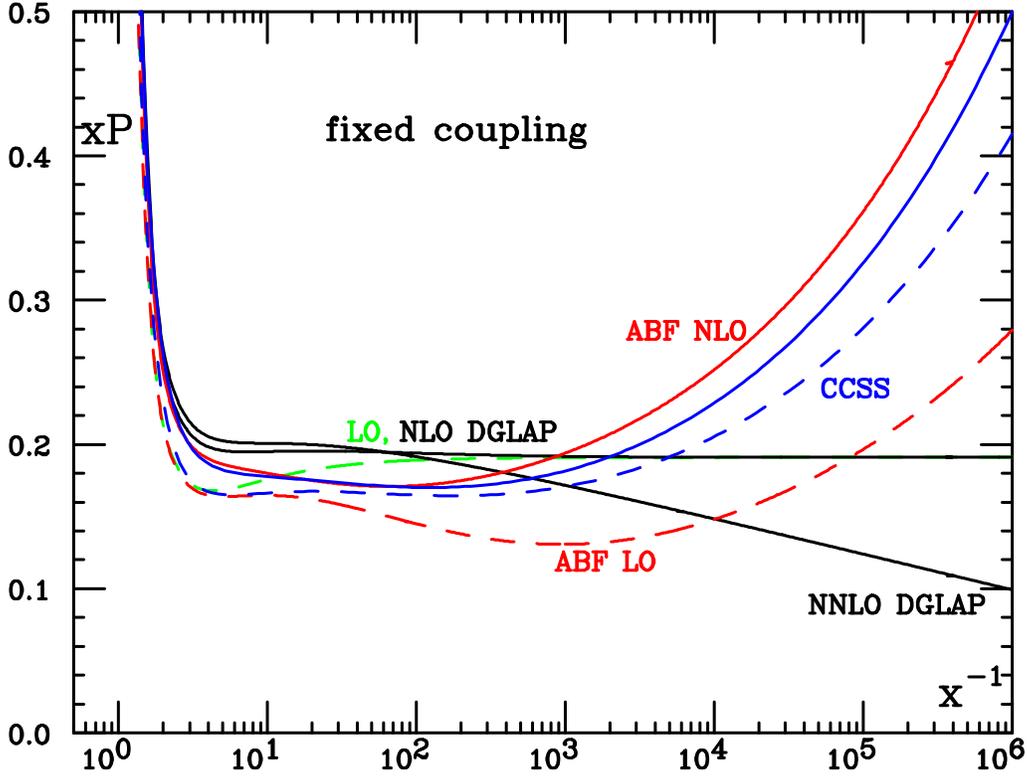}\centering
  \caption{The fixed coupling
    ($\beta_0=0$) $xP_{gg}(x)$ splitting
 function, evaluated with $\alpha_s=0.2$  and $n_f=0$. 
The dashed curves are LO for DGLAP, NL$x$+LO for CCSS and symmetrized LO DL for ABF,
 while the solid curves are NLO and NNLO for DGLAP, NL$x$+NLO for CCSS
    and symmetrized NLO DL for ABF. }
  \label{fig:Pggb0}
\vspace{-.3cm}
\end{figure}
In fig.~\ref{fig:chi} we compare the solution, $\omega$, to the on-shell constraint,
eq.~(\ref{implicit}) for the RGI CCSS result,   and the
solution $\chi_{\rm DIS}$ of eq.~(\ref{symm}) for the symmetrized NLO DL ABF
result. The pure L$x$ and NL$x$ (BFKL) and next-to-leading $\ln Q^2$
(DGLAP) are also shown. All curves are determined  with frozen
coupling ($\beta_0=0$), and with
$n_f=0$, in order to avoid complications related to the diagonalization of
the DGLAP anomalous dimension matrix and to the choice of scheme for the
quark parton distribution.
The resummed CCSS and ABF results are very close, in that they
coincide by construction at the momentum conservation points $M=\frac{1}{2}$
and $M=2$, and differ only in the treatment of NLO DGLAP terms. In
comparison to DGLAP, the resummed kernels have a minimum,
related to the fact that both collinear and anticollinear logs are
resummed.
 In comparison to BFKL, which has a minimum at LO but
not NLO, the resummed kernels always have a perturbatively stable  minimum,
characterized by a lower intercept than leading--order BFKL: specifically,
when $\alpha_s=0.2$, $\lambda\sim0.3$  instead of $\lambda\sim0.5$. This
corresponds to a softer small $x$ rise of the associated splitting
function. 

The fixed--coupling resummed splitting functions up to NLO are shown in 
figure~\ref{fig:Pggb0}, along with the unresummed DGLAP splitting
functions up to NNLO.\footnote{Starting from NLO one needs also to
  specify a factorisation scheme. Small-$x$ results are most
  straightforwardly obtained in the $Q_0$ scheme, while fixed-order
  splitting functions are quoted in the $\overline{\mathrm{MS}}$ scheme (for
  discussions of the relations between different schemes see
  \cite{Catani:1994sq,Ball:1995tn,Altarelli:2000mh,Ciafaloni:2005cg}).} 
 In the CCSS approach the splitting function is determined
by explicitly solving eq.~(\ref{eqGGF}) with the kernel corresponding to
figure~\ref{fig:chi}, and
then applying the numerical deconvolution procedure of
\cite{Ciafaloni:2000cb}. 
 For ${n_{\! f}}=0$ the NLO DGLAP splitting function has the
property that it vanishes at small $x$ --- this makes it relatively
straightforward to combine not just LO DGLAP but also NLO DGLAP with
the NLLx resummation. Both  the CCSS NL$x$+LO and NL$x$+NLO
curves are shown in figure~\ref{fig:Pggb0}.
On the other hand, in the ABF approach the splitting function is the 
inverse Mellin transform of the anomalous dimension obtained using
duality eq.~(\ref{dualdef}) from the symmetrized DL $\chi$
kernel. Hence, the LO and NLO resummed result respectively reproduce all
information contained in the LO and NLO $\chi$ and $\gamma$  kernel
with the additional constraint of collinear-anticollinear symmetry.
Both the ABF LO and NLO results are shown in figure~\ref{fig:Pggb0}.

\begin{figure}[htbp]
 \includegraphics[width=0.85\textwidth]{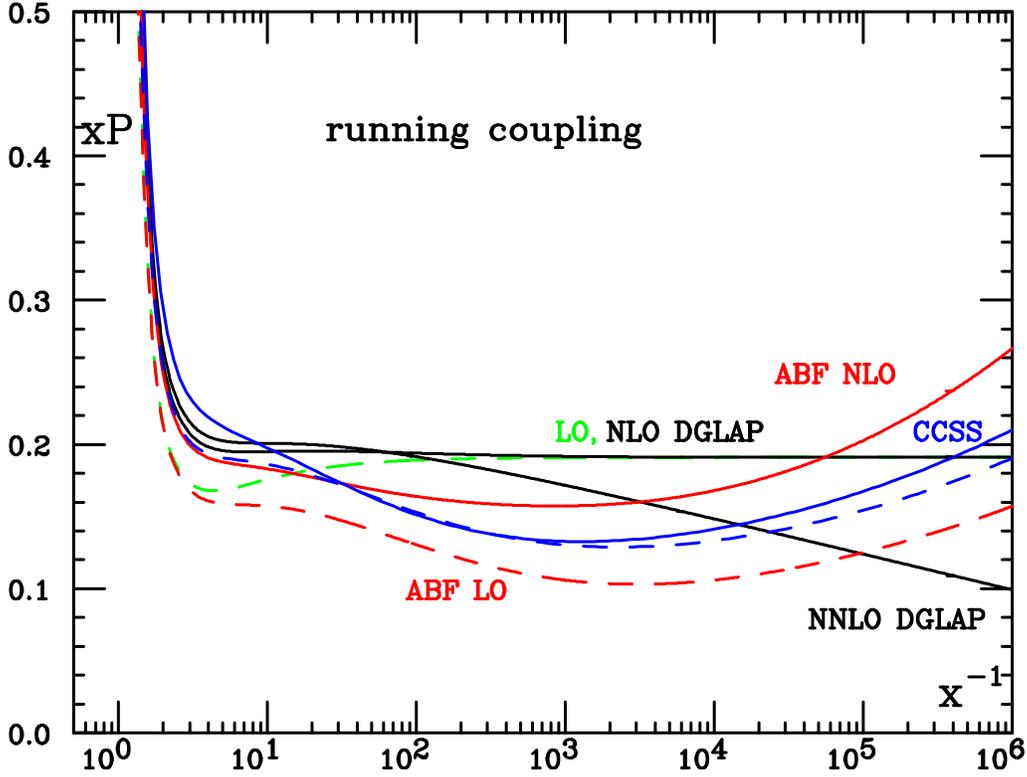}\centering
  \caption{The  running coupling
     $xP_{gg}(x)$ splitting
 function, evaluated with  $\alpha_s=0.2$ and $n_f=0$. The various
     curves correspond to 
     the same cases as in figure~\ref{fig:Pggb0}. }
  \label{fig:Pgg}
\end{figure}
In comparison to unresummed results, the resummed splitting functions
display the characteristic rise at small $x$ of fixed-coupling
leading-order BFKL
resummation, though the small $x$ rise is rather milder ($\sim x^{-0.3}$
instead of $\sim x^{-0.5}$ for $\alpha_s=0.2$).  
At large $x$ there is good agreement between the resummed results and
the corresponding LO (dashed) or NLO (solid) DGLAP curves.
At small $x$ the difference between the ABF LO and CCSS NL$x$+LO
(dashed) curves is mostly due to the inclusion in CCSS of BFKL NL$x$
terms, as well as to differences in the
symmetrization procedure. When comparing CCSS NL$x$+NLO with ABF NLO
this difference is reduced, and , being only due the way the
symmetrization is implemented, it might be taken as 
an estimate of the intrinsic ambiguity of the
fixed--coupling resummation procedure.
At intermediate $x$  the NLO resummed splitting functions
is of a similar order of magnitude as the NLO DGLAP result even down to quite
small $x$, but with a somewhat different shape, characterized  by a shallow
dip at $x\sim 10^{-2}$, until the small $x$ rise sets in for $x \sim
10^{-3}$. It has been suggested~\cite{Ciafaloni:2003kd}  that in
the small ${\alpha_\mathrm{s}} $ limit this dip can be explained as a
consequence of the interplay between
the $-{\alpha_\mathrm{s}} ^3 \ln x$ NNLO term of $xP_{gg}$ (also present in the
resummation) and the first positive resummation effects which start
with an ${\alpha_\mathrm{s}}^4 \ln^3 1/x$ term. 
The unstable small $x$ drop of the  NNLO DGLAP result appears to
be a consequence of the unresummed $\frac{\alpha_s^3}{N^2}$ 
double  pole in the NNLO anomalous dimension.

The running-coupling resummed splitting functions are displayed in
figure~\ref{fig:Pgg}. Note that the unresummed curves are the same as
in the fixed coupling case since their dependence on $\alpha_s$ is
just through a prefactor of $\alpha_s^k$, whereas in the resummed case
there is an interplay between the running of the coupling and the
structure of the small-$x$ logs. All the resummed curves display a
considerable softening of the small $x$ behaviour in comparison to their
fixed-coupling counterparts, due to the softening of the leading small
$x$ singularity in the running-coupling case~\cite{Ciafaloni:1999yw,Altarelli:2001ji}.
As a consequence, the various resummed results are closer to each
other than in the fixed-coupling case, and also closer to the
unresummed LO and NLO DGLAP results. The resummed perturbative
expansion appears to be stable, subject to moderate theoretical
ambiguity, and qualitatively close to NLO DGLAP.

\begin{figure}[htbp]
 \includegraphics[width=0.5\textwidth]{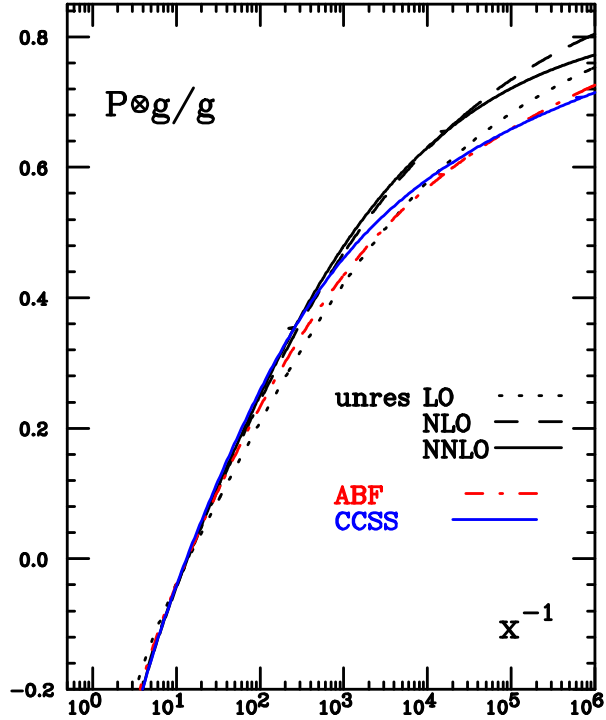}\centering
  \caption{Convolution of resummed and fixed-order $P_{gg}$ splitting
    functions with a toy gluon distribution,
    eq.~(\ref{eq:toy_gluon_for_pgg_test}),  normalised to
    the gluon distribution itself, with $\alpha_s=0.2$ and  ${n_{\!
    f}} = 0$. The resummed CCSS and ABF curves are obtained using
    respectively the
    CCSS NL$x$+NLO and the ABF NLO splitting function shown in fig.~\ref{fig:Pgg}.
}
  \label{fig:conv}
\end{figure}
Finally, to appreciate the impact of resummation it is useful to investigate
not only the properties of the splitting function, but also its
convolution with a physically reasonable gluon distribution. We take
the following toy gluon
\begin{equation}
  \label{eq:toy_gluon_for_pgg_test}
  xg(x) = x^{-0.18} (1-x)^5\,,
\end{equation}
and show in fig.~\ref{fig:conv} the result of its convolution with
various splitting functions of fig.~\ref{fig:Pgg}.
The differences between resummed and unresummed results,
and between the CCSS and ABF  resummations are partly washed out by
the convolution, even though the difference between the unresummed LO
and NLO DGLAP results is clearly visible. In particular,
differences between the fixed-order and resummed
convolution start to become significant only for $x \lesssim 10^{-2}-10^{-3}$,
even though resummation effects started to be visible in the splitting
functions at somewhat larger $x$.

It should be kept in mind that it is
only the $gg$ entry of the singlet splitting function matrix that has
so far been investigated at this level of detail and that the other
entries may yet reserve surprises.

\subsubsection{Explicit solution of the BFKL equation by Regge exponentiation}
\label{sec:pdf,res,sx,num}

The CCSS approach of section~\ref{sec:pdf,res,sx,rgi} exploits a
numerical solution of the BFKL equation 
in which the gluon
Green's function is represented on a grid in
$x$ and $k$. This method  provides an efficient 
determination of the azimuthally averaged
Green's function and splitting functions  --- for percent accuracy, up to
$Y=30$, it runs in a few seconds --- for a wide range of physics
choices, e.g. pure NL$x$, various NL$x$+NLO schemes. 
Here we will discuss an alternative
framework suitable to solve numerically the NLL 
BFKL integral equation~\cite{Andersen:2003an}, based on Monte Carlo
generation of events, which can also be applied to the study of
different resummation schemes and DIS, but so far has been
investigated for simpler NLL BFKL kernels and Regge--like
configurations. This method 
has the advantage that it automatically provides information about
azimuthal decorrelations as well as the pattern of final-state emissions.

This appproach relies on the fact that, as shown in 
Ref.~\cite{Andersen:2003an},
it is possible to trade the simple and double poles in 
$\epsilon$, present in $D = 4 + 2 \epsilon$ dimensional regularisation, by 
a logarithmic dependence on an effective gluon mass $\lambda$. This $\lambda$ 
dependence numerically cancels out when the full NLL BFKL evolution is taken 
into account for a given center--of--mass energy, a consequence of the 
infrared finiteness of the full kernel. The introduction of this mass scale, 
differently to the original work of Ref.~\cite{Fadin:1998py} was performed 
without angular averaging the NLL kernel.

With such reguralisation of the infrared divergencies it is then convenient to 
iterate the NLL BFKL equation for the $t$--channel partial wave, generating, 
in this way, multiple poles in the complex $\omega$--plane. The positions 
of these singularities are set at different values of the gluon Regge 
trajectory depending on the transverse momenta of the Reggeized gluons 
entering the emission vertices. At this point it is possible to Mellin 
transform back to energy space and obtain an iterated form for the solution of 
the NLL BFKL equation:
\begin{align}
\label{solution1}
f({\bf k}_a ,{\bf k}_b, {\rm Y}) 
&= e^{\omega_0^\lambda \left({\bf k}_a\right) {\rm Y}} \frac{}{}\delta^{(2)} ({\bf k}_a - {\bf k}_b)\\
&+\sum_{n=1}^{\infty} \prod_{i=1}^{n} 
\int d^2 {\bf k}_i \int_0^{y_{i-1}} d y_i \left[\frac{\theta\left({\bf k}_i^2 - \lambda^2\right)}{\pi {\bf k}_i^2} \xi\left({\bf k}_i\right) +\widetilde{\mathcal{K}}_r \left({\bf k}_a+\sum_{l=0}^{i-1}{\bf k}_l,
{\bf k}_a+\sum_{l=1}^{i}{\bf k}_l\right)\frac{}{}\right]\nonumber\\
&  \times  e^{
\omega_0^\lambda\left({\bf k}_a+\sum_{l=1}^{i-1} {\bf k}_l
\right)(y_{i-1}-y_i)}\  e^{\omega_0^\lambda\left({\bf
  k}_a+\sum_{l=1}^i {\bf k}_l\right)y_n}\delta^{(2)} \left(\sum_{l=1}^{n}{\bf
k}_l  
+ {\bf k}_a - {\bf k}_b \right), \nonumber
\end{align}
where the strong ordering in longitudinal components of the parton emission is 
encoded in the nested integrals in rapidity with an upper limit set by the 
logarithm of the total energy in the process, $y_0 = {\rm Y}$. The first term 
in the expansion corresponds to two Reggeized gluons propagating in the 
$t$--channel with no additional emissions.
The exponentials carry the dependence on the 
Regge gluon trajectory, {\it i.e.} 
\begin{eqnarray}
\label{trajectory}
\omega_0^\lambda \left({\bf q}\right) &=& 
- \bar{\alpha}_s \ln{\frac{{\bf q}^2}{\lambda^2}}
+ \frac{\bar{\alpha}_s^2}{4}\left[\frac{\beta_0}{2 N_c}\ln{\frac{{\bf q}^2}{\lambda^2}}\ln{\frac{{\bf q}^2 \lambda^2}{\mu^4}}+\left(\frac{\pi^2}{3}-\frac{4}{3}-\frac{5}{3}\frac{\beta_0}{N_c}\right)\ln{\frac{{\bf q}^2}{\lambda^2}}+6 \zeta(3)\right],
\end{eqnarray}
corresponding to no--emission probabilities between two consecutive effective
vertices. Meanwhile, the real emission is built out of two parts, the first
one:
\begin{eqnarray}
\xi \left({\rm X}\right) &\equiv& \bar{\alpha}_s +  
\frac{{\bar{\alpha}_s}^2}{4}\left(\frac{4}{3}-\frac{\pi^2}{3}+\frac{5}{3}\frac{\beta_0}{N_c}-\frac{\beta_0}{N_c}\ln{\frac{{\rm X}}{\mu^2}}\right),  
\end{eqnarray}
which cancels the singularities present in the trajectory order by order in 
perturbation theory, and the second one: $\tilde{\cal K}_r$, which, although 
more complicated in structure, does not generate $\epsilon$ singularities when 
integrated over the full phase space of the emissions, for details see 
Ref.~\cite{Andersen:2003an}.

\begin{figure}[tbp]
\centering
\vspace{5cm}
\includegraphics{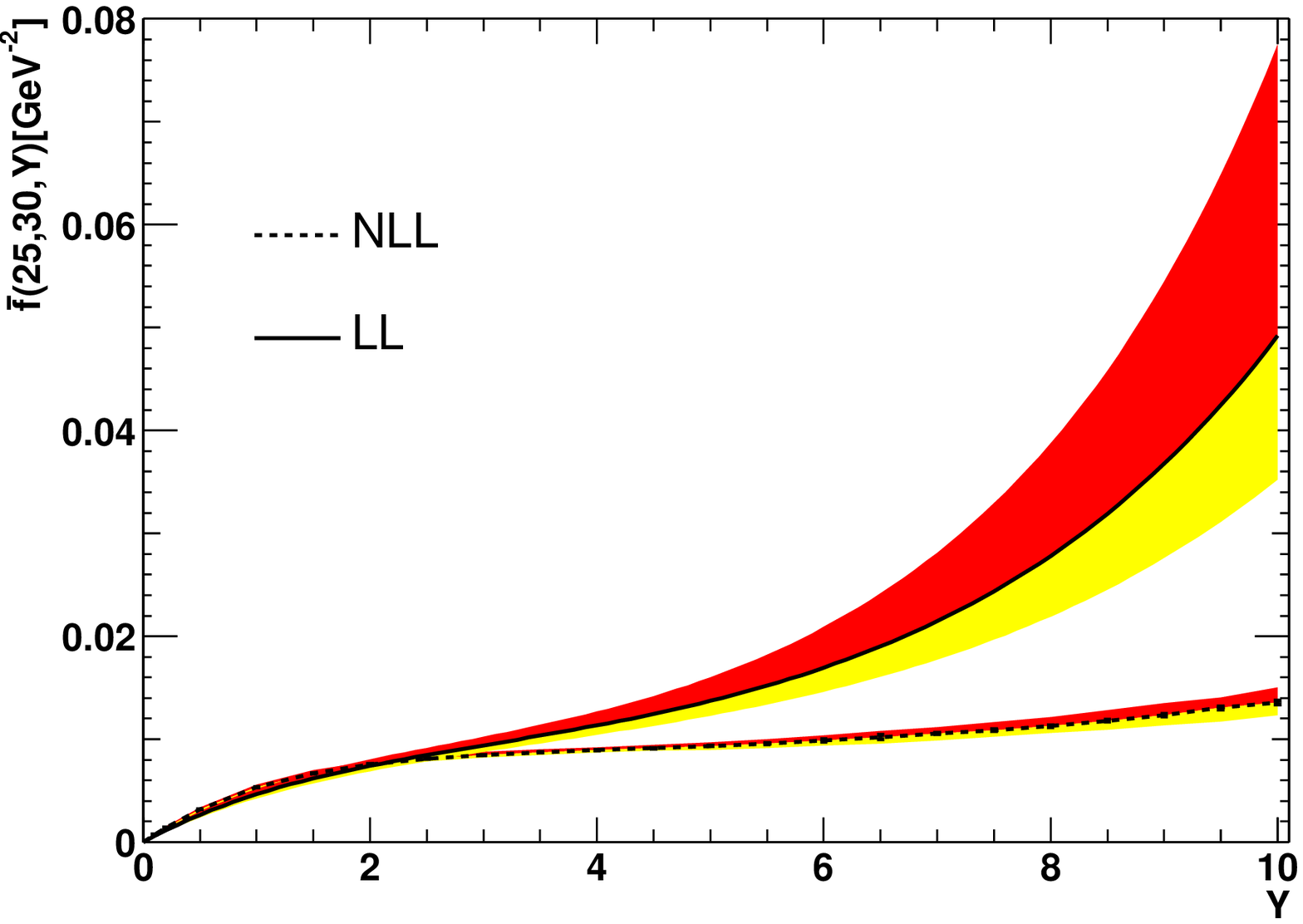}
\hspace{0.2cm}
\includegraphics{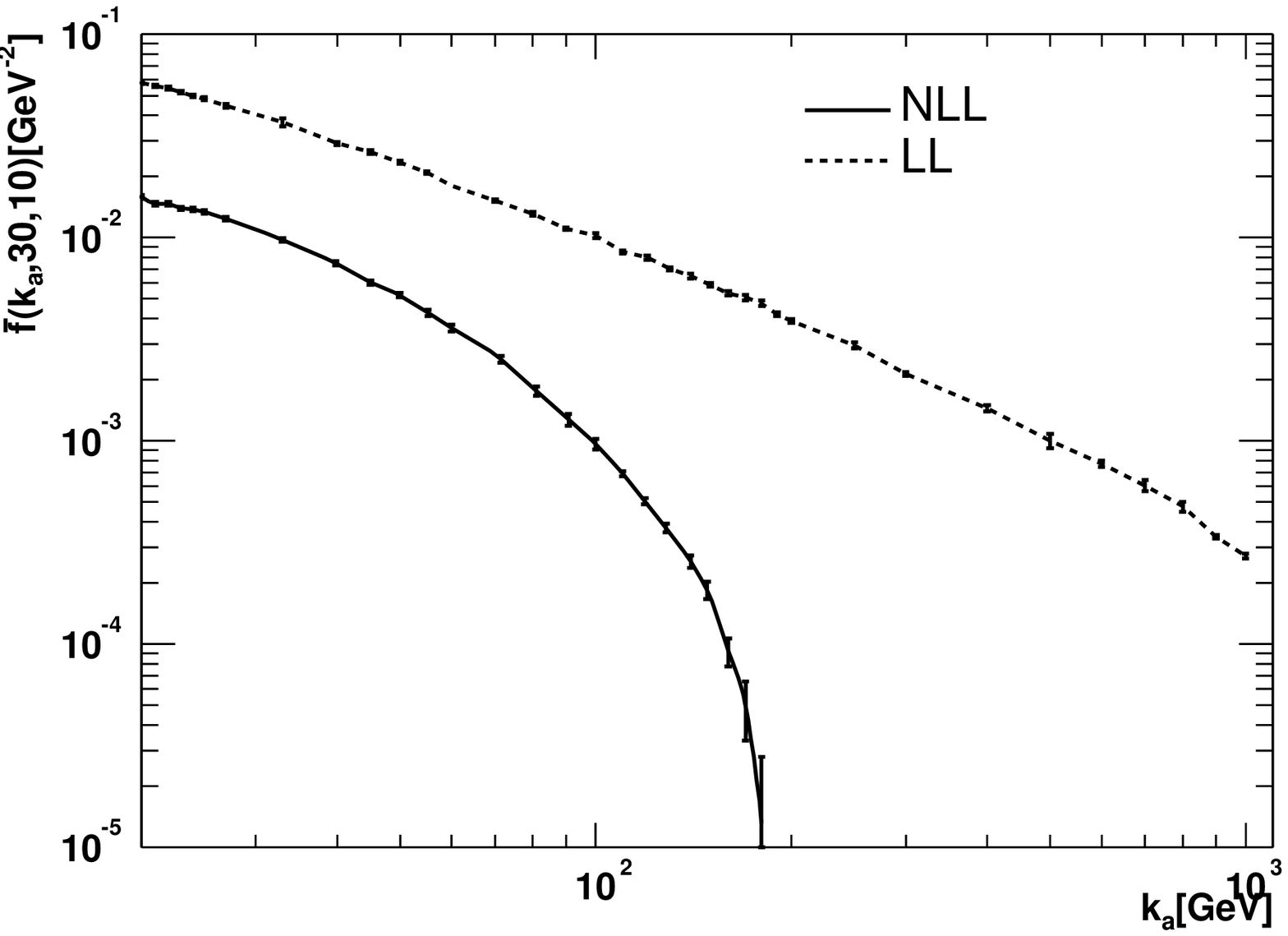}
\caption{Analysis of the gluon Green's function as obtained from the NLL BFKL
  equation. The plot to the left shows the evolution in rapidity of the gluon
  Green's function at LL and NLL for fixed $k_a=25$~GeV and $k_b=30$~GeV. The
  plot on the right hand side shows the dependence on $k_a$ for fixed
  $k_b=30$~GeV and $Y=10$.}
\label{JAfigure1}
\end{figure}

The numerical implementation and analysis of the solution as 
in Eq.~(\ref{solution1}) was performed in 
Ref.~\cite{Andersen:2003wy}. As in previous studies the intercept at NLL 
was proved to be lower than at leading--logarithmic (LL) accuracy. 
In this approach the kernel is not expanded on a set of functions derived
from the LL eigenfunctions, and there are no instabilities in energy
associated with a choice of functions breaking the $\gamma
\leftrightarrow 1 - \gamma$ symmetry, with $\gamma$ being the variable
Mellin--conjugate of the transverse momenta. This is explicitly shown at the
left hand side of Fig.~\ref{JAfigure1} where the coloured bands correspond to
uncertainties from the choice of renormalisation scale.
Since the exponential growth at NLL is slower than at LL, there is little
overlap between the two predictions, and furthermore these move apart for
increasing rapidities. The NLL corrections to the intercept amount to roughly
50\% and are stable with increasing rapidities.

In transverse momentum space the NLL corrections are stable when the two
transverse scales entering the forward gluon Green's function are of similar
magnitude. 
However, when the ratio between these scales departs largely from unity, the
perturbative convergence is poor, driving, as it is well--known, the gluon
Green's function into an oscillatory behaviour with regions of negative
values along the period of oscillation. This behaviour is demonstrated in
the second plot of Fig~\ref{JAfigure1}.

\begin{figure}[tbp]
\vspace{4cm}
\includegraphics{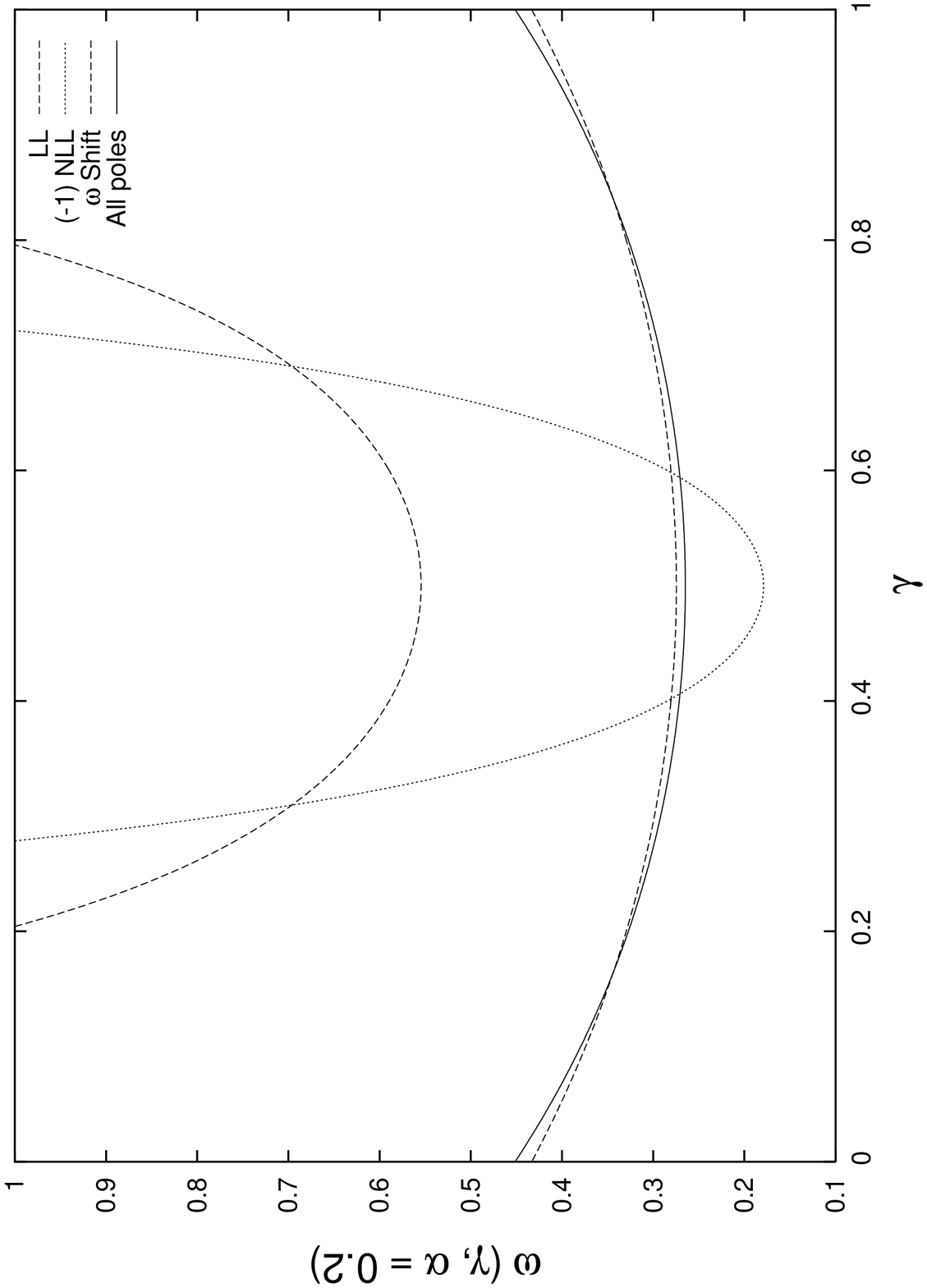}
\hspace{1cm}
\includegraphics{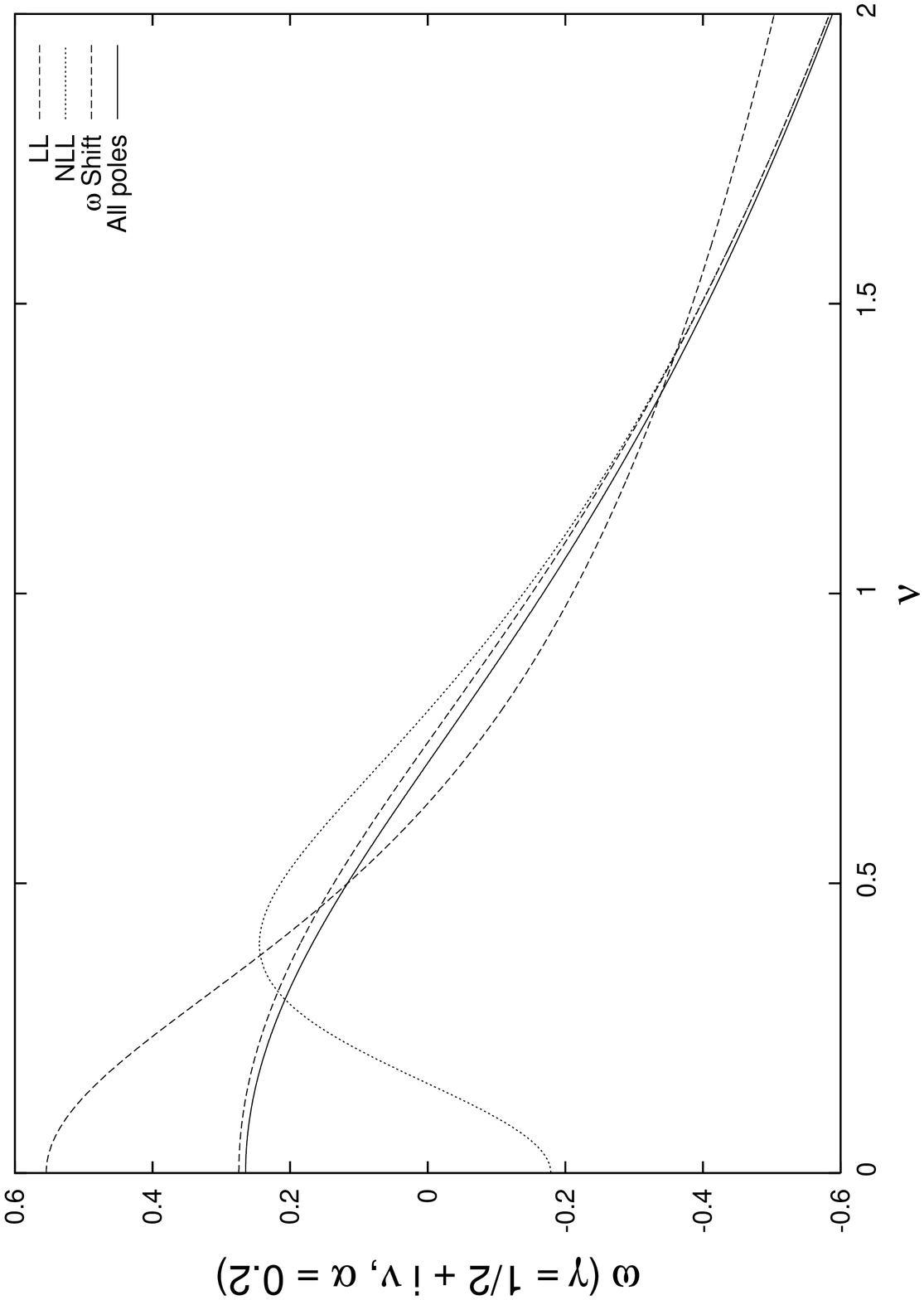}
  \caption{The $\gamma$--representation of the LL and NLL scale 
invariant kernels together with the collinearly--improved kernel by an 
$\omega$--shift and the ``all--poles'' resummation.}
\label{Agus1}
\end{figure}

\begin{figure}[tbp]
\centering
\vspace{5cm}
\includegraphics{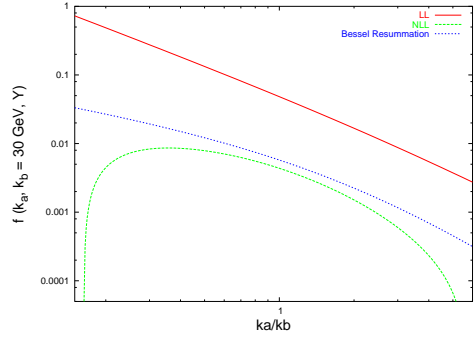}
  \caption{The behaviour of the NLL gluon Green's function using the 
Bessel resummation.}
\label{Agus2}
\end{figure}

The way the perturbative expansion of the BFKL kernel is improved by
simultaneous resummation of energy and collinear logs has been
discussed in sections~\ref{sec:pdf,res,sx,rgi},\ref{sec:pdf,res,sx,dual}.
In particular, the original 
approach based on the introduction in the NLL BFKL kernel of an all order 
resummation of terms compatible with renormalisation group evolution 
described in ref.~\cite{Salam:1998tj} (and incorporated in the CCSS
approach of section~\ref{sec:pdf,res,sx,rgi}) can be implemented in the iterative 
method here
explained~\cite{Vera:2005jt} (the method of
ref.~\cite{Salam:1998tj} was combined 
with the imposition of a veto in rapidities
in refs.~\cite{Schmidt:1999mz,Forshaw:1999xm,Chachamis:2004ab}). 
The main idea is that the solution to the $\omega$--shift proposed in
ref.~\cite{Salam:1998tj}
\begin{eqnarray}
\label{shiftGS}
\omega &=& {\bar \alpha}_s \left(1+\left({\rm a}+\frac{\pi^2}{6}\right){\bar \alpha}_s\right) \left(2 \psi(1)-\psi\left(\gamma+\frac{\omega}{2}-{\rm b}\,{\bar \alpha}_s \right)-\psi\left(1-\gamma+\frac{\omega}{2}-{\rm b}\,{\bar \alpha}_s \right)\right)\nonumber\\
&+& {\bar \alpha}_s^2 \left(\chi_1 \left(\gamma\right) 
+\left(\frac{1}{2}\chi_0\left(\gamma\right)-{\rm b}\right)\left(\psi'(\gamma)+\psi'(1-\gamma)\right)-\left({\rm a}+\frac{\pi^2}{6}\right)\chi_0(\gamma)\right),
\end{eqnarray}
can be very accurately approximated by the sum of the approximated 
solutions to the shift at each of the poles in $\gamma$ of the LL eigenvalue 
of the BFKL kernel. This provides an effective ``solution'' of 
Eq.~(\ref{shiftGS}) of the form~\cite{Vera:2005jt}
\begin{eqnarray}
\label{All-poles}
\omega &=& \bar{\alpha}_s \chi_0 (\gamma) + \bar{\alpha}_s^2 \chi_1 (\gamma) 
+ \left\{\sum_{m=0}^{\infty} \left[\left(\sum_{n=0}^{\infty}
\frac{(-1)^n (2n)!}{2^n n! (n+1)!}\frac{\left({\bar \alpha}_s+ {\rm a} \,{\bar \alpha}_s^2\right)^{n+1}}{\left(\gamma + m - {\rm b} \,{\bar \alpha}_s\right)^{2n+1}}\right) \right. \right. \nonumber\\
&-&\left.\left.\frac{\bar{\alpha}_s}{\gamma + m} - \bar{\alpha}_s^2 \left(\frac{\rm a}{\gamma +m} + \frac{\rm b}{(\gamma + m)^2}-\frac{1}{2(\gamma+m)^3}\right)\right]+ \left\{\gamma \rightarrow 1-\gamma\right\}\right\}, 
\end{eqnarray}
where $\chi_0$ and $\chi_1$ are, respectively, the LL and NLL scale invariant 
components of the kernel in $\gamma$ representation with the collinear limit
\begin{eqnarray}
\chi_1 \left(\gamma \right) \simeq 
\frac{\rm a}{\gamma}+\frac{\rm b}{\gamma^2}-\frac{1}{2 \gamma^3}, \, \,
{\rm a} = \frac{5}{12}\frac{\beta_0}{N_c} -\frac{13}{36}\frac{n_f}{N_c^3}
-\frac{55}{36}, \, \, 
{\rm b} = -\frac{1}{8}\frac{\beta_0}{N_c} -\frac{n_f}{6 N_c^3} -\frac{11}{12}.
\end{eqnarray}
The numerical solution to Eq.~(\ref{shiftGS}) and the value of 
expression~(\ref{All-poles}) are compared in Fig.~\ref{Agus1}.
The stability of the perturbative expansion is recovered in all regions of 
transverse momenta with a prediction for the intercept of 0.3 at NLL for 
$\bar{\alpha}_s = 0.2$, a result valid up to the introduction of scale 
invariance breaking terms. The implementation of expression~(\ref{All-poles}) 
in transverse momentum space is simple given that 
the transverse components decouple from the longitudinal in this form of the 
collinear resummation~\cite{Vera:2005jt}. 
The prescription is to remove the term 
$-\frac{\bar{\alpha}_s^2}{4} \ln^2{\frac{q^2}{k^2}}$ 
from the real emission kernel, ${\cal K}_r \left(\vec{q},\vec{k}\right)$, 
and replace it with
\begin{eqnarray}
\label{presc2}
\left(\frac{q^2}{k^2}\right)^{-{\rm b}{\bar \alpha}_s 
\frac{\left|k-q\right|}{k-q}}
\sqrt{\frac{2\left({\bar \alpha}_s+ {\rm a} \,{\bar \alpha}_s^2\right)}{\ln^2{\frac{q^2}{k^2}}}} 
J_1 \left(\sqrt{2\left({\bar \alpha}_s+ {\rm a} \,{\bar \alpha}_s^2\right) 
\ln^{2}{\frac{q^2}{k^2}}}\right) 
- {\bar \alpha}_s - {\rm a} \, {\bar \alpha}_s^2
+ {\rm b} \, {\bar \alpha}_s^2 \frac{\left|k-q\right|}{k-q}
\ln{\frac{q^2}{k^2}},
\end{eqnarray}
with $J_1$ the Bessel function of the first kind. This prescription does not 
affect angular dependences and generates a well--behaved gluon Green's 
function as can be seen in Fig.~\ref{Agus2} where the oscillations in the 
collinear and anticollinear regions of phase space are consistently removed.
At present, work is in progress to study the effect of the running of the 
coupling in this analysis when the Bessel resummation is introduced in  
the iterative procedure of Ref.~\cite{Andersen:2003an}.

\begin{figure}[tbp]
\centering
\vspace{5cm}  
\includegraphics{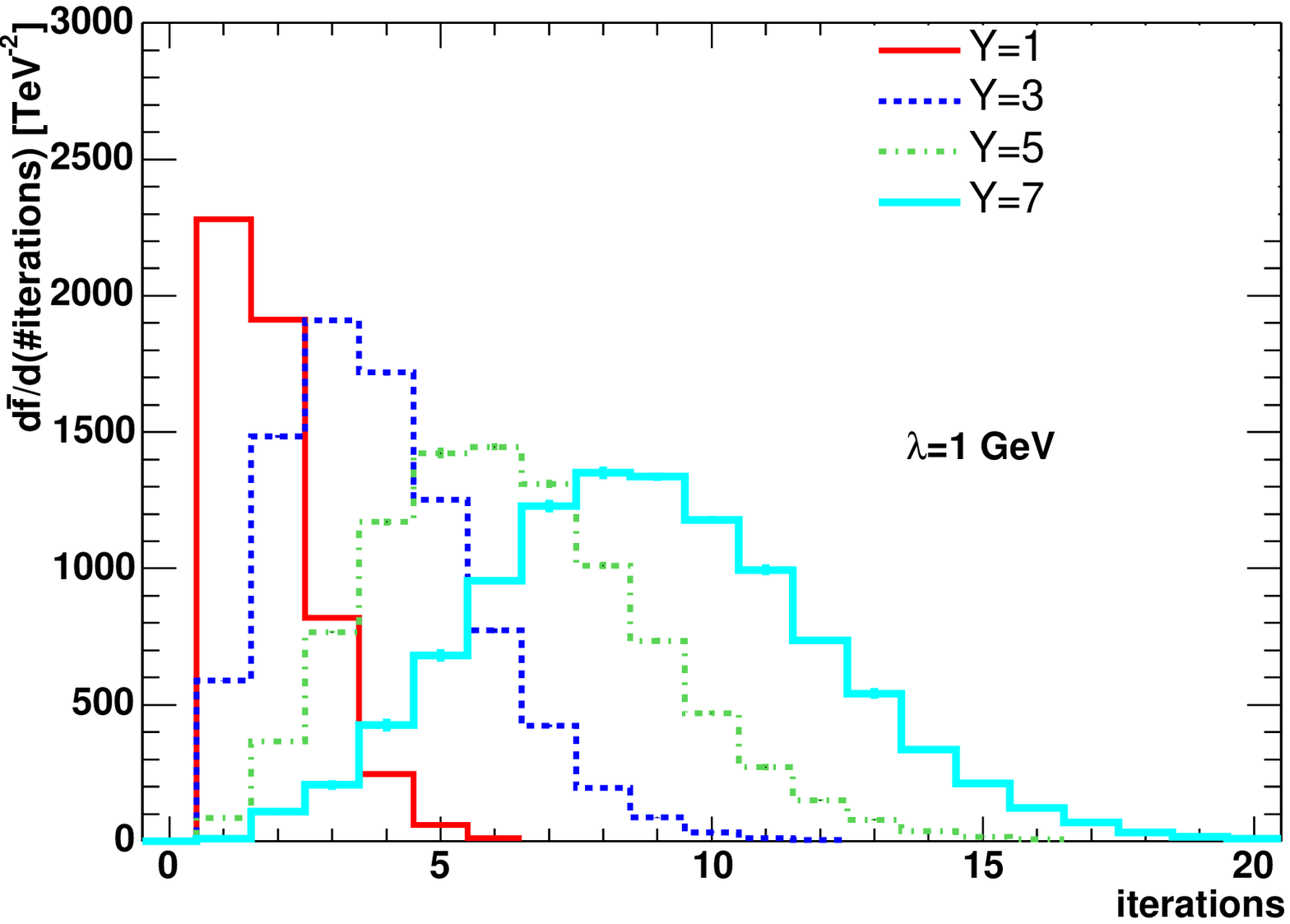}
\hspace{0.2cm}
\includegraphics{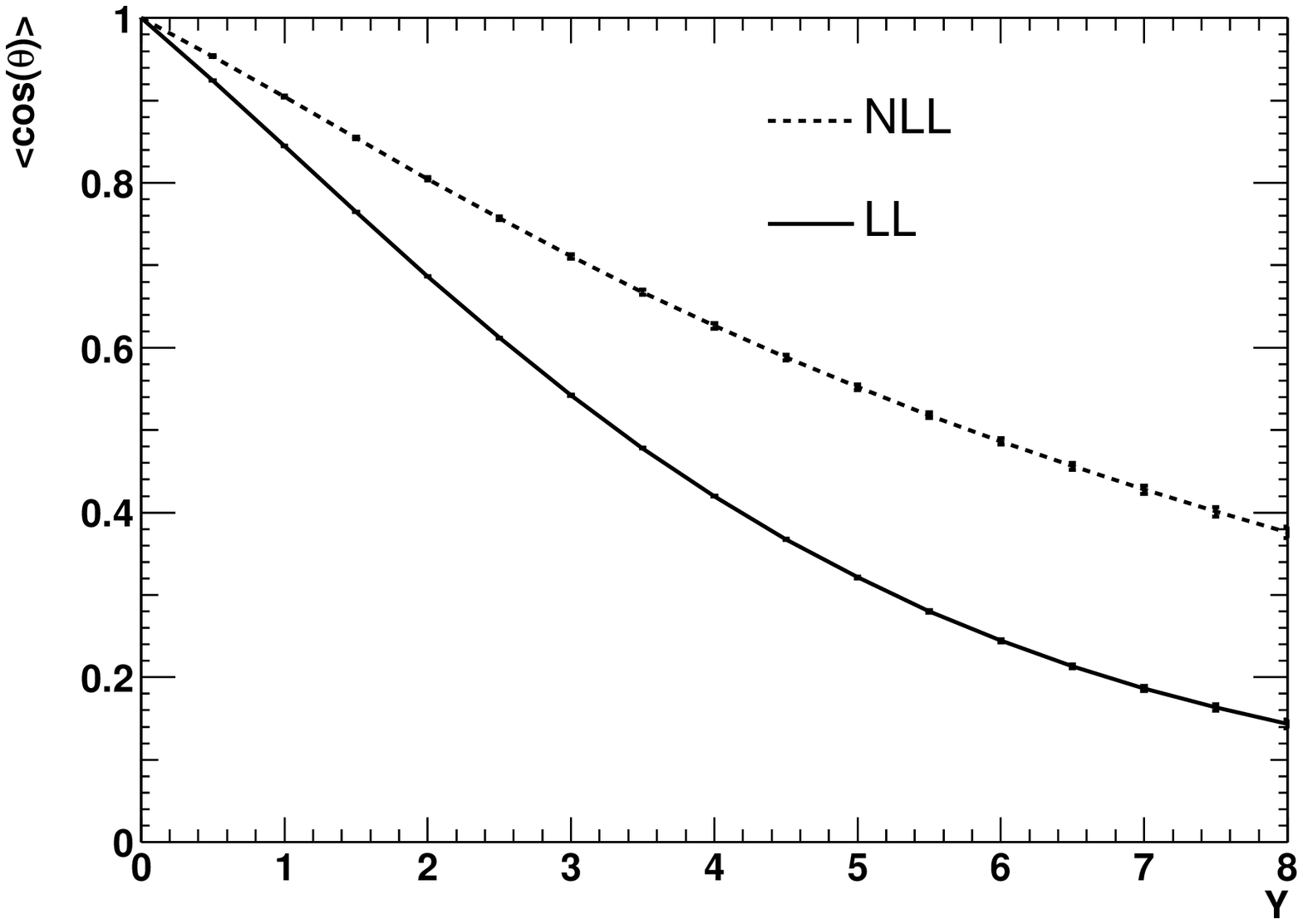}
\caption{Distribution in the number of iterations and angular dependence 
of the NLL gluon Green's function.}
\label{JAfigure2}
\end{figure}

A great advantage of the iterative method here described is that the solution
to the NLL BFKL equation is generated integrating the phase space using a
Monte Carlo sampling of the different parton configurations. This
allows for an investigation 
of the diffusion properties of the BFKL kernel as shown in
ref.~\cite{Andersen:2004tt},  and provides a
good handle on the average multiplicities and angular dependences of the
evolution. Multiplicities can be extracted from the Poisson--like
distribution in the number of iterations of the kernel needed to reach a
convergent solution, which is obtained numerically at the left hand side of
Fig.~\ref{JAfigure2} for a fixed value of the $\lambda$ parameter.  On the
right hand side of the figure a study of the azimuthal angular correlation of
the gluon Green's function is presented at $Y=5$. This decorrelation will
directly impact the prediction for the azimuthal angular decorrelation of two
jets with a large rapidity separation, in a fully inclusive jet sample
(i.e.~no rapidity gaps).  The increase of the angular correlation when the
NLL terms are included is a characteristic feature of these corrections. This
study is possible using this approach because the NLL kernel is treated in
full, without angular averaging, so there is no need to use a Fourier
expansion in angular variables.

%% file: acknowledgments.tex
\section*{Acknowledgments}
A.~Glazov thanks 
       E.~Rizvi, M.~Klein, M.~Cooper-Sarkar and C.~Pascaud for help and many 
       fruitful discussions. 
C.~Gwenlan, A.~Cooper-Sarka and C.~Targett-Adams thank 
       M.~Klein and R.~Thorne for providing the ${F_{L}}$ predictions, 
       as well as for useful discussions.
T.~Carli, G.~Salam and F.~Siegert thank 
       Z.~Nagy, M.~H.~Seymour, T.~Sch\"orner-Sadenius, P.~Uwer and  M.~Wobisch
       for useful discussions on the grid technique,  
       A.~Vogt for discussion on moment-space techniques and 
       Z.~Nagy for help and support with NLOJET++.
R.~Thorne thanks 
       S.~Alekhin for collaboration on the project of obtaining the benchmark 
       parton distributions, for providing his benchmark partons and for many 
       useful exchanges.
J.~Huston and J.~Pumplin thank
        W.K.~Tung and D.~Stump for collaboration on the presented research work. \\

S.~Moch acknowledges
        partial support by the Helmholtz Gemeinschaft under contract VH-NG-105 and 
        by DFG Sonderforschungsbereich Transregio 9, 
        Computergest\"utzte Theoretische Physik.
J.~Bl\"umlein acknowledges
        partial support by DFG Sonderforschungsbereich Transregio 9, 
        Computergest\"utzte Theoretische Physik.
C.~Gwenlan acknowledges
        support by PPARC.
F.~Siegert acknowledges 
        support by the CERN Summer Student Program.
R.S.~Thorne acknowledges 
        support by the Royal Society as University Research Fellow.
J.~Huston and J.~Pumplin acknowledge 
        support by the National Science Foundation.
J.R.~Andersen acknowledges 
        support from PPARC under contract
        PPA/P/S/2003/00281. 
A.~Sabio Vera acknowledges 
        suppoprt from the Alexander von Humboldt foundation.